\documentclass[a4paper]{report}
\usepackage{cmap}
\usepackage{color}
\usepackage[cp1256]{inputenc}
\usepackage[LAE]{fontenc}
\usepackage[english,arabic]{babel}
\usepackage{appendix}
\usepackage{tabularx}

\usepackage{fix-cm}  
\usepackage{graphicx}
\usepackage{epstopdf}
\usepackage{indentfirst}
\usepackage{epsfig}
\usepackage{pdfpages}

\Novocalize

\usepackage{authblk}

\begin{document}
\title{{\Huge {\bf
الفيزياء الاساسية
}\\
${}$
\\
${\rm Fundamental~Physics}$
\\
${}$
}
{\Large {\bf
\\\
\\
}}}
\author{{\Large 
باديس يدري
}
\and
{\Large
${\rm Badis~Ydri}$
}
\and
معهد الفيزياء, جامعة باجي مختار, عنابة, الجزائر
\and
${\rm BM~Annaba~University,Algeria}$
\and
جانفي
 $2016$
 \and
 ${\rm January~2016}$
}
\date{}
\maketitle
\begin{abstract}
محاضرات و تمارين محلولة فى
 الميكانيك الكلاسيكى, 
 الترموديناميك و الميكانيك الاحصائى,
و الميكانيك الكمومى, 
لاقسام الليسانس و الماستر فيزياء فى الجامعات الجزائرية.
\end{abstract}
{\selectlanguage{english}
\begin{abstract}
This book includes my lectures, together with their problem sets and solutions, on $1)$ classical mechanics (one semester), $2)$ thermodynamics and statistical mechanics (one semester), 
and $3)$ quantum mechanics (one semester), which I have been giving to graduate students of theoretical physics at Annaba University since 2010 . 
\end{abstract}
}



{\Huge
\renewcommand{\abstractname}{
اهداء
}
\begin{abstract}
\begin{center}
الي روح الاب الكريم 
\\
ساعد يدري
 \\
 $1943-2015$
\end{center}
\end{abstract}
}

\setcounter{page}{4} 
\tableofcontents

\part*{{\fontsize{50}{50}\selectfont
الميكانيك الكلاسيكي
}}
\addcontentsline{toc}{part}{$I$
الميكانيك الكلاسيكي
} 

{\selectlanguage{english}

\chapter*{
Classical Mechanics
}
This part is based primarily on the standard treatise \cite{Goldstein}. Only fundamental topics are discussed: $1)$ Variational principles and Lagrangian equations, and $2)$ 
Hamiltonian mechanics. In the second chapter we also dicuss the symplectic formalism, canonical transformations and Hamilton-Jacobi equation. See the following chapters: $2$, $8$, $9$ and $10$ of 
Goldstein \cite{Goldstein}. The discussion of the action and the related principle of least action found in  \cite{Landauc}, and the discussion of non-holonomic 
constraints found in \cite{Greiner}, were particularly very useful. Many exercises are taken from \cite{Greiner}, which offers also pedagogical solutions, but also many exercises were 
taken from \cite{Goldstein,Landauc}. The solutions to some of Goldstein's exercises, found in \cite{Reid}, was also consulted.
}

\chapter*{
مبادئ التغاير و معادلات لاغرانج
}
\addcontentsline{toc}{chapter}{$1$
مبادئ التغاير و معادلات لاغرانج
}

\thispagestyle{headings}
\section*{
ميكانيك جملة جسيمات نقطية
}
\addcontentsline{toc}{section}{
ميكانيك جملة جسيمات نقطية
} 
نعتبر
جملة من الجسيمات النقطية ذات  اشعةالموضع
 $\vec{r}_i$
 و الكتل
  $m_i$. 
  قانون نيوتن الثاني للحركة بالنسبة للجسيم رقم 
  $i$
  يعطي ب
\begin{eqnarray}
\vec{F}_i=\vec{F}_i^{(e)}+\sum_{j}\vec{F}_{ji}=\frac{d\vec{p}_i}{dt}.
\end{eqnarray} 
كالعادة تعرف كمية الحركة بدلالة السرعة ب
\begin{eqnarray}
\vec{p}_i=m_i\vec{v}_i=m_i\frac{d\vec{r}_i}{dt}.
\end{eqnarray} 
القوة الخارجية المؤثرة علي الجسيم $i$ هي 
$\vec{F}_i^{(e)}$
 و القوة الداخلية المؤثرة علي الجسيم  $i$ والناجمة عن الجسيم 
 $j$ 
 هي
 $\vec{F}_{ji}$.
 لدينا
 $\vec{F}_{ii}=0$
 و
 $\vec{F}_{ij}=-\vec{F}_{ji}$.
 يمكن كتابة قانون نيوتن الثاني علي الشكل
\begin{eqnarray}
\vec{F}_i=\vec{F}_i^{(e)}+\sum_{j}\vec{F}_{ji}=m_i\frac{d^2\vec{r}_i}{dt^2}.\label{newton}
\end{eqnarray}
بالجمع علي كل الجسيمات نحصل علي 
\begin{eqnarray}
0=\sum_i\vec{F}_i=\sum_i\vec{F}_i^{(e)}=\sum_i m_i\frac{d^2\vec{r}_i}{dt^2}=M\frac{d^2\vec{R}}{dt^2}.
\end{eqnarray}
الكتلة الكلية $M$ معرفة ب 
$M=\sum_i m_i$
و
شعاع موضع مركز كتلة الجملة
$\vec{R}$ 
يعرف ب
\begin{eqnarray}
\vec{R}=\frac{1}{M}\sum_im_i\vec{r}_i.
\end{eqnarray}
اذن القوي الداخلية لانها تخضع لقانون نيوتن الثالث ليس لها اي تأثير علي جركة الجملة.القوة الخارجية الكلية تعطي بدلالة كمية الحركة الكلية ب
\begin{eqnarray}
\vec{F}^{(e)}=M\frac{d\vec{P}}{dt}=M\frac{d^2\vec{R}}{dt^2}.
\end{eqnarray}
اذن يمكن ان نستنتج مباشرة قانون انحفاظ كمية الحركة: اذا انعدمت القوة الخارجية الكلية فان كمية الحركة الكلية تبقي منحفظة في الزمن.

لنحسب الان العمل الذي تقوم به القوي 
$\vec{F}_i^{(e)}$
و
$\vec{F}_{ji}$
في تحريك الجملة من حالة ابتدائية
$1$
الي حالة نهائية
$2$. 
لدينا
\begin{eqnarray}
W_{12}=\sum_i\int_1^2\vec{F}_id\vec{s}_i=\sum_i\int_1^2\vec{F}_i^{(e)}d\vec{s}_i+\sum_{i,j}\int_1^2\vec{F}_{ji}d\vec{s}_i.
\end{eqnarray}
لدينا من جهة
\begin{eqnarray}
W_{12}=\sum_i\int_1^2\vec{F}_id\vec{s}_i&=&\sum_i\int_1^2m_i\frac{d\vec{v}_i}{dt}\vec{v}_i dt\nonumber\\
&=&\sum_i\int_1^2d(\frac{1}{2}m_iv_i^2)\nonumber\\
&=&T_2-T_1.
\end{eqnarray}
الطاقة الحركية الكلية تعرف ب
\begin{eqnarray}
T=\sum_i\frac{1}{2}m_iv_i^2.
\end{eqnarray}
نفترض ان القوي الخارجية
$\vec{F}_i^{(e)}$
محافظة اي انها مشتقة من طاقات كامنة
$V_i$ 
بحيث
\begin{eqnarray}
\vec{F}_i^{(e)}=-\vec{\nabla}_iV_i.
\end{eqnarray}
اذن نحسب
\begin{eqnarray}
\sum_i\int_1^2\vec{F}_i^{(e)}d\vec{s}_i=-\sum_i\int_1^2\vec{\nabla}_iV_i d\vec{s}_i=-\sum_iV_i|_1^2.
\end{eqnarray}
ايضا نفترض ان القوي الداخلية 
$\vec{F}_{ji}$
محافظة اي 
مشتقة من طاقات كامنة
$V_{ij}$
بحيث
\begin{eqnarray}
\vec{F}_{ji}=-\vec{\nabla}_iV_{ij}.
\end{eqnarray}
لان
$\vec{F}_{ij}=-\vec{F}_{ji}$,
يجب ان نأخذ
$V_{ij}$ 
دالة في المسافة
$|\vec{r_i}-\vec{r}_j|$
فقط, اي ان
$V_{ij}=V_{ji}$. 
يمكننا ايضا التحقق من ان القوة
$\vec{F}_{ij}$
هي تقع بمحاذاة الخط الرابط بين الجسيمان
$i$
و
$j$.
نعرف شعاع الفرق ب 
$\vec{r}_{ij}=\vec{r}_i-\vec{r}_j$. 
لدينا اذن
\begin{eqnarray}
\vec{\nabla}_iV_{ij}=-\vec{\nabla}_jV_{ij}=\vec{\nabla}_{ij}V_{ij}.
\end{eqnarray}
يمكننا الان ان نحسب
\begin{eqnarray}
\sum_{i,j}\int_1^2\vec{F}_{ji}d\vec{s}_i&=&-\frac{1}{2}\sum_{i,j}\int_1^2(\vec{\nabla}_iV_{ij}d\vec{s}_i+\vec{\nabla}_jV_{ij}d\vec{s}_j)\nonumber\\
&=&-\frac{1}{2}\sum_{i,j}\int_1^2\vec{\nabla}_{ij}V_{ij}(d\vec{s}_i-d\vec{s}_j)\nonumber\\
&=&-\frac{1}{2}\sum_{i,j}\int_1^2\vec{\nabla}_{ij}V_{ij}d\vec{r}_{ij}\nonumber\\
&=&-\frac{1}{2}\sum_{i\neq j}V_{ij}|_1^2.
\end{eqnarray}
اذن العمل المنجز يعطي ب
\begin{eqnarray}
W_{12}&=&-V_2+V_1.
\end{eqnarray}
الطاقة الكامنة الكلية تعطي اذن ب
\begin{eqnarray}
V=\sum_iV_i+\frac{1}{2}\sum_{i\neq j}V_{ij}.
\end{eqnarray}
من النتائج
$W_{12}=T_2-T_1$ 
و
$W_{12}=-V_2+V_1$
نستنتج ان الطاقة الكلية
$T+V$ 
هي منحفظة.


كالعادة نعرف العزم الحركي الكلي ب
\begin{eqnarray}
\vec{L}=\sum_i\vec{r}_i{\bf x} \vec{p}_i.
\end{eqnarray}
الاشتقاق بالنسبة للزمن يعطي
\begin{eqnarray}
\frac{d\vec{L}}{dt}&=&\sum_i\vec{r}_i{\bf x} \frac{d\vec{p}_i}{dt}\nonumber\\
&=&\sum_i\vec{r}_i{\bf x} \vec{F}_i^{(e)}+\sum_{i\ne j}\vec{r}_i{\bf x}\vec{F}_{ji}\nonumber\\
&=&\sum_i\vec{r}_i{\bf x} \vec{F}_i^{(e)}+\frac{1}{2}\sum_{i\ne j}\vec{r}_{ij}{\bf x}\vec{F}_{ji}.
\end{eqnarray}
بافتراض ان القوي الداخلية بين اي جسيمين, بالاضافة الي كونها متساوية في الشدة و متعاكسة في الاتجاه, تقع بمحاذاة الخط الرابط بين الجسيمين نحصل مباشرة علي
$\vec{r}_{ij}{\bf x}\vec{F}_{ji}=0$\footnote{
يعرف هذا الشرط بالقانون القوي للفعل و رد الفعل. ايضا القوي التي تحقق هذا الشرط هي قوي مركزية.
}.
في هذه الحالة اشتقاق العزم الحركي الكلي بالنسبة للزمن يعطي عزم الدوران الخارجي الكلي اي
\begin{eqnarray}
\frac{d\vec{L}}{dt}
&=&N_i^{(e)}.
\end{eqnarray}
نستنتج مباشرة قانون انحفاظ العزم الحركي: اذا انعدم عزم الدوران الخارجي الكلي فان العزم الحركي الكلي يبقي منحفظا في الزمن.
\section*{
القيود الهولونومية و مبدأ العمل الافتراضي لدالمبارت
}
\addcontentsline{toc}{section}{
القيود الهولونومية و مبدأ العمل الافتراضي لدالمبارت
} 
خلاصة الفقرة السابقة هو معادلات الحركة
\begin{eqnarray}
\vec{F}_i^{(e)}+\sum_j\vec{F}_{ji}=m_i\frac{d^2\vec{r}_i}{dt^2}.\label{newton0}
\end{eqnarray}
الهدف الان هو حل هذه المعادلات من اجل ايجاد اشعة الموضع 
$\vec{r}_i$
كدوال في الزمن. هذه المهمة صعبة  جدا  في الواقع و تتعقد اكثر اذا كانت جملة
الجسيمات خاضعة لقيود علي الحركة.
 القيود علي الحركة هي قوي لا يمكن التعبير عنها مباشرة لكن فقط نعرف تأثيرها الاجمالي علي الحركة. نعتبر هنا حالة القيود الهولونومية التي يعبر عنها بمعادلات من الشكل
\begin{eqnarray}
f(\vec{r}_1,\vec{r}_2,\vec{r}_3,...,t)=0.\label{hol}
\end{eqnarray}
اذن اشعة الموضع
$\vec{r}_i$
ليست كلها مستقلة خطيا وهذا الربط الخطي يمكن ان يتغير من لحظة زمنية الي اخري. نأخذ كمثال علي القيود الهولونومية حركة الجسم الصلب. حركة الجسيمات في هذه الحالة مقيدة بحيث
تبقي المسافة بين الجسيمات ثابتة في الزمن. في هذا المثال نعبر عن هذا القيد الهولونومي بالمعادلة
(حيث
$c_{ij}$ 
هي ثوابت
)

\begin{eqnarray}
(\vec{r}_i-\vec{r}_j)^2-c_{ij}^2=0.\label{sphere}
\end{eqnarray}
كمثال اخر علي القيود الهولونومية حركة جسيم بمحاذاة اي منحني او علي سطح حيث تعطي القيود في هذه الحالة بمعادلة المنحني او السطح.

القيود التي لا يمكن التعبير عنها بمعادلات من الشكل
$(\ref{hol})$
هي قيود غير هولونومية. مثال علي ذلك حركة جزيئات غاز في وعاء: جدران الوعاء هي قيود غير هولونومية. ايضاحركة جسيم علي سطح كرة تحت تأثير حقل ثقالي يعبر عنها بالمعادلات غير الهولونومية
\begin{eqnarray}
\vec{r}^2-a^2\geq 0.
\end{eqnarray}

كما ذكرنا انفا فان وجود قيود هولونومية يعني ان اشعة الموضع
$\vec{r}_i$
 ليست كلها مستقلة خطيا. هذا يعني بالخصوص ان معادلات الحركة
 $(\ref{newton0})$
  ليست كلها مستقلة خطيا. هذه الصعوبة سيتم حلها بادخال الاحداثيات المعممة التي تختزل درجات الحرية المستقلة خطيا للجملة. من الجهة الاخري فان وجود قيود هولونومية يعني وجود قوي مجهولة لا نعرف الا تأثيرها في تقييد حركة الجملة.
   من الواضح انه يجب تحديد هذه القوي بالضبط او التخلص منها نهائيا في الحل. سنتبع في الاتي الطريق الثاني عبر مبدأ دالامبارت.
   
   نفترض ان الجملة تحتوي علي
    $N$ 
    جسيم
    و انها خاضعة ل
    $k$
    قيد هولونومي. اذن يوجد في الجملة
    $3N-k$
    درجة حرية مستقلة خطيا نرمز لها ب
    $q_i$ 
    و نسميها بالاحداثيات المعممة. يمكن اذن ان نعبر عن اشعة الموضع
    $\vec{r}_i$ 
    بدلالة الاحداثيات المعممة
    $q_i$
    و الزمن كالتالي

\begin{eqnarray}
&&\vec{r}_1=\vec{r}_1(q_1,q_2,....,q_{3N-k},t)\nonumber\\
&&.\nonumber\\
&&.\nonumber\\
&&.\nonumber\\
&&\vec{r}_N=\vec{r}_N(q_1,q_2,....,q_{3N-k},t).
\end{eqnarray}
سوف نعتبر الان ازاحات افتراضية متناهية في الصغر
$\delta\vec{r}_i$ 
التي هي ازاحات متسقة مع القيود المفروضة علي الجملة في اللحظة الزمنية
$t$.
عند مقارنة الازاحة الافتراضية
$\delta \vec{r}_i$
مع الازاحة الحقيقية
$d\vec{r}_i$,
التي تحدث خلال مجال زمني
$dt$
و التي يمكن ان تتغير خلالها قوي القيود المفروضة علي الجملة, فانه لدينا من جهة
\begin{eqnarray}
d\vec{r}_i=\frac{\partial \vec{r}_i}{\partial t}dt+\sum_{j=1}^{3N-k}\frac{\partial \vec{r}_i}{\partial q_j}dq_j.
\end{eqnarray}
اما من الجهة الاخري فانه خلال ازاحة افتراضية لدينا
\begin{eqnarray}
\delta \vec{r}_i=\sum_{j=1}^{3N-k}\frac{\partial \vec{r}_i}{\partial q_j}\delta q_j.
\end{eqnarray}
لاحظ اختفاء الحد الاول الناجم عن التغير في الزمن لان الازاحة الافتراضية تنشأ من تغيير مسار الحركة  بمجمله بطريقة متسقة مع القيود المفروضة. انظر الي الشكل 
$1$.

يمكن ان نكتب معادلة الحركة
$(\ref{newton0})$
علي الشكل
$\vec{F}_i-{d\vec{p}_i}/{dt}=0$ 
حيث
 $\vec{p}_i=m_id\vec{r}_i/dt$.
اذن الجسيم رقم 
$i$ 
هو في حالة توازن تحت تأثير القوة الكلية
$\vec{F}_{i~{\rm eff}}=\vec{F}_i-{d\vec{p}_i}/{dt}$.
من الواضح ايضا ان العمل الافتراضي لهذه القوة في الازاحة الافتراضية
$\delta\vec{r}_i$
 ينعدم. بالجمع علي جميع الجسيمات نحصل علي

\begin{eqnarray}
\sum_i(\vec{F}_i-\frac{d\vec{p}_i}{dt})\delta\vec{r}_i=0.
\end{eqnarray} 
نفكك القوة 
$\vec{F}_i$
الي القوة المطبقة 
$\vec{F}_i^{(a)}\equiv \vec{F}_i^{(e)}$ 
و قوة القيود
التي نرمز لها الان ب
 $\vec{f}_i$,
  اي ان
$\vec{F}_i=\vec{F}_i^{(a)}+\vec{f}_i$. 
اذن لدينا
\begin{eqnarray}
\sum_i(\vec{F}_i^{(a)}-\frac{d\vec{p}_i}{dt})\delta\vec{r}_i+\sum_i\vec{f}_i\delta\vec{r}_i=0.
\end{eqnarray}
نقتصر الان علي تلك الجمل الفيزيائية التي ينعدم فيها العمل الافتراضي المنجز من قبل قوي القيود. مثال ذلك الجسم الصلب. في هذه الحالة المسافة 
$r_{ij}$
بين الجسيمات
تبقي ثابتة في الزمن, وبالتالي فان التفاضل
$d\vec{r}_{ij}$
لا يمكن ان يكون الا عموديا علي
$\vec{r}_{ij}$,
اي
 عموديا علي القوي الداخلية 
 $\vec{F}_{ij}$,
و منه  فان عمل القوي الداخلية ينعدم. من الجهة الاخري فان التفاضل الافتراضي 
  $\delta\vec{r}_{ij}$
هو بالتعريف شعاع مماس للمشعب الذي يمثل القيود, الذي هو هي هذه الحالة الكرة
(\ref{sphere}),
اي
انه هو ايضا عمودي علي 
$\vec{r}_{ij}$, 
و منه فان العمل  الافتراضي للقوي الداخلية ينعدم ايضا.
 اذن في حالة الجسم الصلب ينعدم العمل الافتراضي الذي تنجزه قوي القيود التي تعطي في هذه الحالة بالقوي الداخلية.
 
 نحصل اذن,
 من اجل الجمل الفيزيائية التي ينعدم فيها العمل الافتراضي المنجز من قبل قوي القيود, علي مبدأ العمل الافتراضي لدالمبارت
\begin{eqnarray}
\sum_i(\vec{F}_i^{(a)}-\frac{d\vec{p}_i}{dt})\delta\vec{r}_i=0.
\end{eqnarray}
لاحظ ان قوي القيود لا تظهر  صراحة في هذه المعادلة وتأثيرهايقتصر فقط علي جعل
الازاحات الافتراضية ليست كلها مستقلة خطيا.

\section*{
معادلات لاغرانج
}
\addcontentsline{toc}{section}{
معادلات لاغرانج
} 
لنحسب الان العمل الافتراضي بدلالة الاحداثيات المعممة. لدينا
\begin{eqnarray}
\sum_i\vec{F}_i^{(a)}\delta\vec{r}_i&=&\sum_{i,j}\vec{F}_i^{(a)}\frac{\partial \vec{r}_i}{\partial q_j}\delta q_j\nonumber\\
&=&\sum_jQ_j\delta q_j.
\end{eqnarray}
ال
$Q_j$
هي مركبات القوة المعممة و هي معرفة كالتالي
\begin{eqnarray}
Q_j=\sum_i \vec{F}_i^{(a)}\frac{\partial \vec{r}_i}{\partial q_j}.
\end{eqnarray}
لاحظ انه كما ان الاحداثيات المعممة لا تحمل بالضرورة و حدة الطول فان القوة المعممة لا تحمل بالضرورة و حدة القوة.

نحسب ايضا
\begin{eqnarray}
\sum_i\frac{d\vec{p}_i}{dt}\delta\vec{r}_i&=&\sum_{i,j}m_i\frac{d^2\vec{r}_i}{dt^2}\frac{\partial \vec{r}_i}{\partial q_j}\delta q_j\nonumber\\
&=&\sum_{i,j}m_i\bigg[\frac{d}{dt}\bigg(\frac{d\vec{r}_i}{dt}\frac{\partial \vec{r}_i}{\partial q_j}\bigg)-\frac{d\vec{r}_i}{dt}\frac{d}{dt}\bigg(\frac{\partial \vec{r}_i}{\partial q_j}\bigg)\bigg]\delta q_j\nonumber\\
&=&\sum_{i,j}m_i\bigg[\frac{d}{dt}\bigg(\vec{v}_i\frac{\partial \vec{r}_i}{\partial q_j}\bigg)-\vec{v}_i\frac{\partial \vec{v}_i}{\partial q_j}\bigg]\delta q_j.
\end{eqnarray}
بالتعويض
بالنتيجة
${\partial \vec{v}_i}/{\partial\dot{q}_j}={\partial\vec{r}_i}/{\partial q_j}$
نحصل علي
\begin{eqnarray}
\sum_i\frac{d\vec{p}_i}{dt}\delta\vec{r}_i&=&\sum_{i,j}m_i\bigg[\frac{d}{dt}\bigg(\vec{v}_i\frac{\partial \vec{v}_i}{\partial \dot{q}_j}\bigg)-\vec{v}_i\frac{\partial \vec{v}_i}{\partial q_j}\bigg]\delta q_j\nonumber\\
&=&\sum_{j}\bigg[\frac{d}{dt}\bigg(\frac{\partial T}{\partial \dot{q}_j}\bigg)-\frac{\partial T}{\partial q_j}\bigg]\delta q_j.
\end{eqnarray}
الطاقة الحركية الكلية تعطي ب
$T=\sum_i\frac{1}{2}m_iv_i^2$.
اذن مبدأ دالمبارت يصبح
\begin{eqnarray}
\sum_i(\vec{F}_i^{(a)}-\frac{d\vec{p}_i}{dt})\delta\vec{r}_i&=&-\sum_{j}\bigg[Q_j-\frac{d}{dt}\bigg(\frac{\partial T}{\partial \dot{q}_j}\bigg)+\frac{\partial T}{\partial q_j}\bigg]\delta q_j=0.
\end{eqnarray}
لان الاحداثيات المعممة
$q_i$
يمكن اختيارها,  من اجل القيود الهولونومية, بحيث تكون مستقلة خطيا,  يمكننا ان نستخلص مباشرة من النتيجة اعلاه معادلات الحركة
\begin{eqnarray}
-Q_j+\frac{d}{dt}\bigg(\frac{\partial T}{\partial \dot{q}_j}\bigg)-\frac{\partial T}{\partial q_j}=0.
\end{eqnarray}
في المعادلة اعلاه
$j=1,...,n$
حيث
$n=3N-k$
هو عدد الاحداثيات المعممة المستقلة خطيا اي عدد درجات الحرية. من اجل القوي المشتقة من كمون لدينا
$\vec{F}_i^{(a)}=-\vec{\nabla}_iV$
و بالتالي
\begin{eqnarray}
Q_j=-\frac{\partial V}{\partial q_j}.
\end{eqnarray}
اذن نحصل علي معادلات الحركة
\begin{eqnarray}
\frac{d}{dt}\bigg(\frac{\partial L}{\partial \dot{q}_j}\bigg)-\frac{\partial L}{\partial q_j}=0.
\end{eqnarray}
هذه هي معادلات لاغرانج للحركة حيث 
 $L$
 هي اللاغرانجية المعرفة ب
\begin{eqnarray}
L=T-V.
\end{eqnarray}

\section*{حساب التغايرات
}
\addcontentsline{toc}{section}{
حساب التغايرات
} 
نعتبر دالة
$f$
في متغير
 $y$
 الذي هو نفسه دالة في متغير
 $x$. 
 الدالة 
 $f$ 
 يمكن ايضا ان تتعلق بالمشتقة
 $\dot{y}=dy/dx$
 و ايضا ب
 $x$
 .
 يلعب 
 $x$
 هنا دور الزمن و  يلعب 
 $y$ 
دور الموضع. 
نعطي الان التكامل
\begin{eqnarray}
I=\int_{x_1}^{x_2}f(y,\dot{y},x)dx.
\end{eqnarray}
التكامل 
$I$
هو مثال علي ما يسمي بالداليات التي هي دوال يكون فيها المتغير دالة و ليس عدد.
 التكامل
 $I$
 هو اذن دالة, ليست في متغير واحد, لكن في طريق او مسار  بمجمله 
  $y=y(x)$
 الذي 
 يربط نقطتين
 $(x_1,y_1=y(x_1))$
 و
 $(x_2,y_2=y(x_2))$.
نسمي حساب تغيرات, اي حساب تفاضل, الداليات بحساب التغايرات.

السؤال هو: ماهي القيمة المستقرة لهذا التكامل? اي ماهو الطريق
$y_s=y_s(x)$
الذي من اجله ياخذ التكامل قيمة مستقرة اي يأخذ قيمة اصغرية او اعظمية او يكون نقطة انعطاف.

نعتبر مجموعة
الطرق المجاورة و القريبة جدا من الطريق المستقرة
$y_s=y_s(x)$
و التي يمكن ترقيمها بوسيط
$\alpha$
كالتالي
\begin{eqnarray}
y(x)\equiv y(x;\alpha)=y(x;0)+\alpha \eta(x)~,~y_s(x)\equiv y(x;0).
\end{eqnarray}
لان جميع الطرق تنطلق من
$(x_1,y_1=y(x_1))$
و تلتقي في
$(x_2,y_2=y(x_2))$
 لدينا
 \begin{eqnarray}
\eta(x_1)=\eta(x_2)=0.\label{con}
\end{eqnarray}
 يصبح التكامل 
$I$
من اجل هذه المجموعة من الطرق 
دالة
عادية في الوسيط 
$\alpha$
 اي
 \begin{eqnarray}
I(\alpha)=\int_{x_1}^{x_2}f(y(x;\alpha),\dot{y}(x;\alpha),x)dx.
\end{eqnarray}
 القيمة المستقرة للدالة
 $I$
 تعطي اذن بالشرط
 
 \begin{eqnarray}
\frac{dI(\alpha)}{d\alpha}|_{\alpha=0}=0.
\end{eqnarray}
 نقوم بحساب الاشتقاق بشكل عادي كالتالي
  \begin{eqnarray}
\frac{d}{d\alpha}I(\alpha)&=&\int_{x_1}^{x_2}\frac{d}{d\alpha}f(y(x;\alpha),\dot{y}(x;\alpha),x)dx\nonumber\\
&=&\int_{x_1}^{x_2}\bigg[\frac{\partial f}{\partial y}\frac{ dy}{d\alpha}+\frac{\partial f}{\partial \dot{y}}\frac{ d\dot{y}}{d\alpha}\bigg]dx\nonumber\\
&=&\int_{x_1}^{x_2}\bigg[\frac{\partial f}{\partial y}\frac{ dy}{d\alpha}+\frac{\partial f}{\partial \dot{y}}\frac{ d^2{y}}{d\alpha dt}\bigg]dx\nonumber\\
&=&\int_{x_1}^{x_2}\bigg[\frac{\partial f}{\partial y}\frac{ dy}{d\alpha}+\frac{d}{dx}\bigg(\frac{\partial f}{\partial \dot{y}}\frac{ d{y}}{d\alpha}\bigg)-\frac{d}{dx}\bigg(\frac{\partial f}{\partial \dot{y}}\bigg)\frac{dy}{d\alpha}\bigg]dx.
\end{eqnarray}
من الواضح اننا
استعملنا التكامل بالتجزئة 
 للانتقال الي الخط الاخير. ايضا ينعدم الحد الثاني بالشرط
 $(\ref{con})$. 
 نحصل اذن علي
  \begin{eqnarray}
\frac{d}{d\alpha}I(\alpha)
&=&\int_{x_1}^{x_2}\bigg[\frac{\partial f}{\partial y}-\frac{d}{dx}\bigg(\frac{\partial f}{\partial \dot{y}}\bigg)\bigg]\frac{ dy}{d\alpha}dx.
\end{eqnarray}
القيمة المستقرة للدالة
$I$
 تعطي اذن ب
 \begin{eqnarray}
\int_{x_1}^{x_2}\bigg[\frac{\partial f}{\partial y}-\frac{d}{dx}\bigg(\frac{\partial f}{\partial \dot{y}}\bigg)\bigg]\eta(x)dx=0.
\end{eqnarray}
 نستخدم الان النتيجة الاساسية التالية  من حساب التفاضل
  \begin{eqnarray}
\int_{x_1}^{x_2}M(x)\eta(x)dx=0\Rightarrow M(x)=0.\label{calculus}
\end{eqnarray}
 القيمة المستقرة للدالة
$I$
 تعطي اذن بمعادلة الحركة
 \begin{eqnarray}
 \frac{\partial f}{\partial y}-\frac{d}{dx}\bigg(\frac{\partial f}{\partial \dot{y}}\bigg)=0.
\end{eqnarray}
\section*{
مبدأ الفعل الاصغري لهاميلتون
}
\addcontentsline{toc}{section}{
مبدأ الفعل الاصغري لهاميلتون
}
في الفقرات السابقة
 قمنا باشتقاق
 معادلات لاغرانج انطلاقا من اعتبارات تتعلق بالازاحة الافتراضية للجملة حول حالتها اللحظية باستعمال مبدأ العمل الافتراضي لدالمبارت الذي هو مبدأتفاضلي. في هذه الفقرة
 سوف نعيد اشتقاق معادلات لاغرانج انطلاقا من اعتبارات تتعلق بالتغييرات الافتراضية للحركة الاجمالية للجملة
 حول الحركة الحقيقية 
  بين لحظتين زمنيتين 
  $t_1$
 و
 $t_2$
 باستعمال المبدأ التكاملي لهاميلتون المعروف بمبدأ الفعل الاصغري\footnote{
 اذا اردنا دقة اكثر فان مبدأ الفعل الاصغري يختلف عن مبدأ هاميلتون الذي نناقشه هنا. انظر غولدشتاين الفصل 
 $8$
 الباب
 $6$. 
 مبدأ الفعل الاصغري يستخدم التغاير
 $\Delta$
  عوض التغاير
  $\delta$
  الذي يشترط فيه: 
  $(1)$
  ابتداء كل الطرق في  نفس اللحظة 
  $t_1$
  و انتهائها في نفس اللحظة
  $t_2$,
   $(2)$
  انعدام الانتقال الافتراضي 
  $\delta q(t)$
  في اللحظتين الزمنيتين 
  $t_1$
  و
  $t_2$. 
  كلا الشرطين غير متحققين من اجل
  $\Delta$.
 }.
 

الحالة اللحظية للجملة في لحظة زمنية
$t$ 
توصف ب
 $n$
 احداثية معممة
 $q_1$, $q_2$,...,$q_n$,
 و تسمي ايضا بتمثيلة الجملة في اللحظة
 $t$.
  هذه الحالة هي اذن نقطة في فضاء التمثيلات الذي هو فضاء ذو 
 $n$
 بعد تعطي فيه المحاور بالضبط بالاحداثيات المعممة
 $q_i$.
 مع تقدم الزمن تتغير الجملة و تتحرك النقطة
 $(q_1,q_2,...,q_n)$
 في فضاء التمثيلات مختطة منحني يسمي طريق حركة الجملة.

مبدأ الفعل الاصغري لهاميلتون هو مبدأ اقل عمومية من مبدأ دالمبارت لانه يطبق فقط علي الجمل التي تكون فيها كل القوي, و منها قوي القيود, مشتقة من كمون معمم 
$U$.
الكمون المعمم هو كمون يمكن ان يتعلق, بالاضافة الي الاحداثيات المعممة, علي السرعات المعممة و ايضا علي الزمن اي
$U=U(q_i,\dot{q}_i,t)$.
القوي المعممة
 في هذه الحالة يمكن ان نحصل عليها من 
 $U$
 ب
\begin{eqnarray}
Q_j=-\frac{\partial U}{\partial q_j}+\frac{d}{dt}\bigg(\frac{\partial U}{\partial \dot{q}_j}\bigg).\label{generalizedF}
\end{eqnarray}
هذه الجمل تسمي مونوجينية و تبقي من اجلها معادلات لاغرانج صالحة بلاغرانجية معطاة كالعادة ب
$L=T-U$. 
هذه الجمل تصبح محافظة اذا كان الكمون يتعلق فقط بالاحداثيات.

 يمكن ان نبين, 
من اجل الجمل الخاضعة لقيود هولونومية,
ان مبدأ الفعل الاصغري لهاميلتون هو شرط ضروري و كافي من اجل معادلات لاغرانج. في مايلي فاننا سنبين من اجل الجمل المونوجينية
ان مبدأ هاميلتون هو شرط كافي لمعادلات لاغرانج. اذن مبدأ هاميلتون يمكن اخذه المسلمة الاساسية للميكانيك عوضا
عن قوانين نيوتن من اجل الجمل المونوجينية اي لما تكون كل القوي, باستثناء قوي القيود, مشتقة من كمون معمم.

نعرف الفعل بين لحظتين زمنيتين
$t_1$ 
و
$t_2$
بالتكامل
\begin{eqnarray}
I[q]=\int_{t_1}^{t_2}L dt.
\end{eqnarray}
اللاغرانجية
$L$
هي دالة في الاحداثيات و السرعات المعممة
$q_i$
و
$\dot{q}_i$
و كذلك في الزمن
$t$,
اي
$L=L(q_1,q_2,...,q_n,\dot{q}_1,\dot{q}_2,...,\dot{q}_n,t)$, 
 اما الفعل فهو دالية.
 
 من الواضح ان
الفعل يبقي ثابت تحت تأثير اي تحويل للاحداثيات المعممة التي نستخدمها من اجل التعبير عن
$L$
و بالتالي فان معادلات الحركة المشتقة من 
$I$ 
تبقي صامدة تحت تاثير اي تحويل نقطي للاحداثيات.

يتلخص مبدأ الفعل الاصغري لهاميلتون في الاتي: 
{\bf 
يبلغ التكامل
$I$
 قيمته المستقرة, اي يبلغ قيمته الصغري 
او العظمي او يبلغ نقطة انعطاف, من اجل الطريق الحقيقية للحركة. 
}

من الناحية التقنية فاننا نعبر عن هذا المبدأ كالتالي: ان اي تغيير من الرتبة الاولي في طريق الجملة حول طريق الحركة الحقيقية ينجم عنه تغيير من الرتبة الثانية في الفعل
$I$, 
و بالتالي فان كل الطرق المجاورة و 
التي تختلف عن الطريق الحقيقية بازاحة متناهية في الصغر لها نفس الفعل. هذه اذن مسألة تغايرية من اجل دالية الفعل $I$  الذي يتعلق بدالة واحدة التي هي اللاغرانجية $L$.  نكتب مبدأ هاميلتون كالتالي
\begin{eqnarray}
\frac{\delta }{\delta q_i}I[q]=\frac{\delta }{\delta q_i}\int_{t_1}^{t_2}L(q_1,q_2,...,q_n,\dot{q}_1,\dot{q}_2,...,\dot{q}_n,t) dt.
\end{eqnarray}
نعتبر مجموعة الطرق
$q_i(t)$
في فضاء التمثيلات الرابطة بين الحالتين اللحظتين
$(q_1(t_1),...,q_n(t_1))$
و
$(q_1(t_2),...,q_n(t_2))$,
و التي لها نفس فعل الطريق الحقيقية 
$q_{i}^{(s)}(t)$
بين هاتين الحالتين
.
هذه الطرق يمكن ترقيمها بوسيط 
$\alpha$
كالتالي
$q_i(t)\equiv q_i(t,\alpha)=q_i(t,0)+\alpha \eta_i(t)$
حيث
$\alpha=0$
يرفق بالطريق الحقيقية للحركة اي
$q_i(t,0)=q_i^{(s)}(t)$,
 و
$\eta_i$ 
هي دوال كيفية في الزمن
$t$
تنعدم في النقاط الحدية
$t_1$ 
و
$t_2$ 
و مستمرة,
و كذلك نفترض ان مشتقاتها الاولي و الثانية مستمرة.
 من اجل هذه المجموعة من الطرق قان الفعل يصبح دالة في
 $\alpha$
 معطاة ب
\begin{eqnarray}
I(\alpha)=\int_{t_1}^{t_2}L(q_i(t,\alpha),\dot{q}_i(t,\alpha),t) dt.
\end{eqnarray}
نعرف الازاحة الافتراضية
$\delta q_i$
ب
\begin{eqnarray}
\delta q_i=\bigg(\frac{\partial q_i}{\partial \alpha}\bigg)|_{\alpha=0} d\alpha=\eta_i d\alpha.
\end{eqnarray}
بالمقابل التغيير المتناه في الصغر للفعل يعرف ب

\begin{eqnarray}
\delta I&=&\bigg(\frac{d I}{d \alpha}\bigg)|_{\alpha=0} d\alpha.
\end{eqnarray}
نحسب
\begin{eqnarray}
\frac{d I}{d\alpha}&=&\int_{t_1}^{t_2}\bigg(\frac{\partial L}{\partial q_i}\frac{\partial q_i}{\partial \alpha}+\frac{\partial L}{\partial \dot{q}_i}\frac{\partial \dot{q}_i}{\partial \alpha}\bigg)dt\nonumber\\
&=&\int_{t_1}^{t_2}\bigg(\frac{\partial L}{\partial q_i}\frac{\partial q_i}{\partial \alpha}+\frac{\partial L}{\partial \dot{q}_i}\frac{\partial}{\partial t}\frac{\partial {q}_i}{\partial \alpha}\bigg)dt\nonumber\\
&=&\int_{t_1}^{t_2}\bigg(\frac{\partial L}{\partial q_i}\frac{\partial q_i}{\partial \alpha}+\frac{\partial L}{\partial \dot{q}_i}\frac{d}{d t}\frac{\partial {q}_i}{\partial \alpha}\bigg)dt\nonumber\\
&=&\int_{t_1}^{t_2}\bigg(\frac{\partial L}{\partial q_i}\frac{\partial q_i}{\partial \alpha}-\frac{d}{dt}\bigg(\frac{\partial L}{\partial \dot{q}_i}\bigg)\frac{\partial {q}_i}{\partial \alpha}\bigg)dt+\bigg(\frac{\partial L}{\partial \dot{q}_i}\frac{\partial {q}_i}{\partial \alpha}\bigg)_{t_1}^{t_2}.
\end{eqnarray}
الحد الاخير ينعدم لان كل الطرق المعتبرة تمر بالنقاط
 $(t_1,y_i(t_1,0)$
 و
 $(t_2,y_i(t_2,0))$.
 اذن نحصل علي
\begin{eqnarray}
\delta I&=&\int_{t_1}^{t_2}\bigg(\frac{\partial L}{\partial q_i}-\frac{d}{dt}\bigg(\frac{\partial L}{\partial \dot{q}_i}\bigg)\bigg)\delta q_idt.
\end{eqnarray}
مبدأ هاميلتون يعطي ب
\begin{eqnarray}
\frac{\delta I}{d\alpha}&=&\bigg(\frac{d I}{d \alpha}\bigg)|_{\alpha=0} =0.
\end{eqnarray}
هذه تؤدي الي معادلات الحركة
\begin{eqnarray}
\int_{t_1}^{t_2}\bigg(\frac{\partial L}{\partial q_i}-\frac{d}{dt}\bigg(\frac{\partial L}{\partial \dot{q}_i}\bigg)\bigg)\eta_idt=0.
\end{eqnarray}
هذه العلاقة صالحة من اجل كل الدوال
$\eta_i$.
اذن باستعمال النتيجة الاساسية لحساب التفاضل
$(\ref{calculus})$ 
نحصل علي
\begin{eqnarray}
\frac{\partial L}{\partial q_i}-\frac{d}{dt}\bigg(\frac{\partial L}{\partial \dot{q}_i}\bigg)=0.
\end{eqnarray}
نكتب مبدأ الفعل الاصغري لهاميلتون علي الشكل النهائي
\begin{eqnarray}
\frac{\delta I}{\delta q_i}&=&\frac{\partial L}{\partial q_i}-\frac{d}{dt}\bigg(\frac{\partial L}{\partial \dot{q}_i}\bigg)=0.
\end{eqnarray}
هذه  هي معادلات لاغرانج.
\newpage
\section*{
تمارين
}
\addcontentsline{toc}{section}{
تمارين
} 
\paragraph{
تمرين
$1$:
}
\begin{itemize}
\item
بين ان
${\partial \vec{v}_i}/{\partial\dot{q}_j}={\partial\vec{r}_i}/{\partial q_j}$.
\item
احسب الطاقة الحركية بدلالة الاحداثيات و السرعات المعممة.
\end{itemize}

\paragraph{
تمرين
$2$:
}
النواس المضاعف هو جملة مكونة من كتلتين
$m_1$
و
$m_2$
موصولتين بخيط صلب طوله
$l_2$ 
و معلقة الي السقف بخيط صلب اخر طوله
$l_1$
مربوط ايضا بالكتلة 
$m_1$.
انظر الي الشكل
$2$.
ماهي الشروظ الهولونومية التي تخضع لها هاته الجملة و ماهو عدد درجات الحرية. احسب لاغرانجية هاته الجملة و اشتق معادلات لاغرنج للحركة.

\paragraph{
تمرين
$3$:
}
نعطي اللاغرانجية
\begin{eqnarray}
L^{'}=\frac{1}{2}m(a\dot{x}^2+2b\dot{x}\dot{y}+c\dot{y}^2)-\frac{1}{2}K(ax^2+2bxy+cy^2).
\end{eqnarray}
احسب معادلات الحركة. ماهي الجملة الفيزيائية الموصوفة بهذه اللاغرانجية. استنتج اللاغرانجية
$L=T-V$
المرفقة بهذه الجملة.
\paragraph{
تمرين
$4$:
}
نعطي اللاغرانجية
\begin{eqnarray}
L=\frac{1}{12}m^2\dot{x}^4+m\dot{x}^2V(x)-V^2(x).
\end{eqnarray}
احسب معادلات لاغرانج للحركة. ما هو التفسير الفيزيائي لهذه المعادلات.
\paragraph{
تمرين
$5$:
}
بين ان معادلات لاغرانج صامدة تحت تأثير التحويلات النقطية
\begin{eqnarray}
q_i\longrightarrow s_i: q_i=q_i(s_j,t).
\end{eqnarray}
\paragraph{
تمرين
$6$:
}

بين انه من اجل القوي المشتقة من كمون فان القوة المعممة تعطي ب
\begin{eqnarray}
Q_j=-\frac{\partial V}{\partial q_j}.
\end{eqnarray}
\paragraph{
تمرين
$7$:
}
اكتب لاغرانجية جسيم حر يتحرك بسرعة 
$\vec{v}$
بالنسبة لمعلم عطالي
$K$. 
بين ان لاغرانجية الجسيم الحر بالنسبة
لمعلم عطالي
$K^{'}$
يتحرك بسرعة
$\vec{V}$
بالنسبة ل
$K$
يؤدي الي نفس معادلات الحركة.

\paragraph{
تمرين
$8$:
}
طول اي قوس متناه في الصغر في المستوي يعطي ب
\begin{eqnarray}
ds=\sqrt{dx^2+dy^2}.
\end{eqnarray}
بين ان
 اقصر طريق بين نقطتين
$(x_1,y_1)$
 و
 $(x_2,y_2)$
 في المستوي هو المستقيم الرابط بين هاتين النقطتين
 .
 
 اعد نفس السؤال بالنسبة لسطح الكرة. طول قوس متناه في الصغر علي سطح الكرة يعطي ب
\begin{eqnarray}
ds=\sqrt{d\theta^2+\sin^2\theta d\phi^2}.
\end{eqnarray}
\paragraph{
تمرين
$9$:
}
اكتب لاغرانجية هزاز توافقي و معادلات حركته. نفترض الان اننا لا نعرف كيف ان نحل معادلات الحركة و نعرف فقط
ان الحركة اهتزازية بدور
$T=2\pi/\Omega$
 حيث
 $\Omega$
 هو التواتر الزاوي او النبض. موضع الهزاز كدالة في الزمن
 $x(t)$
 يمكن اذن وصفه بسلسة فورييه من الشكل

\begin{eqnarray}
x(t)=\sum_{j=0}a_j\cos j\Omega t.
\end{eqnarray}
نأخذ الطرق في فضاء التمثيلات بين اللحظتين
$t_1=0$
و
$t_2=T$
التي تعطي بالدوال اعلاه. احسب فعل الهزاز علي هذه الطرق بدلالة الوسائط
$a_j$.
بين ان القيمة المستقرة للفعل تعطي ب
\begin{eqnarray}
\Omega =\sqrt{\frac{k}{m}}~,~a_j=0~,~\forall j\neq 1.
\end{eqnarray}

\paragraph{
تمرين
$10$:
}
النواس الكروي هو كتلة نقطية معلقة الي السقف بخيط صلب يمكنها ان تهتز في الفضاء علي سطح كرة. ماهي معادلات القيود و الاحداثيات المعممة في هذه الحالة. احسب لاغرانجية الجملة 
و معادلات الحركة.

\paragraph{
تمرين
$11$:
}
ينحدر قرص منزلقا علي مستوي مائل. عين الاحداثيات المعممة الضرورية لوصف حالة الجملة بالكامل. عين القيود علي الحركة في حالة انحدار
القرص دائرا علي المستوي المائل بدون انزلاق.

\paragraph{
تمرين
$12$:
}
ما هي القيود علي الحركة من اجل الجمل التالية:
\begin{itemize}
 \item 
 جسيم يتحرك علي قطع ناقص.
 \item
 جسيم يتحرك علي كرة.
 \item
جسم صلب مشكل من ثلاث جسيمات.
\item
جسم يتزحلق علي مستوي مائل بزاوية
$\alpha$.
\item
جسم يتحرك علي مستقيم يدور بسرعة زاوية ثابتة
$\Omega$. 
\end{itemize}

\paragraph{
تمرين
$13$:
}
عجلة تتحرك دائرة علي مستوي بدون انزلاق. نفترض ان
العجلة لا يمكنها ان تسقط. احسب معادلات القيود. هل القيود هولونومية ام لا.

\paragraph{
تمرين
$14$:
}
نعتبر جملة مشكلة من كتلتين
$M_1$
و
$M_2$
معلقتين الي بكرتين متراكزتين
 نصف قطريهما
 $R_1$
 و
 $R_2$ 
 علي التوالي. بين ان العمل الافتراصي لقوي القيود ينعدم عند حالة التوازن. استخدم مبدأ العمل
 الافتراضي لدالمبارت لتعيين حالة توازن الجملة.
 
\paragraph{
تمرين
$15$:
}
كتلتان 
$m_1$
و
$m_2$
مرتبطتان بحبل وتتحركان علي مستويين مائلين بزاويتين
$\alpha$
و 
$\beta$ 
علي التوالي. الحبل طوله
$l$ 
و يتحرك بدون احتكاك عبر بكرة تفصلها عن الكتلتين 
المسافتين 
$l_1$
و 
$l_2$
علي التوالي. انظر الي الشكل
$6$.
استعمل مبدأ العمل الافتراضي لدالمبارت لحساب تسارع الجملة. عين المسافة
$l_1$
او المسافة
$l_2$
كدالة في الزمن.

\paragraph{
تمرين
$16$:
}

النواس النابض هو كتلة 
$m$
معلقة الي السقف بنابض ثابت مرونته
$k$
تحت تأثير الحقل الثقالي. ماهي الاحداثيات المعممة في هذه المسألة. احسب لاغرنجية الجملة و اشتق معادلات لاغرنج للحركة.

\paragraph{
تمرين
$17$:
}

يتحرك حجران مربوطان بخيط صلب طوله
$l$
علي مستوي مائل بزاوية 
$\alpha$.
ماهي الاحداثيات المعممة في هذه الحالة. احسب لاغرانجية الجملة. حل معادلات الحركة صراحة.

\paragraph{
تمرين
$18$:
}
جسيم كروي يتحرك داخل انبوب يدور في المستوي
$xy$
حول المحور
$z$
 بسرعة زاوية ثابته
 $\Omega$. 
 اشتق معادلات لاغرنج للحركة. حل معادلات الحركة.

 \paragraph{
تمرين
$19$:
}
نعتبر جملة ذات درجة حرية واحدة 
$q$. 
بين انه اذا كانت لاغرانجية الجملة لا تتعلق صراحة بالزمن اي اذا كانت
$L=L(q,\dot{q})$
فان معادلة لاغرانج للحركة يمكن كتابتها علي الشكل

\begin{eqnarray}
\dot{q}\frac{\partial L}{\partial \dot{q}}-L={\rm constant}.\nonumber
\end{eqnarray}
ملحوظة: اشتق هذه المعادلة بالنسبة للزمن.
\paragraph{
تمرين
$20$:
}

نفترض ان لاغرانجية جملة تتعلق بالتسارع المعمم
$\ddot{q}$
بالاضافة الي الاحداثية و السرعة المعممتين
$q$
و
$\dot{q}$
و الزمن
اي
$L=L(q,\dot{q},\ddot{q},t)$. 
نعتبر الطرق في فضاء التمثيلات التي تربط الحالتين 
$1=(t_1,q_1,\dot{q}_1)$
و
$2=(t_2,q_2,\dot{q}_2)$.
اذن نعتبر فقط التغييرات الافتراضية حول الحركة الحقيقية التي تنعدم في النقطتين
$1$
و
$2$
اي
\begin{eqnarray}
\delta q(t_1)=\delta q(t_2)=0~,~\delta\dot{q}(t_1)=\delta\dot{q}(t_2)=0.\nonumber
\end{eqnarray}
اشتق معادلات لاغرانج في هذه الحالة.

ملحوظة: استعمل مبدأ الفعل الاصغري لهاميلتون
$\delta I=0$
حيث ان الفعل يعرف ب
$I=\int dt L(q,\dot{q},\ddot{q},t)$.
\paragraph{
تمرين
$21$:
}

نعتبر جسم في حالة انزلاق بدون احتكاك و تحت تأثير قوة الثقالة علي منحني
$y=y(x)$
في المستوي الشاقولي. ألجسم يبدأ بالانزلاق في اللحظة
$t=0$ 
في
النقطة
$(x=0,y=y_0)$
و يصل في اللحظة
$t=T$
الي الارض عند النقطة
$(x=x_0,y=0)$. 
\begin{itemize}
 \item 
 باستعمال قانون انحفاظ الطاقة بين النقطة
 $(x=0,y=y_0)$
 و نقطة كيفية
 $(x,y)$
 علي المنحني
 $y=y(x)$
 بين ان الزمن 
 $T$
 يعطي بالمعادلة
\begin{eqnarray}
T=\int_0^{x_0}\sqrt{\frac{1+y^{'2}}{2g(y_0-y)}} dx~,~y^{'}=\frac{dy}{dx}.\nonumber
\end{eqnarray}
\item
عين المنحني
$y=y(x)$
الذي يكون من اجله الزمن 
$T$
اصغري. استعمل معادلة لاغرانج علي الشكل الذي وجدناه في التمرين الاول.

ملحوظة: يمكن ان تستخدم تغيير المتغير
\begin{eqnarray}
y^{'}=-\cot\frac{\theta}{2}.\nonumber
\end{eqnarray}
\end{itemize}

\newpage
\section*{
حلول
}
\addcontentsline{toc}{section}{
حلول
}

\paragraph{
تمرين
$1$:
}
\begin{itemize}
 \item 
السرعة بدلالة الاحداثيات و السرعات المعممة تعطي ب
\begin{eqnarray}
\vec{v}_i=\frac{d\vec{r}_i}{dt}=\frac{\partial \vec{r}_i}{\partial t}+\sum_j\frac{\partial \vec{r}_i}{\partial q_j}\dot{q}_j.
\end{eqnarray}
بالاشتقاق الجزئي بالنسبة ل
$\dot{q}_j$
نحصل علي العلاقة المرغوب فيها.
\item
\begin{eqnarray}
T=M_0+\sum_j M_j\dot{q}_j+\frac{1}{2}\sum_{j,k}M_{jk}\dot{q}_j\dot{q}_k.
\end{eqnarray}
\begin{eqnarray}
M_0=\sum_i\frac{1}{2}m_i\big(\frac{\partial \vec{r}_i}{\partial t}\big)^2~,~M_j=\sum_im_i\frac{\partial \vec{r}_i}{\partial t}.\frac{\partial \vec{r}_i}{\partial q_j}~,~M_{jk}=\sum_i m_i \frac{\partial \vec{r}_i}{\partial q_j}.\frac{\partial \vec{r}_i}{\partial q_k}.
\end{eqnarray}
\end{itemize}

\paragraph{
تمرين
$2$:
}
 احداثيات الكتلة الاولي هي
\begin{eqnarray}
x_1=l_1\sin\theta_1~,~y_1=-l_1\cos\theta_1.
\end{eqnarray}
احداثيات الكتلة الثانية هي
\begin{eqnarray}
x_2=x_1+l_2\sin\theta_2~,~y_2=y_1-l_2\cos\theta_2.
\end{eqnarray}
نلاحظ ان
\begin{eqnarray}
x_1^2+y_1^2=l_1^2.
\end{eqnarray}
\begin{eqnarray}
(x_2-x_1)^2+(y_2-y_1)^2=l_2^2.
\end{eqnarray}
هذه هي معادلات القيود الهولونومية في هذه الحالة. اذن عدد درجات الحرية هو 
$4-2=2$.
الاحداثيات المعممة في هذه الحالة
 هي الزاويتين
$\theta_1$
و
$\theta_2$.

من اجل حساب اللاغرانجية علينا حساب الطاقة الحركية و الطاقة الكامنة. سرعة الكتلة الاولي هي
\begin{eqnarray}
v_1^2=\dot{x}_1^2+\dot{y}_1^2=l_1^2\dot{\theta}_1^2.
\end{eqnarray}
سرعة الكتلة الثانية هي
\begin{eqnarray}
v_2^2=\dot{x}_2^2+\dot{y}_2^2=l_1^2\dot{\theta}_1^2+l_2^2\dot{\theta}_2^2+2l_1l_2\dot{\theta}_1\dot{\theta}_2\cos(\theta_1-\theta_2).
\end{eqnarray}
الطاقة الحركية للجملة هي

\begin{eqnarray}
T&=&\frac{1}{2}m_1v_1^2+\frac{1}{2}m_2 v_2^2\nonumber\\
&=&\frac{1}{2}(m_1+m_2)l_1^2\dot{\theta}_1^2+\frac{1}{2}m_2l_2^2\dot{\theta}_2^2+m_2l_1l_2\dot{\theta}_1\dot{\theta}_2\cos(\theta_1-\theta_2).
\end{eqnarray}
نحسب الان الطاقة الكامنة. قوي الثقالة المؤثرة علي الجسيمين الاول و الثاني هي
$\vec{F}_1=m_1\vec{g}$
و
$\vec{F}_2=m_2\vec{g}$.
 في هذه الحالةالطاقة الكامنة تساوي ناقص عمل قوة الثقالة. اذن
 \begin{eqnarray}
V&=&m_1g.y_1+m_2g.y_2\nonumber\\
&=&-(m_1+m_2)gl_1\cos\theta_1-m_2gl_2\cos\theta_2.
\end{eqnarray}
 اذن لاغرانجية النواس المضاعف تعطي ب
 \begin{eqnarray}
L&=&\frac{1}{2}(m_1+m_2)l_1^2\dot{\theta}_1^2+\frac{1}{2}m_2l_2^2\dot{\theta}_2^2+m_2l_1l_2\dot{\theta}_1\dot{\theta}_2\cos(\theta_1-\theta_2)\nonumber\\
&+&(m_1+m_2)gl_1\cos\theta_1+m_2gl_2\cos\theta_2.
\end{eqnarray}
 معادلات الحركة هي
 
 \begin{eqnarray}
&&\frac{d}{dt}\big(\frac{\partial L}{\partial \dot{\theta}_1}\big)-\frac{\partial L}{\partial\theta_1}=0\Leftrightarrow \frac{d}{dt}\bigg[(m_1+m_2)l_1^2\dot{\theta}_1+m_2l_1l_2\dot{\theta}_2\cos(\theta_1-\theta_2)\bigg]\nonumber\\
&&-\bigg[-m_2l_1l_2\dot{\theta}_1\dot{\theta}_2\sin(\theta_1-\theta_2)-(m_1+m_2)gl_1\sin\theta_1\bigg]=0.
\end{eqnarray}
 
 \begin{eqnarray}
&&\frac{d}{dt}\big(\frac{\partial L}{\partial \dot{\theta}_2}\big)-\frac{\partial L}{\partial\theta_2}=0\Leftrightarrow \frac{d}{dt}\bigg[m_2l_2^2\dot{\theta}_2+m_2l_1l_2\dot{\theta}_1\cos(\theta_1-\theta_2)\bigg]\nonumber\\
&&-\bigg[-m_2l_1l_2\dot{\theta}_1\dot{\theta}_2\sin(\theta_1-\theta_2)-m_2gl_2\sin\theta_2\bigg]=0.
\end{eqnarray}
\paragraph{
تمرين
$3$:
} 
معادلات الحركة تعطي ب
\begin{eqnarray}
&&\frac{d}{dt}\big(\frac{\partial L^{'}}{\partial \dot{x}}\big)-\frac{\partial L^{'}}{\partial x}=0\Leftrightarrow m(a\ddot{x}+b\ddot{y})+K(ax+by)=0.
\end{eqnarray}
 \begin{eqnarray}
&&\frac{d}{dt}\big(\frac{\partial L^{'}}{\partial \dot{y}}\big)-\frac{\partial L^{'}}{\partial y}=0\Leftrightarrow m(b\ddot{x}+c\ddot{y})+K(bx+cy)=0.
\end{eqnarray}
نعرف المتغيرات
\begin{eqnarray}
u_1=ax+by~,~u_2=bx+cy.
\end{eqnarray}
معادلات الحركة تأخذ اذن الشكل
\begin{eqnarray}
m\ddot{u}_1+Ku_1=0~,~m\ddot{u}_2+Ku_2=0.
\end{eqnarray}
هذه معادلات حركة هزازان توافقيان 
$u_1$
و
$u_2$
حيث ان كل هزاز هو
عبارة عن نابض ذو كتلة
$m$
و
ثابت
$K$.
الطاقة الحركية و الطاقة الكامنة للنابض
$u$
 تعطي ب
 \begin{eqnarray}
 T=\frac{1}{2}m\dot{u}^2~,~V=\frac{1}{2}Ku^2.
\end{eqnarray}
اذن لاغرانجية الجملة
$L=T-V$
تعطي ب
\begin{eqnarray}
 L=\frac{1}{2}m(\dot{u}_1^2+\dot{u}_2^2)-\frac{1}{2}K(u_1^2+u_2^2).
\end{eqnarray}
الجملة الفيزيائية هي  اذن عبارة عن هزاز توافقي في بعدين.
\paragraph{
تمرين
$5$:
}
لدينا من جهة
\begin{eqnarray}
\frac{\partial L}{\partial s_i}&=&\sum_j\frac{\partial L}{\partial q_j}\frac{\partial q_j}{\partial s_i}+\sum_j\frac{\partial L}{\partial \dot{q}_j}\frac{\partial \dot{q}_j}{\partial s_i}\nonumber\\
&=&\sum_j\frac{\partial L}{\partial q_j}\frac{\partial q_j}{\partial s_i}+\sum_j\frac{\partial L}{\partial \dot{q}_j}\frac{\partial }{\partial s_i}\sum_k\bigg(\frac{\partial q_j}{\partial s_k}\dot{s}_k+\frac{\partial q_j}{\partial t}\bigg)\nonumber\\
&=&\sum_j\frac{\partial L}{\partial q_j}\frac{\partial q_j}{\partial s_i}+\sum_{j,k}\frac{\partial L}{\partial \dot{q}_j}\bigg(\frac{\partial^2 q_j}{\partial s_i \partial s_k}\dot{s}_k+\frac{\partial^2 q_j}{\partial s_i\partial t}\bigg).
\end{eqnarray}
من جهة اخري لدينا
\begin{eqnarray}
\frac{\partial L}{\partial \dot{s}_i}&=&\sum_j\frac{\partial L}{\partial \dot{q}_j}\frac{\partial \dot{q}_j}{\partial \dot{s}_i}\nonumber\\
&=&\sum_j\frac{\partial L}{\partial \dot{q}_j}\frac{\partial {q}_j}{\partial {s}_i}.
\end{eqnarray}
اي ان
\begin{eqnarray}
\frac{d}{dt}\big(\frac{\partial L}{\partial \dot{s}_i}\big)&=&\sum_j\frac{d}{dt}\big(\frac{\partial L}{\partial \dot{q}_j}\big)\frac{\partial \dot{q}_j}{\partial \dot{s}_i}+\sum_j\frac{\partial L}{\partial \dot{q}_j}\frac{d}{dt}\big(\frac{\partial {q}_j}{\partial {s}_i}\big).
\end{eqnarray}
اذن اذا كان لدينا معادلات لاغرانج
\begin{eqnarray}
&&\frac{d}{dt}\big(\frac{\partial L}{\partial \dot{q}_i}\big)-\frac{\partial L}{\partial q_i}=0,
\end{eqnarray}
فانه يترتب عليه مباشرة معادلات لاغرانج
\begin{eqnarray}
&&\frac{d}{dt}\big(\frac{\partial L}{\partial \dot{s}_i}\big)-\frac{\partial L}{\partial s_i}=0.
\end{eqnarray}
\paragraph{
تمرين
$7$:
}
بالنسبة للمعلم
$K$ 
لدينا
\begin{eqnarray}
L=\frac{1}{2}m\vec{v}^2.
\end{eqnarray}
بالنسبة للمعلم
$K^{'}$ 
لدينا
\begin{eqnarray}
L^{'}&=&\frac{1}{2}m\vec{v}^{'2}\nonumber\\
&=&L+\frac{1}{2}m\vec{V}^2+m\vec{v}\vec{V}\nonumber\\
&=&L+\frac{d F}{dt}.
\end{eqnarray}
\begin{eqnarray}
F=\frac{1}{2}m\vec{V}^2 t+m\vec{r}.\vec{V}.\label{func}
\end{eqnarray}
نحسب الان
\begin{eqnarray}
\frac{\partial L^{'}}{\partial r_i}=\frac{\partial L}{\partial r_i}+\frac{\partial}{\partial r_i}\big(\frac{dF}{dt}\big).
\end{eqnarray}
\begin{eqnarray}
\frac{\partial L^{'}}{\partial \dot{r}_i}&=&\frac{\partial L}{\partial \dot{r}_i}+\frac{\partial}{\partial \dot{r}_i}\big(\frac{dF}{dt}\big)\nonumber\\
&=&\frac{\partial L}{\partial \dot{r}_i}+\frac{\partial F}{\partial {r}_i}.
\end{eqnarray}
المعادلة الاخيرة تؤدي الي
\begin{eqnarray}
\frac{d}{dt}\big(\frac{\partial L^{'}}{\partial \dot{r}_i}\big)&=&\frac{d}{dt}\big(\frac{\partial L}{\partial \dot{r}_i}\big)+\frac{d}{dt}\big(\frac{\partial F}{\partial {r}_i}\big)\nonumber\\
&=&\frac{\partial L}{\partial {r}_i}+\frac{d}{dt}\big(\frac{\partial F}{\partial {r}_i}\big).
\end{eqnarray}
اذن نحصل علي

\begin{eqnarray}
\frac{d}{dt}\big(\frac{\partial L^{'}}{\partial \dot{r}_i}\big)-\frac{\partial L^{'}}{\partial r_i}&=&\frac{d}{dt}\big(\frac{\partial F}{\partial {r}_i}\big)-\frac{\partial}{\partial r_i}\big(\frac{dF}{dt}\big)\nonumber\\
&=&0.
\end{eqnarray}
هذه النتيجة تبقي صالحة من اجل كل الدوال 
$F=F(r_i,t)$
 القابلة للاشتقاق و ليس فقط من اجل الدالة
 $(\ref{func})$
 .

 \paragraph{
تمرين
$8$:
}
طول اي منحني رابط بين النقطتين 
$(x_1,y_1)$
 و
 $(x_2,y_2)$
  يعطي ب

\begin{eqnarray}
I&=&\int_{1}^2ds\nonumber\\
&=&\int_1^2\sqrt{dx^2+dy^2}\nonumber\\
&=&\int_{x_1}^{x_2}dx f(y,\dot{y}).
\end{eqnarray}
\begin{eqnarray}
f(y,\dot{y})=\sqrt{1+\dot{y}^2}
\end{eqnarray}
نحسب مباشرة
\begin{eqnarray}
\frac{\partial f}{\partial y}=0~,~\frac{\partial f}{\partial \dot{y}}=\frac{\dot{y}}{\sqrt{1+\dot{y}^2}}.
\end{eqnarray}
معادلة الحركة هي اذن
\begin{eqnarray}
\frac{\dot{y}}{\sqrt{1+\dot{y}^2}}=c\Leftrightarrow \dot{y}=a=\frac{c}{\sqrt{1-c^2}}.
\end{eqnarray}
 $a$
 و
 $c$
 هي ثوابت تكامل. بالتكامل مرة اخري نحصل علي
 \begin{eqnarray}
y=ax+b.
\end{eqnarray}
 هذه هي معادلة المستقيم. الثوابت
  $a$
 و
 $c$
 تعين من شرط مرور المستقيم بالنقطتين
$(x_1,y_1)$
 و
 $(x_2,y_2)$. 
 اذن اقصر طريق رابط بين نقطتين في المستوي هو المستقيم.

 بالنسبة لحالة الكرة لدينا
 \begin{eqnarray}
f(\phi,\dot{\phi},\theta)=\sqrt{1+\sin^2\theta\dot{\phi}^2}~,~\dot{\phi}=\frac{d\phi}{d\theta}.
\end{eqnarray}
معادلة القيم المستقرة تعطي ب
\begin{eqnarray}
\frac{\sin^2\theta \dot{\phi}}{\sqrt{1+\sin^2\theta \dot{\phi}^2}}=c.
\end{eqnarray}
يمكن كتابة هذه المعادلة علي الشكل
\begin{eqnarray}
\dot{\phi}=-\frac{\dot{\rho}}{\sqrt{1-\rho^2}}~,~\rho=a\cot \theta.
\end{eqnarray}
اذن الحل المستقر يعطي ب
(باهمال ثابت تكامل اضافي)
\begin{eqnarray}
\sin\phi=-a\cot\theta.
\end{eqnarray}
هذه معادلات الدوائر الكبري اي دوائر علي سطح الكرة.
 \paragraph{
تمرين
$9$:
}
الفعل يعطي ب
\begin{eqnarray}
I&=&\int_0^T L dt\nonumber\\
&=&\int_0^T\big(\frac{1}{2}m\dot{x}^2-\frac{1}{2}kx^2\big)dt\nonumber\\
&=&\frac{1}{2}m\int_0^T x^2(t)-\frac{1}{2}k\int_0^T \dot{x}^2(t).
\end{eqnarray}
نحسب

\begin{eqnarray}
\int_0^T x^2(t)&=&\sum_{j=0}\sum_{k=0}a_ja_k\int_0^T \cos j\Omega t\cos k\Omega t dt\nonumber\\
&=&\sum_{j=0}\sum_{k=0}a_ja_k\frac{T}{2}\delta_{jk}\nonumber\\
&=&\frac{T}{2}\sum_{j=0}a_j^2.
\end{eqnarray}
من جهة اخري لدينا

 \begin{eqnarray}
x(t)=\sum_{j=0}a_j\cos j\Omega t\Rightarrow \dot{x}(t)=-\Omega\sum_{j=0}ja_j\sin j\Omega t.
\end{eqnarray}
 اذن نحسب
 \begin{eqnarray}
\int_0^T \dot{x}^2(t)&=&\Omega^2\sum_{j=0}\sum_{k=0}jk a_ja_k\int_0^T \sin j\Omega t\sin k\Omega t dt\nonumber\\
&=&\Omega^2\sum_{j=0}\sum_{k=0}jk a_ja_k\frac{T}{2}\delta_{jk}\nonumber\\
&=&\frac{T\Omega^2}{2}\sum_{j=0}j^2a_j^2.
\end{eqnarray}
الفعل يصبح
 \begin{eqnarray}
I=\frac{\pi}{2}\sum_{j=0}\big(m\Omega j^2-\frac{k}{\Omega}\big)a_j^2.
\end{eqnarray}
القيمة المستقرة تعطي بالشرط
\begin{eqnarray}
\delta I=0\Rightarrow \pi\sum_{j=0}\big(m\Omega j^2-\frac{k}{\Omega}\big)a_j\delta a_j=0.
\end{eqnarray}
الحل يعطي ب

\begin{eqnarray}
\big(m\Omega j^2-\frac{k}{\Omega}\big)a_j=0~,~\forall j.
\end{eqnarray}
قليل من التأمل يعطي الحل النهائي 
\begin{eqnarray}
\Omega =\sqrt{\frac{k}{m}}~,~a_j=0~,~\forall j\neq 1.
\end{eqnarray}

\paragraph{
تمرين
$10$:
}
شعاع الموضع لانه يقع علي سطح كرة يجب ان يحقق
\begin{eqnarray}
\vec{r}^2=L^2.
\end{eqnarray}
هذه هي معادلة القيد. عدد درجات الحرية هو اذن
$2$.  
مرة اخري لان شعاع الموضع يقع علي سطح كرةيمكننا كتابته علي الشكل
\begin{eqnarray}
\vec{r}=L\big(\sin\theta\cos\phi \hat{i}+\sin\theta\sin\phi \hat{j}+\cos\theta \hat{k}\big).
\end{eqnarray}
يمكن اخذ الزاويتين
$\theta$
و
$\phi$
كاحداثيات معممة. 

نحسب السرعة و الطاقة الحركية و الطاقة الكامنة
\begin{eqnarray}
\vec{v}=L\dot{\theta}\big(\cos\theta\cos\phi \hat{i}+\cos\theta\sin\phi \hat{j}-\sin\theta \hat{k}\big)+L\dot{\phi}\sin\theta\big(-\sin\phi \hat{i}+\cos\phi \hat{j}\big).
\end{eqnarray}

\begin{eqnarray}
T=\frac{1}{2}mL^2\dot{\theta}^2+\frac{1}{2}mL^2\dot{\phi}^2\sin^2\theta.
\end{eqnarray}

\begin{eqnarray}
V=-mgL\cos\theta.
\end{eqnarray}
لاغرانجية النواس الكروي تعطي اذن ب
\begin{eqnarray}
L=\frac{1}{2}mL^2\dot{\theta}^2+\frac{1}{2}mL^2\dot{\phi}^2\sin^2\theta+mgL\cos\theta.
\end{eqnarray}
معادلات الحركة تعطي ب
\begin{eqnarray}
\ddot{\theta}=-\frac{1}{L}(g-L\dot{\phi}^2\cos\theta)\sin\theta.
\end{eqnarray}
\begin{eqnarray}
\frac{d}{dt}(\dot{\phi}\sin^2\theta)=0.
\end{eqnarray}

\paragraph{
تمرين
$11$:
}

حالة الجملة تعين بالكامل باعطاء المسافة
$l$
التي يقطعها القرص علي المستوي المائل و الزاوية
$\alpha$
التي يدور بها القرص حول محور دورانه. الاحداثيات المعممة هي اذن
$l$
و
$\alpha$.
انظر الي الشكل
$3$
.

عند انحدار القرص علي المستوي دائرا بدون انزلاق فان انتقال نقطة التماس
$dl$
خلال زمن
$dt$
يساوي ضرب نصف قطر القرص و الانتقال
الزاوي
$d\alpha$
خلال الزمن
$dt$. 
اي
\begin{eqnarray}
dl=Rd\alpha\Leftrightarrow v=R\dot{\alpha}.
\end{eqnarray}
هذا هو قيد الدوران بدون انزلاق او زحلقة. من الواضح انه قيد هولونومي.

\paragraph{
تمرين
$12$:
}

\begin{itemize}
 \item 
 \begin{eqnarray}
 \frac{x^2}{a^2}+\frac{y^2}{b^2}=1.
\end{eqnarray}
 \item
 \begin{eqnarray}
x=r\sin\theta\cos\phi~,~y=r\sin\theta\sin\phi~,~z=r\cos\theta\Rightarrow x^2+y^2+z^2=r^2.
\end{eqnarray}
 \item
 \begin{eqnarray}
 (\vec{r}_i-\vec{r}_j)^2=c_{ij}^2.
\end{eqnarray}
\item
\begin{eqnarray}
x=-l\cos\alpha~,~y=-l\sin\alpha\Rightarrow \frac{y}{x}=\tan\alpha.
\end{eqnarray}
\item
\begin{eqnarray}
x=r\cos\Omega t~,~y=r\sin\Omega t\Rightarrow \frac{y}{x}=\tan\Omega t.
\end{eqnarray}
\end{itemize}

\paragraph{
تمرين
$13$:
}

نحتاج لتحديد حالة الجملة الي احداثيات مركز ثقل العجلة في المستوي,
$x_w$
و
$y_w$,
الي الزاوية
$\psi$
التي تحدد اتجاه العجلة, و الي زاوية دوران العجلة
$\phi$.
انظر الي الشكل
$4$.

مركبات السرعة
$\vec{v}$
هي
\begin{eqnarray}
\dot{x}_w=-v\sin\psi~,~\dot{y}_w=v\cos\psi.
\end{eqnarray}
من الجهة الاخري فان شرط الدوران بدون انزلاق يعطي ب
\begin{eqnarray}
v=R\dot{\phi}.
\end{eqnarray}
بالتعويض نحصل علي معادلات القيد
\begin{eqnarray}
dx_w=-R\sin\psi d\phi~,~dy_w=R\cos\psi d\phi.
\end{eqnarray}
هذه معادلات لا يمكن مكاملتها حتي نحل المسالة. اذن هذه القيود غير هولونومية.

\paragraph{
تمرين
$14$:
}
 
 قوي القيود في هذه الحالة هي قوي التوتر قي الخيوط
 $\vec{T}_1$
 و
 $\vec{T}_2$. 
 انظر الي الشكل
 $5$. 
 دوران البكرات بزاوية
 $\delta \phi$
 يقابل انتقال الكتل بمسافة تعطي ب
\begin{eqnarray}
\delta y_1=R_1\delta\phi_1~,~\delta y_2=-R_2\delta\phi.
\end{eqnarray}   
 العمل الافتراضي لقوي التوتر يعطي اذن ب
 \begin{eqnarray}
\delta W&=&\vec{T}_1\delta \vec{r}_1+\vec{T}_2\delta \vec{r}_2\nonumber\\
&=&T_1\delta y_1+T_2\delta y_2\nonumber\\
&=&(T_1R_1-T_2R_2)\delta\phi.
\end{eqnarray}
 لكن عند التوازن تتساوي عزوم قوي التوتر. اذن عند التوازن ينعدم العمل الافتراضي لقوي التوتر.
 
 مبدأ العمل الافتراضي لدالمبارت عند التوازن يأخذ الشكل
 
 \begin{eqnarray}
 \sum_i \vec{F}_i^{(a)}\delta \vec{r}_i=0.
\end{eqnarray}
 القوي المطبقة في هذه المسألة هي قوي الثقالة. اذن المعادلة اعلاه تأخذ الشكل
 \begin{eqnarray}
m_1g\delta y_1+m_2g\delta y_2=0.
\end{eqnarray}
 حالة التوازن تعطي اذن ب
 \begin{eqnarray}
 m_1R_1=m_2R_2.
\end{eqnarray}
\paragraph{
تمرين
$15$:
}

مبدأ العمل الافتراضي لدالمبارت يأخذ الشكل
 
 \begin{eqnarray}
 \sum_i (\vec{F}_i^{(a)}-\dot{\vec{p}_i})\delta \vec{r}_i=0.
\end{eqnarray}
نكتب هذه المعادلة علي الشكل
\begin{eqnarray}
(m_1\vec{g}-m_1\ddot{\vec{l}}_1)\delta\vec{l}_1+(m_2\vec{g}-m_2\ddot{\vec{l}}_2)\delta\vec{l}_2=0.
\end{eqnarray}
بالاسقاط نحصل علي
\begin{eqnarray}
(m_1{g}\sin\alpha-m_1\ddot{{l}}_1)\delta{l}_1+(m_2{g}\sin\beta-m_2\ddot{{l}}_2)\delta{l}_2=0.
\end{eqnarray}
القيد علي الحركة في هذه الحالة هو
\begin{eqnarray}
l=l_1+l_2\Rightarrow \delta{l}_1=-\delta{l}_2.
\end{eqnarray}
نحصل  اذن 
علي
\begin{eqnarray}
\ddot{l}_1=\frac{m_1\sin\alpha-m_2\sin\beta}{m_1+m_2}g.
\end{eqnarray}

\paragraph{
تمرين
$16$:
}

احداثيات الكتلة 
$m$
في الشكل
$7$
تعطي ب

\begin{eqnarray}
x=r\sin\phi~,~y=r\cos\phi.
\end{eqnarray}
الاحداثيات المعممة هي 
$r$, 
لان طول النابض غير ثابت في هذه المسألة, و 
$\phi$.

الطاقة الحركية تعطي ب
\begin{eqnarray}
T=\frac{1}{2}m(\dot{r}^2+r^2\dot{\phi}^2).
\end{eqnarray}
ليكن 
$r_0$
طول النابض في حالة التوازن. الطاقة الكامنة تعطي ب
\begin{eqnarray}
V&=&-m\vec{g}\vec{r}+\frac{1}{2}k(r-r_0)^2\nonumber\\
&=&-mgr\cos\phi+\frac{1}{2}k(r-r_0)^2.
\end{eqnarray}
لاغرانجية الجملة تعطي ب
\begin{eqnarray}
L=\frac{1}{2}m(\dot{r}^2+r^2\dot{\phi}^2)+mgr\cos\phi-\frac{1}{2}k(r-r_0)^2.
\end{eqnarray}
معادلة الحركة بالنسبة ل
$\phi$:
\begin{eqnarray}
mr\ddot{\phi}=-mg\sin\phi-2m\dot{r}\dot{\phi}.
\end{eqnarray}
الحد الثاني هو قوة كوريوليس الناجمة عن تعلق طول النواس بالزمن. معادلة الحركة بالنسبة ل
$r$:
\begin{eqnarray}
m\ddot{r}=mr\dot{\phi}^2+mg\cos\phi-k(r-r_0).
\end{eqnarray}
الحد الاخير هو قوة هوك.

\paragraph{
تمرين
$17$:
}

الاحداثيات النسبية في الشكل
$8$ 
تعطي ب
\begin{eqnarray}
x=l\cos\alpha~,~y=l\sin\alpha.
\end{eqnarray}
هناك قيد هولونومي واحد و بالتالي لدينا درجة حرية واحدة. 
الاحداثية المعممة هي الزاوية 
$\alpha$. 
لاغرانجية الجملة تعطي ب
\begin{eqnarray}
L&=&\frac{1}{2}ml^2\dot{\alpha}^2+m\vec{g}\vec{r}\nonumber\\
&=&\frac{1}{2}ml^2\dot{\alpha}^2-mgl\sin\alpha.
\end{eqnarray}
معادلات لاغرانج للحركة

\begin{eqnarray}
\ddot{\alpha}+\frac{g}{l}\cos\alpha=0.
\end{eqnarray}
بضرب طرفي هذه المعادلة ب
$\dot{\alpha}$
يمكن مكاملة هذه المعادلة مرة من اجل الحصول علي
\begin{eqnarray}
\dot{\alpha}=\sqrt{2(c-\frac{g}{l}\sin\alpha)}.
\end{eqnarray}
$c$
هو ثابت تكامل. بالمكاملة مرة ثانية باستعمال فصل المتغيرات نحصل علي
\begin{eqnarray}
t-t_0=\int_{\alpha_0}^{\alpha}\frac{d\alpha}{\sqrt{2(c-\frac{g}{l}\sin\alpha)}}.
\end{eqnarray}

\paragraph{
تمرين
$18$:
}

 لاغرانجية الجملة تعطي ب
 \begin{eqnarray}
L=\frac{1}{2}m(\dot{r}^2+\Omega^2r^2).
\end{eqnarray}
 معادلات لاغرانج للحركة تعطي ب
  \begin{eqnarray}
  \ddot{r}-\Omega^2r=0.
\end{eqnarray}
الحل يعطي ب
 \begin{eqnarray}
 r=A\exp(\Omega t)+B\exp(-\Omega t).
\end{eqnarray}

\paragraph{
تمرين
$21$:
}
\begin{itemize}
 \item [$(1$]

مبدأ انحفاظ الطاقة بين النقطة
$(0,y)$
و نقطة كيفية
$(x,y)$ 
يعطي ب
\begin{eqnarray}
 mgy_0=\frac{1}{2}mv^2+mgy~,~v^2=(\frac{dx}{dt})^2+(\frac{dy}{dt})^2.
\end{eqnarray}
نحصل اذن علي
\begin{eqnarray}
dt=\sqrt{\frac{dx^2+dy^2}{2g(y_0-y)}}\Rightarrow T=\int_0^{x_0}\sqrt{\frac{1+y^{'2}}{2g(y_0-y)}}dx.
\end{eqnarray}
\item [$(2$]
الزمن
$T$
هو من الشكل
\begin{eqnarray}
T=\int_0^{x_0}f(y,y^{'},x)dx.
\end{eqnarray}
الدالة
$f$
لا تتعلق صراحة بالزمن
$x$
و تعطي ب
\begin{eqnarray}
f=\sqrt{\frac{1+y^{'2}}{2g(y_0-y)}}.
\end{eqnarray}
معادلة لاغرانج تأخذ الشكل

\begin{eqnarray}
\frac{\partial f}{\partial y}-\frac{d}{dx}\frac{\partial f}{\partial y^{'}}=0\Rightarrow y^{'}\frac{\partial f}{\partial y^{'}}-f=1/c.
\end{eqnarray}
الحساب يعطي
\begin{eqnarray}
\frac{1}{c^2}=\frac{1}{2g(y_0-y)(1+y^{'2})}.
\end{eqnarray}
نقوم بتغيير المتغير

\begin{eqnarray}
y^{'}=-\cot\frac{\theta}{2}.\label{gh1}
\end{eqnarray}
نحصل علي
\begin{eqnarray}
y=y_0-\frac{c^2}{2g}\sin^2\frac{\theta}{2}.\label{gh0}
\end{eqnarray}
بالاشتقاق
\begin{eqnarray}
y^{'}=-\frac{c^2}{2g}\sin \theta .\theta^{'}.\label{gh2}
\end{eqnarray}
من
$(\ref{gh1})$
و
$(\ref{gh2})$
نحصل علي
\begin{eqnarray}
x=\frac{c^2}{2g}\int \cos^2\frac{\theta}{2} d\theta=\frac{c^2}{4g}(\theta-\sin\theta).\label{gh3}
\end{eqnarray}
المعادلتان
$(\ref{gh0})$
و
$(\ref{gh3})$
تعرفان دويري
\footnote{.${\rm cycloid}$}. 
في اللحظة الابتدائية لدينا
$\theta=0$. 
في اللحظة النهائية لدينا
\begin{eqnarray}
0=y_0-\frac{c^2}{2g}\sin^2\frac{\theta_0}{2}~,~x_0=\frac{c^2}{4g}(\theta_0-\sin\theta_0).
\end{eqnarray}
اي
\begin{eqnarray}
\frac{x_0}{y_0}=\frac{\theta_0-\sin\theta_0}{1-\cos\theta_0}.
\end{eqnarray}
$\theta_0$
هي القيمة الاعظمية للزاوية 
$\theta$.
\end{itemize}
\chapter*{
الميكانيك الهاميلتوني
}
\addcontentsline{toc}{chapter}{$2$
الميكانيك الهاميلتوني
} 

\section*{
قوانين الانحفاظ
}
\addcontentsline{toc}{section}{
قوانين الانحفاظ
} 
نعتبر جملة مشكلة من جسيمات نقطية تتفاعل فيما بينها عبر قوي مشتقة من كمون يتعلق فقط بالموضع. نحسب
\begin{eqnarray}
\frac{\partial L}{\partial \dot{x}_i}&=&\frac{\partial T}{\partial \dot{x}_i}-\frac{\partial V}{\partial \dot{x}_i}\nonumber\\
&=&m_ix_{i}-0\nonumber\\
&=&p_{ix}.
\end{eqnarray} 
هذه هي بالضبط كمية حركة الجسيم 
$i$
في الاتجاه
$x$. 

 اذن نعرف كمية الحركة المعممة او كمية الحركة المرافقة او كمية الحركة القانونية 
$p_i$
 المرفقة بالاحداثية المعممة 
 $q_i$
 بالعبارة
 \begin{eqnarray}
p_{i}=\frac{\partial L}{\partial \dot{q}_i}.
\end{eqnarray} 
مثل ما ان الاحداثيات المعممة لا تحمل بالضرورة ابعاد الطول فان كميات الحركة المعممة لا تحمل بالضرورة ابعاد
كمية الحركة.

نعرف الان مفهوم الاحداثية المهملة او الاحداثية الدورية علي انها الاحداثية 
$q_i$
التي لا تدخل في اللاغرانجية
$L$
رغم ان
$L$
يمكن ان يتعلق ب
$\dot{q}_i$. 
في هذه الحالة معادلات لاغرانج تؤدي الي
\begin{eqnarray}
\frac{\partial L}{\partial q_i}-\frac{d}{dt}\big(\frac{\partial L}{\partial \dot{q}_i}\big)=0\Rightarrow \frac{d}{dt}{p}_i=0\Rightarrow p_i={\rm constant}.
\end{eqnarray} 
اذن كمية الحركة المعممة المرفقة باحداثية معممة مهملة تكون منحفظة في الزمن. اي انها ثابت للحركة.
 هذا هو شرط الانحفاظ الاكثر عمومية في الميكانيك التحليلي.

 من اجل الجمل المحافظة يكون الكمون دالة في الاحداثيات  المعممة فقط. في هذه الحالة القوة المعممة
 $Q_i$
 المرفقة بالاحداثية المعممة الدورية
 تنعدم لان الكمون لا يتعلق ب
 $q_i$,
 
 علاوة علي ذلك,
اذا كانت الاحداثية المعممة الدورية 
 $q_i$
 هي بحيث 
$dq_i$
 يقابل انسحاب للجملة في الاتجاه 
$\vec{n}$, 
فان انعدام القوة المعممة
$Q_i$
يكافئ انعدام القوة العادية في الاتجاه
$\vec{n}$,
و
 انحفاظ كمية الحركة المعممة 
$p_i$
 يكافئ انحفاظ كمية الحركة العادية في الاتجاه 
  $\vec{n}$
 .
اي انه في هذه الحالة القوة المعممة و كمية الحركة المعممة هما بالضبط القوة و كمية الحركة العاديين  في الاتجاه
  $\vec{n}$.
نحصل اذن علي قانون انحفاظ كمية الحركة لما تبقي حالة الجملة صامدة تحت تأثير الانسحابات. نقول ان الانسحابات هي تناظرات
للجملة.

بالمثل اذا كانت الاحداثية المعممة 
الدورية
$q_i$
 هي بحيث 
$dq_i$
 يقابل دوران للجملة حول محور 
$\vec{n}$, 
  فان انعدام القوة المعممة
$Q_i$
يكافئ انعدام عزم الدوران حول المحور
$\vec{n}$,
و
 انحفاظ كمية الحركة المعممة 
$p_i$
 يكافئ انحفاظ العزم الحركي في الاتجاه 
  $\vec{n}$
 .
نحصل اذن في هذه الحالة علي قانون انحفاظ العزم الحركي لما تبقي حالة الجملة صامدة تحت تأثير الدورانات. نقول
في هذه الحالة ان الدورانات هي تناظرات للجملة.

  قوانين الانحفاظ, اي وجود احداثيات دورية, 
  تكون دائما مرتبطة بوجود تناظرات معينة تميز حالة الجملة. مثلا وجود احداثية دورية انسحابية يعني ان الجملة تبقي صامدة
 تحت تأثير الانسحابات في الاتجاه المقابل للاحداثية الدورية مما ينتج عنه انحفاظ كمية الحركة في هذا الاتجاة. بالمثل
  فان
  وجود احداثية دورية دورانية يعني ان الجملة تبقي
  صامدة تحت تأثير الدورانات في الاتجاه المقابل للاحداثية الدورية 
   و هذا ينتج عنه انحفاظ العزم الحركي في هذا الاتجاه.

  يمكن البرهان علي قانون انحفاظ الطاقة باستعمال معادلات لاغرانج كالتالي. نحسب
  \begin{eqnarray}
  \frac{dL}{dt}&=&\sum_i\frac{\partial L}{\partial q_i}\frac{dq_i}{dt}+\sum_i\frac{\partial L}{\partial \dot{q}_i}\frac{d\dot{q}_i}{dt}+\frac{\partial L}{\partial t}\nonumber\\
  &=&\sum_i\frac{d}{dt}\big(\frac{\partial L}{\partial \dot{q}_i}\big)\dot{q}_i+\sum_i\frac{\partial L}{\partial \dot{q}_i}\frac{d\dot{q}_i}{dt}+\frac{\partial L}{\partial t}\nonumber\\
  &=&\sum_i\frac{d}{dt}\big(\frac{\partial L}{\partial \dot{q}_i}\dot{q}_i\big)+\frac{\partial L}{\partial t}.
\end{eqnarray} 
  نستنتج اذن
  \begin{eqnarray}
  \frac{dh}{dt}+\frac{\partial L}{\partial t}=0.
\end{eqnarray} 
  $h$ 
  هو بالضبط دالة الطاقة او الهاميلتونية و تعطي ب
  \begin{eqnarray}
  h(q,\dot{q},t)=\sum_i\dot{q}_i\frac{\partial L}{\partial \dot{q}_i}-L.\label{hJ}
\end{eqnarray} 
  اذن اذا لم تتعلق اللاغرانجية
  $L$
  صراحة
  بالزمن فان الهاميلتونية تكون منحفظة في الزمن. في هذه
  الحالة
  $h$ 
  هو ثابت للحركة يسمي ثابت جاكوبي.
  

 من اجل الجمل المحافظة تأخذ اللاغرانجية الشكل العام التالي
 
 \begin{eqnarray}
 L=L_0(q,t)+L_1(q,\dot{q},t)+L_2(q,\dot{q},t).\label{form}
\end{eqnarray} 
$L_1$
و
$L_2$
هي دوال 
متجانسة من الدرجة الاولي و الثانية علي التوالي في السرعات المعممة 
$\dot{q}_i$.

 نقول عن دالة
$f(x,y,...)$
انها دالة متجانسة من الرتبة 
$q$
في المتغيرات
$x$, $y$,...
اذا تحقق الشرط
\begin{eqnarray}
f(t x,t y,...)=t^q f(x,y,...).
\end{eqnarray}
نعرف
$x^{'}=tx$, $y^{'}=ty$,...
نحسب
\begin{eqnarray}
\frac{df(x^{'},y^{'},...)}{dt}=\frac{dx^{'}}{dt}\frac{\partial f}{\partial x^{'}}+\frac{dy^{'}}{dt}\frac{\partial f}{\partial y^{'}}+...\Leftrightarrow qt^{q-1}f(x,y,...)=x\frac{\partial f}{\partial (tx)}+
y\frac{\partial f}{\partial (ty)}+...
\nonumber\\
\end{eqnarray}
من اجل 
$t=1$
نحصل علي مبرهنة اولر
\begin{eqnarray}
  \sum_ix_i\frac{\partial f}{\partial x_i}=qf.
\end{eqnarray} 
  الشكل 
  $(\ref{form})$
  تأخذه ايضا لاغرانجيات جمل اخري كثيرة ليست بالضرورة محافظة. 
  بتطبيق مبرهنة اولر 
  علي 
  $h$
  المعطي بالمعادلة
  $(\ref{hJ})$
   نحصل علي
 \begin{eqnarray}
  h=L_2-L_0.
\end{eqnarray} 
 من الجهة الاخري تأخذ الطاقة الحركية
دائما الشكل
\begin{eqnarray}
T=T_0(q)+T_1(q,\dot{q})+T_2(q,\dot{q}).
\end{eqnarray} 
من الواضح اذن انه لدينا
\begin{eqnarray}
 L_0=T_0-V~,~L_1=T_1~,~L_2=T_2.
\end{eqnarray} 
اذن
\begin{eqnarray}
  h=T_2-T_0+V.
\end{eqnarray} 
 بالاضافة الي هذا, اذا كان تغيير المتغيرات
$\vec{r}_i\longrightarrow q_i$
لا يتعلق بالزمن فان
$T=T_2$
و بالتالي
\begin{eqnarray}
  h=T+V.
\end{eqnarray} 
هذه  بالفعل هي طاقة الجملة.

\section*{
تحويل لوجوندر  و معادلات هاميلتون
}
\addcontentsline{toc}{section}{
تحويل لوجوندر و معادلات هاميلتون
} 

مرة اخري نفترض جملة فيزيائية خاضعة لقيود هولونومية 
$f_j(q_i,\dot{q}_i,t)=0$
و قوي مونوجينية اي قوي مشتقة من كمون معمم يتعلق بالاضافة الي الاحداثيات المعممة  علي السرعات المعممة 
 اي
$U=U(q_i,\dot{q}_i,t)$
و تعطي بالعلاقة

\begin{eqnarray}
Q_j=-\frac{\partial U}{\partial q_j}+\frac{d}{dt}\bigg(\frac{\partial U}{\partial \dot{q}_j}\bigg).
\end{eqnarray}
من اجل جملة تحتوي علي 
$n$
درجة حرية لدينا
$n$
معادلة للحركة تعطي بالضبط بمعادلات لاغرانج
\begin{eqnarray}
\frac{\partial L}{\partial q_i}-\frac{d}{dt}\bigg(\frac{\partial L}{\partial \dot{q}_i}\bigg)=0.
\end{eqnarray}
هذه معادلات تفاضلية من الرتبة الثانية حلها يتطلب تقديم
$2n$
شرط ابتدائي. كمثال علي الشروط الابتدائية يمكن تقديم ال
$n$
قيمة
للموضع 
$q_i$ 
و ال
$n$ 
قيمة
للسرعة
$\dot{q}_i$
 في اللحظة الابتدائية
 $t_0$.
 
 حالة او تمثيلة الجملة هي نقطة
 $(q_1,...,q_2)$
 في فضاء التمثيلات  ذو ال
 $n$
 بعد
 تختط خلال الزمن مسار يحدده بالضبط حل معادلة لاغرانج.


في الصياغة الهاميلتونية للميكانيك تعطي معادلات الحركة بمعادلات تفاضلية من الرتبة الاولي تعرف بمعادلات هاميلتون. لان عدد
الشروط الابتدائية الضرورية يجب ان يبقي نفسه يساوي
$2n$, كما في الصياغة اللاغرانجية,
 فان عدد المعادلات التفاضلية من الرتبة الاولي الضرورية لوصف حالة الجملة يجب ان يعطي ب
 $2n$
  معادلة
 اي انه يجب ان نعمل ب
 $2n$
 متغير. من الطبيعي جدا ان نأخذ نصف هذه المتغيرات ال
 $n$
 احداثية معممة
 $q_i$
 اما من اجل النصف الاخر
 فنأخذ ال
 $n$
 كمية حركة معممة
 $p_i$
 التي تعرف ب
\begin{eqnarray}
p_i=\frac{\partial L(q_j,\dot{q}_j,t)}{\partial \dot{q}_i}.
\end{eqnarray}
 يعرف الزوج
$(q_i,p_i)$
 بالمتغيرات القانونية.
 حالة  او تمثيلة الجملة في الصياغة الهاميلتونية تعطي بنقطة
  $(q_1,q_2,...,q_n,p_1,p_2,...,p_n)$
  في فضاء ذو
  $2n$
  بعد يعرف بالفضاء الطوري للجملة اين تعطي المحاور بالاحداثيات و كميات الحركة المعممة
  $q_i$
  و
  $p_i$.
  معادلات هاميلتون هي معادلات حركة النقطة
 $(q_1,q_2,...,q_n,p_1,p_2,...,p_n)$
  في الفضاء الطوري.

الانتقال من الصياغة اللاغرانجية الي الصياغة الهاميلتونية  يتطلب تغيير متغيرات من الشكل
\begin{eqnarray}
(q_i,\dot{q}_i,t)\longrightarrow (q_i,p_i,t).
\end{eqnarray}
هذا مثال علي ما يعرف باسم تحويل لوجوندر.
 
 قبل ان نواصل نذكر بتعريف تحويل لوجوندر لدالة
 $f(x)$ 
 في متغير
 $x$.
 نفترض ان الدالة محدبة اي انها تحقق الشرط
 \begin{eqnarray}
 \frac{d^2f}{dx^2}>0.
\end{eqnarray}
 هذا الشرط يمكن ايضا ان نعبر عليه كالتالي: الدالة الميل
  \begin{eqnarray}
 s(x)=\frac{df}{dx}
\end{eqnarray}
 هي دالة رتيبة, لانها تتزايد فقط, في 
$x$.
  اذن هناك قيمة واحدة ل
  $s$ 
  من اجل كل نقطة
  $x$
  اي ان الدالة
$s=s(x)$
هي
  مفردة القيمة و تقبل العكس لتعطي دالة مفردة القيمة
  $x=x(s)$.

  اذن يمكن الابتداء من الميل
  $s$
  كمتغير مستقل, نستعمل الدالة العكسية
   $x=x(s)$
   للحصول علي القيمة الوحيدة ل 
   $x$
   المقابلة للميل
   $s$,
   ثم نعوض بهذه القيمة ل
   $x$ 
   في الدالة
   $f$ 
   لنحصل
   علي
   $f(x(s))$.
    تحويل لوجوندر 
    $g(s)$
    للدالة
    $f(x)$ 
    هي نقطة تقاطع المستقيم المماس للدالة
    في النقطة
    $x=x(s)$
    مع محور العينات
    اي

     \begin{eqnarray}
     f(x(s))=sx(s)-g(s)\Leftrightarrow g(s)=sx(s)-f(x(s)).
\end{eqnarray}
    انظر الشكل 
    $10$. 
    تحويل لوجوندر هو تطبيق للثنائية بين النقاط و المستقيمات: الدالة
    $f$ 
    يمكن اعطائها بمجموعة النقاط
    $(x,y)$
    او بمجموعة
    الازواج 
    $(s,-g)$
    المشكلة من 
     الميول 
     $s$
     و نقاط التقاطع
     $-g$
     .
     يمكن تعريف تحويل لوجوندر ايضا بعملية التعظيم
  \begin{eqnarray}
     g(s)={\rm max}_x(sx-f(x)).
\end{eqnarray}    
  نعتبر الان دالة
  $f(x,y)$
  في متغيرين
  $x$
  و
  $y$.
 التفاضل التام للدالة
 $f$ 
 يعطي ب
 \begin{eqnarray}
df=u dx+v dy~,~u=\frac{\partial f}{\partial x}~,~v=\frac{\partial f}{\partial y}.
\end{eqnarray}
 تحويل لوجوندر من المتغيرات
 $(x,y)$
 الي المتغيرات
 $(u,y)$
 يحول
 الدالة
 $f(x,y)$
 الي الدالة
 $g(u,y)$
 المعرفة ب 
 \begin{eqnarray}
g=ux-f.
\end{eqnarray}
نحسب التفاضل
\begin{eqnarray}
dg=xdu-v dy\equiv \frac{\partial g}{\partial u}du+\frac{\partial g}{\partial y} dy.
\end{eqnarray}
نحصل اذن علي
\begin{eqnarray}
x=\frac{\partial g}{\partial u}~,~v=-\frac{\partial g}{\partial y}.
\end{eqnarray}

كما قلنا فان الانتقال من الصياغة اللاغرانجية الي الصياغة الهاميلتونية يكافئ  تحويل لوجوندر من المتغيرات
$(q_i,\dot{q}_i,t)$
الي المتغيرات
$(q_i,p_i,t)$. 
اذن عوض اللاغرانجية
$L=L(q_i,\dot{q}_i,t)$
التي هي دالة في
 $q_i$,$\dot{q}_i$ 
 و
 $t$
 سوف نعمل, في الصياغة الهاميلتونية, بما يسمي
 بالهاميلتونية
 $H$
 التي هي دالة في 
 $q_i$, $p_i$ 
 و
 $t$ 
 معرفة بتحويل لوجوندر
\begin{eqnarray}
H(q_i,p_i,t)=\sum_i \dot{q}_ip_i-L(q_i,\dot{q}_i,t).
\end{eqnarray}
نحسب من جهة
\begin{eqnarray}
dH&=&\frac{\partial H}{\partial q_i}dq_i+\frac{\partial H}{\partial p_i}dp_i+\frac{\partial H}{\partial t}dt.
\end{eqnarray}
من الجهة الاخري نحسب
 \begin{eqnarray}
dH
&=&\dot{q}_idp_i+p_id\dot{q}_i-\frac{\partial L}{\partial \dot{q}_i}d\dot{q}_i-\frac{\partial L}{\partial q_i}dq_i-\frac{\partial L}{\partial t}dt\nonumber\\
&=&\dot{q}_idp_i-\frac{\partial L}{\partial q_i}dq_i-\frac{\partial L}{\partial t}dt\nonumber\\
&=&\dot{q}_idp_i-\dot{p}_idq_i-\frac{\partial L}{\partial t}dt.
\end{eqnarray}
بالمقارنة نحصل علي معادلات هاميلتون للحركة
\begin{eqnarray}
\dot{q}_i=\frac{\partial H}{\partial p_i}~,~-\dot{p}_i=\frac{\partial H}{\partial q_i}.
\end{eqnarray}
نحصل ايضا علي
\begin{eqnarray}
-\frac{\partial L}{\partial t}=\frac{\partial H}{\partial t}.
\end{eqnarray}
من اجل قسم كبير من الجمل الفيزيائية و الاحداثيات المعممة لدينا الاتي متحقق:
\begin{itemize}
\item{}
اللاغرانجية تكتب علي الشكل
$L(q_i,\dot{q}_i,t)=L_0(q_i,t)+L_1(q_i,\dot{q}_i,t)+L_2(q_i,\dot{q}_i,t)$ 
حيث
 $L_2$ 
 هي دالة متجانسة من الدرجة الثانية في
 $\dot{q}_i$
 و
 $L_1$
هي دالة متجانسة من الدرجة الاولي في 
 $\dot{q}_i$. 
 في هذه الحالة نحسب
\begin{eqnarray}
\dot{q}_ip_i&=&\dot{q}_i\frac{\partial L_1}{\partial \dot{q}_i}+\dot{q}_i\frac{\partial L_2}{\partial \dot{q}_i}=L_1+2L_2.
\end{eqnarray}
اذن
\begin{eqnarray}
H=L_2-L_0.
\end{eqnarray}
\item{}
عموما تأخذ الطاقة الحركية الشكل
$T=T_2(q_i,\dot{q}_i,t)+T_1(q_i,\dot{q}_i,t)+T_0(q_i,t)$.
اذا كانت المعادلات التي تعرف الاحداثيات المعممة لا تتعلق بالزمن صراحة اي
$\vec{r}_i=\vec{r}_i(q_1,q_2,...,q_n)$
فان
$\vec{v}_i=\sum_j\dot{q}_j{\partial \vec{r}_i}/{\partial q_j} $
و بالتالي
$T=T_2$
حيث
$T_2$
هي دالة في 
$q_i$
و
$\dot{q}_i$
تربيعية في 
$\dot{q}_i$.
من الجهة الاخري اذا كانت الطاقة الكامنة لا تتعلق بالسرعات المعممة 
$\dot{q}_i$
فان
$L_2=T$, $L_1=0$ 
و
$L_0=-V$.
اذن نحصل علي
\begin{eqnarray}
H=T+V.
\end{eqnarray}
هذه هي الطاقة الكلية للجملة.
\end{itemize}
يمكن ان نبرهن بدون صعوبة باستعمال معادلات هاميلتون ان
\begin{eqnarray}
\frac{dH}{dt}=\frac{\partial H}{\partial t}.
\end{eqnarray}
اذن اذا كانت الطاقة الكامنة لا تتعلق بالزمن صراحة فان
$L$
لا يتعلق بالزمن صراحة و بالتالي فان 
$H$
لا يتعلق بالزمن صراحة اي ان الهاميلتونية 
$H$
هي منحفظة في الزمن.

\section*{
معادلات هاميلتون من حساب التغاير: مبدأ هاميلتون المعدل
}
\addcontentsline{toc}{section}{
معادلات هاميلتون من حساب التغاير: مبدأ هاميلتون المعدل
} 
كما بينا في السابق فان معادلات لاغرانج للحركة يمكن اشتقاقها من مبدأ هاميلتون التكاملي الذي يأخذ الشكل
\begin{eqnarray}
\delta I=\delta \int_{t_1}^{t_2} dt L(q_i,\dot{q}_i,t)=0.\label{hamiltonI}
\end{eqnarray}
بالطبع فان حساب التغاير يتم علي طرق معرفة في فضاء التمثيلات بين النقطتين
$(q_1(t_1),...,q_n(t_1))$
و
$(q_1(t_2),...,q_n(t_2))$.

معادلات هاميلتون تخص حركة حالة الجملة في الفضاء الطوري و بالتالي فان المبدأ التغايري الذي يمكن ان يؤدي الي هاته المعادلات يجب بالضرورة صياغته في
الفضاء الطوري.

بالتعويض بالمعادلة
$L(q_i,\dot{q}_i,t)=\sum_i \dot{q}_ip_i -H(q_i,p_i,t)$
في مبدأ هاميلتون اعلاه, ثم اعادة تفسير الطرق التي يحسب عليها التغاير علي انها الطرق في الفضاء الطوري التي تربط بين النقطتين
$(q_1(t_1),...,q_n(t_1),p_1(t_1),...,p_n(t_1))$
و
$(q_1(t_2),...,q_n(t_2),p_1(t_2),...,p_n(t_2))$,
نحصل علي مبدأ هاميلتون المعدل الذي يعطي ب
\begin{eqnarray}
\delta I=\delta \int_{t_1}^{t_2} dt \bigg(\sum_i \dot{q}_ip_i -H(q_i,p_i,t)\bigg)=0.
\end{eqnarray}
هذا مبدأ تغايري في فضاء ذو 
$2n$
بعد من نفس شكل مبدأ هاميلتون
 $(\ref{hamiltonI})$
 اي من الشكل
 \begin{eqnarray}
\delta I=\delta \int_{t_1}^{t_2} dt {\cal L}(q_i,\dot{q}_i,p_i,\dot{p}_i,t)=0.
\end{eqnarray}
معادلات لاغرانج من اجل هذا المبدأ هي مباشرة معطاة ب
 \begin{eqnarray}
\frac{d}{dt}\bigg(\frac{\partial{\cal L}}{\partial \dot{q}_i}\bigg)-\frac{\partial{\cal L}}{\partial {q}_i}=0\Leftrightarrow \frac{d}{dt}\big({p}_i\big)+\frac{\partial{H}}{\partial {q}_i}=0.
\end{eqnarray}

\begin{eqnarray}
\frac{d}{dt}\bigg(\frac{\partial{\cal L}}{\partial \dot{p}_i}\bigg)-\frac{\partial{\cal L}}{\partial {p}_i}=0\Leftrightarrow \frac{d}{dt}\big(0\big)-\dot{q}_i+\frac{\partial{H}}{\partial {p}_i}=0.
\end{eqnarray}
كما نري نحصل مباشرة علي معادلات هاميلتون للحركة.

\section*{
التحويلات القانونية
}
\addcontentsline{toc}{section}{
التحويلات القانونية
} 

نبدأ بالتذكير بالاحداثيات المعممة الدورية. الاحداثية المعممة 
$q_i$
هي احداثية دورية اذا لم تتعلق الهاميلتونية
$H=H(q_i,p_i)$
بها. اذن في هذه الحالة, باستعمال معادلات هاميلتون, نجد ان كمية الحركة المعممة المقابلة
$p_i$
هي منحفظة في الزمن. لدينا اذن
\begin{eqnarray}
-\dot{p}_i=\frac{\partial H}{\partial q_i}=0\Rightarrow p_i=\beta_i={\rm constant}.
\end{eqnarray}
 الملاحظة الاساسية الاولي هنا هي كالتالي :  حل معادلات هاميلتون هو عملية سهلة في حالة الاحداثيات الدورية.
 
 من الجهة الاخري ان اختيار الاحداثيات المعممة و كميات الحركة المعممة هو عملية كيفية في مجملها لان
 هناك عدد غير منته من الاختيارات الممكنة. لانه دائما يهمنا حل معادلات هاميلتون فانه من مصلحتنا اختيار
 احداثيات معممة 
 $Q_i$
 و كميات حركة معممة  
$P_i$
يكون من اجلها بعض او كل الاحداثيات المعممة 
$Q_i$
دورية. هذه المجموعة الجديدة
$(Q_i,P_i)$
يجب هي الاخري ان تحقق معادلات هاميلتون بهاميلتونية جديدة
$K(Q_i,P_i)$
مغايرة عموما للهاميلتونية الاصلية
$H(q_i,p_i)$.
من اجل هذا السبب بالضبط فان التحويل
$(q_i,p_i)\longrightarrow (Q_i,P_i)$
يسمي تحويل قانوني.

 التحويل القانوني هو تعميم لتغيير المتغيرات في فضاء التمثيلات المعطي بالتحويل النقطي
$q_i\longrightarrow Q_i=Q_i(q_i,t)$.
التحويل القانوني هو في الحقيقة تغيير متغيرات في الفضاء الطوري من الشكل
\begin{eqnarray}
q_i\longrightarrow Q_i=Q_i(q_j,p_j,t)~,~p_i\longrightarrow P_i=P_i(q_j,p_j,t).
\end{eqnarray}
نفترض ان الزوج 
$(q_i,p_i)$
يحل معادلات هاميلتون بهاميلتونية
$H=H(q_i,p_i)$
اي
\begin{eqnarray}
\dot{q}_i=\frac{\partial H}{\partial p_i}~,~-\dot{p}_i=\frac{\partial H}{\partial q_i}.
\end{eqnarray}
كما بينا اعلاه فان هذه المعادلات يمكن اشتقاقها من مبدأ هاميلتون المعدل:
\begin{eqnarray}
\delta\int_{t_1}^{t_2}(p_i\dot{q}_i-H(q,p,t))=0.
\end{eqnarray}
كما قلنا قبل قليل فان التحويل
$q_i\longrightarrow Q_i=Q_i(q_j,p_j,t)$, $p_i\longrightarrow P_i=P_i(q_j,p_j,t)$ 
هو تحويل قانوني لاننا نفترض ان الزوج الجديد
$(ِQ_i,P_i)$
يحل ايضا معادلات هاميلتون لكن بهاميلتونية جديدة
$K(Q,P,t)$
اي ان المتغيرات
الجديدة
 $Q_i$
 و
 $P_i$
 هي متغيرات قانونية.
 لدينا اذن معادلات هاميلتون الجديدة
\begin{eqnarray}
\dot{Q}_i=\frac{\partial K}{\partial P_i}~,~-\dot{P}_i=\frac{\partial K}{\partial Q_i}.
\end{eqnarray}
من الواضح انه يمكننا ان نشتق هذه المعادلات من مبدأ هاميلتون المعدل 
\begin{eqnarray}
\delta\int_{t_1}^{t_2}(P_i\dot{Q}_i-K(Q,P,t))=0.
\end{eqnarray}
اذن يجب ان يكون لدينا
\begin{eqnarray}
\delta\int_{t_1}^{t_2}(p_i\dot{q}_i-H(q,p,t))=\delta\int_{t_1}^{t_2}(P_i\dot{Q}_i-K(Q,P,t))=0.
\end{eqnarray}
او بالمقابل
\begin{eqnarray}
\lambda (p_i\dot{q}_i-H(q,p,t))=P_i\dot{Q}_i-K(Q,P,t)+\frac{dF}{dt}.
\end{eqnarray}
$\lambda$
هي ثابت و 
$F$
هي دالة في احداثيات الفضاء الطوري ذات مشتقة ثانية مستمرة. التحويلات القانونية التي لها
$\lambda\neq 1$
نسميها التحويلات القانونية الممتدة اما التي لها
$\lambda=1$
فنسميها اختصارا بالتحويلات القانونية. 

الثابت 
$\lambda$
ناجم عن تحويل قانوني خاض جدا يسمي بالتحويل السلمي الذي يعرف كالتالي:
\begin{eqnarray}
q_i\longrightarrow Q_i=\mu q_i~,~p_i\longrightarrow P_i=\nu p_i.
\end{eqnarray}
$\mu$
و
$\nu$
هي ثوابت. 
معادلات هاميلتون الجديدة تؤدي الي معادلات هاميلتون القديمة مع الحل
\begin{eqnarray}
K(Q_i,P_i)=\lambda H(q_i,p_i)~,~\lambda=\mu\nu.
\end{eqnarray}
اي
\begin{eqnarray}
\lambda (p_i\dot{q}_i-H(q,p,t))=P_i\dot{Q}_i-K(Q,P,t).
\end{eqnarray}
يمكننا دائما اختيار
$\lambda=1$
باستعمال تحويل سلمي مناسب. نفترض مثلا انه لدينا تحويل قانوني
$(q_i,p_i)\longrightarrow (Q_i^{'},P_i^{'})$
 مع
 $\lambda\neq 1$.
 نعتبر التحويل السلمي
 $(Q_i,P_i)\longrightarrow (Q_i^{'}=\mu Q_i, P_i^{'}=\nu P_i)$
  مع 
  $\lambda=\mu\nu$.
  التحويل القانوني
  $(q_i,p_i)\longrightarrow (Q_i^{'},P_i^{'})$
  هو تركيب للتحويل القانوني
  $(q_i,p_i)\longrightarrow (Q_i^{},P_i^{})$
  و التحويل السلمي
  $(Q_i,P_i)\longrightarrow (Q_i^{'}, P_i^{'})$.
يمكننا التحقق بسهولة ان التحويل القانوني
$(q_i,p_i)\longrightarrow (Q_i,P_i)$
له
$\lambda=1$.

اذن يمكننا التركيز بالكامل علي التحويلات القانونية
التي لها 
$\lambda=1$
دون فقدان اي عمومية في تناولنا للتحويلات القانونية. من اجل هذه التحويلات لدينا
\begin{eqnarray}
p_i\dot{q}_i-H(q,p,t)=P_i\dot{Q}_i-K(Q,P,t)+\frac{dF}{dt}.\label{canonical}
\end{eqnarray}
التحويلات القانونية التي لا تتعلق بالزمن صراحة اي
$Q_i=Q_i(q_j,p_j)$ 
و
$P_i=P_i(q_j,p_j)$
تسمي بالتحويلات القانونية المحدودة.


الدالة 
$F$
هي دالة في احداثيات الفضاء الطوري
$q_i$, $Q_i$, $p_i$
و
$P_i$
 بالاضافة الي الزمن اي انها دالة في 
 $4n+1$
 متغير.
باستعمال
 $Q_i=Q_i(q_j,p_j,t)$
 و
 $P_i=P_i(q_j,p_j,t)$
 و معكوساتها نري ان
 $F$
 هي في الواقع دالة في 
 $2n+1$
 متغير.
الدالة
$F$ 
تسمح لنا بتعيين الشكل المضبوط للتحويل القانوني فقط عندما نأخد نصف متغيراتها من الاحداثيات الطورية القديمة
$(q_i,p_i)$
و النصف الاخر من الاحداثيات الطورية الجديدة
$(Q_i,P_i)$
.
في هذه الحالة فان
$F$ 
تلعب دور مولد التحويل القانوني. اذن لدينا اربعة انواع فقط من التحويلات القانونية  معينة بالدوال المولدة التالية
 
 \begin{eqnarray}
F=F_1(q_i,Q_i,t).
\end{eqnarray} 
\begin{eqnarray}
F=F_2(q_i,P_i,t).
\end{eqnarray}
 \begin{eqnarray}
F=F_3(p_i,Q_i,t).
\end{eqnarray}
\begin{eqnarray}
F=F_4(p_i,P_i,t).
\end{eqnarray}
نناقش فيما تبقي ببعض التفصيل الحالتين الاولي و الثانية.
\paragraph{
الحالة الاولي:
}
في هذه الحالة
\begin{eqnarray}
F=F_1(q_i,Q_i,t).
\end{eqnarray} 
نحسب
\begin{eqnarray}
p_i\dot{q}_i-H=P_i\dot{Q}_i-K+\frac{\partial F_1}{\partial t}+\frac{\partial F_1}{\partial q_i}\dot{q}_i+\frac{\partial F_1}{\partial Q_i}\dot{Q}_i.
\end{eqnarray}
لان 
 $q_i$ 
 و
 $Q_i$
 مستقلان خطيا نحصل علي
\begin{eqnarray}
p_i=\frac{\partial F_1}{\partial q_i}~,~P_i=-\frac{\partial F_1}{\partial Q_i}.
\end{eqnarray}
\begin{eqnarray}
K=H+\frac{\partial F_1}{\partial t}.
\end{eqnarray}
\paragraph{
الحالة الثانية:
}
في هذه الحالة فان الدالة المولدة يجب ان تكون دالة في
الاحداثيات المعممة القديمة
$q_i$
و 
كميات الحركة المعممة الجديدة
$P_i$. 
بالمقارنة
بالحالة الاولي فان 
$P_i$
هنا يلعب دور
$Q_i$
هناك. بالتالي فانه في المعادلة 
$(\ref{canonical})$
يجب تعويض
$P_i\dot{Q}_i$
ب
$Q_i\dot{P}_i$.
 يمكن تحقيق ذلك باختيار الدالة المولدة كالتالي
\begin{eqnarray}
F=F_2(q_i,P_i,t)-Q_iP_i.
\end{eqnarray}
نحسب الان
\begin{eqnarray}
p_i\dot{q}_i-H=-Q_i\dot{P}_i-K+\frac{\partial F_2}{\partial t}+\frac{\partial F_2}{\partial q_i}\dot{q}_i+\frac{\partial F_2}{\partial P_i}\dot{P}_i.
\end{eqnarray}
مرة اخري لان 
$q_i$
و
$P_i$ 
هما مستقلان خطيا نحصل علي
\begin{eqnarray}
p_i=\frac{\partial F_2}{\partial q_i}~,~Q_i=\frac{\partial F_2}{\partial P_i}.
\end{eqnarray}
\begin{eqnarray}
K=H+\frac{\partial F_2}{\partial t}.
\end{eqnarray}
من اجل الحالتين الثالثة و الرابعة نكتب
\begin{eqnarray}
F=F_3(p_i,Q_i,t)+q_ip_i.
\end{eqnarray}
\begin{eqnarray}
F=F_4(p_i,P_i,t)+q_ip_i-Q_iP_i.
\end{eqnarray}
التحويلات القانونية التي تكون دالتها المولدة لا تتعلق بالزمن هي بالضبط التحويلات القانونية المحدودة و من اجلها لدينا
 \begin{eqnarray}
\frac{\partial F}{\partial t}=0\Rightarrow K=H.
\end{eqnarray}
\section*{
 الصياغة السمبليكتية, اقواس بواسون و مبرهنة ليوفيل
}
\addcontentsline{toc}{section}{
الصياغة السمبليكتية, اقواس بواسون و مبرهنة ليوفيل
} 

\paragraph{
الشرط السمبليكتي
:}
يمكن كتابة التحويلات القانونية علي شكل  اخر مختلف, لكن مكافئ للدوال المولدة, باستعمال الصياغة السمبليكتية
\footnote{.${\rm symplectic}~{\rm formulation}$}
لمعادلات هاميلتون. اولا نعرف الشعاع
$\eta$ 
في 
$2n$
بعد
 المشكل من الاحداثيات المعممة
 $q_i$
 و
 كميات الحركة المعممة
 $p_i$,
 و الشعاع
 $\xi$
 المعرف
 ايضا في
 $2n$
بعد
 المشكل من الاحداثيات المعممة
 $Q_i$
 و
 كميات الحركة المعممة
 $P_i$
  اي
 \begin{eqnarray}
\eta=\left(\begin{array}{c}
q_i\\
p_i
\end{array}\right)~,~\xi=\left(\begin{array}{c}
Q_i\\
P_i
\end{array}\right).
\end{eqnarray} 
 هذه اشعة معرفة في الفضاء الطوري. معادلات التحويل القانوني
 المحدود
 $Q_i=Q_i(q_j,p_j)$
 و
 $P_i=P_i(q_j,p_j)$
 يمكن كتابتها علي الشكل
 \begin{eqnarray}
\xi=\xi(\eta).
\end{eqnarray}  
معادلات هاميلتون في المتغيرات
$\eta$
تعطي ب
\begin{eqnarray}
\dot{\eta}=J\frac{\partial H}{\partial \eta}.\label{hamiltonsymp}
\end{eqnarray}
المصفوفة
$J$
هي
$2n{\rm x}2n$
و تعطي ب
\begin{eqnarray}
J=\left(\begin{array}{cc}
0&{\bf 1}_{n}\\
-{\bf 1}_n&0
\end{array}\right).
\end{eqnarray}
معادلات هاميلتون في المتغيرات
$\xi$
تعطي ب
\begin{eqnarray}
\dot{\xi}&=&J\frac{\partial H}{\partial \xi}.
\end{eqnarray}
نعرف المصفوفة
$M$
ب
\begin{eqnarray}
M_{ij}=\frac{\partial \xi_i}{\partial \eta_j}.
\end{eqnarray}
لدينا
\begin{eqnarray}
\dot{\xi}_i&=&M_{ij}\dot{\eta}_j\nonumber\\
&=&M_{ij}J_{jk}\frac{\partial H}{\partial \eta_k}\nonumber\\
&=&M_{ij}J_{jk}M_{lk}\frac{\partial H}{\partial \xi_l}\nonumber\\
&=&(MJM^T)_{il}\frac{\partial H}{\partial \xi_l}.
\end{eqnarray}
 بالمقارنة
 نحصل اذن علي
\begin{eqnarray}
MJM^T=J.
\end{eqnarray}
هذا هو الشرط السمبليكتي
 و المصفوفة
 $M$
 هي مصفوفة سمبليكتية
.
ان الشرط السمبليكتي هو شرط ضروري و كافي من اجل كل التحويلات القانونية حتي تلك التي تتعلق بالزمن و ليس فقط من اجل التحويلات القانونية المحدودة التي اعتبرناها اعلاه. يمكن ان نبرهن ان الشرط السمبليكتي يستلزم وجود دالة مولدة. 
ايضا يمكن استعمال الصياغة السمبليكتية للبرهان علي ان مجموعة التحويلات القانونية تشكل زمرة. 

\paragraph{
التحويلات القانونية المتناهية في الصغر:
}

يمكن ان نبرهن ان التحويلات القانونية تحقق الشرط السمبليكتي كالتالي. اولا لانه لدينا بنية زمرة فان اي تحويل قانوني يمكن تفكيكه كالاتي
\begin{eqnarray}
\eta=\left(\begin{array}{c}
q_i\\
p_i
\end{array}\right)\longrightarrow \xi(\eta,t_0)=\left(\begin{array}{c}
Q_i(q,p,t_0)\\
P_i(q,p,t_0)
\end{array}\right)\longrightarrow \xi(\eta,t)=\left(\begin{array}{c}
Q_i(q,p,t)\\
P_i(q,p,t)
\end{array}\right).\label{decomp}
\end{eqnarray}
الحد الاول يحقق الشرط السمبليكتي لانه تحويل قانوني محدود لا يتعلق بالزمن. مرة اخري لانه لدينا بنية زمرة فان التحويل الثاني يمكن تركيبه من تحويلات قانونية متناهية في الصغر اي نقسم المجال
$t-t_0$
الي مجالات صغيرة متناهية في الصغر
$dt$
و نعتبر فقط التحويل القانوني في كل مجال.

 نبدأ بتعريف التحويلات القانونية المتناهية في الصغر. اولا نلاحظ ان
الدالة
$F_2=q_iP_i$
تولد التحويل القانوني الذي يؤثر كالتطابق
\footnote{.${\rm identity}$}. بالفعل يمكن ان نبرهن في هذه الحالة علي ان
$Q_i=q_i$, $P_i=p_i$ 
و
$K=H$.
التحويل القانوني المتناه في الصغر يقابل اذن
\begin{eqnarray}
F_2=q_iP_i+\epsilon G(q_j,P_j,t).
\end{eqnarray}
نحسب
\begin{eqnarray}
P_i=p_i-\epsilon\frac{\partial G}{\partial q_i}~,~Q_i=q_i+\epsilon\frac{\partial G}{\partial P_i}=q_i+\epsilon\frac{\partial G}{\partial p_i}.
\end{eqnarray}
اي انه يمكننا ان نفكر في 
$G$
علي انها دالة في 
$q$
و
$p$,
عوض
$q$
و
$P$,
و
الزمن.
 الدالة
  $G$ 
  هي الدالة المولدة للتحويل القانوني المتناه في الصغر. لدينا
\begin{eqnarray}
\delta p_i=P_i-p_i=-\epsilon\frac{\partial G}{\partial q_i}~,~\delta q_i=Q_i-q_i=\epsilon\frac{\partial G}{\partial p_i}.\label{ICT}
\end{eqnarray}
يمكن ان نكتب هذه المعادلات
$(\ref{ICT})$
علي الشكل المتراص
\begin{eqnarray}
\delta\eta=\xi-\eta=\epsilon J\frac{\partial G}{\partial \eta}.\label{infinitesimal}
\end{eqnarray}
ايضا يمكن ان نحسب من اجل التحويل القانوني المتناه في الصغر اعلاه
\begin{eqnarray}
M=\frac{\partial \xi}{\partial \eta}&=&1+\frac{\partial}{\partial\eta}\delta\eta\nonumber\\
&=&1+\epsilon J\frac{\partial^2 G}{\partial\eta\partial\eta}.
\end{eqnarray}
المصفوفة
${\partial^2 G}/{\partial\eta\partial\eta}$
هي مصفوفة متناظرة بمركبات معطاة ب

\begin{eqnarray}
\big(\frac{\partial^2 G}{\partial\eta\partial\eta}\big)_{ij}=\frac{\partial^2 G}{\partial\eta_i\partial\eta_j}.
\end{eqnarray}
يمكن ان نتحقق الان مباشرة من الشرط السمبليكتي من اجل التحويلات القانونية المتناهية في الصغر كالتالي

\begin{eqnarray}
MJM^T&=&\bigg(1+\epsilon J\frac{\partial^2 G}{\partial\eta\partial\eta}\bigg)J\bigg(1-\epsilon \frac{\partial^2 G}{\partial\eta\partial\eta}J\bigg)\nonumber\\
&=&J.
\end{eqnarray}
اذا اخترنا
$\epsilon=dt$
فان التحويل القانوني المتناه في الصغر اعلاه هو بالضبط التحويل
$\xi(\eta,t_0)\longrightarrow \xi(\eta,t)$
مع
$t=t_0+dt$
.
هذا التحويل يحقق اذن الشرط السمبليكتي و بالتالي فان التحويل القانوني الذي يظهر في الحد الثاني ل
$(\ref{decomp})$,
 و الذي هو تركيب لتحويلات قانونية متناهية في الصغر من النوع 
 $\epsilon=dt$
 , يحقق الشرط السمبليكتي و هو المراد. 

\paragraph{
اقواس بواسون:
}
نعرف اقواس بواسون لدالتين
$u$ 
و
 $v$
 علي الفضاء الطوري بالنسبة للمتغيرات
 $q_i$ 
 و
  $p_i$
  بالعلاقة

\begin{eqnarray}
[u,v]_{\eta}&=&\sum_i\bigg(\frac{\partial u}{\partial q_i}\frac{\partial v}{\partial p_i}-\frac{\partial u}{\partial p_i}\frac{\partial v}{\partial q_i}\bigg)\nonumber\\
&=&\bigg(\frac{\partial u}{\partial \eta}\bigg)^TJ~\frac{\partial v}{\partial \eta}.\label{poisson}
\end{eqnarray}
نحسب مباشرة ما يسمي باقواس بواسون الاساسية التي تعطي ب
\begin{eqnarray}
[\eta,\eta]_{\eta}&=&J.
\end{eqnarray}
هذه العلاقة تأخذ بدلالة المركبات الشكل
\begin{eqnarray}
[q_i,q_j]_{\eta}=0~,~[p_i,p_j]_{\eta}=0~,~[q_i,p_j]_{\eta}=-[p_i,q_j]_{\eta}=\delta_{ij}.
\end{eqnarray}
نحسب الان
\begin{eqnarray}
[u,v]_{\eta}&=&\frac{\partial u}{\partial \eta_i}J_{ij}\frac{\partial v}{\partial \eta_j}\nonumber\\
&=&\frac{\partial u}{\partial \xi_k}\frac{\partial \xi_k}{\partial \eta_i}J_{ij}\frac{\partial \xi_l}{\partial \eta_j}\frac{\partial v}{\partial \xi_l}\nonumber\\
&=&\frac{\partial u}{\partial \xi_k}(MJM^T)_{kl}\frac{\partial v}{\partial \xi_l}\nonumber\\
&=&[u,v]_{\xi}.
\end{eqnarray}
اي ان اقواس بواسون هي صامدة تحت تأثير التحويلات القانونية.  هذا الشرط  مكافئ تماما للشرط السمبليكتي. ايضا
يمكن استعمال خاصية الصمود هذه للبرهان علي ان الشرط السمبليكتي يستلزم وجود دالة
مولدة للتحويل القانوني.
 كما ستري هناك اشياء اخري, بالاضافة الي اقواس بواسون, تبقي صامدة تحت تأثير التحويلات القانونية.

 نعتبر الان دالة 
$u$
في المتغيرات 
$q_i$, $p_i$
و الزمن
اي
$u=u(q_i,p_i,t)$.
 باستعمال معادلات هاميلتون فان المشتقة التامة في الزمن للدالة
$u$
تعطي ب

\begin{eqnarray}
\frac{du}{dt}&=&\sum_i\bigg(\frac{\partial u}{\partial q_i}\dot{q}_i+\frac{\partial u}{\partial p_i}\dot{p}_i\bigg)+\frac{\partial u}{\partial t}\nonumber\\
&=&\sum_i\bigg(\frac{\partial u}{\partial q_i}\frac{\partial H}{\partial p_i}-\frac{\partial u}{\partial p_i}\frac{\partial H}{\partial q_i}\bigg)+\frac{\partial u}{\partial t}\nonumber\\
&=&[u,H]_{\eta}+\frac{\partial u}{\partial t}.\label{l0}
\end{eqnarray}
هذه هي معادلة حركة الدالة
$u$. 
يمكن الحصول علي معادلات هاميلتون كحالة خاصة كالتالي. اذا اخترنا
$u=q_i,p_i$
نحصل مباشرة علي
$\dot{q}_i=[q_i,H]_{\eta}$, $\dot{p}_i=[p_i,H]_{\eta}$.
باستعمال الكتابة السمبليكتية نحصل اذن علي
\begin{eqnarray}
 \dot{\eta}=[\eta,H]_{\eta}=J\frac{\partial H}{\partial \eta}
 \end{eqnarray}
 التي هي عبارة علي معادلات هاميلتون
 $(\ref{hamiltonsymp})$.

يمكن ايضا التعبير عن التحويلات القانونية المتناهية في الصغر 
$(\ref{infinitesimal})$
باستعمال اقواس بواسون. باختيار
 $u=\eta$ 
 و
 $v=G$
 في
 $(\ref{poisson})$
 نحصل علي
 \begin{eqnarray}
 [\eta,G]_{\eta}=J\frac{\partial G}{\partial \eta},
 \end{eqnarray}
التحويل القانوني المتناه في الصغر
$(\ref{infinitesimal})$
يمكن اذن كتابته علي الشكل
\begin{eqnarray}
\delta\eta=\epsilon[\eta,G]_{\eta}.
\end{eqnarray}
اذا اخترنا مثلا
 $\epsilon=dx$ 
 و
 $G=p_j$
نحصل علي
$\delta q_i=dx[q_i,p_j]_{\eta}=\delta_{ij}dx$
و
$\delta p_i=dx[p_i,p_j]_{\eta}=0$
اي ان
الانسحاب في الاتجاه
$j$
تولده كمية الحركة
$p_j$.

كمثال ثاني نعتبر الاتي. 
اذا اخترنا
$\epsilon=dt$
و
$G=H$
نحصل علي
$\delta\eta=\dot{\eta}dt=d\eta$
اي
\begin{eqnarray}
\epsilon=dt~,~G=H\Rightarrow \delta\eta=\dot{\eta}dt=d\eta.
\end{eqnarray}
اذن الهاميلتونية هي مولدة حركة الجملة اي التطور في الزمن.  هذه النتيجة المهمة جدا يمكن الوصول اليها  ايضا كالتالي. اذا اخترنا
$G=H$
في التحويل القانوني المتناه في الصغر
$(\ref{ICT})$
نستنتج مباشرة ان 
\begin{eqnarray}
\delta p_i=\epsilon\dot{p}_i~,~\delta q_i=\epsilon\dot{q}_i\Rightarrow \epsilon=dt.
\end{eqnarray}
اي ان الهاميلتونية هي الدالة المولدة للحركة المتناهية في الصغر اي للانسحابات في الزمن المتناهية في الصغر. بصيغه اخري  نقول ان حركة الجملة في الزمن هي تحويل قانوني تولده الهاميلتونية.

\paragraph{
مبرهنة ليوفيل:
}

 كما قلنا اعلاه هناك اشياء اخري, بالاضافة الي اقواس بواسون,  تبقي صامدة تحت تأثير التحويلات القانونية و منها التحويلات القانونية التي تحرك الجملة في الزمن. اهم هذه الامور الاخري هو الحجم في الفضاء الطوري.

 الحجم المتناه في الصغر في الفضاء الطوري يعطي بدلالة الاحداثيات 
 $\eta_i$
 بعنصر الحجم
 \begin{eqnarray}
dV_{\eta}=d^{2n}\eta=dq_1...dq_ndp_1...dp_n.
\end{eqnarray}
تحت تأثير التحويل القانوني 
$\eta\longrightarrow \xi$
يتحول الحجم
$dV_{\eta}$
الي الحجم
$dV_{\xi}$
الذي يعطي ب
 \begin{eqnarray}
 dV_{\xi}=d^{2n}\xi=dQ_1...dQ_ndP_1...dP_n.
\end{eqnarray}
هذان الحجمان
$dV_{\eta}$
و
$dV_{\xi}$
مرتبطان كما هو معروف
بالمحدد الجاكوبي للتحويل القانوني 
$\eta\longrightarrow \xi$.
 بالضبط لدينا
\begin{eqnarray}
 d^{2n}\xi&=&|{\rm det}\frac{\partial \xi_i}{\partial \eta_j}|d^{2n}\eta\nonumber\\
 &=&|{\rm det} M_{ij}|d^{2n}\eta.
\end{eqnarray}
اي
\begin{eqnarray}
 dV_{\xi}=||M||dV_{\eta}.
\end{eqnarray}
من الجهة الاخري فان الشرط السمبليكتي
$MJM^{T}=J$
يؤدي الي
\begin{eqnarray}
 {\rm det}J&=&{\rm det}(MJM^T)\nonumber\\
 &=&{\rm det}M.{\rm det}J.{\rm det}M^T\nonumber\\
 &=&{\rm det} J.({\rm det}M)^2.
\end{eqnarray}
نستنتج اذن
$|M|^2=1$
و بالتالي
\begin{eqnarray}
 dV_{\xi}=dV_{\eta}.
\end{eqnarray}
اذن الحجم المتناه في الصغر صامد تحت تأثير التحويلات القانونية. هذا يستلزم مباشرة ان حجم اي منطقة في الفضاء الطوري هو صامد تحت تأثير التحويلات القانونية. لدينا اذن
التكامل الصامد
 \begin{eqnarray}
 V_{\eta}&=&\int dV_{\eta}\nonumber\\
 &=&\int d^{2n}\eta\nonumber\\
 &=&\int dq_1...dq_ndp_1...dp_n.
\end{eqnarray}
يعرف هذا التكامل باسم التكامل الصامد لبوانكريه.

نعتبر الان حجم متناه في الصغر 
$dV_{\eta}$
 في الفضاء الطوري
  يحتوي علي
  $dN_{\eta}$
  نقطة
  $(q_1,...,q_n,p_1,...,p_n)$.
  كل نقطة من هذه النقاط تعرف حالة معينه للجملة الفيزيائية في لحظة ابتدائية
  $t_0$
  تختلف فيما بينها فقط باختلاف الشروط الابتدائية.  
  كثافة الحالات تعرف ب
  \begin{eqnarray}
 \rho=\frac{dN_{\eta}}{dV_{\eta}}.
\end{eqnarray}
 كما بينا اعلاه فان الحجم لا يتغير في الزمن اذا كانت النقاط تتطورتحت تأثير معادلات هاميلتون.
  اذن مع مرور الزمن فان الحجم 
  $dV_{\eta}$
  يمكن ان يتغير شكله لكن لا يمكن ان تتغير قيمته. من الواضح ان عدد الحالات 
  $dN_{\eta}$
  داخل الحجم
   $dV_{\eta}$
   يبقي ايضا ثابت في الزمن لان كل الحركة اللاحقة للجملة تحددها بشكل فريد
   المواقع الابتدائية في الفضاء الطوري و بالتالي فان كل النقاط داخل الحجم 
  $dV_{\eta}$
  في اللحظة
  $t_0$
   تتحرك مع بعضها البعض لتحتل الحجم الجديد
   $dV_{\eta}$
   في اللحظة
   $t$
   . نستنتج اذن ان كثافة الحالات يجب ان تكون ثابته في الزمن اي
    \begin{eqnarray}
 \frac{d\rho}{dt}=0.
\end{eqnarray}
   هذه هي مبرهنة ليوفيل. 
   الدالة 
   $\rho$
   هي
   دالة في الفضاء الطوري في الاحداثيات
   $q_i$, $p_i$
   و الزمن. باستعمال النتيجة
   $(\ref{l0})$
   لدينا
 \begin{eqnarray}
\frac{d\rho}{dt}
&=&[\rho,H]_{\eta}+\frac{\partial \rho}{\partial t}.
\end{eqnarray}
مبرهنة ليوفيل تأخذ اذن الشكل المكافئ
\begin{eqnarray}
\frac{\partial \rho}{\partial t}
&=&-[\rho,H]_{\eta}.
\end{eqnarray}

\paragraph{
التفسير الايجابي و التفسير السلبي للتحويلات القانونية
:
}
التحويلات القانونية يمكن تفسيرها اما ايجابيا او سلبيا. في التفسير السلبي للتحويل القانوني ننتقل من الفضاء الطوري 
$\eta$ 
باحداثيات
 $q_i$
 و
$p_i$  
الي الفضاء الطوري
$\xi$
باحداثيات
$Q_i$
و
$P_i$.
اذن الجملة في اللحظة 
$t$
يمكن ان توصف
بالتمثيلة
$A=(q_i,p_i)$
و ايضا بالتمثيلة المحولة
$A^{'}=(Q_i,P_i)$.
بعبارة اخري اي دالة
$u$
في متغيرات الجملة
تأخد نفس القيمة
$u(A)=u(A^{'})$
في الفضائين
$\eta$ 
و
$\xi$
رغم
ان
التعلق الدالي
ل
 $u$
 علي
 $q_i$
 و
  $p_i$
  يختلف عموما عن تعلقها الدالي علي المتغيرات
  $Q_i$
  و
  $P_i$.

في التفسير الايجابي للتحويل القانوني
فان الاحداثيات
$Q_i$ 
و
 $P_i$ 
هي احداثيات نقطة اخري
$B$
في نفس الفضاء الطوري
مغايرة للنقطة
$A$. 
اذن التحويل القانوني يحرك الجملة من النقطة
$A=(q_i,p_i)$ 
الي النقطة
$B=(Q_i,P_i)$
بمعني انه يسمح لنا
 بالتعبير عن التمثيلة
$B$ 
بدلالة
التمثيلة
$A$
و العكس. اذن من هذا المنظور فان قيمة
الدالة
$u$ 
تتغير
عند الانتقال من 
 $A$
 الي 
 $B$
 رغم
 ان تعلقها الدالي  علي المتغيرات
 $q_i$
 و
  $p_i$
  هو نفسه علي المتغيرات
  $Q_i$
  و
  $P_i$.
  التغير
  $\partial u$
  في قيمة
  الدالة عند الانتقال من 
  $A$
 الي 
 $B$
 يعطي ب
\begin{eqnarray}
\partial u &=&u(B)-u(A)\nonumber\\
&=&u(\eta+\delta\eta)-u(\eta)\nonumber\\
&=&\frac{\partial u}{\partial \eta}\delta \eta\nonumber\\
&=&\epsilon\frac{\partial u}{\partial \eta}J\frac{\partial G}{\partial \eta}\nonumber\\
&=&\epsilon [u,G]_{\eta}.
\end{eqnarray} 
من اجل الهاميلتونية فان الامور معقدة قليلا لان الهاميلتونية تتغير تحت تأثير التحويل القانوني  
في حالة تعلق الدالة المولدة علي الزمن كالتالي
$K=H+{\partial F_2}/{\partial t}=H+\epsilon {\partial G}/{\partial t}$. 
اذن حتي في التفسير السلبي للتحويل القانوني فان الهاميلتونية تتغير من
$H(A)$
الي
$K(A^{'})$
عند الانتقال من
$A$ 
الي
 $A^{'}$,
 اما  في التفسير الايجابي فان الهاميلتونية تتغير كما في الاعلي من
 $H(A)$
الي
$H(B)$
عند الانتقال من
$A$ 
الي
 $B$.
 في هذه الحالة نعرف
 $\partial H$ 
 علي انه الفرق في قيمة
 الهاميلتونية بين التفسيرين اي
\begin{eqnarray}
\partial H&=&(H(B)-H(A))-(K(A^{'})-H(A))\nonumber\\
&=&H(B)-K(A^{'}).
\end{eqnarray}
هذا التعريف ينطبق علي التعريق السابق في حالة اذا لم تتغير الدالة تحت تأثير التحويلات القانونية. 
نحسب الان
\begin{eqnarray}
\partial H
&=&H(B)-H(A^{'})- \epsilon\frac{\partial G}{\partial t}\nonumber\\
&=&H(B)-H(A)-\epsilon\frac{\partial G}{\partial t}\nonumber\\
&=&\epsilon[H,G]_{\eta}-\epsilon\frac{\partial G}{\partial t}\nonumber\\
&=&-\epsilon\frac{dG}{dt}.
\end{eqnarray}
الخلاصة الاساسية هنا هي كالتالي: اذا كانت الدالة المولدة
$G$
هي ثابت للحركة فان التحويل القانوني المتناه في الصغر المقابل لها لا يغير من قيمة الهاميلتونية اي انه يترك الهاميلتونية صامدة. اذن ثوابت الحركة هي بالضبط مولدات التحويلات القانونية المتناهية في الصغر التي تترك الهاميلتونية صامدة.

\section*{
معادلة هاميلتون - جاكوبي
}
\addcontentsline{toc}{section}{
معادلات هاميلتون - جاكوبي
} 

نعتبر تحويل قانوني من الاحداثيات
$(q_i,p_i)$
الي الاحداثيات
$(Q_i,P_i)$
حيث نريد ان تكون 
$Q_i$ 
و
$P_i$
 ثوابت في الزمن اي
$Q_i=\beta_i$
و
$P_i=\alpha_i$. 
هذا التحويل القانوني يمكن تحقيقه بافتراض انعدام الهاميلتونية المحولة  
 $K(Q,P,t)$. 
 لان
$K(Q,P,t)=H(q,p,t)+{\partial F}/{\partial t}$
يجب اذن ان يكون لدينا
\begin{eqnarray}
H(q,p,t)+\frac{\partial F}{\partial t}=0.
\end{eqnarray} 
من الملائم اخذ الدالة المولدة 
$F$
من النوع الثاني اي
$F=F_2(q_i,P_i,t)$.
باستخدام معادلة التحويل
$p_i={\partial F_2}/{\partial q_i}$
يمكن ان نكتب المعادلة اعلاه علي الشكل
\begin{eqnarray}
H(q_1,q_2,...,q_n,\frac{\partial F_2}{\partial q_1},\frac{\partial F_2}{\partial q_2},...,\frac{\partial F_2}{\partial q_n},t)+\frac{\partial F_2}{\partial t}=0.
\end{eqnarray} 
هذه هي معادلة هاميلتون - جاكوبي. هذه معادلة تفاضلية جزئية من الرتبة الاولي في 
$n+1$ 
متغير
$q_1$, ....,$q_n$
و
$t$  
من اجل الدالة المولدة
$F_2$.
نرمز الي الحل ب
$F_2=S=S(q_1,...,q_n,\alpha_1,...,\alpha_n,\alpha_{n+1},t)$ 
و نسميه بدالة هاميلتون الرئيسية. من الواضح ان الاعداد
$\alpha_i$
هي ثوابت التكامل. من الواضح ايضا انه اذا كانت
$S$
حل فان
$S+\alpha$ 
هي ايضا حل. بعبارة اخري فان هناك ثابت تكامل  , 
و ليكن هو
$\alpha_{n+1}$,
يظهر فقط مضافا الي 
$S$
و بالتالي هو غير مهم تماما في الحل لانه
يختفي عند اخذ الاشتقاقات الجزئية. اذن الحل
$F_2=S$
يأخذ الشكل
\begin{eqnarray}
F_2=S=S(q_1,...,q_n,\alpha_1,...,\alpha_n,t).
\end{eqnarray} 
هذا الحل يسمي بالحل التام لمعادلة
هاميلتون - جاكوبي. 
كميات الحركة
 $P_i$
 التي افترضنا انها ثوابت في الزمن في بداية هذه الفقرة يمكن اخذها اذن و بدون اي مشاكل مساوية 
 للثوابت 
 $\alpha_i$
 اي
 \begin{eqnarray}
P_i=\alpha_i.
\end{eqnarray}
نكتب المعادلة
$p_i={\partial F_2}/{\partial q_i}$
الان علي الشكل
\begin{eqnarray}
p_i=\frac{\partial S(q,\alpha,t)}{\partial q_i}.
\end{eqnarray}
في اللحظة الابتدائية 
$t_0$ 
تربط هذه المعادلة بين 
القيم الابتدائية ل
 $q_i$ 
 و
 $p_i$ 
و
 $\alpha_i$.
 اذن يمكن تعيين 
 ثوابت التكامل
  $\alpha_i$
  بدلالة 
  القيم الابتدائية ل
  $q_i$ 
  و
  $p_i$ 
  انطلاقا من هذه المعادلة.
  
  من الجهة الاخري فان المعادلة
  $Q_i={\partial F_2}/{\partial P_i}$
  تكتب الان علي الشكل
  \begin{eqnarray}
Q_i=\beta_i=\frac{\partial S(q,\alpha,t)}{\partial {\alpha}_i}.
\end{eqnarray}
في اللحظة الابتدائية 
$t_0$ 
هذه المعادلة تسمح لنا بتعيين
$\beta_i$
بدلاله
  القيم الابتدائية ل
  $q_i$ 
  و
   $\alpha_i$.
   هذه المعادلة يمكن قلبها للحصول علي
  $q_i$
  بدلالة
  $\alpha_i$, 
$\beta_i$
و الزمن اي
 \begin{eqnarray}
q_i=q_i(\alpha,\beta,t).\label{s1}
\end{eqnarray}
بالتعويض في المعادلة
$p_i={\partial S(q,\alpha,t)}/{\partial q_i}$ 
نحصل علي 
$p_i$
بدلالة
 $\alpha_i$, $\beta_i$
 و الزمن
 اي 
 \begin{eqnarray}
p_i=p_i(\alpha,\beta,t).\label{s2}
\end{eqnarray}
 تشكل المعادلتان
$(\ref{s1})$
و
$(\ref{s2})$
مع بعضهما البعض
الحل التام لمعادلات
هاميلتون. نستنتج اذن ان ايجاد دالة هاميلتون الرئيسية
$S=S(q,\alpha,t)$,
التي هي عبارة عن الدالة المولدة للتحويل القانوني الذي يأخذنا لاحداثيات ثابتة في الزمن,
عبر حل معادلة
هاميلتون - جاكوبي هو مكافئ لايجاد حل لمعادلات هاميلتون للحركة. بعبارة اخري فان معادلات هاميلتون للحركة مكافئة تماما لمعادلة 
هاميلتون - جاكوبي.


المعني الفيزيائي لدالة هاميلتون الرئيسية يمكن توضيحه اكثر كالتالي. نحسب
\begin{eqnarray}
\frac{dS}{dt}&=&\frac{\partial S}{\partial q_i}\dot{q}_i+\frac{\partial S}{\partial t}\nonumber\\
&=&p_i\dot{q}_i-H\nonumber\\
&=&L.
\end{eqnarray}
اي ان
$S$
هو الفعل:
\begin{eqnarray}
S=\int L dt +{\rm constant}.
\end{eqnarray}
اذا كانت الهاميلتونية لا تتعلق بالزمن صراحة فان 
معادلة
هاميلتون - جاكوبي تصبح من الشكل
\begin{eqnarray}
H(q_i,\frac{\partial S}{\partial q_i})+\frac{\partial S}{\partial t}=0.
\end{eqnarray}
يمكن فصل الزمن بافتراض حل من الشكل
\begin{eqnarray}
S(q_i,\alpha_i,t)=W(q_i,\alpha_i)-\alpha_1t.
\end{eqnarray}
تصبح معادلة
هاميلتون - جاكوبي  من الشكل
\begin{eqnarray}
H(q_i,\frac{\partial W}{\partial q_i})=\alpha_1.\label{hcharacteristic}
\end{eqnarray}
اي ان 
$\alpha_1$
هي ثابت للحركة يساوي الطاقة في كل الحالات التي تكون فيها الهاميلتونية هي دالة الطاقة.
الدالة
$W$
تسمي دالة هاميلتون المميزة. هذه الدالة تولد التحويل القانوني الذي تصبح تحت تأثيره كل الاحداثيات دورية اي انها لا تظهر في الهاميلتونية المحولة. 

لنعتبر التحويل القانوني
$(q_i,p_i)\longrightarrow (Q_i,P_i)$
الذي تكون فيه كميات الحركة المعممة الجديدة
$P_i$
ثوابت حركة تساوي
 $\alpha_i$
 مع دالة مولدة
 $W(q_i,P_i)$ 
 لا تتعلق صراحة بالزمن و بالتالي
 $K(Q_i,P_i)=H(q_i,p_i)$. 
نفترض ان الهاميلتونية
  $H(q_i,p_i)$
  هي
ثابت حركةتساوي
$\alpha_1$.
 كما في السابق يجب ان يكون لدينا
 $p_i={\partial W}/{\partial q_i}$
 و
 $Q_i={\partial W}/{\partial P_i}={\partial W}/{\partial \alpha_i}$
 و بالتالي فان المطلب
 $H(q_i,p_i)=\alpha_1$
 هو مكافئ
 ل
 $(\ref{hcharacteristic})$. 
نلاحظ انه تحت تأثير هذا التحويل القانوني 
$K(Q_i,P_i)=P_1$
اي ان
الهاميلتونية المحولة
لا تتعلق
بالاحداثيات المعممة الجديدة
$Q_i$
و بالتالي فهي كلها دورية.
 بالاضافة
الي هذا يمكن ايضا ان نستخلص من معادلات هاميلتون ان
  $Q_1=t+\beta_1$ 
  و
   $Q_i=\beta_i$
   من اجل
   $i\neq 1$
   اي ان كل الاحداثيات المعممة الجديدة باستثناء الاولي هي ثوابت حركة.
\newpage

\section*{
تمارين
}
\addcontentsline{toc}{section}{
تمارين
} 

\paragraph{
تمرين
$1$:
}
نعتبر حركة نواس بسيط ذو كتلة
$m$
و طول 
$l$.
. احسب كمية الحركة المعممة و هاميلتونية الجملة ثم اشتق معادلات هاميلتون للحركة.

\paragraph{
تمرين
$2$:
}
نعتبر جسيم يتحرك في المستوي تحت تأثير قوة مركزية: اي قوة مشتقة من 
كمون لا يتعلق الا بالمسافة 
$r=|\vec{r}|$. 
\begin{itemize}
\item
عين الاحداثيات المعممة و احسب لاغرانجية الجملة و معادلات لاغرنج للحركة.
\item 
احسب كميات الحركة المعممة و هاميلتونية الجملة.
\item
احسب معادلات هاميلتون للحركة. هل هناك احداثيات دورية في هذه الحالة. ماذا تستنتج?
\end{itemize}

\paragraph{
تمرين
$3$:
}
جسيم ذو كتلة
$m$
يتحرك في ثلاث ابعاد يخضع لكمون 
$V(x,y,z)$. 
\begin{itemize}
 \item 
 اكتب هاميلتونية الجسيم في الاحداثيات الديكارتية.
 \item
 اشتق هاميلتونية الجملة في الاحداثيات الاسطوانية.
 \item
احسب  لاغرانجية, كميات الحركة المعممة و هاميلتونية الجسيم
 في الاحداثيات الكروية.
\end{itemize}

\paragraph{
تمرين
$4$:
}

نعتبر جسيم في حالة سقوط حر في حقل ثقالي منتظم
$\vec{g}$.
\begin{itemize}
 \item 
عين هاميلتونية الجسيم و بين انها ثابت للحركة اي انها منحفظة في الزمن.
\item
صف الفضاء الطوري في هذه الحالة. ما هو مسار الجسيم في هذا الفضاء.
\item
احسب عدد الحالات في الفضاء الطوري التي لها كمية حركة
$p_1\leq p\leq p_2$
و طاقة
$E_1\leq E\leq E_2$.

ملحوظة: عدد الحالات يجب ان يكون متناسبا مع المساحة
$F$
في الفضاء الطوري المحددة ب
$p_1\leq p\leq p_2$
و
$E_1\leq E\leq E_2$.
\item
ماذا يحدث لعدد الحالات المحسوب في السؤال السابق تحت تأثير معادلات هاميلتون.
\end{itemize}

\paragraph{
تمرين
$5$:
}
\begin{itemize}
 \item 
 اكتب هاميلتونية هزاز توافقي في بعد واحد كتلته
 $m$
 و تواتره الزاوي
 $\Omega$.
 \item
 الدالة المولدة لتحويل قانوني
 $(q,p)\longrightarrow (Q,P)$
 من النوع الاول تعطي ب
 \begin{eqnarray}
F_1(q,Q)=\frac{1}{2}m\Omega q^2\cot Q.
\end{eqnarray}
احسب التحويل القانوني صراحة.
\item
احسب الهاميلتونية الجديدة 
$K(Q,P)$.
ماذا تلاحظ.
\item 
حل معادلات هاميلتون الجديدة. استنتج معادلات مسار الهزاز التوافقي. 
\end{itemize}
\paragraph{
تمرين
$6$:
}
نواس طوله
$l$
و كتلته
$m$
يسمح له بالاهتزاز بحيث تتحرك نقطة تعليقه
علي قطع مكافئ
$y=ax^2$.
انظر الشكل
$13$.
احسب هاميلتونية الجملة و معادلات هاميلتون للحركة. اعد السؤال في التقريب التربيعي. حل معادلات الحركة و عين تواتر الحركة.

\paragraph{
تمرين
$7$:
}
لاغرانجية جملة تعطي ب
\begin{eqnarray}
L=\dot{q}_1^2+\frac{\dot{q}_2^2}{a+bq_1^2}+k_1q_1^2+k_2\dot{q}_1\dot{q}_2.
\end{eqnarray}
احسب هاميلتونية الجملة و معادلات هاميلتون للحركة.

\paragraph{
تمرين
$8$:
}
نواس كتلته
$m$
و طوله
$l$
يهتز في مستوي بحيث تتحرك نقطة تعليقه حركة دائرية منتظمة  نصف قطرها
$a$
بتواتر زاوي
$\gamma$.
\begin{itemize}
 \item 
 احسب لاغرانجية الجملة و حاول تبسيطها.
 \item 
 احسب هاميلتونية الجملة.
 \item
 اشتق معادلات هاميلتون للحركة.
\end{itemize}

\paragraph{
تمرين
$9$:
}
\begin{itemize}
 \item 
لاغرانجية جملة تعطي ب
\begin{eqnarray}
L=\frac{m}{2}(\dot{q}^2\sin^2\Omega t+\dot{q} q\Omega \sin 2\Omega t +q^2\Omega^2).
\end{eqnarray}
احسب الهاميلتونية المرفقة بهذه اللاغرانجية. هل هي منحفظة.
\item 
اكتب اللاغرانجية اعلاه بدلالة المتغير
\begin{eqnarray}
Q=q\sin\Omega t.
\end{eqnarray}
احسب الهاميلتونية المرفقة بهذه اللاغرانجية و هل هي منحفظة.
\end{itemize}
\paragraph{
تمرين
$10$:
}
نعطي التحويل
\begin{eqnarray}
Q=\log(\frac{1}{q}\sin p)~,~P=q\cot p.
\end{eqnarray}
بين مباشرة ان هذا التحويل قانوني.

\paragraph{
تمرين
$11$:
}
نعطي التحويل التالي
\begin{eqnarray}
Q_1=q_1~,~Q_2=p_2~,~P_1=p_1-2p_2~,~P_2=-2q_1-q_2.
\end{eqnarray}
بين ان هذا التحويل هو تحويل قانوني و عين دالته المولدة.

\paragraph{
تمرين
$12$:
}
\begin{itemize}
\item
احسب الطاقة الحركية و الطاقة الكامنة ثم استنتج لاغرانجية نواس بسيط كتلته
$m$
و طوله
$l$ 
(
الشكل 
$9$
)
و اشتق معادلات حركته.
\item 
نعتبر الان جملة النواس البسيط مع كتلة
  $m^{'}$
  موضوعة 
في نقطة تعليقه  بحيث يمكنها الحركة علي خط افقي مستقيم في المستوي الذي يهتز فيه النواس. انظر الشكل
$12$.
احسب في هذه الحالة لاغرانجية الجملة ثم استنتج معادلات لاغرانج للحركة.
\item
اشتق هاميلتونية الجملة و معادلات هاميلتون للحركة.
\end{itemize}

\newpage
\section*{
حلول
}
\addcontentsline{toc}{section}{
حلول
} 

\paragraph{
تمرين
$1$:
}
موضع, سرعة و طاقة حركة النواس تعطي ب
\begin{eqnarray}
\vec{r}=l (\sin\theta \hat{i}+\cos\theta \hat{j}).
\end{eqnarray}
\begin{eqnarray}
\vec{v}=l\dot{\theta} (\cos\theta \hat{i}-\sin\theta \hat{j}).
\end{eqnarray}
\begin{eqnarray}
T=\frac{1}{2}ml^2\dot{\theta}^2.
\end{eqnarray}
الطاقة الكامنة للنواس معطاة بناقص عمل قوة الثقالة اي
\begin{eqnarray}
V=-W=-mg\hat{j}\vec{r}=-mgl\cos\theta.
\end{eqnarray}
لاغرانجية الجملة تعطي اذن ب
\begin{eqnarray}
L=\frac{1}{2}ml^2\dot{\theta}^2+mgl\cos\theta.
\end{eqnarray}
نحسب الان كمية الحركة المعممة ب
\begin{eqnarray}
p_{\theta}=\frac{\partial L}{\partial \dot{\theta}}=ml^2\dot{\theta}.
\end{eqnarray}
هاميلتونية الجملة تحسب كالتالي
\begin{eqnarray}
H&=&\dot{\theta}p_{\theta}-L\nonumber\\
&=&\frac{1}{2}ml^2\dot{\theta}^2-mgl\cos\theta\nonumber\\
&=&\frac{p_{\theta}^2}{2ml^2}-mgl\cos\theta.
\end{eqnarray}
معادلات هاميلتون تعطي ب
\begin{eqnarray}
\dot{\theta}=\frac{\partial H}{\partial p_{\theta}}=\frac{p_{\theta}}{ml^2}~,~-\dot{p}_{\theta}=\frac{\partial H}{\partial \theta}=mgl\sin\theta.
\end{eqnarray}

\paragraph{
تمرين
$2$:
}

الاحداثيات المعممة هي 
$r$, $\phi$.
شعاع الموضع يعطي ب
\begin{eqnarray}
\vec{r}=r\vec{u}_r~,~\vec{u}_r=\cos\phi \hat{i}+\sin\phi \hat{j}.
\end{eqnarray}
شعاع السرعة اذن
\begin{eqnarray}
\vec{v}=\dot{r}\vec{u}_r+r\dot{\phi}\vec{u}_{\phi}~,~\vec{u}_{\phi}=-\sin\phi \hat{i}+\cos\phi \hat{j}.
\end{eqnarray}
الطاقة الحركية و اللاغرانجية
\begin{eqnarray}
T=\frac{1}{2}m\vec{v}^2=\frac{1}{2}m(\dot{r}^2+r^2\dot{\phi}^2).
\end{eqnarray}
\begin{eqnarray}
L=T-V=\frac{1}{2}m\vec{v}^2=\frac{1}{2}m(\dot{r}^2+r^2\dot{\phi}^2)-V(r).
\end{eqnarray}
معادلات لاغرانج للحركة
\begin{eqnarray}
r: m\ddot{r}-mr\dot{\phi}^2+\frac{\partial V}{\partial r}=0.
\end{eqnarray}
\begin{eqnarray}
\phi: \frac{d}{dt}(mr^2\dot{\phi})=0.
\end{eqnarray}
كميات الحركة المعممة
\begin{eqnarray}
p_r=\frac{\partial L}{\partial \dot{r}}=m\dot{r}.
\end{eqnarray}
\begin{eqnarray}
p_{\phi}=\frac{\partial L}{\partial \dot{\phi}}=mr^2\dot{\phi}.
\end{eqnarray}
الهاميلتونية تعطي اذن ب
\begin{eqnarray}
H=\dot{r}p_r+\dot{\phi}p_{\phi}-L=\frac{1}{2m}p_r^2+\frac{1}{2mr^2}p_{\phi}^2+V(r).
\end{eqnarray}
معادلات هاميلتون للحركة
\begin{eqnarray}
r: \dot{r}=\frac{\partial H}{\partial p_r}=\frac{p_r}{m}~,~-\dot{p}_r=\frac{\partial H}{\partial r}=-\frac{p_{\phi}^2}{mr^3}+\frac{\partial V}{\partial r} \Rightarrow m\ddot{r}-mr\dot{\phi}^2+\frac{\partial V}{\partial r}=0.
\end{eqnarray}
\begin{eqnarray}
\phi: \dot{\phi}=\frac{\partial H}{\partial p_{\phi}}=\frac{p_{\phi}}{mr^2}~,~-\dot{p}_{\phi}=\frac{\partial H}{\partial \phi}=0\Rightarrow  \frac{d}{dt}(mr^2\dot{\phi})=0.
\end{eqnarray}
الاحداثية المعممة 
$\phi$
هي احداثية دورية و بالتالي فان كمية الحركة المعممة المقابلة يجب ان تكون منحفظة و هو ما بيناه اعلاه. لان 
$\phi$
تصف دوران الجملة فان 
$p_{\phi}$
يجب ان يكون مرتبط بالعزم الحركي للجملة. بالفعل

\begin{eqnarray}
\vec{L}=m\vec{r}{\rm x} \vec{v}=mr^2\dot{\phi}\vec{u}_r{\rm x}\vec{u}_{\phi}=p_{\phi}\hat{k}.
\end{eqnarray}

\paragraph{
تمرين
$3$:
}
هاميلتونية الجملة في الاحداثيات الديكارتية
\begin{eqnarray}
H=\frac{p_x^2}{2m}+\frac{p_y^2}{2m}+\frac{p_z^2}{2m}+V(x,y,z).
\end{eqnarray}
هاميلتونية الجملة في الاحداثيات الاسطوانية
\footnote{
اشعة الوحدة الاسطوانية:
$\vec{u}_{\rho}=\cos\phi \hat{i}+\sin\phi\hat{j}$,  $\vec{u}_{\phi}=-\sin\phi \hat{i}+\cos\phi\hat{j}$
و
$\hat{k}$.
}
\begin{eqnarray}
H=\frac{p_{\rho}^2}{2m}+\frac{p_{\phi}^2}{2m\rho^2}+\frac{p_z^2}{2m}+V(\rho,\phi,z).
\end{eqnarray}
هاميلتونية الجملة في الاحداثيات الكروية
\footnote{
اشعة الوحدة الكروية:
$\vec{u}_r=\sin\theta\cos\phi \hat{i}+\sin\theta\sin\phi\hat{j}+\cos\theta \hat{k}$,  $\vec{u}_{\theta}=\cos\theta\cos\phi \hat{i}+\cos\theta\sin\phi\hat{j}-\sin\theta \hat{k}$
و
$\hat{u}_{\phi}=-\sin\phi \hat{i}+\cos\phi \hat{j}$.
}
\begin{eqnarray}
H=\frac{p_r^2}{2m}+\frac{p_{\theta}^2}{2mr^2}+\frac{p_{\phi}^2}{2mr^2\sin^2\theta}+V(r,\theta,\phi).
\end{eqnarray}

\paragraph{
تمرين
$4$:
}
هاميلتونية الجملة تعطي ب
(
مع
$p=m\dot{z}$)
\begin{eqnarray}
H=\frac{p^2}{2m}+mg z.
\end{eqnarray}
باستعمال معادلات هاميلتون يمكننا ان نبين
\begin{eqnarray}
\frac{dH}{dt}=\frac{\partial H}{\partial t}=0\Rightarrow H=E.
\end{eqnarray}
ثابت حركة الجملة
$E$
هو طاقة الجملة. 

الفضاء الطوري
  ذو بعدين في هذه الحالة مع محاور تعطي بالموضع
  $z$
  و
  كمية الحركة
  $p$. 
  مسار الجملة في هذا الفضاء يعطي ب
  \begin{eqnarray}
E=\frac{p^2}{2m}+mg z\Rightarrow z=\frac{1}{mg}(E-\frac{p^2}{2m}).
\end{eqnarray}
هذه معادلة قطع مكافئ.

 عدد الحالات 
  التي لها كمية حركة
$p_1\leq p\leq p_2$
و طاقة
$E_1\leq E\leq E_2$
 هو متناسب مع المساحة
$F$
في الفضاء الطوري المحددة ب
$p_1\leq p\leq p_2$
و
$E_1\leq E\leq E_2$.  
هذه المساحة تحسب عن طريق التكامل التالي
 \begin{eqnarray}
F=\int_{p_1}^{p_2}dp\int_{\frac{1}{mg}(E_1-\frac{p^2}{2m})}^{\frac{1}{mg}(E_2-\frac{p^2}{2m})}dz=\frac{E_2-E_1}{mg}\int_{p_1}^{p_2}dp=\frac{E_2-E_1}{mg}(p_2-p_1).
\end{eqnarray}
تحت تأثير معادلات هاميلتون فان كمية الحركة تنسحب كالتالي
\begin{eqnarray}
\dot{p}=-mg \Rightarrow p_i^{'}=p_i-mg t~,~i=1,2.
\end{eqnarray}
   المساحة 
 $F$
  تتغير اذن الي
  \begin{eqnarray}
F^{'}=\frac{E_2-E_1}{mg}(p_2^{'}-p_1^{'}).
\end{eqnarray}
انظر الي الشكل
$11$.
  بالتعويض نحصل علي
  $F^{'}=F$ 
  اي ان عدد الحالات لا يتغير تحت تأثير معادلات هاميلتون. هذا مثال علي مبرهنة ليوفيل.
  
\paragraph{
تمرين
$5$:
}
هاميلتونية هزاز توافقي في بعد واحد كتلته
 $m$
 و تواتره الزاوي
 $\Omega$ 
 تعطي ب
 \begin{eqnarray}
H=\frac{1}{2m}p^2+\frac{1}{2}m\Omega^2q^2.
\end{eqnarray}
  لدينا من اجل الدوال المولدة من النوع الاول
  \begin{eqnarray}
p&=&\frac{\partial F_1}{\partial q}\nonumber\\
&=&m\Omega q\cot Q.
\end{eqnarray}

 \begin{eqnarray}
P&=&-\frac{\partial F_1}{\partial Q}\nonumber\\
&=&\frac{1}{2}m\Omega q^2\frac{1}{\sin^2 Q}.
\end{eqnarray}
نحصل علي التحويل القانوني
\begin{eqnarray}
q=\sqrt{\frac{2P}{m\Omega}}\sin Q~,~p=\sqrt{2m\Omega P }\cos Q.\label{ct}
\end{eqnarray}
نحسب الهاميلتونية
\begin{eqnarray}
K(Q,P)&=&H\big(q(Q,P),p(Q,P)\big)\nonumber\\
&=&\Omega P.
\end{eqnarray}
اذن الاحداثية المعممة الجديدة
$Q$
هي احداثية دورية و منه نستنتج ان كمية الحركة المعممة الجديدة
$P$
هي ثابت للحركة.

حل معادلة هاميلتون الجديدة المتبقية نحصل علي
\begin{eqnarray}
\dot{Q}=\frac{\partial K}{\partial P}=\Omega\Rightarrow Q=\Omega t +Q_0.
\end{eqnarray}
بالتعويض في 
$(\ref{ct})$
نحصل علي معادلات مسار الهزاز التوافقي المعروفة.

\paragraph{
تمرين
$6$:
}
نأخذ المبدأ مركز القطع المكافي 
$y=ax^2$. 
احداثيات نقطة التعليق هي
$x$
و
$y=ax^2$.
احداثيات النواس هي اذن
\begin{eqnarray}
x_m=x+l\sin\theta~,~y_m=y-l\cos\theta.
\end{eqnarray}
الطاقة الحركية تعطي ب
\begin{eqnarray}
T&=&\frac{1}{2}m(\dot{x}_m^2+\dot{y}_m^2)+\frac{1}{2}M(\dot{x}^2+\dot{y}^2)\nonumber\\
&=&\frac{1}{2}(m+M)(1+4a^2x^2)\dot{x}^2+\frac{1}{2}ml^2\dot{\theta}^2+ml\dot{\theta}\dot{x}(\cos\theta+2ax\sin\theta)\nonumber\\
&=&\frac{1}{2}A(x)\dot{x}^2+\frac{1}{2}B\dot{\theta}^2+C(x,\theta)\dot{\theta}\dot{x}.
\end{eqnarray}
الطاقة الكامنة تعطي ب
\begin{eqnarray}
V=mg(-l\cos\theta+y)+Mg y.
\end{eqnarray}
نشتق الان كميات الحركة المعممة
\begin{eqnarray}
P_x=\frac{\partial T}{\partial \dot{x}}=A\dot{x}+C\dot{\theta}~,~P_{\theta}=\frac{\partial T}{\partial \dot{\theta}}=B\dot{\theta}+C\dot{x}.
\end{eqnarray}
قلب هذه العلاقات يعطي
\begin{eqnarray}
\dot{x}=\frac{1}{\Delta}(BP_x-CP_{\theta})~,~\dot{\theta}=\frac{1}{\Delta}(-CP_x+AP_{\theta}),
\end{eqnarray}
المحدد يعطي ب
\begin{eqnarray}
\Delta&=&AB-C^2\nonumber\\
&=&m^2l^2(\sin^2\theta+4a^2x^2\cos^2\theta-2ax\sin 2\theta)+Mml^2(1+4a^2x^2).
\end{eqnarray}
بالتعبير عن الطاقة الحركية بدلالة كميات الحركة المعممة نحصل علي
\begin{eqnarray}
T
&=&\frac{B}{2\Delta}P_{x}^2+\frac{A}{2\Delta}P_{\theta}^2-\frac{C}{\Delta}P_xP_{\theta}.
\end{eqnarray}
هاميلتونية الجملة تعطي ب
\begin{eqnarray}
H=\frac{B}{2\Delta}P_{x}^2+\frac{A}{2\Delta}P_{\theta}^2-\frac{C}{\Delta}P_xP_{\theta}+a(m+M)gx^2-mgl\cos\theta.
\end{eqnarray}
معادلات هاميلتون تعطي ب
\begin{eqnarray}
\dot{x}=\frac{\partial H}{\partial P_x}=\frac{B}{\Delta}P_x-\frac{C}{\Delta}P_{\theta}~,~\dot{\theta}=\frac{\partial H}{\partial P_{\theta}}=-\frac{C}{\Delta}P_x+\frac{A}{\Delta}P_{\theta}.
\end{eqnarray}
\begin{eqnarray}
-\dot{P}_x=\frac{\partial H}{\partial x}~,~-\dot{P}_{\theta}=\frac{\partial H}{\partial {\theta}}.
\end{eqnarray}
حساب هاتين المعادلتين الاخيرتين سهل لكن طويل جدا.

في التقريب التربيعي المعادلات تتبسط الي حد كبير. نجد في الاخير معادلات الحركة
\begin{eqnarray}
\ddot{x}+l\ddot{\theta}+g\theta=0~,~\ddot{x}+\frac{ml}{m+M}\ddot{\theta}+2agx=0.
\end{eqnarray}
الحل هو دوال اهتزازية بنفس التواتر الزاوي 
$\gamma$
اي
\begin{eqnarray}
x=A\exp(i\gamma t)~,~\theta=B\exp(i\gamma t).
\end{eqnarray}
بالتعويض في معادلات الحركة بهذا الاقتراح نحصل علي التواتر الزاوي
\begin{eqnarray}
\gamma^2=\frac{g(1+2al)\pm\sqrt{g^2(1+2al)^2-8al g^2\frac{M}{M+m}}}{2\frac{M}{M+m}l}.
\end{eqnarray}

\paragraph{
تمرين
$7$:
}
نحسب كميات الحركة المعممة
\begin{eqnarray}
p_1=\frac{\partial L}{\partial \dot{q}_1}=2\dot{q}_1+k_2\dot{q}_2~,~p_2=\frac{\partial L}{\partial \dot{q}_2}=k_2\dot{q}_1+\frac{2}{a+bq_1^2}\dot{q}_2.
\end{eqnarray}
قلب هذه المعادلات يعطي ب
\begin{eqnarray}
\dot{q}_1=\frac{1}{\Delta}(\frac{2}{a+bq_1^2}p_1-k_2p_2)~,~\dot{q}_2=\frac{1}{\Delta}(-k_2p_1+2p_2).
\end{eqnarray}
المحدد يعطي ب
\begin{eqnarray}
\Delta=\frac{4}{a+bq_1^2}-k_2^2.
\end{eqnarray}
نحسب
\begin{eqnarray}
\dot{q}_1^2+\frac{\dot{q}_2^2}{a+bq_1^2}+k_2\dot{q}_1\dot{q}_2=\frac{p_1^2}{\Delta (a+bq_1^2)}+\frac{p_2^2}{\Delta}-\frac{k_2p_1p_2}{\Delta}.
\end{eqnarray}
الهاميلتونية تعطي ب
\begin{eqnarray}
H&=&p_1\dot{q}_1+p_2\dot{q}_2-L\nonumber\\
&=&\frac{p_1^2}{\Delta (a+bq_1^2)}+\frac{p_2^2}{\Delta}-\frac{k_2p_1p_2}{\Delta}-k_1q_1^2.
\end{eqnarray}
معادلات هاميلتون للحركة

\begin{eqnarray}
\dot{q}_1&=&\frac{\partial H}{\partial p_1}=\frac{2p_1}{\Delta(a+bq_1^2)}-\frac{k_2p_2}{\Delta}.
\end{eqnarray}
\begin{eqnarray}
\dot{q}_2&=&\frac{\partial H}{\partial p_2}=\frac{2p_2}{\Delta}-\frac{k_2p_1}{\Delta}.
\end{eqnarray}
\begin{eqnarray}
-\dot{p}_1&=&\frac{\partial H}{\partial q_1}.
\end{eqnarray}

\begin{eqnarray}
-\dot{p}_2&=&\frac{\partial H}{\partial q_2}=0.
\end{eqnarray}
فقط المعادلة الثالثة تحتاج الي حساب طويل نوعا ما.

\paragraph{
تمرين
$8$:
}

نأخذ مبدأ الاحداثيات مركز الدائرة. احداثيات نقطة التعليق هي
\begin{eqnarray}
x_s=a\cos \gamma t~,~y_s=-a\sin \gamma t.
\end{eqnarray}
احداثيات الكتلة
$m$ 
هي
\begin{eqnarray}
x_m=a\cos \gamma t+l\sin\theta~,~y_m=-a\sin \gamma t+l\cos\theta.
\end{eqnarray}
نحسب الطاقة الحركية
\begin{eqnarray}
T=\frac{1}{2}ml^2\dot{\theta}^2+\frac{1}{2}ma^2\gamma^2+ma\gamma l\dot{\theta}\sin(\theta-\gamma t).
\end{eqnarray}
نحسب الطاقة الكامنة
\begin{eqnarray}
V=-mga\sin \gamma t+mgl\cos\theta.
\end{eqnarray}
لاغرانجية الجملة تعطي ب
\begin{eqnarray}
L=\frac{1}{2}ml^2\dot{\theta}^2+mal\gamma^2\sin(\theta-\gamma t)+mgl\cos\theta.
\end{eqnarray}
في هذه المعادلة الاخيرة اهملنا الحدود الثابتة و التي تتعلق بالزمن فقط و التي هي عبارة عن مشتقة تامة.

كمية الحركة المعممةتعطي ب

\begin{eqnarray}
p_{\theta}=\frac{\partial L}{\partial \dot{\theta}}=ml^2\dot{\theta}.
\end{eqnarray}
الهاميلتونية تعطي ب

\begin{eqnarray}
H&=&\dot{\theta}p_{\theta}-L\nonumber\\
&=&\frac{p_{\theta}^2}{2ml^2}-mal\gamma^2\sin(\theta-\gamma t)-mgl\cos\theta.
\end{eqnarray}
معادلات هاميلتون للحركة
\begin{eqnarray}
\dot{\theta}=\frac{\partial H}{\partial p_{\theta}}=\frac{p_{\theta}}{ml^2}.
\end{eqnarray}
\begin{eqnarray}
-\dot{p}_{\theta}=\frac{\partial H}{\partial {\theta}}=-mal\gamma^2\cos(\theta-\gamma t)+mgl\sin\theta.
\end{eqnarray}

\paragraph{
تمرين
$9$:
}
كمية الحركة المعممة المرفقة ب
$q$
تعطي ب
\begin{eqnarray}
p=\frac{\partial L}{\partial \dot{q}}=\frac{m}{2}(2\dot{q}\sin^2\Omega t+q\Omega\sin 2\Omega t).
\end{eqnarray}
القلب هنا بسيط لانه لدينا متغير واحد. الهاميلتونية اذن تعطي ب
\begin{eqnarray}
h&=&p\dot{q}-L\nonumber\\
&=&\frac{m}{2}\bigg[\frac{1}{m^2\sin^2\Omega t}\big(p-\frac{m\Omega q\sin 2\Omega t}{2}\big)^2-q^2\Omega^2\bigg].
\end{eqnarray}
هذه الهاميلتونية غير منحفظة.

اللاغرانجية بدلالة المتغير
$Q$
تعطي ب
\begin{eqnarray}
L=\frac{m}{2}\big(\dot{Q}^2+\Omega^2Q^2\big).
\end{eqnarray}
هذه هاميلتونية هزاز توافقي. كمية الحركة المعممة المرفقة بالمتغير
$Q$
تعطي ب
\begin{eqnarray}
P=\frac{\partial L}{\partial \dot{Q}}=m\dot{Q}=\frac{p}{\sin\Omega t}.
\end{eqnarray}
الهاميلتونية في هذه الحالة منحفظة تعطي ب
\begin{eqnarray}
H=\frac{m}{2}\big(\frac{P^2}{m^2}-\Omega^2Q^2\big).
\end{eqnarray}
\paragraph{
تمرين
$10$:
}

نعرف ان 
$q$
و
$p$
تحققان معادلات هاميلتون. 
نحسب  اذن اولا
\begin{eqnarray}
\dot{q}=\frac{\partial H}{\partial p}&=&\frac{\partial P}{\partial p}\frac{\partial H}{\partial P}+\frac{\partial Q}{\partial p}\frac{\partial H}{\partial Q}\nonumber\\
&=&-\frac{q}{\sin^2 p}\frac{\partial H}{\partial P}+\cot p\frac{\partial H}{\partial Q}.
\end{eqnarray}
\begin{eqnarray}
-\dot{p}=\frac{\partial H}{\partial q}&=&\frac{\partial P}{\partial q}\frac{\partial H}{\partial P}+\frac{\partial Q}{\partial q}\frac{\partial H}{\partial Q}\nonumber\\
&=&\cot p \frac{\partial H}{\partial P}-\frac{1}{q} \frac{\partial H}{\partial Q}.
\end{eqnarray}
قلب هذه المعادلات نحصل علي
\begin{eqnarray}
\frac{\partial H}{\partial P}&=&-\frac{\dot{q}}{q}+\dot{p}\cot p.
\end{eqnarray}
\begin{eqnarray}
\frac{\partial H}{\partial Q}&=&-\dot{q}\cot p +\frac{q}{\sin^2 p}\dot{p}.
\end{eqnarray}
من الجهة الاخري فاننا نتوقع ان يكون التحويل القانوني محدود و بالتالي 
$K=H$. 
اذن نحسب من الجهة الاخري
\begin{eqnarray}
\frac{\partial H}{\partial P}=\dot{Q}&=&\frac{\partial Q}{\partial q}\dot{q}+\frac{\partial Q}{\partial p}\dot{p}.
\end{eqnarray}
\begin{eqnarray}
\frac{\partial H}{\partial Q}=-\dot{P}&=&-\frac{\partial P}{\partial q}\dot{q}-\frac{\partial P}{\partial p}\dot{p}.
\end{eqnarray}
بالمقارنة نجد ان المعادلات تتطابق.
\paragraph{
تمرين
$11$:
}
من الواضح ان التحويل القانوني محدود اي ان الدالة المولدة لا تتعلق بالزمن و
$K=H$. 
الدالة المولدة يجب ان تحقق
\begin{eqnarray}
p_i\dot{q}_i-H=P_i\dot{Q}_i-K+\frac{dF}{dt}.
\end{eqnarray}
نأخذ الدالة المولدة من النوع الاول
اي
$F=F_1(q_i,Q_i)$
لكن نبنيها علي
 شكل تحويل لوجوندر لدالة
 $F_{13}$
 تتعلق ب
 $p_1$, $q_2$
 و
 $Q_i$
 اي
 
\begin{eqnarray}
F=F_1(q_i,Q_i)=q_1p_1+F_{13}(p_1,q_2,Q_i).
\end{eqnarray}
باستعمال التحويل القانوني يمكن كتابة الشرط اعلاه كما يلي
\begin{eqnarray}
p_2\dot{q}_2=(p_1-2p_2)\dot{q}_1+(-2q_1-q_2)\dot{p}_2+\frac{\partial F_{13}}{\partial p_1}\dot{p}_1+\frac{\partial F_{13}}{\partial Q_1}\dot{q}_1+\frac{\partial F_{13}}{\partial q_2}\dot{q}_2+
\frac{\partial F_{13}}{\partial Q_2}\dot{p}_2+q_1\dot{p}_1.\nonumber\\
\end{eqnarray}
يجب ان يكون لدينا
\begin{eqnarray}
\frac{\partial F_{13}}{\partial p_1}=-q_1=-Q_1.
\end{eqnarray}
\begin{eqnarray}
\frac{\partial F_{13}}{\partial q_2}=p_2=Q_2.
\end{eqnarray}
\begin{eqnarray}
\frac{\partial F_{13}}{\partial Q_1}=-p_1+2p_2=-p_1+2Q_2.
\end{eqnarray}
\begin{eqnarray}
\frac{\partial F_{13}}{\partial Q_2}=2q_1+q_2=2Q_1+q_2.
\end{eqnarray}
حل المعادلات التفاضلية الاربعة اعلاه يعطي ب
\begin{eqnarray}
F_{13}=2Q_1Q_2+q_2Q_2-p_1Q_1.
\end{eqnarray}

\paragraph{
تمرين
$12$:
}
\begin{itemize}
\item
واضح.
\item 
موضع و سرعة و طاقة حركة الكتلة
$m$
تعطي ب
\begin{eqnarray}
\vec{r}=(x+l\sin\theta)\vec{i}-l\cos\theta \vec{j}.
\end{eqnarray}
\begin{eqnarray}
\vec{v}=(\dot{x}+l\dot{\theta}\cos\theta)\vec{i}+l\dot{\theta}\sin\theta \vec{j}.
\end{eqnarray}
\begin{eqnarray}
T=\frac{1}{2}m(\dot{x}^2+l^2\dot{\theta}^2+2l\dot{x}\dot{\theta}\cos\theta).
\end{eqnarray}
طاقة حركة الكتلة
$m^{'}$
تعطي ب
\begin{eqnarray}
T=\frac{1}{2}m^{'}\dot{x}^2.
\end{eqnarray}
لاغرانجية الجملة
\begin{eqnarray}
L=\frac{1}{2}(m+m^{'})\dot{x}^2+\frac{1}{2}m(l^2\dot{\theta}^2+2l\dot{x}\dot{\theta}\cos\theta)+mgl\cos\theta.
\end{eqnarray}
الحصول علي معادلات لاغرانج امر بسيط.
\item
اشتقاق هاميلتونية الجملة و معادلات هاميلتون للحركة امر طويل نسبيا لكن يبقي بسيط.
\end{itemize}

\renewcommand\thefigure{\thepart.\arabic{figure}}    
\setcounter{figure}{0}   

\newpage
{\selectlanguage{english}

\begin{figure}[htbp]
\begin{center}
\includegraphics[width=12.0cm,angle=-0]{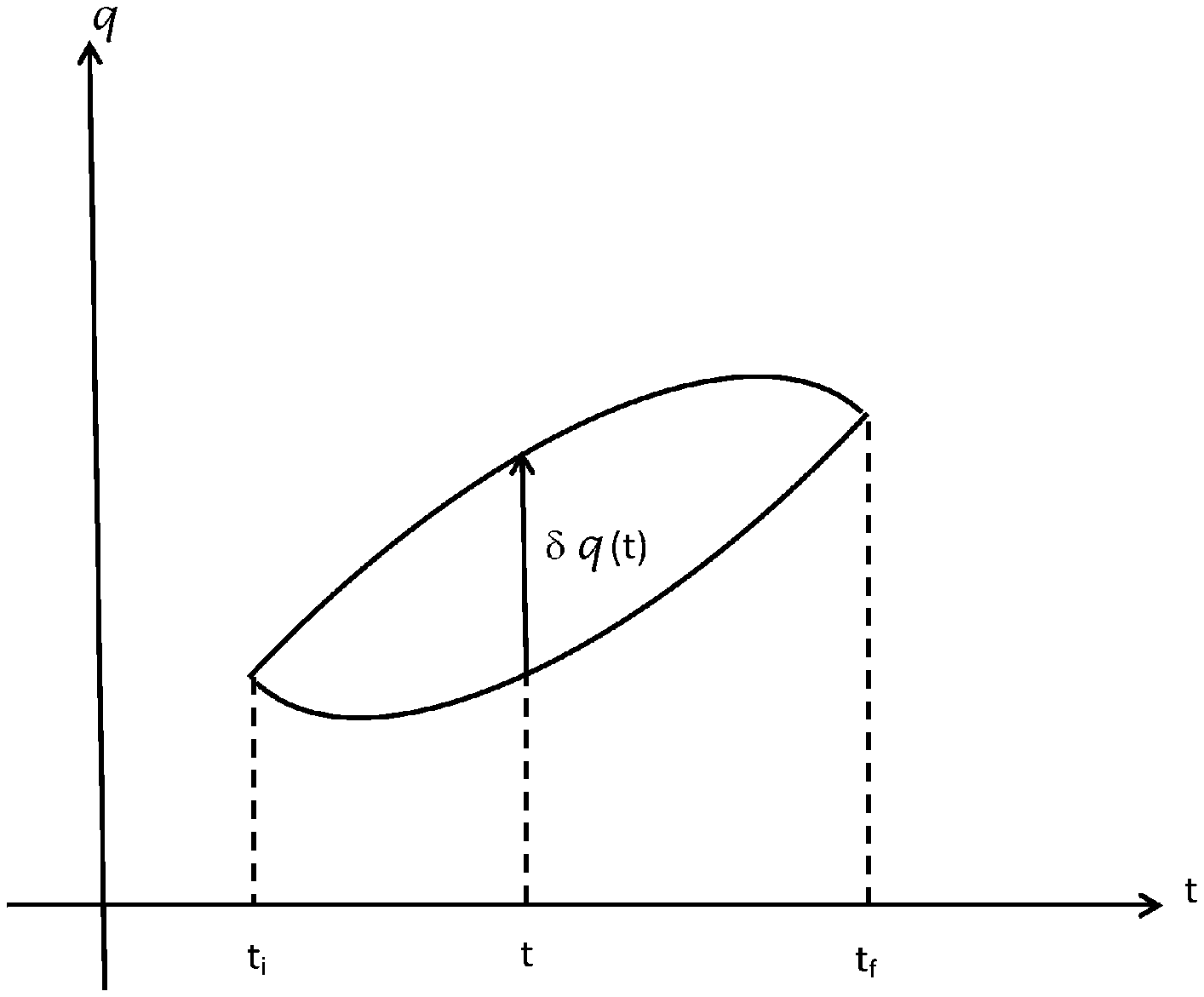}
\end{center}
\caption{
}
\end{figure}

\begin{figure}[htbp]
\begin{center}
\includegraphics[width=10.0cm,angle=-0]{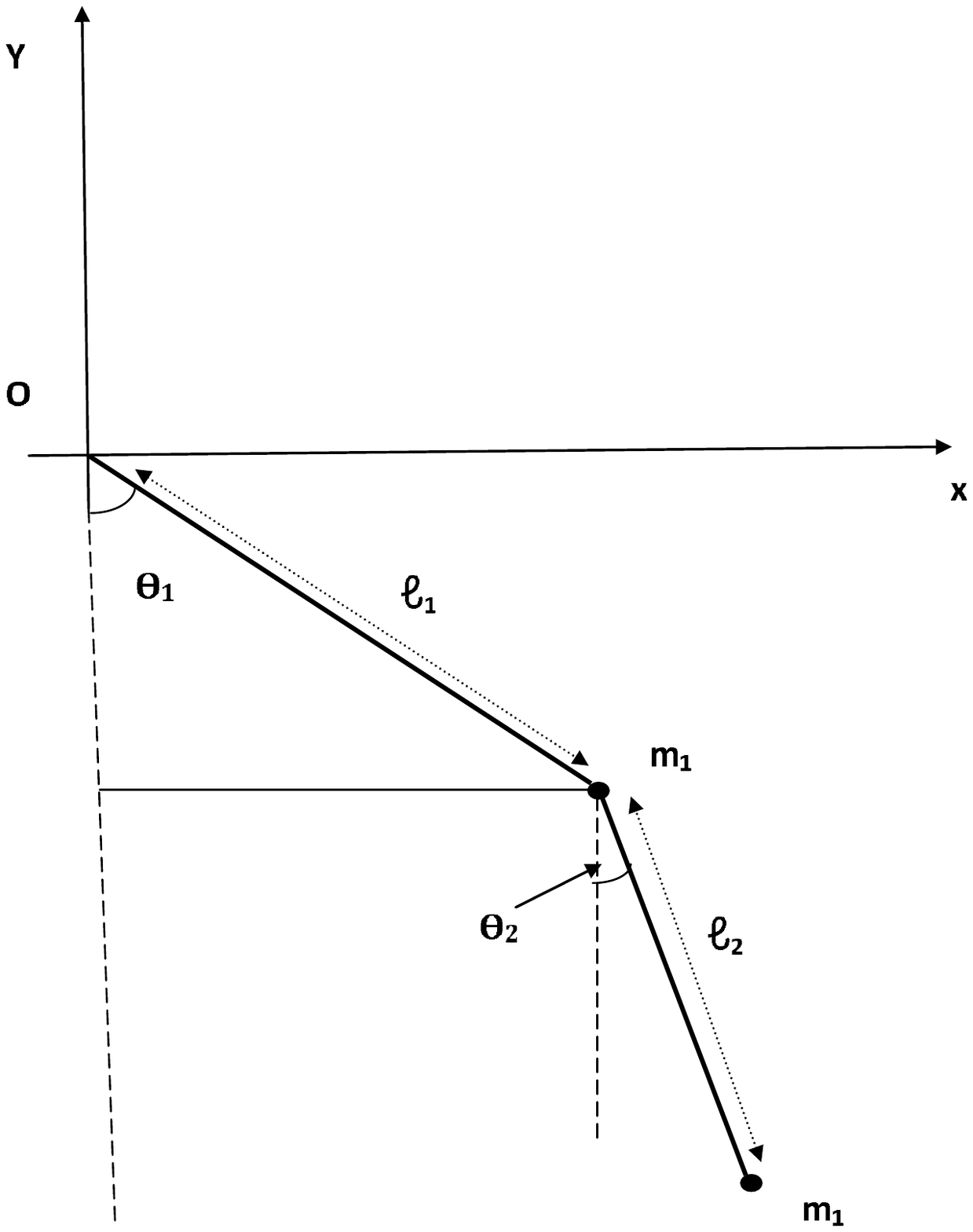}
\end{center}
\caption{}
\end{figure}

\begin{figure}[htbp]
\begin{center}
\includegraphics[width=10.0cm,angle=-0]{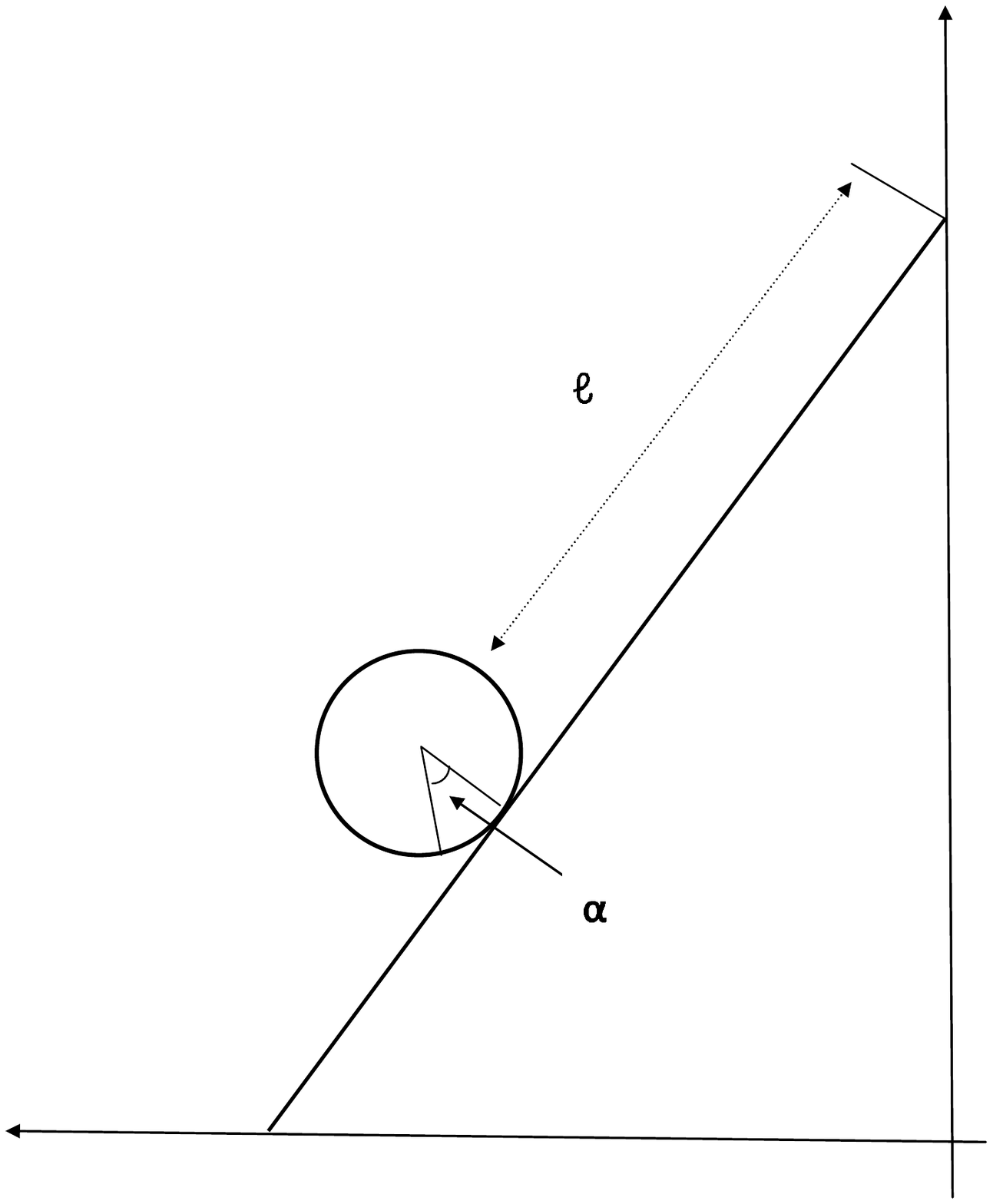}
\end{center}
\caption{}
\end{figure}

\begin{figure}[htbp]
\begin{center}
\includegraphics[width=11.0cm,angle=-0]{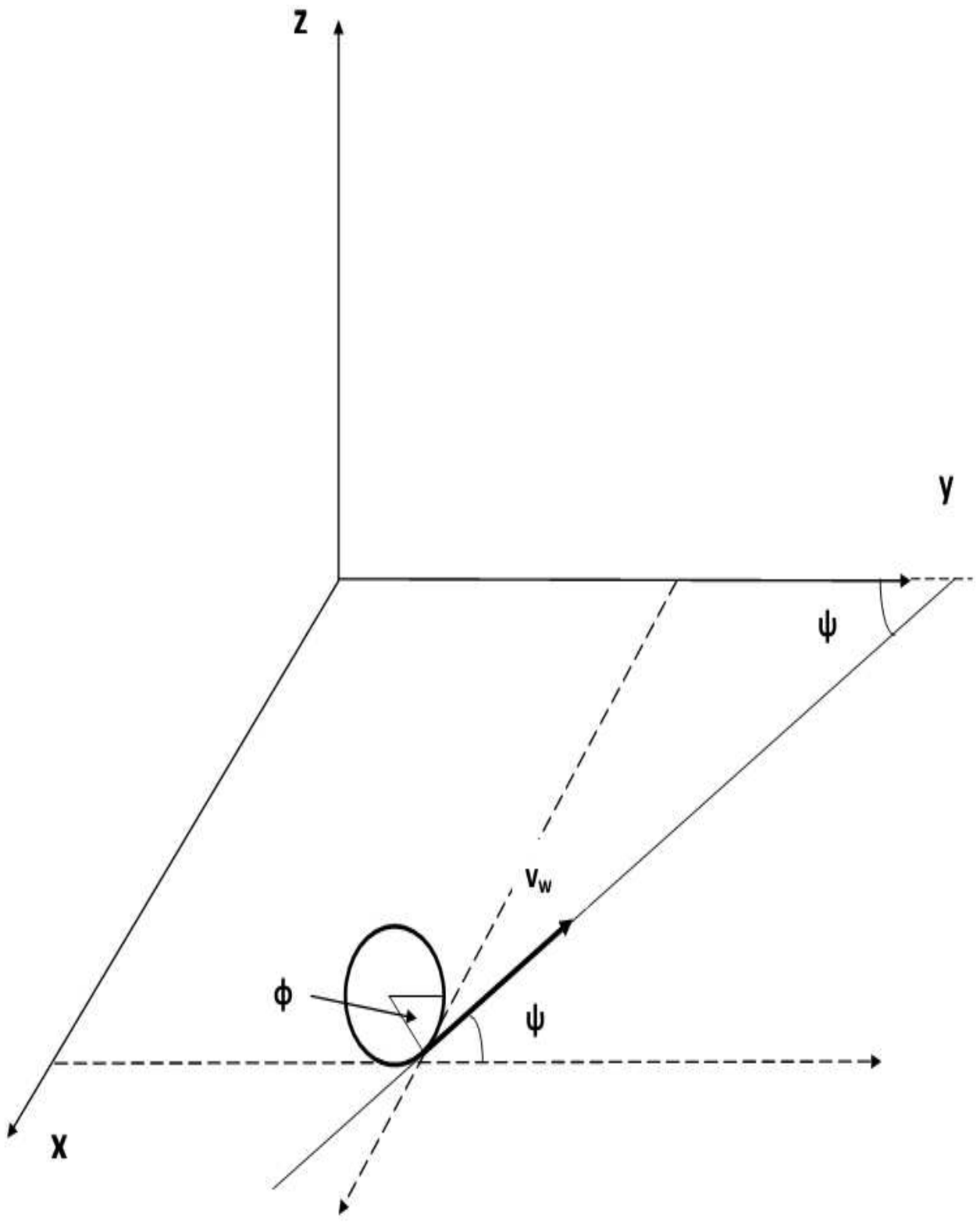}
\end{center}
\caption{}
\end{figure}

\begin{figure}[htbp]
\begin{center}
\includegraphics[width=10.0cm,angle=-0]{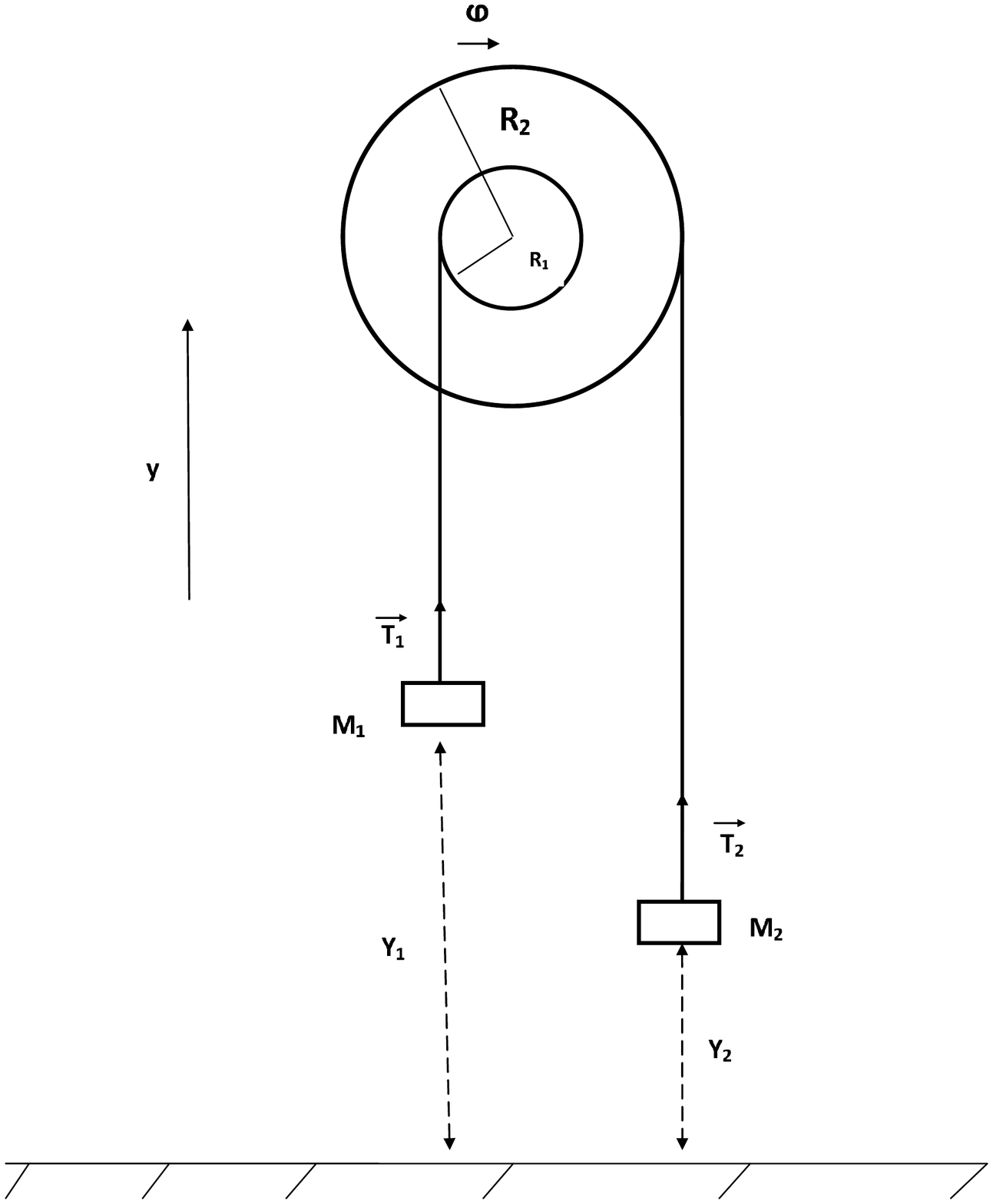}
\end{center}
\caption{}
\end{figure}

\begin{figure}[htbp]
\begin{center}
\includegraphics[width=10.0cm,angle=-0]{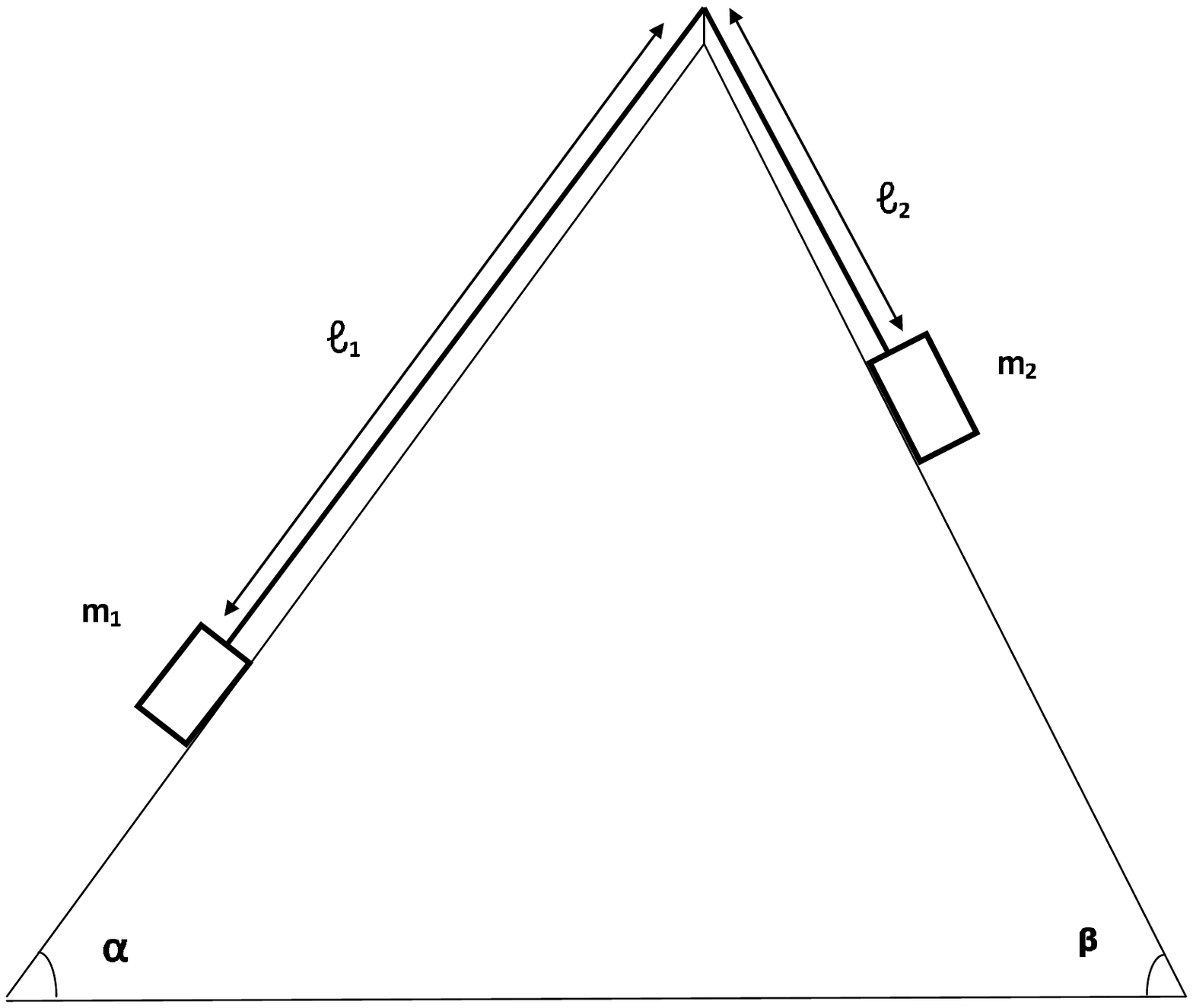}
\end{center}
\caption{}
\end{figure}

\begin{figure}[htbp]
\begin{center}
\includegraphics[width=15.0cm,angle=-0]{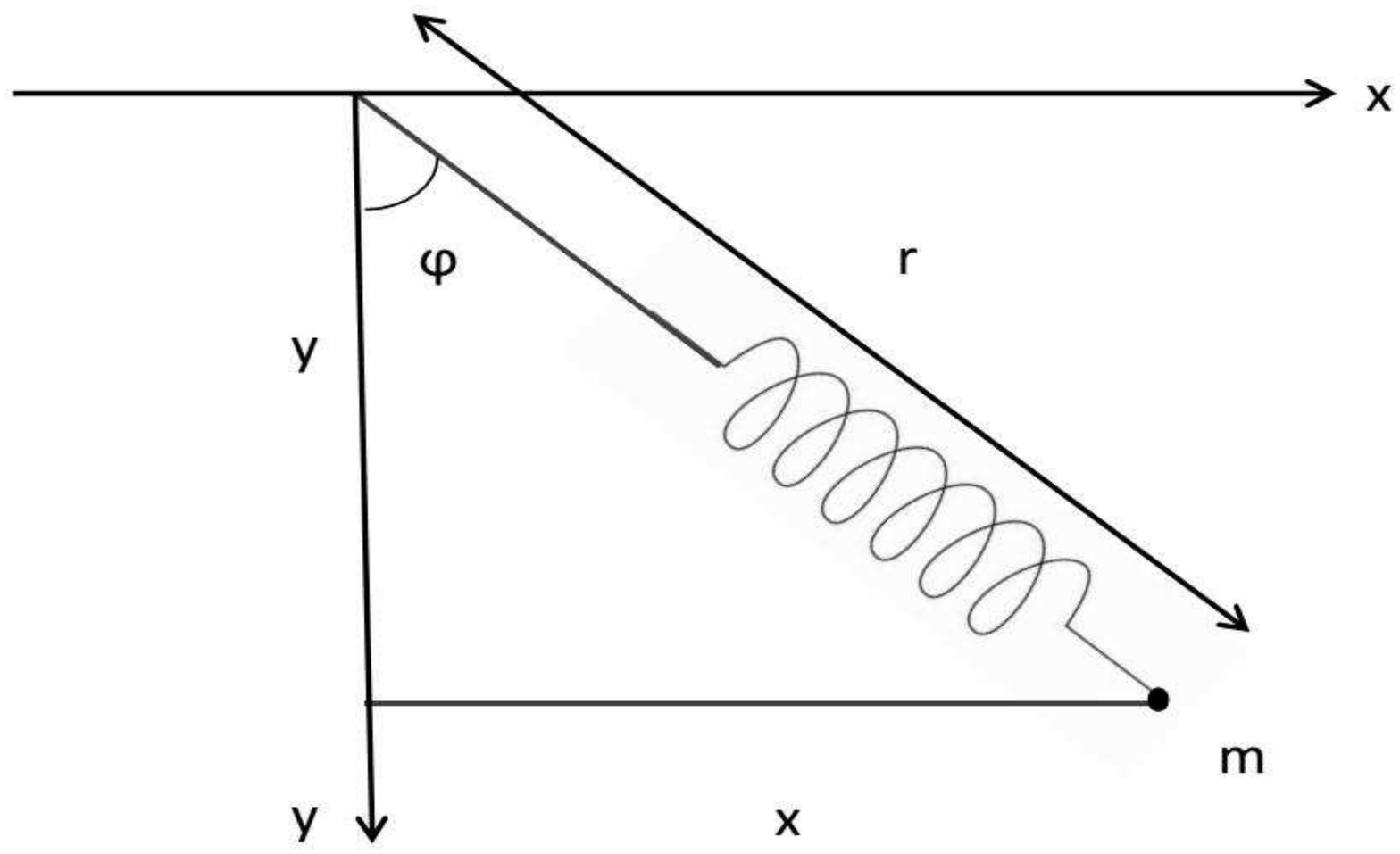}
\end{center}
\caption{}
\end{figure}

\begin{figure}[htbp]
\begin{center}
\includegraphics[width=10.0cm,angle=-0]{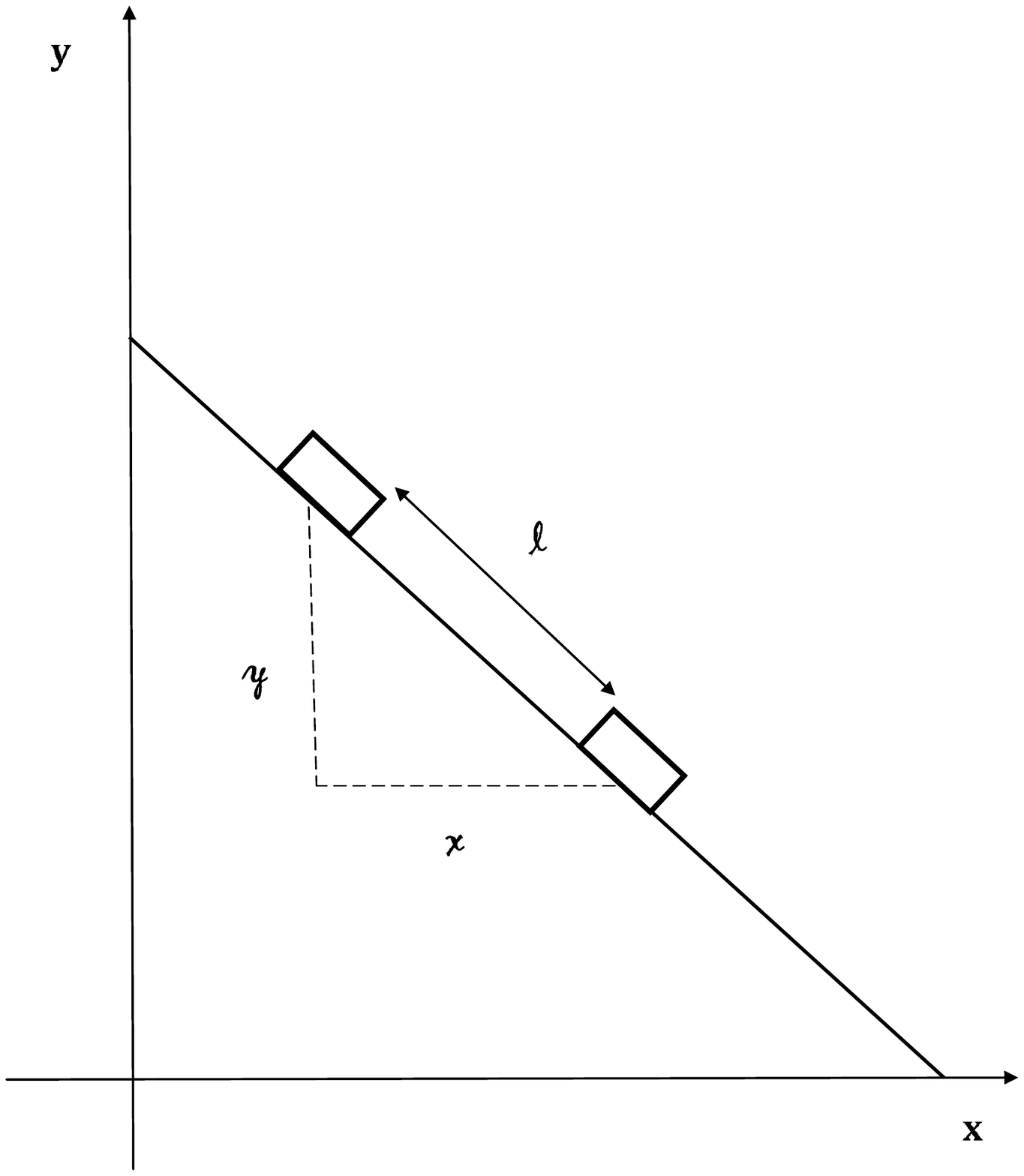}
\end{center}
\caption{}
\end{figure}

\begin{figure}[htbp]
\begin{center}
\includegraphics[width=12.0cm,angle=-0]{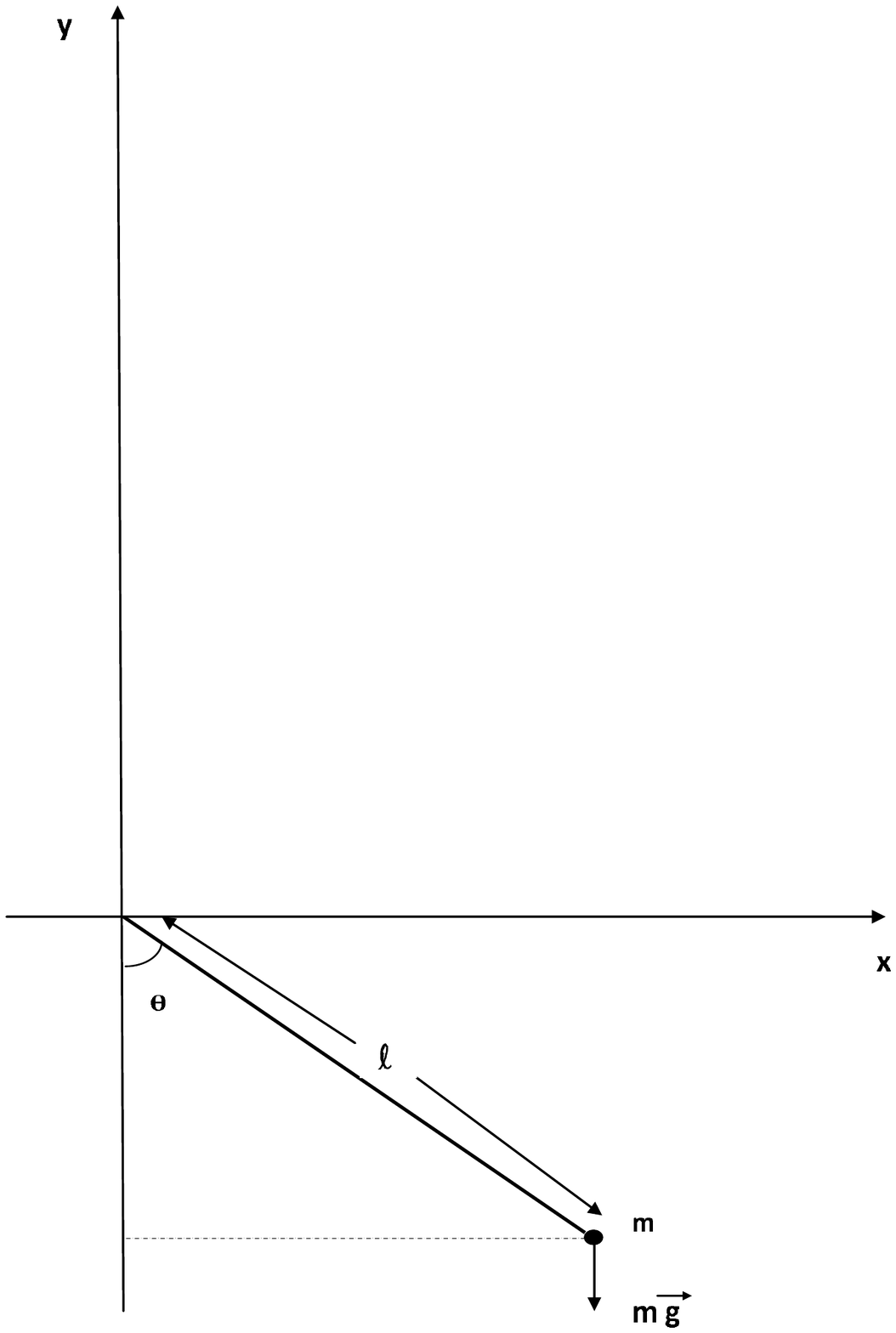}
\end{center}
\caption{}
\end{figure}

\begin{figure}[htbp]
\begin{center}
\includegraphics[width=11.0cm,angle=-0]{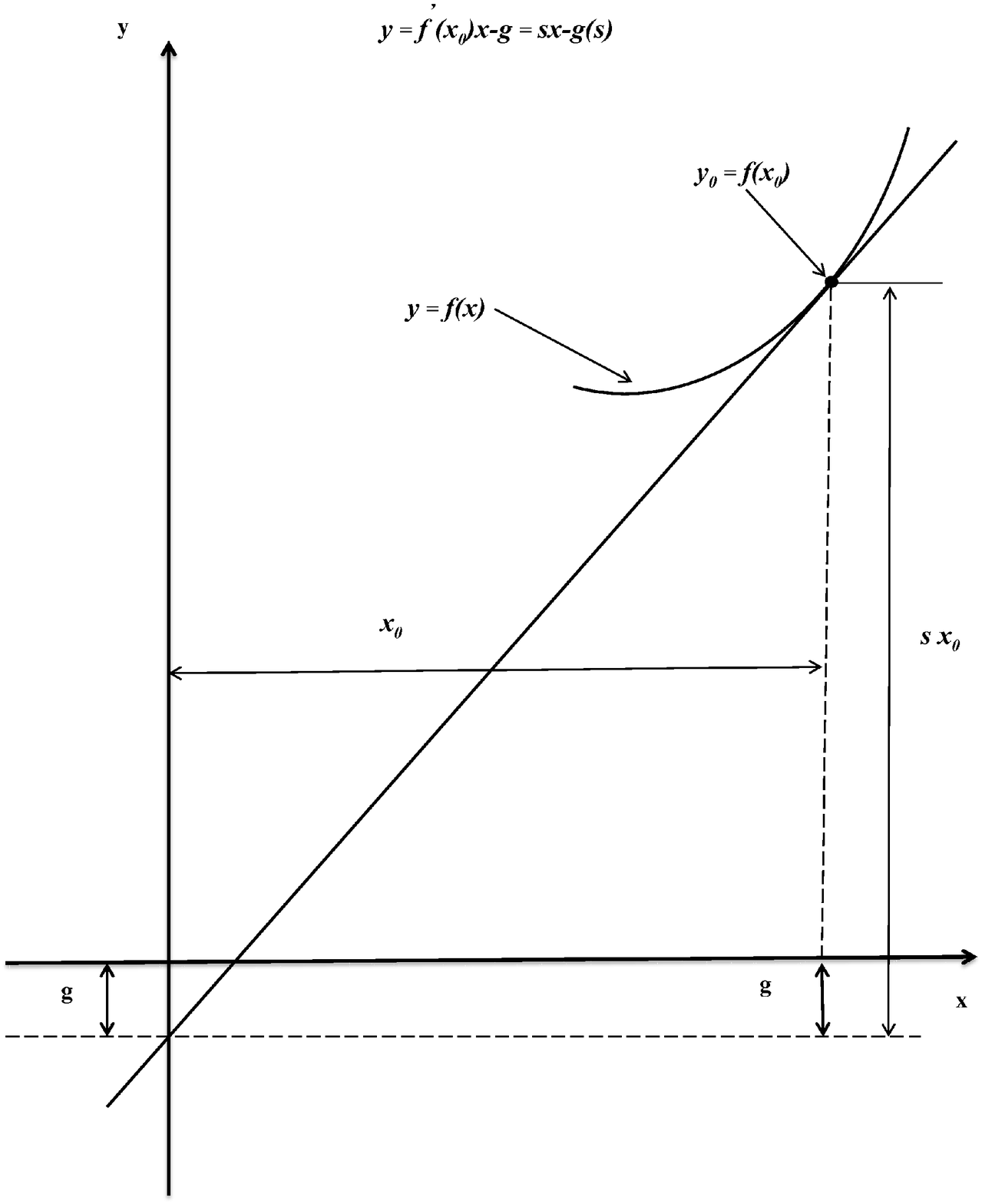}
\end{center}
\caption{}
\end{figure}

\begin{figure}[htbp]
\begin{center}
\includegraphics[width=15.0cm,angle=-0]{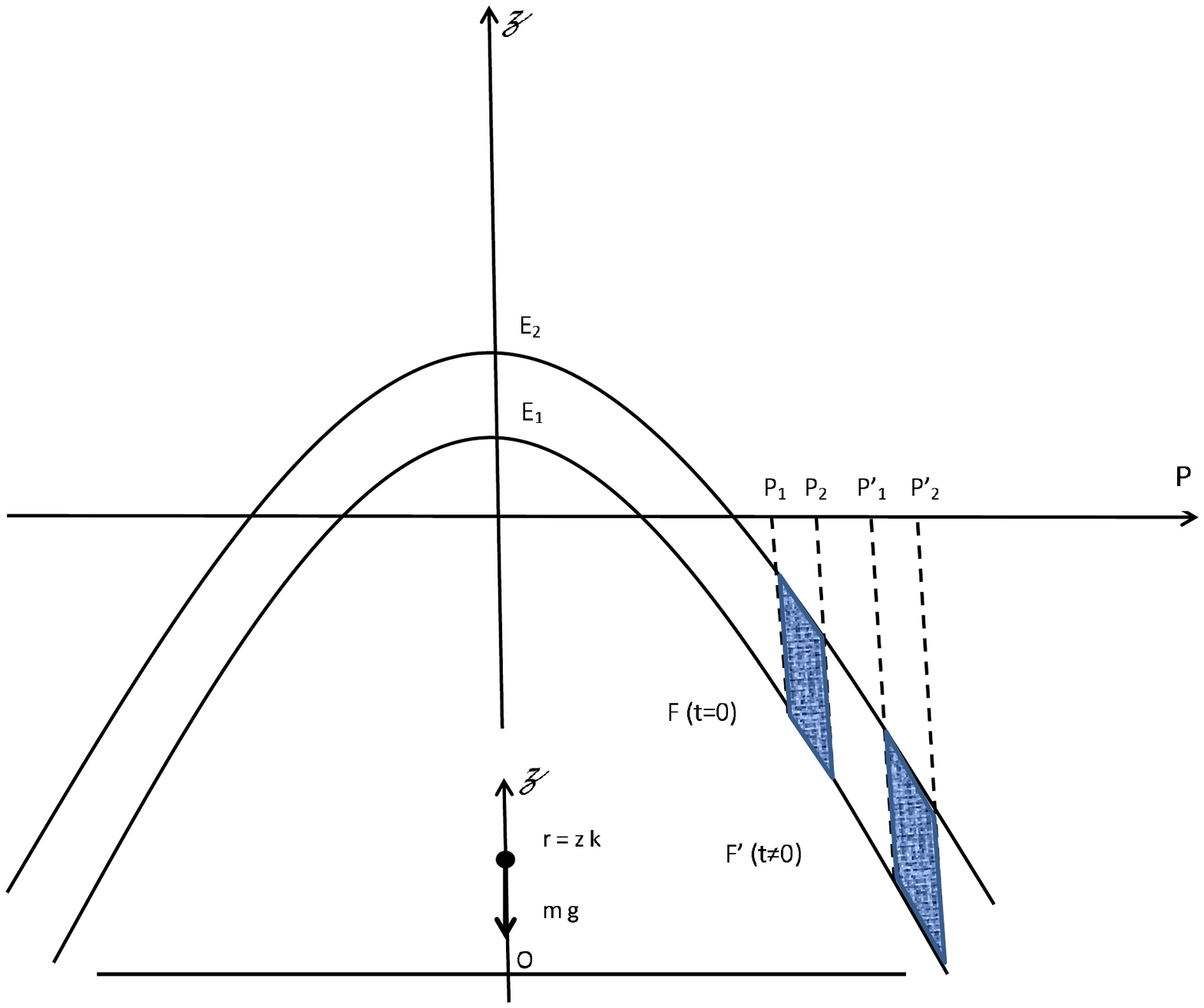}
\end{center}
\caption{}
\end{figure}

\begin{figure}[htbp]
\begin{center}
\includegraphics[width=11.0cm,angle=-0]{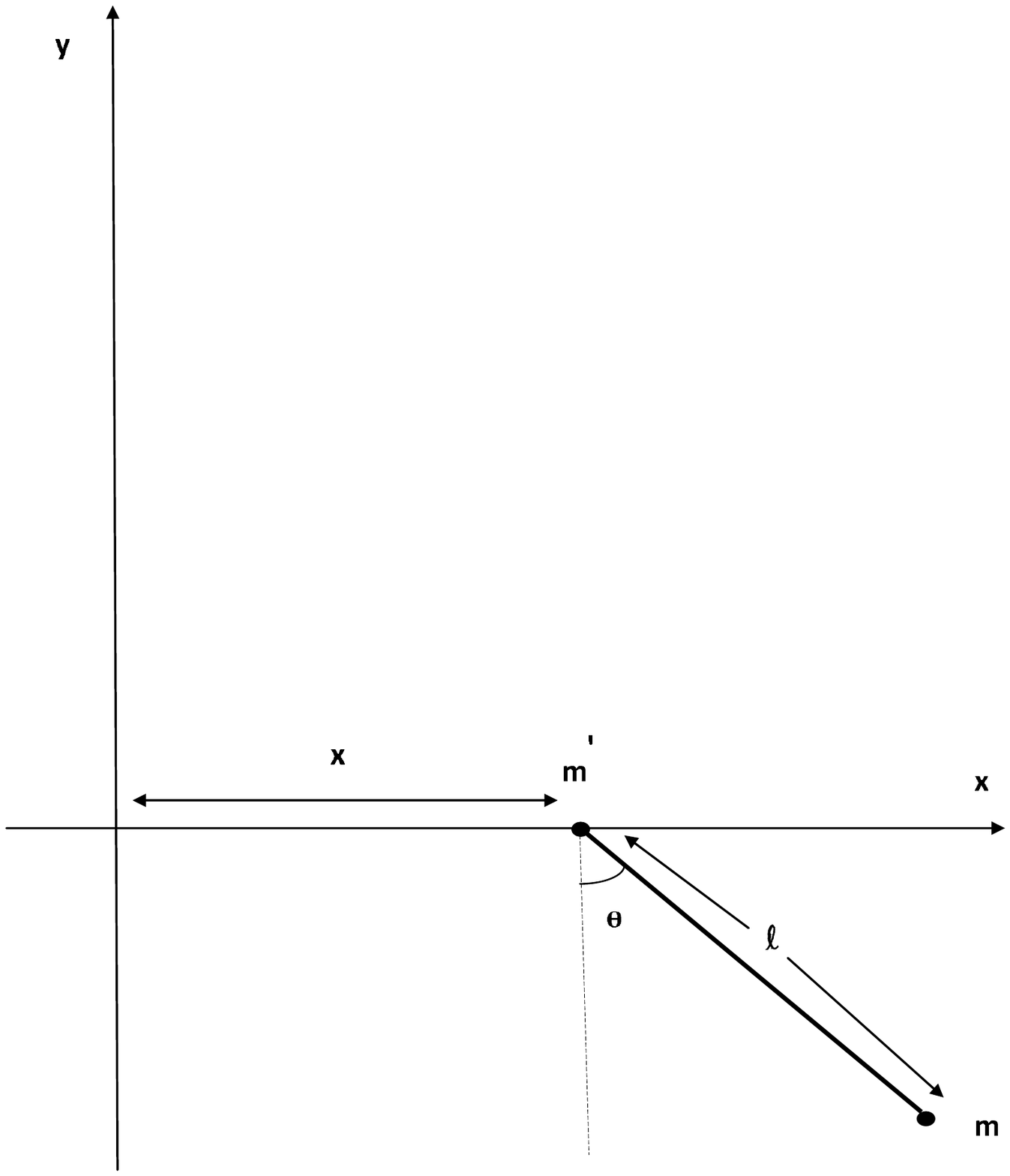}
\end{center}
\caption{}
\end{figure}

\begin{figure}[htbp]
\begin{center}
\includegraphics[width=11.0cm,angle=-0]{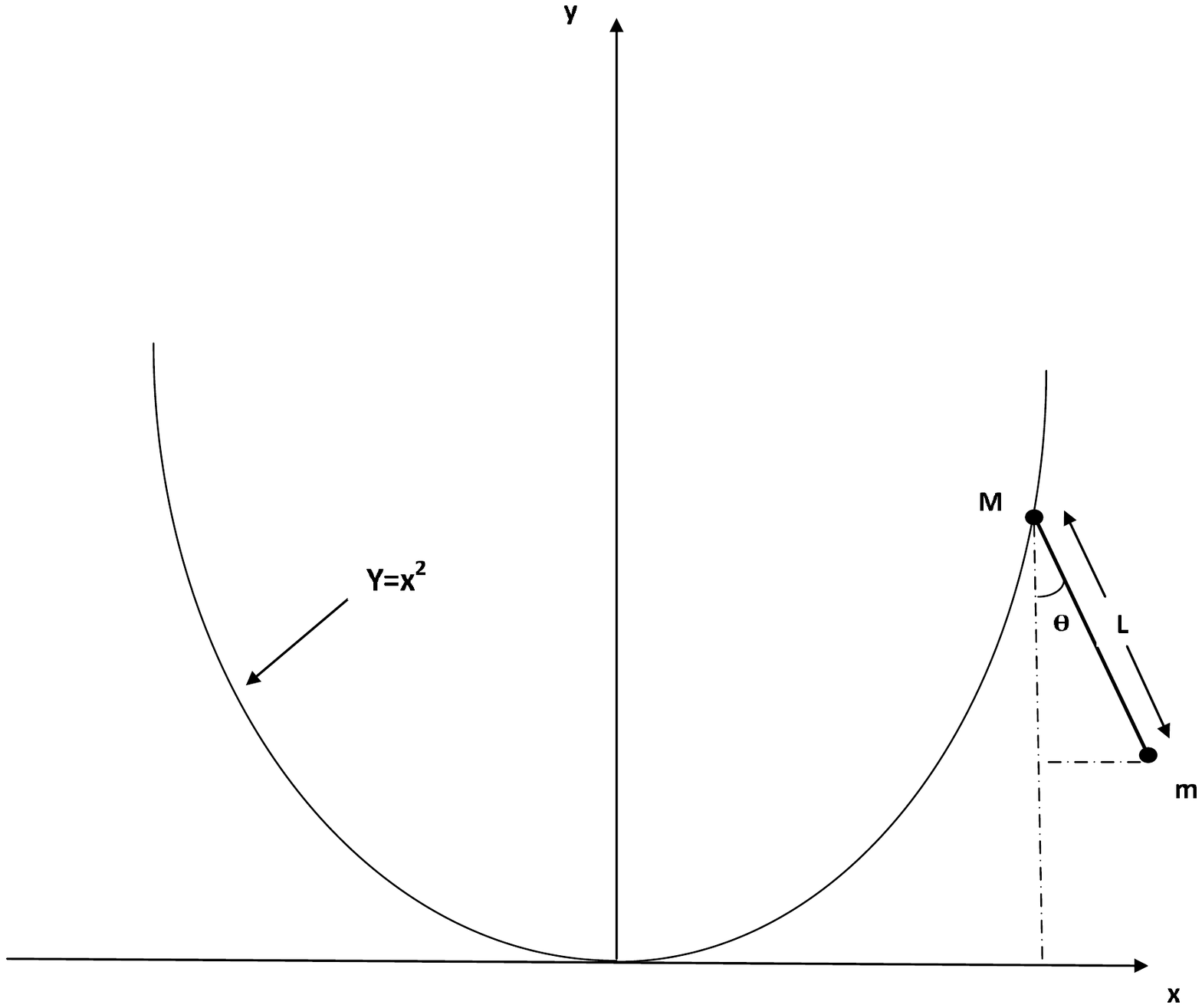}
\end{center}
\caption{}
\end{figure}

}

\part*{{\fontsize{50}{50}\selectfont 
الترموديناميك و الميكانيك الاحصائي
}}
\addcontentsline{toc}{part}{$II$
 الترموديناميك و الميكانيك الاحصائي
} 

{\selectlanguage{english}

\chapter*{
Thermodynamics and Statistical Mechanics
}
This part is based primarily on the books \cite{Huang,Ngo}. Huang is very elegant and precise while Ngo and Ngo (in french) is very pedagogical. Again, only fundamental topics are 
discussed: $1)$ Thermodynamics, and $2)$ 
The micocanonical ensemble. In the first chapter, we have concentrated on the discussion of fundamentals of thermodynamics, like the first and second principles and the
thermoynamical state functions, and avoided discussion of applications due to lack of time. In the second chapter, we have given an introduction to statistical mechanics, by providing 
a systematic exposition of the micocanonical ensemble, and its application to the ideal gaz.  See chapters $1$ and $6$ of Huang, and chapters $1$-$5$ of Ngo and Ngo. The treatise of Landau \cite{Landaus} was also consulted extensively. 
Most exercises are taken from \cite{Ngo}, which offers also pedagogical solutions, and some exercises were 
taken from \cite{Huang}. 
}

\chapter*{
مقدمة في الترموديناميك
}
\addcontentsline{toc}{chapter}{$1$
مقدمة في الترموديناميك
} 

\section*{
مقدمة
}
\addcontentsline{toc}{section}{
مقدمة 
} 

الترموديناميك هو فرع من الفيزياء يهتم بوصف المادة ظواهريا
\footnote{.${\rm phenomenological}$} 
وبالخصوص يهتم بدراسة خصائص الجمل الفيزيائية الماكروسكوبية المتوازنة حراريا و دراسة التبادلات في الطاقة التي تحدث بينها اي بين شكلي الطاقة الاساسيين : كمية الحرارة و العمل الميكانيكي. 

الترموديناميك ناجح جدا في وصف 
العالم الماكرسكوبي دون الرجوع الي المبادي الاولية
\footnote{.${\rm first}~{\rm principles}$}
للديناميك الجزيئي الاستثناء ربما هي النظرية الحركية للغازات التي يمكن اشتقاق  الترموديناميك الخاص  بها مباشرة من النظرية الذرية. الميكانيك الاحصائي هو النظرية الاساسية التي تحاول بناء الترموديناميك انطلاقا من الفيزياء التي تحكم العالم الجزيئي و الذري وبالتالي فهي تعتبر اصل الترموديناميك و تفسيره في ان معا.

الميكانيك الاحصائي اذن هو همزة الوصل بين العالم الميكروسكوبي و العالم الماكروسكوبي و هو بالضبط احصائي لان عدد الجسيمات الميكروسكوبية المشكلة لاي جملة فيزيائية هو عدد كبير جدا. وجود هذا العدد الكبير من الجسيمات هو الاصل الذي يسمح لنا باشتقاق القوانين البسيطة والانيقة للترموديناميك, التي تصف خصائص الجمل الفيزيائية الماكروسكوبية المتوازنة, ابتداءا من قوانين الميكانيك الاحصائي.

 نبدأ  في هذا الفصل دراستنا بالترموديناميك, و نؤجل الميكانيك الاحصائي للفصول القادمة, حتي نعرف عن قرب اللغة و العالم اللذان يتعامل بها و معه الميكانيك الاحصائي قبل الغوص في الرياضيات التي تحكمه و تؤسسه.

 \section*{
تعاريف عامة
}
\addcontentsline{toc}{section}{
تعاريف عامة
} 

\paragraph{
الجمل الترموديناميكية
:}

  الجملة الترموديناميكية هي جزء ماكروسكوبي من الكون محدود بسطح مغلق قد يكون اعتباري. هذه الجملة قد تكون معزولة او مغلوقة اذا لم يكن هناك اي تفاعلات لها مع الوسط الخارجي امااذا كان هناك تفاعل
مع الوسط الخارجي فالجملة تسمي مفتوحة. 

الجملة الماكروسكوبية 
\footnote{.${\rm macroscopic}$}
هي جملة ذات ابعاد كبيرة جدا بالمقارنة مع ابعاد مكوناتها الجزيئية, الذرية, النووية او الاولية. عموما عدد هذه المكونات الميكروسكوبية
\footnote{.${\rm microscopic}$}
هو عدد هائل مثلا من نفس رتبة عدد
افوقادرو
\footnote{.$َ{\rm Avogadro}$}
\begin{eqnarray}
 {\cal N}=6.021{\rm x} 10^{23}.
\end{eqnarray}
كما اشرنا اليه اعلاه فان الترموديناميك علي خلاف الميكانيك الاحصائي لا يهتم بهذه المكونات الميكروسكوبية للجملة الترموديناميكية و يحاول فقط وصف الجملة الماكروسكوبية ياستعمال متغيرات
ماكروسكوبية مثل:
الحجم
$V$, 
الضغط 
$P$,
درجة الحرارة
$T$
و عدد الجسيمات
$N$.

\paragraph{
الجمل المتجانسة
:}
الجملة الترموديناميكية تسمي متجانسة اذا كان تركيز مكوناتها لا يتغير من نقطة الي اخري في الجملة. اي ان الجملة المتجانسة تتكون من حالة طورية واحدة او  حالة طورية منتظمة. في الحالة العكسية فان الجملة تسمي غير متجانسة.

\paragraph{
المقادير التكثيفية %
 و المقادير التمديدية
:}
نعتبر جملتين ترموديناميكيتين متماثلتين
$1$
و
$2$. 
نفترض ان كل جملة هي جملة 
متجانسة محتواة داخل حجم
$V$
و مشكلة
من 
$r$
مركبة عدد مولاتها 
$n_1$, $n_2$,....,$n_r$ 
علي التوالي. الجملة المشكلة من اضافة الجملتين اعلاه الي بعضهما البعض هي ايضا جملة متجانسة حجمها 
 $2V$
 و مشكلة من نفس المركبات 
مع عدد مولات يساوي
$2n_1$,$2n_2$,....,$2n_r$ 
علي التوالي. ليكن
$X$
احد المقادير الفيزيائية الذي يأخذ القيمة 
$X_0$ 
في الجملتين 
$1$
و
$2$
 كل علي حدة. 
 
 اذا كان المتغير
 $X$ 
 يأخذ في الجملة الكلية
 $1+2$
 نفس القيمة
 $X_0$
 فاننا نقول ان
 $X$ 
هو مقدار تكثيفي
 \footnote{.${\rm intensive}$}.
 مثال:
 الضغط و درجة الحرارة هي مقادير تكثيفية. 
اما اذا كان المتغير
 $X$ 
 يأخذ في الجملة الكلية
 $1+2$
 القيمة
 $2X_0$
 فاننا نقول ان
 $X$ 
هو مقدار تمديدي
 \footnote{.${\rm extensive}$}.
 مثال:
 الحجم و عدد الجسيمات هي مقادير تمديدية.
 
 اذن المقادير التمديدية هي المقادير التي تتناسب مع كمية المادة في الجملة اما المقادير التكثيفية فهي المقادير التي لا تتعلق
 بكمية المادة في الجملة.

\paragraph{
الحالات الماكروسكوبية
:
}
تعين حالة الجملة الماكروسكوبية ترموديناميكا بالكامل باعطاء قيم محددة للمتغيرات الماكروسكوبية
$V$, $P$, $T$, $N$.
بالتالي فان الحالة الماكروسكوبية هي صورة معينة للجملة مرفقة بقيمة محددة للمتغيرات الماكروسكوبية.

\paragraph{
 التوازن الترموديناميكي
:}

تكون الجملة الفيزيائية الماكروسكوبية في حالة توازن ترموديناميكي اذا كانت المتغيرات الماكروسكوبية الواصفة لهذه الجملة
لا تتغير مع الزمن. في الحالة العكسية تكون الجملة غير متوازنة. كل جملة غير متوازنة تتطور في الزمن بصورة او باخري الي ان تبلغ حالة توازنها. الزمن الذي تستغرقه الجملة حتي تبلغ التوازن يسمي زمن
الاسترخاء.

\paragraph{
معادلة الحالة
:}
معادلة حالة الجملة الترموديناميكية هي علاقة تربط بين متغيراتها الترموديناميكية في حالة توازنها. مثلا اذا كانت المتغيرات الترموديناميكية لجملة ما هي
$V$, $P$, $T$
فان معادلة الحالة هي علاقة من الشكل
\begin{eqnarray}
f(P,V,T)=0.
\end{eqnarray}
اذن هناك متغيران فقط مستقلان خطيا من بين الثلاثة. اذا مثلنا حالة الجملة بنقطة في الفضاء الثلاثي
$P-V-T$ 
فان معادلة الحالة تعرف سطح في هذا الفضاء معطي بالضبط ب
$f=0$
 حيث كل نقطة منه هي عبارة عن حالة توازن ممكنة للجملة.
 
 اسقاط سطح معادلة الحالة علي المستوي
 $P-V$
 يعطي ما يعرف بالمخطط
 $P-V$. 
 كل نقطة من هذا المخطط تمثل حالة توازن.
 
\paragraph{
الغاز المثالي:
}
نأخذ هنا مسألة الغاز المثالي كمثال.
من الناحية التجربيبة كل الغازات الحقيقية تتصرف بنفس الشكل عندما تكون مميهة
\footnote{.${\rm diluted}$}.
هذا التصرف الكوني\footnote{.${\rm universal}$}
يمكن وصفه بغاز مثالي.  اذن الغاز المثالي هو الغاز الذي تكون جزيئاته ذات تفاعلات متبادلة مهملة. مثلا الهيليوم تحت ضغط منخفض هو غاز مثالي. حالة الغاز المثالي تحدد باعطاء
$4$
متغيرات
$V$, $P$, $T$
و
$N$.
معادلة حالة الغاز المثالي هي معادلة
بويل
\footnote{.${\rm Boyl}$}
\begin{eqnarray}
 PV=n RT~,~PV=N k T~,~n=\frac{N}{{\cal N}}~,~k={\cal N}R.
\end{eqnarray}
$n$
هو عدد المولات,  
$k$
هو ثابت بولتزمان و $R$  هو ثابت الغازات المثالية اللذان يعطيان ب
\begin{eqnarray}
 k=1.38{\rm x}10^{-23} J/K~,~R=8.315 J/K{\rm mole}.
\end{eqnarray}
يمكن استخدام معادلة حالة الغاز المثالي لتعريف ترموماتر وبالتالي اعطاء تعريف كوني لسلم درجة الحرارة.

\paragraph{العمل
و كمية الحرارة
:}
الجمل الفيزيائية الماكروسكوبية تتبادل الطاقة مع الوسط الخارجي بطريقتين مختلفتين هما العمل و كمية الحرارة.
\begin{itemize}
 \item العمل
 $W$: 
 هو تبادل ماكروسكوبي للطاقة 
 علي شكل ميكانيكي.
اذا كان 
  التبادل في الطاقة ينجم عنه تغير في المتغيرات الترموديناميكية ما عدا درجة الحرارة فاننا نقول ان هناك تبادل  عمل. 
  
  اذا كانت المتغيرات الترموديناميكية هي
  $P$ , $V$ 
  و
  $T$
  فان العمل
  $dW$
  خلال تحول ترموديناميكي متناه في الصغر يتغير خلاله الحجم ب
  $dV$
  يعطي ب
  \begin{eqnarray}
dW=PdV.
\end{eqnarray}
  \item
  كمية الحرارة
  $Q$:
  هو تبادل ميكروسكوبي للطاقة علي شكل حراري. اذا كان التبادل في الطاقة ينجم عنه تغير
  في درجة الحرارة فقط مع ثبوت باقي المتغيرات الترموديناميكية علي قيمها فاننا نقول ان هناك تبادل لكمية حرارة.
  
  كمية الحرارة
  $dQ$
  الممتصة من قبل جملة متجانسة مما يؤدي الي ارتفاع درجة الحرارة ب
  $dT$
  مع عدم القيام باي عمل تعطي ب
  \begin{eqnarray}
dQ=C dT,
\end{eqnarray}
حيث
$C$ 
هي ما يسمي بالسعة الحرارية.
\end{itemize}
نقيس العمل و كمية الحرارة بالجول. كمية الحرارة تقاس ايضا بالكالوري, الذي هو كمية الحرارة الضرورية لرفع درجة حرارة
$1$ 
غرام من الماء تحت
$1$
ضغط جوي من 
$14.5$
درجة مئوية
الي
$15.5$ 
درجة مئوية,
ويعرف ب
\begin{eqnarray}
1{\rm ca}=4.18 J.
\end{eqnarray} 

 \paragraph{
الخزان الحراري
:}

  الخزان الحراري هو جملة ترموديناميكية لا تتغير درجة حرارتها تحت تأثير اي تبادل  لاي قيمة منتهية لكمية الحرارة.

\section*{
التحويلات الترموديناميكية
}

\addcontentsline{toc}{section}{
التحويلات الترموديناميكية
}
\paragraph{
التحويلات الترموديناميكية
:}
التحويل الترموديناميكي هو عملية تغير للحالة الماكروسكوبية لجملة ترموديناميكية. يمكن للتحويل ان يكون شبه ساكن
\footnote{.${\rm quasi-static}$},
عكسي
\footnote{.${\rm reversible}$},
او غير عكسي
\footnote{.${\rm irreversible}$}.
خلال التحويل شبه الساكن تبقي الجملة في حالة توازن ترموديناميكي تقريبي اما
التحويل العكسي فهو تحويل شبه ساكن يمكن دائما الرجوع بالجملة فيه الي حالتها الابتدائية عكس التحويل غير العكسي الذي لا يمكن الرجوع فيه بالجملة الي حالتها الابتدائية.

\paragraph{
المتغيرات الخارجية
:}

المتغيرات الترموديناميكية ما عدا درجة الحرارة تسمي المتغيرات الخارجية
\footnote{.${\rm external}~{\rm variables}$},
و اي تغير فيها يمكن ان يولد عمل ميكانيكي.
 لتكن
 $S$
 جملة ترموديناميكية في حالة ابتدائية معينة
 ثم نجري عليهاتحويل ترموديناميكي عن طريق تغيير بعض المتغيرات الخارجية لهذه الجملة. اذا كان تطور الجملة بحيث
 انها تبقي دائما في حالة توازن ترموديناميكي فان التحويل الذي قمنا به يسمي
 شبه ساكن. بالاضافة الي ذلك اذا كان يمكن الرجوع بالجملة الي حالتها الابتدائية بالقيام
 بتحويل عكسي للجملة فان التحويل يسمي تحويل عكسي. و اذا كان لا يمكن
 الرجوع الي الحالة الابتدائية فان التحويل يسمي غير عكسي.

 في حالة التحويلات شبه الساكنة و العكسية يجب علي المتغيرات الخارجية ان تتغير ببطء كاف بالمقارنة مع زمن
 الاسترخاء للجملة الترموديناميكية. انظر المثال ادناه.
\paragraph{التحويلات الادياباتيكية
:}
كما ذكرنا سابقا فان الجملة المعزولة هي اي جملة لا تتفاعل مع العالم الخارجي. الجملة المعزولة حراريا هي الجملة التي لا تتبادل كمية حرارة مع العالم الخارجي. يمكن تحقيق
هذا الامر عبر عزل الجملة بجدار ادياباتيكي
\footnote{.${\rm adiabatic}$}
 لا يسمح بنقل كمية الحرارة. جميع التحويلات الترموديناميكية التي تخضع لها الجملة في هذه الحالة تسمي تحويلات ادياباتيكية. 
 
علي العكس من الجدار الادياباتيكي الذي لا يسمح بنقل كمية الحرارة هناك الجدار الدياتارم
\footnote{.${\rm diatherme}$}
الذي هو ناقل مثالي لكمية الحرارة.
 
 \paragraph{
 التحويلات الايزوحرارية
:}
\footnote{.${\rm isothermic}$}
 هي تحويلات ترموديناميكية متساوية الحرارة اي تحدث عند نفس درجة الحرارة.

\paragraph{
العمل في التحويلات شبه الساكنة
:} 
نعتبر غاز محتجز داخل اسطوانة ادياباتيكية اي كاتمة للحرارة. احدي قاعدتي الاسطوانة عبارة عن مكبس يتحرك بدون  اي احتكاك مع جدران الاسطوانة. لما يتحرك
المكبس فان حجم الغاز و هو متغير خارجي يتغير. انظر الي الشكل 
$1$.
نفترض الان
\begin{itemize}
 \item 
 ان سرعة تحرك المكبس ضعيفة جدا.
 \item
 ان الزمن اللازم من اجل الانتقال من الوضعية
 $x$
 الي الوضعية
 $x+dx$
 هو اكبر بكثير من الزمن اللازم للجملة للقيام بتبادل حراري مع الوسط الخارجي للوصول الي التوازن. اذا افترضنا
 مثلا ان زمن استرخاء الجملة هو
 $1/1000~{\rm sec}$
 فانه يكفي ان ينتقل المكبس من
 $x$
 الي 
 $x+dx$
 في زمن قدره
 $1/10~{\rm sec}$
 حتي يمكن اعتبار التحول كأنه شبه ساكن الي درجة كبيرة لانه في كل لحظة ستكون الجملة  في حالة توازن ترموديناميكي.
\end{itemize}
تحت هذه الظروف فان التحول من 
$x$ 
الي 
$x+dx$
هو تحول شبه ساكن.

في اللحظة
$t=0$
المكبس في الوضعية
$x$. 
لان الجملة في حالة توازن فان الضغط متساو من كل الجهات. القوة الكلية التي تؤثر علي 
المكبس هي
$PA$
حيث
$A$ 
هي مساحة المكبس.
حتي يتحرك المكبس مسافة
$dx$
فان
الضغط داخل الاسطوانة يجب ان يكون اكبر بقليل من الضغط الخارجي. العمل المقدم
من الغاز هو
$PAdx$. اذن
العمل المقدم من الوسط الخارجي هو

\begin{eqnarray}
dW=-PdV.
\end{eqnarray}

  \section*{
المبدأ الصفر للترموديناميك
}
\addcontentsline{toc}{section}{
المبدأ الصفر للترموديناميك
}
 اذا كانت
 $1$
 و
 $2$
  جملتان ترموديناميكيتان
ماكروسكوبيتان
متوازنتان كل علي حدة
مع جملة
ترموديناميكية ماكروسكوبية ثالثة
$3$
فان
$1$
و
$2$
متوازنتان فيما بينهما. بعبارة اخري مكافئة:
كل جملتين ترموديناميكيتين ماكروسكوبيتين متوازنتين لهما 
نفس درجة الحرارة.

\section*{
الطاقة الداخلية و المبدأ الاول للترموديناميك 
}
\addcontentsline{toc}{section}{
الطاقة الداخلية و المبدأ الاول للترموديناميك
} 
\paragraph{
الطاقة الداخلية:
}
  من المعروف ان الطاقة الكلية لجملة ترموديناميكية معزولة, والتي تساوي مجموع الطاقة الحركية و الطاقة الكامنة, هي ثابت للحركة اي انها تبقي منحفظة
في الزمن. مبدأ انحفاظ الطاقة هذا هو قانون كوني يؤدي تطبيقه في الترموديناميك الي المبدأ الاول للترموديناميك. في الترموديناميك
الطاقة الكلية للجملة تسمي بالطاقة الداخلية
\footnote{.${\rm internal}~{\rm energy}$}
و نرمز اليها ب
$U$. 
اذن من اجل 
جملة ترموديناميكية معزولة فان التغير في الطاقة الداخلية يكون منعدم اي
\begin{eqnarray}
dU=0.
\end{eqnarray} 
اذا كانت الجملة غير معزولة اي انها تتفاعل مع الوسط الخارجي فان الطاقة الكلية تتغير بمقدار
$dU$.
خلال هذا التحويل الترموديناميكي فان الجملة تتبادل مع الوسط الخارجي عمل
$dW$
و كمية حرارة
$dQ$
حيث
\begin{eqnarray}
dU=dQ+dW.
\end{eqnarray} 
اذا كان العمل 
$dW$
و
و كمية الحرارة
$dQ$
موجبين فان الوسط الخارجي هو الذي يقوم بالعمل و الجملة تمتص كمية الحرارة. اما اذا كان
العمل و كمية الحرارة سالبين فان الجملة هي التي تقوم بالعمل
$dW$
و الوسط الخارجي هو الذي يمتص كمية الحرارة
$dQ$.

\paragraph{
المبدأ الاول للترموديناميك: 
}
ينص علي ان التغير في الطاقة الداخلية للجملة هو نفسه من اجل كل التحويلات الترموديناميكية التي تربط
بين نفس الحالة الابتدائية و نفس الحالة النهائية. هذا يعني بالخصوص ان
$U$
هو دالة حالة 
\footnote{.${\rm state}~{\rm function}$}
بمعني ان 
$dU$
لا يتعلق بالطريق المتبع بين حالة ابتدائية و حالة نهائية و يتعلق فقط بالحالتين الابتدائية
و النهائية. بعبارة اخري توجد
دالة
$U$ 
تفاضلها هو بالضبط
$dU$
اي 
ان
$dU$ 
هو تفاضل تام او ان

\begin{eqnarray}
\int_C dU=U_f -U_i.
\end{eqnarray} 
من الواضح ان العمل
$dW$
و كمية الحرارة
$dQ$
لا يتمتعان بالخواص اعلاه لانها ليست بدوال حالة. اذن العمل و كمية الحرارة يتعلقان بالطريق المتبع. فقط في حالات خاصة يكون فيها العمل او كمية الحرارة دوال حالة و بالتالي التغير فيهما 
لا يتعلق بالطريق المتبع و يتعلق فقط بالحالتين الابتدائية و النهائية. مثال ذلك التحويلات الادياباتيكية الكاتمة للحرارة 
اي التي تحدث بدون تبادل لكمية الحرارة و بالتالي في هذه الحالة
\begin{eqnarray}
dQ=0~,~dU=dW.
\end{eqnarray} 
اذا اعتبرنا ان المتغيرات الترموديناميكية
 للجملة هي
 $P$, $V$
 و
 $T$ 
 مع معادلة حالة
 $f(P,V,T)=0$
 فان اثنين فقط من المتغيرات هي مستقلة خطيا. لنأخذ هنا
 $P$
 و
 $V$ 
 كمتغيرات ترموديناميكية مستقلة خطيا. اذن الطاقة الداخلية للجملة هي دالة في 
 $P$
 و
 $V$
 اي
 \begin{eqnarray}
U=U(P,V).
\end{eqnarray} 
نحسب 
 \begin{eqnarray}
dU=(\frac{\partial U}{\partial V})_P dV +(\frac{\partial U}{\partial P})_V dP .
\end{eqnarray}  
لان 
$dU$
هو تفاضل تام فانه لدينا مباشرة النتيجة
\begin{eqnarray}
\frac{\partial}{\partial P}\bigg[(\frac{\partial U}{\partial V})_P\bigg]_V=\frac{\partial}{\partial V}\bigg[(\frac{\partial U}{\partial P})_V\bigg]_P.
\end{eqnarray}  
في الترموديناميك تعتبر الطاقة الداخلية مقدار تمديدي. 
هذا يعني ان طاقة جملة
$1+2$
مشكلة من
جملتين
$1$
و
$2$
طاقتيهما
$U_1$
و
$U_2$
علي التوالي هي
\begin{eqnarray}
U_{1+2}=U_1+U_2.
\end{eqnarray} 
 لكن هذا الامر هو فقط تقريب صالح بالنسبة للجمل الماكروسكوبية ذات 
التفاعلات الضعيفة لان الطاقة
$U_{1+2}$
يجب ان تعطي في الحقيقة بالمعادلة
\begin{eqnarray}
U_{1+2}=U_1+U_2+U_{12}.
\end{eqnarray} 
$U_{12}$
هي طاقة التفاعل التي يمكن اهمالها من اجل الجمل الماكروسكوبية ذات التفاعلات
الضعيفة.

الطاقة الداخلية هي مقدار تمديدي فقط في النهاية الترموديناميكية
\footnote{.${\rm thermodynamical}~{\rm limit}$}
المعرفة ب
\begin{eqnarray}
N\longrightarrow \infty~,~V\longrightarrow \infty~:~\frac{N}{V}={\rm constant}.
\end{eqnarray} 
في هذه النهاية اثار السطح تصبح مهملة امام اثار الحجم. اي ان الطاقة
تصبح متناسبة طرديا مباشرة مع ابعاد الجملة
$N$
او
$V$.

\section*{
المبدأ الثاني للترموديناميك
}
\addcontentsline{toc}{section}{
المبدأ الثاني للترموديناميك
}
هناك بعض التحويلات الترموديناميكية التي تحقق المبدأ الاول للترموديناميك  لكنها
لا يمكن ان تحدث بصورة تلقائية في الطبيعة. 

كمثال نعتبر معدن ساخن موضوع داخل ماء بارد. المشاهد عادة ان الحرارة تنتقل من المعدن الساخن
الي الماء البارد. لكن مع الاعتماد فقط علي المبدأ الاول للترموديناميك يمكن للحرارة ان تنتقل من البارد الي
الساخن بحيث يزداد المعدن سخونة و يزداد الماء برودة. لكن هذا غير مشاهد في 
الطبيعة. المبدأ الثاني للترموديناميك يهدف الي توضيح الاتجاهات التي تنتقل فيها الحرارة.

\paragraph{
بيان كلوسيوس:
}\footnote{.${\rm Clausius}~{\rm statement}$}
لا يوجد تحويل ترموديناميكي نتيجته الوحيدة تكون نقل كمية حرارة من جسم بارد
الي جسم ساخن. يمكن نقل الحرارة من جسم بارد الي جسم ساخن ببذل عمل معين.
\paragraph{
بيان كلفن- بلانك:
}\footnote{.${\rm Kelvin}-{\rm Planck}~{\rm statement}$}
لا توجد تحويلات ترموديناميكية تكو ن نتيجتها الوحيدة هو استخراج
كمية حرارة من خزان حراري وحيد ذو درجة حرارة ثابته و تحويله بالكامل الي عمل.

كمثال ناخذ غاز مثالي يتمدد بطريقة عكسية ايزو حرارية اي عند نفس درجة الحرارة
$T$.
من اجل غاز مثالي الطاقة الداخلية لا تتعلق الا بدرجة الحرارة 
(
انظر الي التمرينات
)
و بالتالي 
$dU=0$
لان
$T$
ثابته في هذه الحالة. اذن في هذه الحالة
$-dQ=dW$.
لان الغاز يتمدد فانه يبرد و بالتالي
$dQ>0$
اي ان الغاز يمتص حرارة للحفاظ علي ثبات درجة الحرارة 
و منه نستنتج ان
$dW<0$
اي ان الغاز هو الذي يقوم بالعمل. نلاحظ انه في هذه الحالة كل كمية الحرارة تم
تحويلها الي عمل لكن ليس هذه النتيجة الوحيدة  لهذا التحويل الترموديناميكي لان الغاز في حالته 
النهائية يحتل حجم اكبر.

من المهم ان نقنع انفسنا ان بيان كلوسيوس هو مكافئ تماما لبيان
كلفن- بلانك.
\section*{
دورة كارنو
}
\addcontentsline{toc}{section}{
دورة كارنو
}

دورة كارنو
\footnote{.${\rm Carnot}~{\rm cycle}$}
هي مجموعة متتابعة من التحويلات الترموديناميكية شبه الساكنة و العكسية التي يستعمل فيها
خزانيين حراريين بدرجتي حرارة
$T_1$
و
$T_2<T_1$
.
 لنعتبر غاز في اسطوانة قاعدتها عبارة عن مكبس يتحرك بدون احتكاك. الطريق الذي يتبعه الغاز في المستوي
 $P-V$
 هو كالاتي
 (انظر الي الشكل
 $2$
 ): 
 
 \begin{itemize}
  \item 
  الحالة الابتدائية للغاز
  (النقطة
  $A$)
  تكون بحجم
  $V_1$,
  درجة
  حرارة
  $T_1$
  و ضغط
  $P_1$.

  نضع الجملة علي اتصال بترموستات ذو درجة حرارة
  $T_1$.
  الترموستات هو عبارة عن خزان حراري كبير يسمح بالحفاظ علي درجة حرارة ثابته للغاز لان كل تبادل للحرارة مع الجملة المعتبرة لا يغير من درجة حرارتها.
  
  نقوم بتغيير حالة الغاز بطريقة عكسية ايزوحرارية حتي يصبح الضغط
  $P_2$
  عند النقطة
  $B$
  و ذلك عن طريق تغيير شبه ساكن للضغط المطبق علي المكبس.
  الحجم النهائي يصبح
  $V_2>V_1$
  اي ان الغاز يتمدد و بالتالي يبرد. حتي يحافظ الغاز علي درجة حرارة
  ثابته يمتص كمية
  حرارة
  $Q_1>0$
  من المنبع الحار.
  
  \item
  نعزل الغاز بجدار ادياباتيكي كاتم للحرارة يمنع اي تبادل لكمية الحرارة. نقوم بترك
  الغاز يتمدد بطريقة
  عكسية ادياباتيكية حتي
  تصبح درجة الحرارة مساوية
  ل
  $T_2<T_1$ 
  التي هي درجة حرارة المنبع البارد عند
  النقطة
  $C$.
   يصبح الحجم
  $V_3$
  و الضغط
  $P_3$
  عند النقطة
  $C$.
  \item 
  نضع الجملة علي اتصال بترموستات ذو درجة حرارة
  $T_2$
  حتي نحافظ علي درجة حرارته مساوية ل
  $T_2$.
  نقوم بضغط الغاز من الحجم
  $V_3$
  الي الحجم
  $V_4<V_3$
  و من الضغط
  $P_3$
  الي الضغط
  $P_4>P_3$.
  هذه العملية هي ايضا عكسية و ايزوحرارية اي ان درجة الحرارة تبقي ثابتة
  مساوية ل
  $T_2$.
  الغاز المضغوط
  يسخن
  و بالتالي حتي
  نحافظ
  علي درجة حرارة ثابته يجب علي الغاز
 ان يعطي كمية حرارة
  $Q_2<0$
  الي الوسط الخارجي. النقطة
  $D$ 
  تقع علي الخط
  الادياباتيكي
  المار ب
  $A$.
 \item
 نغلق الدورة بعملية ضغط ثابتة تكون ادياباتيكية و عكسية من
 $D$
 الي
 $A$.
 خلال هذا المقطع تتغير درجة حرارة الغاز من 
 $T_2$
 الي
 $T_1$.
 \end{itemize}
خلال دورة كارنو لدينا حسب المبدأ الاول للترموديناميك
\begin{eqnarray}
\Delta U=0.
\end{eqnarray} 
كمية الحرارة المنقولة خلال دورة كارنو هي
\begin{eqnarray}
ِQ=Q_1+Q_2.
\end{eqnarray} 
العمل المتبادل بين الغاز و الوسط الخارجي خلال
$ABC$ 
هو
\begin{eqnarray}
W_1=-\int_{ABC}PdV<0.
\end{eqnarray} 
لان الحجم خلال هذا الجزء يتزايد باستمرار و بالتالي
$dV>0$.
$W_1$
هي المساحة تحت المنحني
$ABC$.
لان
$W_1<0$ 
فان الجملة هي التي تقوم بالعمل.
 
 العمل المتبادل بين الغاز و الوسط الخارجي خلال
$CDA$ 
هو
\begin{eqnarray}
W_2=-\int_{CDA}PdV>0.
\end{eqnarray} 
لان الحجم خلال هذا الجزء ينكمش باستمرار و بالتالي
$dV<0$.
$W_2$
هي المساحة تحت المنحني
$CDA$.
لان
$W_2>0$ 
فان الوسط الخارجي هو الذي يقوم بالعمل. نعتبر الحالتين التاليتين:
\begin{itemize}
 \item  الامكانية الاولي :
\begin{eqnarray}
|W_1|>|W_2|.
\end{eqnarray}
اذن العمل الكلي في هذه الحالة هو سالب اي
\begin{eqnarray}
W=W_1+W_2<0.
\end{eqnarray}
اي ان الجملة هي التي تقوم بالعمل اي توفر عمل للوسط الخارجي اي انها تتصرف كمحرك حراري لانها حولت طاقة حرارية الي عمل ميكانيكي. باستعمال المبدأ الاول للترموديناميك كالتالي
(
مع
$Q_2^{'}=-Q_2>0$, $W_1^{'}=-W_1>0$)

\begin{eqnarray}
\Delta U=W+Q=0\Rightarrow -W=Q>0.
\end{eqnarray}
هذا مكافئ ل
\begin{eqnarray}
W_1^{'}-W_2=Q_1-Q_2^{'}.
\end{eqnarray}
يمكن ان نبرهن بشكل عام جدا انه اذا كان
$-W>0$
فان
$Q_2^{'}>0$
و
$Q_1>0$.

\item 
الامكانية الثانية: هناك ايضا الامكانية التي يكون فيها العمل الكلي
$W$
 موجب اي ان الوسط الخارجي هو الذي يقوم بالعمل.

يمكن ايضا ان نبرهن بشكل عام جدا انه اذا كان
$-W<0$
و
$Q_2^{'}<0$
فان
$Q_1<0$.
اذن الحرارة يمتصها الان الغاز من المنبع البارد بينما يتخلص من الحرارة باعطائها للمنبع الحار  و من اجل تحقيق كل هذا يجب ان نوفر عمل للغاز من الوسط الخارجي.
 اي انه في هذه الحالة فان دورة كارنو تعمل في الاتجاه المخالف و تتصرف
كبراد لاننا نستعمل عمل ميكانيكي لاحداث تدرج في درجة الحرارة.

\end{itemize}
من الواضح ان مردود محرك كارنو الحراري هو النسبة بين العمل الذي قامت به الجملة (الغاز) و كمية الحرارة المأخوذة من المبع الحار اي
\begin{eqnarray}
\eta=\frac{W_1^{'}-W_2}{Q_1}=1-\frac{Q_2^{'}}{Q_1}.
\end{eqnarray}
 نذكر الان بعض النتائج المهمة بدون اي برهان:
\begin{itemize}
 \item 
 مبرهنة كارنو:
 لا يوجد
 محرك حراري يعمل بين درجتي حرارة
 $T_1$
 و
 $T_2$
 هو اكثر فعالية من محرك كارنو الحراري.
 \item
 لازمة كارنو:
 كل دورات كارنو التي تعمل بين درجتي حرارة
 $T_1$
 و
 $T_2$
 لها نفس المردود.
 \item
 اذا كانت الدورة تحتوي علي تحويلات غير عكسية فان المردود سيكون اقل.
\end{itemize}
دورة كارنو تسمح لنا بتعريف السلم المطلق لدرجات الحرارة عبر العلاقة التجريبية:
\begin{eqnarray}
\frac{Q_1}{Q_2^{'}}=\frac{T_1}{T_2}.\label{key}
\end{eqnarray}
انظر الي التمرينات.
\section*{
مبرهنة كلوسيوس, 
الانتروبي و 
 المبدأ الثاني للترموديناميك
}
\addcontentsline{toc}{section}{
مبرهنة كلوسيوس, 
الانتروبي و 
 المبدأ الثاني للترموديناميك
} 

 نعيد كتابة العلاقة
 $(28)$
 اعلاه علي الشكل
\begin{eqnarray}
\frac{Q_1}{T_1}+\frac{Q_2}{T_2}=0.
\end{eqnarray}
يمكن تعميم هذه المعادلة لكل الدورات شبه الساكنة كالاتي
\begin{eqnarray}
\oint \frac{dQ}{T}=0.
\end{eqnarray}
التكامل مأخوذ علي طول الدورة و 
$dQ$
هي كمية الحرارة المتبادلة بطريقة شبه ساكنة في نقطة الدورة اين تكون درجة الحرارة مساوية ل
$T$.
هذه النتيجة هي جزء من ما يسمي بمبرهنة كلوسيوس.

\paragraph{
مبرهنة كلوسيوس:
}
في اي تحويل ترموديناميكي دوري
${\cal O}$
فان المتراجحة التالية صحيحة:
\begin{eqnarray}
\oint \frac{dQ}{T}\leq 0.
\end{eqnarray}
اذا كان التحويل الترموديناميكي عكسي فان
\begin{eqnarray}                                                                                                                                                                                                                                                                                                
\oint \frac{dQ}{T}= 0.
\end{eqnarray}
البرهان يجري كالاتي. نقسم التحويل
${\cal O}$
الي 
$n\longrightarrow \infty$
تحويل متناه في الصغر
 حيث تكون درجة الحرارة تقريبا ثابته في كل خطوة. اذن نتصور ان الجملة في كل خطوة
 $i$
 هي علي اتصال
 بمخزن حراري ذو درجة حرارة
 $T_i$
 اي انها تمتص كمية حرارة
 $Q_i$
 في كل خطوة من اجل الحفاظ علي ثبات درجة الحرارة عند
 $T_i$. 
 
 نبني  
 $n$
 دورة لكارنو
 $\{C_1,C_2,...,C_n\}$
 حيث  كل
 $C_i$
هي بحيث
 \begin{enumerate}
  \item 
  تعمل بين درجتي الحرارة
  $T_i$
  و
  $T_o\geq T_i$
  من اجل كل
  $i$.
  \item
  تمتص كمية الحرارة
  $Q_i^o$
  من $T_o$.
  \item
  تتخلص من كمية الحرارة
  $Q_i$
  ل
  $T_i$.
 \end{enumerate}
نعتبر التحويل الترموديناميكي
${\cal O}+\{C_1,C_2,...,C_n\}$
حيث الخطوة
$i$
مشتركة بين
${\cal O}$
و
$C_i$
لكن في اتجاهين متعاكسين. اذن كمية الحرارة الكلية المتبادلة خلال هذا التحويل هي
\begin{eqnarray}
Q_o=\sum_{i=1}^nQ_i^o.
\end{eqnarray}
لكن السلم المطلق لدرجة الحرارة يعطي
\begin{eqnarray}
\frac{Q_i^o}{Q_i}=\frac{T_o}{T_i}.
\end{eqnarray}
اذن كمية الحرارة الكلية المتبادلة هي
\begin{eqnarray}
Q_o=T_o\sum_{i=1}^n\frac{Q_i}{T_i}.
\end{eqnarray}
هذه هي كمية الحرارة الكلية الممتصة من الخزان الحراري
$T_o$
و التي حولت حسب المبد الاول للترموديناميك
($\Delta U=W_o+ Q_o=0$)
 بالكامل الي عمل من دون نتائج اخري. باستعمال
 المبدأ الثاني
 للترموديناميك حسب بيان كلفن- بلانك فانه
 لا توجد تحويلات ترموديناميكية تكو ن نتيجتها الوحيدة هو استخراج
كمية حرارة من خزان حراري وحيد ذو درجة حرارة ثابته و تحويله بالكامل الي عمل.  اذن الوسط الخارجي يجب ان يوفر عمل اي ان 
$W_o>0$
و بالتالي فان
$Q_o\leq 0$
و هو يكافئ
\begin{eqnarray}
\sum_{i=1}^n\frac{Q_i}{T_i}\leq 0.
\end{eqnarray}
و هذا ما نريد. 

اذا كان التحويل عكسي فاننا يمكننا ان نعيد نفس الخطوات من اجل التحويل العكسي
 $-{\cal O}$
 الذي نعوض فيه
 $Q_i$ ب
 $-Q_i$
لنحصل علي

\begin{eqnarray}
-\sum_{i=1}^n\frac{Q_i}{T_i}\leq 0.
\end{eqnarray}
من المعادلتين اعلاه نحصل من اجل التحويلات العكسية مباشرة علي

\begin{eqnarray}
\sum_{i=1}^n\frac{Q_i}{T_i}=0.
\end{eqnarray}
و هذا يكمل البرهان علي مبرهنة كلوسيوس.

\paragraph{
لازمة كلوسيوس و تعريف الانتروبي:
}
التكامل
\begin{eqnarray}
\oint \frac{dQ}{T},
\end{eqnarray}
لا يتعلق بالطريق المتبع و يتعلق فقط بالحالتين الابتدائية و النهائية. البرهان سهل جدا يعتمد علي تطبيق مباشر لمبرهنة كلوسيوس. لنعتبر حالتين
$A$
و 
$B$
و ليكن
$I$
و
$II$
طريقين مختلفين بين
$A$
و 
$B$.
ليكن
$II^{'}$
الطريق العكسي ل
$II$. 
باستعمال مبرهنة كلوسيوس لدينا 
\begin{eqnarray}
\int_I \frac{dQ}{T}+\int_{II^{'}}\frac{dQ}{T}=0.
\end{eqnarray}
اذن مباشرة نستنتج
\begin{eqnarray}
\int_I \frac{dQ}{T}=\int_{II}\frac{dQ}{T}.
\end{eqnarray}
يمكننا اذن ان نعرف دالة حالة جديدة 
$S$
هي الانتروبي
\footnote{.${\rm entropy}$}
بالتفاضل التام
\begin{eqnarray}
dS=\frac{dQ}{T}.
\end{eqnarray}
هذه العلاقة تعرف الانتروبي خلال تحويل عكسي متناه في الصغر. 
من التعريف اعلاه من
 الواضح ان انتروبي اي حالة ترموديناميكية 
 $A$
 هو الانتروبي
 خلال اي تحويل عكسي
 يربط بين حالة ابتدائية
 $O$ 
 و الحالة
 $A$ 
و هو معرف 
 فقط الي غاية ثابت تجميعي كيفي بالعلاقة
\begin{eqnarray}
S(A)=\int_O^A \frac{dQ}{T}.
\end{eqnarray}
في المقابل فان الفرق في الانتروبي بين حالتين
$A$
و
$B$
هو معرف بدقة بالعلاقة
\begin{eqnarray}
S(A)-S(B)=\int_B^A \frac{dQ}{T}.
\end{eqnarray}
الانتروبي هو مقدار تمديدي فقط في النهاية الترموديناميكية مثله مثل الطاقة الداخلية, و هو مقياس اللانظام او الفوضي في الجملة الترموديناميكية. 
\paragraph{
المبدأ الثاني للترموديناميك (مرة اخري)
:
}
انتروبي جملة معزولة حراريا لا يمكنه الا ان يزيد اي ان
\begin{eqnarray}
\Delta S\geq 0.
\end{eqnarray}
من اجل الجمل الترموديناميكية العكسية فان
$\Delta S=0$
اما من اجل الجمل  الترموديناميكية غير العكسية فان
$\Delta S>0$.

البرهان يجري كالاتي. لنعتبر حالتين ترموديناميكيتين
$A$
و
$B$
و ليكن
$R$
طريق عكسي 
و 
$I$
طريق غير عكسي يربطان بين الحالتين
$A$
و
$B$.
من اجل الطريق
$R$
لدينا من التعريف

\begin{eqnarray}
S(B)-S(A)=\int_R \frac{dQ}{T}.
\end{eqnarray}
نعتبر التحويل الدوري المشكل من 
$I$
و عكس
$R$.
باستعمال مبرهنة كلوسيوس لدينا مباشرة
\begin{eqnarray}
\int_I\frac{dQ}{T}-\int_R \frac{dQ}{T}\leq 0.
\end{eqnarray}
اي
\begin{eqnarray}
S(B)-S(A)\geq \int_I \frac{dQ}{T}.
\end{eqnarray}
بصفة عامة لدينا
\begin{eqnarray}
S(B)-S(A)\geq \int \frac{dQ}{T}.
\end{eqnarray}
اذا اعتبرنا الان جملة معزولة حراريا اي لا تتبادل اي كمية حرارة مع الوسط الخارجي فان
$dQ=0$
و بالتالي نحصل مباشرة من النتيجة اعلاه علي
\begin{eqnarray}
S(B)-S(A)\geq 0.
\end{eqnarray}
اذن انتروبي جملة معزولة حراريا لا يتناقص ابدا.

علي الرغم من ان انتروبي جملة معزولة لا يمكنه الا ان يتزايد فان انتروبي الجمل غير المعزولة يمكنه ان يتناقص. اي ان تناقص الانتروبي لا يمكنه ان يتم
الا بتبادل طاقة بين الجملة و الوسط الخارجي.
\paragraph{
التحويلات العكسية و غير العكسية (مرة اخري)
:
}
الحالة الماكروسكوبية 
$X$
لجملة معزولة حراريا و متوازنة ترموديناميكيا تتعين بالكامل اذا علمنا الطاقة
$E$,
الحجم
$V$
و عدد الجسيمات الميكروسكوبية 
$N_i$
المكونة لها. 
نفترض من اجل التبسيط ان كل الجسيمات الميكروسكوبية هي من نفس النوع. نكتب 
$X=(E,V,N)$. 
الانتروبي
$S$
هو دالة في
$X$. 
الفضاء الرباعي ذو الاحداثيات
$E$, $V$, $N$
و
$S$
هو فضاء الحالة لهذه الجملة الترموديناميكية. انظر الي الشكل $3$.

اي نقطة من هذا الفضاء تمثل حالة توازن معينه للجملة الترموديناميكية. اي مسار داخل هذا
الفضاء يمثل تحول ترموديناميكي شبه ساكن لانه عبارة عن توالي لحالات توازن ترموديناميكية. اذا كان التحول الترموديناميكي يتم بانتروبي ثابت فالتحول عكسي و هو يوافق خط افقي مستقيم داخل فضاء الحالة. في حالة اذا كان التحول
يتم بانتروبي متزايد فانه تحول غير عكسي.
كل تحول عكسي
هو تحول شبه ساكن لكن العكس غير صحيح.

في الواقع توجد تحولات غير عكسية ليست شبه ساكنة. في هذه الحالة يحدث التحويل بتوالي حالات لا توازن و بالتالي فان هذه الحالات لا تنتمي الي الفضاء الرباعي
$(S,X)$. 
حتي يتم وصف هذا التحويل ندخل متغيرات جديدة 
$Y$
التي تتعلق بطبيعة هذا التحويل. تطور الجملة يتم الان داخل الفضاء
$(S,X,Y)$.

\section*{
المبدأ الثالث للترموديناميك
}
\addcontentsline{toc}{section}{
المبدأ الثالث للترموديناميك
} 

ذكرنا قبل قليل ان تعريف انتروبي حالة ترموديناميكية كيفية
 $A$
 يعتمد علي وجود تحويلات عكسية تربط
 $A$
 باي حالة مرجعية مختارة
 $O$.
  من اجل معادلات الحالة التي تكافئ 
   سطح حالة مشكل من ورقة 
   \footnote{.${\rm sheet}$}
   واحدة فان كل الحالات الترموديناميكية تكون مرتبطة فيما بينها بتحويلات عكسية لانها تقع 
  كلها علي هذه الورقة. بعبارة اخري فان الطريق العكسي الذي يربط
   $A$
   و
  $O$
يوجد دائما في هذه الحالة. 
  
  اذا اعتبرنا من الجهة الاخري مادتين مختلفتين او مادة واحدة بطورين مختلفين فان معادلة الحالة تكافي سطح
  حالة  قد يكون مشكل من اكثر من ورقة واحدة غير متصلة. في هذه الحالة فان الطريق العكسي
  الذي يربط بين
 $A$
   و
  $O$
  قد لا يوجد وبالتالي فان الفرق في الانتروبي لا يمكن تعريفه في هذه الحالة. اذن المبدأ الثاني للترموديناميك لا يمكن ان يعين بصورة وحيدة الفرق 
  في الانتروبي  بين حالتين
  $A$
  و
  $B$
  اذا كانت 
  $A$ 
  تخص مادة او طور
  و
  $B$
  تخص
  مادة او طور اخر. المبدأ الثالث للانتروبي, الذي صاغه نارنست
\footnote{.${\rm Nernst}$}
في
$1905$,
يجعل تعريف الانتروبي وحيد في كل الحالات و من ضمنها الحالات المذكورة انفا. هذ المبدأ ينص علي الاتي:
  
  انتروبي اي جملة ترموديناميكية في حالة توازن يساوي الصفر عند درجة حرارة الصفر المطلق:
\begin{eqnarray}
S(0)=0.
\end{eqnarray}
  
\section*{
الدوال الترموديناميكية
}
\addcontentsline{toc}{section}{
الدوال الترموديناميكية
} 
اول و اهم الدوال الترموديناميكية هي الطاقة الداخلية المعرفة ب
\begin{eqnarray}
dU=dW+dQ=-PdV+TdS.
\end{eqnarray}
$U$
هي مقدار تمديدي و بالتالي فهي دالة في المقادير التمديدية
$S$, $V$
و
$N$
اي
\begin{eqnarray}
U=U(S,V,N).
\end{eqnarray}
لان
$U$
هي دالة حالة و باعتبار
$ٍS$
و
$V$
كمتغيرات مستقلة
فان
\begin{eqnarray}
dU=\big(\frac{\partial U}{\partial S}\big)_{V,N} dS+\big(\frac{\partial U}{\partial V}\big)_{S,N} dV.
\end{eqnarray}
بالمقارنه نحصل علي
\begin{eqnarray}
T=\big(\frac{\partial U}{\partial S}\big)_{V,N}~,~-P=\big(\frac{\partial U}{\partial V}\big)_{T,N}.
\end{eqnarray}
اذا سمحنا ايضا لعدد الجسيمات بالتغير فاننا نحصل من الجهة الاخري علي
\begin{eqnarray}
dU=\big(\frac{\partial U}{\partial S}\big)_{V,N} dS+\big(\frac{\partial U}{\partial V}\big)_{S,N} dV+\big(\frac{\partial U}{\partial N}\big)_{V,S} dN.
\end{eqnarray} 
نعرف الكمون الكيميائي علي انه هو المتغير المرفق بالتغير  في عدد الجسيمات اي
\begin{eqnarray}
\mu=\big(\frac{\partial U}{\partial N}\big)_{V,S}.
\end{eqnarray}
اي ان التغير في الطاقة الداخلية يعطي  في العموم ب
\begin{eqnarray}
dU=-PdV+TdS+\mu dN.
\end{eqnarray}
الطاقة الداخلية هي دالة متجانسة ذات رتبة
$1$
و بالتالي من اجل اي عدد حقيقي
$\lambda$
لدينا
\begin{eqnarray}
U(\lambda S,\lambda V,\lambda N)=\lambda U(S,V,N).
\end{eqnarray}
بالاشتقاق بالنسبة الي 
$\lambda$
ثم وضع
$\lambda=1$
نحصل علي صيغة اولر 
\footnote{.${\rm Euler}$}
للطاقة الداخلية:
\begin{eqnarray}
U(S,V,N)=-PV+TS+\mu N.
\end{eqnarray}
نحصل علي الدوال الترموديناميكية الاخري عن طريق تحويلات لوجوندر للطاقة الداخلية. نعرف الطاقة الحرة لهلمولتز
\footnote{.${\rm Helmholtz}~{\rm free}~{\rm energy}$}
علي انها تحويل لوجوندر للطاقة الداخليةبالنسبة للمتغيرات
$T\leftrightarrow S$
المعرف ب
\begin{eqnarray}
F=F(T,V,N)=U(S,V,N)-TS=-PV+\mu N.
\end{eqnarray}
\begin{eqnarray}
dF=dU-dT.S-T.dS=-PdV+\mu dN-S dT.
\end{eqnarray}
نعرف الكمون الترموديناميكي (او الطاقة الحرة) لجيبس
\footnote{.${\rm Gibbs}~{\rm thermodynamic}~{\rm potential}$.}
علي انها تحويل لوجوندر للطاقة الحرة لهلمولتز بالنسبة للمتغيرات
$P\leftrightarrow V$
المعرف ب
\begin{eqnarray}
G=G(T,P,N)=F(T,V,N)+PV=\mu N.
\end{eqnarray}
\begin{eqnarray}
dG=-S dT+V dP+\mu dN.
\end{eqnarray}
نعرف الانتالبي
\footnote{.${\rm enthalpy}$}
علي انه تحويل لوجوندر للطاقة الداخلية بالنسبة للمتغيرات 
$P\leftrightarrow V$
المعرف ب
\begin{eqnarray}
H=H(S,P,N)=U(S,V,N)+PV=TS+\mu N.
\end{eqnarray}
\begin{eqnarray}
dH=dU+dP.V+P.dV=TdS+VdP+\mu dN.
\end{eqnarray}
نختم هذا الفصل بالمبرهنات المفيدة التالية:
\paragraph{
مبرهنة $1$:
}
من اجل جملة معزولة ميكانيكيا عند درجة حرارة ثابته فان الطاقة الحرة لهلمولتز لا تتزايد ابدا. حالة التوازن هي الحالة التي تكون فيها الطاقة الحرة لهلمولتز اصغرية.

البرهان كما يلي. نعتبر تحويل ايزوحراري بين حالتين ترموديناميكيتين
$A$
و
$B$.
من المبدأ الثاني لدينا
\begin{eqnarray}
\int_A^B\frac{dQ}{T}\leq S(B)-S(A).
\end{eqnarray}
لان التحويل ايزوحراي فان
$T$
ثابته و بالتالي
\begin{eqnarray}
\frac{\Delta Q}{T}\leq \Delta S.
\end{eqnarray}
$\Delta Q$
هي كمية الحرارة الممتصة خلال التحويل. لكن من المبدأ الاول لدينا
\begin{eqnarray}
-\Delta W=-\Delta U+\Delta Q\leq -\Delta U +T\Delta S=-\Delta F.
\end{eqnarray}
$\Delta W$
هو العمل الذي قامت به الجملة و
$F$
هي الطاقة الحرة لهلمولتز. اذن نحصل علي
\begin{eqnarray}
\Delta F\leq \Delta W.
\end{eqnarray}
من اجل التحويلات المعزولة ميكانيكيا لدينا
$\Delta W=0$
و بالتالي
\begin{eqnarray}
\Delta F\leq 0.
\end{eqnarray}
اذن الطاقة الحرة لهلمولتز لا تتزايد ابدا في هذه الحالة و  في حالة التوازن (اي من اجل التحويلات العكسية) فان
$\Delta F=0$.

\paragraph{
مبرهنة 
$2$:
}
من اجل جملة محفوظة عند درجة حرارة ثابته و ضغظ ثابت فان الكمون الترموديناميكي لجيبس لا يتزايد ابدا. حالة التوازن هي الحالة التي يكون فيها الكمون الترموديناميكي لجيبس اصغري.

البرهان سهل جدا. من اجل درجة حرارة ثابته لدينا
\begin{eqnarray}
\Delta F\leq \Delta W.
\end{eqnarray}
من اجل ضغط ثابت نحصل مباشرة علي
\begin{eqnarray}
\Delta G\leq 0.
\end{eqnarray}
\newpage

\section*{
تمارين
}
\addcontentsline{toc}{section}{
تمارين
} 

\paragraph{
تمرين
$1$:
}
\begin{itemize}
 \item 
في تجربة جول
\footnote{.${\rm Joule}$}
نسمح لغاز مثالي بالتمدد الحر في الفراغ من الحجم
$V_1$
و درجة الحرارة
$T_1$
الي الحجم
$V_2>V_1$
و درجة الحرارة
$T_2$.
تجريبيا نلاحظ ان
$T_1=T_2$.
بين ان الطاقة الداخلية لغاز مثالي لا تتعلق الا بدرجة الحرارة.
\item
نعتبر الان تمدد ايزوحراري عكسي من الحالة
$(T_1,V_1)$
الي الحالة
$(T_2,V_2)$. 
احسب الفرق في الانتروبي.
\item
هل تمدد جول هو تحويل عكسي.
احسب الفرق في الانتروبي في تجربة جول.
\end{itemize}
\paragraph{
تمرين
$2$:
}
الحراة النوعية تحت حجم او ضغط ثابت هي معطاة كالاتي
\begin{eqnarray}
C_v=(\frac{dQ}{dT})_V~,~C_p=(\frac{dQ}{dT})_P.
\end{eqnarray}
\begin{itemize}
 \item 
 باستعمال المبدأ الاول للترموديناميك استخرج معادلات ال
 $dQ$
 ثم عبر عن 
 $C_v$
 و
 $C_p$
 بدلالة الطاقة الداخلية
 $U$
 و درجة الحرارة
 $T$.
 \item
 برهن علاقة ماير
 \footnote{.${\rm Mayer}$}
 للغازات المثالية
 \begin{eqnarray}
C_p-C_v=nR.
\end{eqnarray}
\end{itemize}

 \paragraph{
تمرين
$3$:
}
دورة كارنو تمتص كمية حرارة
$Q_2^{'}>0$
عند درجة حرارة
$T_2$
و تتخلص من كمية حرارة
$Q_1^{'}$
عند درجة الحرارة
$T_1<T_2$. 
كمية الحرارة الكلية المتبادلة هي 
$Q=Q_2^{'}-Q_1^{'}$
 و بالتالي فان العمل الكلي هو
$W=-Q$
اي ان المردود هو 
\begin{eqnarray}
\eta=\frac{W}{Q_2^{'}}=1-\frac{Q_1^{'}}{Q_2^{'}}.
\end{eqnarray}
 تعريف درجة الحرارة المطلقة يعطي بالعلاقة التجريبية
 \begin{eqnarray}
\frac{Q_1^{'}}{Q_2^{'}}=\frac{T_1}{T_2}.
\end{eqnarray}
  من الواضح انه لان
$0\leq \eta\leq 1$
فان درجة الحرارة المطلقة هي دائما اكبر او يساوي من الصفر.

 بين انه باستعمال سلسلة من دورات كارنو التي تؤدي كلها نفس العمل
 $W$
 والتي 
 تمتص فيها كل دورة كمية الحرارة التي تتخلص منها الدورة السابقة يمكن الحصول علي سلم منتظم لدرجات الحرارة المطلقة.
 
  \paragraph{
تمرين
$4$:
}
 نعتبر 
 $1$
 مول من غاز مثالي محتوي داخل اسطوانة مغلقة دياتارم, اي ذات جدران ناقلة مثالية للحرارة,  احدي قاعدتيها عبارة
 عن مكبس متحرك. نضع الاسطوانه داخل خزان كبير مملوء بسائل درجة حرارته
 $T$. 
 \begin{itemize}
  \item 
 في البداية ضغط الغاز
 يساوي
 $P_1$
 و حجمه يساوي
 $V_1$.
 نترك الغاز يتمدد بطريقة شبه ساكنة حتي يصبح حجمه
 $V_2$
 و ضغطه
 $P_2$.
 احسب العمل المقدم من الغاز الي الوسط الخارجي.
 \item
 لنفترض الان ان التمدد كان بطريقة غير عكسية اين يتم تغيير ضغط الغاز من
 $P_1$
 الي القيمة
 $P_2$
 بغته. احسب العمل في هذه الحالة.
 \item
 لنفترض ان التمدد كان بطريقة غير عكسية اولا من الضغط
 $P_1$
 الي
 الضغط
 $P_3<P_1$
 ثم
 من الضغط
 $P_3$
 الي الضغط
 $P_2<P_3$.
 احسب العمل في هذه الحالة. ماذا يمكنك ان تستنتج. خذ مثلا
 $P_1=3~{\rm atm}$, $P_2=1~{\rm atm}$, $P_3=2~{\rm atm}$.
\item
لنفترض الان ان جدران الاسطوانة ادياباتيكية, اي عازلة مثالية للحرارة, و نفترض ان تمدد الغاز يتم عبر تحول عكسي. اوجد
العلاقة بين الضغط و الحجم في هذا التحول. استخدم
\begin{eqnarray}
 \gamma=\frac{C_p}{C_v}.
\end{eqnarray}
\item
لنفترض ان الغاز المثالي هو غاز ثنائي الذرة
\footnote{.${\rm diatomic}$}
و بالتالي فان 
$ \gamma={7}/{5}$.
اذا كان الحجم النهائي للغاز يساوي
مرة و نصف حجمه الابتدائي, احسب درجة الحرارة النهائية 
$T_2$
بدلالة درجة الحرارة الابتدائية
$T_1$.
خذ مثلا
$T_1=300~{\rm K}$.
 \end{itemize}
 \paragraph{
تمرين
$5$:
}

\begin{itemize}
 \item
 الطاقة الداخلية لاي جملة هي دالة في
 $T$
و
$V$.
 ما هو الشرط الذي يجب ان تحققه المشتقة الجزئية
 $(\partial U/\partial V)_T$
 حتي يكون الانتروبي
 $S=S(T,V)$
 دالة حالة.
 \item
 ماذا يمكن ان نستنتجه من اجل الغاز المثالي.
 \item
 احسب
 $S(T,V)$
 في حالة الغاز المثالي. 
 \end{itemize}
 \paragraph{
تمرين
$6$:
}
 \begin{itemize}
  \item 
نضع سائل (ماء مثلا) داخل اسطوانة شاقولية ذات غطاء عبارة عن مكبس متحرك. لما نجذب علي المكبس فان الفراغ بين السائل و المكبس يمتلئ ببخار مشبع.
ضغط البخار المشبع يتعلق بدرجة حرارة السائل فقط لا غير. الجملة الترموديناميكية سائل زائد مكبس تغمر في ترموستات ذو درجة حرارة
$T$. 
صف المنحنيات ذات درجة الحرارة الثابته اي الايزوحراريات للجملة سائل زائد بخار داخل الفضاء
$P-V$.
\item
نهتم بمجال درجات الحرارة اين يتواجد البخار و السائل في نفس الوقت. ليكن
$V_1$
و
$V_2$
حجمي السائل و البخار في وحدة الكتل و لتكن
$U_1$
و
$U_2$
طاقتي السائل و البخار في وحدة الكتل. المقادير
$P$, $V_i$, $U_i$
هي دوال تابعة لدرجة الحرارة فقط. الكتلة الكلية للمادة المحتواة داخل الاسطوانة
هي
$m=m_1+m_2$.
نعتبر تحويل ايزوحراري متناه في الصغر في اثناءه كتلة
$dm$
من السائل تتبخر.
\begin{itemize}
\item 
احسب التغير
$dV$
في الحجم و التغير
$dU$
في الطاقة الكلية للجملة. استنتج كمية الحرارة
$\lambda=dQ/dm$
اللازمة من اجل تبخر و حدة من كتلة السائل.

\item
استنتج معادلة
كلابيرون
\footnote{.${\rm Clapeyron}$}
التي تربط بين
$P$, $T$, $\lambda$, $V_1$
و
$V_2$.
\item
افترض ان السائل هو ماء و البخار هو غاز مثالي. باعتبار
ان
$V_2>>V_1$
استخرج القانون الرابط بين
$P$, $\lambda$
و
$T$.
\end{itemize}
 \end{itemize}

 \paragraph{
تمرين
$7$:
}
\begin{itemize}
\item
بين انه اذا كانت
$x$, $y$
و
$z$
مرتبطة فيما بينها بمعادلة حالة فان
\begin{eqnarray}
(\frac{\partial x}{\partial y})_z (\frac{\partial y}{\partial x})_z=1~,~ (\frac{\partial x}{\partial y})_z(\frac{\partial y}{\partial z})_x=-(\frac{\partial x}{\partial z})_y .
\end{eqnarray} 
\item
لتكن 
$f$
دالة في اثنين فقط من المتغيرات. بين ان
\begin{eqnarray}
(\frac{\partial x}{\partial y})_f (\frac{\partial y}{\partial z})_f=(\frac{\partial x}{\partial z})_f.
\end{eqnarray} 
\end{itemize}
\paragraph{
تمرين
$8$:
}
\begin{itemize}
 \item 
 باستعمال المبدأ الثاني للترموديناميك في معادلة
 ال
 $dQ$
 الاولي استخرج العلاقة
 \begin{eqnarray}
(\frac{\partial P}{\partial T})_V =\frac{1}{T}\bigg[P+(\frac{\partial U}{\partial V})_T\bigg].
\end{eqnarray}
اعد كتابة معادلة ال
$dQ$
الاولي باستعمال هذه العلاقة.
\item
اعد كتابة معادلة ال
$dQ$
الثانية بالمرور عبر نفس الخطوات. نحصل هكذا علي ما يسمي بمعادلات ال
$TdS$.
\item 
اعد كتابة معادلات ال
$TdS$
باستعمال معاملات التمدد الحراري
$\alpha$, 
الانضغاطية الايزوحرارية
$\kappa_T$
و الانضغاطية الادياباتيكية
$\kappa_S$
المعرفة ب
\begin{eqnarray}
\alpha=\frac{1}{V}(\frac{\partial V}{\partial T})_P~,~\kappa_T=-\frac{1}{V}(\frac{\partial V}{\partial P})_T~,~\kappa_S=-\frac{1}{V}(\frac{\partial V}{\partial P})_S.
\end{eqnarray}
هذه المعاملات هي التي تقاس تجريبيا.
\item
احسب
$C_p-C_v$
و
$\gamma=C_p/C_v$.
\end{itemize}
 \paragraph{
تمرين
$9$:
}
\begin{itemize}
 \item  
 انطلاقا من المبدأ الاول للترموديناميك
 $dU=-PdV+TdS$
 استخرج علاقات ماكسويل
 \begin{eqnarray}
(\frac{\partial U}{\partial V})_S=-P~,~(\frac{\partial U}{\partial S})_V=T.
\end{eqnarray}
\item
انطلاقا من تعريف التغير في الطاقة الحرة لهلمولتز
$dF=-PdV-SdT$,
 التغير في الكمون الترموديناميكي
 $dG=-SdT+VdP$
 و التغير في الانتالبي
 $dH=TdS+VdP$
 استخرج علاقات ماكسويل الستة الاخري. 
 \end{itemize}
 
 \paragraph{
تمرين
$10$:
}

تتميز مادة بالخواص التالية
\begin{itemize}
\item
 العمل خلال تحويل ايزوحراري
 $T_0$
 يعطي ب
 \begin{eqnarray}
W
&=&-RT_0 \ln \frac{V}{V_0}.
\end{eqnarray}
\item
الانتروبي يعطي ب
\begin{eqnarray}
S
&=&R\frac{V}{V_0} \big(\frac{T}{T_0}\big)^a.
\end{eqnarray}
\end{itemize}
حيث
$T_0$, $V_0$
و
$a$
ثوابت.
\begin{itemize}
\item 
احسب الطاقة الحرة لهلمولتز.
\item
احسب معادلة الحالة.
\item
احسب العمل من اجل تحويل ايزوحراري
$T$
كيفي.
\end{itemize}

\paragraph{
تمرين
$11$:
}

نعتبر تحويل ترموديناميكي عكسي دوري مشكل من ستة قطع مستقيمة في المخطط
$T-S$
 كالاتي:
\begin{itemize}
 \item
 تحويل ايزوحراري عند درجة الحرارة 
 $T_1$
 من الانتروبي
 $S_1$ 
 الي الانتروبي
 $S_2>S_1$.
 \item
 تحويل عند انتروبي ثابت معطي ب
 $S_2$
 من درجة الحرارة
$T_1$ 
 الي درجة الحرارة
 $T_3>T_1$.
 \item
 تحويل ايزوحراري عند درجة الحرارة
 $T_3$
 من الانتروبي 
 $S_2$
 الي الانتروبي
 $S_3<S_2$.
 \item
 تحويل عند انتروبي ثابت معطي ب
 $S_3$
 من درجة الحرارة
 $T_3$
 الي
 درجة الحرارة
 $T_2>T_3$.
 \item
 تحويل ايزوحراري عند درجة الحرارة
 $T_2$
 من الانتروبي
 $S_3$
 الي الانتروبي
 $S_1<S_3$.
 \item
 تحويل عند انتروبي ثابت معطي ب
 $S_1$
 من درجة الحرارة
 $T_2$
 الي درجة الحرارة
 $T_1<T_2$.
\end{itemize}
احسب العمل و كمية الحرارة الممتصة ثم استنتج المردود. بين ان دورة كارنو التي تعمل بين درجتي الحرارة 
الاعلي و الاخفض لها مردود اعلي.

\newpage
\section*{
حلول
}
\addcontentsline{toc}{section}{
حلول
} 

\paragraph{
تمرين
$1$:
}
\begin{itemize}
 \item 
لان الغاز يتمدد بشكل حر في الفراغ فان الضغط عليه صفر منذ بداية التحول و بالتالي فان العمل ينعدم اي
\begin{eqnarray}
\Delta W=0.
\end{eqnarray}
لان درجة الحرارة لا تتغير فان كمية الحرارة المتبادلة مع الوسط الخارجي تنعدم اي
\begin{eqnarray}
\Delta Q=0.
\end{eqnarray}
اذن
\begin{eqnarray}
\Delta U=0\leftrightarrow U_1=U_2.
\end{eqnarray}
لان 
$U$
هي دالة حالة يمكن ان تتعلق فقط ب
$T$
و
$V$,
 و لان
 $U$
 هي نفسها من اجل
 $(T_1,V_1)$
 و
 $(T_2=T_1,V_2>V_1)$
  فان
 $U$
 لا تتعلق ب
 $V$
 و تتعلق فقط ب
 $T$.
 \item
 لان الغاز مثالي فان
 $U=U(T)$
 و بالتالي فانه خلال التحويل الايزوحراري لدينا
 $\Delta U=0$
 اي
 $\Delta Q=-\Delta W$. 
 نحسب اذن
 \begin{eqnarray}
\Delta W=-\int PdV=-RT \ln \frac{V_2}{V_1}\Rightarrow \Delta Q=RT \ln \frac{V_2}{V_1}.
\end{eqnarray}
 لان التحويل عكسي ايزوحراري فان الفرق في انتروبي الغاز هو يعطي ب
\begin{eqnarray}
(\Delta S)_{\rm gas}=\int \frac{dQ}{T}=\frac{\Delta Q}{T}=R \ln \frac{V_2}{V_1}.
\end{eqnarray}
$\Delta Q$
هي كمية الحرارة الممتصة من الغاز اي ان
$-\Delta Q$
هي كمية الحرارة التي يفقدها الخزان الحراري
$T$. 
الفرق في انتروبي الخزان الحراري هو اذن
\begin{eqnarray}
(\Delta S)_{\rm reservoir}=\int \frac{dQ}{T}=-\frac{\Delta Q}{T}=-R \ln \frac{V_2}{V_1}.
\end{eqnarray}
الفرق في الانتروبي الكلي ينعدم كما يجب بالنسبة لتحويل عكسي. يمكن استخدام العمل المقدم, الذي يمكن تخزينه في نابض مثلا, لعكس التحويل.
 \item 
 تمدد جول هو تحويل غير عكسي و بالتالي لا يمكن تطبيق العلاقة
 $dS=dQ/T$. 
 لكن لان الانتروبي هو دالة حالة لا تتعلق الا بالحالتين الابتدائية و النهائية فان الفرق في انتروبي الغاز ما زال يعطي ب
 \begin{eqnarray}
(\Delta S)_{\rm gas}=R \ln \frac{V_2}{V_1}.
\end{eqnarray}
في هذه الحالة لانه لا يوجد تبادل حراري بين الغاز و الخزان فان انتروبي الخزان يتعدم اي
\begin{eqnarray}
(\Delta S)_{\rm reservoir}=0.
\end{eqnarray}
الانتروبي الكلي اكبر من الصفر يعطي ب
 \begin{eqnarray}
(\Delta S)_{\rm total}=R \ln \frac{V_2}{V_1}.
\end{eqnarray}
تم تضييع كمية العمل
$W$
لان التحويل عكسي.
 \end{itemize}

\paragraph{
تمرين
$2$:
}
الطاقة الداخلية هي مقدار تمديدي يتعلق بدرجة الحرارة و الحجم اي

\begin{eqnarray}
U=U(T,V).
\end{eqnarray}
اذن
\begin{eqnarray}
dU=(\frac{\partial U}{\partial T})_V dT+(\frac{\partial U}{\partial V})_T dV.
\end{eqnarray}
لدينا ايضا
\begin{eqnarray}
dU=dW+dQ=-PdV+dQ.
\end{eqnarray}
من هاتين المعادلتين نستنتج
\begin{eqnarray}
dQ=(\frac{\partial U}{\partial T})_V dT+\bigg[P+(\frac{\partial U}{\partial V})_T\bigg] dV.
\end{eqnarray}
هذه هي معادلة ال
$dQ$
الاولي.
تحت حجم ثابت نحصل علي
\begin{eqnarray}
dQ=(\frac{\partial U}{\partial T})_V dT.
\end{eqnarray}
اذن
\begin{eqnarray}
C_v=(\frac{dQ}{dT})_V=(\frac{\partial U}{\partial T})_V.
\end{eqnarray}
اذا اخترنا
$T$
و
$P$
كمتغيرات مستقلة في الطاقة الداخلية عوض
$T$
و
$V$
فاننا نحصل علي
\begin{eqnarray}
dU&=&(\frac{\partial U}{\partial T})_P dT+(\frac{\partial U}{\partial P})_T dP\nonumber\\
&=&-P dV+dQ\nonumber\\
&=&-P\bigg[(\frac{\partial V}{\partial T})_P dT+(\frac{\partial V}{\partial P})_T dP\bigg]+dQ.
\end{eqnarray}
اذن
\begin{eqnarray}
dِQ&=&\bigg[(\frac{\partial U}{\partial T})_P+P(\frac{\partial V}{\partial T})_P\bigg] dT+\bigg[(\frac{\partial U}{\partial P})_T+P(\frac{\partial V}{\partial P})_T \bigg] dP.
\end{eqnarray}
هذه هي معادلة ال
$dQ$
الثانية.
تحت ضغط ثابت
\begin{eqnarray}
dِQ&=&\bigg[(\frac{\partial U}{\partial T})_P+P(\frac{\partial V}{\partial T})_P\bigg] dT.
\end{eqnarray}
اذن
\begin{eqnarray}
C_p=(\frac{dQ}{dT})_P=(\frac{\partial U}{\partial T})_P+P(\frac{\partial V}{\partial T})_P.
\end{eqnarray}
يمكن كتابة هذه المعادلة علي الشكل
\begin{eqnarray}
C_p=(\frac{dQ}{dT})_P=(\frac{\partial H}{\partial T})_P.
\end{eqnarray}
$H$
هو الانتالبي
\begin{eqnarray}
H=U+PV.
\end{eqnarray}
يمكن استخراج معادلة ال
$dQ$
الاخيرة بنفس الطريقة لنجد
\begin{eqnarray}
dِQ&=&(\frac{\partial U}{\partial P})_VdP+\bigg[P+(\frac{\partial U}{\partial V})_P\bigg] dV.
\end{eqnarray}
بالنسبة للغازات المثالية لدينا
\begin{eqnarray}
PV=nRT\Rightarrow V=\frac{nRT}{P}.
\end{eqnarray}
\begin{eqnarray}
(\frac{\partial V}{\partial T})_P=\frac{nR}{P}\Rightarrow P(\frac{\partial V}{\partial T})_P=nR.
\end{eqnarray}
بالتالي
\begin{eqnarray}
C_p-C_v&=&(\frac{\partial U}{\partial T})_P+P(\frac{\partial V}{\partial T})_P-(\frac{\partial U}{\partial T})_V\nonumber\\
&=&nR,
\end{eqnarray}
لان الطاقة الطاقة الداخلية لا تتعلق الا بدرجة الحرارة بالنسبة الي غاز مثالي اي
\begin{eqnarray}
(\frac{\partial U}{\partial T})_P=(\frac{\partial U}{\partial T})_V.
\end{eqnarray}
\paragraph{
تمرين
$3$:
}
مباشرة لدينا في كل دورة
$n$:
\begin{eqnarray}
-W=Q_{n+1}^{'}-Q_n^{'}.
\end{eqnarray}
ايضا
\begin{eqnarray}
\frac{Q_{n+1}^{'}}{Q_n^{'}}=\frac{T_{n+1}}{T_n}\Rightarrow \frac{T_{n+1}}{Q_{n+1}^{'}}=\frac{T_{n}}{Q_{n}^{'}}=x.
\end{eqnarray}
$x$
لا يتعلق ب 
$n$.
 باستخدام العلاقة الاخيرة في العلاقة الاولي نحصل علي
 \begin{eqnarray}
 T_{n+1}=T_n-xW.
\end{eqnarray}
باختيار
$T_1= 0~{\rm K}$ 
و
$xW=-1~{\rm K}$ 
نحصل علي سلم منتظم لدرجة الحرارة المطلقة.

 \paragraph{
تمرين
$4$:
}
\begin{itemize}
 \item

من اجل غاز مثالي لدينا

\begin{eqnarray}
 PV=RT.
\end{eqnarray}
العمل في الحالة الاولي التي هي عبارة عن تحويل عكسي ايزوحراري هو
\begin{eqnarray}
 W=-\int P dV=-RT\int_{V_1}^{V_2}\frac{dV}{V}=-RT\ln \frac{V_2}{V_1}=RT\ln \frac{P_2}{P_1}.
\end{eqnarray}
\item
العمل في الحالة الثانية التي هي عبارة عن تحويل غير عكسي يتم تغيير الضغط فيه بغته من
$P_1$
الي
$P_2$
و بالتالي فان الضغط  يساوي
$P_2$
خلال كل التحويل هو
\begin{eqnarray}
 W=-\int P dV=-P_2\int_{V_1}^{V_2}dV=RT(\frac{P_2}{P_1}-1).
\end{eqnarray}
\item
في الحالة الثالثة نركب تحويلين شبيهين بالتحويل الثاني. اذن لدينا
\begin{eqnarray}
&&P_1\longrightarrow P_3~:~W=RT(\frac{P_3}{P_1}-1)\nonumber\\
&&P_3\longrightarrow P_2~:~W=RT(\frac{P_2}{P_3}-1).
\end{eqnarray}
اذن العمل الكلي في الحالة الثالثة هو
\begin{eqnarray}
W=RT(\frac{P_3}{P_1}-1)+RT(\frac{P_2}{P_3}-1).
\end{eqnarray}
العمل بالقيمة المطلقة هو اعظمي في التحويل شبه الساكن العكسي.
\item
الحالة الرابعة هي تحويل عكسي ادياباتيكي اي
\begin{eqnarray}
 dQ=0\Rightarrow dU=dW=-P dV.
\end{eqnarray}
من اجل غاز مثالي
\begin{eqnarray}
U=U(T)~,~dU=C_v dT=C_v\frac{V}{R}dP+C_v\frac{P}{R}dV.
\end{eqnarray}
من المعادلتين اعلاه نحصل علي
\begin{eqnarray}
V dP=-P(1+\frac{R}{C_v})dV=-P\gamma dV.
\end{eqnarray}
المكاملة تعطي مباشرة
\begin{eqnarray}
PV^{\gamma}={\rm constant}.
\end{eqnarray}
\item
لدينا مباشرة
\begin{eqnarray}
P_1V_1^{\gamma}=P_2V_2^{\gamma}\Rightarrow T_1V_1^{\gamma -1}=T_2V_2^{\gamma -1}.
\end{eqnarray}
اذن
\begin{eqnarray}
 T_2=T_1 \bigg(\frac{V_1}{V_2}\bigg)^{\gamma -1}.
\end{eqnarray}
\end{itemize}
\paragraph{
تمرين
$5$:
}
\begin{itemize}
 \item
 ننطلق من
 \begin{eqnarray}
 dU=dW+dQ=-PdV+TdS\Rightarrow dS=\frac{1}{T}(\frac{\partial U}{\partial T})_VdT+\frac{1}{T}\big(P+(\frac{\partial U}{\partial V})_T\big)dV.\nonumber\\
\end{eqnarray}
حتي تكون 
$S$
دالة حالة يجب ان يكون لدينا
\begin{eqnarray}
(\frac{\partial S}{\partial T})_V=\frac{1}{T}(\frac{\partial U}{\partial T})_V\Rightarrow \frac{\partial^2 S}{\partial V\partial T}=\frac{1}{T}\frac{\partial^2 U}{\partial V\partial T}.
 \end{eqnarray}
\begin{eqnarray}
(\frac{\partial S}{\partial V})_T=\frac{1}{T}\big(P+(\frac{\partial U}{\partial V})_T\big)\Rightarrow \frac{\partial^2 S}{\partial T\partial V}=-\frac{1}{T^2}\big(P+(\frac{\partial U}{\partial V})_T\big)+ \frac{1}{T}\big((\frac{\partial P}{\partial T})_V+\frac{\partial^2 U}{\partial T\partial V}\big).
 \end{eqnarray}
 بمقارنة المعادلتين اعلاه نحصل علي العلاقة
 \begin{eqnarray}
(\frac{\partial U}{\partial V})_T=-P+T(\frac{\partial P}{\partial T})_V.
\end{eqnarray}
\item
من اجل الغاز المثالي لدينا معادلة الحالة
\begin{eqnarray}
P=\frac{RT}{V}\Rightarrow T(\frac{\partial P}{\partial T})_V=P.
\end{eqnarray}
اذن
\begin{eqnarray}
(\frac{\partial U}{\partial V})_T=0\Rightarrow U=U(T).
\end{eqnarray}
الطاقة الداخلية لغاز مثالي لا تتعلق الا بدرجة حرارته.
\item
نحسب من اجل الغاز المثالي
\begin{eqnarray}
(\frac{\partial S}{\partial T})_V=\frac{1}{T}(\frac{\partial U}{\partial T})_V=\frac{C_v}{T}.
\end{eqnarray}
\begin{eqnarray}
(\frac{\partial S}{\partial V})_T=\frac{1}{T}\big(P+(\frac{\partial U}{\partial V})_T\big)=\frac{P}{T}=\frac{R}{V}.
\end{eqnarray}
بالتالي

\begin{eqnarray}
dS&=&\frac{C_v}{T} dT+\frac{R}{V}dV\nonumber\\
&=&C_v d\ln T+R d\ln V.
\end{eqnarray}
المكاملة هنا سهلة و نحصل مباشرة علي
\begin{eqnarray}
S&=&S_0+C_v \ln T+R \ln V\nonumber\\
&=&S_0+C_v\ln TV^{\frac{R}{C_v}}\nonumber\\
&=&S_0+C_v\ln T V^{\gamma -1}.
\end{eqnarray}

 \end{itemize}
 
 \paragraph{
تمرين
$6$:
}
 \begin{itemize}
  \item 
انظر الي الشكل
$4$. 
الخط المستقيم يوافق التوازن بين السائل و البخار من اجل درجة حرارة معينه. خلال هذا التوازن فان تناقص الحجم
 لا يؤدي الي تغير في الضغط لكن يؤدي فقط الي تغير في كتلة السائل. تحت حجم معين فانه لا يتبقي الا السائل في الاسطوانة و اي تناقص في الحجم هنا يؤدي الي تزايد في الضغط. فوق حجم معين فانه لا يوجد الا بخار في الاسطوانة و اي تزايد في الحجم هنا يؤدي الي تناقص في الضغط. 
 
 اذا زدنا درجة الحرارة فان ضغط البخار المشبع يزداد و يضيق الخط المستقيم الموافق للتوازن. اذا تعدت درجة الحرارة 
 $T$
 درجة حرارة حرجة
 $T_c$
 فان البخار فقط هو الذي يتبقي في الاسطوانة مهما كان الحجم.
\item
\begin{itemize}
\item 
الحجم الكلي و الطاقة الداخلية الكلية يعطيان بالمعادلات
\begin{eqnarray}
V=m_1V_1+m_2V_2.
\end{eqnarray}
\begin{eqnarray}
U=m_1U_1+m_2U_2.
\end{eqnarray}
اذا كانت
$dm$
هي كتلة السائل التي تبخرت خلال التحويل الايزوحراري المتناه في الصغر فان التغير في الحجم و التغير في الطاقة الداخلية يعطيان ب
\begin{eqnarray}
V+dV=(m_1-dm)V_1+(m_2+dm)V_2\Rightarrow dV=(V_2-V_1)dm.
\end{eqnarray}
\begin{eqnarray}
U+dU=(m_1-dm)U_1+(m_2+dm)U_2\Rightarrow dU=(U_2-U_1)dm.
\end{eqnarray}
كمية الحرارة في وحدة الكتل تعطي اذن ب
\begin{eqnarray}
\lambda=\frac{dQ}{dm}&=&\frac{dU}{dm}+P\frac{dV}{dm}\nonumber\\
&=&U_2-U_1+P(V_2-V_1).
\end{eqnarray}
\item
من النتائج اعلاه لدينا مباشرة
\begin{eqnarray}
dU=(U_2-U_1)dm=\frac{U_2-U_1}{V_2-V_1}dV\Rightarrow (\frac{\partial U}{\partial V})_T=\frac{U_2-U_1}{V_2-V_1}.
\end{eqnarray}
لكننا نعرف ان
\begin{eqnarray}
(\frac{\partial U}{\partial V})_T=T(\frac{\partial P}{\partial T})_V-P.
\end{eqnarray}
اذن نحصل علي
\begin{eqnarray}
\frac{U_2-U_1}{V_2-V_1}=T\frac{dP}{dT}-P\Rightarrow \frac{dP}{dT}=\frac{1}{T}\frac{\lambda}{V_2-V_1}.
\end{eqnarray}
هذه هي معادلة كلابيرون.
\item
اذا كان حجم البخار اكبر بكثير من حجم السائل اي
$V_2>>V_1$
نحصل علي
\begin{eqnarray}
\frac{dP}{dT}=\frac{1}{T}\frac{\lambda}{V_2}.
\end{eqnarray}
اذا افترضنا ايضا ان البخار هو غاز مثالي فان
\begin{eqnarray}
V_2=\frac{nRT}{P}.
\end{eqnarray}
اذن
\begin{eqnarray}
\frac{dP}{P}=\frac{\lambda}{nR}\frac{dT}{T^2}\Rightarrow P=P_0\exp(-\frac{\lambda}{nRT}).
\end{eqnarray}
\end{itemize}
 \end{itemize}

 \paragraph{
تمرين
$7$:
}
\begin{itemize}
\item
 كمثال نعتبر 
$x=T$, $y=V$
و 
$z=P$.
 نأخذ
 $T$
 و
 $V$ 
هي المتغيرات المستقلة ونعبر عن 
$P$ 
بدلالتهما.  بالتالي
\begin{eqnarray}
dP&=&(\frac{\partial P}{\partial T})_V dT+(\frac{\partial P}{\partial V})_T dV\nonumber\\
&=&(\frac{\partial P}{\partial T})_V\bigg((\frac{\partial T}{\partial V})_P dV+(\frac{\partial T}{\partial P})_V dP \bigg)+(\frac{\partial P}{\partial V})_T dV.
\end{eqnarray}
نستنتج اذن
\begin{eqnarray}
(\frac{\partial P}{\partial T})_V (\frac{\partial T}{\partial P})_V=1~,~ (\frac{\partial P}{\partial T})_V(\frac{\partial T}{\partial V})_P=-(\frac{\partial P}{\partial V})_T .
\end{eqnarray}
\item 
كمثال نعتبر
$f=S$. 
لدينا
$f=f(x,y)$. 
نأخذ
$x$ 
كدالة في
$f$
و
$y$. 
نحصل مباشرة علي العلاقات
\begin{eqnarray}
(\frac{\partial f}{\partial x})_y (\frac{\partial x}{\partial f})_y=1.
\end{eqnarray}
\begin{eqnarray}
(\frac{\partial x}{\partial f})_y (\frac{\partial f}{\partial y})_x=-(\frac{\partial x}{\partial y})_f.
\end{eqnarray}
من الجهة الاخري
اذا اعتبرنا
$f$
دالة في 
$x$ 
و
$z$
فانه
يجب ان نعبر عن
$y$
بدلالة
$x$
و
$z$.
نحصل علي العلاقة

\begin{eqnarray}
(\frac{\partial f}{\partial y})_x (\frac{\partial y}{\partial z})_x=(\frac{\partial f}{\partial z})_x.
\end{eqnarray}
نحسب الان
\begin{eqnarray}
(\frac{\partial x}{\partial y})_f (\frac{\partial y}{\partial z})_f (\frac{\partial z}{\partial x})_f &=& -(\frac{\partial f}{\partial y})_x (\frac{\partial z}{\partial f})_x. (\frac{\partial f}{\partial z})_y (\frac{\partial x}{\partial f})_y. (\frac{\partial f}{\partial x})_z (\frac{\partial y}{\partial f})_z\nonumber\\
&=&-(\frac{\partial z}{\partial y})_x.(\frac{\partial x}{\partial z})_y.(\frac{\partial y}{\partial x})_z\nonumber\\
&=&1.
\end{eqnarray}
\end{itemize}

 \paragraph{
تمرين
$8$:
}
\begin{itemize}
 \item 
معادلة ال
$dQ$
الاولي تعطي ب
\begin{eqnarray}
dQ&=&(\frac{\partial U}{\partial T})_V dT+\bigg[P+(\frac{\partial U}{\partial V})_T\bigg] dV\nonumber\\
T dS&=&C_v dT+\bigg[P+(\frac{\partial U}{\partial V})_T\bigg] dV.
\end{eqnarray}
لان الانتروبي  هو دالة حالة فانه لدينا مباشرة
\begin{eqnarray}
\frac{\partial}{\partial V}_T\frac{C_v}{T}=\frac{\partial}{\partial T}_V\bigg[P+(\frac{\partial U}{\partial V})_T\bigg].
\end{eqnarray}
نحصل علي
\begin{eqnarray}
(\frac{\partial P}{\partial T})_V =\frac{1}{T}\bigg[P+(\frac{\partial U}{\partial V})_T\bigg].
\end{eqnarray}
بالتعويض فان معادلة ال
$dQ$
الاولي تصبح
\begin{eqnarray}
T dS&=&C_v dT+T  (\frac{\partial P}{\partial T})_V dV.
\end{eqnarray}
\item
معادلة
$dQ$
الثانية هي
\begin{eqnarray}
dِQ&=&\bigg[(\frac{\partial U}{\partial T})_P+P(\frac{\partial V}{\partial T})_P\bigg] dT+\bigg[(\frac{\partial U}{\partial P})_T+P(\frac{\partial V}{\partial P})_T \bigg] dP\nonumber\\
T dS&=&C_p dT+\bigg[(\frac{\partial U}{\partial P})_T+P(\frac{\partial V}{\partial P})_T \bigg] dP.
\end{eqnarray}
بالمرور عبر نفس الخطوات يمكن ان نبين ان
\begin{eqnarray}
-T(\frac{\partial V}{\partial T})_P=(\frac{\partial U}{\partial P})_T+P(\frac{\partial V}{\partial P})_T .
\end{eqnarray}
اذن معادلة ال
$dQ$
الثانية تأخذ الشكل
\begin{eqnarray}
T dS&=&C_p dT-T(\frac{\partial V}{\partial T})_P dP.
\end{eqnarray}
\item 
معادلة ال
$dQ$
الثانية نعبر عليها مباشرة بدلالة معامل التمدد الحراري
$\alpha$
كالتالي
\begin{eqnarray}
T dS&=&C_p dT-\alpha TV dP.
\end{eqnarray}
معادلة 
$dQ$
الاولي تحتاج الي عمل اكثر باستعمال نتائج التمرين السابق. لدينا
\begin{eqnarray}
(\frac{\partial P}{\partial T})_V &=&-(\frac{\partial P}{\partial V})_V (\frac{\partial V}{\partial T})_V\nonumber\\
&=&\frac{\alpha}{\kappa_T}.
\end{eqnarray}
اذن معادلة
$dQ$
الاولي يمكن ان نعبر عليها بدلالة معامل التمدد الحراري
$\alpha$
و معامل الانضغاطية الايزوحرارية كالتالي
\begin{eqnarray}
T dS&=&C_v dT+\frac{\alpha}{\kappa_T} T dV.
\end{eqnarray}
\item
بمساواة معادلتي ال
$dQ$
اعلاه نحصل مباشرة علي
\begin{eqnarray}
C_p dT-\alpha TV dP=C_v dT+\frac{\alpha}{\kappa_T} T dV.
\end{eqnarray}
نأخذ كمتغيرات مستقلة
$P$
و
$V$. 
نصل الي المعادلة
\begin{eqnarray}
\bigg((C_p-C_v)(\frac{\partial T}{\partial V})_P-\frac{\alpha}{\kappa_T}T\bigg)dV +\bigg((C_p-C_v)(\frac{\partial T}{\partial P})_V-\alpha TV\bigg)dP =0.
\end{eqnarray}
منه نستنتج ان
\begin{eqnarray}
C_p-C_v=\frac{\alpha^2}{\kappa_T}TV.
\end{eqnarray}
من الجهة الاخري فانه لدينا من اجل التحويلات الادياباتيكية
\begin{eqnarray}
C_p=\alpha TV (\frac{\partial P}{\partial T})_S~,~C_v=-\frac{\alpha T}{\kappa_T} (\frac{\partial V}{\partial T})_S.
\end{eqnarray}
اذن لدينا
\begin{eqnarray}
\gamma&=&\frac{C_p}{C_v}\nonumber\\
&=&-V\kappa_T (\frac{\partial P}{\partial T})_S(\frac{\partial T}{\partial V})_S\nonumber\\
&=&-V\kappa_T (\frac{\partial P}{\partial V})_S\nonumber\\
&=&\frac{\kappa_T}{\kappa_S}.
\end{eqnarray}

\end{itemize}

 \paragraph{
تمرين
$10$:
}
\begin{itemize}
\item 
الطاقة الحرة لهلمولتز تعطي ب
\begin{eqnarray}
dF&=&-PdV-SdT\nonumber\\
&=&dW-SdT.
\end{eqnarray}
خلال التحويل الايزوحراري 
$T_0$
لدينا
\begin{eqnarray}
dF&=&dW\Rightarrow F(T_0,V)=W=-RT_0\ln \frac{V}{V_0}.
\end{eqnarray}
 من جهة اخري خلال التحويل تحت الحجم الثابت
 $V_0$
 لدينا
 \begin{eqnarray}
dF&=&-SdT=-R\frac{V}{V_0}\big(\frac{T}{T_0}\big)^adT\Rightarrow F(T,V)=-\frac{R}{a+1}\frac{V}{V_0}T_0\big(\frac{T}{T_0}\big)^{a+1}+f(V).\nonumber\\
\end{eqnarray}
 بالتعويض ب
 $T=T_0$
في المعادلة الاخيرة نحصل علي
 \begin{eqnarray}
F(T_0,V)=-\frac{R}{a+1}\frac{V}{V_0}T_0+f(V).
\end{eqnarray}
بالمقارنة بالمعادلة السابقة نحصل علي
\begin{eqnarray}
f(V)=-RT_0\ln \frac{V}{V_0}+\frac{R}{a+1}\frac{V}{V_0}T_0.
\end{eqnarray}
 تعطي الطاقة الحرة اذن بالمعادلة
 \begin{eqnarray}
F(T,V)=-RT_0\ln \frac{V}{V_0}+\frac{R}{a+1}\frac{V}{V_0}T_0\big(1-\frac{T^{a+1}}{T_0^{a+1}} \big).
\end{eqnarray}
\item
معادلة الحالة يمكن ان نحصل عليها من علاقة ماكسويل
 \begin{eqnarray}
 -P&=&(\frac{\partial F}{\partial V})_T\nonumber\\
&=&-RT_0 \frac{1}{V}+\frac{R}{a+1}\frac{T_0}{V_0}\big(1-\frac{T^{a+1}}{T_0^{a+1}} \big).
\end{eqnarray}
\item
العمل  من اجل درحة حرارة ثابتة
$T$
كيفية
يحسب مباشرة كالعادة بالعلاقة
\begin{eqnarray}
W&=&-\int P dV\nonumber\\
&=&\int \bigg[-RT_0 \frac{1}{V}+\frac{R}{a+1}\frac{T_0}{V_0}\big(1-\frac{T^{a+1}}{T_0^{a+1}} \big)\bigg]dV\nonumber\\
&=&-RT_0 \ln \frac{V}{V_0}+\frac{R}{a+1}\frac{T_0}{V_0}(V-V_0)\big(1-\frac{T^{a+1}}{T_0^{a+1}} \big).
\end{eqnarray}

 \end{itemize}
 
 \paragraph{
تمرين
$11$:
}
التحويل عكسي دوري اذن
\begin{eqnarray}
\Delta U=0.
\end{eqnarray}
العمل اذن
 \begin{eqnarray}
\Delta W=-\Delta Q.
\end{eqnarray}
لان التحويل عكسي لدينا الاتي
\begin{eqnarray}
d Q=T dS.
\end{eqnarray}
\begin{eqnarray}
\Delta Q=\Delta Q_1 +0+\Delta Q_3+0+\Delta Q_2+0.
\end{eqnarray}
\begin{eqnarray}
\Delta Q_1=T_1(S_2-S_1)=B.
\end{eqnarray}
\begin{eqnarray}
\Delta Q_3=T_3(S_3-S_2).
\end{eqnarray}

\begin{eqnarray}
\Delta Q_2=T_2(S_1-S_3).
\end{eqnarray}

\begin{eqnarray}
\Delta Q_3+\Delta Q_2=-A-B.
\end{eqnarray}
اذن نحصل علي
\begin{eqnarray}
\Delta Q =-A\Rightarrow \Delta W=A.
\end{eqnarray}
كمية الحرارة الممتصة
\begin{eqnarray}
\Delta Q =-\Delta Q_3-\Delta Q_2=A+B.
\end{eqnarray}
اذن المردود هو
\begin{eqnarray}
\eta =\frac{A}{A+B}.
\end{eqnarray}
دورة كارنو التي تعمل بين
$T_1$
و
$T_2$
تعطي عمل اكبر اي
$A$
يزداد لكن
$B$
يبقي نفسه اذن المردود يزداد.

\chapter*{
مدخل الي الميكانيك الاحصائي الكلاسيكي:  المجموعة الميكروقانونية
}
\addcontentsline{toc}{chapter}{$2$
مدخل الي الميكانيك الاحصائي الكلاسيكي:  المجموعة الميكروقانونية
} 

\section*{
الحالات الميكروسكوبية
}
\addcontentsline{toc}{section}{
الحالات الميكروسكوبية
} 
نحدد علي المستوي الماكروسكوبي حالة جملة ترموديناميكية عن طريق اعطاء
قيم محددة لكل المتغيرات الترموديناميكية المستقلة خطيا و هذا ما يحدد الحالة الماكروسكوبية 
\footnote{${\rm macrostate}.$}
للجملة. هذه الحالة الماكروسكوبية الواحدة تقابل عدد هائل من
الحالات الميكروسكوبية
\footnote{${\rm microstates}.$}
للجملة التي يتم في كل واحدة منها تحديد حالة كل المكونات الجزئية
او الذرية اوالنووية لهذه الجملة باستعمال درجات الحرية  المعروفة باسم الاعداد الكمية, اي باستعمال الميكانيك الكمي, رغم انه في بعض الاحيان القليلة تكون درجات الحرية الكلاسيكية كافية.

لان عدد 
المكونات الذرية كبير جدا فان استعمال التقنيات الاحصائية لدراسة الحالات الميكروسكوبية  امر لابد منه و ايضا يمكن ان نري تقريبا بوضوح لماذا يمكن لعدد كبير من الحالات
الميكروسكوبية  ان يقابل حالة ماكروسكوبية واحدة. مثلا فان عدد الحالات الميكروسكوبية 
$\Omega(E)$
التي تقابل الحالة الماكروسكوبية التي تكون فيها طاقة الجملة تساوي
$E$,
بافتراض ان هناك
$n$
 درجة حرية في الجملة,
يعطي بالعلاقة التقريبية
\begin{eqnarray}
  \Omega(E)\sim E^n.
 \end{eqnarray}
 من اجل جملة مشكلة من
 $N$
 جسيم سلمي حر داخل علبة مكعبة فان
 $n=3N$.
 
 عدد الحالات الميكروسكوبية
 $\Omega(E)$
 يتعلق ايضا و بقوة علي طبيعة الجسيمات المكونة للجملة هل هي متطابقة ام لا و  ايضا علي سبين هذه الجسيمات هل هو  عدد صحيح (بوزونات)
 او عدد نصف صحيح (فرميونات),
و هذا العدد يلعب دور مهم جدا كما سنري في الميكانيك الاحصائي, كما ان حسابه هو عملية ليست بالبسيطة عموما.

\section*{
مثال: نموذج ايزينغ - المشاء العشوائي
}
\addcontentsline{toc}{section}{
مثال: نموذج ايزينغ - المشاء العشوائي
} 

نموذج ايزينغ
\footnote{${\rm Ising}~{\rm model}.$}
في بعد واحد يتشكل 
من
$N$
ذرة ذات سبين
$1/2$
علي شبكة خطية. في غياب حقل مغناطيسي خارجي فان احتمال ان تكون المركبة
$S_3$
للسبين مساوية ل
$+1/2$
و
$-1/2$
هو
$p=1/2$
و
$q=1/2$
علي التوالي. عدد الحالات الميكروسكوبية الكلي للجملة هو
\begin{eqnarray}
  \Omega(N)\sim 2^N.
 \end{eqnarray}
اذا كان
$n_1$
هو عدد الذرات التي سبينها علوي و 
$n_2$
هو عدد الذرات التي سبينها سفلي فان 
مركبة السبين الكلية هي
 \begin{eqnarray}
  S_3=\sum_{i=1}^N S_{3,i}=\frac{1}{2}(n_1-n_2)=\frac{1}{2}(2n_1-N).
 \end{eqnarray}
 مركبة السبين الكلية هذه تلعب نفس دور الانتقال
 $x=(n_1-n_2)l$
 في مسألة المشاء العشوائي كما سنري لاحقا.  الحالة الماكروسكوبية للجملة محددة في هذه الحالة بقيمة
 $S_3$.
 عبارة
 $n_1$ 
 بدلالة
 $S_3$
 تعطي ب
 \begin{eqnarray}
 n_1=\frac{2S_3+N}{2}.
 \end{eqnarray}
 عدد الحالات الميكروسكوبية التي لها سبين
 $S_3$
 يساوي الي عدد الطرق التي يمكننا فيها اختيار
 $n_1$
 سبين من بين ال
 $N$
 سبين. هذا العدد يعطي بعدد التبديلات الكلي
 $N!$
 مقسوم علي جداء عددي التبديلات الجزئيين
 $n_1!$
 و
 $n_2!$
 لان الترتيب بين السبينات العلوية او السفلية فيما بينها غير مهم. اذن

 \begin{eqnarray}
 \Omega(n_1,N)=C_{n_1}^N=\frac{N!}{n_1!n_2!}=\frac{N!}{n_1!(N-n_1)!}.
 \end{eqnarray}
 هذا هو عدد الحالات الميكروسكوبية التي تحتوي علي
 $n_1$
 سبين علوي و
 $n_2$
 سبين سفلي. احتمال كل واحدة من هذه الحالات هو بالضبط
 $p^{n_1}q^{n_2}$
 لان
 $p$
 هو احتمال ان يكون السبين علوي و
 $q$
 هو احتمال ان يكون السبين سفلي. اذن احتمال ان نحصل علي
 $n_1$
 سبين علوي
 من بين ال
 $N$
 سبين
 هو
 
 \begin{eqnarray}
 W_N(n_1)=\frac{N!}{n_1!(N-n_1)!}p^{n_1}q^{N-n_1}.
 \end{eqnarray}
 هذا هو توزيع الاحتمال ثنائي الحد. هذه التسمية راجعة الي الخاصية
 \begin{eqnarray}
 \sum_{n_1=0}^NW_N(n_1)=(p+q)^N=1.
 \end{eqnarray}
 نلاحظ ان
 \begin{eqnarray}
  \Omega(n_1,N)=\Omega(N)W_N(n_1).
 \end{eqnarray}
 النتائج اعلاه تطبق ايضا و بالكامل علي مسألة المشاء العشوائي في بعد واحد. نعتبر جسيم يتحرك علي
 خط مستقيم انطلاقا من
 $x=0$
 اما الي اليمين خطوة واحدة تساوي
 $+a$
باحتمال
$p$
او الي اليسار خطوة واحدة تساوي
$-a$
باحتمال
$q$.
بعد
$N$
خطوة موضع المشاء هو
$x=ma$
حيث
$m=n_1-n_2$
و
$N=n_1+n_2$.
من الواضح ان
$-N\leq m\leq +N$
وان
$m$
زوجي اذا كان 
$N$
زوجي و العكس. الحالة الماكروسكوبية تقابل
$m$
ثابت
اما الحالة الميكروسكوبية فتقابل اعطاء
الخطوة الاولي: يمين او يسار,
الخطوة الثانية: يمين او يسار وهكذا الي غاية اخر خطوة. عدد الحالات الميكروسكوبية التي
فيها
$n_1$
خطوة الي اليمين و
$n_2$
جطوة الي اليسار هو
$\Omega(n_1,N)$
و احتمال ان يقوم المشاء ب 
$n_1$
خطوة الي اليمين و
$n_2$
خطوة الي اليسار هو
$W_N(n_1)$.

نحسب الان القيمة المتوسطة 
$<n_1>$,
 التشتت او التفاوت
 \footnote{${\rm dispersion}~{\rm or}~{\rm variance}.$}
 في المتوسط
 $<\Delta n_1^2>$
 و الانحراف المعياري
$\sigma_{n_1}$. 
لدينا
 \begin{eqnarray}
  <n_1>&=&\sum_{n_1}W_N(n_1)n_1\nonumber\\
  &=&\sum_{n_1}\frac{N!}{n_1!(N-n_1)!}p^{n_1}q^{N-n_1}n_1\nonumber\\
  &=&p\frac{\partial}{\partial p}\sum_{n_1}\frac{N!}{n_1!(N-n_1)!}p^{n_1}q^{N-n_1}\nonumber\\
  &=&p\frac{\partial}{\partial p}(p+q)^{N}\nonumber\\
  &=&Np.
 \end{eqnarray}
مباشرة نستنتج ايضا القيم المتوسطة التالية
\begin{eqnarray}
  <n_2>
  &=&Nq.
 \end{eqnarray}
\begin{eqnarray}
  <x>
  &=&N(p-q)a~,~<S_3>=N(p-q)\frac{1}{2}.
 \end{eqnarray}
 التشتت او التفاوت معرف ب
\begin{eqnarray}
  <\Delta n_1^2>&=&<(n_1-<n_1>)^2>\nonumber\\
  &=&<n_1^2>-<n_1>^2.
 \end{eqnarray}
 التشتت يقيس مربع عرض توزيع الاحتمال الذي تخضع له قيم المتغير
 $n_1$
اي
 $W_N(n_1)$.
 اذن
 $\sqrt{<\Delta n_1^2>}$
 يقيس مدي تشتت فيم المتغير
 $n_1$
 حول القيمة المتوسطة.
 الانحراف المعياري هو بالضبط هذا الجذر التربيعي للتشتت اي
 \begin{eqnarray}
  \sigma_{n_1}=\sqrt{<\Delta n_1^2>}.
 \end{eqnarray} 
 نحسب
 \begin{eqnarray}
  <n_1^2>  &=&(p\frac{\partial}{\partial p})^2\sum_{n_1}\frac{N!}{n_1!(N-n_1)!}p^{n_1}q^{N-n_1}\nonumber\\
  &=&Np(Np+q).
 \end{eqnarray}
اذن
\begin{eqnarray}
 \sigma_{n_1}^2&=&Npq.
 \end{eqnarray}
 العرض النسبي لتوزيع الاحتمال
 $W_N(n_1)$
  يعطي اذن ب
  
 \begin{eqnarray}
 \frac{\sigma_{n_1}}{<n_1>}&=&\frac{\sqrt{Npq}}{Np}=\sqrt{\frac{q}{p}}\frac{1}{\sqrt{N}}.
 \end{eqnarray}
 اذن العرض النسبي يقترب من الصفر مثل
 $1/\sqrt{N}$.
 اي ان القيمة المتوسطة هي القيمة الاكثر احتمالا. يمكن رؤية هذا الامر بسهولة اكثر في النهاية
 $N\longrightarrow\infty$ 
 التي يصبح فيها توزيع الاحتمال
 ثنائي الحدين توزيع غوس للاحتمال. يمكن ان نحسب ايضا التشتتات
 
 \begin{eqnarray}
 \sigma_{x}^2&=&4Npqa^2~,~\sigma_{S_3}^2=Npq.
 \end{eqnarray}
 
\section*{
انتروبي المعلومات و مسلمات الميكانيك الاحصائي
}
\addcontentsline{toc}{section}{
انتروبي المعلومات و مسلمات الميكانيك الاحصائي
} 
 \paragraph{انتروبي المعلومات:}
  هو انتروبي يقيس كمية المعلومات  غير المتوفرة لنا او الناقصة عن حالة  جملة احتمالية. 
 
 نأخذ كمثال ملموس جملة مشكلة من
 $N$
 علبة و كرة واحدة. الكرة تعبر عن جسيم مثلا و العلب تمثل الحالات التي يمكن ان يتواجد فيها الجسيم. نفترض ان الكرة موجودة بالضرورة في احدي العلب. يمكن ان يكون صحيحا احد الامرين:
 \begin{itemize}
  \item 
  كل العلب متساوية الاحتمال
  \footnote{.${\rm equiprobable}$}.
  اي ان احتمال وجود الكرة في احدي العلب هو
  $1/N$.
  \item
  العلبة التي توجد بداخلها الكرة عليها علامة تدل علي وجود الكرة بداخلها.
 \end{itemize}
من الواضح جدا انه لدينا معلومات اكثر حول حالة الجملة في الحالة الثانية. اما في الحالة الاولي فان قصورنا عن معرفة يقينية بالعلبة التي توجد فيها الكرة من بين ال
$N$
امكانية يعكس نقص معلوماتنا عن الجملة. انتروبي المعلومات
$I$
هو دالة ارتياب 
تقيس بالضبط كمية المعلومات الناقصة عن الجملة. هذا الانتروبي يجب ان يحقق الاتي:
\begin{enumerate}
 \item اولا:
 $I$
 يجب ان يكون دالة في 
 $N$
 اي
 \begin{eqnarray}
  I=I(N).
 \end{eqnarray}
\item
ثانيا:
اذا زاد عدد العلب فان الارتياب
$I$
يزداد لان كمية المعلومات الناقصة عن الجملة يزداد. بعبارة اخري فان كمية المعلومات المعروفة عن الجملة تتناقص. اذن
\begin{eqnarray}
I(M)>I(N)~,~M>N.
 \end{eqnarray}
 \item
 ثالثا: اذا كانت هناك علبة واحدة فاننا نعرف كل شيء عن الجملة. اي ان كمية المعلومات الناقصة تساوي صفر في هذه الحالة اي
 \begin{eqnarray}
I(1)=0.
 \end{eqnarray}
 \item
 رابعا: اذا قسمنا كل علبة الي
 $M$
 خانة متساوية الاحتمال فانه في الاجمال يكون لدينا
 $NM$
 حجرة متساوية الاحتمال. في هذه الحالة كمية المعلومات الناقصة عن الجملة تعطي ب
 $I(NM)$.
 
 من ناحية اخري كان بالامكان ان نجد العلبة ثم الخانة التي بها الكرة. كما قلنا سابقا فان كمية المعلومات الناقصة عند محاولتنا معرفة العلبة التي بها
 الكرة هي
 $I(N)$.
 بالمثل فان كمية المعلومات الناقصة عند محاولتنا معرفة الخانة التي بها الكرة يجب ان يكون
 $I(M)$.
 اذا افترضنا ان كمية المعلومات هي مقدار اضافي
\footnote{.${\rm additive}$}
فان كمية المعلومات الناقصة الكلية هو المجموع
$I(N)+I(M)$.
كون كمية المعلومات هي مقدار اضافي يعني انه اذا عرفت المعلومات الخاصة بجملة شيئا فشيئا بدون تكرار فان كمية المعلومات الاجمالية تساوي الي مجموع كميات
المعلومات المحصل عليها في كل مرحلة. في هذا المثال عرفنا في المرحلة الاولي ان كمية المعلومات الناقصة هي
$I(N)$ 
ثم عرفنا ان كمية المعلومات الناقصة في المرحلة الثانية هي
$I(M)$.
بالتالي فان كمية المعلومات الناقصة الكلية هي المجموع
$I(N)+I(M)$.

اذن من جهة وجدنا ان كمية المعلومات الناقصة هي
$I(NM)$
و من الحهة الاخري وجدنا ان كمية المعلومات الناقصة هي
$I(N)+I(M)$.
نستنتج مباشرة ان
  \begin{eqnarray}
I(NM)=I(N)+I(M).
 \end{eqnarray}
\end{enumerate}
من المعادلات الاربعة اعلاه يمكننا ان نستنتج ان انتروبي المعلومات يجب ان يعطي بالعلاقة
 \begin{eqnarray}
I(N)=C\ln N.
 \end{eqnarray}
اذا اخذنا 
$C=k$,
حيث 
$k$
هو
ثابت بولتزمان, فان انتروبي المعلومات
$I$
يصبح بالضبط, كما سنبين لاحقا,
الانتروبي الاحصائي
$S$
حيث 
$N$
هو عدد الحالات الميكروسكوبية التي يمكن ان تكون فيها الجملة.

\paragraph{
حساب عدد الحالات الميكروسكوبية- 
توزيع
$N$
 جسم مختلف علي 
$r$
علبة مختلفة
:}

في المثال اعلاه افترضنا ان كل العلب متساوية الاحتمال. لنفترض الان ان العلب غير متساوية الاحتمال. ليكن
$P_i$
احتمال ان تحتل الكرة العلبة
$1\leq i\leq N$.
لدينا
\begin{eqnarray}
P_i\geq 0~,~\sum_{i=1}^NP_i=1.
 \end{eqnarray}
تعرف الاحتمالات
 $P_i$
 في المجموعة التالية.
نعتبر انه لدينا
$N$
جملة متطابقة حيث كل جملة هي عبارة عن كرة موضوعة في علبة من بين 
$N$
علبة.
لتكن
$N_i$
عدد الجمل التي تكون فيها الكرة في العلبة
$i$.
من الواضح انه لما
$N\longrightarrow \infty$
لدينا
\begin{eqnarray}
P_i=\frac{N_i}{N}.
 \end{eqnarray}
اي 
$N_i=NP_i$
هو عدد الجمل التي تكون فيها الكرة في العلبة
$i$. 
لدينا
\begin{eqnarray}
\sum_{i=1}^NN_i=\sum_{i=1}^NNP_i=N.
 \end{eqnarray}
علينا تحديد عدد الحالات الميكروسكوبية المختلفة التي تحقق الشروط الماكروسكوبية 
$=N$
ثابت
و
$=N_i$
ثابتة
.
  هذه المسألة مكافئة لمسألة توزيع
$N$
جسم مختلف, هنا ال
$N$
جملة المعتبرة اعلاه, علي
$r=N$
علبة مختلفة, هنا العلبة تقابل مجموعة الجمل التي فيها الكرة في نفس الوضعية, من دون ان يكون للترتيب داخل العلبة اية اهمية.

اولا لدينا
$N!$
تبديلة مختلفة لل
$N$
جملة. لكن
$N_i=NP_i$
جملة فيها الكرة في العلبة
$i$
و بالتالي فهي جمل متطابقة اي ليس  للترتيب اي اهمية
.
اذن ال
$N_i!$
تبديلة لهذه الجمل تؤدي كلها الي نفس الحالة الميكروسكوبية. عدد الحالات الميكروسكوبية هو اذن
\begin{eqnarray}
{\cal N}=\frac{N!}{\prod_{i=1}^rN_i!}.\label{verify}
 \end{eqnarray}
كل هذه الحالات هي متساوية الاحتمال و بالتالي كمية المعلومات الناقصة عن مجموع ال
$N$
جملة هي
\begin{eqnarray}
I_N=k\ln \frac{N!}{\prod_{i=1}^N(NP_i)!}=k\bigg(\ln N!-\sum_{i=1}^N \ln (NP_i)!\bigg).
 \end{eqnarray}
 نستعمل علاقة ستيرلينغ
 \footnote{.${\rm Stirling}$}
 \begin{eqnarray}
\ln n!=n\ln n -n~,~n\longrightarrow\infty.
 \end{eqnarray}
 نحصل علي
 \begin{eqnarray}
I_N&=&k\bigg(N\ln N -N-\sum_{i=1}^N NP_i \ln NP_i+\sum_{i=1}^NNP_i\bigg)\nonumber\\
&=&-kN\sum_{i=1}^NP_i\ln P_i.
 \end{eqnarray}
 لان كل جملة من ال
 $N$
 جملة
 هي متساوية الاحتمال فان كمية المعلومات الناقصة لكل جملة هي
 $I_N/N$
 اي
 \begin{eqnarray}
I
&=&-k\sum_{i=1}^NP_i\ln P_i.
 \end{eqnarray}
 \paragraph{
 التوازن الاحصائي و الانتروبي الاحصائي:
 }
التوازن الاحصائي يوافق الحالة اين تكون جميع الحالات الميكروسكوبية
متساوية الاحتمال. في هذه الحالة
\begin{eqnarray}
P_i=\frac{1}{\Omega(E)},
 \end{eqnarray}
 حيث
$\Omega(E)$ 
هو عدد الحالات الميكروسكوبية التي لها طاقة
$E$. 
نحصل اذن من اجل جملة معزولة  متوازنة احصائيا علي انتروبي معلومات
$I$
مساو للانتروبي الاحصائي
$S$
المعرف, كما سنبين لاحقا, ب
\begin{eqnarray}
S
=k\ln \Omega(E).
 \end{eqnarray}
   سنبين ايضا لاحقا ان الانتروبي الاحصائي هو نفسه الانتروبي الترموديناميكي الذي عرفناه في الفصل السابق.
لان الانتروبي الاحصائي لا يعرف الا في حالة التوازن فان انتروبي المعلومات هو اذن تعميم للانتروبي الاحصائي للوضعيات الخارجة عن التوازن.

\paragraph{
المسلمة الاولي للميكانيك الاحصائي
:}
استخدمنا في الفقرة السابقة, بدون ان نذكر ذلك صراحة, المسلمة الاولي للميكانيك الاحصائي التي نناقشها الان.

{\bf
النص: 
من اجل جملة معزولة في حالة توازن احصائي فان كل الحالات الميكروسكوبية المسموح بها هي متساوية الاحتمال.
}

اذن اذا كان
$\Omega(E)$
هو عدد الحالات الميكروسكوبية التي تحقق الشرط الماكروسكوبي
$=E$
ثابت
حيث
$E$
هي طاقة الجملة فان احتمال ان تكون الجملة في احدي هذه الحالات الميكروسكوبية هو
$1/\Omega(E)$.
هذا منطقي جدا لان مع معرفتنا لطاقة الجملة فقط لا يوجد اي سبب مسبق يجعلنا نفضل حالة ميكروسكوبية ما علي اخري.
اذا لم يتحقق هذا الامر فان الجملة ليست في حالة توازن و سوف تتطور في الزمن الي ان تبلغ التوازن.

يمكن صياغة هذه المسلمة بدلالة كمية المعلومات الناقصة او انتروبي المعلومات
$I$
كالتالي. نعتبر جملة معزولة و ليكن
$P_i$
احتمال احتلال الحالة الميكروسكوبية
$i$.
انتروبي المعلومات لهذه الجملة معرف ب
\begin{eqnarray}
I
&=&-k\sum_{i} P_i\ln P_i.
 \end{eqnarray}
الاحتمالات
$P_i$
تحقق الشرط
\begin{eqnarray}
\sum_iP_i=1.
 \end{eqnarray}
نريد ايجاد القيمة العظمي ل
$I$
 مع شرط انحفاظ الاحتمال اعلاه. من اجل اجراء هذه العملية نستخدم طريقة مضروبات لاغرانج. نعرف الدالة
 $F$
 ب
 \begin{eqnarray}
F&=&I-\lambda(\sum_i P_i-1)\nonumber\\
&=&-k\sum_{i} P_i\ln P_i-\lambda(\sum_i P_i-1).
 \end{eqnarray}
 المتغير
 $\lambda$
هو بالضبط مضروب لاغرانج. شرط القيم القصوي, اصغرية او اعظمية, بالنسبة للمتغير
$P_i$
 يعطي كالعادة ب
 \begin{eqnarray}
\frac{\partial F}{\partial P_i}&=&\frac{\partial I}{\partial P_i}-\lambda\nonumber\\
&=&-k\ln P_i-k-\lambda\nonumber\\
&=&0\Rightarrow P_i=\exp(-1-\lambda).
 \end{eqnarray}
 باستخدام الان قانون انحفاظ الاحتمال نحصل علي قيمة مضروب لاغرانج
  \begin{eqnarray}
\exp(1+\lambda)=\Omega(E).
 \end{eqnarray}
 اي ان
 \begin{eqnarray}
P_i=\exp(-1-\lambda)=\frac{1}{\Omega(E)}.
 \end{eqnarray}
 نحصل اذن علي توزيع الاحتمال الخاص بالتوازن: كل الحالات الميكروسكوبية متساوية الاحتمال. نلاحظ ايضا ان
  \begin{eqnarray}
\frac{\partial^2 F}{\partial P_i^2}=\frac{\partial^2 I}{\partial P_i^2}<0.
 \end{eqnarray}
 اي عند التوازن فان انتروبي المعلومات اعظمي وبالتالي فان كمية المعلومات المتوفرة عن الجملة اصغرية. نقول ان الفوضي او اللانظام اعظمي و ان الانتروبي هو مقياس الفوضي.

 \paragraph{
المسلمة الثانية للميكانيك الاحصائي- الفرضية الارجودية
:}
استخدمنا  ايضا في الفقرة السابقة, و ايضا بدون ان نذكر ذلك صراحة, المسلمة الثانية للميكانيك الاحصائي التي تعرف ايضا تحت مسمي
 الفرضية الارجودية
\footnote{.${\rm ergodic}~{\rm hypothesis}$}
  و التي نتناولها الان بالنقاش
.

 الحالة الميكروسكوبية التي تتواجد فيها الجملة في اي لحظة زمنية تتغير مع تطور الجملة في الزمن بسبب التفاعلات التي تخضع لها الجملة. اذا لاحظنا الجملة لزمن غير منته فان الزمن الذي
 تقضيه الجملة في كل حالة ميكروسكوبية هو نفسه بالنسبة لكل الحالات و هو مقتضي المسلمة الاولي اعلاه. 
 
 عوض اعتبار جملة واحدة و تتبع تطورها خلال الزمن و هو امر قد يكون صعبا لاسباب واضحة فاننا نعتبر مجموعة من الجمل المتطابقة في لحظة معينة. تشكل هذه المجموعة
 بحيث ان احتمال الحصول علي احد هذه الجمل المتطابقة في حالة
 ميكروسكوبية معينة هو نفسه مهما كانت الحالة الميكروسكوبية. اذا كانت المجموعة مشكلة من 
 $N\longrightarrow \infty$
 جملة متطابقة فان
 $N/\Omega(E)$
 هو عدد الجمل في اي حالة ميكروسكوبية لان احتمال الحصول علي اي حالة ميكروسكوبية هو نفسه معطي ب
 $1/\Omega(E)$:
 الحالات متساوية الاحتمال. الفرضية الارجودية تنص علي الاتي.

 {\bf 
 النص:
 المتوسط في الزمن لمتغير ما يساوي متوسط هذا المتغير مأخوذ علي مجموعة من الجمل المتطابقة  التي لها الخواص المذكورة اعلاه.
 }
 
 اذا كان
 $y=y(t)$
 هو المتغير قيد الدراسة فان المتو سط في الزمن و المتوسط علي مجموعة مشكلة من
 $N$
جملة متطابقة في اللحظة
 $t$
 يعطيان علي التوالي بالعلاقات التالية
   \begin{eqnarray}
   <y>={\rm lim}_{t^{'}\longrightarrow\infty}\frac{1}{t^{'}}\int_0^{t^{'}}y(t^{'})dt^{'}.
 \end{eqnarray}
  \begin{eqnarray}
   <y>_t=\frac{1}{N}\sum_{i=1}^Ny_i(t).
 \end{eqnarray}
 حسب الفرضية الارجودية فانه يجب ان يكون لدينا
   \begin{eqnarray}
   <y>=<y>_t.
 \end{eqnarray}
 \section*{
المجموعة الميكروقانونية
}
\addcontentsline{toc}{section}{
المجموعة الميكروقانونية
} 

\paragraph{
تعريف
:}

المسلمة الاولي للميكانيك الاحصائي التي تنص علي تساوي احتمال الحالات الميكروسكوبية تؤدي مباشرة الي ان اي جملة معزولة في حالة توازن
ترموديناميكي يجب ان تنتمي الي مجموعة احصائية مميزة باحتمال ثابت. هذه المجموعة هي ما يعرف باسم المجموعة الميكروقانونية
\footnote{${\rm microcanonical}~{\rm ensemble}.$}.
 ان استعمال المجموعة الميكروقانونية في التطبيق معقد عموما و بالتالي فاننا نستعمل مكانها تقريبات مثل المجموعة القانونية و المجموعة
 القانونية الكبري.

 من المعلوم ان طاقة جملة معزولة, 
 و لتكن
 $E_0$
 هذه الطاقة,
  ثابته بالضرورة.  في العموم هناك دائما ارتياب في معرفة قيمة الطاقة معطي بالخطا 
$\delta E<<E_0$.
 من الواضح ان هذا الارتياب راجع الي الاخطاء التجريبية و لكن ايضا هو راجع الي التأثيرات الفيزيائية الناجمة 
عن الميكانيك الكمي مثل مبدأ الارتياب لهايزنبرغ
\footnote{.${\rm Heisenberg}~{\rm uncertainty}~{\rm principle}$}
الذي ينص في احد بنوده علي ان الارتياب في الطاقة متناسب عكسا مع المدة المحددة التي يجري فيها القياس 
علي الجملة. طاقة الجملة هي اذن في مجال بين
$E_0$
و
$E_0+\delta E$
اما الحالات الميكروسكوبية المسموح بها للجملة فهي التي لها طاقة 
$E$
حيث
 \begin{eqnarray}
   E_0\leq E\leq E_0+\delta E~,~\delta E<<E_0.
 \end{eqnarray}
 كما فعلنا في السابق عوض اعتبار
 جملة واحدة و اتباع تطورها في الزمن نعتبر مجموعة من الجمل المتطابقة مع الجملة الاصلية من الناحية الماكروسكوبية لا يمكن التمييز بينها باجراء قياسات علي المقادير 
 الماكروسكوبية. كل جملة من هذه المجموعة هي في حالة
 ميكروسكوبية تحقق الشرط اعلاه. 
 نفترض التوازن الاحصائي و بالتالي كل الحالات
 الميكروسكوبية هي متساوية الاحتمال. هذه المجموعة الاحصائية تسمي المجموعة الميكروقانونية.
 
ليكن
$\Omega(E)$
عدد الحالات الميكروسكوبية التي لها طاقة بين
$E$
و
$E+\delta E$.
انتروبي الجملة يعطي بالمعادلة
 \begin{eqnarray}
  S=k\ln \Omega(E).
 \end{eqnarray}
 ليكن
 $\Phi(E)$
 عدد الحالات الميكروسكوبية التي لها طاقة اقل او تساوي من
 $E$.
 من المعروف ان الطاقة في الميكانيك
 الكمي هي في العموم متغير متقطع لكن الفرق في الطاقة
 $\Delta E$
 بين مستويين هو بحيث
 $\Delta E<<\delta E<<E$.
 يمكننا اذن ان نفترض و هو تقريب ممتاز ان
 $E$
 هو متغير مستمر و بالتالي فان
 $\Phi(E)$ 
 هي دالة مستمرة. هذه الدالة تتزايد بسرعة شديدة مع
 $E$. 
  يمكننا ايضا ان نعرف كثافة الحالات الميكروسكوبية اي عدد الحالات في و حدة الطاقة بالعلاقة
  \begin{eqnarray}
  \rho(E)=\frac{d\Phi(E)}{dE}.
 \end{eqnarray}
 اي ان
 \begin{eqnarray}
  \Phi(E)=\int_0^E dE^{'}\rho(E^{'}).
 \end{eqnarray}
 $\rho(E)$
 مثل
 $\Omega(E)$
 دالة تترايد بشدة مع الطاقة.
 $\Phi(E)$
 هي مساحة السطح تحت منحني الدالة
 $\rho(E^{'})$
 بين المحور
 $E^{'}=0$
 و المحور
 $E^{'}=E$.
لان
$\delta E<<E$
فان عدد الحالات
الميكروسكوبية
$\Omega(E)$
يساوي كثافة الحالات الميكروسكوبية
$\rho(E)$
مضروبة في
$\delta E$
اي
\begin{eqnarray}
  \Omega(E)=\rho(E)\delta E.
 \end{eqnarray}
 ليكن
 $n$
 عدد درجات حرية الجملة و
 $\epsilon$
  الطاقة  المتوسطة من اجل درجة حرية واحدة. الطاقة
 $E$
 تعطي   
 اذن بالعلاقة
 \begin{eqnarray}
 E=n\epsilon.
 \end{eqnarray}
  لدينا اذن
 \begin{eqnarray}
  S=k\bigg[\ln \rho(E)\epsilon+\ln \frac{\delta E}{\epsilon}\bigg].
 \end{eqnarray}
  في العموم يتعلق عدد الحالات الميكروسكوبية
 $\Omega(E)$
  بالطاقة
   $E$
  كالاتي
 \begin{eqnarray}
  \Omega(E)\sim E^n.
 \end{eqnarray}
 اذن
 $\rho(E)$
 يتصرف مثل 
 $E^n$
 و بالتالي يتصرف الحد الاول في الانتروبي اعلاه مثل
 $n  \ln E$,
 اما الحد الثاني فانه يتصرف مثل
 $\ln n$.
اذن لان 
 $n$ 
 كبير جدا فان الحد الاول في الانتروبي
  يهيمن بالكامل علي قيمة الانتروبي. نحصل اذن علي العلاقة
  \begin{eqnarray}
  S=k\ln \rho(E).
 \end{eqnarray}
 \paragraph{
اشتقاق الترموديناميك
:}
في المجموعة الميكروقانونية كل جملة تحتوي علي 
$N$
جسيم
و لها حجم
$V$
و طاقة 
بين
$E$
و
$E+\delta E$.
اذا افترضنا ان الميكانيك الكلاسيكي قابل للتطبيق فان
عدد الحالات الميكروسكوبية
$\Omega(E)\delta E$
التي لها طاقة بين
$E$
و
$E+\delta E$
هو متناسب مع الحجم الذي تحتله المجموعة الميكروقانونية في الفضاء الطوري اي
 \begin{eqnarray}
 \Omega(E)\sim \Gamma(E)=\int_{E\leq H\leq E+\delta E} d^{3N}pd^{3N}q.
 \end{eqnarray}
بالمقابل فان عدد الحالات الميكروسكوبية
$\Phi(E)$
التي لها طاقة اقل او تساوي من
$E$
متناسب مع الحجم في الفضاء الطوري المحتوي داخل سطح الطاقة ذو الطاقة
$E$
اي

 \begin{eqnarray}
 \Phi(E)\sim \Sigma(E)=\int_{H\leq E} d^{3N}pd^{3N}q.
 \end{eqnarray}
 من الواضح ان
 \begin{eqnarray}
 \Gamma(E)= \Sigma(E+\delta E)-\Sigma(E).
 \end{eqnarray}
 من اجل
 $\delta E\longrightarrow 0$
 نحصل علي
$ \Gamma(E)= \delta E \rho(E)$
حيث نعرف الان كثافة الحالات
$\rho(E)$
بالعلاقة
\begin{eqnarray}
 \rho(E)= \frac{\partial\Sigma (E)}{\partial E}.
 \end{eqnarray}
كما بينا اعلاه يمكن ان نعرف الانتروبي باحدي العلاقات
 \begin{eqnarray}
  S=k\ln \Gamma(E).\label{entropymC}
 \end{eqnarray}
\begin{eqnarray}
  S=k\ln \rho(E).
 \end{eqnarray}
 هذا التعريف للانتروبي الاحصائي  يؤدي, كما سنبين
 في التمرينات, الي الانتروبي الترموديناميكي بكل خواصه المعروفة مثل الخاصية التمديدية و المبدأ الثاني للترموديناميك. 
 
 يمكننا الان اشتقاق  كل الترموديناميك  انطلاقا من المجموعة  الميكرقانونية باستعمال هذا التعريف للانتروبي كالاتي. نحتاج اولا الي تعريف 
 التحويلات الترموديناميكية شبه الساكنة في هذا الاطار. هذه التحويلات تقابل هنا التغيرات البطيئة جدا في 
 الطاقة و الحجم الناجمة عن تفاعلات الجملة مع الوسط الخارجي. خلال هذه التحويلات فان المجموعة الميكروقانونية تمثل بمجموعة
 من النقاط موزعة بانتطام ( مسلمة تساوي الاحتمال) في حجم يتحرك ببطء شديد في الفضاء الطوري حيث في 
 كل لحظة لدينا مجموعة ميكروقانونية. التغير المتناه في الصغر في الانتروبي خلال هذه التحويلات الترموديناميكية يعطي ب
 \begin{eqnarray}
  dS(E,V)=\big(\frac{\partial S}{\partial E}\big)_VdE+\big(\frac{\partial S}{\partial V}\big)_EdE.
 \end{eqnarray}
 نعرف درجة الحرارة و الضغط بالعلاقات
 \begin{eqnarray}
\big(\frac{\partial S}{\partial E}\big)_V=\frac{1}{T}~,~\big(\frac{\partial S}{\partial V}\big)_E=\frac{P}{T}.
 \end{eqnarray}
 نحصل مباشرة علي المبدأ الاول للترموديناميك
 \begin{eqnarray}
 dE=-PdV+TdS.
 \end{eqnarray}
 اذن للحصول علي الترموديناميك انطلاقا من المجموعة الميكروقانونية نتبع الخطوات التالية:
 \begin{itemize}
  \item 
  احسب كثافة الحالات
  $\rho(E)$
  انطلاقا من الهاميلتونية.
  \item
  احسب الانتروبي باستعمال العلاقة
  \begin{eqnarray}
  S=k\ln \rho(E).
 \end{eqnarray}
 \item
 اقلب الدالة
 $S=S(E,V)$
 من اجل حساب
 $E$
 بدلالة
 $S$
 و
 $V$.
 النتيجة هي بالضبط الطاقة الداخلية
 اي
 \begin{eqnarray}
  U=E(S,V).
 \end{eqnarray}
 \item
 احسب باقي المقادير الترمودياميكية باستعمال العلاقات
 \begin{eqnarray}
  T=\big(\frac{\partial U}{\partial S}\big)_V.
 \end{eqnarray}
 \begin{eqnarray}
  P=-\big(\frac{\partial U}{\partial V}\big)_S.
 \end{eqnarray}
 \begin{eqnarray}
  F=U-TS.
 \end{eqnarray}
 
 \begin{eqnarray}
  G=U+PV-TS.
 \end{eqnarray}
 \begin{eqnarray}
  C_v=\big(\frac{\partial U}{\partial T}\big)_S.
 \end{eqnarray}
 \end{itemize}

 \section*{
التوازن الترموديناميكي
}
\addcontentsline{toc}{section}{
التوازن الترموديناميكي
} 
كل جملة معزولة تتطور في الزمن الي ان تبلغ حالة توازنها اين تصيح الحالات الميكروسكوبية متساوية الاحتمال و يصبح الانتروبي- انتروبي المعلومات-
اعظمي. هذا هو ما ينص عليه المبدأ الثاني للترموديناميك في حالته الميكروسكوبية. الانتروبي الاحصائي الذي  هو مقياس
اللانظام في الجملة هو متناسب مع لوغاريتم عدد الحالات الميكروسكوبية. اذن اللانظام يصبح اعظمي عند التوازن و المقصود به ان عدد الحالات المسموح بها للجملة 
يصبح اعظمي عند التوازن.

عدد الحالات الميكروسكوبية المسموح بها لجملة ماكروسكوبية هو في العموم دالة في الطاقة و ايضا في الحجم و في عدد الجسيمات اي
\begin{eqnarray}
 \Omega=\Omega(E,V,N).
 \end{eqnarray}
 بالتالي
 \begin{eqnarray}
 S=S(E,V,N).
 \end{eqnarray}
 نعتبر جملتين ماكروسكوبيتين 
 $1$
 و
 $2$.
 المتغيرات الترموديناميكية هي
 $E_1$, $V_1$, $N_1$
بالنسبة للجملة
 $1$
 و
 $E_2$, $V_2$, $N_2$
 بالنسبة للجملة
 $2$.
 الجملة الكلية
 $1+2$
 هي جملة معزولة بجدران ادياباتيكية لها طاقة
 $E_0$,
 حجم
 $V_0$
 و عدد جسيمات
 $N_0$
 كلها متغيرات ثابتة. ليكن
 $\Omega_T(E_0,V_0,N_0)$
 عدد الحالات الميكروسكوبية المسموح بها للجملة
 $1+2$
 بدون اي شروط قسرية علي الجملتين
 $1$
 و
 $2$ 
 اي اننا نأخذ بعين الاعتبار كل التمثيلات الممكنة للجملتين
 $1$
 و
 $2$.
\paragraph{ 
التوازن الحراري:
} 
 نفترض اولا ان الجملتين 
 $1$
 و
 $2$
 مفصولتين بجدار دياتارم ثابت و غير نفاذ للجسيمات
 \footnote{.${\rm impermeable}$}
 . اذن
 $N_1$, $N_2$, $V_1$, $V_2$
 تبقي ثابتة لكن هناك تبادل للحرارة بين الجملتين
 $1$
 و
 $2$.
نفترض ان طاقة التفاعل بين
الجملتين
$1$
و
$2$
هي مهملة بالمقارنة مع الطاقات الداخلية
$E_1$
و
$E_2$.
اذن مبدأ انحفاظ الطاقة يعطي مباشرة
$E_0=E_1+E_2$.
ايضا نستنتج مباشرة ان عدد الحالات الميكروسكوبية المسموح بها للجملة الكلية  هو
\begin{eqnarray}
 \Omega(E_0,E_1)=\Omega_1(E_1)\Omega_2(E_2)=\Omega(E_1)\Omega_2(E_0-E_1).
 \end{eqnarray}
الجداء راجع الي ان كل حالة ميكروسكوبية لاي من الجملتين
$1$
او
$2$
يمكن ان يرفق بكل الحالات الميكروسكوبية المسموح بها للجملة الاخري. عدد الحالات الميكروسكوبية الكلي المسموح بها للجملة الكلية هو
\begin{eqnarray}
\Omega_T(E_0)=\sum_{E_1} \Omega(E_0,E_1).
 \end{eqnarray}
اذن الاحتمال
$P(E_1)$
حتي تكون طاقة الجملة
$1$
تساوي
$E_1$
هو

\begin{eqnarray}
P(E_1)=\frac{\Omega(E_0,E_1)}{\Omega_T(E_0)}.
 \end{eqnarray}
 الانتروبي المرفق بعدد الحالات 
 $ \Omega(E_0,E_1)$
 هو
 
 \begin{eqnarray}
 S(E_0,E_1)&=&k\ln \Omega_T(E_0)\nonumber\\
 &=&S_1(E_1)+S_2(E_0-E_1).
 \end{eqnarray}
 اذن الانتروبي هو مقدار تمددي كما يجب و كما ينص عليه الترموديناميك. 
 الجملتان  
 $1$
 و
 $2$
تتبادلان الحرارة عبر الجدار الدياتارم الي غاية ان يحصل توازن حراري اي لما يصبح
$\Omega(E_0,E_1)$
او
$S(E_0,E_1)$
اعظمي او لما يكون الاحتمال
$P(E_1)$
اعظمي. لدينا
\begin{eqnarray}
\frac{dS}{dE_1}&=&\frac{dS_1}{dE_1}+\frac{dS_2}{dE_2}\frac{dE_2}{dE_1}\nonumber\\
&=&\frac{dS_1}{dE_1}-\frac{dS_2}{dE_2}\nonumber\\
&=&0\Rightarrow \frac{dS_1}{dE_1}=\frac{dS_2}{dE_2}.
 \end{eqnarray}
يمكن ان نبين ان هذه القيمة القصوي هي قيمة اعظمية باستعمال  ايجابية السعة الحرارية. 
نعرف درجة الحرارة المطلقة  ب
$1/T=dS/dE$.
 بصفة عامة
\begin{eqnarray}
(\frac{\partial S}{\partial E})_{V,N}&=&\frac{1}{T}.
 \end{eqnarray}
 اذن عند التوازن الحراري نحصل علي تساوي درجة حرارة الجملتين
 $1$
 و
 $2$
 اي
 \begin{eqnarray}
T_1=T_2.
 \end{eqnarray}
 هذا هو المبدأ الصفر للترموديناميك.

عند تطور الجملة في الزمن فان عدد الحالات الميكروسكوبية يتغير من عدد ابتدائي
$\Omega_i$
الي عدد نهائي
$\Omega_f$.
اذا كان
$\Omega_f>\Omega_i$
فان التحول غير عكسي و اذا كان
$\Omega_f=\Omega_i$
فان التحول عكسي. 

 نعتبر تحويل شبه ساكن متناه في الصغر, اي ان الجملة  الكلية تبقي دائما في حالة توازن احصائي مثلا خلال تحويل عكسي, 
 حيث  ايضا يبقي الحجم و عدد الجسيمات في كل جملة  
 ثابتا. من الواضح انه خلال هذا التحويل لا يوجد عمل ميكانيكي  و بالتالي فان
 $\Delta E_1=Q_1$
 حيث
 $Q_1$
 هي كمية الحرارة المتبادلة. من الجهة الاخري فان
  درجة الحرارة
 $T_1$
 تبقي ثابتة لان التحول متناه في الصغر. اذن
 
 \begin{eqnarray}
\Delta S_1=\frac{\Delta E_1}{T_1}=\frac{Q_1}{T_1}.
 \end{eqnarray}
 حسب المبدأ الثاني للترموديناميك فانه خلال التحول الذي ادي الي التوازن الترموديناميكي فان 
 الانتروبي الكلي لا يمكن الا ان يزداد لان الجملة معزولة اي
 \begin{eqnarray}
\frac{dS}{dt}>0\Rightarrow (\frac{1}{T_1}-\frac{1}{T_2})\frac{dE_1}{dt}>0.
 \end{eqnarray}
 اذا كان
 $T_2>T_1$
 فان
 $dE_1/dt>0$
 اي ان الحرارة تنتقل من 
 الجملة
 $2$
 الي الجملة
 $1$
 اي من الساخن الي البارد.
 
 \paragraph{ 
التوازن الحراري الميكانيكي:
} 
 نفترض الان ان الجملتين 
 $1$
 و
 $2$
 مفصولتين بجدار دياتارم متحرك بدون احتكاك و غير نفاذ للجسيمات
 . اذن
  هناك تبادل للحرارة و ايضا للعمل الميكانيكي بين الجملتين
 $1$
 و
 $2$.
نفترض ان المقادير الاتية تبقي ثابتة خلال التبادلات التي تحدث بين الجملتين: 

 \begin{eqnarray}
E_0=E_1+E_2~,~V_0=V_1+V_2~,~N_0=N_1+N_2.
 \end{eqnarray}
 ايضا لان الجدار غير نفاذ فان 
 $N_1$
 و
 $N_2$
 ثابتين. القيمة القصوي  للدالة
 $S(E_0,E_1,V_1,N_1)$
 تحقق الشرطين
 
 \begin{eqnarray}
\frac{\partial S}{\partial E_1}=0~,~\frac{\partial S}{\partial V_1}=0.
 \end{eqnarray}
 لكن
 \begin{eqnarray}
S(E_0,E_1,V_1,N_1)=S_1(E_1,V_1,N_1)+S_2(E_2,V_2,N_2).
 \end{eqnarray}
بالمرور عبر نفس الخطوات من الفقرة السابقة نحصل مباشرة علي
 \begin{eqnarray}
(\frac{\partial S_1}{\partial E_1})_{V_1,N_1}=(\frac{\partial S_2}{\partial E_2})_{V_2,N_2}\Rightarrow \frac{1}{T_1}=\frac{1}{T_2}.
 \end{eqnarray}
 
 \begin{eqnarray}
(\frac{\partial S_1}{\partial V_1})_{E_1,N_1}=(\frac{\partial S_2}{\partial V_2})_{E_2,N_2}\Rightarrow \frac{P_1}{T_1}=\frac{P_2}{T_2}.
 \end{eqnarray}
 كما في السابق يمكن ان نبين ان هذه القيمة القصوي هي قيمة اعظمية. 
 اذن التوازن الحراري الميكانيكي يعطي بتساوي درجة الحرارة و ضغط الجملتين
 $1$
 و
 $2$
اي
\begin{eqnarray}
T_1=T_2~,~P_1=P_2.
 \end{eqnarray}
\paragraph{ 
الضغط:
} 
يمكن ان نعبر عن الضغط بدلالة الطاقة الداخلية كالاتي. ننطلق من
$dE(S,V,N)=0$
اي من
\begin{eqnarray}
(\frac{\partial  E}{\partial S})_{V,N}\frac{dS}{dV}=-(\frac{\partial E}{\partial V})_{S,N}.
 \end{eqnarray}
بالتالي
\begin{eqnarray}
P=(\frac{\partial S}{\partial V})_{E,N}(\frac{\partial E}{\partial S})_{V,N} =-(\frac{\partial E}{\partial V})_{S,N}.
 \end{eqnarray}
 اذن المشتقة الجزئية للطاقة بالنسبة للحجم مع ثبات الانتروبي و عدد الجسيمات- تحول ادياباتيكي عكسي- هي تساوي ناقص الضغط. من المستحسن ان نري هذه 
 النتيجة ايضا من خلال مثال.
 
 نعتبر غاز مثالي داخل اسطوانة مغلقة بمكبس ذي سطح
 $A$.
 في اللحظة الابتدائية يكون الغاز في حالة ميكروسكوبية
 $i$
 ذات طاقة
 $E=\epsilon_i$. 
 رأينا من خلال مثال العلبة المكعبة ان طاقة  اي حالة ميكروسكوبية تتعلق بالحجم الذي يحتله الغاز اي ان
 $\epsilon_i=\epsilon_i(V)$.
  من اجل انتقال 
  $dx$
  موجب للمكبس يزداد حجم الغاز بكمية
  $dV$. 
  شروط الحركة هي بحيث ان التحويل الترموديناميكي هو تحويل 
  ادياباتيكي عكسي اي لا يوجد تبادل للحرارة و كل التبادلات الطاقوية تكون علي شكل عمل ميكانيكي. الجملة اذن تحافظ عل نفس اعداد الاحتلال و بالتالي تبقي في نفس الحالة الميكروسكوبية
  $i$.
  لكن طاقة
  الحالة الميكروسكوبية تصبح
  \begin{eqnarray}
\epsilon_i(V+dV)=\epsilon_i(V)+(\frac{\partial \epsilon_i}{\partial V})_{S,N}dV.
 \end{eqnarray}
 التغير في الطاقة الداخلية للغاز يعطي ب
 \begin{eqnarray}
\Delta E&=&E(V+dV)-E(V)\nonumber\\
&=&\epsilon_i(V+dV)-\epsilon_i(V)\nonumber\\
&=&(\frac{\partial \epsilon_i}{\partial V})_{S,N}dV.
 \end{eqnarray}
 هذه الطاقة تساوي العمل المقدم من المكبس اي تساوي
  \begin{eqnarray}
W=-P_i Adx=-P_idV.
 \end{eqnarray}
  $P_i$
  هو الضغط الذي يجب تطبيقه علي يمين المكبس حتي يكون التحويل عكسي ادياباتيكي. بمطابقة المعادلتين اعلاه نحصل مباشرة علي
  \begin{eqnarray}
P_i =-(\frac{\partial \epsilon_i}{\partial V})_{S,N}.
 \end{eqnarray}
  هذه العبارة
  تعطي قيمة الضغط الخاصة بالحالة الميكروسكوبية
  $i$.
  للحصول علي قيمة الضغط
  $P$
  الذي نقيسه فعليا علي المستوي الماكروسكوبي علينا ان نأخذ القيمة المتوسطة كالاتي.
  
 كما فعلنا في السابق مرات متعددة نعتبر مجموعة 
  $\{{\cal M}\}$
 من الجمل المتطابقة علي المستوي الماكروسكوبي. ليكن
 ${\cal P}_i$
 احتمال الحصول علي الحالة الميكروسكوبية
 $i$. 
 الطاقة المتوسطة علي المجموعة
 $\{{\cal M}\}$
  هي
  \begin{eqnarray}
<E> =\sum_{\cal M}{\cal P}_i\epsilon_i.
 \end{eqnarray}
 لان الجملة ماكروسكوبية فان تقلبات
 \footnote{${\rm fluctuations}.$}
 الطاقة
 $E$
 حول القيمة المتوسطة
 $<E>$
 هي في الغالب مهملة. هذه القيمة المتوسطة هي ايضا القيمة الاكثر احتمالا. اذن
 $E\simeq <E>$
 و نحصل علي
 
 \begin{eqnarray}
E =\sum_{\cal M}{\cal P}_i\epsilon_i.
 \end{eqnarray}
 بالمثل فان
 \begin{eqnarray}
P&\simeq &<P>\nonumber\\
&=&\sum_{\cal M}{\cal P}_iP_i.
 \end{eqnarray}
 لكن لان التحويل ادياباتيكي اي ان اعداد الاحتلال ثابتة فان احتمال التواجد في اي حالة ميكروسكوبية
 $i$
 هو ثابت و بالتالي
  \begin{eqnarray}
dE &=&\sum_{\cal M}d{\cal P}_i\epsilon_i+\sum_{\cal M}{\cal P}_id\epsilon_i\nonumber\\
&=&\sum_{\cal M}{\cal P}_id\epsilon_i\nonumber\\
&=&-\sum_{\cal M}{\cal P}_iP_idV\nonumber\\
&=&-PdV.
 \end{eqnarray}
 و هذا هو المطلوب و المعروف.
 \section*{
الغاز المثالي الكلاسيكي
}
\addcontentsline{toc}{section}{
الغاز المثالي الكلاسيكي
} 

 عندما تكون درجة حرارة غاز مثالي بعيدة عن الصفر المطلق فان الغاز يتصرف تقريبا بطريقة كلاسيكية. اذن يمكننا في هذه الحالة 
 تطبيق قوانين الميكانيك الكلاسيكي علي جملة الغاز المثالي المشكلة
 من
 $N$
 جسيم, جزئ او ذرة, داخل حيز من الفضاء 
 حجمه
 $V$.
  بالتعريف فان التفاعل بين جسيمات الغاز المثالي ضعيفة جدا و يمكن اهمالها وبالتالي فان طاقة الجملة
  تهيمن عليها الطاقة الحركية للجسيمات. الهاميلتونية تعطي في هذه الحالة ب
  \begin{eqnarray}
  H=\sum_{i=1}^N\frac{\vec{p}_i^2}{2m}.
 \end{eqnarray}
 $m$
 هي كتلة ذرات الغاز ونفترض ايضا ان ذرات الغاز هي جسيمات سلمية و بالتالي فان السبين
 الخاص بها ينعدم. الفضاء الطوري اذن هو ذو
 $6N$
 بعد تعطي فيه المحاور باشعة
 الموضع
 $\vec{r}_i$
 و اشعة 
 كمية الحركة
 $\vec{p}_i$.
 لتكن
 $\Omega(E_0)$
 عدد الحالات الميكروسكوبية التي لها طاقة بين
 $E_0$
 و
 $E_0+\delta E$. 
 من المعروف من الميكانيك الكمي ان كل حالة ميكروسكوبية تمثل في الفضاء الطوري بنقطة تحتل خلية, تعرف بخلية هايزنبرغ
 \footnote{.${\rm Heisenberg}~{\rm cell}$}
 , حجمها هو
 \begin{eqnarray}
  V_{\rm cell}=h^{{n}/{2}},
 \end{eqnarray}
 حيث
 $h$
 هو ثابت بلانك
 \footnote{${\rm Planck}.$}
 و
 $n$
 هو عدد درجات الحرية. هذه القيمة ترجع الي مبدأ الارتياب لهايزنبرغ
 $\Delta x\Delta p\sim h$.
 اذن عدد الحالات الميكروسكوبية
 $\Omega$
 هو حاصل قسمة الحجم  في الفضاء الطوري الذي تحتله الحالات الميكروسكوبية التي لها 
 طاقة بين
 $E_0$
 و
 $E_0+\delta E$.
  علي حجم خلية هايزنبرغ واحدة. نكتب اذن
\begin{eqnarray}
  \Omega(E_0)&=&\frac{\nu}{h^{3N}}\nonumber\\
  &=&\frac{1}{h^{3N}}\int_{E_0\leq H\leq E_0+\delta E} d^{3N}\vec{r}d^{3N}\vec{p}.
 \end{eqnarray}
 \begin{eqnarray}
 d^{3N}\vec{r}&=&\prod_{i=1}^Ndx_idy_idz_i~,~d^{3N}\vec{p}=\prod_{i=1}^Ndp_{xi}dp_{yi}dp_{zi}.
 \end{eqnarray}
 لان الطاقة لا تتعلق الا بكميات الحركة فان التكامل اعلاه ينقسم الي حجم في فضاء ال
 $\vec{r}$
 و حجم في فضاء ال
 $\vec{p}$
اي
\begin{eqnarray}
 \Omega(E_0)&=&\frac{1}{h^{3N}}\int d^{3N}\vec{r} \int_{E_0\leq H\leq E_0+\delta E} d^{3N}\vec{p}\nonumber\\
 &=&\frac{V^N}{h^{3N}} \int_{E_0\leq H\leq E_0+\delta E} d^{3N}\vec{p}.
 \end{eqnarray}
 نحسب اولا عدد الحالات الميكروسكوبية
 $\Phi(E_0)$
 التي لها طاقة اقل او تساوي من
 $E_0$.
 بالمرور عبر نفس الخطوات اعلاه نجد ان 
 هذا العدد يعطي بالتكامل
 \begin{eqnarray}
 \Phi(E_0)
 &=&\frac{V^N}{h^{3N}} \int_{H\leq E_0} d^{3N}\vec{p}\nonumber\\
&=&\frac{V^N}{h^{3N}} \int_{\sum_{i=1}^N\frac{\vec{p}_i^2}{2m}\leq E_0} \prod_{i=1}^Nd^{3}\vec{p}_i.
 \end{eqnarray}
لكن
\begin{eqnarray}
\sum_{i=1}^N\frac{\vec{p}_i^2}{2m}= E_0
 \end{eqnarray}
 هي معادلة كرة ذات نصف قطر
 $R=\sqrt{2mE_0}$
 في 
 $3N$
 بعد حيث تلعب مركبات كميات الحركة دور الاحداثيات الديكارتية علي هذه الكرة. المسألة اذن هي مسألة حساب
 حجم كرة في
 $3N$
 بعد. نحن نعرف ان الحجم في ثلاث ابعاد يعطي ب
 \begin{eqnarray}
V_3=\int_{x_1^2+x_2^2+x_3^2\leq R^2} dx_1dx_2dx_3=\frac{4}{3}\pi R^3.
 \end{eqnarray}
 في 
 $n$
 بعد يعطي الحجم ب
 \begin{eqnarray}
V_n=\int_{x_1^1+x_2^2+...+x_n^2\leq R^2} dx_1dx_2dx_3...dx_{n}
.
 \end{eqnarray}
 من الواضح ان هذا الحجم يجب ان يكون متناسب مع
 $R^n$
اي
\begin{eqnarray}
V_n=C_n R^n.
 \end{eqnarray}
 منه نحصل علي
 \begin{eqnarray}
dV_n=dx_1dx_2dx_3...dx_{n}=nC_n r^{n-1}dr.
 \end{eqnarray}
 علينا الان ان نحسب
 $C_n$.
 نبدأ من التكامل
 \begin{eqnarray}
\int e^{-x^2} dx=\sqrt{\pi}.
 \end{eqnarray}
 نرفع طرفي هذه المعادلة للقوة
 $n$
 لنحصل علي
  \begin{eqnarray}
\int e^{-\sum_{i=1}^nx_i^2} dx_1...dx_n=\pi^{n/2}.
 \end{eqnarray}
 اي
  \begin{eqnarray}
\int e^{-r^2} nC_n r^{n-1}dr=\pi^{n/2}.
 \end{eqnarray}
 نجري تغيير المتغير
 $y=r^2$
 لنحصل علي
  \begin{eqnarray}
\frac{n}{2}C_n\int y^{\frac{n}{2}-1}e^{-y}dy=\pi^{n/2}.
 \end{eqnarray}
 يمكننا الان ان نستخدم التكامل المعروف
 \begin{eqnarray}
\int y^{\alpha}e^{-y}dy=\Gamma(\alpha+1)=\alpha!.
 \end{eqnarray}
 الدالة
 غاما هي تعميم لدالة المعاملي للمتغيرات السالبة و غير الصحيحة و المركبة. من اجل قيم صحيحة للوسيط
 $\alpha$
 فان التكامل اعلاه يعطي بالضبط
 $\alpha!$
 اي ان
 $\Gamma(\alpha+1)=\alpha!$
 من اجل القيم الصحيحة ل
 $\alpha$.
  لكن التكامل اعلاه معرف من اجل جميع قيم
  $\alpha$:
  موجبة او سالبة, صحيحة او غير صحيحة, حقيقية او مركبة و ناتج التكامل هو بالتعريف الدالة غاما التي تعمم
  المعاملي لكل هذه المجالات.
  
 باستخدام الدالة غاما نحصل علي الثابت
 $C_n$
 بسهولة.  بالفعل لدينا مباشرة
  \begin{eqnarray}
\frac{n}{2}C_n\Gamma(\frac{n}{2})=\pi^{n/2}\Rightarrow \frac{n}{2}C_n(\frac{n}{2}-1)!=\pi^{n/2}.
 \end{eqnarray}
 حتي من اجل القيم غير الصحيحة فان المعاملي يحقق الخاصية
 \begin{eqnarray}
\frac{n}{2}(\frac{n}{2}-1)!=(\frac{n}{2})!.
 \end{eqnarray}
الخلاصة ان عدد درجات الحرية
 $n$,
الثابت
$C_n$,
حجم الكرة في
 $n$
 بعد, و عدد الحالات الميكروسكوبية 
 $\Phi(E_0)$
 التي لها طاقة اقل او تساوي من
 $E_0$
تعطي اذن ب
\begin{eqnarray}
n=3N.
 \end{eqnarray}
 \begin{eqnarray}
C_n=\frac{\pi^{n/2}}{(\frac{n}{2})!}.
 \end{eqnarray}
 \begin{eqnarray}
V_n= (2mE_0)^{n/2}\frac{\pi^{n/2}}{({n}/{2})!}.
 \end{eqnarray}
 \begin{eqnarray}
\Phi(E_0)= \frac{V^n}{h^{3N}}(2mE_0)^{n/2}\frac{\pi^{n/2}}{({n}/{2})!}.
 \end{eqnarray}
 جسيمات الغاز لها خاصية كمية اخري لا يمكن اهمالها وهي كونها جسيمات متطابقة. لانه لدينا
 $N$
 جسيم متطابق فان هناك
 $N!$
 تبديلة ممكنة لهذه الجسيمات توافق
 $N!$
 تمثيلة متطابقة للجملة. اي ان عدد الحالات الميكروسكوبية التي حصلنا عليها هو اكبر ب
 $N!$
 مرة من عدد الحالات الميكروسكوبية التي هي فعلا مختلفة. اذن  عدد الحالات الميكروسكوبية 
 $\Phi(E_0)$
 التي لها طاقة اقل او تساوي من
 $E_0$
 هو في الواقع معطي ب
  \begin{eqnarray}
\Phi(E_0)= \frac{1}{N!}\frac{V^n}{h^{3N}}(2mE_0)^{n/2}\frac{\pi^{n/2}}{({n}/{2})!}.
 \end{eqnarray}
في المعادلة الاخيرة اعلاه المعاملات
$N!$
و
$h^{3N}$
هي معاملات راجعة للتأثيرات الكمية لا يمكن الحصول عليها بالاعتماد علي الميكانيك الكلاسيكي فقط. ايضا من المعروف ان تكميم جملة مشكلة من جسيم واحد حر داخل
علبة مكعبة حجمها
$V$
يؤدي الي قيم مكممة للطاقة بفسحة تعطي ب
\begin{eqnarray}
\epsilon_0=\frac{\pi^2\hbar^2}{2mV^{2/3}}.
 \end{eqnarray}
 بالتعويض في عدد الحالات
 $\Phi(E)$
 نحصل علي
  \begin{eqnarray}
\Phi(E_0)= \frac{1}{N!}\frac{\pi^{3N/2}}{2^{3N}({3N}/{2})!}\big(\frac{E_0}{\epsilon_0}\big)^{3N/2}.
 \end{eqnarray}
 بالاشتقاق نحصل علي كثافة
  الحالات الميكروسكوبية
 $\rho(E_0)$
 كما يلي
  \begin{eqnarray}
\rho(E_0)&=&\frac{d\Phi(E_0)}{dE_0}\nonumber\\
&=& \frac{1}{N!}\frac{\pi^{3N/2}}{2^{3N}({3N}/{2}-1)!}\big(\frac{E_0}{\epsilon_0}\big)^{3N/2}\frac{1}{E_0}.
 \end{eqnarray}
 باستخدام هذه الكثافة يمكن ان نحصل علي عدد الحالات الميكروسكوبية 
 $\Omega(E_0)$
 التي لها طاقة
 $E_0$
 بارتياب
 $\delta E$
 كما يلي
  \begin{eqnarray}
\Omega(E_0)&=&\rho(E_0)\delta E\nonumber\\
&=& \frac{1}{N!}\frac{\pi^{3N/2}}{2^{3N}({3N}/{2}-1)!}\big(\frac{E_0}{\epsilon_0}\big)^{3N/2}\frac{\delta E}{E_0}.
 \end{eqnarray}
 نحسب الان انتروبي الغاز المثالي. لدينا
  \begin{eqnarray}
\frac{S}{k}&=&\ln\Omega(E_0)\nonumber\\
&=&\ln \rho(E_0)\delta E\nonumber\\
&=& \ln\bigg(\frac{1}{N!}\frac{\pi^{3N/2}}{2^{3N}({3N}/{2}-1)!}\big(\frac{E_0}{\epsilon_0}\big)^{3N/2}\frac{\delta E}{E_0}\bigg)\nonumber\\
&=&-\ln N!-\ln(3N/2-1)!+(3N/2)\ln\pi/4+(3N/2)\ln(E_0/\epsilon_0)+\ln\delta E/E_0.\nonumber\\
 \end{eqnarray}
 نلاحظ ايضا ان
 \begin{eqnarray}
\ln\rho(E_0)
&=&-\ln N!-\ln(3N/2-1)!+(3N/2)\ln\pi/4+(3N/2)\ln(E_0/\epsilon_0)+\ln 1/E_0.\nonumber\\
 \end{eqnarray}
  \begin{eqnarray}
\ln\Phi(E_0)
&=&-\ln N!-\ln(3N/2)!+(3N/2)\ln\pi/4+(3N/2)\ln(E_0/\epsilon_0).\nonumber\\
 \end{eqnarray}
 لان
 $N$
 كبير جدا فان الحدود الاخيرة في عبارتي
 $\ln\Omega(E_0)$
 و
 $\ln\rho(E_0)$
 يمكن اهمالها و نحصل اذن علي النتيجة التي ذكرناها سابقا الاتية
  \begin{eqnarray}
\frac{S}{k}&=&\ln\Omega(E_0)\simeq \ln\rho(E_0)\simeq \ln\Phi(E_0).
 \end{eqnarray}
 اي ان لوغاريتم المنحني
 $\rho(E_0)$
 يساوي الي لوغاريتم المساحة
 $\Phi(E_0)$
 تحت  هذا المنحني
 و هو يساوي الي لوغاريتم المساحة 
 $\Omega(E_0)\delta E$
 المرتكزة حول القيمة الاكثر احتمالا للطاقة
 $E_0$. 
 نستخدم الان علاقة ستيرلينغ لتبسيط العلاقة اعلاه للانتروبي كالاتي
  \begin{eqnarray}
S&=&\frac{3Nk}{2}\ln\frac{2E_0}{3N\epsilon_0}+\alpha+NS_0=S(E_0,V,N).
 \end{eqnarray}
  \begin{eqnarray}
S_0=\frac{3k}{2}(1+\ln\frac{\pi}{4})~,~\alpha=k(-N\ln N+N).
 \end{eqnarray}
 الثابت
 $\alpha$
 هو المساهمة في قيمة الانتروبي الناجمة عن تطابق الجسيمات. تعلق الانتروبي بالحجم محتوي في الفسحة الطاقوية
 $\epsilon_0$.
 
 كما شرحنا في الفقرة السابقة نحصل علي الطاقة الداخلية لجملة الغاز المثالي الكلاسيكي عن طريق قلب العلاقة
 $S=S(E_0,V,N)$.
بعد حساب بسيط نحصل  علي
\begin{eqnarray}
E_0=\frac{3h^2}{4\pi m}\frac{N^{5/3}}{V^{2/3}}\exp\big(\frac{2S}{3Nk}- \frac{5}{3}\big).
 \end{eqnarray}
 درجة حرارة الغاز المثالي تعطي ب
 \begin{eqnarray}
T&=&(\frac{\partial E_0}{\partial S})_{V,N}\nonumber\\
&=&\frac{2E_0}{3NK}\Rightarrow E_0=\frac{3}{2}NkT.
 \end{eqnarray}
 السعة الحرارية تحت حجم ثابت تعطي ب
  \begin{eqnarray}
C_v&=&(\frac{\partial E_0}{\partial T})_{V,N}\nonumber\\
&=&\frac{3Nk}{2}.
 \end{eqnarray}
ضغط الغاز المثالي يعطي ب

  \begin{eqnarray}
P&=&-(\frac{\partial E_0}{\partial V})_{S,N}\nonumber\\
&=&\frac{2E_0}{3V}\Rightarrow PV=NkT.
 \end{eqnarray}
 في النهاية نحصل اذن علي معادلة حالة الغاز المثالي المعروفة.

  \section*{
مسائل اضافية
}
\addcontentsline{toc}{section}{
مسائل اضافية
} 

\paragraph{
توزيع غوس
للاحتمال:
}

التوزيع ثنائي الحدين للاحتمال يعطي ب
\begin{eqnarray}
 W_N(n_1)=\frac{N!}{n_1!(N-n_1)!}p^{n_1}q^{N-n_1}.
 \end{eqnarray}
 نهتم بالنهاية
 $N\longrightarrow\infty$
 اين نحصل علي توزيع غوس
 \footnote{.${\rm Gauss}$}
 للاحتمال
 .
 من اجل القيم الكبيرة ل
 $n_1$
 فان
 $W_N(n_1)$
 يمكن اعتبارها دالة مستمرة للمتغير المستمر
 $n_1$
 علي الرغم من ان القيم الصحيحة
 ل
 $n_1$
 هي فقط التي تحمل اي معني فيزيائي.
 \begin{itemize}
 \item
بين المعني الفيزيائي للاعداد
$p$, $q$, $n_1$
و
$N$
من اجل حركة المشاء العشوائي. بين ايضا المعني الفيزيائي ل
$W_N(n_1)$.
  \item 
  احسب الدالة
  $\ln W_N(n_1)$,
  المشتقة الاولي
  $B_1=d\ln W_N(n_1)/dn_1$
  و المشتقة الثانية
  $B_2=d^2\ln W_N(n_1)/dn_1^2$.
  \item
  احسب القيمة
  $n_1=\bar{n_1}$
  التي يكون عندها توزيع الاحتمال
  $W_N(n_1)$
  اعظمي.
 \item
 احسب تحليل تايلور للدالة
 $\ln W_N(n_1)$
 و من ثم احسب الاحتمال
 $W_N(n_1)$.
 \item
 احسب توزيع غوس للاحتمال
 $P(x)$ 
 المعرف ب
 \begin{eqnarray}
  P(x)dx=W_n(n_1)dn_1.
 \end{eqnarray}
تذكر ان
$m=n_1-n_2$, $N=n_1+n_2$
و ان
$l$
هو طول خطوة المشاء العشوائي بينما
$x=ml$
هو موضع المشاء العشوائي.
  
 \end{itemize}

\paragraph{
توزيع بواسون
للاحتمال:
}
 نعتبر مرة اخري التوزيع ثنائي الحدين للاحتمال المعطي ب
\begin{eqnarray}
 W_N(n_1)=\frac{N!}{n_1!(N-n_1)!}p^{n_1}q^{N-n_1}.
 \end{eqnarray}
 نعتبر الان 
 النهاية
$N\longrightarrow\infty$
و
$p\longrightarrow 0$
من اجل القيم الصغيرة ل
$n_1$
اين نحصل علي توزيع بواسون
\footnote{.${\rm Poisson}$}
للاحتمال.
\begin{itemize}

\item
احسب
المعامل
$C_N^{n_1}=N!/n_1!(N-n_1)!$
في النهاية
$N\longrightarrow\infty$
و
$p\longrightarrow 0$
من اجل القيم الصغيرة ل
$n_1$
اي
$n_1<<N$.
\item
بين انه في هذه النهاية
$q^{N-n_1}=e^{-Np}$
ثم احسب الاحتمال
$W_N(n_1)$
في هذه الحالة.
\item
احسب القيم المتوسطة
$<n_1>$
و
$<n_1^2>$
ثم احسب التشتت في المتوسط المعرف ب
$\sigma^2=<n_1^2>-<n_1>^2$.
\end{itemize}

\paragraph{
الانتروبي الاحصائي و المبدأ الثاني للترموديناميك:
}
بين ان التعريف
\begin{eqnarray}
  S=k\ln \Gamma(E)
 \end{eqnarray}
للانتروبي الاحصائي يؤدي الي الانتروبي الترموديناميكي بكل خواصه المعروفة مثل $-1$ الانتروبي هو مقدار تمديدي و
$-2$
الانتروبي يحقق المبدأ الثاني للترموديناميك. بين ايضا ان درجة الحرارة تعطي ب
\begin{eqnarray}
  \frac{1}{T}=\frac{\partial S}{\partial E}.
 \end{eqnarray}
 \paragraph{
 مبرهنة التقسيم المتساوي للطاقة:
}
 نأخذ جسم ذو سبين صفر و كتلة
 $m$
 داخل علبة حجمها
 $V$.
 الجملة معزولة ذات طاقة بين
 $E$
 و
 $E+\delta E$.
 \begin{itemize}
  \item
  عرف الفضاء الطوري في هذه الحالة.
  اكتب هاميلتونية الجملة. ماهو عدد الحالات
  $\Omega(E)$
  المسموح بها للجملة. احسب
  $p_i\partial H/\partial p_i$.
  
  \begin{itemize}
  \item
  بين ان القيمة المتوسطة ل
  $p_i\partial H/\partial p_i$
  تكتب علي الشكل
  \begin{eqnarray}
  <p_i\frac{\partial H}{\partial p_i}>=\frac{\frac{d}{d E}\int_{H\leq E}p_i\frac{\partial H}{\partial p_i} d^3xd^3p}{\frac{d}{dE}\int_{H\leq E}d^3xd^3p}.
 \end{eqnarray}
 \item
 بين ان
 
 \begin{eqnarray}
 \int_{H\leq E}p_i\frac{\partial (H-E)}{\partial p_j}=V\delta_{ij}\int_{H\leq E}(E-H)d^3p.
 \end{eqnarray}
 \item
 برهن ان
 \begin{eqnarray}
\frac{d}{d\alpha} \int_{f(\alpha)}^{g(\alpha)}F(\alpha,x)dx= \int_{f(\alpha)}^{g(\alpha)}\frac{\partial}{\partial \alpha}F(\alpha,x)dx+\frac{\partial g(\alpha)}{\partial\alpha}F(\alpha,g(\alpha))-\frac{\partial f(\alpha)}{\partial\alpha}
F(\alpha,f(\alpha)).
\end{eqnarray}
 \item
 احسب
 $<p_i\partial H/\partial p_i>$
 بدلالة عدد الحالات
 $\Phi(E)$ 
 التي لها طاقة اقل من
 $E$.
 ماذا يمكن ان نستنتج بالنسبة للقيمة المتوسطة للطاقة
 $<H>$.
 عبر عن
 $<H>$
 بدلالة الانتروبي ثم بدلالة درجة الحرارة.
 \item
 ماذا يمكن ان نستنتج بالنسبة الي حالة الهزاز التوافقي.
 
  \end{itemize}
 \end{itemize}

  \paragraph{
تناقض جيبس:
}
\begin{itemize}
 \item 
 بالنسبة لغاز مثالي فان الانتروبي يعطي ب
  \begin{eqnarray}
  S=\frac{3Nk}{2}\ln \frac{2E}{3N\epsilon_0}+NS_0+\alpha.
\end{eqnarray}
استخرج من هذه النتيجة المعادلة الاتية
\begin{eqnarray}
  S=Nk\ln V\epsilon^{3/2}+NS_0^{'}+\alpha(N),
\end{eqnarray}
حيث
$\epsilon$
هي طاقة جزئ واحد من الغاز. تذكر ان
\begin{eqnarray}
S_0=\frac{3k}{2}(1+\ln\frac{\pi}{4})~,~\alpha=k(-N\ln N+N).
\end{eqnarray}
استخرج ايضا العلاقات التي تعطي
$\alpha(N)$
و
$S_0^{'}$.
برهن ايضا العلاقة التالية (معادلة ساكور - تترود
\footnote{${\rm Sakur-Tetrode}~{\rm equation}.$}
)
\begin{eqnarray}
  S=Nk\ln \frac{V}{N}\epsilon^{3/2}+\frac{3}{2}Nk(\frac{5}{3}+\ln \frac{4\pi m}{3h^2}).
\end{eqnarray}
استخدم ايضا

\begin{eqnarray}
 \epsilon_0=\frac{\pi^2\hbar^2}{2mV^{2/3}}.
\end{eqnarray}
\item
نأخذ جملة معزولة عن الوسط الخارجي مشكلة من غازين مفصولين بحاجز. نفترض ان درجة الحرارة
ثابتة اي
$T_1=T_2=T$
و ان كتل جزيئات الغاز متساوية اي
$m_1=m_2=m$.
احسب الانتروبي
$S_1$
و الانتروبي
$S_2$
بدلالة الحجمين
$V_1$
و
$V_2$
و اعداد الجزيئات
$N_1$
و
$N_2$.
نفترض ان الحاجز الفاصل يرفع بعد فترة. احسب التغير في الانتروبي. ماذا تستنتج.
\itemنفترض
ان جزيئات الغاز الاول متطابقة مع جزيئات الغاز الثاني. احسب التغير في الانتروبي في هذه
الحالة. ماذا تستنتج.
\end{itemize}
 
 \newpage
  \section*{
تمارين
}
\addcontentsline{toc}{section}{
تمارين
} 

\paragraph{
تمرين
$1$:
}
\begin{itemize}
 \item 
نعتبر جسيم ذو كتلة
$m$
يتحرك داخل علبة مكعبة طول ضلعها
$L$. 
حل معادلة شرودينغر 
\footnote{${\rm Schrodinger}~{\rm equation}.$}
لايجاد قيم الطاقة المسموح بها. افترض الشروط الحدية التي تنعدم فيها دالة الموجة علي الجدران.
\item
احسب درجة انحلال المستويات الطاقوية ال
$10$
الاولي.
\item 
نضع واحد مول من الهيليوم داخل العلبة. الضغط
${\rm pa}$~
$P=10^5$
و درجة الحرارة
${\rm K}$~
$T=273$.
نفترض ان الهيليوم غاز مثالي. أحسب
$L$
و كتلة ذرة واحدة من الهيليوم. احسب الفسحة الطاقوية
$\epsilon_0$.

استخدم النتيجة الاحصائية: الطاقة المتوسطة لذرة واحدة من الغاز هي
$<E>=3kT/2$
من اجل اعطاء تقدير تقريبي لرتبة عظم الاعداد الكمية
$n_x,n_y$
و
$n_z$.
\item
وضح في اطار المثال اعلاه الفرق بين العمل الميكانيكي و كمية الحرارة من الناحية الميكروسكوبية.
\item
نفترض الان انه لدينا ثلاث جسيمات داخل المكعب بحيث ان طاقة الجملة تساوي الي
$18\epsilon_0$.
ماهي التمثيلات الطاقوية المسموح بها للجملة في الحالات التالية:
\begin{itemize}
\item
الجسيمات متمايزة.
\item
الجسيمات عبارة عن بوزونات متطابقة ذات سبين  يساوي صفر.
\item
الجسيمات عبارة عن بوزونات متطابقة ذات سبين يساوي واحد.
\item
الجسيمات عبارة عن فرميونات متطابقة ذات سبين يساوي
نصف.
\end{itemize}
احسب في كل مرة عدد الحالات الميكروسكوبية المسموح بها.
\end{itemize}

\paragraph{
تمرين
$2$:
}
\begin{itemize}
 \item 
 ماهي طاقة جسيم
 يتحرك في بعد واحد بين 
 حائطين عاكسين واقعين في
$x=0$
و
$x=L$.
\item
نعتبر ثلاث جسيمات غير متفاعلة فيما بينها تتحرك في بعد واحد بين حائطين عاكسين واقعين في
$x=0$
و
$x=L$.

\begin{itemize}
\item
كيف تميز الحالة الميكروسكوبية للجملة اذا كانت الجسيمات لها سبينات
$s_1$,$s_2$
و
$s_3$
علي التوالي.
\item
لتكن
طاقة الجملة تساوي
$E=27\epsilon_0$
حيث
$\epsilon_0=\pi^2\hbar^2/2mL^2$.
ماهي التمثيلات الطاقوية التي يمكن ان تكون فيها الجملة.
\item
احسب عدد الحالات الميكروسكوبية اذا كانت الجسيمات متمايزة بدون سبين.
\item
احسب عدد الحالات الميكروسكوبية اذا كانت الجسيمات متطابقة ذات سبين
$s=0,1,1/2$.
\end{itemize}
 \end{itemize}
 
 \paragraph{
تمرين
$3$:
}
 نعتبر جسيم متحرك في بعد واحد بين حائطين عاكسين مفصولين بمسافة
 $L$.
 \begin{itemize}
  \item 
 احسب عدد الحالات
 $\Phi(E)$
 التي لها طاقة اقل او يساوي من
 $E$
 و عدد الحالات
 $\Omega(E)$
 التي لها طاقة
 $E$ 
 بارتياب
 $\delta E$
 اذا كانت حركة الجسيم كمية.
 \item
 اعد السؤال السابق بافتراض ان حركة الجسيم كلاسيكية.
 \end{itemize}
 \paragraph{
تمرين
$4$:
}
احسب عدد الحالات الميكروسكوبية المسموح بها لجملة جسيم حر متحرك داخل
مكعب ذو حجم
$V$. 
افترض ان الجسيم ذو سبين صفر يتحرك بصورة كلاسيكية. ماهو عدد الحالات اذا كان الجسيم ذو سبين
نصف. ماهو عدد الحالات اذا كان الجسيم ذو سبين واحد.
 
 \paragraph{
تمرين
$5$:
}
نعتبر حركة ثلاثة جسيمات حرة بين حائطين عاكسين علي مسافة تساوي
$L$.
نفترض ان حركة الجسيمات هي حركة كلاسيكية و بالتالي فانه يمكننا ان نهمل تطابق الجسيمات.
\begin{itemize}
 \item 
 صف الفضاء الطوري للجملة و اكتب الهاميلتونية.
 \item
 احسب عدد الحالات
 $\Phi(E)$
 التي لها طاقة اقل او تساوي من
$ E$.
 \item
 احسب عدد الحالات
 $\Omega(E)$
 التي لها طاقة
 $E$
 بارتياب
 $\delta E$.
 \item
 اذا افترضنا ان الجسيمات هي عبارة عن فرميونات فماذا يصبح عدد الحالات
 $\Omega(E)$.
 \item
 ماذا يصبح عدد الحالات اذا كانت الجسيمات ذات سبين
 $s$.
 \item
 ماهو عدد الحالات
 $\Omega(E)$
 اذا افترضنا الان ان الجسيمات متطابقة.
\end{itemize}
\paragraph{
تمرين
$6$:
}
تحقق من المعادلة 

\begin{eqnarray}
{\cal N}=\frac{N!}{\prod_{i=1}^rN_i!}
 \end{eqnarray}
صراحة من اجل
$N=2$
و
$N=3$.
عين في كل مرة الحالات الميكروسكوبية.
\paragraph{
تمرين
$7$:
}
نعتبر جملة عبارة عن حجر نرد. ناقش  العلاقة بين المتوسط في الزمن و المتوسط علي المجموعة الاحصائية في هذه الحالة.

\paragraph{
تمرين
$8$:
}
نعتبر جملة غاز مثالي كلاسيكي مشكلة من 
$N=3$
جسيمات حرة تتحرك في بعد واحد علي قطعة مستقيمة طولها
$L$.
احسب عدد الحالات الميكروسكوبية
$\Phi(E)$
و
$\Omega(E)$
و اشتق عبارة الانتروبي 
$ٍ$.
اشتق عبارة الطاقة الداخلية
$E$
و باقي المقادير الترموديناميكية مثل درجة الحرارة و الضغط و كذا معادلة الحالة.

\paragraph{
تمرين
$9$:
}
\begin{itemize}
 \item 
 صف حركة هزاز توافقي في بعد واحد و احسب سعة
 حركة الهزاز بدلالة الطاقة.
 \item
 احسب عدد الحالات الميكروسكوبية للهزاز التي لها طاقة اقل او تساوي من
 $E$.
 \item
 ما هو عدد الحالات الميكروسكوبية التي لها طاقة
 $E$
 بارتياب
 $\delta E$.
 \item
 احسب
 \begin{eqnarray}
  p\frac{\partial H}{\partial p}+x\frac{\partial H}{\partial x},
 \end{eqnarray}
ثم بين ان
 \begin{eqnarray}
 \int_0^E H dp dx=\int_0^E(E-H)dp dx.
 \end{eqnarray}
 \item
 القيمة المتوسطة لمتغير ديناميكي
 $f$
 مأخوذة علي الحالات الميكروسكوبية التي لها طاقة
 $E$
 بارتياب
 $\delta E$
 يعطي بالعلاقة
 \begin{eqnarray}
 <f>=\frac{\frac{d}{dE}\int_0^E fdpdx}{\frac{d}{dE}\int_0^Edp dx}.
 \end{eqnarray}
 احسب القيمة المتوسطة
 $<H>$.
\end{itemize}

\paragraph{
تمرين
$10$:
}
جملة مغلقة مشكلة من مكعبين ملتصقين عبر جدار ادياباتيكي كاتم للحرارة. جملة المكعبين معزولة عن
الوسط الخارجي بجدار ادياباتيكي.
\begin{itemize}
 \item 
في الحالة الابتدائية المكعب الاول يحتوي علي جسيمين غير متطابقين بطاقة كلية تساوي
$E_I=12\epsilon_0$
و المكعب الثاني يحتوي علي جسيم واحد بطاقة كلية تساوي
$E_{II}=9\epsilon_0$.
احسب عدد الحالات الميكروسكوبية المسموح بها للجملة.
 \item
في الحالة النهائية نرفع الحاجز الادياباتيكي الفاصل بين المكعبين. نحصل اذن علي ثلاث جسيمات حرة داخل متوازي اسطح بطاقة
اجمالية تساوي
$E=21\epsilon_0$.
ما هي الطاقات المسموح بهل للجملة في هذه الحالة. احسب عدد الحالات
و ماذا تستنتج.
 \end{itemize}
\paragraph{
تمرين
$11$:
}
\begin{itemize}
 \item 
نعتبر جملة معزولة مشكلة من مكعبين متلاصقين طول ضلع كل واحد منهما هو
$L$.
المكعبان مفصولان بجدار كاتم للحرارة و غير نفاذ للجسيمات. المكعب الاول يحتوي علي جسيمين طاقتهما
$E_I=12\epsilon_0$
و المكعب الثاني يحتوي ايضا  علي جسيمين طاقتهما
$E_{II}=18\epsilon_0$.
نفترض ان الجسيمات متمايزة.

 احسب عدد الحالات الميكروسكوبية المسموح بها للمكعب الاول و عدد الحالات الميكروسكوبية المسوح بها للمكعب الثاني. ما هو عدد الحالات الاجمالية المسموح 
 بها للجملة الكلية.
 \item
 نفترض الان ان الجدار الفاصل بين المكعبين هو جدار غير نفاذ للجسيمات لكنه دياتارم و بالتالي فانه يسمح 
 بتبادل الطاقة بين المكعبين. الجملة الكلية تصبح غير متوازنة ترموديناميكيا و بالتالي فان
 طاقة كل مكعب يمكنها ان تتغير الي ان تبلغ الجملة الكلية التوازن من جديد. خلال كل هذا التحول فان الطاقة الكلية 
 $E=E_I+E_{II}=30\epsilon_0$
 تبقي دائما منحفظة و يمكن لها ان تتوزع علي المكعبين بطرق مختلفة.
 \begin{itemize}
 \item
 ماهي التمثيلات الطاقوية الممكنة.
 \item
 احسب عدد الحالات الميكروسكوبية المسموح بها للجملة في كل تمثيلة طاقوية. ما هو العدد الكلي للحالات.
 \item
 ما هو احتمال ان تكون الجملة في اي من الحالات التي وجدناها. ما هو احتمال ان تكون طاقة المكعب الاول تساوي
 $6\epsilon_0$, 
 $9\epsilon_0$, $12\epsilon_0$
 و
 $15\epsilon_0$.
 \end{itemize}

\end{itemize}

\paragraph{
تمرين
$12$:
}
دالة غاما تعرف بالتكامل
\begin{eqnarray}
\Gamma(t)=\int_0^{\infty}x^{t-1} e^{-x} dx.
 \end{eqnarray}

\begin{itemize}
 \item 
 بين انه من اجل 
 $t$
 موجب لدينا
 $\Gamma(t+1)=t\Gamma(t)$.
\item
بين انه من اجل قيم طبيعية ل
$t$ 
اي
$t=n$
فان
$\Gamma(n+1)=n!$.
\item
احسب
$(1/2)!$
باجراء التكامل اعلاه ثم استنتج قيمة
$(3/2)!$,
الخ.
\item
احسب حجم الكرة في 
$n$
بعد المعطي بالتكامل
\begin{eqnarray}
\int_{x_1^2+...+x_n^2\leq R^2}dx_1...dx_n.
 \end{eqnarray}
 \end{itemize}

\newpage
 \section*{
حلول
}
\addcontentsline{toc}{section}{
حلول
} 

\paragraph{
 مبرهنة التقسيم المتساوي للطاقة:
}
 \begin{itemize}
  \item
  الفضاء الطوري ستة ابعاد: 
  $\vec{r}=(x_1,x_2,x_3)$, $\vec{p}=(p_1,p_2,p_3)$.
تعطي  الهاميلتونية ب 
  \begin{eqnarray}
 H=\frac{p_1^2}{2m}+\frac{p_2^2}{2m}+\frac{p_3^2}{2m}.
 \end{eqnarray}
 عدد الحالات
  $\Omega(E)$
  المسموح بها للجملة
  يعطي ب
  \begin{eqnarray}
 \Omega(E)&=&\frac{1}{h^3}\int_{E\leq H\leq E+\delta E}d^3xd^3p\nonumber\\
 &=&\frac{d\Phi(E)}{dE}.\delta E\nonumber\\
 &=&\frac{d}{dE}\bigg(\frac{1}{h^3}\int_{ H\leq E}d^3xd^3p\bigg).\delta E\nonumber\\
 &=&\frac{V}{h^3}\frac{d}{dE}\bigg(\int_{ H\leq E}d^3p\bigg).\delta E.
 \end{eqnarray}
 نحسب ايضا
  \begin{eqnarray}
 \sum_i p_i\frac{\partial H}{\partial p_i}=2H.
 \end{eqnarray}
  
  \begin{itemize}
  \item
  القيمة المتوسطة لدالة 
  $f$
  تتعلق فقط بكمية الحركة محسوبة علي الحالات الميكروسكوبية التي لها طاقة
  $E$
  بارتياب
  $\delta E$
  يعطي ب
  
  \begin{eqnarray}
 <f>&=&\frac{\int_{E\leq H\leq E+\delta E} f(p)d^3xd^3p}{\int_{E\leq H\leq E+\delta E}d^3xd^3p}.
 \end{eqnarray}
من الواضح اذن من هذا التعريف و باستعمال نتيجة السؤال السابق ان
\begin{eqnarray}
 <f>&=&\frac{\frac{d}{dE}\int_{H\leq E} f(p)d^3p}{\frac{d}{dE}\int_{H\leq E}d^3p}.
 \end{eqnarray}
 \item
 نحسب
 \begin{eqnarray}
 \int_{H\leq E}p_i\frac{\partial (H-E)}{\partial p_j}d^3xd^3p&=&V\int_{H\leq E}\bigg[\frac{\partial}{\partial p_j}(p_i(H-E))-(H-E)\frac{\partial p_i}{\partial p_j}\bigg]d^3p\nonumber\\
 &=&V\delta_{ij}\int_{H\leq E}(E-H)d^3p.
 \end{eqnarray}
 \item
 نعتبر الدالة
 \begin{eqnarray}
 G(f(\alpha),g(\alpha),\alpha)&=&\int_{f(\alpha)}^{g(\alpha)}F(\alpha,x)dx.
 \end{eqnarray}
 اذن
 \begin{eqnarray}
\frac{d}{d\alpha} &=&\frac{\partial G}{\partial \alpha}+\frac{\partial G}{\partial f}\frac{\partial f}{\partial\alpha}+\frac{\partial G}{\partial g}\frac{\partial g}{\partial\alpha}\nonumber\\
&=& \int_{f(\alpha)}^{g(\alpha)}\frac{\partial}{\partial \alpha}F(\alpha,x)dx+\frac{\partial g(\alpha)}{\partial\alpha}F(\alpha,g(\alpha))-\frac{\partial f(\alpha)}{\partial\alpha}
F(\alpha,f(\alpha)).\nonumber\\
\end{eqnarray}

 \item
 لدينا
 \begin{eqnarray}
 <p_i\frac{\partial H}{\partial p_j}>&=&\frac{\frac{d}{dE}\int_{H\leq E} p_i\frac{\partial H}{\partial p_j}d^3p}{\frac{d}{dE}\int_{H\leq E}d^3p}\nonumber\\
 &=&\frac{\frac{d}{dE}\int_{H\leq E} p_i\frac{\partial (H-E)}{\partial p_j}d^3p}{\frac{d}{dE}\int_{H\leq E}d^3p}\nonumber\\
 &=&\frac{-\delta_{ij}\frac{d}{dE}\int_{H\leq E} (H-E)d^3p}{\frac{d}{dE}\int_{H\leq E}d^3p}.
 \end{eqnarray}
 اذن يجب ان نحسب
 \begin{eqnarray}
 \frac{d}{dE}\int_{H\leq E} (H-E)d^3p&=&\frac{1}{2m}\frac{d}{dE}\int_{0\leq p^2\leq 2mE} (p^2-2mE)p^2dp d\Omega\nonumber\\
 &=&\frac{1}{2}\frac{d}{d\alpha}\int_{0}^{\alpha} (x-\alpha)\sqrt{x}dx d\Omega.
 \end{eqnarray}
 باستعمال
 نتيجة السؤال السابق حيث
  $f(\alpha)=0$
  و
  $g(\alpha)=\alpha$
  نحصل علي
  \begin{eqnarray}
 \frac{d}{dE}\int_{H\leq E} (H-E)d^3p
 &=&\frac{1}{2}\int_{0}^{\alpha} (-1)\sqrt{x}dx d\Omega\nonumber\\
 &=&-\int_{H\leq E} d^3p.
 \end{eqnarray}
 بالتعويض في المعادلة اعلاه نحصل علي
 \begin{eqnarray}
 <p_i\frac{\partial H}{\partial p_j}>
 &=&\frac{\delta_{ij}\frac{d}{dE}\int_{H\leq E} d^3p}{\frac{d}{dE}\int_{H\leq E}d^3p}\nonumber\\
 &=&\frac{\delta_{ij}\Phi(E)}{d\Phi(E)/dE}.
 \end{eqnarray}
 مباشرة نستنتج
 \begin{eqnarray}
 <2H>
 =\frac{3}{d\ln\Phi(E)/dE}=\frac{3k}{2dS/dE}=\frac{3}{2}kT.
 \end{eqnarray}

 \item
بالنسبة للهزاز التوافقي لدينا
 \begin{eqnarray}
 H=\frac{1}{2m}\sum_ip_i^2+\frac{k}{2}\sum_ix_i^2.
 \end{eqnarray}
 اذن
 \begin{eqnarray}
 <2H>&=&\sum_i(<p_i\frac{\partial H}{\partial p_i}>+<x_i\frac{\partial H}{\partial x_i}>).
 \end{eqnarray}
 نحصل الان علي
 \begin{eqnarray}
 <2H>&=6kT.
 \end{eqnarray}
  \end{itemize}
 \end{itemize}

  \paragraph{
تناقض جيبس:
}
\begin{itemize}
 \item  
  انتروبي غاز مثالي يعطي ب
  \begin{eqnarray}
  S=\frac{3Nk}{2}\ln \frac{2E}{3N\epsilon_0}+NS_0+\alpha.
\end{eqnarray}
 من الجهة الاخري فان الطاقة الداخلية لغاز مثالي تعطي ب
 \begin{eqnarray}
 E=\frac{3}{2}NkT.
\end{eqnarray}
مباشرة طاقة جزئ واحد هي 
\begin{eqnarray}
\epsilon=\frac{E}{N}=\frac{3}{2}kT.
\end{eqnarray}
يمكن ان نستخرج اذن بسهولة المعادلة الاتية
\begin{eqnarray}
  S=Nk\ln V\epsilon^{3/2}+NS_0^{'}+\alpha(N),
\end{eqnarray}
\begin{eqnarray}
 S_0^{'}=S_0+\frac{3}{2}k\ln\frac{4m}{3\pi^2\hbar^2}=\frac{3k}{2}(1+\ln\frac{4\pi m}{3 h^2})~,\alpha(N)=\alpha=k(-N\ln N+N).
\end{eqnarray}
يمكن ايضا ان نستخرج بسهولة 
معادلة ساكور - تترود

\begin{eqnarray}
  S=Nk\ln \frac{V}{N}\epsilon^{3/2}+\frac{3}{2}Nk(\frac{5}{3}+\ln \frac{4\pi m}{3h^2}).
\end{eqnarray}
\item
لان درجات الحرارة متساوية فان
$\epsilon_1=\epsilon_2=\epsilon$
و لان الكتل متساوية فان
$(S_0^{'})_1=(S_0^{'})_2=S_0^{'}$.
في الحالة الابتدائية نحصل اذن علي الانتروبيات
\begin{eqnarray}
  S_1=N_1k\ln V_1\epsilon^{3/2}+N_1S_0^{'}+\alpha(N_1).
\end{eqnarray}
\begin{eqnarray}
  S_2=N_2k\ln V_2\epsilon^{3/2}+N_2S_0^{'}+\alpha(N_2).
\end{eqnarray}
في الحالة النهائية فان
$V=V_1+V_2$
و
$N=N_1+N_2$.
الانتروبي يعطي ب
\begin{eqnarray}
  S=Nk\ln V\epsilon^{3/2}+NS_0^{'}+\alpha(N).
\end{eqnarray}
نلاحظ ان
$\alpha(N)=\alpha(N_1)+\alpha(N_2)$
لان هذا المعامل ناجم عن تطابق الجسيمات (تأثير كمي) اي عن
$N_1!N_2!$
(لان الجزئ الاول يختلف عن الجزئ الثاني)
و ليس عن
$(N_1+N_2)!$.
اذن
\begin{eqnarray}
  S=N_1k\ln V\epsilon^{3/2}+N_1S_0^{'}+\alpha(N_1)+N_2k\ln V\epsilon^{3/2}+N_2S_0^{'}+\alpha(N_2).
\end{eqnarray}
التغير في الانتروبي هو
\begin{eqnarray}
  \Delta S=N_1k\ln \frac{V}{V_1}+N_2k\ln \frac{V}{V_2}> 0.
\end{eqnarray}
هذا يسمي انتروبي الخلط
\footnote{${\rm entropy}~{\rm of}~{\rm mixing}.$}
.
$\Delta S>0$
لان التفاعل غير عكسي.

هذه النتيجة مؤكدة تجريبيا من اجل الغازات المختلفة. لكن من اجل الغازات المتطابقة فانها تؤدي الي ما يسمي بتناقض جيبس: انتروبي
خليط من الغاز مشكل من جسيمات متطابقة هو مختلف عن الصفر معطي بالمعادلة اعلاه و هذا من جهة مخالف للتجربة, و من جهة اخري اذا كان هذا صحيح فان هذا يعني ان انتروبي 
الغاز يتعلق بتاريخه و لما كان 
ممكنا ان يكون الانتروبي دالة تتعلق فقط بالحالة الترموديناميكية التي تتواجد فيها 
الجملة.   في الحقيقة  لا يمكن تعريف الانتروبي اصلا في هذه الحالة
لانه يمكننا تصور عدد كيفي من الحواجز الفاصلة داخل الاناء الذي يحتوي الغاز و بالتالي فان انتروبي الغاز يمكن ان يكون اي عدد كبير نريده.  

\item
نفترض الان ان الجزئ الاول و الجزئ الثاني متطابقان. في هذه الحالة عندما نلغي الحاجز
الفاصل فان لا شيئ يحدث. التفاعل هو عكسي في هذه الحالة. بالفعل في هذه الحالة
$\alpha(N)=\alpha(N_1+N_2)$
و بالتالي فان
 التغير في الانتروبي هو

\begin{eqnarray}
 \Delta S=N_1k\ln \frac{V}{V_1}+N_2k\ln \frac{V}{V_2}+N_1k\ln \frac{N_1}{N}+N_2k\ln \frac{N_2}{N}.
\end{eqnarray}
لكن
\begin{eqnarray}
 \frac{V_1}{N_1}=\frac{V_2}{N_2}=\frac{V}{N}=\frac{kT}{P}.
\end{eqnarray}
اذن

\begin{eqnarray}
 \Delta S=0.
\end{eqnarray}
\end{itemize}

\paragraph{
تمرين
$1$:
}
\begin{itemize}
 \item 
معادلة شرودينغر تعطي ب
\begin{eqnarray}
\bigg[-\frac{\hbar^2}{2m}\bigg(\frac{\partial^2}{\partial x^2}+\frac{\partial^2}{\partial y^2}+\frac{\partial^2}{\partial z^2}\bigg)+V(x,y,z)\bigg]\psi(x,y,z)=E\psi(x,y,z).
 \end{eqnarray}
 الجسم حر داخل المكعب اذن
 $V=0$.
 نستخدم فصل المتغيرات اي
 $\psi(x,y,z)=\psi_x(x)\psi_y(y)\psi_z(z)$.
 نحصل مباشرة علي المعادلات و الحلول
 \begin{eqnarray}
E=E_x+E_y+E_z=\frac{\hbar^2}{2m}(\Omega_x^2+\Omega_y^2+\Omega_z^2).
 \end{eqnarray}
 
 \begin{eqnarray}
\frac{d^2\psi_x}{dx^2}+\Omega_x^2\psi_x=0\Rightarrow \psi_x(x)=A\cos\Omega_x x+B\sin\Omega_x x ~,~\Omega_x^2=\frac{2mE_x}{\hbar^2}.
 \end{eqnarray}
 \begin{eqnarray}
\frac{d^2\psi_y}{dy^2}+\Omega_y^2\psi_y=0\Rightarrow \psi_y(y)=A\cos\Omega_y y+B\sin\Omega_y y ~,~\Omega_y^2=\frac{2mE_y}{\hbar^2}.
 \end{eqnarray}
 \begin{eqnarray}
\frac{d^2\psi_z}{dz^2}+\Omega_z^2\psi_z=0\Rightarrow \psi_z(z)=A\cos\Omega_z z+B\sin\Omega_z z ~,~\Omega_z^2=\frac{2mE_z}{\hbar^2}.
 \end{eqnarray}
 اذا افترضنا الشروط الحدية 
 $\psi_x(0)=\psi_x(L)=0$, $\psi_y(0)=\psi_y(L)=0$, $\psi_z(0)=\psi_z(L)=0$ 
 فاننا نحصل مباشرة علي التواترات الزاوية
 \begin{eqnarray}
\Omega_x=\frac{\pi n_x}{L}~,~\Omega_y=\frac{\pi n_y}{L}~,~\Omega_z=\frac{\pi n_z}{L}.
 \end{eqnarray}
 قيم الطاقة المسوح بها هي اذن
  \begin{eqnarray}
E=\frac{\hbar^2\pi^2}{2mL^2}(n_x^2+n_y^2+n_z^2).
 \end{eqnarray}
 \item
 نعرف الفسحة الطاقوية ب
  \begin{eqnarray}
\epsilon_0=\frac{\hbar^2\pi^2}{2mL^2}.
 \end{eqnarray}
 المستويات الطاقوية العشرة الاولي هي كالاتي
 \begin{itemize}
 \item
 $E=3\epsilon_0$
 يوافق الحالة
 $(1,1,1)$.
 \item
 $E=6\epsilon_0$
 يوافق الثلاث حالات
 $(1,1,2),(1,2,1),(2,1,1)$.
 \item
 $E=9\epsilon_0$
 يوافق الثلاث حالات
 $(1,2,2),(2,1,2),(2,2,1)$.
 \item
 $E=11\epsilon_0$
 يوافق الثلاث حالات
 $(1,1,3),(1,3,1),(3,1,1)$.
 \item
 $E=12\epsilon_0$
 يوافق الحالة
 $(2,2,2)$.
 \item
 $E=14\epsilon_0$
 يوافق الست حالات
 $(1,2,3)$
 و تبديلاتها
 .
\item
 $E=17\epsilon_0$
 يوافق الثلاث حالات
 $(2,2,3),(2,3,2),(3,2,2)$.
 \item
 $E=18\epsilon_0$
 يوافق الثلاث حالات
 $(1,1,4),(1,4,1),(4,1,1)$.
 \item
 $E=19\epsilon_0$
 يوافق الثلاث حالات
 $(1,3,3),(3,1,3),(3,3,1)$.
 \item
 $E=21\epsilon_0$
 يوافق الست حالات
 $(1,2,4)$
 و تبديلاتها
 .
 \end{itemize}
 \item
 من معادلة الحالة
 $PV=nRT$
 نحسب
  \begin{eqnarray}
L=\bigg(\frac{RT}{P}\bigg)^{1/3}=\bigg(\frac{8.315.273}{10^5}\bigg)^{1/3}=0.3m.
 \end{eqnarray}
 كتلة واحد مول من الهيليوم هو
 $4$
 غرام
 و بالتالي فان كتلة ذرة واحدة من الهيليوم هي
 \begin{eqnarray}
m=\frac{4.10^{-3}}{6.022.10^{23}}.
 \end{eqnarray}
 الفسحة الطاقوية تعطي ب
 \begin{eqnarray}
\epsilon_0=\frac{h^2}{8mL^2}=\frac{(6.63.10^{-34})^2}{8.m.(0.3)^2}=91.91.10^{-42}\frac{m^2kg}{s^2}.
 \end{eqnarray}
 الطاقة المتوسطة لذرة هيليوم واحدة هي
 \begin{eqnarray}
<E>=\frac{3}{2}kT=565.11 J.
 \end{eqnarray}
رتبة عظم الاعداد الكمية
 $n_x,n_y$
 و
 $n_z$
 تعطي اذن ب
 \begin{eqnarray}
n_x,n_y,n_z\sim \sqrt{\frac{<E>}{\epsilon_0}}\sim 10^{21}.
 \end{eqnarray}
 \item
 العمل الميكانيكي يوافق تغير في الحجم
 $V$ 
 الذي يؤدي الي تغير في الفسحة الطاقوية.  بالفعل اذا تناقص الحجم  
 $V$
 فان
 الفسحة الطاقوية تكبر و العكس. اذن التغير في الحجم
 يؤدي الي تغير في طاقة الجملة عبر تغير المستويات الطاقوية مع الحفاظ علي اعداد الاحتلال
 \footnote{${\rm occupation}~{\rm numbers}.$}
 ثابتة.
 
 من الجهة الاخري فان التغير في الانتروبي يؤدي الي تغير في طاقة الجملة,
 الذي هو عبارة هنا عن كمية حرارة,  عبر تغير اعداد الاحتلال مع 
 الحفاظ علي المستويات الطاقوية ثابتة.
 
\item
لدينا ثلاث امكانيات عند توزريع الجسيمات علي المستويات الطاقوية المختلفة للحصول علي الطاقة
$E=18\epsilon_0$:

\begin{itemize}
\item
يمكن وضع الجسيم الاول علي المستوي
$E=3\epsilon_0$,
الجسيم الثاني علي المستوي
$E=6\epsilon_0$
و الجسيم الثالث علي المستوي
$E=9\epsilon_0$.
\item
يمكن وضع الجسيم الاول علي المستوي
$E=3\epsilon_0$,
الجسيم الثاني علي المستوي
$E=3\epsilon_0$
و الجسيم الثالث علي المستوي
$E=12\epsilon_0$.
\item
يمكن وضع الجسيم الاول علي المستوي
$E=6\epsilon_0$,
الجسيم الثاني علي المستوي
$E=6\epsilon_0$
و الجسيم الثالث علي المستوي
$E=6\epsilon_0$.
\end{itemize}
الان عدد الحالات الميكروسكوبية المختلفة يتعلق علي طبيعة و سبين الجسيمات كالاتي:
\paragraph{
الجسيمات متمايزة
:}
\begin{itemize}
\item 
عند توزيع ثلاث جسيمات متمايزة علي 
$E=3\epsilon_0$,
$E=6\epsilon_0$
و
$E=9\epsilon_0$
فانه لدينا 
$3!$
تبديلة ممكنة و مختلفة للجسيمات و 
$1.3.3$
امكانية راجعة الي انحلال المستويات الطاقوية. اذن هناك
$6.9=54$
حالة ميكروسكوبية في هذه الحالة.
\item 
عند توزيع ثلاث جسيمات متمايزة علي 
$E=3\epsilon_0$,
$E=3\epsilon_0$
و
$E=12\epsilon_0$
فانه لدينا 
$3!/2!$
تبديلة ممكنة و مختلفة للجسيمات و 
$1.1.1$
امكانية راجعة الي انحلال المستويات الطاقوية. اذن هناك
$3.1=3$
حالات ميكروسكوبية في هذه الحالة.
\item 
عند توزيع ثلاث جسيمات متمايزة علي 
$E=6\epsilon_0$,
$E=6\epsilon_0$
و
$E=6\epsilon_0$
فانه لدينا 
$3!/3!$
تبديلة ممكنة و مختلفة للجسيمات و 
$3.3.3$
امكانية راجعة الي انحلال المستويات الطاقوية. اذن هناك
$1.27=27$
حالة ميكروسكوبية في هذه الحالة.
\end{itemize}
\paragraph{
الجسيمات بوزونات متطابقة
:}
\begin{itemize}
\item 
عند توزيع ثلاث بوزونات متطابقة علي 
$E=3\epsilon_0$,
$E=6\epsilon_0$
و
$E=9\epsilon_0$
فانه
علينا ان نقسم العدد المحصل سابقا علي
$3!$
لان التبديلات الان غير مهمة بسبب ان الجسيمات متطابقة. اذن نحصل علي
$54/6=9$
حالة ميكروسكوبية مختلفة.
\item 
عند توزيع ثلاث بوزونات متطابقة علي 
$E=3\epsilon_0$,
$E=3\epsilon_0$
و
$E=12\epsilon_0$
فانه 
علينا ان نقسم العدد المحصل سابقا علي
$3$
لان التبديلة الان غير مهمة بسبب ان الجسيمات متطابقة. اذن نحصل علي
$3/3=1$
حالة ميكروسكوبية في هذه الحالة.
\item 
عند توزيع ثلاث بوزونات متطابقة علي 
$E=6\epsilon_0$,
$E=6\epsilon_0$
و
$E=6\epsilon_0$
فاننا ننطلق من ال
$27$
حالة ميكروسكوبية التي حصلنا عليها سابقا. من بين هذه الامكانيات الحالات الميكروسكوبية التي هي 
فعلا مختلفة هي كما يلي. كل الجسيمات تقع علي نفس الحالة الكمية علي المستوي الطاقوي
$E=6\epsilon_0$:
$3$
حالات. جسيمان يقعان علي نفس الحالة الكمية علي المستوي
$E=6\epsilon_0$:
$6$
حالات. كل جسيم يقع علي حالة مختلفة علي المستوي
$E=6\epsilon_0$:
حالة واحدة. اذن هناك
$10$
حالات مختلفة.

\end{itemize}
\paragraph{
الجسيمات بوزونات متطابقة ذات سبين يساوي واحد
:}
في هذه الحالة لدينا ثلاث حالات سبين مختلفة كل مرة. اذن لدينا الاتي:
\begin{itemize}
\item 
عند توزيع ثلاث بوزونات متطابقة علي 
$E=3\epsilon_0$,
$E=6\epsilon_0$
و
$E=9\epsilon_0$
فانه لدينا
$9$
حالات ميكروسكوبية مختلفة. عند اضافة السبين يصبح عدد الحالات
$9.3.3.3=243$
حالة ميكروسكوبية.
\item 
عند توزيع ثلاث بوزونات متطابقة علي 
$E=3\epsilon_0$,
$E=3\epsilon_0$
و
$E=12\epsilon_0$
فانه لدينا 
حالة ميكروسكوبية واحدة. عند اضافة السبين فان الجسيمين علي المستوي
$E=3\epsilon_0$
لهما
$6$
حالات سبين مختلفة اما الجسيم الاخر فان له 
$3$
حالات سبين مختلفة. اذن يصبح لدينا
$6.3=18$
حالة ميكروسكوبية.
\item 
من اجل ثلاث بوزونات متطابقة  موزعة علي 
$E=6\epsilon_0$,
$E=6\epsilon_0$
و
$E=6\epsilon_0$
كان لدينا
$3+6+1=10$
امكانيات. من اجل الامكانيات الثلاث الاولي يمكن لكل الجسيمات ان يكون لها نفس السبين (ثلاث حالات), جسيمان لهما نفس السبين (ست حالات) او كل جسيم له سبين مختلف (حالة واحدة). اذن
$3$
تصبح
$3.10$
حالة. من اجل الامكانيات الستة التالية فان الجسيمين اللذين يقعان علي نفس الحالة الكمية لهما
$6$
حالات سبين مختلفة اما الجسيم الثالث فله
$3$
حالات سبين مختلفة. اذن
$6$
تصبح
$6.3.6$.
من اجل الامكانية الاخيرة
لدينا
$3.3.3$
حالة سبين مختلفة. اذن لدينا في الاجمال
$3.10+6.3.6+1.3.3.3=165$
حالة ميكروسكوبية.
\end{itemize}
\paragraph{
الجسيمات فرميونات متطابقة ذات سبين يساوي نصف
:}
في هذه الحالة لدينا حالتي سبين مختلفة كل مرة. ايضا الفرميونات لا يمكن ان تحتل نفس الحالة الكمية مبدأ الاستبعاد لباولي
\footnote{${\rm Pauli}$.}
. اذن لدينا الاتي:
\begin{itemize}
\item 
عند توزيع ثلاث فرميونات متطابقة علي 
$E=3\epsilon_0$,
$E=6\epsilon_0$
و
$E=9\epsilon_0$
فانه لدينا
$9$
حالات ميكروسكوبية مختلفة. عند اضافة السبين يصبح عدد الحالات
$9.2.2.2=72$
حالة ميكروسكوبية.
\item 
عند توزيع ثلاث فرميونات متطابقة علي 
$E=3\epsilon_0$,
$E=3\epsilon_0$
و
$E=12\epsilon_0$
فانه لدينا 
حالة ميكروسكوبية واحدة. عند اضافة السبين فان الجسيمين علي المستوي
$E=3\epsilon_0$
لهما
حالة سبين واحدة لان السبين يجب ان يكون مقترنا هنا حسب مبدأ الاستبعاد لباولي اما الجسيم الاخر فان له 
حالتي سبين مختلفتين. اذن يصبح لدينا
$1.2=2$
حالة ميكروسكوبية.
\item 
من اجل ثلاث فرميونات متطابقة  موزعة علي 
$E=6\epsilon_0$,
$E=6\epsilon_0$
و
$E=6\epsilon_0$
 يمكن ان يكون لدينا
$6+1=7$
امكانيات. 
من اجل الامكانيات الستة الاولي فان الجسيمين اللذين يقعان علي نفس الحالة الكمية لهما
حالة سبين واحدة لان السبين لا يمكن الا ان يكون مقترنا هنا حسب مبدأ الاستبعاد لباولي اما الجسيم الثالث فله
حالتي سبين مختلفتين. اذن
$6$
تصبح
$6.2$.
من اجل الامكانية الاخيرة
لدينا
$2.2.2$
حالة سبين مختلفة. اذن لدينا في الاجمال
$6.2+1.2.2.2=20$
حالة ميكروسكوبية.
\end{itemize}

\end{itemize}

\paragraph{
تمرين
$2$:
}
\begin{itemize}
 \item 
 \begin{eqnarray}
  E=\epsilon_0 n^2~,~\epsilon_0=\frac{\hbar^2\pi^2}{2mL^2}.
 \end{eqnarray}
\item
\begin{itemize}
\item
\begin{eqnarray}
 |n_1s_1m_1>|n_2s_2m_2>|n_3s_3m_3>.
\end{eqnarray}

\item
توجد امكانيتان: 
$(A)$ $|9\epsilon_0>|9\epsilon_0>|9\epsilon_0>$~,~$(B)$ $|25\epsilon_0>|\epsilon_0>|\epsilon_0>$.
\item
اذا كانت الجسيمات متمايزة بدون سبين فان عدد الحالات هو كما يلي: 
$(A)$ $1$~,~$(B)$ $3$.
\item
 اذا كانت الجسيمات متطابقة ذات سبين
 $s=0$
 فان عدد الحالات 
هو كما يلي:
$(A)$ $1$~,~$(B)$ $1$.

 اذا كانت الجسيمات متطابقة ذات سبين
 $s=1$
 فان عدد الحالات 
هو كما يلي:
$(A)$ 
$10$:
كل الجسيمات لها نفس السبين
$(3)$,
جسيمان لهما نفس السبين و الاخر مختلف
$(6)$,
سبينات الجسيمات كلها مختلفة
$(1)$~, ~$(B)$ $18$: 
الجسيمان اللذان يقعان علي نفس المستوي لهما سته حالات سبين و الجسيم الاخير له ثلاث حالات سبين.

 اذا كانت الجسيمات متطابقة ذات سبين
 $s=1/2$
 فان عدد الحالات 
هو كما يلي:
$(A)$ 
$0$:
مبدأ الاستبعاد لباولي
~, ~$(B)$ $2$: 
الجسيمان اللذان يقعان علي نفس المستوي لهما حالة سبين و احدة و الجسيم الاخير له حالاتي سبين.
\end{itemize}
 \end{itemize}

 \paragraph{
تمرين
$3$:
}

 \begin{itemize}
  \item 
  نعرف ان الطاقة تعطي ب
  \begin{eqnarray}
   E=\frac{\pi^2\hbar^2 n^2}{2mL^2}.
  \end{eqnarray}
  مباشرة عدد الحالات يعطي ب
\begin{eqnarray}
   \Phi(E)=n=\frac{L}{\pi\hbar}\sqrt{2mE}.
  \end{eqnarray}
  اذن
\begin{eqnarray}
   \Omega(E)=\frac{d\Phi}{dE}\delta E=n=\frac{L}{2\pi\hbar}\sqrt{\frac{2m}{E}}\delta E.
  \end{eqnarray}
 \item
 في الميكانيك الكلاسيكي لدينا
\begin{eqnarray}
   \Phi(E)&=&\frac{1}{h}\int_{H\leq E} dx dp\nonumber\\
   &=&\frac{2L}{h}\int_0^{p\leq \sqrt{2mE}}dp\nonumber\\
   &=&\frac{2L}{h}\sqrt{2mE}.
  \end{eqnarray}
  هذه نفس العبارة التي حصلنا عليها في الميكانيك الكمي.
 \end{itemize}
 \paragraph{
تمرين
$4$:
}
نحسب
\begin{eqnarray}
   \Phi(E)&=&\frac{1}{h^3}\int_{H\leq E} dxdydz dp_xdp_ydp_z\nonumber\\
   &=&\frac{V}{h^3}\int_{p_x^2+p_y^2+p_z^2\leq 2mE}dp_xdp_ydp_z\nonumber\\
   &=&\frac{V}{h^3}\frac{4}{3}\pi(2mE)^{3/2}.
  \end{eqnarray}
\begin{eqnarray}
  \Omega(E)=\frac{d \Phi(E)}{dE}\delta E.
  \end{eqnarray}
  عند اضافة سبين
  $s$
  فاننا نضرب عدد الحالات ب
  $2s+1$.
 
 \paragraph{
تمرين
$5$:
}
\begin{itemize}
 \item 
الفضاء الطوري له سته ابعاد هي
$x_1,p_1,x_2,p_2,x_3,p_3$.
الهاميلتونية تعطي ب
\begin{eqnarray}
 H=\frac{p_1^2}{2m}+\frac{p_2^2}{2m}+\frac{p_3^2}{2m}.
  \end{eqnarray}
 \item
 
\begin{eqnarray}
   \Phi(E)&=&\frac{1}{h^3}\int_{H\leq E} dx_1dx_2dx_3 dp_1dp_2dp_3\nonumber\\
   &=&\frac{L^3}{h^3}\int_{p_1^2+p_2^2+p_3^2\leq 2mE}dp_1dp_2dp_3\nonumber\\
   &=&\frac{L^3}{h^3}\frac{4}{3}\pi(2mE)^{3/2}.
  \end{eqnarray}
 \item
 \begin{eqnarray}
  \Omega(E)&=&\frac{d \Phi(E)}{dE}\delta E\nonumber\\
  &=&\frac{4\pi L^3m}{h^3}(2mE)^{1/2}\delta E.
  \end{eqnarray}
 \item
نضرب في اثنين.
 \item
نضرب في
 $2s+1$.
 \item
 نقسم علي
 $3!$.
\end{itemize}

\paragraph{
تمرين
$8$:
}
نحصل علي
\begin{eqnarray}
 \Phi(E)=\frac{L^3}{h^3}\frac{4}{3}\pi\sqrt{8m^3}E^{3/2}=\frac{\pi}{6}(E/\epsilon_0)^{3/2}.
\end{eqnarray}
\begin{eqnarray}
 \Omega(E)=4\pi\frac{L^3}{h^3}\sqrt{2m^3}E^{1/2}\delta E=\frac{\pi}{4}(E/\epsilon_0)^{1/2}\delta E/\epsilon_0.
\end{eqnarray}
نتصور ان
$N=3$
عدد كبير بحيث يمكننا ان نحسب الانتروبي من العبارة
\begin{eqnarray}
 \frac{S}{k}=\ln\Phi\Rightarrow S=k\ln\frac{\pi}{6}+\frac{3k}{2}\ln \frac{E}{\epsilon_0}.
\end{eqnarray}
الطاقة الداخلية تعطي بالعبارة
\begin{eqnarray}
 E=\epsilon_0e^{-\frac{2}{3}\ln\frac{\pi}{6}}e^{\frac{2S}{3k}}.
\end{eqnarray}
نحسب المقادير الترموديناميكية

\begin{eqnarray}
T=(\frac{\partial E}{\partial S})_{V,N}=\frac{2}{3k}E\Rightarrow E=\frac{3}{2}kT.
\end{eqnarray}
\begin{eqnarray}
P=-(\frac{\partial E}{\partial V})_{S,N}=-(\frac{\partial \epsilon_0}{\partial V})_{S,N}\frac{1}{\epsilon_0}.E=\frac{2}{L}E\Rightarrow PL=2E=3kT.
\end{eqnarray}

  \paragraph{
تمرين
$9$:
}
\begin{itemize}
 \item 
عدد ابعاد الفضاء الطوري هو
$2$
باحداثيات معطاة ب
$x$
و
$p$.
الهاميلتونية تعطي ب
\begin{eqnarray}
  H=\frac{p^2}{2m}+\frac{1}{2}kx^2.
  \end{eqnarray}
  معادلات هاميلتون
  \begin{eqnarray}
  \frac{\partial H}{\partial p}=\frac{p}{m}=\dot{x}.
  \end{eqnarray}
  \begin{eqnarray}
  \frac{\partial H}{\partial x}=kx=-\dot{p}.
  \end{eqnarray}
اذن  الحركة دورية بتواتر زاوي
\begin{eqnarray}
  \Omega=\sqrt{\frac{k}{m}}.
  \end{eqnarray}
 الحركة دورية تعطي صراحة بالحلول
 \begin{eqnarray}
  x=A\cos (\Omega t+\phi)~,~p=m\dot{x}=-mA\Omega\sin(\Omega t+\phi).
  \end{eqnarray}
  نحسب بالتالي
  \begin{eqnarray}
  H=\frac{A^2k}{2}\Rightarrow A=\sqrt{\frac{2H}{m\Omega^2}}.
  \end{eqnarray}
  
 \item
 \begin{eqnarray}
   \Phi(E)&=&\frac{1}{h}\int_{H\leq E} dx dp\nonumber\\
   &=&\frac{1}{h}\int_{p^2/b^2+x^2/a^2\leq 1}dx dp.
  \end{eqnarray}
  نحصل اذن علي مساحة قطع ناقص بانصاف محاور
  \begin{eqnarray}
   a=\sqrt{\frac{2E}{k}}~,~b=\sqrt{2mE}.
  \end{eqnarray}
  مساحة القطع الناقص
  $x^2/a^2+y^2/b^2=1$
  تحسب كالاتي
  \begin{eqnarray}
   S&=&\int_{-a}^adx\int_{-b\sqrt{1-x^2/a^2}}^{b\sqrt{1-x^2/a^2}}dy\nonumber\\
   &=&2ba\int_{-a}^a\frac{dx}{a}\sqrt{1-x^2/a^2}\nonumber\\
   &=&2ab\int_{-\pi/2}^{\pi/2}\cos\phi d\phi\nonumber\\
   &=&ab\pi.
  \end{eqnarray}
  اذن عدد الحالات هو
  \begin{eqnarray}
   \Phi(E)&=&\frac{E}{\hbar \Omega}.
  \end{eqnarray}
 \item
\begin{eqnarray}
   \Omega(E)&=&\frac{\delta E}{\hbar \Omega}.
  \end{eqnarray}
  \item
  \begin{eqnarray}
   p\frac{\partial H}{\partial p}+x\frac{\partial H}{\partial x}=2H.
  \end{eqnarray}
  اذن
  \begin{eqnarray}
 \int_0^E 2H dpdx&=&\int_0^E  p\frac{\partial H}{\partial p}dpdx+\int_0^Ex\frac{\partial H}{\partial x}dpdx\nonumber\\
 &=&\int_0^E  p\frac{\partial (H-E)}{\partial p}dpdx+\int_0^Ex\frac{\partial (H-E)}{\partial x}dpdx\nonumber\\
 &=&-2\int_0^E(H-E)dp dx.
  \end{eqnarray}
  اذن
  \begin{eqnarray}
 <H>&=&\frac{\frac{d}{dE}\int_0^E H dpdx}{\frac{d}{dE}\int_0^E dpdx}\nonumber\\
 &=&\frac{\frac{d}{dE}\int_0^E (E-H) dpdx}{\frac{d}{dE}\int_0^E dpdx}\nonumber\\
 &=&\frac{\int_0^E  dpdx}{\frac{d}{dE}\int_0^E dpdx}\nonumber\\
 &=&\frac{\Phi}{{d\Phi}/{dE}}\nonumber\\
 &=&\frac{1}{{d\ln\Phi}/{dE}}\nonumber\\
 &=&\frac{k}{{dS}/{dE}}\nonumber\\
 &=&kT.
  \end{eqnarray}
  استخدمنا اعلاه العلاقة
  \begin{eqnarray}
 \frac{d}{dx}\int_0^{h(x)}f(x,y)dy=\int_0^{h(x)}\frac{\partial}{\partial x}f(x,y)dy+h^{'}(x)f(x,h(x)).
  \end{eqnarray}
\end{itemize}
\paragraph{
تمرين
$10$:
}

\begin{itemize}
 \item 
 الطاقة داخل مكعب تعطي ب
 \begin{eqnarray}
  E=\epsilon_0(n_x^2+n_y^2+n_z^2).
  \end{eqnarray}
بالنسبة للمكعب الاول هناك تمثيلتان طاقويتان:
\begin{itemize}
\item $:{\bf A}$
الجسيمان علي المستوي الطاقوي
$E=6\epsilon_0$.
لان هذا المستوي منحل ثلاث مرات لدينا اذن
$3.3=9$
حالة ميكروسكوبية ممكنة.
\item $:{\bf B}$
احد الجسيمان علي المستوي
$E=3\epsilon_0$
و الاخر علي المستوي
$E=9\epsilon_0$.
لان 
$E=3\epsilon_0$
غير منحل و
$E=9\epsilon_0$
منحل ثلاث مرات لدينا
$3.2=6$
حالة ميكروسكوبية حيث
$2$ 
راجع الي تمايز الجسيمات
.
\end{itemize}
عدد الحالات الميكروسكوبية في المكعب الاول هو
$\Omega_I=9+6=15$.

بالنسبة للمكعب الثاني الجسيم يغع علي المستوي
$E=9\epsilon_0$.
اذن هناك ثلاث حالات ميكروسكوبية اي
$\Omega_{II}=3$.

لان الجدار الفاصل ادياباتيكي فان عدد الحالات الكلي هو
$\Omega_I.\Omega_{II}=15.3=45$.

 \item
 
 لما ننزع الجدار الفاصل فان الحجم يصبح متوازي اسطح. الطاقة تعطي الان ب
 \begin{eqnarray}
  E&=&\frac{\pi^2\hbar^2}{2m}(n_x^2/L_x^2+n_y^2/L_y^2+n_z^2/L_z^2)\nonumber\\
  &=&\epsilon_0(n_x^2/4+n_y^2+n_z^2).
  \end{eqnarray}
هناك طاقات تساوي
$\epsilon_0 p/4$
حيث 
$p$
عدد طبيعي لا يقبل القسمة علي
$4$
و لانه لدينا ثلاث جسيمات (عدد فردي) فان هذه الطاقات لا يمكن ان تجمع ل
$21\epsilon_0$.
يتبقي لنا الطاقات التالية:
\begin{itemize}
\item $E=3\epsilon_0$: $(n_x,n_y,n_z)=(2,1,1)$.
\item $E=6\epsilon_0$: $(n_x,n_y,n_z)=(2,1,2),(2,2,1),(4,1,1)$.
\item $E=9\epsilon_0$: $(n_x,n_y,n_z)=(2,2,2),(4,1,2),(4,2,1)$.
\item $E=11\epsilon_0$: $(n_x,n_y,n_z)=(2,3,1),(2,1,3)$.
\item $E=12\epsilon_0$: $(n_x,n_y,n_z)=(4,2,2)$.
\item $E=14\epsilon_0$: $(n_x,n_y,n_z)=(2,3,2),(2,2,3)$.
\item $E=18\epsilon_0$: $(n_x,n_y,n_z)=(2,4,1),(2,1,4)$.
\end{itemize}
اذن هناك ثلاث تمثيلات طاقوية ممكنة:
\begin{itemize}
\item $:{\bf A}$
جسيمان علي المستوي الطاقوي
$E=6\epsilon_0$
و الاخر علي المستوي الطاقوي
$E=9\epsilon_0$.
الجسيم علي المستوي
$E=9\epsilon_0$
له ثلاث امكانيات, الجسيمان علي المستوي
$E=6\epsilon_0$
لهما
تسعة امكانيات و تمايز الجسميات يؤدي الي معامل ضرب يساوي ثلاثة. اذن هناك
$3.9.3=81$
حالة ميكروسكوبية ممكنة.
\item $:{\bf B}$
جسيم علي المستوي
$E=3\epsilon_0$,
الجسيم الثاني علي المستوي
$E=6\epsilon_0$
و الثالث علي المستوي
$E=12\epsilon_0$.
الجسيم
علي المستوي
$E=3\epsilon_0$
له امكانية واحدة, الجسيم
علي المستوي
$E=6\epsilon_0$ 
له ثلاث امكانيات, الجسيم علي
المستوي
$E=12\epsilon_0$
له امكانية واحدة و تمايز الجسيمات يؤدي الي
$3!$.
اذن لدينا
$6.1.3.1=18$
حالة ميكروسكوبية.
\item  $:{\bf C}$
جسيمان علي المستوي الطاقوي
$E=9\epsilon_0$
و الاخر علي المستوي الطاقوي
$E=3\epsilon_0$.
الجسيمان علي المستوي
$E=9\epsilon_0$
لهما تسعة امكانيات, الجسيم علي المستوي
$E=3\epsilon_0$
له
 امكانية واحدة و تمايز الجسميات يؤدي الي معامل ضرب يساوي ثلاثة. اذن هناك
$3.9.1=27$
حالة ميكروسكوبية ممكنة.
\end{itemize}
عدد الحالات الاجمالي هو
$81+18+27$.
عدد الحالات يزداد عند رفع الحاجز الادياباتيكي.
\end{itemize}
\paragraph{
تمرين
$11$:
}
\begin{itemize}
 \item 
 بالنسبة للعلبة الاولي لدينا تمثيلتان طاقويتان: $(A)$
 الجسيمان علي المستوي الطاقوي
 $E=6\epsilon_0$ 
 المنحل ثلاث مرات. اذن هناك 
 $3.3=9$
 حالة ميكروسكوبية. $(B)$
 جسيم علي المستوي الطاقوي
 $E=9\epsilon_0$
 المنحل ثلاث مرات و جسيم علي المستوي الطاقوي
 $E=3\epsilon_0$
 غير المنحل. اذن هناك
 $3.2=6$
 حالة ميكروسكوبية. عدد الحالات الاجمالية بالنسبة للمكعب الاول هو
 $\Omega_I=9+6=15$.
 
 بالنسبة للعلبة الثانية لدينا  ايضا تمثيلتان طاقويتان: $(A)$
 الجسيمان علي المستوي الطاقوي
 $E=9\epsilon_0$ 
 المنحل ثلاث مرات. اذن هناك 
 $3.3=9$
 حالة ميكروسكوبية. $(B)$
 جسيم علي المستوي الطاقوي
 $E=6\epsilon_0$
 المنحل ثلاث مرات و جسيم علي المستوي الطاقوي
 $E=12\epsilon_0$
 غير المنحل. اذن هناك
 $3.2=6$
 حالة ميكروسكوبية. عدد الحالات الاجمالية بالنسبة للمكعب الثاني هو
 $\Omega_{II}=9+6=15$.
 
 عدد الحالات الكلي هو
 $\Omega=\Omega_I.\Omega_{II}=15.15=225$
 حالة ميكروسكوبية.
 
 \item

 \begin{itemize}
 \item
 يمكن ان تتوزع الطاقة كما يلي:
 $(A)$ 
 $E_I=6\epsilon_0$~,~$E_{II}=24\epsilon_0$
 و العكس.
 $(B)$
 $E_I=9\epsilon_0$~,~$E_{II}=21\epsilon_0$
 و العكس.
 $(C)$
 $E_I=12\epsilon_0$~,~$E_{II}=18\epsilon_0$
و العكس.
$(D)$
 $E_I=15\epsilon_0$~,~$E_{II}=15\epsilon_0$.

 \item
 $(A)$:
 العلبة الاولي:
 الجسيمان  علي المستوي الطاقوي
 $E=3\epsilon_0$ (حالة واحدة).
 العلبة الثانية: 
الجسيمان علي المستوي الطاقوي
$E=12\epsilon_0$ (حالة واحدة),
جسيم علي المستوي الطاقوي
$E=18\epsilon_0$
المنحل ثلاث مرات
و الجسيم الاخر علي المستوي الطاقوي
$E=6\epsilon_0$
المنحل ثلاث مرات
(تسعة حالات),
جسيم عي المستوي الطاقوي
$E=21\epsilon_0$
المنحل ستة مرات و الجسيم الاخر علي المستوي الطاقوي
$E=3\epsilon_0$ ($12$ حالة).
اذن هناك 
$\Omega_A=\Omega_1.\Omega_2.2=1.31.2=62$
حالة.

$(B)$:
 العلبة الاولي:
 جسيم علي المستوي الطاقوي
 $E=3\epsilon_0$
 و جسيم علي المستوي
 الطاقوي
 $E=6\epsilon_0$ 
 المنحل ثلاث مرات
 ( ستة حالات).
 العلبة الثانية: 
جسيم علي المستوي الطاقوي
$E=3\epsilon_0$ 
و
جسيم علي المستوي الطاقوي
$E=18\epsilon_0$
المنحل ثلاث مرات
(ستة حالات),
جسيم عي المستوي الطاقوي
$E=9\epsilon_0$
المنحل ثلاثة مرات و الجسيم الاخر علي المستوي الطاقوي
$E=12\epsilon_0$ (ستة حالات).
اذن هناك 
$\Omega_B=\Omega_1.\Omega_2.2=6.12.2=144$
حالة.

$(C)$:
 العلبة الاولي:
 جسيم علي المستوي الطاقوي
 $E=3\epsilon_0$
 و جسيم علي المستوي
 الطاقوي
 $E=9\epsilon_0$ 
 المنحل ثلاث مرات
 ( ستة حالات), الجسيمان علي المستوي الطاقوي
 $E=6\epsilon_0$
 المنحل ثلاث مرات
 (تسعة حالات).
 العلبة الثانية: 
جسيم علي المستوي الطاقوي
$E=12\epsilon_0$ 
و
جسيم علي المستوي الطاقوي
$E=6\epsilon_0$
المنحل ثلاث مرات
(ستة حالات),
الجسيمان عي المستوي الطاقوي
$E=9\epsilon_0$
المنحل ثلاثة مرات 
 (تسعة حالات).
اذن هناك 
$\Omega_C=\Omega_1.\Omega_2.2=15.15.2=450$
حالة.

$(D)$:
 العلبة الاولي:
 جسيم علي المستوي الطاقوي
 $E=6\epsilon_0$
 المنحل ثلاث مرات 
 و جسيم علي المستوي
 الطاقوي
 $E=9\epsilon_0$ 
 المنحل ثلاث مرات
 ( $18$ حالة), 
جسيم علي المستوي الطاقوي
$E=3\epsilon_0$ 
و
جسيم علي المستوي الطاقوي
$E=12\epsilon_0$
(حالتان).
 العلبة الثانية: نفس التعداد.
اذن هناك 
$\Omega_D=\Omega_1.\Omega_2=20.20=400$
حالة.

 \item
 العدد الاجمالي للحالات الميكروسكوبية هو
 \begin{eqnarray}
  \Omega=\Omega_A+\Omega_B+\Omega_C+\Omega_D=62+144+450+400=1056.
 \end{eqnarray}
حسب مسلمة تساوي الاحتمال فان احتمال الحصول علي اي حالة ميكروسكوبية هو
\begin{eqnarray}
  P=1/1056.
 \end{eqnarray}
 نحسب ايضا الاحتمالات
\begin{eqnarray}
&&  P(E_I=6\epsilon_0)=31/1056~,~P(E_I=9\epsilon_0)=72/1056\nonumber\\
&&P(E_I=15\epsilon_0)=400/1056~,~P_I(E_I=12\epsilon_0)=225/1056.\nonumber\\
 \end{eqnarray}
 \end{itemize}

\end{itemize}
  
\paragraph{
تمرين
$12$:
}

\begin{itemize}
 \item
المكاملة بالتجزئة.
\item
استخدم الخاصية التي برهنا عليها في السؤال السابق.
\item
المكاملة بالتجزئة و تغيير المتغير
$x=az^2$
يؤديان الي العلاقات
\begin{eqnarray}
 (1/2)!=\Gamma(3/2)&=&\frac{1}{2}\int_0^{\infty}x^{-1/2}e^{-x} dx\nonumber\\
 &=&-a^{3/2}\frac{d}{da}\int_{-\infty}^{+\infty}dz e^{-az^2}\nonumber\\
 &=&-a^{3/2}\frac{d}{da}(\sqrt{\pi}a^{-1/2})\nonumber\\
 &=&\frac{1}{2}\sqrt{\pi}.
\end{eqnarray}
اذن
\begin{eqnarray}
 (3/2)!=(1/2)\Gamma(1/2)=\sqrt{\pi}/4.
\end{eqnarray}

\item
انظر الي المحاضرة.
 \end{itemize}
\renewcommand\thefigure{\thepart.\arabic{figure}}    
\setcounter{figure}{0}   

\newpage

{\selectlanguage{english}

\begin{figure}[htbp]
\begin{center}
\includegraphics[width=10.0cm,angle=-0]{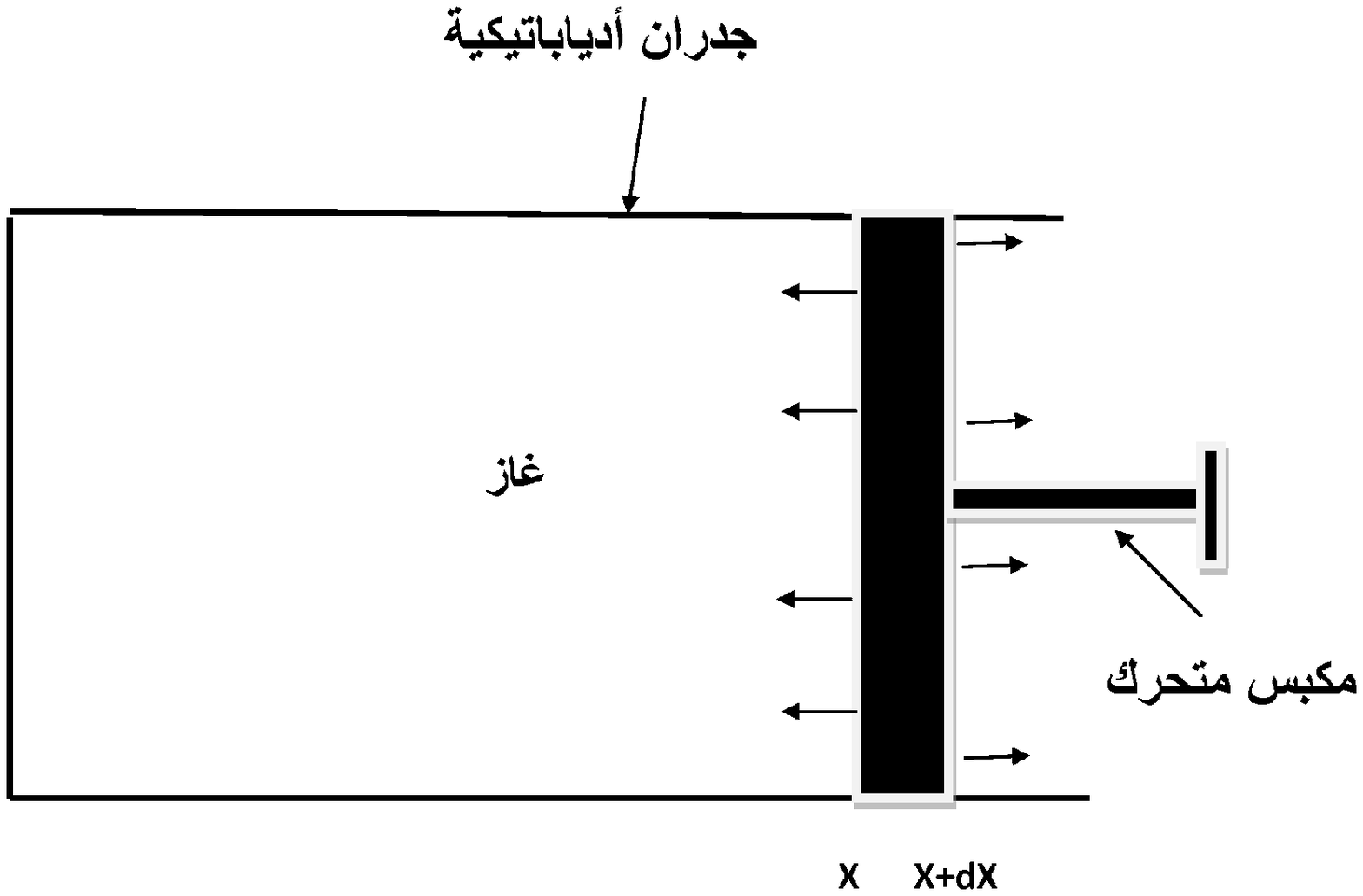}
\end{center}
\caption{}
\end{figure}
\begin{figure}[htbp]
\begin{center}
\includegraphics[width=10.0cm,angle=-0]{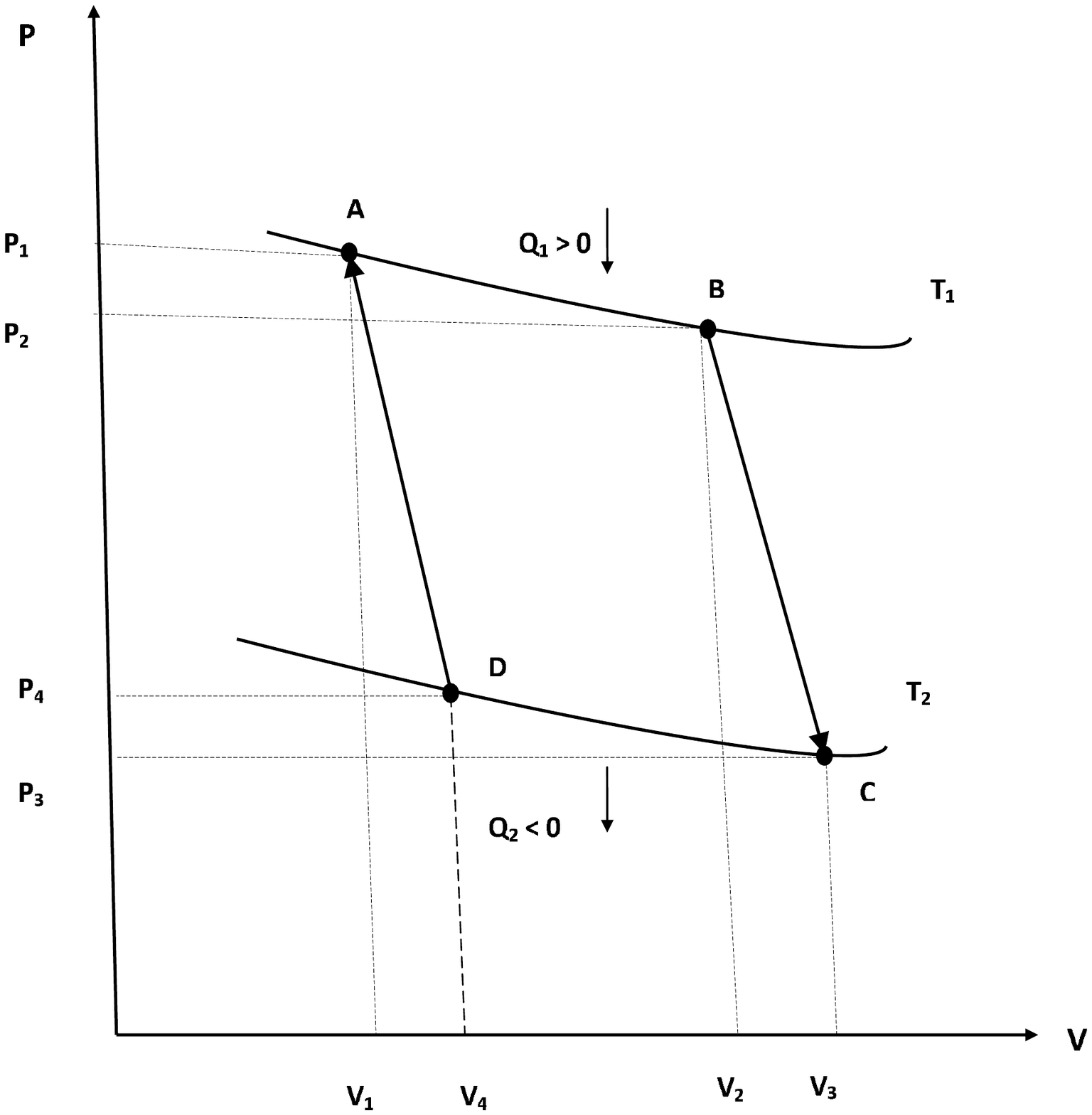}
\end{center}
\caption{}
\end{figure}
\begin{figure}[htbp]
\begin{center}
\includegraphics[width=10.0cm,angle=-0]{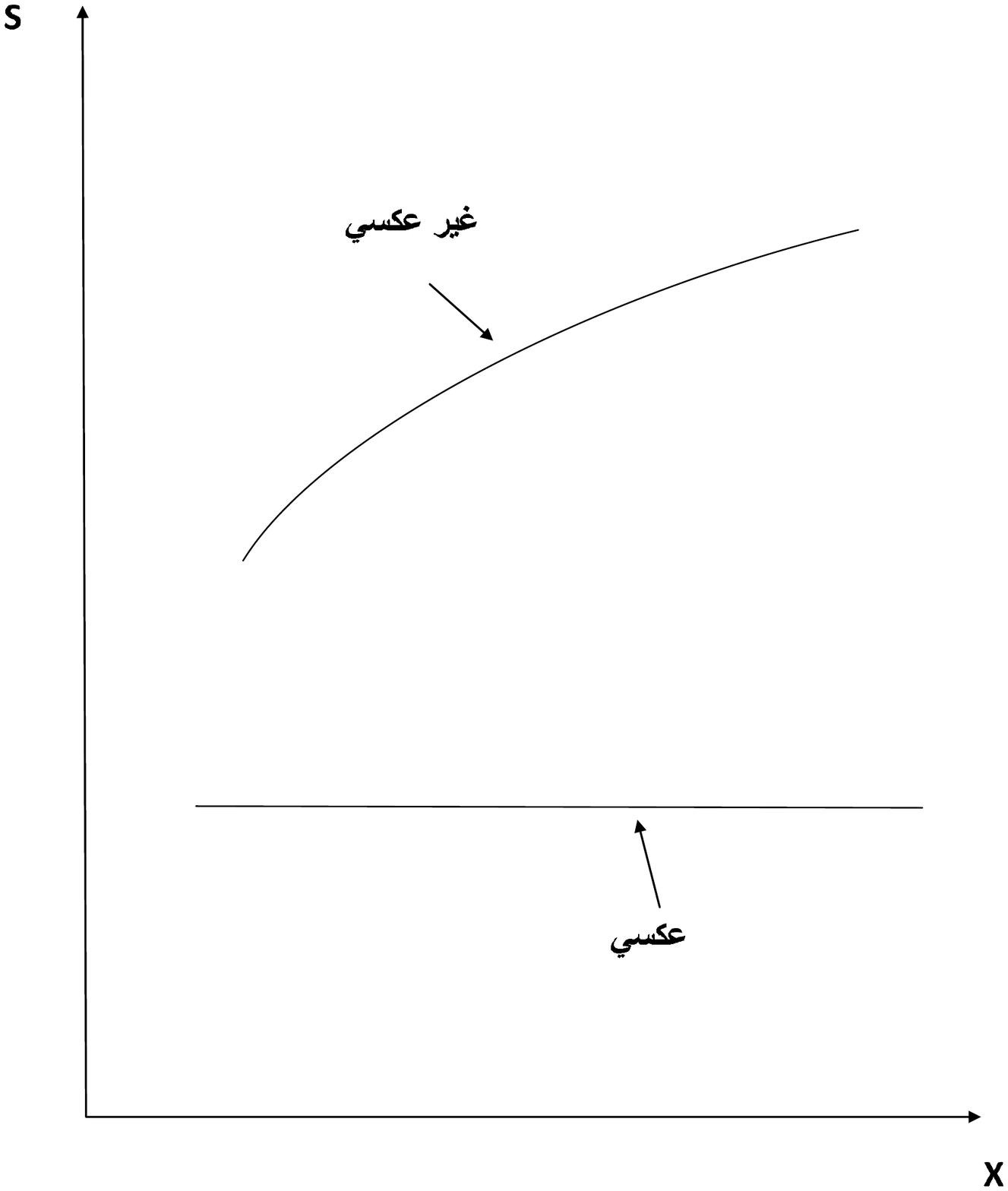}
\end{center}
\caption{}
\end{figure}
\begin{figure}[htbp]
\begin{center}
\includegraphics[width=10.0cm,angle=-0]{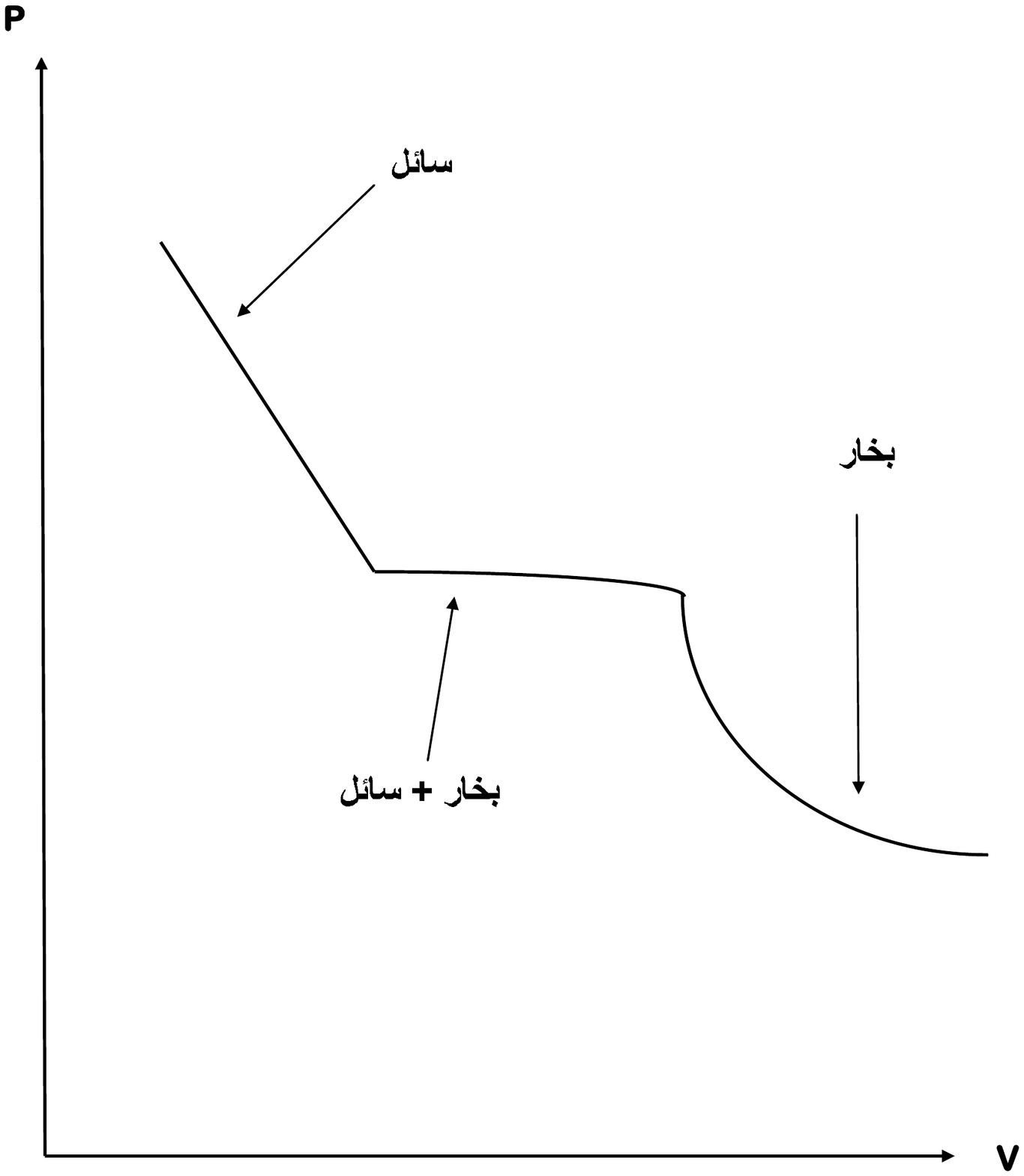}
\end{center}
\caption{}
\end{figure}

}
 

\part*{{\fontsize{50}{50}\selectfont
الميكانيك الكمومي
}}
\addcontentsline{toc}{part}{$III$
الميكانيك الكمومي
} 

{\selectlanguage{english}

\chapter*{
Quantum Mechanics
}
This part is based primarily on the books \cite{Sakurai,Griffiths}. Sakurai is the historical texbook on quantum mechanics for the older generations of  graduate students in the US, while Griffiths is the most
 pedagogical textbook on quantum mechanics out there.  This course on quantum mechanics should only be taken by graduate students who have already taken a modern physics course. Again, due to lack of time,
 only fundamental topics are 
discussed: $1)$ An introduction to quantum mechanics, $2)$ Theory of perturbations and $3)$ Theory of scattering. In the first chapter, a quick introduction to the subject 
is given,
 which includes: Dirac quantization, Hilbert spaces, wave functions, rotational invriance, exact solutions of the Schrodinger equation, etc. In the second chapter, a systematic expostion
 of the theory of perturbation theory, both time-independent and time-dependent, is presented. This is a very long chapter with many applications such as the Hydrogen atom and emission
  and absorption of radiation. In the third chapter, the theory of quantum scattering is discussed in great detail. See in particular chapters $5$ and $7$ of Sakurai, and chapters 
  $6$, $9$ and $11$ of Griffiths. The treatise of Landau et al and Tannoudji et al \cite{Landaus,tannoudji} were also consulted occasionally. 
Most exercises are taken from \cite{Sakurai,Griffiths}. The solutions Manual for Griffiths is found in \cite{GriffithsS}.
}










\chapter*{
مدخل الي الميكانيك الكمومي
}
\addcontentsline{toc}{chapter}{$1$
مدخل الي الميكانيك الكمومي
}

\thispagestyle{headings}
\section*{
التكميم القانوني و معادلة شرودينغر
}
\addcontentsline{toc}{section}{
{\bf
التكميم القانوني و معادلة شرودينغر
}
} 
\subsection*{
علاقات التبادل القانونية
}
\addcontentsline{toc}{subsection}{
علاقات التبادل القانونية
} 

نعتبر جسيم يتحرك في ثلاثة ابعاد تحت تأثير كمون
$V$.
في الميكانيك الكلاسيكي حالة الجسيم تعطي بالنقطة
$(\vec{x},\vec{p})$ 
في الفضاء الطوري
\footnote{.${\rm phase~space}$}
حيث
$\vec{x}$ 
هو شعاع الموضع و
$\vec{p}$ 
هو شعاع كمية الحركة اي
$\vec{p}=m\dot{\vec{x}}$.
نحصل علي 
$x_i$ 
و
$p_i$
من معادلات هاميلتون للحركة
\begin{eqnarray}
\dot{p}_i=-\frac{\partial H}{\partial x_i}~,~\dot{x}_i=\frac{\partial H}{\partial p_i}.
\end{eqnarray}
الزوج
$(x_i,p_i)$
يعرف بالمتغيرات القانونية
\footnote{.${\rm canonical~variables}$}
. الدالة
$H$
هي دالة علي الفضاء الطوري, اي
$H=H(x_i,p_i)$,
 تعرف بالهاميلتونية 
 \footnote{.${\rm hamiltonian}$}
 و تتطابق مع الطاقة الكلية للجملة. اذن 
\begin{eqnarray}
H&=&T+V=\frac{\sum_i{p}_ip_i}{2m}+V({x}_i).
\end{eqnarray}
 يمكن صياغة معادلات هاميلتون بدلالة اقواس بواسون
 \footnote{.${\rm Poisson~brackets}$}
 . قوس بواسون لاي دالتين
 $u$
 و 
$v$
بالنسبة للمتغيرات القانونية
$x_i$
و
$p_i$
يعرف ب

\begin{eqnarray}
[u,v]_{\rm P.B}&=&\sum_i\bigg(\frac{\partial u}{\partial x_i}\frac{\partial v}{\partial p_i}-\frac{\partial u}{\partial p_i}\frac{\partial v}{\partial x_i}\bigg).
\end{eqnarray}
اقواس بواسون الاساسية تعطي ب
\begin{eqnarray}
[x_i,x_j]_{\rm P.B}=0~,~[p_i,p_j]_{\rm P.B}=0~,~[x_i,p_j]_{\rm P.B}=\delta_{ij}.
\end{eqnarray}
لتكن 
$Q$
دالة في المتغيرات القانونية
$x_i$, $p_i$ 
و الزمن اي
$Q=Q(x_i,p_i,t)$.
المشتقة الكلية بالنسبة للزمن للدالة
$Q$ 
تعطي ب
\begin{eqnarray}
\frac{dQ}{dt}&=&[Q,H]_{\rm P.B}+\frac{\partial Q}{\partial t}.\label{Hclassical}
\end{eqnarray}
هذه هي معادلة حركة الدالة
$Q$. 
معادلات هاميلتون يمكن الحصول عليها كحالة خاصة. بالفعل اذا اخترنا
$Q=x_i,p_i$
نحصل مباشرة علي
$\dot{x}_i=[x_i,H]_{\rm P.B}$, $\dot{p}_i=[p_i,H]_{\rm P.B}$
و هي بالضبط معادلات هاميلتون اعلاه
. 

تكميم 
\footnote{.${\rm quantization}$}
هذه الجملة الكلاسيكية يعطي الجملة الكمومية المقابلة. حسب ديراك 
\footnote{.${\rm Dirac}$}
فانه يمكننا الحصول علي الجملة الكمومية انطلاقا من الجملة الكلاسيكية عن طريق تعويض اقواس بواسون بمبدلات
\footnote{.${\rm commutators}$}
كالتالي
\begin{eqnarray}
[,]_{\rm P.B}\longrightarrow \frac{1}{i{\hbar}}[,].\label{diracprescription}
\end{eqnarray}
هذا هو مبدأ التوافق
\footnote{.${\rm correspondence}~{\rm principle}$}
. 
بعبارة اخري فاننا نحصل علي الجملة الكمومية عن طريق تعويض الموضع
 $x_i$
 بمؤثر الموضع
 $\hat{x}_i$
 و كمية الحركة
 $p_i$
 بمؤثر كمية الحركة 
 $\hat{p}_i$
 بحيث تصبح اقواس بواسون الاساسية معطاة بالمبدلات التالية
\begin{eqnarray}
[\hat{x}_i,\hat{x}_j]=0~,~[\hat{p}_i,\hat{p}_j]=0~,~[\hat{x}_i,\hat{p}_j]_{\rm P.B}=i{\hbar}\delta_{ij}.
\end{eqnarray}
هذه هي علاقات التبادل الاساسية او القانونية.  من الواضح ان المبدل
 $[A,B]$
 معرف ب
 $A.B-B.A$.
 ايضا لان
 $\hat{x}_i$
 و
 $\hat{p}_i$
 هي مؤثرات و ليست اعداد فانها يجب ان تؤثر علي فضاء ما
  ${\cal H}$
  يعرف باسم فضاء هيلبرت
  \footnote{.${\rm hilbert~space}$}.
  فضاء هيلبرت هو فضاء شعاعي مركب يمكن ان يكون, و هذا متحقق في هذه الحالة, ذو بعد لا نهائي.
 

\subsection*{
معادلة هايزنبرغ
}
\addcontentsline{toc}{subsection}{
معادلة هايزنبرغ
} 
قياسا علي الهاميلتونية الكلاسيكية التي هي دالة في 
$x_i$
و
$p_i$
فان الهاميلتونية الكمومية هي دالة في المؤثرات
 $\hat{x}_i$
 و
  $\hat{p}_i$ 
  نحصل عليها كالاتي. لان المؤثرات 
   $\hat{x}_i$ 
   تتبادل فيما بينها و
   ايضا
   $\hat{p}_i$
   تتبادل فيما بينها فان الهاميلتونية الكمومية هي مؤثر علي فضاء هيلبرت
   ${\cal H}$
   يعطي ب
\begin{eqnarray}
\hat{H}=\frac{\sum_i\hat{p}_i^2}{2m}+V(\hat{x}_i).
\end{eqnarray}
بالمثل فان اي دالة  كلاسيكية, اي دالة علي الفضاء الطوري
$Q=Q(x_i,p_i)$,
تعوض بعد التكميم بمؤثر 
$\hat{Q}=\hat{Q}(t)$
علي فضاء هيلبرت
${\cal H}$ 
يعطي تطوره في الزمن 
بالمقابل الكمومي  لمعادلة الحركة
$(\ref{Hclassical})$,
الذي يحصل عليه بوصفة التكميم
$(\ref{diracprescription})$,
اي ب

\begin{eqnarray}
i\hbar \frac{d\hat{Q}}{dt}&=&[\hat{Q},\hat{H}].
\end{eqnarray}
هذه هي معادلة هايزنبرغ للحركة
\footnote{.${\rm Heisenberg}$}. 
كما سنري هذه المعادلة مكافئة تماما لمعادلة شرودينغر
\footnote{.${\rm Schrodinger}$}. 

نعتبر الان المؤثر الاحادي
$U=U(t,t_0)$
علي فضاء هيلبرت
${\cal H}$,
اي المؤثر الذي يحقق 
$UU^+=U^+U=1$,
و الذي
يتعلق بالزمن بحيث

\begin{eqnarray}
\hat{Q}(t)=U(t,t_0)\hat{Q}(t_0)U(t,t_0)^+.\label{evo}
\end{eqnarray}
المؤثر الاحادي
$U(t,t_0)$
يعرف باسم مؤثر التطور. 
من الواضح ان
$\hat{Q}(t_0)$
يتطابق مع المؤثر
$\hat{Q}(t)$
في اللحظة الزمنية
$t_0$. 
اذن
$\hat{Q}(t_0)$
لا يتعلق بالزمن اي
${d\hat{Q}(t_0)}/{dt}=0$.
ايضا يجب ان يكون لدينا
$[U(t_0,t_0),\hat{Q}(t_0)]=0$ 
من اجل اي مؤثر
$\hat{Q}(t_0)$ 
علي فضاء هيلبرت
${\cal H}$ 
و بالتالي 
$U(t_0,t_0)={\bf 1}$. 
اذن
المؤثر
$U(t,t_0)$
يحمل  بالكامل كيفية تيعية $\hat{Q}(t)$
للزمن. بالفعل يمكن ان نحسب من جهة
\begin{eqnarray}
i\hbar \frac{d\hat{Q}}{dt}&=&i\hbar \frac{d U}{d t}\hat{Q}_0U^++i\hbar U\hat{Q}_0\frac{dU^+}{dt}.
\end{eqnarray}
من الجهة الاخري نحسب
\begin{eqnarray}
[\hat{Q},\hat{H}]=
-\hat{H}U\hat{Q}_0U^++U\hat{Q}_0U^+\hat{H}.
\end{eqnarray}
اذن نحصل علي
\begin{eqnarray}
i\hbar\frac{dU}{dt}=-\hat{H}U~,~i\hbar\frac{dU^+}{dt}=U^+\hat{H}.
\end{eqnarray}
الهاميلتونية في هذه الحالة, حالة الجسيم الحر في ثلاث ابعاد, لا تتعلق بالزمن و بالتالي نحصل مباشرة علي
\begin{eqnarray}
U=U(t,t_0)=e^{\frac{i}{\hbar}\hat{H}(t-t_0)}.
\end{eqnarray}
\subsection*{
معادلة شرودينغر
}
\addcontentsline{toc}{subsection}{
معادلة شرودينغر
} 
هناك فرق شاسع في الميكانيك الكمومي  بين الملاحظات
\footnote{.${\rm observables}$}, 
التي هي عبارة عن الكميات الفيزيائية التي يمكن ان تقاس في التجربة, و اشعة حالة
\footnote{.${\rm state~vectors}$}
الجملة التي تحدد حالة الجملة في الزمن. 
كما ذكرنا انفا الملاحظات يعبر عنها بمؤثرات  تؤثر علي فضاء هيلبرت
${\cal H}$
مثل مؤثرات الموضع
$\hat{x}_i$
و مؤثرات كمية الحركة 
$\hat{p}_i$.
هذه المؤثرات يجب ان تكون ايضا هرميتية
\footnote{.${\rm hermitian}$}
اي
$\hat{Q}^+=\hat{Q}$
لان الكميات الفيزيائية المقاسة يجب ان تكون بالضرورة حقيقية. اشعة الحالة يعبر عليها من الجهة الاخري بعناصر من فضاء هيلبرت
${\cal H}$ 
و بالتالي فان الملاحظات يمكن ان تؤثر عليها لتنتج اشعة حالة اخري. 

في ما يسمي بترميز ديراك
\footnote{.${\rm dirac~notation}$}
نرمز لاشعة الحالة بالكات
\footnote{.${\rm ket}$}
$|\psi(t_0)>$ 
 الذي  يمكن ايضا ان نشترط فيه ان يكون منظم اي
$<\psi(t_0)|\psi(t_0>=1$.


اذن في الملخص حصلنا علي ملاحظات يعبر عنها بمؤثرات هرميتية تتعلق بالزمن
$\hat{Q}(t)$
و اشعة حالة 
$|\psi(t_0)>$  
ثابتة في الزمن. هذه هي  بالضبط وجهة نظر هايزنبرغ. من وجهة نظر شرودينغر فان الملاحظات هي التي تصبح ثابتة في الزمن معطاة ب
 $\hat{Q}(t_0)$
 اما اشعة الحالة فانها تتعلق بالزمن تعطب ب
 $|\psi(t)>$
 الذي هو يساوي
 $|\psi(t_0)>$
 في اللحظة الزمنية
 $t_0$.
 بعبارة اخري ادق يعطي شعاع الحالة في وجهة نظر شرودينغر بالكات
\begin{eqnarray}
|\psi(t)>=U(t,t_0)^+|\psi_0>.
\end{eqnarray}
المؤثر الاحادي
$U(t,t_0)$
هو بالضبط مؤثر التطور المعرف في المعادلة
$(\ref{evo})$.
من الواضح مباشرة ان تطور شعاع الحالة
 $|\psi(t)>$
 في الزمن يعطي ب
\begin{eqnarray}
i\hbar\frac{d}{dt}|\psi(t)>&=&U^+\hat{H}U|\psi(t)>\nonumber\\
&=&\hat{H}|\psi(t)>.
\end{eqnarray} 
هذه هي معادلة شرودينغر. من الواضح ايضا ان القيم المتتظرة 
\footnote{.${\rm expectation~values}$}
من وجهتي نظر هايزنبرغ و شرودينغر هي متساوية لانها تعبر عن قياس فيزيائي يجب ان يكون نفسه بالضرورة اي
\begin{eqnarray}
<\psi(t)|\hat{Q}_0|\psi(t)>=<\psi_0|\hat{Q}(t)|\psi_0>.
\end{eqnarray}

\thispagestyle{headings}
\section*{
فضاء هيلبرت
}
\addcontentsline{toc}{section}{
{\bf
فضاء هيلبرت
} }
\subsection*{
اشعة الحالة
}
\addcontentsline{toc}{subsection}{
اشعة الحالة
} 

تلعب فكرة فضاء هيلبرت دورا محوريا قي الميكانيك الكمومي. في الواقع فان اهم مكونات الميكانيك الكمومي,
اشعة الحالة و المؤثرات التي تؤثر عليها,  كلاهما مرتبط ارتباطا وثيقا بفضاء هيلبرت خاصة الجملة.  بالفعل فان
مجموعة كل اشعة الحالة تشكل فضاء هيلبرت بينما يعبر عن الملاحطات بمؤثرات هرميتية تؤثر علي فضاء هيلبرت.

فضاء هيلبرت 
${\cal H}$
هو فضاء شعاعي مركب ذو ابعاد غير متناهية في اغلب الاحيان ممنوح جداء داخلي
\footnote{.${\rm inner~product}$}
. باتباع ديراك نرمز لاشعة فضاء هيلبرت ب
$|\psi>$ 
و نسميها كاتس (مفردها كات
\footnote{.${\rm ket}$}). 
في اساس معين, نفترضه متقطع 
\footnote{.${\rm discrete}$}
للتبسيط, نكتب اشعة الحالة علي شكل الاشعة العمودية 

\begin{eqnarray}
|\psi>=\sum_n a_n|e_n>.\label{expa}
\end{eqnarray}
رمزنا لعناصر الاساس ب
$|e_n>$
حيث 
$n$
يأخذ قيم من
$0$
الي
$\infty$.
الافادة بان فضاء هيلبرت
${\cal H}$ 
هو فضاء مركب يكافئ بالضبط المطلب بان المركبات
$a_n$ 
هي اعداد مركبة.

ليكن
$|\phi>$ 
شعاع حالة اخر بمركبات
$b_n$
اي
$|\phi>=\sum_nb_n|e_n>$. 
الجداء الداخلي بين 
$|\psi>$ 
و
 $|\phi>$ 
 و الذي يرمز له ب
 $<\phi|\psi>$ 
 يعرف ب
\begin{eqnarray}
<\phi|\psi>=\sum_nb_n^*a_n.
\end{eqnarray}
بالمثل فان الجداء الداخلي بين
$|\phi>$
و
$|\psi>$ 
و الذي يرمز له ب
$<\psi|\phi>$ 
يعرف ب
\begin{eqnarray}
<\psi|\phi>=\sum_na_n^*b_n.
\end{eqnarray}
من هذه التعريفات نلاحظ مباشرة ان
$<\phi|\psi>=<\psi|\phi>^*$.
الجداء الداخلي يعمم الجداء السلمي في الفضاءات الشعاعية الحقيقية.

من التعريف اعلاه من الواضح ان الاساس
$\{|e_n>\}$
هو متعامد و متجانس اي
\begin{eqnarray}
<e_n|e_m>=\delta_{nm}.
\end{eqnarray}
الجداء الداخلي $<\phi|\psi>$
يمكن ايضا ان يفهم علي انه قيمة الدالة الخطية
$<\phi|$ 
في النقطة (الشعاع)
$|\psi>$ 
من فضاء هيلبرت 
${\cal H}$. 
بعبارة اخري
\begin{eqnarray}
<\phi|&:&{\cal H}\longrightarrow {\bf C}\nonumber\\
&&|\psi>\longrightarrow <\phi|\psi>.
\end{eqnarray}
مجموعة كل الدوال الخطية 
 $<\phi|$ 
 تعرف فضاء هيلبرت اخر 
 ${\cal H}^*$
 الذي هو ثنوي
 \footnote{.${\rm dual}$}
 ل
 ${\cal H}$. 
العناصر
 $<\phi|$
 التي تعرف في ترميز ديراك بالبراس
 (مفرد برا
 \footnote{.${\rm bra}$}
 )
 تعطي بالاشعة الافقية
\begin{eqnarray}
<\phi|=\sum_nb_n^*<e_n|.
\end{eqnarray}
المجموعة 
$\{<e_n|\}$
هي اساس في الفضاء الثنوي
 ${\cal H}^*$ 
 و هي بالتالي ثنوية للاساس
 $\{|e_n>\}$.
 في معني اخر مضبوط فان البرا
 $<\phi|$ 
 هو المرافق الهرميتي للكات
 $|\phi>$ 
 اي
 $<\phi|=(|\phi>)^+$. 

من الواضح ان طويلة شعاع الحالة
$|\psi>$
يجب ان تعرف بدلالة الجداء الداخلي
$<\psi|\psi>$
الذي هو دائما عدد حقيقي موجب.  بالتأكيد فان الطويلة تساوي 
$\sqrt{<\psi|\psi>}$.
الحالتان
$|\psi>$
و
$a|\psi>$ 
من اجل اي عدد مر كب
$a$ 
تمثل نفس الحالة الفيزيائية. بعبارة اخري يمكننا دائما ان ننظم الشعاع
$|\psi>$
بحيث
$<\psi|\psi>=1$.
 عناصر فضاء هيلبرت 
 ${\cal H}$
 هي اذن
 اشعة قابلة للتنظيم اي
\begin{eqnarray}
{\rm if}~|\psi>\in{\cal H}~{\rm then}~<\psi|\psi><\infty.
\end{eqnarray}
باستعمال الشرط
$<e_n|e_m>=\delta_{nm}$ 
يمكن ان نحسب المركبات
 $a_n$
 خاصة شعاع الحالة
 $|\psi>$.
  نجد
\begin{eqnarray}
a_n=<e_n|\psi>.
\end{eqnarray}
النشر
$(\ref{expa})$
يأخذ اذن الشكل
\begin{eqnarray}
|\psi>=\sum_n|e_n><e_n|\psi>.
\end{eqnarray}
بعبارة اخري يجب ان يكون لدينا علاقة الاكتمال
\footnote{.${\rm completeness~relation}$}
\begin{eqnarray}
\sum_n|e_n><e_n|={\bf 1}.\label{compl1}
\end{eqnarray}
الكمية
$|e_n><e_n|$ 
هي مؤثر نحصل عليه من الجداء الخارجي
\footnote{.${\rm outer~product}$}
بين الكات
$|e_n>$
و البرا
$<e_n|$. 
هذا المؤثر هو ايضا مسقط
\footnote{.${\rm projector}$}
.
\subsection*{
الملاحظات
}
\addcontentsline{toc}{subsection}{
الملاحظات
} 

 تمثل الملاحظات خاصة الجملة بمؤثرات هرميتية علي فضاء هيلبرت
${\cal H}$. 
 اي مؤثر
$\hat{Q}$
يؤثر علي
 ${\cal H}$ 
 هو تحويل خطي لانه يأخذ شعاع حالة
 $|\psi>$
 الي شعاع حالة
 اخر نرمز له ب 
  $\hat{Q}|\psi>$
  اي
\begin{eqnarray}
\hat{Q}&:&{\cal H}\longrightarrow {\cal H}\nonumber\\
&&|\psi>\longrightarrow \hat{Q}|\psi>.
\end{eqnarray}
هذه الدالة خطية لانه
$\hat{Q}(a|\psi>+b|\phi>)=a\hat{Q}|\psi>+b|\hat{Q}|\psi>$
من اجل اي عددين مركبين
$a$
و
$b$.
اذن المؤثر
$\hat{Q}$
يمكن تمثيله بالمصفوفة غير المتناهية الابعاد التي تعطي مركباتها في الاساس
$\{|e_n>\}$ 
ب
$<e_n|\hat{Q}|e_m>$
اي
\begin{eqnarray}
\hat{Q}=\sum_n\sum_m<e_n|\hat{Q}|e_m>|e_n><e_m|.
\end{eqnarray}
المؤثر الهرميتي هو المؤثر الذي يحقق الشرط الاضافي
$\hat{Q}^+=\hat{Q}$
حيث
$\hat{Q}^+$ 
هو المرافق الهرميتي ل
$\hat{Q}$
الذي يعرف ب
$<\phi|\hat{Q}|\psi>=<\psi|\hat{Q}^+|\phi>^*$.
بعبارة اخري
مركبات اي مؤثر هرميتي تحقق 
$<e_n|\hat{Q}^+|e_m>=(<e_m|\hat{Q}|e_n>)^*$.
القيمة المنتظرة ل 
$\hat{Q}$
في شعاع الحالة
$|\psi>$
تعطي بالجداء الداخلي بين
$\hat{Q}|\psi>$
و
$|\psi>$
اي
\begin{eqnarray}
<\hat{Q}>=<\psi|\hat{Q}|\psi>.
\end{eqnarray}
لان
$\hat{Q}^+=\hat{Q}$
فان القيمة المنتظرة
$<\hat{Q}>$
يجب ان تكون حقيقية. متوسط نتائج عدة قياسات للمؤثر
$\hat{Q}$
التي تجري علي جمل متطابقة, اي محضرة بنفس الطريقة, هي بالضبط القيمة المنتظرة
$<\hat{Q}>$. 
بالمقابل لان نتيجة اي قياس هي عدد حقيقي فان القيمة المنتظرة لمؤثر يمثل ملاحظ يجب ان تكون حقيقية و بالتالي فان المؤثر يجب ان يكون هرميتي.

نقول عن 
شعاع حالة منظم
$|\psi>$
انه شعاع ذاتي للمؤثر الهرميتي
$\hat{Q}$
مرفق بالقيمة الذاتية
$\lambda$
اذا كان

\begin{eqnarray}
\hat{Q}|\psi>=\lambda|\psi>.
\end{eqnarray}
القيمة المنتظرة ل
$\hat{Q}$
في 
$|\psi>$ 
تساوي
$\lambda$. 
علاوة علي ذلك فان الانحراف المعياري ل
$\hat{Q}$
في
$|\psi>$ 
المعرف ب
$\sigma^2=<\psi|(\hat{Q}-<\hat{Q}>)^2|\psi>$
هو صفر. بعبارة اخري الشعاع الذاتي
$|\psi>$
هو شعاع يقيني
\footnote{.${\rm determinate}$}
للجملة بمعني ان نتائج كل القياسات التي تجري علي مجموعة من الجمل المتطابقة المحضرة بنفس الطريقة في الحالة
 $|\psi>$
 تعطي نفس القيمة
 $\lambda$.

مجموعة كل القيم الذاتية خاصة
$\hat{Q}$
تسمي طيف المؤثر. الطيف يمكن ان يكون منحل اي يمكن ان توجد حالتان او اكثر مرفقة بنفس القيمة الذاتية. في حالة مؤثر هرميتي ذي طيف 
متقطع فان القيم الذاتية تكون حقيقية و اشعتها الذاتية متعامدة فيما بينها. اذن اذا كانت
$|\lambda_n>$
هي الاشعة الذاتية ل
$\hat{Q}$
المرفقة ب
$\lambda_n$
فانه يجب ان يكون لدينا
\begin{eqnarray}
<\lambda_m|\lambda_n>=\delta_{mn}.
\end{eqnarray}
وجود الانحلال يعني ان هناك قيم ذاتية مرفقة بفضاءات جزئية منحلة. الفضاء الجزئي المنحل المرفق بالقيمة الذاتية
$\lambda_n$,
و الذي يسمي ايضا بالفضاء الذاتي
ل
$\hat{Q}$
المرفق ب
$\lambda_n$ 
,
يحتوي علي كل الاشعة الذاتية المرفقة ب
$\lambda_n$. 
 نستعمل طريقة التعميد ل
 غرام-شميت
 \footnote{.${\rm Gram-Schmidt}$}
 من اجل ايجاد الاشعةالذاتية المتعامدة داخل كل فضاء ذاتي. 
a

من اجل فضاء هيلبرت منته فان مجموعة الاشعة الذاتية
 $|\lambda_n>$ 
 لاي مؤثر هرميتي
  $\hat{Q}$
  هي مجموعة مكتملة. بعبارة اخري فان اي شعاع حالة
  $|\psi>$ 
  داخل فضاء هيلبرت يمكن كتابته علي شكل تركيب خطي ل
  $|\lambda_n>$
  اي
\begin{eqnarray}
|\psi>=\sum_{n}c_n|\lambda_n>.
\end{eqnarray}
المجموع علي 
$n$
هو
محدود علي مجموعة منتهية من
${\bf N}$. 
الخاصية اعلاه هي مكافئة تماما لعلاقة الاكتمال
\begin{eqnarray}
\sum_n|\lambda_n><\lambda_n|={\bf 1}.\label{compl2}
\end{eqnarray}
البرهان علي خاصية الاكتمال في حالة الفضاءات المنتهية لا يعمم الي حالة الفضاءات غير المنتهية. مع ذلك فاننا نأخذ خاصية الاكتمال, التي هي
خاصية محورية في الميكانيك الكمومي, كمسلمة كما فعل ديراك. من الناحية التقنية هذا يعني انه يجب علينا ان نحدد اصناف المؤثرات الهرميتية التي يمكن ان تمثل الملاحطات في النظرية. اذن
علاقة الاكتمال
$(\ref{compl1})$
هي فقط مسلمة. بالمثل فان علاقة الاكتمال
$(\ref{compl2})$
التي يكون فيها المجموع علي
$n$
غير محدد هي ايضا مسلمة.

\thispagestyle{headings}
\section*{
الاطياف المستمرة و الدوال الموجية
}
\addcontentsline{toc}{section}{
{\bf 
الاطياف المستمرة و الدوال الموجية
}} 
\subsection*{
مؤثر الموضع و الدوال الموجية
}
\addcontentsline{toc}{subsection}{
مؤثر الموضع و الدوال الموجية
} 
في حالة المؤثرات الهرميتية ذات الاطياف المستمرة, اي المؤثرات التي تملأ قيمها الذاتية مجال مستمر, فان الاشعة الذاتية المرافقة غير قابلة للتنظيم. بعبارة اخري هذه الاشعة الذاتية المرافقة لا تقع
في فضاء هيلبرت و بالتالي لا يمكن ان تمثل حالات فيزيائية. في مثل هذه الوضعية فقط التركيب الخطي لهذه الاشعة الذاتية يمكن ان يعبر عن حالات فيزيائية.

كمثال نأخذ مؤثرات الموضع و كمية الحركة 
$\hat{x}$ 
و
$\hat{p}$ 
في بعد واحد. علاقات التبادل القانونية تكتب علي الشكل 
\begin{eqnarray}
[\hat{x},\hat{p}]=i\hbar.
\end{eqnarray}
الشعاع الذاتي
$|x>$ 
لمؤثر الموضع
$\hat{x}$
المرفق بالقيمة الذاتية
$x$ 
يعرف ب
\begin{eqnarray}
\hat{x}|x>=x|x>.
\end{eqnarray}
مؤثر الموضع هو مؤثر هرميتي. اذن من الطبيعي ان نفترض ان القيم الذاتية ل
$\hat{x}$
حقيقية. نحسب ايضا
\begin{eqnarray}
<x^{'}|\hat{x}|x>=x<x^{'}|x>=x^{'}<x^{'}|x>.
\end{eqnarray}
يمكن ان  نستنتج مباشرة ان
\begin{eqnarray}
<x^{'}|x>=\delta(x^{'}-x).\label{Dortho}
\end{eqnarray}
الاشعة الذاتية
$|x>$
متعامدة لكنها غير قابلة للتنطيم بالمعني العادي. لكننا نقول انها قابلة للتنظيم حسب ديراك بمعني ان
$<x|x>=\delta(0)$. 
المعادلة
$(\ref{Dortho})$
 تسمي شرط التعامد و التجانس لديراك. يمكن اذن ان نشتق علاقة الاكتمال
\begin{eqnarray}
\int dx |x><x|={\bf 1}.
\end{eqnarray}
اي شعاع حالة
 $|\psi>$ 
 يمكن ان ينشر في الاساس
 $|x>$ 
 علي الشكل
\begin{eqnarray}
|\psi>=\int dx^{'}\psi(x^{'})|x^{'}>.
\end{eqnarray}
نحسب
\begin{eqnarray}
\psi(x)=<x|\psi>.
\end{eqnarray}
هذه هي دالة موجة الجملة الموافقة لشعاع الحالة
$|\psi>$. 
شرط القابلية للتنظيم
$<\psi|\psi> <\infty$
يصبح شرط قابلية التكامل للمربع
\begin{eqnarray}
<\psi|\psi>=\int dx|\psi(x)|^2<\infty.
\end{eqnarray}
مجموعة كل الدوال
$\psi(x)$
علي مجال
 $[a,b]$
 التي تحقق شرط قابلية التكامل للمربع تشكل فضاء هيلبرت يسمي
 $L_2(a,b)$.  
 
 نكتب ايضا الجداء الداخلي لشعاعي حالة
 $|\psi>$
 و
 $|\phi>$
 في اساس الموضع علي الشكل

\begin{eqnarray}
<\phi|\psi>=\int dx\phi(x)^*\psi(x).
\end{eqnarray}
\subsection*{
مؤثر كمية الحركة و الانسحابات
}
\addcontentsline{toc}{subsection}{
مؤثر كمية الحركة و الانسحابات
} 
قبل تقطير مؤثر كمية الحركة
$\hat{p}$
نعتقد انه من المفيد ان نبدأ بتقديم مفهوم الانسحاب غير المتناه في الصغر
$U(dx)$.
هذا يعطي ب
\begin{eqnarray}
U(dx)|x>=|x+dx>.
\end{eqnarray}
تأثير
$U(dx)$
علي شعاع الحالة
$|\psi>$
هو
\begin{eqnarray}
U(dx)|\psi>&=&\int dx^{'}\psi(x^{'})U(dx)|x^{'}>\nonumber\\
&=&\int dx^{'} \psi(x^{'})|x^{'}+dx>\nonumber\\
&=&\int dx^{'}\psi(x^{'}-dx)|x^{'}>.\label{translation}
\end{eqnarray}
في المعادلة اعلاه افترضنا ان التكامل علي
 $x$
 هو من
 $-\infty$ 
 الي
 $\infty$.
 بافتراض ان شعاع الحالة المنسحب
 $U(x)|\psi>$
 هو منظم يمكن ان نتحقق من ان المؤثر
 $U$ 
 احادي اي ان
 $U^+U={\bf 1}$. 
 هذا المؤثر يحقق ايضا
 $U(0)={\bf 1}$, $U^{-1}(dx)=U(-dx)$
 و
 $U(dx_1)U(dx_2)=U(dx_1+dx_2)$. 
 المؤثر الاحادي
 $U(dx)$
 يمكن دائما نشره حول مؤثر الوحدة كالتالي
 
\begin{eqnarray}
U(dx)={\bf 1}-i K dx.
\end{eqnarray}
يعرف المؤثر الهرميتي
$K$
بمولد الانسحاب و نحسبه كالتالي. نبدأ من
\begin{eqnarray}
[\hat{x},U(dx)]|x>=dx|x+dx>.
\end{eqnarray}
هذا يمكن اعادة كتابته علي الشكل
\begin{eqnarray}
-i[\hat{x},K]|x>=|x>+O(dx).
\end{eqnarray}
بعبارة اخري
\begin{eqnarray}
[\hat{x},K]=i.
\end{eqnarray}
مولد الانسحاب
$K$
يمكن اذن مطابقته مع مؤثر كمية الحركة
$\hat{p}$
تقسيم
$\hbar$
اي
\begin{eqnarray}
K=\frac{\hat{p}}{\hbar}.
\end{eqnarray} 
 يمكن بناء انسحاب منته
$U( x)$
انطلاقا من الانسحابات غير المتناهية في الصغر علي الشكل التالي. نعتبر
$N$
انسحاب متعاقب كلهاغير متناهية في الصغر
$U(dx)$. 
الانسحاب المنته
$U( x)$
هو بالضبط تركيب هذه الانسحابات غير المتناهية في الصغر اي
 $U(x)=U(dx)U(dx)..U(dx)$ 
 حيث
 $dx= x/N$. 
 اذن
 $U( x)=({\bf 1}-iK dx)^N=e^{-iK x}$.


المعادلة
$(\ref{translation})$
يمكن وضعها علي الشكل 
\begin{eqnarray}
({\bf 1}-i\frac{\hat{p}}{\hbar}dx)|\psi>=\int dx^{'}(\psi(x^{'})-dx\frac{\partial \psi(x^{'})}{\partial x^{'}})|x^{'}>.
\end{eqnarray}
استعملنا في هذه المعادلة الاشتقاق الجزئي عوض الاشتقاق التام لان شعاع الحالة
$|\psi>$,
و بالتالي دالة الموجة
$\psi(x)$,
يمكن ان تتعلق ايضا علي الزمن الذي حافظنا عليه مثبت هنا. بالمقابل يمكن ان نكتب
\begin{eqnarray}
\hat{p}|\psi>=\int dx^{'}\frac{\hbar}{i}\frac{\partial \psi(x^{'})}{\partial x^{'}}|x^{'}>.
\end{eqnarray}
او
\begin{eqnarray}
<x|\hat{p}|\psi>=\frac{\hbar}{i}\frac{\partial \psi(x)}{\partial x}.
\end{eqnarray}
اذن في اساس الموضع يأخذ مؤثر كمية الحركة الشكل
\begin{eqnarray}
<x|\hat
{p}|x^{'}>=-\frac{\hbar}{i}\frac{\partial}{\partial x^{'}}\delta(x-x^{'}).
\end{eqnarray}
ليكن
 $|p>$
 الشعاع الذاتي المرفق بالقيمة الذاتية
 $p$
 اي
\begin{eqnarray}
\hat{p}|p>=p|p>.
\end{eqnarray}
في اساس الموضع تكتب هذه المعادلة علي الشكل
\begin{eqnarray}
\frac{\hbar}{i}\frac{\partial}{\partial x}<x|p>=p<x|p>.
\end{eqnarray}
حلول هذه المعادلة تأخذ الشكل
$<x|p>=Ae^{i\frac{p}{\hbar}x}$
حيث 
$p$
و
$A$
مركبان و هي دوال لا تحقق شرط قابلية التكامل للمربع. لكن من اجل
$p$ 
حقيقي تصبح هذه الدوال محققة لشرط قابلية التكامل للمربع خاصة ديراك بمعني
\begin{eqnarray}
<p^{'}|p>&=&\int_{-\infty}^{\infty} dx <x|p>^*<x|p>\nonumber\\
&=&|A|^2\int_{-\infty}^{\infty}dx~ e^{i\frac{p-p^{'}}{\hbar}x}\nonumber\\
&=&|A|^22\pi\hbar\delta(p-p^{'}).
\end{eqnarray}
اذن الاختيار
$A=1/\sqrt{2\pi\hbar}$
يعطي الاشعة الذاتية
\begin{eqnarray}
<x|p>=\frac{1}{\sqrt{2\pi\hbar}}e^{i\frac{p}{\hbar}x}.
\end{eqnarray}
هذه حالات متعامدة و متجانسة حسب ديراك اي
\begin{eqnarray}
<p^{'}|p>=\delta(p-p^{'}).
\end{eqnarray}
\subsection*{
معادلة شرودينغر في فضاء المواضع
}
\addcontentsline{toc}{subsection}{
معادلة شرودينغر في فضاء المواضع
} 
اي شعاع حالة
$|\psi>$
يمكن نشره علي الشكل
\begin{eqnarray}
|\psi>=\int dp <p|\psi>|p>.
\end{eqnarray}
او
\begin{eqnarray}
|\psi>=\int dx <x|\psi>|x>.
\end{eqnarray}
دالة الموجة في فضاء الموضع
هي 
$\psi(x)=<x|\psi>$
بينما دالة الموضع في فضاء كمية الحركة هي
$\tilde{\psi}(p)=<p|\psi>$.
هذه الدوال مرتبطة كما يلي
\begin{eqnarray}
\psi(x)= \int \frac{dp}{\sqrt{2\pi\hbar}}\tilde{\psi}(p)e^{i\frac{px}{\hbar}}.
\end{eqnarray}
\begin{eqnarray}
\tilde\psi(p)= \int \frac{dx}{\sqrt{2\pi\hbar}}{\psi}(x)e^{-i\frac{px}{\hbar}}.
\end{eqnarray}
اذن هذه الدوال الموجية هي تحويلات فوريي
\footnote{.${\rm Fourier~transforms}$}
بالنسبة لبعضها البعض.

معادلة شرودينغر في فضاء الموضع تصبح
\begin{eqnarray}
<x|i\hbar\frac{d}{dt}|\psi(t)>=\int dx^{'}<x|\hat{H}|x^{'}>\psi(t,x^{'}).
\end{eqnarray}
بالمقابل لدينا
\begin{eqnarray}
i\hbar\frac{\partial }{\partial t}\psi(t,x)=\bigg(-\frac{\hbar^2}{2m}\frac{\partial^2}{\partial x^2}+V(x)\bigg)\psi(t,x).
\end{eqnarray}
اعلاه استعملنا النتيجة
\begin{eqnarray}
\int dx^{'}<x|\hat{p}^2|x^{'}>\psi(t,x^{'})=-\hbar^2\frac{\partial^2}{\partial x^2}\psi(t,x).
\end{eqnarray}

\thispagestyle{headings}
\section*{ 
القياس
}
\addcontentsline{toc}{section}{
{\bf
القياس
}} 
\subsection*{
التفسير الاحصائي
}
\addcontentsline{toc}{subsection}{
التفسير الاحصائي
} 

الاصلان المحوريان للميكانيك الكمومي هما معادلة شرودينغر و التفسير الاحصائي. معادلة شرودينغر تسمح لنا بحساب
تطور دالة الموجة في الزمن بينما يسمح لنا التفسير الاحصائي لبورن
\footnote{.${\rm Born}$}
بحساب احتمالات مختلف النتائج الممكنة لاي قياس.

نحن يهمنا قياس كمية فيزيائية معينة
$Q(x,p)$. 
المؤثر الهرميتي المرفق بهذه الكمية هو
$\hat{Q}=\hat{Q}(\hat{x},\hat{p})$.
نفترض ان
$\hat{Q}$
لديه طيف متقطع
$q_n$
مرفق بالدوال الذاتية
$\psi_n(x)=<x|\psi_n>$.
نفترض ايضا ان
$|\psi_n>$ 
تحقق شرط التعامد و التجانس
$<\psi_m|\psi_n>=\delta_{mn}$ 
و انها تشكل اساس مكتمل اي
$\sum_n|\psi_n><\psi_n|={\bf 1}$.
دالة موجة الجملة
$\psi(x)=<x|\psi>$
يمكن ان تنشر حسب مبدأ التركيب الخطي في الاساس
$\{\psi_n(x)\}$
كالاتي

\begin{eqnarray}
|\psi>=\sum_n c_n|\psi_n>.
\end{eqnarray}
بالاضافة نفترض ان
$|\psi>$
منظمة اي
\begin{eqnarray}
<\psi|\psi>=1\leftrightarrow~\sum_n|c_n|^2=1.
\end{eqnarray}
نذكر ايضا ان المركبات
$c_n$
تعطي ب
\begin{eqnarray}
c_n=<\psi_n|\psi>.
\end{eqnarray}
القيمة 
المنتظرة
للمؤثر
$\hat{Q}$
في الحالة
$|\psi>$
تعطي ب
\begin{eqnarray}
<\hat{Q}>=<\psi|\hat{Q}|\psi>=\sum_n|c_n|^2q_n.
\end{eqnarray}
ينص التفسير الاحصائي علي ان قياس الملاحظ
$Q(x,p)$
في الحالة
$\psi(x)$
يعطي
القيم الذاتية
$q_n$
للمؤثر الهرميتي
$\hat{Q}$
باحتمالات
تعطي ب
 $|c_n|^2=|<\psi_n|\psi>|^2$
 حيث
 $|\psi_n>$
 هي الشعاع الذاتي ل
 $\hat{Q}$ 
 المرفق ب
  $q_n$.

كمثال نأخذ
$\hat{Q}=\hat{x}$.
في هذه الحالة الاشعة الذاتية
هي
$|x>$ 
بحيث
$<x^{'}|x>=\delta(x^{'}-x)$
و
$\int dx |x><x|={\bf 1}$. 
شعاع الحالة
$|\psi>$ 
يمكن نشره علي الشكل
$|\psi>=\int dx \psi(x)|x>$
حيث
$\psi(x)=<x|\psi>$. 
اذن
احتمال ايجاد الجملة في النقطة
$x$ 
بخطأ
$dx$
يعطي ب
$|<x|\psi>|^2dx=|\psi(x)|^2dx$. 

بالمثل اذا اخذنا
$\hat{Q}=\hat{p}$
فاننا نجد ان احتمال ايجاد الجملة بكمية حركة
$p$
مع خطأ
$dp$
يعطي ب
$|<p|\psi>|^2dp=|\tilde{\psi}(p)|^2dp$.

\subsection*{
انهيار الدالة الموجية
}
\addcontentsline{toc}{subsection}{
انهيار الدالة الموجية
} 
قياس ملاحظ
$Q(x,p)$
في الحالة
$\psi(x)$
مرة واحدة
يعطي حتما نتيجة مؤكدة ما
مثلا القيمة الذاتية
$q_n$ 
ل
$\hat{Q}$.
 اي قياس ثان يجري مباشرة بعد القياس الاول يلزم عنه بدون اي شك الحصول علي نفس القيمة الذاتية
$q_n$. 
هذا الامر يرجع الي كون الدالة الموجية
$\psi(x)$
تنهار بعد القياس الاول علي الحالة الذاتية
$\psi_n(x)$ 
المرفقة بالقيمة الذاتية
$q_n$ 
و تكرار القياس مباشرة بعد القياس الاولي سوف يؤدي حتما الي نفس النتيجة. الخلاصة هي ان عملية القياس و انهيار دالة الموجة في اعقاب عملية القياس 
يختلف اختلافا جدريا عن التطورالاحادي لدالة الموجة الذي توفره معادلة شرودينغر.

\subsection*{
علاقات الارتياب
}
\addcontentsline{toc}{subsection}{
علاقات الارتياب
} 

الانحراف المعياري في قياسات اي مؤثر هرميتي
$\hat{A}$
في شعاع الحالة
$|\psi>$
يعطي ب
\begin{eqnarray}
\sigma_A^2&=&<\psi|(\hat{A}-<\hat{A}>)^2|\psi>\nonumber\\
&=&<\psi_A|\psi_A>~,~|\psi_A>=(\hat{A}-<\hat{A}>)|\psi>.
\end{eqnarray}
بالمثل فان
الانحراف المعياري في قياسات اي مؤثر هرميتي
$\hat{B}$
في شعاع الحالة
$|\psi>$
يعطي ب
\begin{eqnarray}
\sigma_B^2&=&<\psi|(\hat{B}-<\hat{B}>)^2|\psi>\nonumber\\
&=&<\psi_B|\psi_B>~,~|\psi_B>=(\hat{B}-<\hat{B}>)|\psi>.
\end{eqnarray}
باستعمال متراجحة شوارز
\footnote{.${\rm Schwarz}$}
نحصل علي
\begin{eqnarray}
\sigma_A^2\sigma_B^2=<\psi_A|\psi_A><\psi_B|\psi_B>\geq |<\psi_A|\psi_B>|^2.
\end{eqnarray}
نحسب
\begin{eqnarray}
<\psi_A|\psi_B>=\frac{1}{2}<\psi|[\hat{A},\hat{B}]|\psi>+\frac{1}{2}<\psi|[\hat{A}-<\hat{A}>,\hat{B}-<\hat{B}>]_+|\psi>.
\end{eqnarray}
حد المبدل هو عدد مركب تخيلي بينما حد المبدل المضاد
هو عدد حقيقي. اذن
\begin{eqnarray}
|<\psi_A|\psi_B>|^2&=&\frac{1}{4}|<\psi|[\hat{A},\hat{B}]|\psi>|^2+\frac{1}{4}|<\psi|[\hat{A}-<\hat{A}>,\hat{B}-<\hat{B}>]|\psi>|^2\nonumber\\
&\geq &\frac{1}{4}|<\psi|[\hat{A},\hat{B}]|\psi>|^2.
\end{eqnarray}
نحصل علي علاقة الارتياب
\begin{eqnarray}
\sigma_A^2\sigma_B^2\geq \frac{1}{4}|<\psi|[\hat{A},\hat{B}]|\psi>|^2.\label{uncertainty}
\end{eqnarray}
من اجل
$\hat{A}=\hat{x}$
و
$\hat{B}=\hat{p}$ 
نحصل علي
$\sigma_x^2\sigma_p^2\geq \frac{\hbar^2}{4}$ 
اي
\begin{eqnarray}
\sigma_x\sigma_p\geq \frac{\hbar}{2}.
\end{eqnarray}
في العموم لدينا علاقة ارتياب من اجل كل زوج من المؤثرات غير المتلائمة
\footnote{.${\rm incompatible~observables}$}
اي من اجل كل زوج من الملاحظات التي لا تتبادل
\footnote{.${\rm do~not~commute}$}. 
المؤثرات الهرميتية غير المتلائمة لا يمكن تقطيرها في ان معا و بالتالي لا توجد مجموعة مكتملة
\footnote{.${\rm complete~set}$}
مشتركة من الاشعة الذاتية. في المقابل المؤثرات الهرميتية المتلائمة, اي التي تتبادل, لها مجموعة مكتملة مشتركة من الاشعة الذاتية.

لنعتبر الان سعاع حالة متعلق بالزمن
 $|\psi(t)>$
 يتطور في الزمن حسب معادلة شرودينغر. القيمة المنتظرة لمؤثر هرميتي
 $\hat{Q}$
 في
 $|\psi(t)>$,
 اي
 $<\hat{Q}>=<\psi(t)|\hat{Q}|\psi(t)>$,
 تتطور في الزمن حسب
\begin{eqnarray}
i\hbar\frac{d}{dt}<\hat{Q}>=<[\hat{Q},\hat{H}]>.
\end{eqnarray}
الان نختار في علاقة الارتياب
$(\ref{uncertainty})$
المؤثرات
$\hat{A}=\hat{Q}$
و
$\hat{B}=\hat{H}$.
نحصل اذن علي

\begin{eqnarray}
\sigma_Q^2\sigma_H^2\geq \frac{1}{4}|<[\hat{Q},\hat{H}]>|^2.
\end{eqnarray}
بعبارة اخري
\begin{eqnarray}
\sigma_Q\sigma_H\geq \frac{\hbar}{2}|\frac{d<\hat{Q}>}{dt}|.
\end{eqnarray}
نعرف
\begin{eqnarray}
\sigma_t=\frac{\sigma_Q}{|\frac{d<\hat{Q}>}{dt}|}.
\end{eqnarray}
اذن نجد

\begin{eqnarray}
\sigma_t\sigma_H\geq \frac{\hbar}{2}.
\end{eqnarray}
هذه هي علاقة الارتياب زمن - طاقة. الكمية
$\sigma_t$
هي كمية الزمن
التي في خلالها تتغير القيمة المنتظرة ل
$\hat{Q}$
بوحدة انحراف معياري. اذن اذا جعلنا الارتياب في الطاقة صغير جدا فان كمية الزمن اللازمة من اجل ان يتغير الملاحظ بصورة محسوسة تكون كبيرة جدا.

\thispagestyle{headings}
\section*{ 
الصمود تحت تأثير الدورانات
}
\addcontentsline{toc}{section}{
{\bf 
الصمود تحت تأثير الدورانات
}} 
\subsection*{
العزوم الحركية
}
\addcontentsline{toc}{subsection}{
العزوم الحركية
} 
يعرف العزم الحركي الزاوي ب
\begin{eqnarray}
\vec{L}=\vec{r}x \vec{p}.
\end{eqnarray}
بدلالة المركبات لدينا
\begin{eqnarray}
L_1=x_2p_3-x_3p_2~,~L_2=x_3p_1-x_1p_3~,~L_3=x_1p_2-x_2p_1.
\end{eqnarray}
في الميكانيك الكمومي نقوم بالتعويضات التالية
\begin{eqnarray}
x_i\longrightarrow \hat{x}_i~,~p_i\longrightarrow \hat{p}_i~:~[\hat{x}_i,\hat{p}_j]=i\hbar \delta_{ij}.
\end{eqnarray}
اذن مؤثرات العزم الحركي الزاوي تعطي ب
\begin{eqnarray}
\hat{L}_1=\hat{x}_2\hat{p}_3-\hat{x}_3\hat{p}_2~,~\hat{L}_2=\hat{x}_3\hat{p}_1-\hat{x}_1\hat{p}_3~,~\hat{L}_3=\hat{x}_1\hat{p}_2-\hat{x}_2\hat{p}_1.
\end{eqnarray}
نحسب
\begin{eqnarray}
[\hat{L}_1,\hat{L}_2]&=&\hat{x}_2[\hat{p}_3,\hat{x}_3]\hat{p}_1+\hat{x}_1[\hat{x}_3,\hat{p}_3]\hat{p}_2\nonumber\\
&=&i\hbar\hat{L}_3.
\end{eqnarray}
بالمثل نحسب
\begin{eqnarray}
[\hat{L}_3,\hat{L}_1]=i\hbar\hat{L}_2~,~[\hat{L}_2,\hat{L}_3]=i\hbar\hat{L}_1.
\end{eqnarray}
يمكن كتابة علاقات التبادل هذه علي الشكل الموجز
\begin{eqnarray}
[\hat{L}_i,\hat{L}_j]=i\hbar\epsilon_{ijk}\hat{L}_k
\end{eqnarray}
هذه المعادلة تعرف جبرية
\footnote{.${\rm algebra}$}
العزم الحركي التي هي جبرية من جبريات 
لي
\footnote{.${\rm Lie}$}
 تعرف رياضيا بجبرية
$su(2)$.
الرمز
$\epsilon_{ijk}$ 
هو رمز ضد-تناظري بالكلية يعرف باسم
تنسور
\footnote{.${\rm tensor}$}
لفي- سيفيتا
\footnote{.${\rm Levi-Civita}$}
معرف ب
$\epsilon_{123}=\epsilon_{312}=\epsilon_{231}=1$, $\epsilon_{213}=\epsilon_{132}=\epsilon_{321}=-1$ 
و
$\epsilon_{ijk}=0$ 
اذا كان
$i=j$ 
او
$i=k$ 
او
$j=k$. 

تعني علاقات التبادل اعلاه  ان المؤثرات
$\hat{L}_i$
هي مؤثرات غير متلائمة وبالتالي, باستعمال مبدأ الارتياب, لا يمكن تقطيرها
\footnote{.${\rm diagonalized}$}
في وقت واحد. 
اذن لا يوجد شعاع عزم حركي يقيني
\footnote{.${\rm determinate}$}. 
لنعرف مربع العزم الحركي ب
\begin{eqnarray}
\hat{L}^2=\hat{L}_1^2+\hat{L}_2^2+\hat{L}_3^2.
\end{eqnarray} 
هذا المؤثر يتبادل مع المركبات
$\hat{L}_i$. 
بالفعل نحسب
\begin{eqnarray}
[\hat{L}^2,\hat{L}_3]&=&[\hat{L}_1^2,\hat{L}_3]+[\hat{L}_2^2,\hat{L}_3]\nonumber\\
&=&\hat{L}_1[\hat{L}_1,\hat{L}_3]+[\hat{L}_1,\hat{L}_3]\hat{L}_1+\hat{L}_2[\hat{L}_2,\hat{L}_3]+[\hat{L}_2,\hat{L}_3]\hat{L}_2\nonumber\\
&=&0.
\end{eqnarray}
بالمثل نحسب
\begin{eqnarray}
[\hat{L}^2,\hat{L}_2]=0~,~[\hat{L}^2,\hat{L}_1]=0.
\end{eqnarray}
اذن يمكن تقطير 
$\hat{L}^2$
و واحد من مركبات العزم الحركي مثلا 
$\hat{L}_3$
في نفس الوقت. نكتب
\begin{eqnarray}
\hat{L}_3|f>=\mu|f>~,~\hat{L}^2|f>=\lambda|f>.
\end{eqnarray}
نعرف مؤثرات الرفع و الخفض ب
\begin{eqnarray}
\hat{L}_{\pm}=\hat{L}_1\pm i\hat{L}_2.
\end{eqnarray}
نحسب علاقات التبادل
\begin{eqnarray}
[\hat{L}_+,\hat{L}_-]=2\hbar \hat{L}_3~,~[\hat{L}_3,\hat{L}_{\pm}]=\pm \hbar \hat{L}_{\pm}~,~[\hat{L}^2,\hat{L}_{\pm}]=0.
\end{eqnarray}
اذن
\begin{eqnarray}
\hat{L}_3(\hat{L}_{\pm}|f>)=(\mu\pm\hbar)(\hat{L}_{\pm}|f>)~,~\hat{L}^2(\hat{L}_{\pm}|f>)=\lambda(\hat{L}_{\pm}|f>).\label{hjh}
\end{eqnarray}
من الواضح ان
$\hat{L}_{\pm}|f>$
هو شعاع ذاتي ل
$\hat{L}_3$
مقابل للقيمة الذاتية
$\mu\pm\hbar$. 
بعبارة اخري
$\hat{L}_+$
يرفع القيمة الذاتية ل
$\hat{L}_3$
ب
$\hbar$
بينما
 $\hat{L}_-$ 
 يخفض القيمة الذاتية ل
 $\hat{L}_3$
 ب
 $\hbar$.

من العلاقة
$<\hat{L}^2>=<\hat{L}_1^2>+<\hat{L}_2^2>+<\hat{L}_3^2>$ 
نستنتج ان
$\mu^2\leq \lambda$. 
اذن انطلاقا من شعاع ذاتي
 $|f>$
 ل
  $\hat{L}_3$ 
  بقيمة ذاتية
  $\mu$
  نحصل عن طريق التطبيق المتتالي ل
  $\hat{L}_+$
علي الاشعة الذاتية بالقيم الذاتية
$\mu + n \hbar$
حيث 
$n$
هو عدد صحيح موجب. يجب دائما ان يكون لدينا
 $(\mu +n\hbar)^2\leq \lambda$
 و بالتالي
 توجد قيمة اعظمية ل 
 $n$. 
 الشعاع الذاتي المقابل هو الشعاع الذاتي الاعلي و يرمز له ب
 $|f_h>=|l>$
 و يجب ان يحقق
\begin{eqnarray}
\hat{L}_+|l>=0.
\end{eqnarray}
لنرمز ايضا للقيمة الذاتية ل
$\hat{L}_3$ 
المقابلة ل
$|l>$
ب
$\hbar l$
اي
\begin{eqnarray}
\hat{L}_3|l>=\hbar l|l>.
\end{eqnarray}
باستعمال العلاقة
$\hat{L}^2=\hat{L}_-\hat{L}_++\hbar \hat{L}_3+\hat{L}_3^2$
نحصل علي
\begin{eqnarray}
\hat{L}^2|l>=\hbar^2l(l+1)|l>.
\end{eqnarray}
اذن
$\lambda=\hbar^2l(l+1)$. 

بالمثل انطلاقا من شعاع ذاتي
 $|f>$
 ل
 $\hat{L}_3$
 بقيمة ذاتية
 $\mu$
 نحصل عن طريق التطبيق المتتالي ل
 $\hat{L}_-$
 علي الاشعة الذاتية بالقيم الذاتية
  $\mu - n \hbar$
  حيث 
 $n$ 
 هو عدد صحيح موجب. مرة اخري يجب ان يكون لدينا
 $(\mu -n\hbar)^2\leq \lambda$
 و بالتالي يوجد قيمة اعظمية ل
 $n$. 
 الشعاع الذاتي المقابل هو الان الشعاع الادني و يرمز له ب
 $|f_l>=|k>$
 و يجب ان يحقق
\begin{eqnarray}
\hat{L}_-|k>=0.
\end{eqnarray}
لنرمز للقيمة الذاتية ل
$\hat{L}_3$
المرفقة ب
 $|k>$
 ب
 $\hbar k$
 اي
\begin{eqnarray}
\hat{L}_3|k>=\hbar k|k>.
\end{eqnarray}
باستعمال العلاقة 
$\hat{L}^2=\hat{L}_+\hat{L}_--\hbar \hat{L}_3+\hat{L}_3^2$
نحصل علي
\begin{eqnarray}
\hat{L}^2|k>=\hbar^2k(k-1)|k>.
\end{eqnarray}
اذن
$\lambda=\hbar^2k(k-1)$
و بالتالي 
$l(l+1)=k(k-1)$ 
اي
$k=-l$. 
اذن شعاع الحالة الادني ل
$\hat{L}_3$
هو
$|f_l>=|-l>$
بالقيمة الذاتية
$-\hbar l$. 

 نرمز للقيم الذاتية ل
 $\hat{L}_3$
  ب
 $\hbar m$
 حيث
$m$ 
تأخذ 
$N$
قيمة بين
 $-l$ 
 و
 $+l$
 كل قيمتين متتاليتين مفصولتين بوحدة.  اذن
 $l=-l+N$
 اي
  $l=N/2$.
  بعبارة اخري 
   $l$
   يمكنه ان يكون عدد صحيح, مرفق بالعزم الحركي الزاوي مثل الذي عرفناه اعلاه, او ان يكون عدد
   نصف صحيح و هذا ما يقابل السبين
   \footnote{.${\rm spin}$}.
   نرمز للاشعة الذاتية المقابلة ب
  $|lm>$
  حيث
 \begin{eqnarray}
\hat{L}^2|lm>=\hbar^2l(l+1)|lm>~,~L_3|lm>=\hbar m|lm>.
\end{eqnarray}
\begin{eqnarray}
l=0,\frac{1}{2},1,\frac{3}{2},...~,~m=-l,-l+1,...,l-1,l.
\end{eqnarray}
من اجل كل قيمة ل
$l$
لدينا 
  $2l+1$
  حالة ذاتية 
اجمالا في فضاء هيلبرت. من الواضح ان
$|ll>=|l>$
و
$|l-l>=|-l>$. 
 المعادلة
$(\ref{hjh})$
تصبح
\begin{eqnarray}
\hat{L}_3(\hat{L}_{\pm}|lm>)=\hbar(m\pm1)(\hat{L}_{\pm}|lm>)~,~\hat{L}^2(\hat{L}_{\pm}|lm>)=\hbar l(l+1)(\hat{L}_{\pm}|lm>).
\end{eqnarray}
بعبارة اخري
 \begin{eqnarray}
\hat{L}_{\pm}|lm>=A_l^m|lm\pm 1>.
\end{eqnarray}
نحسب
\begin{eqnarray}
|A_l^m|^2&=&<lm|\hat{L}_{\mp}\hat{L}_{\pm}|lm>\nonumber\\
&=&<lm|(\hat{L}^2\mp \hbar \hat{L}_3-\hat{L}_3^2)|lm>\nonumber\\
&=&\hbar^2(l(l+1)-m(m\pm 1)).
\end{eqnarray}
\subsection*{
التوافقيات الدورانية
}
\addcontentsline{toc}{subsection}{
التوافقيات الدورانية
} 
مؤثرات العزم الحركي في اساس الموضع تأخذ الشكل
\begin{eqnarray}
\vec{\hat{L}}=\frac{\hbar}{i}\vec{r}x\vec{\nabla}.
\end{eqnarray}
مؤثر التدرج يعطي ب
\begin{eqnarray}
\vec{\nabla}&=&\vec{i}\frac{\partial}{\partial \hat{x}_1}+\vec{j}\frac{\partial}{\partial \hat{x}_2}+\vec{k}\frac{\partial}{\partial \hat{x}_3}.
\end{eqnarray}
نعرف الاحداثيات الكروية بالمعادلات
\begin{eqnarray}
\hat{x}_1=r\sin\theta\cos\phi~,~\hat{x}_2=r\sin\theta\sin\phi~,~\hat{x}_3=r\cos\theta.
\end{eqnarray}
اشعة وحدة الاحداثيات الكروية 
 $r$, $\theta$
 و
  $\phi$
 هي
\begin{eqnarray}
&&\vec{u}_r=\sin\theta\cos\phi ~\vec{i}+\sin\theta\sin\phi ~\vec{j}+\cos\theta ~\vec{k}\nonumber\\
&&\vec{u}_{\theta}=\cos\theta\cos\phi ~\vec{i}+\cos\theta\sin\phi ~\vec{j}-\sin\theta ~\vec{k}\nonumber\\
&&\vec{u}_{\phi}=-\sin\phi ~\vec{i}+\cos\phi ~\vec{j}.
\end{eqnarray}
في الاحداثيات الكروية يصبح مؤثر التدرج معطي  ب
\begin{eqnarray}
\vec{\nabla}&=&\vec{u}_r\frac{\partial}{\partial r}+\vec{u}_{\theta}\frac{1}{r}\frac{\partial}{\partial \theta}+\vec{u}_{\phi}\frac{1}{r\sin\theta}\frac{\partial}{\partial \phi}.
\end{eqnarray}
نلاحظ ان
$\vec{u}_rx\vec{u}_{r}=0$, $\vec{u}_rx\vec{u}_{\theta}=\vec{u}_{\phi}$ 
و
$\vec{u}_rx\vec{u}_{\phi}=-\vec{u}_{\theta}$.
اذن
\begin{eqnarray}
\vec{\hat{L}}&=&\frac{\hbar}{i}\big(\vec{u}_{\phi}\frac{\partial}{\partial \theta}-\vec{u}_{\theta}\frac{1}{\sin\theta}\frac{\partial}{\partial \phi}\big).
\end{eqnarray}
اي
\begin{eqnarray}
&&\hat{L}_1=\frac{\hbar}{i}\big(-\sin\phi \frac{\partial}{\partial \theta}-\cot\theta \cos\phi \frac{\partial}{\partial \phi}\big)\nonumber\\
&&\hat{L}_2=\frac{\hbar}{i}\big(\cos\phi \frac{\partial}{\partial \theta}-\cot\theta \sin\phi \frac{\partial}{\partial \phi}\big)\nonumber\\
&&\hat{L}_3=\frac{\hbar}{i}\frac{\partial}{\partial \phi}.
\end{eqnarray}
يمكن ان نحسب مباشرة
\begin{eqnarray}
\hat{L}_{\pm}=\pm \hbar e^{\pm i\phi}\big(\frac{\partial}{\partial \theta}\pm i\cot\theta  \frac{\partial}{\partial \phi}\big).
\end{eqnarray}
ايضا
\begin{eqnarray}
\hat{L}_{+}\hat{L}_{-}=-\hbar^2\bigg(\frac{\partial^2}{\partial\theta^2}+i\frac{\partial}{\partial\phi}+(\cot\theta)^2\frac{\partial^2}{\partial\phi^2}+\cot\theta\frac{\partial}{\partial\theta}\bigg).
\end{eqnarray}
اذن
\begin{eqnarray}
\hat{L}^2&=&\hat{L}_{+}\hat{L}_{-}-\hbar\hat{L}_3+\hat{L}_3^2\nonumber\\
&=&-\hbar^2\bigg(\frac{\partial^2}{\partial\theta^2}+\frac{1}{\sin^2\theta}\frac{\partial^2}{\partial\phi^2}+\cot\theta\frac{\partial}{\partial\theta}\bigg)\nonumber\\
&=&-\hbar^2\bigg(\frac{1}{\sin\theta}\frac{\partial}{\partial\theta}(\sin\theta\frac{\partial}{\partial\theta})+\frac{1}{\sin^2\theta}\frac{\partial^2}{\partial\phi^2}\bigg).
\end{eqnarray}
الدوال الذاتية ل
$\hat{L}^2$
هي
$Y_l^m(\theta,\phi)=<\theta|<\phi|lm>$
 و هي تحقق
\begin{eqnarray}
-\hbar^2\bigg(\frac{1}{\sin\theta}\frac{\partial}{\partial\theta}(\sin\theta\frac{\partial}{\partial\theta})+\frac{1}{\sin^2\theta}\frac{\partial^2}{\partial\phi^2}\bigg)Y_l^m=\hbar^2l(l+1)Y_l^m.
\end{eqnarray}
الدوال
 $Y_l^m(\theta,\phi)$ 
 هي ايضا دوال ذاتية ل 
 $\hat{L}_3$
 اي
\begin{eqnarray}
\frac{\hbar}{i}\frac{\partial}{\partial\phi}Y_l^m=\hbar m Y_l^m.
\end{eqnarray}
يمكن الحصول علي الحل الصريح باستعمال طريقة فصل المتغيرات. نكتب
\begin{eqnarray}
Y_l^m(\theta,\phi)=\Theta_l^m(\theta)\Phi_m(\phi).
\end{eqnarray}
نحصل علي المعادلات التفاضلية
\begin{eqnarray}
-\frac{1}{\sin\theta}\frac{d }{d\theta}(\sin\theta\frac{d\Theta_l^m}{d\theta})+\frac{m^2}{\sin^2\theta}\Theta_l^m=l(l+1)\Theta_l^m.
\end{eqnarray}
\begin{eqnarray}
\frac{d}{d\phi}\Phi_m=i m \Phi_m\Leftrightarrow \Phi_m(\phi)=e^{im\phi}.
\end{eqnarray}
من الواضح انه يجب ان يتحقق الشرط
$\Phi_m(\phi+2\pi)=\Phi_m(\phi)$ 
و بالتالي فان 
$m$ 
هو عدد صحيح اي
\begin{eqnarray}
m=0,\pm 1,\pm 2,....
\end{eqnarray}
يمكن وضع المعادلة التفاضلية الاخري علي الشكل
(مع 
$x=\cos\theta$
)
\begin{eqnarray}
\frac{d}{dx}\bigg[(1-x^2)\frac{d\Theta_l^m}{dx}\bigg]+[l(l+1)-\frac{m^2}{1-x^2}]\Theta_l^m=0.
\end{eqnarray}
هذه معادلة لوجوندر
\footnote{.${\rm Legendre}$}.
يعطي الحل القانوني بكثيرات حدود لوجوندر المرفقة
\footnote{.${\rm associated~Legendre~polynomials}$}
$P_l^m(x)$
اي
\begin{eqnarray}
\Theta_l^m(\theta)=A P_l^m(x)~,~x=\cos\theta.
\end{eqnarray}
يمكن اعطاء كثيرات حدود لوجوندر المرفقة بدلالة كثيرات حدود لوجوندر
$P_l(x)$
بالعلاقة

\begin{eqnarray}
P_l^m(x)=(1-x^2)^{\frac{|m|}{2}}\bigg(\frac{d}{dx}\bigg)^{|m|}P_l(x).
\end{eqnarray}
كثيرات حدود لوجوندر
$P_l(x)$ 
تعطي بعلاقة رودريغاز
\footnote{.${\rm Rodrigues}$}
كالاتي
\begin{eqnarray}
P_l(x)=\frac{1}{2^ll!}\bigg(\frac{d}{dx}\bigg)^{l} (x^2-1)^l.
\end{eqnarray}
من الواضح من هذه العلاقة ان
 $l$
 يجب ان يكون عدد صحيح موجب و ان 
  $P_l(x)$
  هو كثير حدود من الدرجة
  $l$
  في
  $x=\cos\theta$. 
  كثيرات حدود لوجوندر المرفقة
  $P_l^m(x)$
  هي كثيرات حدود في 
  $x=\cos\theta$ 
  فقط من اجل
  $m$
  زوجي. 
  من اجل
  $m$ 
  فردي  فان كثيرات الحدود هذه تكون مضروبة في قوة ل
  $\sin\theta$. 
  ايضا اذا كان
  $|m|>l$
  فان
  $P_l^m(x)=0$
  و بالتالي فان القيم المسموح بها ل
  $l$
  و
  $m$
  هي
\begin{eqnarray}
l=0,1,2,...~,~m=-l,-l+1,...,0,....,l-1,l.
\end{eqnarray}
كما في السابق لدينا
$2l+1$
حالة من اجل كل قيمة ل
 $l$. 
 لكن 
 $l$
 الان هو دائما صحيح.

الحل المكتمل يعطي اذن ب
\begin{eqnarray}
Y_l^m(\theta,\phi)=AP_l^m(\cos\theta)e^{im\phi}.
\end{eqnarray}
نفرض شرط التنظيم
\begin{eqnarray}
\int_0^{2\pi}\int_0^{\pi}\sin\theta d\theta d\phi~ |Y_l^m(\theta,\phi)|^2=1.
\end{eqnarray}
نجد
\begin{eqnarray}
A=\epsilon\sqrt{\frac{2l+1}{4\pi}\frac{(l-|m|)!}{(l+|m|)!}}.
\end{eqnarray}
\begin{eqnarray}
\epsilon=(-1)^m~,~m\geq 0~,~\epsilon=1~,~m\leq 0.
\end{eqnarray}
يمكن ايضا ان نتحقق من شرط التعامد و التجانس
\begin{eqnarray}
\int_0^{2\pi}\int_0^{\pi}\sin\theta d\theta d\phi~ [Y_l^m(\theta,\phi)]^*Y_t^s(\theta,\phi)=\delta_{lt}\delta_{ms}.
\end{eqnarray}
\thispagestyle{headings}
\section*{
الحلول المضبوطة لمعادلة شرودينغر
}
\addcontentsline{toc}{section}{
{\bf
الحلول المضبوطة لمعادلة شرودينغر
}} 
\subsection*{
الحالات المستقرة, حالات التصادم و الحالات المرتبطة
}
\addcontentsline{toc}{subsection}{
الحالات المستقرة, حالات التصادم و الحالات المرتبطة
} 

\paragraph{
الحالات المستقرة:
}
تكتب معادلة شرودينغر علي الشكل
\begin{eqnarray}
i\hbar\frac{\partial }{\partial t}\Psi(t,x)=\bigg(-\frac{\hbar^2}{2m}\frac{\partial^2}{\partial x^2}+V(x)\bigg)\Psi(t,x).
\end{eqnarray}
ماهي الحلول
$\psi(t,x)$
 من اجل كمون معين
  $V$.
  نريد حل هذه المسألة بشكل عام. نبدا من فصل المتغيرات
\begin{eqnarray}
\Psi(t,x)=\psi(x)\phi(t).
\end{eqnarray}
نحصل علي
\begin{eqnarray}
\frac{i\hbar}{\phi}\frac{d\phi}{d t}=\frac{1}{\psi}\bigg(-\frac{\hbar^2}{2m}\frac{d^2}{d x^2}+V(x)\bigg)\psi.
\end{eqnarray}
الطرف الايسر لهذه المعادلة هو دالة في الزمن
$t$
فقط اما الطرف الايمن فهو دالة في
$x$
فقط. اذن كلا الطرفين يجب ان يكونا مساويين لثابت
$E$
لا يتعلق ب
$t$
و
$x$. 
لدينا اذن
\begin{eqnarray}
\frac{d\phi}{d t}=-i\frac{E}{\hbar}\phi~\longrightarrow~\phi(t)=e^{-i\frac{Et}{\hbar}}.
\end{eqnarray}
المعادلة الاخري تكتب علي الشكل
\begin{eqnarray}
\bigg(-\frac{\hbar^2}{2m}\frac{d^2}{d x^2}+V(x)\bigg)\psi=E\psi.
\end{eqnarray}
الثابت
$E$
يجب ان يكون حقيقي لانه لا شيئ سوي القيمة الذاتية للهاميلتونية
$H=-\frac{\hbar^2}{2m}\frac{d^2}{dx^2}+V(x)$
مرفق بالدالة الذاتية
$\psi(x)$. 
بعبارة اخري الحل المفصول
$\Psi(t,x)=e^{-i\frac{Et}{\hbar}}\psi(x)$
هو حل يقيني ذو طاقة متعينة تساوي
$E$. 

علاوة علي ذلك فان الحل اعلاه هو حل مستقر لان كثافة الاحتمال لا تتعلق بالزمن اي
 $\rho=\Psi^*(t,x)\Psi(t,x)=\psi^*(x)\psi(x)$. 
 في الحقيقة فان القيمة المنتظرة لاي ملاحظ
 $Q(x,p)$ 
 لا تتعلق ايضا بالزمن. بالفعل
\begin{eqnarray}
<\hat{Q}>=<\Psi|\hat{Q}|\Psi>&=&\int dx \Psi^*(t,x)Q(x,\frac{\hbar}{i}\frac{\partial}{\partial x})\Psi(t,x)\nonumber\\
&=&\int dx \psi(x)Q(x,\frac{\hbar}{i}\frac{d}{d x})\psi(x).
\end{eqnarray}
لتكن
$\psi_n(x)$
الدالة الذاتية
للهاميلتونية
$H$
بالقيمة الذاتية
$E_n$. 
الحل العام لمعادلة شرودينغر هو تركيب خطي للحلول المفصولة
$\Psi_n(t,x)=e^{-i\frac{E_nt}{\hbar}}\psi_n(x)$.
هذا يعطي ب
\begin{eqnarray}
\Psi(t,x)=\sum_nc_n\psi_n(x)e^{-i\frac{E_nt}{\hbar}}.
\end{eqnarray}
هذا هو مبدأ التركيب الخطي الكمومي. المعاملات
$c_n$
يجب تعيينها من الشرط الابتدائي
\begin{eqnarray}
\Psi(0,x)=\sum_nc_n\psi_n(x).
\end{eqnarray}

\paragraph{
حالات التصادم و حالات الارتباط:
}
نعتبر جسيم ذو طاقة
 $E$ 
يتحرك في بعد واحد في كمون
$V(x)$. 
في الميكانيك الكلاسيكي اذا كانت الطاقة
$E$ 
هي اصغر من قيم الكمون
$V(\infty)$
و
$V(-\infty)$
فانه لدينا حالة مرتبطة اي ان الجسيم لا يمكن ان يهرب من الكمون الي اللانهاية. اذا كانت الطاقة
 $E$ 
 اكبر من
 $V(\infty)$ 
 و
 $V(-\infty)$ 
 فانه لدينا حالة تصادم اي ان الجسيم يأتي من اللانهاية, يتفاعل مع الكمون, ثم يرجع مرة اخري الي اللانهاية. حتي نحصل علي حالة تصادم يكفي ان تكون
 $E$ 
 اكبر من 
  $V(\infty)$ 
 او
 $V(-\infty)$.
 
 بالمثل فانه في الميكانيك الكمومي هناك نوعان من الحلول الممكنة لمعادلة شرودينغر. الحالات المرتبطة 
 \footnote{.${\rm bound~states}$}
 و حالات التصادم
 \footnote{.${\rm scattering~states}$}
 . تعرف هذه الحالات ب 
\begin{eqnarray}
&&E<V(-\infty)~{\rm and}~E<V(+\infty)~:~{\rm bound}~{\rm state}\nonumber\\
&&E>V(-\infty)~{\rm or}~E>V(+\infty)~:~{\rm scattering}~{\rm state}.
\end{eqnarray}
من اجل الهزاز التوافقي لدينا فقط حالات مرتبطة اما من اجل الجسيم الحر فلدينا حالات تصادم فقط. في اغلب الحالات فان الكمون
ينعدم في اللانهاية و بالتالي نحصل علي الشرط المبسط
\begin{eqnarray}
&&E<0~:~{\rm bound}~{\rm state}\nonumber\\
&&E>0~:~{\rm scattering}~{\rm state}.
\end{eqnarray}

\subsection*{
الجسيم الحر
}
\addcontentsline{toc}{subsection}{
الجسيم الحر
} 
في هذه الحالة ينعدم الكمون في كل مكان. معادلة شرودينغر المستقلة عن الزمن تصبح
\begin{eqnarray}
-\frac{\hbar^2}{2m}\frac{d^2\psi}{d x^2}=E\psi.
\end{eqnarray}
نعيد كتابة هذه المعادلة علي الشكل
\begin{eqnarray}
\frac{d^2\psi}{dx^2}=-k^2\psi~,~k^2=\frac{2mE}{\hbar^2}.
\end{eqnarray}
من الواضح ان
 $E$ 
 هي الطاقة الحركية للجسيم
 $T=\frac{1}{2}mv^2={p^2}/{2m}$
 و بالتالي
 $E\geq 0$. 
 سرعة و كمية حركة الجسيم تعطي اذن ب
\begin{eqnarray}
v=\sqrt{\frac{2E}{m}}=\frac{p}{m}~,~p=\hbar k.
\end{eqnarray}
الحل العام لمعادلة شرودينغر غير المتعلقة بالزمن يعطي ب
\begin{eqnarray}
\psi(x)=Ae^{ikx}+Be^{-ikx}.
\end{eqnarray}
بالضرب بالمعامل الطوري المتعلق بالزمن
 $e^{-i\frac{Et}{\hbar}}$ 
 نحصل علي
\begin{eqnarray}
\Psi(t,x)=Ae^{i\frac{k}{\hbar}(x-v_{\rm phase}t)}+Be^{-i\frac{k}{\hbar}(x+v_{\rm phase}t)}.
\end{eqnarray}
الحد الاول يمثل موجة منتشرة الي اليمين بسرعة
$v_{\rm phase}$ 
بينما يعبر الحد الثاني عن موجة منتشرة الي اليسار بسرعة
$v_{\rm phase}$.
السرعة الطورية تعطي ب
\begin{eqnarray}
v_{\rm phase}=\frac{E}{\hbar k}=\frac{p}{2m}=\frac{1}{2}v.
\end{eqnarray}
يمكن كتابة الحل اعلاه علي الشكل المكافئ
\begin{eqnarray}
\Psi_k(t,x)=Ae^{i(kx-\frac{\hbar k^2}{2m}t)}.
\end{eqnarray}
\begin{eqnarray}
k=\pm\frac{\sqrt{2mE}}{\hbar}.
\end{eqnarray}
\begin{eqnarray}
&&k>0~,~{\rm wave}~{\rm traveling}~{\rm to}~{\rm the}~{\rm right}\nonumber\\
&&k<0~,~{\rm wave}~{\rm traveling}~{\rm to}~{\rm the}~{\rm left}.
\end{eqnarray}
المشكل الاول الذي لدينا مع هذه الحلول المنتشرة هو انها عبارة عن امواج تنتشر بنصف سرعة الجسيم. المشكل الثاني
هو ان هذه الحلول غير قابلة للتنظيم. اذن ليس لدينا جسيم حر بكمية حركة متعينة.

نحصل علي الحل العام لمعادلة شرودينغر عن طريق اخذ تركيب خطي للحلول المفصولة اعلاه كما يلي 
\begin{eqnarray}
\Psi(t,x)&=&\int_{-\infty}^{\infty}\frac{dk}{\sqrt{2\pi}}\frac{\phi(k)}{A}\Psi_k(t,x)\nonumber\\
&=&\int_{-\infty}^{\infty}\frac{dk}{\sqrt{2\pi}}\phi(k)e^{i(kx-\frac{\hbar k^2}{2m}t)}.
\end{eqnarray}
 يمكن تنظيم هذه الدالة الموجية من اجل اختيارات مناسبة للدوال
 $\phi(k)$.
  تسمي هذه الدالة الموجية بالحزمة الموجية
  \footnote{.${\rm wave~packet}$}
  . يمكن تعيين الدوال
  $\phi(k)$ 
  من الشروط الابتدائية
\begin{eqnarray}
\Psi(0,x)&=&\int_{-\infty}^{\infty}\frac{dk}{\sqrt{2\pi}}\phi(k)e^{ikx}.
\end{eqnarray}
حل هذه الشروط الابتدائية يعطي بمبرهنة بلانشارل 
\footnote{.${\rm Plancherel's~theorem}$}
اي ان
$\phi(k)$
هو تحويل فوريي
\footnote{.${\rm Fourier~transform}$}
ل
$\Psi(0,x)$
يعطي ب
\begin{eqnarray}
\phi(k)&=&\int_{-\infty}^{\infty}\frac{dx}{\sqrt{2\pi}}\Psi(0,x)e^{-ikx}.
\end{eqnarray}
لنلاحظ ان
\begin{eqnarray}
\int dx \Psi^*(t,x)\Psi(t,x)=\int dx \Psi^*(0,x)\Psi(0,x)=\int dk \phi^*(k)\phi(k).
\end{eqnarray}
يبقي ان نتحقق ان سرعة الحزمة الموجية تساوي سرعة الجسيم
$v$. 
سرعة الحزمة الموجية تعرف باسم سرعة المجموعة
\footnote{.${\rm group~velocity}$}
و يمكن ان تكون اكبر من, تساوي او اصغر من السرعة الطورية. نعتبر حزمة موجية عامة 
تعطي ب
\begin{eqnarray}
\Psi(t,x)&=&\int_{-\infty}^{\infty}\frac{dk}{\sqrt{2\pi}}\phi(k)e^{i(kx-\Omega t)}.
\end{eqnarray}
نفترض علاقة تشتت
\footnote{.${\rm dispersion~relation}$}
عامة. اي اننا نفترض ان التواتر الزاوي
$\Omega$
هو دالة كيفية في
 $k$
 بمعني
 $\Omega=\Omega(k)$.
 بالاضافة الي هذا نفترض ان
 $\phi(k)$
 هو متمركز حول القيمة
 $k=k_0$
 اي ان المركبات المختلفة للحزمة الموجية تنتشر تقريبا بنفس السرعة الطورية
 $v_{\rm phase}=\Omega/k$ 
 و بالتالي فان شكل الحزمة الموجية يتغير ببطء.  في الحقيقة فانه فقط في هذه الحالة يكون لمفهوم سرعة المجموعة معني واضح. اذن ننشر
 $\Omega$
 كسلسلة 
 تايلور
 \footnote{.${\rm Taylor~series}$}
 حول
 $k=k_0$
 كما يلي
\begin{eqnarray}
\Omega(k)=\Omega(k_0)+\Omega^{'}(k_0)(k-k_0)+...
\end{eqnarray}
نحسب 
(مع 
$k^{'}=k-k_0$, $\Omega_0=\Omega(k_0)$ 
و
 $\Omega_0^{'}=d\Omega(k)/dk|_{k=k_0}$
)
\begin{eqnarray}
\Psi(t,x)&=&e^{-i\Omega_0t}\int_{-\infty}^{\infty}\frac{dk^{'}}{\sqrt{2\pi}}\phi(k^{'}+k_0)e^{i((k^{'}+k_0)x-\Omega_0^{'}k^{'} t)}\nonumber\\
&=&e^{i(-\Omega_0+k_0\Omega_0^{'})t}\int_{-\infty}^{\infty}\frac{dk^{'}}{\sqrt{2\pi}}\phi(k^{'}+k_0)e^{i(k^{'}+k_0)(x-\Omega_0^{'}t)}.
\end{eqnarray}
في اللحظة
$t=0$
نحصل علي 
\begin{eqnarray}
\Psi(0,x)&=&\int_{-\infty}^{\infty}\frac{dk^{'}}{\sqrt{2\pi}}\phi(k^{'}+k_0)e^{i(k^{'}+k_0)x}.
\end{eqnarray}
اذن
\begin{eqnarray}
\Psi(t,x)&=&e^{i(-\Omega_0+k_0\Omega_0^{'})t}\Psi(0,x-\Omega_0^{'}t).
\end{eqnarray}
كما نريد بالضبط فان شكل الحزمة الموجية لا يتغير و تتحرك الحزمة بسرعة المجموعة
\begin{eqnarray}
v_{\rm group}=\Omega_0^{'}=\frac{d\Omega}{dk}|_{k=k_0}.
\end{eqnarray}
في حالتنا هذه 
$\Omega=\hbar k^2/2m$
و بالتالي تصبح سرعة المجموعة معطاة ب
\begin{eqnarray}
v_{\rm group}=\frac{\hbar k_0}{m}=\frac{p_0}{m}=v_0.
\end{eqnarray}
\subsection*{
الهزاز التوافقي
}
\addcontentsline{toc}{subsection}{
الهزاز التوافقي
} 
لتكن
$x_0$
قيمة اصغرية محلية للكمون
 $V$
 اي
\begin{eqnarray}
V^{'}(x_0)=0.
\end{eqnarray}
بالاضافة يمكن دائما ان نختار, من دون اي فقدان للعمومية, الكمون بحيث
$V(x_0)=0$.
ننشر الان
$V(x)$
كسلسلة تايلور حول
$x_0$
كا يلي
\begin{eqnarray}
V(x)&=&V(x_0)+(x-x_0)V^{'}(x_0)+\frac{1}{2}(x-x_0)^2V^{''}(x_0)+...\nonumber\\
&=&\frac{1}{2}V^{''}(x_0)(x-x_0)^2+...
\end{eqnarray}
هذه هي الطاقة الكامنة لهزاز توافقي بسيط بثابت مرونة
$k=V^{''}(x_0)=m\Omega^2$. 

معادلة شرودينغر غير المتعلقة بالزمن التي تصف الحركة حول القيمة الاصغرية المحلية
$x_0$ 
تعطي اذن ب
\begin{eqnarray}
\bigg(-\frac{\hbar^2}{2m}\frac{d^2}{d x^2}+\frac{1}{2}m\Omega^2x^2\bigg)\psi=E\psi.
\end{eqnarray}
يمكن كتابة هذه المعادلة ايضا علي الشكل
\begin{eqnarray}
\bigg(\frac{\hat{p}^2}{2m}+\frac{1}{2}m\Omega^2\hat{x}^2\bigg)|\psi>=E|\psi>.
\end{eqnarray}
لنذكر ان
$\psi(x)=<x|\psi>$. 
نعرف مؤثرات الرفع و الخفض
$a^+$
و
$a$
ب
\begin{eqnarray}
a^+=\frac{1}{\sqrt{2\hbar m\Omega}}(m\Omega\hat{x}-i\hat{p})~,~a=\frac{1}{\sqrt{2\hbar m\Omega}}(m\Omega\hat{x}+i\hat{p}).
\end{eqnarray}
لان
$[\hat{x},\hat{p}]=i\hbar$ 
نحسب علاقات التبادل
\begin{eqnarray}
[a,a^+]=1.
\end{eqnarray}
يمكننا ان نتحقق  الان مباشرة من ان هاميلتونية الهزاز التوافقي البسيط المعطاة ب
$\hat{H}=\frac{\hat{p}^2}{2m}+\frac{1}{2}m\Omega^2\hat{x}^2$ 
يمكن كتابتها علي الشكل
\begin{eqnarray}
\hat{H}=\hbar\Omega(a^+a+\frac{1}{2}).
\end{eqnarray}
نحسب
\begin{eqnarray}
[\hat{H},a]=-\hbar\Omega a~,~[\hat{H},a^+]=\hbar\Omega a^+.
\end{eqnarray}
باستخدام هذه المعادلات و معادلة شرودينغر غير متعلقة بالزمن
$\hat{H}|\psi>=E|\psi>$ 
نحصل علي
\begin{eqnarray}
\hat{H}a|\psi>=(E-\hbar \Omega)a|\psi>~,~\hat{H}a^+|\psi>=(E+\hbar\Omega)a^+|\psi>.
\end{eqnarray}
بعبارة اخري 
$a|\psi>$
هو شعاع ذاتي ل
$\hat{H}$
بالقيمة الذاتية
$E-\hbar\Omega$
بينما 
$a^+|\psi>$
هو شعاع ذاتي بالقيمة الذاتية
$E+\hbar\Omega$.
اذن 
$a$ 
ينقص الطاقة و لهذا نسمية بمؤثر الخفض بينما 
$a^+$
يزيد الطاقة و لهذا الاسم مؤثر الرفع. نعرف مؤثر العدد ب
\begin{eqnarray}
N=a^+a.
\end{eqnarray}
ليكن
$|n>$
الشعاع الذاتي ل
$N$
المرفق بالقيمة الذاتية
$n$
اي
\begin{eqnarray}
N|n>=n|n>.
\end{eqnarray}
لان 
$N$
مؤثر هرميتي فان القيم الذاتية
$n$
حقيقية
و الاشعة الذاتية
$|n>$
متعامدة. في الحقيقة
$n$
يجب ان يكون موجب لان
$n=|a|n>|^2$. 
علاوة علي ذلك فانه باستخدام علاقات التبادل
$[N,a]=-a$ 
و
$[N,a^+]=a^+$
نحسب
$Na|n>=(n-1)a|n>$
و
$Na^+|n>=(n+1)a^+|n>$.
بعبارة اخري
\begin{eqnarray}
a|n>=c_n|n-1>~,~a^+|n>=d_n|n+1>.
\end{eqnarray}
باشتراط ان الاشعة الذاتية
$|n>$
منظمة
اي 
$<n|n>=1$
نحصل علي
$|c_n|^2={n}$
و
$|d_n|^2={n+1}$. 
اذن باخذ
$c_n$
و
$d_n$
اعداد حقيقية موجبة, من اجل التبسيط, لدينا
\begin{eqnarray}
a|n>=\sqrt{n}|n-1>~,~a^+|n>=\sqrt{n+1}|n+1>.
\end{eqnarray}
القيم المسموح بها للطاقة هي بالتالي معطاة ب
\begin{eqnarray}
E_n=\hbar\Omega(n+\frac{1}{2}).
\end{eqnarray}
من الواضح ان الاشعة الذاتية المرفقة هي بالضبط
$|n>$. 

نستعمل الان النتيجة العامة التالية:
{\bf 
طاقة اي حل قابل للتنظيم لمعادلة شرودينغر غير المتعلقة بالزمن يجب
ان تكون اكبر او تساوي من القيمة الاصغرية للكمون
$V$.}
من اجل حالتنا قيد الدراسة فان القيمة الاصغرية ل
 $V$
 هي صفر ووجدنا ان القيم
  $E_n$
  للطاقة هي دائما اكبر من الصفر لان
  $n\geq 0$.
  بالفعل فان طاقة الحالة الاساسية
  $E_0$ 
  للهزاز التوافقي البسيط هي
\begin{eqnarray}
E_0=\frac{1}{2}\hbar\Omega.
\end{eqnarray}
من الواضح انه انطلاقا من اي شعاع حالة
$|n>$
يمكن الوصول الي شعاع الحالة الاساسية 
$|0>$ 
عن طريق التطبيق المتكرر لمؤثر الخفض
$a$.
هذا يعني بالخصوص ان 
$n$ 
يجب ان يكون عدد طبيعي لانه يساوي عدد المرات التي يجب التأثير فيها ب
$a$
للذهاب من
$|n>$
الي
$|0>$.
نحصل علي شرط التكميم

\begin{eqnarray}
n\in{\bf N}.
\end{eqnarray}
شعاع الحالة الاساسية
$|0>$
يجب ان يحقق الشرط
$a|0>=0$.
يكتب هذا الشرط في فضاء الموضع كالتالي
(
مع
$\psi_0(x)=<x|0>$
)

\begin{eqnarray}
(\frac{d}{dx}+\frac{m\Omega}{\hbar}x)\psi_0(x)=0.
\end{eqnarray}
الحل المنظم يعطي ب
\begin{eqnarray}
\psi_0(x)=(\frac{m\Omega}{\pi\hbar})^{\frac{1}{4}}e^{-\frac{m\Omega}{2\hbar}x^2}.
\end{eqnarray}
يمكن حساب اشعة الحالة
$|n>$ 
بدلالة 
$|0>$
كالتالي
\begin{eqnarray}
&&|1>=a^+|0>\nonumber\\
&&|2>=\frac{a^+}{\sqrt{2}}|1>=\frac{(a^+)^2}{\sqrt{2!}}|0>\nonumber\\
&&|3>=\frac{a^+}{\sqrt{3}}|2>=\frac{(a^+)^3}{\sqrt{3!}}|0>\nonumber\\
&&.\nonumber\\
&&.\nonumber\\
&&|n>=\frac{a^+}{\sqrt{n}}|n-1>=\frac{(a^+)^n}{\sqrt{n!}}|0>.
\end{eqnarray}

\subsection*{
كمون دالة دلتا
}
\addcontentsline{toc}{subsection}{
كمون دالة دلتا
} 
يعطي الكمون في هذه الحالة ب
\begin{eqnarray}
V(x)=-\alpha \delta (x).
\end{eqnarray}
الثابت 
 $\alpha$
 موجب. معادلة شرودينغر غير المتعلقة بالزمن تكتب علي الشكل
\begin{eqnarray}
-\frac{\hbar^2}{2m}\frac{d^2\psi}{dx^2}-\alpha\delta (x)\psi=E\psi.
\end{eqnarray}
\paragraph{
الحالات المرتبطة
($E\leq 0$):
} 
نعرف
\begin{eqnarray}
\kappa=\frac{\sqrt{-2mE}}{\hbar}.
\end{eqnarray}
معادلة شرودينغر غير المتعلقة بالزمن تصبح
\begin{eqnarray}
\frac{d^2\psi}{dx^2}+\frac{2m\alpha}{\hbar^2}\delta (x)\psi=\kappa^2\psi.
\end{eqnarray}
من اجل
$x< 0$
او
$x>0$
لدينا
\begin{eqnarray}
\frac{d^2\psi}{dx^2}=\kappa^2\psi.
\end{eqnarray}
الحل من اجل
$x<0$
يأخذ الشكل
\begin{eqnarray}
\psi(x)=Ae^{-\kappa x}+Be^{\kappa x}.
\end{eqnarray}
في النهاية
$x\longrightarrow -\infty$ 
الحل اعلاه ينفجر ما لم ينعدم
$A$.
اذن يجب ان يكون لدينا
\begin{eqnarray}
\psi(x)=Be^{\kappa x}~,~x<0.
\end{eqnarray}
بالمثل فان الحل من اجل
$x>0$ 
يأخذ الشكل
\begin{eqnarray}
\psi(x)=Fe^{-\kappa x}+Ge^{\kappa x}.
\end{eqnarray}
الان في النهاية
$x\longrightarrow \infty$
الحل ينفجر مالم ينعدم
$G$.
اذن يجب ان يكون لدينا
\begin{eqnarray}
\psi(x)=Fe^{-\kappa x}~,~x>0.
\end{eqnarray}
الدالة الموجية هي دائما مستمرة بينما مشتقتها الاولي
$d\psi(x)/dx$ 
هي دائما مستمرة باستثناء في النقاط التي يتباعد فيها الكمون. اذن من الشرط الحدي الاول نحصل علي
\begin{eqnarray}
F=B.
\end{eqnarray}
لدينا اذن النتيجة
\begin{eqnarray}
&&\psi(x)=Be^{+\kappa x}~,~x\leq 0\nonumber\\
&&\psi(x)=Be^{-\kappa x}~,~x\geq 0.
\end{eqnarray}
نكامل الان طرفي معادلة شرودينغر بين
 $-\epsilon$ 
 و
 $+\epsilon$.
 لدينا
\begin{eqnarray}
\int_{-\epsilon}^{\epsilon}dx\frac{d^2\psi}{dx^2}=-\frac{2m\alpha}{\hbar^2}\int_{-\epsilon}^{\epsilon}dx \delta (x)\psi+\kappa^2\int_{-\epsilon}^{\epsilon}dx\psi.
\end{eqnarray}
نحصل في النهاية
$\epsilon\longrightarrow 0$
علي النتيجة
\begin{eqnarray}
\frac{d\psi}{dx}|_{+\epsilon}-\frac{d\psi}{dx}|_{-\epsilon}=-\frac{2m\alpha}{\hbar^2}\psi(0).\label{discontinuity_psi}
\end{eqnarray}
بعبارة اخري فان المشتقة الاولي للدالة الموجية غير مستمرة في النقطة
$x=0$
حيث يتباعد الكمون. المعادلة اعلاه تعطي النتيجة
\begin{eqnarray}
-2B\kappa=-\frac{2m\alpha}{\hbar^2}B.
\end{eqnarray}
اذن
\begin{eqnarray}
\kappa=\frac{m\alpha}{\hbar^2}.
\end{eqnarray}
طاقة الحالة المرتبطة هي اذن معطاة ب
\begin{eqnarray}
E=-\frac{\hbar^2\kappa^2}{2m}=-\frac{m\alpha^2}{2\hbar^2}.
\end{eqnarray}
تنظيم الدالة الموجية
$\psi(x)$
يعطي
$B=\sqrt{\kappa}$. 
الدالة الموجية للحالة المرتبطة تعطي اذن ب
\begin{eqnarray}
\psi(x)=\sqrt{\kappa}e^{-\kappa |x|}.
\end{eqnarray}
\paragraph{
حالات التصادم
($E\geq 0$):} 
نعرف
\begin{eqnarray}
k=\frac{\sqrt{2mE}}{\hbar}.
\end{eqnarray}
\begin{eqnarray}
\frac{d^2\psi}{dx^2}+\frac{2m\alpha}{\hbar^2}\delta (x)\psi=-k^2\psi.
\end{eqnarray}
الحل من اجل
$x<0$
هو من الشكل
\begin{eqnarray}
\psi(x)=Ae^{ik x}+Be^{-ik x}.
\end{eqnarray}
الحل من اجل
$x>0$
هو من الشكل
\begin{eqnarray}
\psi(x)=Fe^{ik x}+Ge^{-ik x}.
\end{eqnarray}
من شرط استمرارية دالة الموجة نحصل علي
\begin{eqnarray}
A+B=F+G.
\end{eqnarray}
نحسب المشتقات الاولي
\begin{eqnarray}
\frac{d\psi}{dx}|_{+\epsilon}=ik(F-G)~,~\frac{d\psi}{dx}|_{-\epsilon}=ik(A-B).
\end{eqnarray}
من الشرط
$(\ref{discontinuity_psi})$
نحصل علي
\begin{eqnarray}
ik(F-G-A+B)=-\frac{2m\alpha}{\hbar^2}(A+B).
\end{eqnarray}
بالمقابل
\begin{eqnarray}
F-G=(1+2i\beta)A-(1-2i\beta)B~,~\beta=\frac{m\alpha}{\hbar^2k}.
\end{eqnarray}
الثوابت 
$A$
و
$F$
هي سعات الامواج المنتشرة الي اليمين بينما
$B$
و
$G$
هي سعات الامواج المنتشرة الي اليسار. في تجربة تصادم معينة فان الجسيمات تأتي من جهة واحدة مثلا من اليسار. في هذه
الحالة
$A$ 
يقابل الموجة الواردة, 
$B$
يقابل الموجة المنعكسة و
$F$
يقابل الموجة المرسلة
اي المنكسرة بينما
$G=0$. 
نعتبر اذن
\begin{eqnarray}
G=0~,~{\rm scattering}~{\rm from}~{\rm left}.
\end{eqnarray} 
نحصل علي المعاملات
\begin{eqnarray}
B=\frac{i\beta}{1-i\beta}A~,~F=\frac{1}{1-i\beta}A.
\end{eqnarray} 
اذن نحصل علي الدوال الموجية
\begin{eqnarray}
&&\psi_{\rm incid}=Ae^{ikx}\nonumber\\
&&\psi_{\rm refle}=\frac{i\beta}{1-i\beta}Ae^{-ikx}\nonumber\\
&&\psi_{\rm trans}=\frac{1}{1-i\beta}Ae^{ikx}.
\end{eqnarray}
الدوال الموجية الكلية تعطي اذن ب
\begin{eqnarray}
\psi(x)=\psi_{\rm incid}(x)+\psi_{\rm refle}(x)~,~x<0.
\end{eqnarray}
\begin{eqnarray}
\psi(x)=\psi_{\rm trans}(x)~,~x>0.
\end{eqnarray}
نذكر ان
$|\psi(x)|^2$ 
هو احتمال ايجاد الجسيم في النقطة 
$x$. 
بعبارة اخري اذا كان لدينا عدد ضخم من الجسيمات كلها في نفس الحالة
$\psi(x)$
فان الكمية
$|\psi(x)|^2$
تقيس عدد الجسيمات التي توجد في النقطة
 $x$. 
 بالتالي
 $\int dx |\psi_{\rm incid}|^2=|A|^2\int dx$, $\int dx |\psi_{\rm refle}|^2=|B|^2\int dx$ 
 و
$\int dx |\psi_{\rm trans}|^2=
|F|^2\int dx$
هي اعداد الجسيمات الواردة, المنعكسة و المنكسرة علي التوالي التي لها طاقة
$E$. 
رغم ان هذه الاعداد غير منتهية,  لان الدوال الموجية
$\psi_{\rm incid}$, $\psi_{\rm refle}$
و
$\psi_{\rm trans}$
غير قابلة للتنظيم, فان نسبها منتهية.

اذن الاحتمال النسبي لجسيم وارد ان ينعكس يعطي بأخذ نسبة عددالجسيمات الواردة لعدد الجسيمات المنعكسة اي 

\begin{eqnarray}
R=\frac{|B|^2}{|A|^2}=\frac{\beta^2}{1+\beta^2}=\frac{1}{1+\frac{2\hbar^2E}{m\alpha^2}}.
\end{eqnarray}
هذا يسمي معامل الانعكاس. بالمثل فان الاحتمال النسبي لجسيم وارد ان ينكسر يعطي
بأخذ نسبة عدد الجسيمات الواردة لعدد الجسيمات المنكسرة اي
\begin{eqnarray}
T=\frac{|F|^2}{|A|^2}=\frac{1}{1+\beta^2}=\frac{1}{1+\frac{m\alpha^2}{2\hbar^2E}}.
\end{eqnarray}
هذا يسمي معامل الانكسار او الارسال. لدينا
\begin{eqnarray}
R+T=1.
\end{eqnarray}
لنلاحظ انه لما
$E\longrightarrow \infty$
فان
$R\longrightarrow 0$
و
$T\longrightarrow 1$.
اي ان الجسيم الذي له طاقة كافية احتمال مروره عير الكمون اكبر من احتمال انعكاسه.

الدوال الموجية
$\psi_{\rm incid}$, $\psi_{\rm refle}$
و
$\psi_{\rm trans}$
ليست فيزيائية لانها دوال غير قابلة للتنظيم. يجب تعويض هذه الدوال بدوال قابلة للتنظيم, عبارة عن حزم موجية مثل ما فعلنا في حالة الجسيم الحر, و هذا يؤدي بالضرورة الي تعويض الطاقة
$E$
بمجال من القيم المسموحة للطاقة. نعتبر اذن حزم موجية مركزة حول القيمة
 $k$
 للعدد الموجي كي تكون الطاقة مركزة حول
 القيمة
 $E$.
 الحزم الموجية الواردة, المنعكسة و المنكسرة
 يجب ان تحقق نفس الشروط الحدية التي تحققها
 $\psi_{\rm incid}$, $\psi_{\rm refle}$
 و
 $\psi_{\rm trans}$
 علي التوالي. 
 التحليل الذي قمنا به اعلاه بالنسبة ل
 $\psi_{\rm incid}$, $\psi_{\rm refle}$
 و
 $\psi_{\rm trans}$
 يبقي صالحا بالكامل بالنسبة لهذه الحزم الموجية اما
 $R$
 و
 $T$
 فيصبح لهما تفسير
 معاملي الانعكاس والانكسار للجسيمات ذات الطاقة
 $E$.

\subsection*{
الكمون المربع
}
\addcontentsline{toc}{subsection}{
الكمون المربع
} 

نعتبر الان الكمون
\begin{eqnarray}
&&V=-V_0~,~-a<x<a\nonumber\\
&&V=0~,~|x|>0.
\end{eqnarray}
\paragraph{
الحالات المرتبطة
($E<0$):
} 
نعرف
\begin{eqnarray}
\kappa=\frac{\sqrt{-2mE}}{\hbar}.
\end{eqnarray}
لدينا ثلاث مناطق. المنطقة الاولي توافق 
$x<-a$
بينما توافق المنطقة الثالثة
$x>a$. 
في هاته المنطقتين تكتب معادلة شرودينغر علي الشكل
\begin{eqnarray}
\frac{d^2\psi}{dx^2}=\kappa^2\psi.
\end{eqnarray}
الحل العام هو
\begin{eqnarray}
\psi(x)=Ae^{-\kappa x}+Be^{\kappa x}.
\end{eqnarray}
من الواضح ان الحل في المنطقة الاولي هو
\begin{eqnarray}
\psi_{I}(x)=Be^{\kappa x}~,~x<-a.
\end{eqnarray}
بالمثل الحل في المنطقة الثالثة هو
\begin{eqnarray}
\psi_{III}(x)=Fe^{-\kappa x}~,~x>a.
\end{eqnarray}
طاقة اي حل قابل للتنظيم لمعادلة شرودينغر يجب ان تكون اكبر او تساوي من القيمة الاصغرية للكمون. في هذه الحالة هذا يعني ان
 $E>-V_0$.
 اذن في المنطقة الثانية
 اي من اجل
 $-a<x<a$
 تكتب معادلة شرودينغر علي الشكل
\begin{eqnarray}
\frac{d^2\psi}{dx^2}=-l^2\psi~,~l=\frac{\sqrt{2m(E+V_0)}}{\hbar}.
\end{eqnarray}
الحل يعطي ب
\begin{eqnarray}
\psi_{II}(x)=C\sin lx+ D\cos lx.
\end{eqnarray}
لان الكمون زوجي يمكن ان نفترض ان الدالة الموجية اما زوجية او فردية. بافتراض انها زوجية لدينا مباشرة
$C=0$.
نحصل علي
\begin{eqnarray}
\psi_{II}(x)=D\cos lx.
\end{eqnarray}
الشروط الحدية
$\psi_{I}(-a)=\psi_{II}(-a)$, $\psi_{II}(a)=\psi_{III}(a)$ 
تؤدي الي المعادلات
\begin{eqnarray}
B=F.
\end{eqnarray}
\begin{eqnarray}
Be^{-\kappa a}=D\cos la.
\end{eqnarray}
الشروط الحدية
$\psi_{I}^{'}(-a)=\psi_{II}^{'}(-a)$, $\psi_{II}^{'}(a)=\psi_{III}^{'}(a)$ 
تؤدي الي المعادلات
\begin{eqnarray}
\kappa Be^{-\kappa a}=Dl\sin la.
\end{eqnarray}
اذن الطاقات المسموح بها يجب ان تحقق الشرط
\begin{eqnarray}
\tan la =\frac{\kappa}{l}.
\end{eqnarray}
نعرف
\begin{eqnarray}
z=la~,~z_0=\frac{a}{\hbar}\sqrt{2mV_0}.
\end{eqnarray}
نلاحظ ان
$\kappa^2+l^2=2mV_0/\hbar^2$
و بالتالي
$a^2\kappa^2=z_0^2-z^2$.
اذن
\begin{eqnarray}
\tan z=\sqrt{\frac{z_0^2}{z^2}-1}.
\end{eqnarray}
يجب حل هذه المعادلة المتسامية من اجل المجهول
$z$ 
المكافئ للطاقة
$E$ 
بدلالة
$z_0$ 
الذي يقيس حجم البئر.

من اجل بئر عميقة اي
$z_0\longrightarrow \infty$
لدينا 
$\tan z\longrightarrow \infty$.
 بالتالي
 $z=n\pi/2$
 حيث
 $n$
 فردي. اذن في هذه الحالة نقاط تقاطع الدالتين
 $\tan z$
 و
 $\sqrt{\frac{z_0^2}{z^2}-1}$
 تقع في

\begin{eqnarray}
z_n=n\frac{\pi}{2}\leftrightarrow E_n^{'}=E_n+V_0=\frac{\hbar^2\pi^2n^2}{2m(2a)^2}.
\end{eqnarray}
من اجل
$V_0$
منته فان هناك عدد منته من الحلول. في النهاية
 $V_0\longrightarrow\infty$ 
 القيم
  $E_n^{'}$ 
  تصبح طاقات الكمون المربع اللانهائي.

من اجل كمون ضحل و ضيق فانه يكون لدينا عدد اقل من الحالات المرتبطة. بالفعل من اجل كل القيم
$z_0$
التي هي اقل من
$\pi/2$
مهما كانت صغيرة فانه يكون لدينا حالة مرتبطة وحيدة.

\paragraph{
حالات التصادم
($E>0$):}
نعرف
\begin{eqnarray}
k=\frac{\sqrt{2mE}}{\hbar}.
\end{eqnarray}
لدينا الحلول
\begin{eqnarray}
&&\psi_{I}(x)=Ae^{ikx}+Be^{-ikx}~,~x<-a\nonumber\\
&&\psi_{II}(x)=C\sin lx +D\cos lx~,~-a<x<a\nonumber\\
&&\psi_{III}(x)=Fe^{ikx}~,~x>a.
\end{eqnarray}
في المناطق الاولي و الثالثة الجسيم حر. الموجة الواردة متناسبة مع
 $A$, 
 الموجة المنعكسة متناسبة مع
  $B$
  و
  الموجة المنكسرة (المرسلة) متناسبة مع
  $F$.
  استمرارية الدالة الموجية في النقاط
  $x=\pm a$
  يعطي المعادلات

\begin{eqnarray}
&&Ae^{-ika}+Be^{ika}=-C\sin la+D\cos la\nonumber\\
&&Fe^{ika}=C\sin la +D\cos la.\label{hg1}
\end{eqnarray}
استمرارية المشتقة الاولي للدالة الموجية في
$x=\pm a$
تؤدي الي المعادلات
\begin{eqnarray}
&&ik(Ae^{-ika}-Be^{ika})=l(C\cos la+D\sin la)\nonumber\\
&&ik(Fe^{ika})=l(C\cos la -D\sin la).\label{hg2}
\end{eqnarray}
نستعمل المعادلة الثانية من
$(\ref{hg1})$
و المعادلة الثانية من
$(\ref{hg2})$
لايجاد
\begin{eqnarray}
C=(\sin la+\frac{ik}{l}\cos la)e^{ika}F~,~D=(\cos la-\frac{ik}{l}\sin la )e^{ika}F.
\end{eqnarray}
نعوض هذه العبارات في المعادلة الاولي من
$(\ref{hg1})$
و المعادلة الاولي من
$(\ref{hg2})$
لايجاد
\begin{eqnarray}
&&Ae^{-ika}+Be^{ika}=(\cos 2la-\frac{ik}{l}\sin 2la)e^{ika}F\nonumber\\
&&Ae^{-ika}-Be^{ika}=(\cos 2la-\frac{il}{k}\sin 2la)e^{ika}F.
\end{eqnarray}
اذن
\begin{eqnarray}
F=\frac{e^{-2ika}}{\cos 2la -i\frac{k^2+l^2}{2kl}\sin 2l}A.
\end{eqnarray}
\begin{eqnarray}
B=i\frac{l^2-k^2}{2kl}\sin 2la F.
\end{eqnarray}
معامل الانكسار او الارسال هو
\begin{eqnarray}
T=\frac{|F|^2}{|A|^2}=\frac{1}{\cos^2 2la+(\frac{k^2+l^2}{2kl})^2\sin^2 2la}.
\end{eqnarray}
معامل الانعكاس هو
\begin{eqnarray}
R=\frac{|B|^2}{|A|^2}=\frac{(\frac{k^2-l^2}{2kl})^2\sin^2 2la}{\cos^2 2la+(\frac{k^2+l^2}{2kl})^2\sin^2 2la}.
\end{eqnarray}
نتحقق من ان
\begin{eqnarray}
R+T=1.
\end{eqnarray}
\newpage
\section*{
تمارين
}
\addcontentsline{toc}{section}{
{\bf 
تمارين
}} 
\paragraph{
تمرين
$1$:
}
\begin{itemize}
\item[$(1$]
ليكن
 $|f>$
 و
$|g>$
شعاعي حالة في فضاء هيلبرت
${\cal H}$.
برهن علي صحة متراجحة شوارز
\begin{eqnarray}
|<f|g>|^2\leq ~<f|f><g|g>.\nonumber
\end{eqnarray}
\item[$(2$]
بين باستعمال متراجحة شوارز ان الجداء الداخلي
$<f|g>$ 
موجود.
\item[$(3$]
فضاء هيلبرت هو فضاء مركب. بالتعريف الفضاء الشعاعي هو فضاء مغلق تحت تأثير الجمع الشعاعي و الضرب السلمي. اذن اذا كان
 $|f>$
 و
 $|g>$
  اي شعاعي حالة في فضاء هيلبرت فان المجموع
  $|h>=|f>+|g>$
  هو ايضا شعاع حالة في فضاء هيلبرت. بين انه اذا كانت الدالتين الموجيتين 
  $f(x)$
  و
  $g(x)$
  قابلتين للتنظيم فان 
  $h(x)$
  هي ايضا دالة موجية قابل للتنظيم.
\end{itemize}
\paragraph{
تمرين
$2$:
}
\begin{itemize}
\item[$(1$]
مبدأ الارتياب يعطي ب
\begin{eqnarray}
\sigma_A^2\sigma_B^2\geq \frac{1}{4}|<[\hat{A},\hat{B}]|>|^2.\nonumber
\end{eqnarray}
بين ان الشرط الضروري و الكافي من اجل صحة المتراجحة اعلاه يعطي ب
\begin{eqnarray}
(\hat{B}-<\hat{B}>)|\psi>=ia(\hat{A}-<\hat{A}>)|\psi>.\nonumber
\end{eqnarray}
الحالة الموصوفة بالشعاع
$|\psi>$
هي اذن حالة ذات ارتياب اصغري.
\item[$(2$] 
جد حل للشرط اعلاه من اجل
$\hat{A}=\hat{x}$
و
$\hat{B}=\hat{p}$. 
نطم الدالة الموجية المحصل عليها لواحد. استعمل
\begin{eqnarray}
\int_{-\infty}^{\infty}dx e^{-x^2}=\sqrt{\pi}.\nonumber
\end{eqnarray}
\item[$(3$] 
جد الدالة الموجية المقابلة في فضاء كميات الحركة.
\item[$4)$]
من اجل التبسيط نعتير 
$<\hat{x}>=0$.
عين متي تكون كمية حركة الجسيم معرفة جيدا و متي يكون الجسيم متموضعا جيدا في فضاء المواضع.
\end{itemize}
\paragraph{
تمرين
$3$:
}
\begin{itemize}
\item[$(1$]
علم الدالة الموجية
$\psi(t,x)$
ب
$\psi(t,x)=\sqrt{\rho}e^{\frac{iS}{\hbar}}$. 
بالتعويض بهذا الاقتراح في معادلة شرودينغر نحصل علي معادلة الاستمرارية
\begin{eqnarray}
\frac{\partial \rho}{\partial t}+\frac{\partial j }{\partial x}=0.\nonumber
\end{eqnarray}
اكتب تيار الاحتمال
$j$
بدلالة
$\rho$
و
$S$
و ايضا
$\psi$ 
و
$\psi^*$. 
\item[$(2$] 
باستعمال معادلة الاستمرارية تحقق من قانون انحفاظ الاحتمال المعطي ب
\begin{eqnarray}
\frac{dP}{dt}=0~,~P=\int_{-\infty}^{\infty}\rho(t,x)dx.\nonumber
\end{eqnarray}
\item[$(3$]
اربط بين
$<\hat{p}>$
و
$\frac{d<\hat{x}>}{dt}$.
ما هو معني تيار الاحتمال 
 $j$.
\item[$(4$]
نفترض ان الكمون
$V$
مركب اي
 $V=V_0-i\Gamma$.
 عين في هذه الحالة معدل تغير الاحتمال الكلي 
 $\frac{dP}{dt}$.
 ماذا يصف
 $P$.
\item[$(5$]
ما هي المعادلة التي يحققها
$S$ 
و ما هي نهايتها الكلاسيكية
$\hbar\longrightarrow 0$.
ما هو معني
$S$.
\end{itemize}

\paragraph{
تمرين
$4$:
}

\begin{itemize}
\item[$(1$]
نعتبر جسيم ذو سبين 
$1/2$. 
الحالة الذاتية العليا للمؤثرات
$\hat{S}^2$ 
و
$\hat{S}_3$
يرمز لها ب
$|+>\equiv|\frac{1}{2}\frac{1}{2}>$
اما الحالة الذاتية الدنيا فيرمز لها ب
 $|->\equiv|\frac{1}{2}-\frac{1}{2}>$.
 اكتب المؤثرات
 $\hat{S}^2$, $\hat{S}_3$
 و
 $\hat{S}_{\pm}$
 في هذا الاساس. عبر عن
$\hat{S}_i$
بدلالة مصفوفات باولي
\begin{eqnarray}
{\sigma}_1=\left(\begin{array}{cc}
0 &1\\
1 &0
\end{array}\right)~,~{\sigma}_2=\left(\begin{array}{cc}
0 &-i\\
i &0
\end{array}\right)~,~{\sigma}_3=\left(\begin{array}{cc}
1 &0\\
0 &-1
\end{array}\right).\nonumber
\end{eqnarray}
\item[$(2$]
ما هي القيم التي نحصل عليها و ما هي احتمالاتها
اذا قسنا السبين
$\hat{S}_3$
في حالة عامة للجسيم.
\item[$(3$]
ما هي القيم التي نحصل عليها و ما هي احتمالاتها
اذا قسنا السبين
$\hat{S}_1$
في حالة عامة للجسيم.
\item[$(4$]
اذا افترضنا ان الجسيم في اللحظة الابتدائية موجود في الحالة
$|+>$.
ماهي نتائج قياس السبين 
$\hat{S}_1$
و ماهي احتمالاتها. اذا ادي القياس  للنتيجة
$+\hbar/2$
ما هي حالة الجسيم بعد القياس. ما هي نتائج قياس السبين
$\hat{S}_3$
الذي نجريه مباشرة بعد القياس السابق و ماهي احتمالاتها.
\end{itemize}
\paragraph{
تمرين
$5$:
}
نعتبر جملة مشكلة من جسيمين سبينهما 
$1/2$
مثل الالكترون و البروتون في الحالة الاساسية لذرة الهيدروجين. ما هو العزم الحركي الكلي للجملة. انشئ فضاء هيلبرت هذه الجملة.

\paragraph{
تمرين
$6$:
}
نعتبر هزازان توافقيان مستقلان بمؤثرات احداث و تدمير
$a_+,a_+^+$ 
و
$a_-,a_-^+$,
اي
$[a_+,a_+^+]=1$, $[a_-,a_-^+]=1$
و
$[a_+,a_-^+]=0$.
مؤثرات العدد الفردية تعطي ب
$N_+=a_+^+a_+$, $N_-=a_-^+a_-$
اما مؤثر العدد الكلي فيعطي ب
$N=N_++N_-$.
فضاء هيلبرت الكلي هو الجداء التنسوري للفضاءات الهيلبرتية الفردية. اذن اذا كان
$\{|n_+>\}$
هو اساس فضاء هيلبرت للهزاز التوافقي الاول و
$\{|n_->\}$
هو اساس فضاء هيلبرت للهزاز التوافقي الثاني فان
$\{|n_+>|n_->\}$
هو اساس فضاء هيلبرت الكلي. نعرف
\begin{eqnarray}
J_+=\hbar a_+^+a_-~,~J_-=\hbar a_-^+a_+~,~J_3=\frac{\hbar}{2}(a_+^+a_+-a_-^+a_-).\nonumber
\end{eqnarray} 
نحقق من ان
$J_{\pm}$
و
$J_3$ 
تحقق علاقات العزم الحركي. احسب مربع العزم الحركي
$J^2$
بدلالة مؤثر العدد الكلي
$N=N_++N_-$. 
كيف تؤثر 
$J_{\pm}$, $J_3$
و
 $J^2$
 علي الاساس
 $|n_1,n_2>$. 
 جد العلاقة بين الاعداد الكمية
$n_+$
و
$n_-$
من جهة و الاعداد الكمية
 $j$ 
 و
  $m$
 من جهة اخري. ماذا تلاحظ بالنسبة للمجموع
 $n_++n_-$.
 اكتب
 $|j,m>$
 بدلالة
 مؤثرات الانشاء
 $a_+^+$
 و
 $a_-^+$.

\paragraph{
تمرين
$7$:
}
\begin{itemize}
\item[$(1$]
 نقول عن العبارتين
 $D_1(x)$ 
 و
 $D_2(x)$, 
 اللتان تتعلقان بدالة ديراك دلتا, انهما متساويتان اذا تحقق الشرط 
\begin{eqnarray}
\int_{-\infty}^{\infty}dx f(x)D_1(x)=\int_{-\infty}^{\infty}dx f(x)D_2(x).\nonumber
\end{eqnarray}
بين ان
\begin{eqnarray}
\delta(cx)=\frac{1}{|c|}\delta(x).\nonumber
\end{eqnarray}
\item[$(2$]
الدالة الخطوة
$\theta(x)$
تعرف ب
\begin{eqnarray}
&&\theta(x)=1~,~x>0\nonumber\\
&&\theta(x)=0~,~x<0.\nonumber
\end{eqnarray}
بين ان
\begin{eqnarray}
\frac{d\theta}{dx}=\delta(x).\nonumber
\end{eqnarray}
\end{itemize}
\paragraph{
تمرين
$8$:
}
مبرهنة بلانشارل
\footnote{.${\rm Plancherl}$}
تعطي ب
\begin{eqnarray}
f(x)=\frac{1}{\sqrt{2\pi}}\int_{-\infty}^{\infty}F(k)e^{ikx}dk\leftrightarrow F(k)=\frac{1}{\sqrt{2\pi}}\int_{-\infty}^{\infty}f(x)e^{-ikx}dx.\nonumber
\end{eqnarray}
بين ان
\begin{eqnarray}
\delta(x)=\frac{1}{2\pi}\int_{-\infty}^{\infty}e^{ikx} dk.\nonumber
\end{eqnarray}
\paragraph{
تمرين
$9$:
}
نعتبر كمون دالة ديراك. في المنطقتين
 $I$ ($x<0$) 
 و
 $II$ ($x>0$)
 تعطي الدوال الموجية ب
\begin{eqnarray}
&&\psi_I(x)=Ae^{ikx}+Be^{-ikx}~,~x<0\nonumber\\
&&\psi_{II}(x)=Fe^{ikx}+Ge^{-ikx}~,~x>0.\nonumber
\end{eqnarray}
 الشروط الحدية عند
$x=0$
تعطي ب
\begin{eqnarray}
&&F+G=A+B\nonumber\\
&&F-G=A(1+2i\beta)-B(1-2i\beta).\nonumber
\end{eqnarray}
اذن يمكن ان نجد ثابتين بدلالة الثابتين الاخرين. لدينا
\begin{eqnarray}
\beta=\frac{m\alpha}{\hbar^2k}~,~k=\frac{\sqrt{2mE}}{\hbar}.\nonumber
\end{eqnarray}
\begin{itemize}
\item[$(1$]
احسب مصفوفة التصادم
$S$
المعرفة ب
\begin{eqnarray}
\left(\begin{array}{c}
B\\
F
\end{array}\right)=\left(\begin{array}{cc}
S_{11}&S_{12}\\
S_{21}&S_{22}
\end{array}\right)\left(\begin{array}{c}
A\\
G
\end{array}\right).\nonumber
\end{eqnarray}
هذا يعطي السعات الصادرة
$B$ 
و
 $F$,
 اي التي تتحرك بعيدا عن الكمون, بدلالة السعات الواردة 
 $A$ 
 و
  $G$,
  اي التي تتحرك نحو الكمون.
\item[$(2$]
احسب مصفوفة التحويل
$T$
المعرفة ي
\begin{eqnarray}
\left(\begin{array}{c}
F\\
G
\end{array}\right)=\left(\begin{array}{cc}
T_{11}&T_{12}\\
T_{21}&T_{22}
\end{array}\right)\left(\begin{array}{c}
A\\
B
\end{array}\right).\nonumber
\end{eqnarray}
هذا يعطي السعات علي يمين الكمون
$F$ 
و
$G$ 
بدلالة السعات علي يساره
$A$
و
$B$.
\item[$(3$] 
ناقش مصفوفة التصادم و مصفوفة التحويل من اجل كمون كيفي ينعدم لما
$x\longrightarrow\pm \infty$.
\item[$(4$] 
من اجل التصادم من اليسار اكتب معاملات الانكسار و الانعكاس بدلالة
 $S_{ij}$ 
 و
 $T_{ij}$.
\item[$(5$]
بين انه من اجل كمون مشكل من قطعتين غير متصلتين فان مصفوفة التحويل تحقق
\begin{eqnarray}
T=T_2T_1.\nonumber
\end{eqnarray}
 $T_i$ 
 هي مصفوفة التحويل من اجل القطعة
  $i$
  علي حدة.
\end{itemize}

\newpage
\section*{
حلول
}
\addcontentsline{toc}{section}{
{\bf 
حلول
}} 
\paragraph{
تمرين
$1$:
}
\begin{itemize}
\item[$(1$]
نعتبر طويلة الشعاع
$|\psi>=|f>+a|g>$
مع
$a=-<g|f>/<g|g>$.
\item[$(2$] 
باستعمال متراجحة شوارز لدينا
$|<f|g>|\leq \sqrt{<f|f><g|g>}$. 
التكاملات
$<f|f>=\int dx f^*(x)f(x)$
و
$<g|g>=\int dx g^*(x)g(x)$
تقترب من اعداد منتهية لان كل من
$f(x)$ 
و
$g(x)$
 يحقق شرط قابلية التكامل للمربع. اذن الجداء الداخلي 
 $<f|g>=\int dx f^*(x)g(x)$
 يقترب من عدد منته.
\item[$(3$] 
يجب ان نبين ان
$<h|h>=\int dx h^*(x)h(x)$
يقترب من عدد منته. مرة اخري نستعمل متراجحة شوارز.
\end{itemize}
\paragraph{
تمرين
$2$:
}
\begin{itemize}
\item[$(1$] 
لدينا
\begin{eqnarray}
\sigma_A^2\sigma_B^2=<\psi_A|\psi_A><\psi_B|\psi_B>\geq |<\psi_A|\psi_B>|^2.
\end{eqnarray}
تتحقق المساواة اذا كان
$|\psi_B>=c|\psi_A>$.
بالاضافة لدينا
\begin{eqnarray}
|<\psi_A|\psi_B>|^2&=&\frac{1}{4}|<\psi|[\hat{A},\hat{B}]|\psi>|^2+\frac{1}{4}|<\psi|[\hat{A}-<\hat{A}>,\hat{B}-<\hat{B}>]|\psi>|^2\nonumber\\
&\geq &\frac{1}{4}|<\psi|[\hat{A},\hat{B}]|\psi>|^2.
\end{eqnarray}
تتحقق المتراجحة اذا كان
$<\psi|[\hat{A}-<\hat{A}>,\hat{B}-<\hat{B}>]|\psi>=0$. 
هذا يؤدي الي الشرط
$(c+c^{*})<\psi_A|\psi_B>=0$ 
اي ان
$c$
هو عدد تخيلي. نكتب
$c=ia$. 
الشرط الضروري و الكافي هو اذن
\begin{eqnarray}
(\hat{B}-<\hat{B}>)|\psi>=ia(\hat{A}-<\hat{A}>)|\psi>.
\end{eqnarray}
\item[$(2$]
من اجل
 $\hat{A}=\hat{x}$ 
 و
 $\hat{B}=\hat{p}$
 نحصل في اساس الموضع علي المعادلة
\begin{eqnarray}
(\frac{\hbar}{i}\frac{d}{dx}-<\hat{p}>)\psi(x)=ia(x-<\hat{x}>)\psi(x).
\end{eqnarray}
ننزع من
$\psi(x)$
التصرف كموجة مستوية بكتابة
\begin{eqnarray}
\psi(x)=e^{\frac{i<\hat{p}>x}{\hbar}}\phi(x).
\end{eqnarray}
نحصل علي معادلة تفاضلية ل
$\phi(x)$
معطاة ب
\begin{eqnarray}
\frac{d\ln\phi}{dx}=-\frac{a}{\hbar}(x-<\hat{x}>)
\end{eqnarray}
اذن
$\phi$
تعطي ب
\begin{eqnarray}
\phi(x)=Ae^ {-\frac{a}{2\hbar}(x-<\hat{x}>)^2}.
\end{eqnarray}
الحل العام الذي يتميز بارتياب اصغري يعطي ب
\begin{eqnarray}
\psi(x)=Ae^{-\frac{a}{2\hbar}(x-<\hat{x}>)^2} e^{\frac{i<\hat{p}>x}{\hbar}}.
\end{eqnarray}
التنظيم يعطي القيمة
\begin{eqnarray}
A=(\frac{a}{\hbar\pi})^{\frac{1}{4}}.
\end{eqnarray}  
\item[$(3$]
نحسب
\begin{eqnarray}
\tilde{\psi}(k)&=&\int \frac{dx}{\sqrt{2\pi\hbar}}e^{-\frac{ikx}{\hbar}}\psi(x)\nonumber\\
&=&A\int \frac{dx}{\sqrt{2\pi\hbar}}e^{-\frac{a}{2\hbar}[x-\frac{i}{a}(<\hat{p}>-k-ia<\hat{x}>)]^2}e^{-\frac{1}{2a\hbar}(<\hat{p}>-k-ia<\hat{x}>)^2-\frac{a}{2\hbar}<\hat{x}>^2}\nonumber\\
&=&(\frac{1}{\pi a\hbar})^{\frac{1}{4}}e^{-\frac{1}{2a\hbar}(<\hat{p}>-k-ia<\hat{x}>)^2-\frac{a}{2\hbar}<\hat{x}>^2}.
\end{eqnarray}
\item[$(4$]
نعتبر
$<\hat{x}>=0$ 
من اجل التبسيط. كثافة احتمال ايجاد الجسيم بكمية حركة
$k$ 
بارتياب
$<\hat{p}>$
هي غوسية 
\footnote{.${\rm Gaussian}$}
متمركزة في فضاء كمية الحركة حول
$k$
تساوي
$|\tilde{\psi}(k)|^2$. 
عرض هاته الغوسية هو
$d_k^2=a\hbar$
الذي هو متناسب عكسا مع العرض
$d_x^2=\hbar/a$ 
خاصة الغوسية
$\psi(x)$. 

في النهاية
$a\longrightarrow 0$ 
لدينا
$d_k\longrightarrow 0$
اي ان دالة الموجة
$\tilde{\psi}(k)$
تصبح دالة دلتا مركزة علي
 $k$.
 في هذه الحالة
 $d_x\longrightarrow\infty$
 و بالتالي فان دالة الموجة
 $\psi(x)$ 
 هي موجة مستوية ذات كمية حركة
 $k$. 

في النهاية
$a\longrightarrow \infty$
لدينا
$d_x\longrightarrow 0$
و بالتالي
دالة موجة الجسيم
$\psi(x)$
تصبح دالة دلتا مركزة حول 
 $0$
  اي ان الجسيم متموضع بشكل جيد حول النقطة
  $0$. 
 من الجهة الاخري تصبح دالة الموجة في فضاء كمية الحركة
  $\tilde{\psi}(k)$ 
  ثابت مستقل عن
 $k$. 

\end{itemize}
\paragraph{
تمرين
$3$:
}
\begin{itemize}
\item[$(1$]
نحدد دالة الموجة
$\psi(t,x)$
كالتالي
\begin{eqnarray}
\psi(t,x)=\sqrt{\rho}e^{\frac{iS}{\hbar}}.
\end{eqnarray}
سعة الاحتمال المقابل هي
$\rho=\rho(t,x)$
معرفة ب
\begin{eqnarray}
\rho=\psi^*(t,x)\psi(t,x).
\end{eqnarray}
بالتعويض في معادلة شرودينغر نحصل علي
\begin{eqnarray}
\sqrt{\rho}\bigg[\frac{1}{2m}(\frac{\partial S}{\partial x})^2+V+\frac{\partial S}{\partial t}\bigg]-\frac{\hbar^2}{2m}\frac{\partial^2\sqrt{\rho}}{\partial x^2}&=&i\hbar\bigg[\frac{\partial\sqrt{\rho}}{\partial t}+\frac{1}{m}\frac{\partial\sqrt{\rho}}{\partial x}\frac{\partial S}{\partial x}+\frac{1}{2m}\sqrt{\rho}\frac{\partial^2S}{\partial x^2}\bigg]\nonumber\\
&=&\frac{i\hbar}{2\sqrt{\rho}}\bigg[\frac{\partial \rho}{\partial t}+\frac{\partial j }{\partial x}\bigg].\label{schro}
\end{eqnarray}
تيار الاحتمال يعرف ب
\begin{eqnarray}
j=\frac{1}{m}\rho\frac{\partial S}{\partial x}.
\end{eqnarray}
يمكن ان نتحقق من ان
\begin{eqnarray}
j=\frac{\hbar}{2im}[\psi^*\frac{\partial \psi}{\partial x}-\psi\frac{\partial\psi^*}{\partial x}].
\end{eqnarray}
من الواضح ان الطرف الايمن للمعادلة
$(\ref{schro})$
هو عدد تخيلي
بينما الطرف الايسر هو عدد حقيقي. اذن يجب ان ينعدم كل من طرفي هذه المعادلة كل علي حدة. نتحصل بالتالي علي معادلة الاستمرارية
\begin{eqnarray}
\frac{\partial \rho}{\partial t}+\frac{\partial j }{\partial x}=0.
\end{eqnarray}
\item[$(2$]
هذا يعبر عن انحفاظ الاحتمال. بالفعل نحسب
\begin{eqnarray}
\frac{dP}{dt}&=&\int dx \frac{\partial \rho}{\partial t}\nonumber\\
&=&-\int dx\frac{\partial j}{\partial x}\nonumber\\
&=&j(t,-\infty)-j(t,+\infty).
\end{eqnarray}
لان دالة الموجة
$\psi(t,x)$
هي دالة تحقق شرط قابلية التكامل للمربع يجب ان يكون لدينا
$\psi\longrightarrow 0$
و
$\frac{\partial\psi}{\partial x}\longrightarrow 0$ 
لما
$x\longrightarrow\pm\infty$.
اذن
$j(t,x)\longrightarrow 0$
لما
$x\longrightarrow \pm\infty$
اي
$\frac{dP}{dt}=0$.


\item[$(3$]
تيار الاحتمال
$j$
هو, في معني معين, سرعة الجسيم. بالفعل نحسب من جهة
\begin{eqnarray}
\int j(t,x)dx=\frac{<\hat{p}>}{m}.
\end{eqnarray}
نحسب من الجهة الاخري
\begin{eqnarray}
\frac{<\hat{p}>}{m}=\frac{d<\hat{x}>}{dt}.
\end{eqnarray}
هذه المعادلة الاخيرة مثال علي مبرهنة
ايرنفاست
\footnote{.${\rm Ehrenfest's~theorem}$}
التي تنص علي ان القيم المنتظرة للمؤثرات الكمومية تتبع القوانين الكلاسيكية.
\item[$(4$]
في حالة الكمون المركب
$V=V_0-i\Gamma$
نحسب
\begin{eqnarray}
i\hbar\frac{dP}{dt}&=&i\hbar\int \bigg(\frac{\partial \psi^*}{\partial t}.\psi+\psi^*.\frac{\partial \psi}{\partial t}\bigg)dx\nonumber\\
&=&\int \bigg((-V^*\psi^*).\psi+\psi^*.(V\psi)\bigg)dx\nonumber\\
&=&-2i\Gamma P.
\end{eqnarray}
اذن
\begin{eqnarray}
\frac{dP}{dt}&=&-\frac{2\Gamma}{\hbar} P~,~P=P_0e^{-\frac{2\Gamma}{\hbar}t}.
\end{eqnarray}
هذه المعادلة تصف التهافت التلقائي
\footnote{.${\rm spontaneous~decay}$}
لجسيم غير مستقر بعمر
$\tau=\frac{\hbar}{2\Gamma}$.
\item[$(5$]
من خلال وضع الطرف الايسر للمعادلة
$(\ref{schro})$
يساوي صفر نحصل علي معادلة هاميلتون - جاكوبي
\footnote{.${\rm Hamilton-Jacobi~equation}$}
الكمومية. هذه تعطي ب
\begin{eqnarray}
\bigg[\frac{1}{2m}(\frac{\partial S}{\partial x})^2+V+\frac{\partial S}{\partial t}\bigg]-\frac{\hbar^2}{2m}\frac{\partial^2\sqrt{\rho}}{\partial x^2}=0.
\end{eqnarray}
في النهاية
$\hbar\longrightarrow 0$
تصبح هذه المعادلة بالضبط معادلة
هاميلتون - جاكوبي. من الميكانيك التحليلي نعرف ان
$S$
هو الفعل خاصة الجملة و بالتالي
$\frac{\partial S}{\partial x}$ 
هو بالفعل كمية الحركة.

\end{itemize}

\paragraph{
تمرين
$4$:
}
\begin{itemize}
\item[$(1$]
في هذا الاساس
\begin{eqnarray}
|+>=\left(\begin{array}{c}
1\\
0
\end{array}\right)~,~|->=\left(\begin{array}{c}
0\\
1
\end{array}\right).
\end{eqnarray}
نجد

\begin{eqnarray}
\hat{S}^2=\frac{3}{4}\hbar^2\left(\begin{array}{cc}
1 &0\\
0 &1
\end{array}\right)~,~\hat{S}_3=\frac{\hbar}{2}\left(\begin{array}{cc}
1 &0\\
0 &-1
\end{array}\right).
\end{eqnarray}
\begin{eqnarray}
\hat{S}_+=\hbar\left(\begin{array}{cc}
0 &1\\
0 &0
\end{array}\right)~,~\hat{S}_-=\hbar\left(\begin{array}{cc}
0 &0\\
1 &0
\end{array}\right).
\end{eqnarray}
\begin{eqnarray}
\hat{S}_1=\frac{\hbar}{2}\left(\begin{array}{cc}
0 &1\\
1 &0
\end{array}\right)~,~\hat{S}_2=\frac{\hbar}{2}\left(\begin{array}{cc}
0 &-i\\
i &0
\end{array}\right).
\end{eqnarray}
بالتالي
\begin{eqnarray}
\hat{S}_i=\frac{\hbar}{2}{\sigma}_i.
\end{eqnarray}

\item[$(2$] 
الحالة العامة لجسيم ذو سبين نصف هي من الشكل
\begin{eqnarray}
|\chi>=a|+>+b|->=\left(\begin{array}{c}
a\\
b
\end{array}\right).
\end{eqnarray}
شرط التنظيم يكافئي
\begin{eqnarray}
1=|a|^2+|b|^2.
\end{eqnarray}
اذن بقياس
$\hat{S}_3$
سوف نحصل علي
$+\frac{\hbar}{2}$
باحتمال
$|a|^2$
و
 $-\frac{\hbar}{2}$
 باحتمال
 $|b|^2$.
\item[$3)$]
يجب ان نعين الاشعة الذاتية ل
$\hat{S}_1$.
نجد القيم الذاتية
\begin{eqnarray}
+\frac{\hbar}{2}~,~-\frac{\hbar}{2}.
\end{eqnarray}
الاشعة الذاتية المقابلة هي
\begin{eqnarray}
|+>_1=\frac{1}{\sqrt{2}}(|+>+|->)~,~|->_1=\frac{1}{\sqrt{2}}(|+>-|->).
\end{eqnarray}
الحالة العامة
 $|\chi>$
 لجسيم ذو سبين نصف تكتب في الاساس
 $\{|+>_1,|->_1\}$
علي الشكل
 \begin{eqnarray}
|\chi>=\frac{1}{\sqrt{2}}(a+b)|+>_1+\frac{1}{\sqrt{2}}(a-b)|->_1.
\end{eqnarray}
اذن بقياس
$\hat{S}_1$
سوف نحصل علي
$+\frac{\hbar}{2}$ 
باحتمال
$|a+b|^2/2$
و
$-\frac{\hbar}{2}$ 
باحتمال
$|a-b|^2/2$.
\item[$(4$] 
نطبق النتائج السابقة و مبدأ الارتياب.
\end{itemize}
\paragraph{
تمرين
$5$:
}
العزم الحركي المداري لذرة الهيدروجين هو صفر في الحالة الاساسية. العزم الحركي الكلي هو اذن مجموع عزوم السبينات. ليكن
\begin{eqnarray}
\vec{\hat{S}}=\vec{\hat{S}_a}+\vec{\hat{S}_b}.
\end{eqnarray}
كل سبين لديه امكانيتان اما حالة السبين العلوي
$|+>$
او حالة السبين السفلي. لدينا اذن اربع امكانيتان في المجموع هي
\begin{eqnarray}
|+>|+>,|+>|->,|->|+>,|->|->.
\end{eqnarray}
نرمز لهاته الحالات ب
\begin{eqnarray}
|m_1>|m_2>.
\end{eqnarray}
اعلاه
$m_1,m_2=+1/2, -1/2$. 
نلاحظ ان هذه الحالات هي اشعة ذاتية ل
$\hat{S}_3=(\hat{S}_a)_3+(\hat{S}_b)_3$,
اي
\begin{eqnarray}
\hat{S}_3|m_1>|m_2>=\hbar(m_1+m_2)|m_1>|m_2>.
\end{eqnarray}
بعبارة اخري
\begin{eqnarray}
&&\hat{S}_3|+>|+>=\hbar|+>|+>\nonumber\\
&&\hat{S}_3|+>|->=0\nonumber\\
&&\hat{S}_3|->|+>=0\nonumber\\
&&\hat{S}_3|->|->=-\hbar|->|->.
\end{eqnarray}
لدينا حالتان تنعدم فيهما المركبة الثالثة لعزم السبين الكلي. من الواضح انه لا يمكن لهاته الحالتين ان يكون لديهما نفس قيمة عزم السبين
الكلي. حتي نميز بينهما نطبق مؤثر الخفض
 $\hat{S}_-=(\hat{S}_a)_-+(\hat{S}_b)_-$. 
 نجد
\begin{eqnarray}
\hat{S}_-|+>|+>=\hbar(|->|+>+|+>|->).
\end{eqnarray}
\begin{eqnarray}
\hat{S}_-(|->|+>+|+>|->)=2|->|->.
\end{eqnarray}
\begin{eqnarray}
\hat{S}_-|->|->=0.
\end{eqnarray}
هذا يعني ان الحالات
 $|+>|+>$, $(|->|+>+|+>|->)/\sqrt{2}$ 
 و
 $|->|->$
 هي في نفس
 الحالة المتعددة
 \footnote{.${\rm multiplet}$}
  بسبين يساوي
  $s=1$. 
  بعبارة اخري
 \begin{eqnarray}
&&|11>=|+>|+>\nonumber\\
&&|10>=\frac{1}{\sqrt{2}}(|+>|->+|->|+>)\nonumber\\
&&|1-1>=|->|->.
\end{eqnarray}
هذه الحالة المتعددة تسمي بالحالة الثلاثية
\footnote{.${\rm triplet}$}
$s=1$.
الحالة الاخيرة
$(|->|+>-|+>|->)/\sqrt{2}$
توافق سبين
$0$, 
اي
 \begin{eqnarray}
&&|00>=\frac{1}{\sqrt{2}}(|+>|->-|->|+>).
\end{eqnarray}
هذه تسمي بالحالة العازبة
\footnote{.${\rm singlet}$}
$s=0$.

اخيرا نحسب
\begin{eqnarray}
\hat{S}^2=\hat{S}_a^2+\hat{S}_b^2+2(\hat{S}_a)_3(\hat{S}_b)_3+(\hat{S}_a)_+(\hat{S}_b)_-+(\hat{S}_a)_-(\hat{S}_b)_+.
\end{eqnarray}
ايضا نحسب
\begin{eqnarray}
\hat{S}^2|+>|->=\hat{S}^2|->|+>=\hbar^2(|+>|->+|->|+>)
\end{eqnarray}
اذن
\begin{eqnarray}
\hat{S}^2|10>=2\hbar^2|10>~,~\hat{S}^2|00>=0.
\end{eqnarray}
هذا يؤكد ان
$|10>$
لديه سبين
$s=1$ 
و
$|00>$
لديه سبين
$s=0$.
\paragraph{
تمرين
$6$:
}
نجد
\begin{eqnarray}
[J_3,J_{\pm}]=\pm \hbar J_{\pm}~,~[J_+,J_-]=2\hbar J_3.
\end{eqnarray}
\begin{eqnarray}
J^2=\frac{\hbar^2}{2}N(\frac{N}{2}+1).
\end{eqnarray}
التالي نحسب
\begin{eqnarray}
&&J_+|n_+,n_->=\hbar\sqrt{n_-(n_++1)}|n_++1,n_--1>\nonumber\\
&&J_-|n_+,n_->=\hbar\sqrt{n_+(n_-+1)}|n_+-1,n_-+1>\nonumber\\
&&J_3|n_+,n_->=\hbar\frac{n_+-n_-}{2}|n_+,n_->.
\end{eqnarray}
لدينا العلاقات
\begin{eqnarray}
j=\frac{n_++n_-}{2}~,~m=\frac{n_+-n_-}{2}.
\end{eqnarray}
المجموع
$n_++n_-$
دائما مثبت.

لدينا
\begin{eqnarray}
|n_+,n_->\equiv |j,m>&=&\frac{(a_+^+)^{n_+}}{\sqrt{n_+!}}\frac{(a_-^+)^{n_-}}{\sqrt{n_-!}}|0>|0>\nonumber\\
&=&\frac{(a_+^+)^{j+m}}{\sqrt{(j+m)!}}\frac{(a_-^+)^{j-m}}{\sqrt{(j-m)!}}|0>|0>.
\end{eqnarray}
القيمة الذاتية
$n_+$ 
يمكن فهمها علي انها عدد الجسيمات ذات السبين
$1/2$ 
التي هي في حالة السبين العلوي بينما
$n_-$ 
هي عدد الجسيمات ذات السبين 
$1/2$ 
التي هي في حالة السبين السفلي التي تشكل مع بعضها البعض الحالة ذات السبين
 $j$ 
 المعطاة ب
 $|n_+,n_->\equiv |j,m>$.

\paragraph{
تمرين
$7$:
}
\begin{itemize}
\item[$(1$]
نحسب
\begin{eqnarray}
\int_{-\infty}^{\infty} f(x) \delta (cx) dx&=&\int_{-\infty}^{\infty}f(\frac{y}{c})\delta (y)\frac{dy}{|c|}\nonumber\\
&=&\frac{f(0)}{|c|}\nonumber\\
&=&\frac{1}{|c|}\int_{-\infty}^{\infty} f(x) \delta (x) dx.
\end{eqnarray}
القيمة المطلقة تأتي من اشارة التكامل. نستنتج مباشرة ان
$\delta(cx)=\frac{1}{|c|}\delta(x)$.
\item[$(2$]
نحسب
\begin{eqnarray}
\int_{-\infty}^{\infty} f(x) \frac{d\theta}{dx} dx&=&[f(x)\theta(x)]_{-\infty}^{\infty}-\int_{-\infty}^{\infty}df(x)\theta(x)\nonumber\\
&=&[f(x)\theta(x)]_{-\infty}^{\infty}-\int_{0}^{\infty}df(x)\nonumber\\
&=&f(0).
\end{eqnarray}
بالتالي
$\frac{d\theta}{dx}=\delta(x)$.
\end{itemize}
\paragraph{
تمرين
$8$:
}
نختار
$f(x)=\delta (x)$.
نجد
$F(k)=1/\sqrt{2\pi}$. 
بالتعويض نحصل علي النتيجة المرادة:
\begin{eqnarray}
\delta(x)=\frac{1}{2\pi}\int_{-\infty}^{\infty}e^{ikx} dk.
\end{eqnarray}
\paragraph{
تمرين
$9$:
}
\begin{itemize}
\item[$(1$]
نجد
\begin{eqnarray}
S=\frac{1}{1-i\beta}\left(\begin{array}{cc}
i\beta &1\\
1&i\beta
\end{array}\right).
\end{eqnarray}
\item[$(2$]
نجد
\begin{eqnarray}
T=\left(\begin{array}{cc}
1+i\beta &1+i\beta\\
-i\beta&-i\beta
\end{array}\right).
\end{eqnarray}
\item[$(3$]
مازال لدينا الدوال الموجية
\begin{eqnarray}
&&\psi_I(x)=Ae^{ikx}+Be^{-ikx}~,~x\longrightarrow -\infty\nonumber\\
&&\psi_{II}(x)=Fe^{ix}+Ge^{-ikx}~,~x\longrightarrow +\infty.
\end{eqnarray}
في المنطقة
$III$ 
اين
 لا ينعدم الكمون تأخذ الدالة الموجية الشكل العام
\begin{eqnarray}
\psi_{III}=Cf(x)+Dg(x).
\end{eqnarray}
الدالتان
$f$
و
$g$
هما حلان خاصان مستقلان خطيا لمعادلة شرودينغر. لدينا ثابتا تكامل
$C$
و
$D$
لان معادلة شرودينغر هي معادلة تفاضلية من الدرجة الثانية.

لدينا اربعة شروط حدية. شرطان يربطان المنطقتين
$I$ 
و
$III$
و شرطان يربطان المنطقتين
$II$
و
$III$. 
يمكن استعمال شرطان حديان للتخلص من
$C$ 
و
 $D$.
 يتبقي لدينا الثوابت الاربعة
 $A$, $B$, $F$
 و
 $G$.
 الشرطان الحديان المتبقيان يمكن استعمالهما لتعيين ثابتين بدلالة الثابتين الاخرين. يمكننا اذن تعريف المصفوفة
 $S$
 و المصفوفة
 $T$ 
 بنفس الطريقة كما فعلنا في السابق.
\item[$(4$]
اولا نحسب
\begin{eqnarray}
S_{11}=-\frac{T_{21}}{T_{22}}~,~S_{12}=\frac{1}{T_{22}}~,~S_{21}=T_{11}-\frac{T_{12}T_{21}}{T_{22}}~,~S_{22}=\frac{T_{12}}{T_{22}}.
\end{eqnarray}
نجد من اجل
$G=0$
المعاملات
\begin{eqnarray}
R_l=\frac{|B|^2}{|A|^2}=|S_{11}|^2=|\frac{T_{21}}{T_{22}}|^2.
\end{eqnarray}
\begin{eqnarray}
T_l=\frac{|F|^2}{|A|^2}=|S_{21}|^2=|T_{11}-\frac{T_{21}T_{12}}{T_{22}}|^2.
\end{eqnarray}
\item[$(5$]
البرهان بائن تقريبا.
\end{itemize}

\chapter*{
نظرية الاضطرابات
}
\addcontentsline{toc}{chapter}{$2$
نظرية الاضطرابات 
}

\thispagestyle{headings}
\section*{
نظرية الاضطرابات غير المتعلقة بالزمن
}
\addcontentsline{toc}{section}{\bf 
نظرية الاضطرابات غير المتعلقة بالزمن
} 

\subsection*{
الاضطرابات غير المنحلة
}
\addcontentsline{toc}{subsection}{ 
الاضطرابات غير المنحلة
} 
نفترض انه يمكننا ان نحل بالضبط من اجل القيم الذاتية
$E_n^0$
و الدوال الذاتية 
$\psi_n^0$
لكمون
$V^0$.
نسمي الهاميلتونية في هذه الحالة
$H^0$.
لدينا اذن
\begin{eqnarray}
H^0|\psi_n^0>=E_n^0|\psi_n^0>.
\end{eqnarray} 
\begin{eqnarray}
<\psi_n^0|\psi_m^0>=\delta_{nm}.
\end{eqnarray}
هذه هي المسألة غير المضطربة.

الان لتكن
$H$
هاميلتونية اخري يمكن كتابتها علي الشكل
\begin{eqnarray}
H=H^0+\lambda H^1.
\end{eqnarray}
الهاميلتونية
$\lambda H^1$
تسمي الاضطراب حيث
$\lambda$
هو وسيط تحكم يأخذ قيم صغيرة. المسألة المضطربة تعرف ب
\begin{eqnarray}
H|\psi_n>=E_n|\psi_n>.\label{perturb}
\end{eqnarray} 
هدف نظرية الاضطرابات هو ايجاد الحلول التقريبية لمعادلة القيم الذاتية بدلالة الحلول المضبوطة. بعبارة اخري نود
الحصول علي عبارات تقريبية للقيم الذاتية
$E_n$ 
و الدوال الذاتية 
$\psi_n$ 
بدلالة القيم الذاتية غير المضطربة
$E_n^0$
و الدوال الذاتية غير المضطربة
 $\psi_n^0$.  
 نكتب
 $E_n$
 و
 $\psi_n$
 بدلالة
 $E_n^0$
 و
 $\psi_n^0$
 علي الشكل

\begin{eqnarray}
|\psi_n>=|\psi_n^0>+\lambda|\psi_n^1>+\lambda^2|\psi_n^2>+...\label{psin}
\end{eqnarray} 
\begin{eqnarray}
E_n=E_n^0+\lambda E_n^1+\lambda^2 E_n^2+...\label{en}
\end{eqnarray} 
ال
$E_n^1$
و
$|\psi_n^1>$
هي التصحيحات من الرتبة الاولي للقيمة الذاتية
 $E_n$
و الدالة الذاتية
$|\psi_n>$
علي التوالي  
بينما
$E_n^2$ 
و
$|\psi_n^2>$
هي التصحيحات من الرتبة الثانية. بالتعويض ب
$(\ref{psin})$ 
و
$(\ref{en})$
في
$(\ref{perturb})$
نجد

\begin{eqnarray}
\lambda(H^0|\psi_n^1>+H^1|\psi_n^0>)+\lambda^2(H^0|\psi_n^2>+H^1|\psi_n^1>)+O(\lambda^3)&=&\nonumber\\\lambda(E_n^0|\psi_n^1>+E_n^1|\psi_n^0>)+\lambda^2(E_n^0|\psi_n^2>+E_n^1|\psi_n^1>+E_n^2|\psi_n^0>)+O(\lambda^3).
\end{eqnarray}
نظرية الاضطرابات من الرتبة الاولي تعطي ب
\begin{eqnarray}
H^0|\psi_n^1>+H^1|\psi_n^0>&=&E_n^0|\psi_n^1>+E_n^1|\psi_n^0>.\label{1storderper}
\end{eqnarray}
اذن
\begin{eqnarray}
<\psi_n^0|H^0|\psi_n^1>+<\psi_n^0|H^1|\psi_n^0>&=&E_n^0<\psi_n^0|\psi_n^1>+E_n^1<\psi_n^0|\psi_n^0>.
\end{eqnarray}
\begin{eqnarray}
E_n^1=<\psi_n^0|H^1|\psi_n^0>.\label{fundresult1}
\end{eqnarray}
نعيد كتابة
$(\ref{1storderper})$
علي الشكل
\begin{eqnarray}
(H^0-E_n^0)|\psi_n^1>&=&-(H^1-E_n^1)|\psi_n^0>.
\end{eqnarray}
هذه معادلة تفاضلية غير متجانسة من اجل
$\psi_n^1$.
ننشر
$\psi_n^1$
كما يلي
\begin{eqnarray}
|\psi_n^1>=\sum_{m\neq n}c_m^{(n)}|\psi_m^0>.
\end{eqnarray}
الحد الذي يتعلق ب
$|\psi_n^0>$
غائب لان
$(H^0-E_n^0)|\psi_n^0>=0$.
بالتعويض نحصل علي
\begin{eqnarray}
\sum_{m\neq n}c_m^{(n)}(E_m^0-E_n^0)|\psi_m^0>&=&-(H^1-E_n^1)|\psi_n^0>.
\end{eqnarray}
بعبارة اخري
\begin{eqnarray}
c_m^{(n)}(E_m^0-E_n^0)&=&-<\psi_m^0|H^1|\psi_n^0>.
\end{eqnarray}
بالمقابل
\begin{eqnarray}
c_m^{(n)}&=&-\frac{<\psi_m^0|H^1|\psi_n^0>}{E_m^0-E_n^0}.
\end{eqnarray}
بالتالي
\begin{eqnarray}
|\psi_n^1>&=&-\sum_{m\neq n}\frac{<\psi_m^0|H^1|\psi_n^0>}{E_m^0-E_n^0}|\psi_m^0>.
\end{eqnarray}
نظرية الاضطرابات من الرتبة الثانية تعرف اذن بالمعادلة
\begin{eqnarray}
H^0|\psi_n^2>+H^1|\psi_n^1>&=&E_n^0|\psi_n^2>+E_n^1|\psi_n^1>+E_n^2|\psi_n^0>.\label{2ndorderper}
\end{eqnarray}
\begin{eqnarray}
<\psi_n^0|H^0|\psi_n^2>+<\psi_n^0|H^1|\psi_n^1>&=&E_n^0<\psi_n^0|\psi_n^2>+E_n^1<\psi_n^0|\psi_n^1>+E_n^2.\nonumber\\
\end{eqnarray}
نحصل باستعمال النتيجة
$<\psi_n^0|\psi_n^1>=0$ 
علي 
\begin{eqnarray}
E_n^2&=&<\psi_n^0|H^1|\psi_n^1>\nonumber\\
&=&\sum_{m\neq n}\frac{|<\psi_m^0|H^1|\psi_n^0>|^2}{E_n^0-E_m^0}.
\end{eqnarray}
نعيد كتابة
$(\ref{2ndorderper})$
علي الشكل
\begin{eqnarray}
(H^0-E_n^0)|\psi_n^2>&=&-(H^1-E_n^1)|\psi_n^1>+E_n^2|\psi_n^0>.
\end{eqnarray}
هذه معادلة تفاضلية غير متجانسة من اجل
$\psi_n^2$.
ننشر
$\psi_n^2$
علي الشكل
\begin{eqnarray}
|\psi_n^2>=\sum_{m\ne n}d_m^{(n)}|\psi_m^0>.
\end{eqnarray}
نحسب
\begin{eqnarray}
\sum_{m\ne n}d_m^{(n)}(E_m^0-E_n^0)|\psi_m^0>&=&-(H^1-E_n^1)|\psi_n^1>+E_n^2|\psi_n^0>.
\end{eqnarray}
اذن
\begin{eqnarray}
d_k^{(n)}(E_k^0-E_n^0)&=&-<\psi_k^0|(H^1-E_n^1)|\psi_n^1>.
\end{eqnarray}
بعبارة اخري
\begin{eqnarray}
d_k^{(n)}&=&-\frac{<\psi_k^0|(H^1-E_n^1)|\psi_n^1>}{E_k^0-E_n^0}\nonumber\\
&=&\sum_{m\neq n}\frac{<\psi_k^0|H^1|\psi_m^0><\psi_m^0|H^1|\psi_n^0>}{(E_n^0-E_k^0)(E_n^0-E_m^0)}-\frac{<\psi_k^0|H^1|\psi_n^0>
<\psi_n^0|H^1|\psi_n^0>}{(E_n^0-E_k^0)^2}.\nonumber\\
\end{eqnarray}
\subsection*{
حالة الاضطرابات  المنحلة
}
\addcontentsline{toc}{subsection}{ 
حالة الاضطرابات المنحلة
} 
لنفترض الان انه لدينا حالتان غير مضطربتان
$|\psi_a^0>$ 
و
$|\psi_b^0>$
التي لها نفس القيمة غير المضطربة للطاقة
$E^0$.
 اذن لدينا
\begin{eqnarray}
H^0|\psi_a^0>=E^0|\psi_a^0>~,~H^0|\psi_b^0>=E^0|\psi_b^0>~,~<\psi_a|\psi_b>=0.
\end{eqnarray}
اي تركيب خطي ل
 $|\psi_a^0>$
 و
 $|\psi_b^0>$
 هو ايضا حالة ذاتية ل
  $H^0$
  بنفس قيمة الطاقة
  $E^0$. 
  في العموم يرفع الاضطراب
  $\lambda H^1$
  هذا الانحلال بين الحالتين. بعبارة اخري اذا زدنا في قيمة
  الوسيط 
  $\lambda$ 
  فان المستوي الطاقوي
  $E^0$
  سوف ينقسم الي مستوي علوي و مستوي سفلي. بالمقابل اذا خفضنا في قيمة الوسيط
  $\lambda$ 
  فان المستوي العلوي يختزل الي تركيب خطي للحالتين
  $|\psi_a^0>$
  و
  $|\psi_b^0>$
  بينما يختزل
  المستوي السفلي الي تركيب خطي اخر متعامد مع التركيب الخطي الاول. هذه هي التركيبات الخطية
  التي يجب استعمالها في المعادلة
  $(\ref{fundresult1})$
  لحساب التصحيح من الرتبة الاولي للطاقة. المشكل هو ان هذه التركيبات الخطية غير معلومة لنا. اذن نكتب هذه التركيبات الخطية علي الشكل العام
\begin{eqnarray}
|\psi^0>=\alpha |\psi_a^0>+\beta |\psi_b^0>.
\end{eqnarray}
ننطلق من معادلة شرودينغر
\begin{eqnarray}
H|\psi>=E |\psi>.
\end{eqnarray}
ننشر
\begin{eqnarray}
E=E^0+\lambda E^1+\lambda^2 E^2+...
\end{eqnarray}
\begin{eqnarray}
|\psi>=|\psi^0>+\lambda|\psi^1>+\lambda^2|\psi^2>+....
\end{eqnarray}
نظرية الاضطرابات من الرتبة الاولي هي مرة اخري معرفة بالمعادلة
\begin{eqnarray}
H^0|\psi^1>+H^1|\psi^0>&=&E^0|\psi^1>+E^1|\psi^0>.
\end{eqnarray}
اذن بضرب كلا طرفي هذه المعادلة ب
 $<\psi_a^0|$
 نحصل علي
\begin{eqnarray}
&&<\psi_a^0|H^1|\psi^0>=E^1<\psi_a^0|\psi^0>\nonumber\\
&&\alpha <\psi_a^0|H^1|\psi_a^0>+\beta <\psi_a^0|H^1|\psi_b^0>=E^1\alpha.
\end{eqnarray}
نعرف
\begin{eqnarray}
W_{ij}=<\psi_i^0|H^1|\psi_j^0>~,~i,j=a,b.
\end{eqnarray}
لدينا
\begin{eqnarray}
\alpha W_{aa}+\beta W_{ab}=\alpha E^1.\label{dege1}
\end{eqnarray}
بالمثل بالضرب ب
$<\psi_b^0|$
نحصل علي
\begin{eqnarray}
\alpha W_{ba}+\beta W_{bb}=\beta E^1.\label{dege2}
\end{eqnarray}
بضرب هذه المعادلة ب
$W_{ab}$
نحصل علي
\begin{eqnarray}
\alpha |W_{ab}|^2+\beta W_{ab} W_{bb}=\beta W_{ab} E^1.
\end{eqnarray}
بالتعويض ب
$\beta W_{ab}=\alpha E^1-\alpha W_{aa}$ 
نحصل من اجل 
$\alpha\neq 0$ 
علي المعادلة التربيعية
\begin{eqnarray}
(E^1)^2-(W_{aa}+W_{bb})E^1+W_{aa}W_{bb}-|W_{ab}|^2=0.
\end{eqnarray}
الحلان هما
\begin{eqnarray}
E^1_{\pm}=\frac{1}{2}\bigg[W_{aa}+W_{bb}\pm \sqrt{(W_{aa}-W_{bb})^2+4|W_{ab}|^2}\bigg].
\end{eqnarray}
التركيبات الخطية المرفقة بهذين الحلين يمكن ايجادهما بالعودة الي المعادلتين
 $(\ref{dege1})$
 و
 $(\ref{dege2})$ 
 و
 ايجاد المعاملات
 $\alpha$
 و
 $\beta$.
 
 
 ليكن
 $A$ 
 مؤثر هرميتي يتبادل مع
 $H^0$ 
 و
 $H^1$.
 نفترض ان
 $|\psi_a^0>$ 
 و
 $|\psi_b^0>$
 هي ايضا اشعة ذاتية ل
 $A$ 
 بقيم ذاتية مختلفة اي
  \begin{eqnarray}
A|\psi_a^0>=\mu|\psi_a^0>~,~A|\psi_b^0>=\nu|\psi_b^0>~,~\mu\neq \nu.
\end{eqnarray}
لان
$[A,H^1]=0$ 
نحسب
\begin{eqnarray}
0&=&<\psi_a^0|[A,H^1]|\psi_b^0>\nonumber\\
&=&<\psi_a^0|AH^1|\psi_b^0>-<\psi_a^0|H^1A|\psi_b^0>\nonumber\\
&=&(\mu-\nu)W_{ab}.
\end{eqnarray}
لان
$\mu\neq \nu$
نستنتج ان
$W_{ab}=0$.
بعبارة اخري
\begin{eqnarray}
E^1_{+}=W_{aa}~,~E^1_{-}=W_{bb}.
\end{eqnarray}
هذه هي النتيجة التي كنا سوف نحصل عليها لو استعملنا نظرية الاضطرابات غير المنحلة من الرتبة الاولي. الاشارة زائد تقابل
$\alpha=1$
و
$\beta=0$
 اي
 $|\psi^0>=|\psi_{a}^0>$
 بينما تقابل الاشارة ناقص
 $\alpha=0$
 و
 $\beta=1$
 اي
 $|\psi^0>=|\psi_b^0>$. 

الخلاصة انه اذا وجدنا مؤثر هرميتي
$A$ 
يتبادل مع
$H$
فانه يمكننا ان نستخدم الاشعة الذاتية المشتركة كاشعة غير مضطربة و نطبق مباشرة نظرية الاضطرابات غير المنحلة من الرتبة الاولي.

في الاخير من اجل مستوي طاقوي 
 $E^0$
 منحل 
 $n$
 مرة فان التصحيحات من الرتبة الاولي
 $E^1$
 تعطي بالقيم الذاتية
 للمصفوفة
 $W_{ij}=<\psi_i^0|H^1|\psi_j^0>$.
 الاشعة الذاتية المرفقة بهذه القيم الذاتية هي بالضبط التركيبات الخطية التي يمكن ان نستعملها كاشعة حالة غير مضطربة.

\section*{
ذرة الهيدروجين
}
\addcontentsline{toc}{section}{\bf
ذرة الهيدروجين
} 

\subsection*{
المسألة المركزية الكمومية
}
\addcontentsline{toc}{subsection}{
المسألة المركزية الكمومية
}

معادلة شرودينغر في ثلاث ابعاد في اساس الموضع تكتب علي الشكل
\begin{eqnarray}
i\hbar\frac{\partial}{\partial t}\Psi&=&H\psi\nonumber\\
&=&\bigg(\frac{\vec{\hat{p}}^2}{2m}+V(\vec{\hat{r}})\bigg)\Psi\nonumber\\
&=&\bigg(-\frac{\hbar^2}{2m}\vec{\nabla}^2+V(\vec{{r}})\bigg)\Psi.
\end{eqnarray}
الكمون المركزي يعرف ب
\begin{eqnarray}
V(\vec{{r}})=V({r}).
\end{eqnarray}
في هذه الحالة من الافضل ان نعمل في الاحداثيات الكروية.  اللابلاسية
\footnote{.${\rm Laplacian}$}
$\vec{\nabla}^2$
في الاحداثيات الكروية تعطي ب

\begin{eqnarray}
\vec{\nabla}^2&=&\frac{1}{r^2}\frac{\partial}{\partial r}(r^2\frac{\partial}{\partial r})+\frac{1}{r^2\sin\theta}\frac{\partial}{\partial\theta}(\sin\theta\frac{\partial}{\partial\theta})+\frac{1}{r^2\sin^2\theta}\frac{\partial^2}{\partial\phi^2}\nonumber\\
&=&\frac{1}{r^2}\frac{\partial}{\partial r}(r^2\frac{\partial}{\partial r})-\frac{\hat{L}^2}{\hbar^2 r^2}.
\end{eqnarray}
معادلة شرودينغر تصبح
\begin{eqnarray}
i\hbar\frac{\partial}{\partial t}\Psi&=&\bigg(-\frac{\hbar^2}{2mr^2}\frac{\partial}{\partial r}(r^2\frac{\partial}{\partial r})+\frac{\hat{L}^2}{2m r^2}+V({r})\bigg)\Psi.
\end{eqnarray}
نحل هذه المعادلة عن طريق فصل المنغيرات اي
\begin{eqnarray}
\Psi=\Psi(t,\vec{r})=\psi_{nlm}(\vec{r})e^{-\frac{iE_{n}t}{\hbar}}.
\end{eqnarray}
الدالة الموجية 
$\psi_{nlm}(\vec{r})$
تحل المعادلة التفاضلية
\begin{eqnarray}
E_n\psi_{nlm}&=&\bigg(-\frac{\hbar^2}{2mr^2}\frac{\partial}{\partial r}(r^2\frac{\partial}{\partial r})+\frac{\hat{L}^2}{2m r^2}+V({r})\bigg)\psi_{nlm}.
\end{eqnarray}
فصل المتغيرات الثاني يجري كالاتي
\begin{eqnarray}
\psi_{nlm}=\psi_{nlm}(\vec{r})=R_{nl}(r)F_l^m(\theta,\phi).
\end{eqnarray}
نحصل علي
\begin{eqnarray}
\frac{1}{R_{nl}}\frac{d}{dr}\bigg(r^2\frac{dR_{nl}}{dr}\bigg)-\frac{2mr^2}{\hbar^2}(V(r)-E_n)=\frac{1}{\hbar^2}\frac{1}{F_l^m}{\hat{L}^2F_l^m}.
\end{eqnarray}
الطرف الايمن لهذه المعادلة يتعلق ب
$r$
فقط بينما يتعلق الطرف الايسر ب
$\theta$
و
$\phi$
فقط. 
اذن كلا الطرفين يجب ان يكون مساو لثابت نرمز له ب
$l(l+1)$. 
الدالة
$F_l^m$
يجب ان تحقق المعادلة
\begin{eqnarray}
\hat{L}^2F_l^m=\hbar^2l(l+1)F_l^m.
\end{eqnarray}
نستنتج مباشرة ان
$F_l^m$ 
هي بالضبط التوفيقة الكروية
$Y_l^m$
اي
\begin{eqnarray}
F_l^m=Y_l^m(\theta,\phi).
\end{eqnarray}
المعادلة التفاضلية المتبقية تعطي ب
\begin{eqnarray}
\frac{1}{R_{nl}}\frac{d}{dr}\bigg(r^2\frac{dR_{nl}}{dr}\bigg)-\frac{2mr^2}{\hbar^2}(V(r)-E_n)=l(l+1).
\end{eqnarray}
عن طريق تعريف
$u_{nl}(r)=rR_{nl}(r)$
يمكن ان نبين ان المعادلة اعلاه هي مكافئة ل
\begin{eqnarray}
-\frac{\hbar^2}{2m}\frac{d^2u_{nl}}{dr^2}+\bigg[V(r)+\frac{\hbar^2}{2m}\frac{l(l+1)}{r^2}\bigg]u_{nl}=E_nu_{nl}.
\end{eqnarray}
هذه معادلة شرودينغر في بعد واحد بكمون فعلي
معطي ب
\begin{eqnarray}
V_{\rm eff}(r)=V(r)+\frac{\hbar^2}{2m}\frac{l(l+1)}{r^2}.
\end{eqnarray}
الحد الثاني يسمي حد الطرد المركزي وتأثيره العام هو دفع الجسيم بعيدا عن المركز. شرط التنظيم هو
\begin{eqnarray}
\int_0^{\infty}r^2dr |R_{nl}(r)|^2=\int_0^{\infty}dr|u_{nl}(r)|^2=1.
\end{eqnarray}
\subsection*{
كمون كولومب
}
\addcontentsline{toc}{subsection}{
كمون كولومب
} 

 نعتبر الان الكمون الخاص بذرة الهيدروجين المعطي بكمون كولومب
\begin{eqnarray}
V(r)=-\frac{e^2}{4\pi\epsilon_0}\frac{1}{r}.
\end{eqnarray}
معادلة شرودينغر تصبح
\begin{eqnarray}
-\frac{\hbar^2}{2m}\frac{d^2u_{nl}}{dr^2}+\bigg[-\frac{e^2}{4\pi\epsilon_0}\frac{1}{r}+\frac{\hbar^2}{2m}\frac{l(l+1)}{r^2}\bigg]u_{nl}=E_nu_{nl}.
\end{eqnarray}
نحن نبحث عن حالات مرتبطة و بالتالي
$E<0$.
 نعرف
\begin{eqnarray}
\kappa_n=\frac{\sqrt{-2m E_n}}{\hbar}.
\end{eqnarray}
ايضا نستعمل
\begin{eqnarray}
\rho=\kappa_n r~,~\rho_{0n}=\frac{me^2}{2\pi\epsilon_0 \hbar^2\kappa_n}.
\end{eqnarray}
معادلة شرودينغر يمكن اذن وضعها علي الشكل التالي
\begin{eqnarray}
\frac{d^2u_{nl}}{d \rho^2}=\bigg[1-\frac{\rho_{0n}}{\rho}+\frac{l(l+1)}{\rho^2}\bigg]u_{nl}.
\end{eqnarray}
نستعمل طريقة فروبينيوس
\footnote{.${\rm Frobenius}$}
من اجل حل هذه المعادلة التفاضلية. في النهاية 
$\rho\longrightarrow \infty$ 
المعادلة التفاضلية اعلاه تختزل ل
\begin{eqnarray}
\frac{d^2u_{nl}}{d \rho^2}=u_{nl}.
\end{eqnarray}
الحل هو
\begin{eqnarray}
u_{nl}(r)=A e^{-\rho}+Be^{\rho}.
\end{eqnarray}
لما
$\rho\longrightarrow \infty$
 الحد الثاني ينفجر و بالتالي يجب ان نختار
 $B=0$.
نحصل علي
\begin{eqnarray}
u_{nl}(r)=A e^{-\rho}~,~\rho\longrightarrow\infty.
\end{eqnarray}
من الجهة الاخري في النهاية
$\rho\longrightarrow 0$ 
المعادلة التفاضلية اعلاه تصبح
\begin{eqnarray}
\frac{d^2u_{nl}}{d \rho^2}=\frac{l(l+1)}{\rho^2}u_{nl}.
\end{eqnarray}
الحل هو
\begin{eqnarray}
u_{nl}(r)=C \rho^{l+1}+D\rho^{-l}.
\end{eqnarray}
من جديد لما
 $\rho\longrightarrow 0$
 الحد الثاني ينفجر و بالتالي يجب ان نختار
 $D=0$. 
 نحصل علي
\begin{eqnarray}
u_{nl}(r)=C \rho^{l+1}~,~\rho\longrightarrow 0.
\end{eqnarray}
نزيل
 التصرف المقارب
 \footnote{.${\rm asymptotic~behavior}$}
 عند
 $\rho\longrightarrow \infty$ 
 و
 عند
 $\rho\longrightarrow 0$
 باعتبار الاقتراح التالي
\begin{eqnarray}
u_{nl}(r)=\rho^{l+1} e^{-\rho}v_{nl}(\rho).
\end{eqnarray}
نجد ان الدالة
$v_{nl}(\rho)$
يجب ان تحقق المعادلة التفاضلية
\begin{eqnarray}
\rho\frac{d^2v_{nl}(\rho)}{d\rho^2}+2(l+1-\rho)\frac{d v_{nl}}{d \rho}+\bigg(\rho_{0n}-2(l+1)\bigg)v_{nl}=0.
\end{eqnarray}
نعتبر الان السلسلة
\begin{eqnarray}
v_{nl}(\rho)=\sum_{j=0}^{\infty}c_j\rho^j.
\end{eqnarray}
بالتعويض في المعادلة التفاضلية نصل الي النتيجة
\begin{eqnarray}
\sum_{j=0}\bigg[(j+1)(j+2l+2)c_{j+1}+(\rho_{0n}-2(j+l+1))c_j\bigg]\rho^j=0.
\end{eqnarray}
بعبارة اخري
\begin{eqnarray}
c_{j+1}=\frac{2(j+l+1)-\rho_{0n}}{(j+1)(j+2l+2)}c_j.
\end{eqnarray}
من اجل القيم الكبيرة ل
$j$
لدينا
\begin{eqnarray}
c_{j+1}\simeq \frac{2}{j+1}c_j.
\end{eqnarray}
بافتراض ان هذه النتيجة مضبوطة نحصل علي
\begin{eqnarray}
c_j=\frac{2^j}{j!}c_0.
\end{eqnarray}
اذن
\begin{eqnarray}
v_{nl}(\rho)=c_0e^{2\rho}\Leftrightarrow v_n(\rho)=c_0\rho^{l+1}e^{\rho}.
\end{eqnarray}
من الواضح ان هذا لديه التصرف المقارب الخاطئ من اجل
$\rho\longrightarrow\infty$. 
 هذا يعني ان السلسلة يجب ان تنقطع و بالتالي توجد قيمة اعظمية 
 $j_{\rm max}$ 
 ل
 $j$
 بحيث
\begin{eqnarray}
c_{j_{\rm max}+1}=0.
\end{eqnarray}   
بعبارة اخري
$j_{\rm max}$
يجب ان يحقق
\begin{eqnarray}
2(j_{\rm max}+l+1)-\rho_{0n}=0.
\end{eqnarray}
عوض العمل ب
$j_{\rm max}$
نعمل ب
$n$
المعرف ب
\begin{eqnarray}
n=j_{\rm max}+l+1.
\end{eqnarray}
اذن
\begin{eqnarray}
\rho_{0n}=2n.
\end{eqnarray}
من هذا القيد نشتق ان الطاقة يجب ان تكون مكممة كالتالي
\begin{eqnarray}
E_n=-\frac{m}{2\hbar^2}\bigg(\frac{e^2}{4\pi\epsilon_0}\bigg)^2\frac{1}{n^2}.
\end{eqnarray}
الدالة الموجية الفضائية المكتملة هي
\begin{eqnarray}
\psi_{n l m}(r,\theta,\phi)=R_{n l}(r)Y_l^m(\theta,\phi).
\end{eqnarray}
\begin{eqnarray}
R_{n l}(r)=\frac{u_{nl}(r)}{r}=\frac{\rho^{l+1}}{r}e^{-\rho}v_{nl}(\rho).
\end{eqnarray}
الحالة الاساسية توافق
$n=1$
بطاقة
$E_1=-13.6~eV$. 
من الواضح انه في هذه الحالة
$j_{\rm max}=0$ 
و بالتالي
$l=0$,$v_{nl}(\rho)=c_0$
و
$R_{10}=c_0\kappa_1 e^{-\kappa_1 r}$.
الثابت
$\kappa_1$ 
هو مقلوب
ما يسمي بنصف قطر بور
اي
$\kappa_1=\sqrt{-2mE_1}/\hbar=1/a$. 
شرط التنظيم يسمح لنا بتثبيت
$c_0$. 

الحالة المثارة الاولي توافق
$n=2$ 
بطاقة
$E_2=E_1/4$
و
$\kappa_2=\kappa_1/2=\frac{1}{2a}$.
في هذه الحالة
$j_{\rm max}=1-l$
و بالتالي لدينا الامكانيات
$l=1$
و
$l=0$.
من اجل
$l=1$
لدينا
$j_{\rm max}=0$, $v_{nl}(\rho)=c_0$ 
و
بالتالي
$R_{21}=\frac{c_0}{4a^2}re^{-r/2a}$ 
بينما من اجل
$l=0$ 
لدينا
$j_{\rm max}=1$, $v_{nl}(\rho)=c_0+c_1\rho=c_0(1-\rho)$
 و بالتالي
 $R_{20}=\frac{c_0}{2a}(1-\frac{r}{2a})e^{-r/2a}$. 
 من جديد الثابت
 $c_0$
 يعين من شرط التنظيم.

في الحالة العامة الدالة
$v_{nl}(\rho)$
هي كثير حدود من الدرجة
$j_{\rm max}=n-l-1$ 
في
 $\rho$. 
من الواضح انه من اجل كل قيمة معينة ل
$n$
العدد الكمومي
$l$
يمكن ان يأخذ فقط القيم
$0$, $1$,...,$n-1$. 
من اجل كل قيمة ل
$l$
العدد الكمومي
$m$
يمكن ان يأخذ
ال
$2l+1$
قيمة
$-l$,$-l+1$,....,$l-1$,$l$.
اذن من اجل كل قيمة ل
$n$
لدينا
$\sum_{l=0}^{n-1}(2l+1)=n^2$ 
حالة. بعبارة اخري
درجة انحلال المستوي الطاقوي
 $E_n$
 هو
 $n^2$. 
 لنلاحظ ايضا ان كثير الحدود
 $v_{nl}(\rho)$ 
 هو كثير حدود لاغار
 المرافق
 $L_{n-l-1}^{2l+1}(2\rho)$ \footnote{.${\rm associated~Laguerre~polynomial}$}. 
يعرف كثير حدود لاغار المرافق
 $L_{q-p}^p(x)$ 
 بدلالة كثيرات حدود لاغار
 $L_{q}(x)$
 \footnote{.${\rm Laguerre~polynomials}$}
 كالتالي
\begin{eqnarray}
L_{q-p}^p(x)=(-1)^p\bigg(\frac{d}{dx}\bigg)^p L_q(x)~,~L_q(x)=e^x\bigg(\frac{d}{dx}\bigg)^q(e^{-x}x^q).
\end{eqnarray}

\section*{
البنية الدقيقة لذرة الهيدوجين
}
\addcontentsline{toc}{section}{\bf
البنية الدقيقة لذرة الهيدوجين
} 

\subsection*{
التصحيح النسبي
}
\addcontentsline{toc}{subsection}{
التصحيح النسبي
} 
الهاميلتونية الكلاسيكية لذرة الهيدروجين هي
\begin{eqnarray}
H^0=\frac{\vec{p}^2}{2m}-\frac{e^2}{4\pi\epsilon_0}\frac{1}{r}.
\end{eqnarray}
طاقات بور تعطي ب
\begin{eqnarray}
E_n^0=-\frac{\alpha^2}{2n^2}m c^2.
\end{eqnarray}
الاشعة الذاتية المرافقة هي
 $|\psi_{nlm}^0>$.
 الثابت 
  $\alpha$
  يعرف باسم
   ثابت البنية الدقيقة و يعطي ب
\begin{eqnarray}
\alpha=\frac{e^2}{4\pi\epsilon_0\hbar c}=\frac{1}{137.036}.
\end{eqnarray}
هذا هو ثابت الاقتران للتفاعلات الكهرومغناطيسية. هذه هي المسألة غير مضطربة في هذه الحالة.

التصحيح الاول ل
$E_n$
ياتي من الاخذ بعين الاعتبار لحركة النواة و هذا عن طريق تعويض
$m$
بالكتلة المختزلة
$mm_p/(m+m_p)$.
البنية الدقيقة لذرة الهيدروجين تتكون من التصحيح النسبي و التصحيح الناجم عن الاقتران بين عزم السبين و العزم المداري. هذه التصحيحات تكون من الرتبة
 $\alpha^4 mc^2$. 
 التصحيح التالي هو سحب لامب
 \footnote{.${\rm Lamb~shift}$}
 و هو من الرتبة
 $\alpha^5 mc^2$
 و ينجم عن تكميم الحقل المغناطيسي. بعد ذلك يأتي تصحيح البنية فائقة الدقة الذي هو من الرتبة
 $(m/m_p)\alpha^4 m c^2$
 و الذي ينجر عن التفاعل المغناطيسي بين العزوم المغناطيسية
 \footnote{.${\rm dipole~moments}$}
 للالكترون و البروتون. فيمايلي سنحسب فقط البنية الدقيقة لذرة الهيدروجين. 

الطاقة و كمية الحركة النسبيان يعطيان ب
\begin{eqnarray}
E=\gamma m v.
\end{eqnarray}
\begin{eqnarray}
p=\gamma m v.
\end{eqnarray}
المعامل 
$\gamma$
يسمي معامل لورنز
\footnote{.${\rm Lorentz}$}
و يعطي ب
\begin{eqnarray}
\gamma=\frac{1}{\sqrt{1-\frac{v^2}{c^2}}}.
\end{eqnarray}
نحسب
\begin{eqnarray}
E^2=p^2c^2+m^2c^4.
\end{eqnarray}
طاقة السكون
\footnote{.${\rm rest~energy}$}
تعطي ب
\begin{eqnarray}
E_0=m c^2.
\end{eqnarray}
اذن الطاقة الحركية تعطي ب
\begin{eqnarray}
T=E-E_0&=&\sqrt{p^2c^2+m^2c^4}-m c^2\nonumber\\
&=&m c^2\bigg[\sqrt{1+\frac{p^2}{m^2c^2}}-1\bigg]\nonumber\\
&=& m c^2\bigg[\frac{1}{2}\frac{p^2}{m^2c^2}-\frac{1}{8}\frac{p^4}{m^4 c^4}+..\bigg]\nonumber\\
&=&\frac{p^2}{2m}-\frac{p^4}{8m^3c^2}+...
\end{eqnarray}
الاضطراب يعطي بالتالي بالمؤثر 
\begin{eqnarray}
\lambda H^1_r=-\frac{\hat{p}^4}{8m^3c^2}.
\end{eqnarray}
من الواضح ان 
الاضطراب المكتوب في المعادلة اعلاه متناظر كرويا و بالتالي فانه يتبادل مع مربع العزم الحركي
$\hat{L}^2$
و مع المركبة الثالثة للعزم الحركي
$\hat{L}_3$.
 ترفق الاشعة الذاتية المشتركة
$|\psi_{n l m}^0>$,
ل
$H^0$ 
(الهاميلتونية غير المضطربة)
و
$\hat{L}^2$ 
و
$\hat{L}_3$,
بقيم ذاتية مختلفة ل
$\hat{L}^2$ 
و
$\hat{L}_3$
تعطي ب
$\hbar^2 l(l+1)$ 
و
$\hbar m$
من اجل
الحالات ال
$n^2$
التي لها نفس الطاقة
$E_n$. 
اذن
$n$, $l$ 
و
$m$
هي اعداد كمومية جيدة
و بالتالي يمكن استخدام نظرية الاضطراب غير المنحلة من الرتبة الاولي. التصحيح يعطي ب
\begin{eqnarray}
\lambda E^1_{r}&=&\lambda <\psi_{n l m}^0|H^1_r|\psi_{n l m}^0>\nonumber\\
&=&-\frac{1}{8m^3c^2}<\psi_{n l m}^0|\hat{p}^4|\psi_{n l m}^0>\nonumber\\
&=&-\frac{1}{8m^3c^2}(\hat{p}^2|\psi_{n l m}^0>)^+(\hat{p}^2|\psi_{n l m}^0>).
\end{eqnarray}
في المعادلة اعلاه استعملنا الخاصية الهرميتية للمؤثر
$\hat{p}^2$.
معادلة شرودينغر بالنسبة للحالات غير المضطربة تعطي ب
 \begin{eqnarray}
\hat{p}^2|\psi_{n l m}^0>=2m(E_n^0-V)|\psi_{nlm}^0>.
\end{eqnarray}
اذن
\begin{eqnarray}
\lambda E^1_{r}
&=&-\frac{1}{2m c^2}<\psi_{n l m}^0|(E_n^0-V)^2|\psi_{n l m}^0>\nonumber\\
&=&-\frac{1}{2m c^2}\bigg[(E_n^0)^2+2\hbar c\alpha E_n^0<\frac{1}{r}>+\hbar^2 c^2 \alpha^2<\frac{1}{r^2}>\bigg].
\end{eqnarray}
نستعمل النتائج
\begin{eqnarray}
<\frac{1}{r}>=\frac{\alpha}{n^2}\frac{mc}{\hbar}.
\end{eqnarray}
\begin{eqnarray}
<\frac{1}{r^2}>=\frac{\alpha^2m^2c^2}{\hbar^2}\frac{1}{n^3(l+\frac{1}{2})}.
\end{eqnarray}
بالتعويض بهذه العبارات نحصل علي
\begin{eqnarray}
\lambda E^1_{r}
&=&-\frac{(E_n^0)^2}{2m c^2}\bigg[\frac{4n}{l+\frac{1}{2}}-3\bigg].
\end{eqnarray}
\subsection*{
الاقتران  بين السبين و العزم الحركي المداري
}
\addcontentsline{toc}{subsection}{
الاقتران  بين السبين و العزم  الحركي المداري
} 
في معلم السكون خاصة الالكترون يدور البروتون حول الالكترون ويتولد عن ذلك حقل مغناطيسي
$\vec{B}$. 
التيار المتولد عن البروتون
هو
$I=e/T$
حيث
$T$
هو دور المدار. قانون بيو- ساقار
\footnote{.${\rm Biot-Savart~law}$}
ينص علي ان الحقل المغناطيسي المتولد عن هذا التيار متناسب طردا مع
$I$ 
و متناسب عكسا مع
نصف قطر المدار. لدينا بالضبط
\begin{eqnarray}
B=\frac{\mu_0 I}{2 r}=\frac{\mu_0 e}{2 r T}=\frac{ e}{2\epsilon_0 r c^2 T}.
\end{eqnarray}
هذا الحقل المغناطيسي عمودي علي مستوي المدار. اذا كان
$\vec{r}$
هو الشعاع من الالكترون الي البروتون و
$\vec{v}$ 
هي سرعة البروتون فان
 $\vec{B}$ 
 هو في اتجاه
 $\vec{r}x \vec{v}$. 
 بعبارة اخري
 $\vec{B}$
 هو في اتجاه العزم الحركي المداري
 $\vec{L}=m\vec{r}x \vec{v}$
 اي

\begin{eqnarray}
\vec{B}=\frac{e}{4\pi\epsilon_0 r^3 m c^2}\vec{L}.
\end{eqnarray}
من الجهة الاخري الالكترون له سبين
$\vec{S}$
و بالتالي له عزم مغناطيسي
$\vec{\mu}$. 
في الحقيقة العزم المغناطيسي متناسب طردا مع السبين و معامل التناسب يسمي النسبة المغناطيسية
\footnote{.${\rm gyromagnetic~ratio}$}.

كمثال نحسب النسبة المغناطيسية من اجل شحنة خطية
$q$
موزعة بانتظام حول دائرة ذات نصف قطر
$r$.
الدائرة تدور حول محورها بدور
$T$.
العزم المغناطيسي هوالتيار
$q/T$ 
ضرب المساحة
$\pi r^2$
اي

\begin{eqnarray}
\mu=\frac{q\pi r^2}{T}.
\end{eqnarray}
كتلة الشحنة
$q$
تساوي
 $m$ 
 و هي ايضا موزعة بانتظام. العزم المداري (السبين) يساوي عزم العطالة
  $m r^2$
  ضرب التواتر الزاوي
  $2\pi/T$
  اي
\begin{eqnarray}
S=\frac{2\pi m r^2}{T}.
\end{eqnarray}
النسبة المغناطيسية تعطي اذن ب
\begin{eqnarray}
\frac{\mu}{S}=\frac{q}{2 m}.
\end{eqnarray}
العزم المغناطيسي و السبين هما في نفس الجهة. اذن
\begin{eqnarray}
\vec{\mu}=\frac{q}{2 m}\vec{S}.
\end{eqnarray}
في حالة الالكترون فان العزم المغناطيسي يعطي بضعف هذه القيمة اي
\begin{eqnarray}
\vec{\mu}_e=-\frac{e}{ m}\vec{S}.
\end{eqnarray}
الحقل المغناطيسي المتولد عن حركة البروتون يؤثر علي العزم المغناطيسي للالكترون بحيث يحاول ان يجعل اتجاه
$\vec{\mu}_e$ 
محاذي لاتجاه
 $\vec{B}$.
 الهاميلتونية المرافقة تعطي ب
 \begin{eqnarray}
\lambda H^1_{so}&=&-\vec{\mu}_e.\vec{B}\nonumber\\
&=&\frac{e^2}{4\pi\epsilon_0}\frac{1}{m^2c^2r^3}\vec{S}.\vec{L}.
\end{eqnarray}
معلم السكون خاصة الالكترون ليس بمعلم عطالي لانه متسارع. التصحيح
الكينيماتي 
\footnote{.${\rm kinematic}$}
الراجع لهذا التأثير يتلخص في ضرب
$\lambda H^1_{so}$
بمعامل
يساوي
$1/2$.
هذا يسمي بمداورة توماس
\footnote{.${\rm Thomas~precession}$}. 
نحصل علي
 \begin{eqnarray}
\lambda H^1_{so}
&=&\frac{e^2}{8\pi\epsilon_0}\frac{1}{m^2c^2r^3}\vec{S}.\vec{L}.
\end{eqnarray}
نلاحظ ان تصحيح النسبة المغناطيسية للالكترون و مداورة توماس يلغيان بعضهما البعض تماما. 

الهاميلتونية 
$\lambda H^1_{so}$
تصف التفاعل بين عزم السبين و العزم الحركي المداري لذرة الهيدوجين. في الميكانيك الكمومي يتم تعويض 
$\vec{S}$
و
$\vec{L}$
بالمؤثرات
$\vec{\hat{S}}$
و
$\vec{\hat{L}}$. 
مؤثر الهاميلتونية في هذه الحالة لا يتبادل مع
 $\vec{\hat{L}}$
 و
 $\vec{\hat{S}}$. 
 مع ذلك 
 $\lambda H^1_{so}$
 يتبادل مع
  $\hat{L}^2$, $\hat{S}^2$, $\hat{J}^2$
  و
  $\hat{J}_3$
  حيث
  $\vec{\hat{J}}$
  هو العزم الحركي الكلي الذي يعطي ب
\begin{eqnarray}
\vec{\hat{J}}=\vec{\hat{L}}+\vec{\hat{S}}.
\end{eqnarray}
لان
$s=\frac{1}{2}$
فان القيم الذاتية الممكنة ل
$\hat{J}^2$
هي
$\hbar^2 j(j+1)$
حيث

\begin{eqnarray}
j=l+\frac{1}{2}~,~j=l-\frac{1}{2}.
\end{eqnarray}
الاشعة الذاتية المقابلة هي
\begin{eqnarray}
|jj_3>&=&\sum_{m,\sigma}C_{jj_3}^{lms\sigma}|lm>|s\sigma>\nonumber\\
&=&C_{jj_3}^{lj_3-\frac{1}{2}s\frac{1}{2}}|lj_3-\frac{1}{2}>|s\frac{1}{2}>+C_{jj_3}^{lj_3+\frac{1}{2}s\frac{1}{2}}|lj_3+\frac{1}{2}>|s-\frac{1}{2}>.
\end{eqnarray}
المعاملات 
$C_{jj_3}^{lms\sigma}$
هي معاملات كلابش - غوردون
\footnote{.${\rm Clebsch-Gordon~coefficients}$}
 التي تحقق, ضمن امور اخري, العلاقة
 $C_{jj_3}^{lms\sigma}=0$
 باستثناء لما
 $j_3=m+\sigma$. 

الاشعة الذاتية غير المضطربة يمكن اخذها الاشعة الذاتية المشتركة ل
 $H^0$, $\hat{L}^2$, $\hat{S}^2$, $\hat{J}^2$
 و
 $\hat{J}_3$. 
 هذه تعطي ب
\begin{eqnarray}
|\psi_{njj3}>=|R_{nl}>|jj_3>.
\end{eqnarray}
هذا يجب ان يقارن ب
$|\psi_{nlm}>|s\sigma>=|R_{nl}>|lm>|s\sigma>$ 
التي هي عبارة عن الاشعة الذاتية المشتركة ل
$H^0$, 
$\hat{L}^2$, $\hat{L_3}$, $\hat{S}^2$, $\hat{S}_3$. 
الطاقات غير المضطربة المقابلة ل
$|\psi_{njj3}>$
او
$|\psi_{nlm}>|s\sigma>$
ما زالت تعطي بطاقات بور.

نحسب
\begin{eqnarray}
\hat{S}.\hat{L}=\frac{1}{2}(\hat{J}^2-\hat{S}^2-\hat{L}^2).
\end{eqnarray}
القيم الذاتية ل
$\vec{S}.\vec{L}$
المقابلة ل
 $|\psi_{njj3}>$
 تعطي ب
\begin{eqnarray}
\frac{\hbar^2}{2}(j(j+1)-s(s+1)-l(l+1))=\frac{\hbar^2}{2}(j(j+1)-\frac{3}{4}-l(l+1)).
\end{eqnarray}
التصحيح من الرتبة الاولي يعطي ب
\begin{eqnarray}
\lambda E^1_{so}&=&<\psi_{njj_3}|\lambda H^1_{so}|\psi_{njj_3}>\nonumber\\
&=&\frac{e^2}{8\pi\epsilon_0}\frac{1}{m^2c^2}\frac{\hbar^2}{2}(j(j+1)-\frac{3}{4}-l(l+1))<\psi_{njj_3}|\frac{1}{r^3}|\psi_{njj_3}>\nonumber\\
&=&\frac{e^2}{8\pi\epsilon_0}\frac{1}{m^2c^2}\frac{\hbar^2}{2}(j(j+1)-\frac{3}{4}-l(l+1))<R_{nl}|\frac{1}{r^3}|R_{nl}>\nonumber\\
&=&\frac{\alpha \hbar^3}{4 m^2 c}(j(j+1)-\frac{3}{4}-l(l+1))<\frac{1}{r^3}>.
\end{eqnarray}
نستعمل النتيجة
\begin{eqnarray}
<\frac{1}{r^{3}}>=\frac{\alpha^3m^3c^3}{\hbar^3 n^3l(l+1)(l+\frac{1}{2})}.
\end{eqnarray}
اذن

\begin{eqnarray}
\lambda E^1_{so}&=&\frac{(E_n^0)^2}{m c^2}\frac{n}{l(l+1)(l+\frac{1}{2})}\bigg(j(j+1)-\frac{3}{4}-l(l+1)\bigg).
\end{eqnarray}
نحسب بصراحة
\begin{eqnarray}
\lambda E^1_{so}&=&\frac{(E_n^0)^2}{2m c^2}\frac{2n}{l+\frac{1}{2}}\frac{1}{l+1}~,~j=l+\frac{1}{2}.
\end{eqnarray}
\begin{eqnarray}
\lambda E^1_{so}&=&-\frac{(E_n^0)^2}{2m c^2}\frac{2n}{l+\frac{1}{2}}\frac{1}{l}~,~j=l-\frac{1}{2}.
\end{eqnarray}
في الخلاصة تصحيح البنية الدقيقة يعطي ب
\begin{eqnarray}
\lambda E^1_{r}+\lambda E^1_{so}&=&\frac{(E_n^0)^2}{2m c^2}(3-\frac{4n}{l+1})~,~j=l+\frac{1}{2}.
\end{eqnarray}
\begin{eqnarray}
\lambda E^1_{r}+\lambda E^1_{so}&=&\frac{(E_n^0)^2}{2m c^2}(3-\frac{4n}{l})~,~j=l-\frac{1}{2}.
\end{eqnarray}
يمكن كتابة هاتين العبارتين علي الشكل  
\begin{eqnarray}
\lambda E^1_{r}+\lambda E^1_{so}&=&\frac{(E_n^0)^2}{2m c^2}(3-\frac{4n}{j+\frac{1}{2}}).
\end{eqnarray}
المستويات الطاقوية لذرة الهيدروجين تصبح
\begin{eqnarray}
E_{n}&=&E_n^0+\lambda E^1_{r}+\lambda E^1_{so}\nonumber\\
&=&E_n^0\bigg[1+\frac{\alpha^2}{n^2}(\frac{n}{j+\frac{1}{2}}-\frac{3}{4})\bigg].
\end{eqnarray}
الانحلال في
$l$
انكسر لكن الانحلال في
$j$
مازال موجودا.

\section*{
نظرية الاضطرابات المتعلقة بالزمن
}
\addcontentsline{toc}{section}{\bf
نظرية الاضطرابات المتعلقة بالزمن
} 
\subsection*{
 تمثيل ديراك
}
\addcontentsline{toc}{subsection}{
 تمثيل ديراك
} 
نعتبر هاميلتونية متعلقة بالزمن
$H$
يمكن كتابتها علي الشكل
\begin{eqnarray}
H=H_0+V(t).
\end{eqnarray}
الهاميلتونية غير المتعلقة بالزمن
$H_0$
تقابل مسألة تقبل الحل المضبوط اي
\begin{eqnarray}
  H_0|n>=E_n|n>.
\end{eqnarray}
الكمون المتعلق بالزمن 
$V(t)$
نفترض انه صغير بالمقارنة مع
$H_0$. 
 اما شعاع الحالة الابتدائي في اللحظة
$t=0$ 
 فاننا نفترض انه من الشكل
\begin{eqnarray}
|\psi(0)>=\sum_nc_n(0)|n>.
\end{eqnarray}
من اجل
$t>0$
شعاع الحالة يأخذ الشكل العام
\begin{eqnarray}
|\psi(t)>=\sum_nc_n(t)e^{-\frac{iE_n t}{\hbar}}|n>.\label{expp}
\end{eqnarray}
الطور
$e^{-iE_nt/\hbar}$ 
هو الجزء المعتاد المتعلق بالزمن لدالة الموجة الذي نحصل عليه حتي لو كان الكمون
 $V$ 
 غير متعلق بالزمن. المجهول هو سعات الاحتمال
 $c_n(t)$
 التي تتعلق بالزمن فقط لان الكمون
 $V$ 
 غير متعلق بالزمن. هذه مسألة غير مستقرة لان الاضطراب 
 $V(t)$
 يمكن ان يؤدي الي الانتقال بين الحالات الكمومية
 $|n>$.
 كمثال علي ذلك, اذا انطلقنا في اللحظة
 $t=0$
 من الحالة
 $|\psi(0)>=|i>$, 
 اي من حالة ذاتية للهاميلتونية 
  $H_0$,
  فانه في اي لحظة زمنية لاحقة
  $t>0$
  يمكن ان نجد الجملة في حالة اخري
   $|j>$, $j\neq i$.
   من اجل
   $V=0$
   لدينا
   $c_j(t)=c_i(0)\delta_{ij}$
   و بالتالي الجملة تبقي دائما في الحالة
   $|i>$
   بينما من اجل
   $V\neq 0$
   نجد ان
   $c_j(t)\neq 0$
   في العموم
   و بالتالي يوجد احتمال انتقال من الحالة 
   $|i>$
   الي الحالة
   $|j>$. 

من المفيد جدا ان نتخلص من الطور
$e^{-iE_nt/\hbar}$
 الذي هو دائما موجود باية حال سواء كان الاضطراب متعلق بالزمن او لا.
من اجل هذه الغاية ندخل تمثيل او تصور ديراك. نذكر اولا ان شعاع الحالة
$|\psi(t)>$ 
هو شعاع الحالة في تمثيل او تصور شرودينغر. شعاع الحالة في تمثيل او تصور هايزنبرغ يعطي ب
\begin{eqnarray}
|\psi>=e^{\frac{i}{\hbar}Ht}|\psi(t)>.
\end{eqnarray}
 كل مؤثر
$O$
في تمثيل او تصور شرودينغر يعطي في تمثيل او تصور هايزنبرغ  بالمؤثر
\begin{eqnarray}
O(t)=e^{\frac{i}{\hbar}Ht}O e^{-\frac{i}{\hbar}Ht}.
\end{eqnarray}
ما يسمي بتمثيل او تصور ديراك يعرف بشعاع الحالة و المؤثر اللذان يعطيان ب 
\begin{eqnarray}
|\psi(t)>_I=e^{\frac{i}{\hbar}H_0t}|\psi(t)>.
\end{eqnarray}
\begin{eqnarray}
O_I(t)=e^{\frac{i}{\hbar}H_0t}O e^{-\frac{i}{\hbar}H_0t}.
\end{eqnarray}
نحسب مباشرة
\begin{eqnarray}
i\hbar\frac{\partial}{\partial t}|\psi(t)>_I&=&-H_0|\psi(t)>+e^{\frac{i}{\hbar}H_0t}(H_0+V)|\psi(t)>\nonumber\\
&=&V_I(t)|\psi(t)>_I.
\end{eqnarray}
النشر
$(\ref{expp})$ 
يصبح
\begin{eqnarray}
|\psi(t)>_I=\sum_nc_n(t)|n>.\label{exppI}
\end{eqnarray}
اذن نحصل مباشرة علي
\begin{eqnarray}
i\hbar\frac{d{c}_m(t)}{dt}=\sum_n c_n(t)e^{i\Omega_{mn}t}V_{mn}.
\end{eqnarray}
\begin{eqnarray}
\Omega_{mn}=\frac{E_m-E_n}{\hbar}~,~V_{mn}=<m|V|n>.
\end{eqnarray}
\subsection*{
مسائل الجمل ذات الحالتان
}
\addcontentsline{toc}{subsection}{
مسائل الجمل ذات الحالتان
} 
في هذه الحالة لدينا
\begin{eqnarray}
i\hbar\frac{d{c}_1(t)}{dt}=\sum_n c_n(t)e^{i\Omega_{1n}t}V_{1n}=c_1(t)V_{11}+c_2(t)e^{-i\Omega_0 t}V_{12}.
\end{eqnarray}
\begin{eqnarray}
i\hbar\frac{d{c}_2(t)}{dt}=\sum_n c_n(t)e^{i\Omega_{2n}t}V_{2n}=c_1(t)e^{i\Omega_0 t}V_{21}+c_2(t)V_{22}.
\end{eqnarray}
\begin{eqnarray}
\Omega_0=\frac{E_2-E_1}{\hbar}~,~E_2>E_1.
\end{eqnarray}
نفترض ان
\begin{eqnarray}
V_{11}=V_{22}=0.
\end{eqnarray}
في هذه الحالة نحصل علي
\begin{eqnarray}
\frac{d{c}_1(t)}{dt}=-\frac{i}{\hbar}c_2(t)e^{-i\Omega_0 t}V_{12}.
\end{eqnarray}
\begin{eqnarray}
\frac{d{c}_2(t)}{dt}=-\frac{i}{\hbar}c_1(t)e^{i\Omega_0 t}V_{21}.
\end{eqnarray}
نعتبر اضطراب جيبي من الشكل
\begin{eqnarray}
V_{12}=V_{21}^*=\gamma e^{i\Omega t}.
\end{eqnarray}
المعادلات التفاضلية المقترنة المكتوبة اعلاه تصبح
\begin{eqnarray}
\frac{d{c}_1(t)}{dt}=-\frac{i\gamma }{\hbar}c_2(t)e^{-i(\Omega_0-\Omega) t}.\label{lki}
\end{eqnarray}
\begin{eqnarray}
\frac{d{c}_2(t)}{dt}=-\frac{i\gamma}{\hbar}c_1(t)e^{i(\Omega_0-\Omega) t}.
\end{eqnarray}
انطلاقا من هاتين المعادلتين نحصل علي المعادلة التفاضلية من الدرجة الثانية
\begin{eqnarray}
\frac{d^2{c}_1(t)}{dt}+i(\Omega-\Omega_0)\frac{dc_1}{dt}+\frac{\gamma^2 }{\hbar^2}c_1(t)=0.
\end{eqnarray}
نأخذ الاقتراح
\begin{eqnarray}
c_1=e^{i\frac{\Omega-\Omega_0}{2}t}\hat{c}_1.
\end{eqnarray}
نجد
\begin{eqnarray}
\frac{d^2{\hat{c}}_1(t)}{dt}+\bigg[\frac{(\Omega-\Omega_0)^2}{4}+\frac{\gamma^2 }{\hbar^2}\bigg]\hat{c}_1(t)=0.
\end{eqnarray}
بعبارة اخري
\begin{eqnarray}
c_1=e^{i\frac{\Omega-\Omega_0}{2}t}\bigg[A\cos\Omega_r t+B\sin\Omega_r t\bigg].
\end{eqnarray}
التواتر
 $\Omega_r$
 يسمي بتواتر
 رابي
\footnote{.${\rm Rabi}$}
 ويعطي ب
\begin{eqnarray}
\Omega_r^2=\frac{(\Omega-\Omega_0)^2}{4}+\frac{\gamma^2 }{\hbar^2}.
\end{eqnarray}
المعادلة التفاضلية
$(\ref{lki})$
يمكن كتابتها علي الشكل
\begin{eqnarray}
-\frac{i\gamma}{\hbar}c_2=i\frac{\Omega-\Omega_0}{2}e^{-i\frac{\Omega-\Omega_0}{2}t}\bigg[A\cos\Omega_r t+B\sin\Omega_r t\bigg]+
\Omega_r e^{-i\frac{\Omega-\Omega_0}{2}t}\bigg[-A\sin\Omega_r t+B\cos\Omega_r t\bigg].
\end{eqnarray}
نستعمل الشروط الابتدائية
\begin{eqnarray}
c_1(0)=1~,~c_2(0)=0.
\end{eqnarray}
نجد
\begin{eqnarray}
A=1~,~B=-i\frac{\Omega-\Omega_0}{2\Omega_r}.
\end{eqnarray}
بالتالي
\begin{eqnarray}
c_1=e^{i\frac{\Omega-\Omega_0}{2}t}\bigg[\cos\Omega_r t-i\frac{\Omega-\Omega_0}{2\Omega_r}\sin\Omega_r t\bigg].
\end{eqnarray}
\begin{eqnarray}
c_2=-\frac{i}{\Omega_r}\frac{\gamma}{\hbar}e^{-i\frac{\Omega-\Omega_0}{2}t}\sin\Omega_r t.
\end{eqnarray}
 الحالة الابتدائية للجملة هي
$|1>$. 
احتمال ايجاد الجملة في اللحظة
$t$
في الحالة
$|2>$
يعطي ب
\begin{eqnarray}
|c_2|^2=\frac{1}{\Omega_r^2}\frac{\gamma^2}{\hbar^2}\sin^2\Omega_r t=\frac{1}{\Omega_r^2}\frac{\gamma^2}{\hbar^2}\frac{1}{2}(1-\cos 2\Omega_r t).
\end{eqnarray}
هذه العلاقة تسمي علاقة رابي. احتمال ايجاد الجملة في الحالة
$|2>$
يهتز في الزمن بتواتر
$2\Omega_r$.
سعة الاهتزاز تبلغ فيمتها الاعظمية عند
$\Omega=\Omega_0$. 
هذا سلوك او تصرف رنيني.
 
الرنين يعرف ب
\begin{eqnarray}
\Omega=\Omega_0~,~\Omega_r=\frac{\gamma}{\hbar}.
\end{eqnarray}
احتمال الانتقال عند الرنين يصبح
 \begin{eqnarray}
|c_2|^2=\frac{1}{2}(1-\cos 2\Omega_r t).
\end{eqnarray}
من اللحظة
$t=0$
الي اللحظة
$t=\pi\hbar/2\gamma$ 
يزداد احتمال ايجاد الجملة في  الحالة
$|2>$ 
حتي يصبح يساوي  
$1$. 
نذكر ان الطاقة
$E_2$
اكبر من الطاقة
$E_1$.
اذن خلال هذا الزمن الجملة تمتص الطاقة من الاضطراب حتي تصبح الحالة
$|2>$
مسكونة
كليا في اللحظة
 $t=\pi\hbar/2\gamma$ 
بينما تفرغ الحالة
$|1>$
كليا. 
 من اللحظة
 $t=\pi\hbar/2\gamma$ 
 الي اللحظة
 $t=\pi\hbar/\gamma$
الاحتمال
 $|c_2|^2$
 يتناقص 
 و بالتالي تفقد الجملة الطاقة للاضطراب حتي تصبح الحالة
  $|1>$
  في اللحظة
  $t=\pi\hbar/\gamma$
  مسكونة كليا بينما تفرغ الحالة
  $|2>$
  كليا. هذه الدورة بين الامتصاص و الارسال تستمر بدون توقف.

بعيدا عن الرنين القيمة الاعظمية لاحتمال الانتقال 
$|c_2|^2$
تعطي ب
 \begin{eqnarray}
|c_2|^2=\frac{1}{\Omega_r^2}\frac{\gamma^2}{\hbar^2}\sin^2\Omega_r t\leq |c_2|^2_{\rm max}=\frac{\frac{\gamma^2}{\hbar^2}}{\frac{(\Omega-\Omega_0)^2}{4}+\frac{\gamma^2 }{\hbar^2}}.
\end{eqnarray}
هذا منحني رنين بقمة عند
$\Omega=\Omega_0$. 
التواترت التي توافق الاحتمال
$|c_2|_{\rm max}^2=1/2$ 
هي
$\Omega=\Omega_0\pm 2\gamma/\hbar$. 
اذن عرض منحني الرنين  المأخوذ عند نصف القيمة الاعظمية هو
$4\gamma/\hbar$.
 من الواضح ان العرض يصبح اصغر, اي نحصل علي  قمم رنين اضيق, من اجل الكمونات الضعيفة. 

من بين تطبيقان الجمل ذات الحالتان نذكر الرنين المغناطيسي النووي
\footnote{.${\rm nuclear~magnetic~resonance}$}
و المايزر
\footnote{.${\rm MASERS: microwave~amplification~by~stimulated~emission~of~radiation}$}.

\subsection*{
نشر دايزون
}
\addcontentsline{toc}{subsection}{
نشر دايزون
} 

في تمثيل ديراك الذي يسمي ايضا بتمثيل التفاعل تأخذ معادلة شرودينغر الشكل
\begin{eqnarray}
i\hbar\frac{\partial}{\partial t}|\psi(t)>_I
&=&V_I(t)|\psi(t)>_I.
\end{eqnarray}
مؤثر التطور في الزمن في تمثيل التفاعل يعرف ب
\begin{eqnarray}
|\psi(t)>_I=U_I(t,t_0)|\psi(t_0)>_I.
\end{eqnarray}
من الجهة الاخري
\begin{eqnarray}
|\psi(t)>_I&=&e^{\frac{i}{\hbar}H_0 t}|\psi(t)>\nonumber\\
&=&e^{\frac{i}{\hbar}H_0 t}U(t,t_0)e^{-\frac{i}{\hbar}H_0 t_0}|\psi(t_0)>_I.
\end{eqnarray}
اذن
\begin{eqnarray}
U_I(t,t_0)=e^{\frac{i}{\hbar}H_0 t}U(t,t_0)e^{-\frac{i}{\hbar}H_0 t_0}.
\end{eqnarray}
هذا المؤثر يخضغ للمعادلة التفاضلية
\begin{eqnarray}
i\hbar\frac{d}{dt}U_I(t,t_0)
&=&V_I(t)U_I(t,t_0).
\end{eqnarray}
الشرط الابتدائي يعطي ب
\begin{eqnarray}
U_I(t_0,t_0)={\bf 1}.
\end{eqnarray}
الحل يعطي بالمعادلة التكاملية
\begin{eqnarray}
U_I(t,t_0)={\bf 1}-\frac{i}{\hbar}\int_{t_0}^tdt_1V_I(t_1)U_I(t_1,t_0).
\end{eqnarray}
يمكن تكرير هذه المعادلة كما يلي
\begin{eqnarray}
U_I(t,t_0)&=&{\bf 1}-\frac{i}{\hbar}\int_{t_0}^tdt_1V_I(t_1)\bigg[{\bf 1}-\frac{i}{\hbar}\int_{t_0}^{t_1} V_I(t_2)U_I(t_2,t_0)dt_2\bigg]\nonumber\\
&=&{\bf 1}+\bigg(\frac{-i}{\hbar}\bigg)\int_{t_0}^tdt_1V_I(t_1)+\bigg(\frac{-i}{\hbar}\bigg)^2\int_{t_0}^tdt_1\int_{t_0}^{t_1}dt_2V_I(t_1)V_I(t_2)U(t_2,t_0).\nonumber\\
\end{eqnarray}
بالتكرير الي رتبة كيفية نحصل علي 
\begin{eqnarray}
U_I(t,t_0)
={\bf 1}&+&\bigg(\frac{-i}{\hbar}\bigg)\int_{t_0}^tdt_1V_I(t_1)+\bigg(\frac{-i}{\hbar}\bigg)^2\int_{t_0}^tdt_1\int_{t_0}^{t_1}dt_2V_I(t_1)V_I(t_2)+...\nonumber\\
&+&\bigg(\frac{-i}{\hbar}\bigg)^n\int_{t_0}^tdt_1\int_{t_0}^{t_1}dt_2...\int_{t_0}^{t_{n-1}}dt_nV_I(t_1)V_I(t_2)...V_I(t_n)+....
\end{eqnarray}
هذا الحل يعرف بنشر دايزون
\footnote{.${\rm Dyson}$}.

نذكر ان
\begin{eqnarray}
|\psi(t)>_I=\sum_n c_n(t)|n>.
\end{eqnarray}
نحسب مباشرة 
\begin{eqnarray}
c_n(t)=<n|U_I(t,t_0)|\psi(t_0)>_I.
\end{eqnarray}
نحتار الحالة الابتدائية بحيث
\begin{eqnarray}
|\psi(t_0)>_I=|i>\Leftrightarrow |\psi(t_0)>=e^{-\frac{i}{\hbar}E_it_0}|i>.
\end{eqnarray}
سعة احتمال الانتقال تصبح
\begin{eqnarray}
c_n(t)=<n|U_I(t,t_0)|i>=e^{\frac{i}{\hbar}(E_nt-E_it_0)}<n|U(t,t_0)|i>.
\end{eqnarray}
احتمال الانتقال يعطي اذن ب
\begin{eqnarray}
P_{i\longrightarrow n}(t)=|c_n(t)|^2=|<n|U_I(t,t_0)|i>|^2=|<n|U(t,t_0)|i>|^2.
\end{eqnarray}
هذا يعطي ب
\begin{eqnarray}
P_{i\longrightarrow n}(t)=|c_n^{(0)}(t)+c_n^{(1)}(t)+c_n^{(2)}(t)+...|^2.
\end{eqnarray}
\begin{eqnarray}
&&c_n^{(0)}(t)=\delta_{ni}.
\end{eqnarray}
\begin{eqnarray}
c_n^{(1)}(t)&=&\bigg(\frac{-i}{\hbar}\bigg)\int_{t_0}^tdt_1<n|V_I(t_1)|i>\nonumber\\
&=&\bigg(\frac{-i}{\hbar}\bigg)\int_{t_0}^tdt_1e^{i\Omega_{ni}t_1}V_{ni}(t_1).
\end{eqnarray}
\begin{eqnarray}
c_n^{(2)}(t)&=&\bigg(\frac{-i}{\hbar}\bigg)^2\int_{t_0}^tdt_1\int_{t_0}^{t_1}dt_2<n|V_I(t_1)V_I(t_2)|i>\nonumber\\
&=&\bigg(\frac{-i}{\hbar}\bigg)^2\int_{t_0}^tdt_1\int_{t_0}^{t_1}dt_2\sum_m e^{i\Omega_{nm}t_1}e^{i\Omega_{mi}t_2}
V_{nm}(t_1)V_{mi}(t_2).
\end{eqnarray}
\begin{eqnarray}
\Omega_{ij}=\frac{E_i-E_j}{\hbar}~,~V_{ij}(t)=<i|V(t)|j>.
\end{eqnarray}
\subsection*{
قاعدة فيرمي الذهبية
}
\addcontentsline{toc}{subsection}{
قاعدة فيرمي الذهبية
} 

نعتبر اضطراب ثابت في الزمن معطي ب
\begin{eqnarray}
V(t)&=&0~,~t<0\nonumber\\
&=&V~,~t\geq 0.
\end{eqnarray}
المؤثر
$V$ 
يتعلق بالزمن فقط ضمنيا. نحسب (مع
$t_0=0$
)
\begin{eqnarray}
c_n^{(1)}(t)
&=&\bigg(\frac{-i}{\hbar}\bigg)\int_{0}^tdt_1e^{i\Omega_{ni}t}V_{ni}(t)\nonumber\\
&=&V_{ni}\frac{1-e^{i\Omega_{ni}t}}{E_n-E_i}.
\end{eqnarray}
اذن
\begin{eqnarray}
|c_n^{(1)}(t)|^2
&=&\frac{2|V_{ni}|^2}{(E_n-E_i)^2}(1-\cos\Omega_{ni}t).
\end{eqnarray}
اذن من اجل
$n\neq i$
احتمال الانتقال من الرتبة الاولي في اللحظة
$t$
يعطي ب
\begin{eqnarray}
P_{i\longrightarrow n}(t)=\frac{4|V_{ni}|^2}{(E_n-E_i)^2}\sin^2\frac{(E_n-E_i)t}{2\hbar}.
\end{eqnarray}
هذا يتعلق ب
$(1$
المركبة المصفوفية
$V_{ni}$
بين الحالة الابتدائية
$|i>$
و الحالة النهائية
$|n>$
و علي
$(2$ 
الفرق في الطاقة
$E_n-E_i=\hbar\Omega_{ni}$
بين الحالتين. 
الزمن
$t$
هو المجال الزمني الذي يكون خلاله الاضطراب
مشتغل.  من اجل قيمة معينة
ل
 $t$
 ندرس الاحتمال 
 $P_{i\longrightarrow n}(t)$
 كدالة في
 $\Omega_{ni}$
 اي
 $P_{i\longrightarrow n}(t)=P(\Omega_{ni})$.
 القيمة الاعظمية لهذا الاحتمال تقع عند
$\Omega_{ni}=0$
اين يصبح الاحتمال متناسب مع
 $t^2$ 
 و ينعدم عند
 $\Omega_{ni}=2n\pi/t$, 
$n=1,2...$. 
هناك قمم اخري اصغر تظهر عند
$\Omega_{ni}=(2n+1)\pi/t$, $n=1,2...$.
اذن عرض الاحتمال 
$P(\Omega_{ni})$ 
هو
 $1/t$.
بعبارة اخري 
$|c_n^{(1)}|^2$
من اجل القيم الكبيرة للزمن هو غير مهمل فقط من اجل الحالات
$|n>$
التي لها طاقة حول
$E_n$
اي

\begin{eqnarray}
|E_n-E_i|t\sim 2\pi\hbar.
\end{eqnarray}
علاقة الارتياب هذه تعني بالخصوص  انه في النهاية
$t\longrightarrow 0$
نحصل علي قمة عريضة جدا و بالتالي الانتقالات التي لا تحفظ الطاقة تصبح اكثر احتمالا بينما في النهاية
 $t\longrightarrow\infty$ 
 نحصل علي قمة ضيقة جدا و بالتالي الانتقالات التي تحقق
 $E_n\simeq E_i$
 هي الاكثر احتمالا في هذه الحالة.

المساحة تحت المنحني
$P(\Omega_{ni})$
هي اذن متناسبة مع
$t^2x 1/t=t$ 
و التي تساوي الاحتمال الكلي للانتقال من الحالة الابتدائية
$|i>$ 
الي الحالات النهائية
$|n>$
التي لها طاقة متمركزة حول
$E_n$.  
يمكن جعل هذه الفكرة اكثر دقة كالاتي. الاحتمال الكلي هو مجموع احتمالات الانتقال الي الحالات النهائية التي لها طاقة
$E_n\simeq E_i$.
 هذا يعطي ب
\begin{eqnarray}
\sum_{n:E_n\simeq E_i}|c_n^{(1)}|^2.
\end{eqnarray}
نفترض ان الحالات النهائية مستمرة. ندخل كثافة الحالات النهائية
 $\rho(E)$.
  بعبارة اخري 
  $\rho(E)dE$
  هو عدد الحالات التي لها طاقة بين
  $E$
  و
  $E+dE$.
  يمكن اذن تعويض الاحتمال الكلي اعلاه بالتكامل
\begin{eqnarray}
\int dE_n \rho(E_n)|c_n^{(1)}|^2=\int dE_n\rho(E_n)\frac{4|V_{ni}|^2}{(E_n-E_i)^2}\sin^2\frac{(E_n-E_i)t}{2\hbar}. 
\end{eqnarray}
من اجل الازمان الكبيرة يمكن ان نستعمل العلاقة
\begin{eqnarray}
{\rm lim}_{t\longrightarrow\infty}\frac{\sin^2tx}{tx^2}=\pi\delta(x).
\end{eqnarray}
نحصل علي 
\begin{eqnarray}
\sum_{n:E_n\simeq E_i}|c_n^{(1)}|^2=\int dE_n \rho(E_n)|c_n^{(1)}|^2&=&\int dE_n\rho(E_n)\frac{2\pi t}{\hbar}|V_{ni}|^2\delta(E_n-E_i)\nonumber\\
&=&\bigg[\rho(E_n)\frac{2\pi t}{\hbar}\overline{|V_{ni}|}^2\bigg]_{E_n= E_i}.
\end{eqnarray}
الخط علي 
$|V_{ni}|^2$
يدل علي انه يجب ان نأخذ الحالات 
$|n>$ 
التي لها تقريبا نفس الطاقة
$E_n$ 
و   لكن لها ايضا تقريبا نفس 
عناصر المصفوفة
$ V_{ni}$
لانه يمكن للحالات التي لها نفس الطاقة ان تكون لها عناصر مصفوفية
$ V_{ni}$
مختلفة
.
خذ مثلا
$|n>=|\vec{p}>$
في حالة الفعل الكهروضوئي.


معدل الانتقال هو بالضبط احتمال الانتقال في وحدة الزمن. هذا يعرف ب
\begin{eqnarray}
w_{i\longrightarrow [n]}=\frac{d}{dt}\sum_{n:E_n\simeq E_i}|c_n^{(1)}|^2&=&\frac{2\pi }{\hbar}\bigg[\overline{|V_{ni}|}^2\rho(E_n)\bigg]_{E_n= E_i}.
\end{eqnarray}
هذه هي قاعدة فيرمي الذهبية
\footnote{.${\rm Fermi's~golden~rule}$}.
هذه العلاقة يمكن ايضا كتابتها علي الشكل
\begin{eqnarray}
w_{i\longrightarrow n}&=&\frac{2\pi }{\hbar}{|V_{ni}|}^2\rho(E_n)\delta(E_n-E_i)dE_n.
\end{eqnarray}
التصحيح من الرتبة الثانية لسعة الاحتمال هو
\begin{eqnarray}
c_n^{(2)}(t)
&=&\bigg(\frac{-i}{\hbar}\bigg)^2\int_{0}^tdt_1\int_{0}^{t_1}dt_2\sum_m e^{i\Omega_{nm}t_1}e^{i\Omega_{mi}t_2}V_{nm}V_{mi}\nonumber\\
&=&\frac{i}{\hbar}\sum_m\frac{V_{nm}V_{mi}}{E_m-E_i}\int_0^{t}dt_1(e^{i\Omega_{ni}t_1}-e^{i\Omega_{nm}t_1})\nonumber\\
&=&\sum_m\frac{V_{nm}V_{mi}}{E_i-E_m}\bigg[\frac{1-e^{i\Omega_{ni}t}}{E_n-E_i}-\frac{1-e^{i\Omega_{nm}t}}{E_n-E_m}\bigg].
\end{eqnarray}
تصرف الحد الاول في هذه العبارة من اجل الازمان الكبيرة
$t$ 
يشبه تصرف 
$c_n^{(1)}$ 
مع التعويض
$V_{ni}\longrightarrow \sum_m V_{nm}V_{mi}/(E_i-E_m)$.
اذن فقط الحالات
التي لها
$E_n\simeq E_i$
سيكون لها مشاركة معتبرة. الحد الثاني لما
 $E_m\neq E_n$
 و
 $E_m\neq E_i$
 يؤدي الي اهتزاز سريع لا يتزايد مع الزمن
 $t$
 و بالتالي لا يشارك في احتمال الانتقال. 
 في المحصلة معدل الانتقال باضافة التصحيح من الرتبة الثانية يعطي ب 
\begin{eqnarray}
w_{i\longrightarrow [n]}=\frac{d}{dt}\sum_{n:E_n\simeq E_i}|c_n^{(1)}+c_n^{(2)}|^2&=&\frac{2\pi }{\hbar}\bigg[\overline{|V_{ni}+\sum_m\frac{V_{nm}V_{mi}}{E_i-E_m}|}^2\rho(E_n)\bigg]_{E_n= E_i}.
\end{eqnarray}
 الحالة لما
 $E_m\simeq E_i$
 مع
 $V_{nm}V_{mi}\neq 0$
 هي حالة خاصة 
 تؤدي الي نفس العبارة مع التعويض
 $E_i-E_m\longrightarrow E_i-E_m+i\epsilon$
 حيث
 $\epsilon$
 هو عدد حقيقي متناه في الصغر.

كا ذكرنا قبل قليل حد الرتبة الاولي في المعادلة اعلاه يقابل انتقالات تحفظ الطاقة. من الجهة الاخري حد الرتبة الثانية يمكن فهمه علي انه تركيب
انتقالين غير حافظين للطاقة من
$|i>$
الي
$|m>$
ثم من
$|m>$ 
الي
$|n>$.
 هذه الانتقالات تسمي افتراضية لانها لا تحفظ الطاقة علي الرغم انه لدينا انحفاظ للطاقة اجمالي بين
 $|i>$
 و
 $|n>$.

\section*{
امتصاص و ارسال الاشعاع
}
\addcontentsline{toc}{section}{\bf
امتصاص و ارسال الاشعاع
} 

\subsection*{
الاضطراب التوافقي
}
\addcontentsline{toc}{subsection}{
الاضطراب التوافقي
} 
نعتبر الان الاضطراب التوافقي
\begin{eqnarray}
V(t)=V e^{i\Omega t}+
V^+ e^{-i\Omega t}.
\end{eqnarray} 
مرة اخري
$V$
و
$V^+$ 
يتعلقان ضمنيا بالزمن. في اللحظة الابتدائية
$t=0$ 
فقط
الحالة الذاتية
$|i>$ 
ل
$H_0$
تكون مسكونة او مأهولة. نحسب التصحيح من الرتبة الاولي لسعة الاحتمال كما يلي 
 \begin{eqnarray}
c_n^{(1)}(t)&=&\frac{-i}{\hbar}\int_0^t dt_1 e^{i\Omega_{ni}t}V_{ni}(t)\nonumber\\
&=&\frac{1}{\hbar}\bigg[V_{ni}\frac{1-e^{i(\Omega_{ni}+\Omega)t}}{\Omega_{ni}+\Omega}+V^+_{ni}\frac{1-e^{i(\Omega_{ni}-\Omega)t}}{\Omega_{ni}-\Omega}\bigg].
\end{eqnarray} 
نذكر انه من اجل الاضطراب الثابت تحصلنا علي
\begin{eqnarray}
c_n^{(1)}(t)
&=&\frac{1}{\hbar}\bigg[V_{ni}\frac{1-e^{i\Omega_{ni}t}}{\Omega_{ni}}\bigg].
\end{eqnarray}
بعبارة اخري التغيير الوحيد هو
\begin{eqnarray}
\Omega_{ni}\longrightarrow \Omega_{ni}\pm\Omega.
\end{eqnarray}
اذن من اجل الازمان الكبري
$t$
فان الاحتمال
$|c_n^{(1)}|^2$ 
يكون ذو قيمة معتبرة فقط في الحالتين المستبعدتين لبعضهما البعض
\begin{eqnarray}
&&\Omega_{ni}+\Omega=0\Leftrightarrow E_n=E_i-\hbar\Omega.\label{stimulated}
\end{eqnarray}
\begin{eqnarray}
&&\Omega_{ni}-\Omega=0\Leftrightarrow E_n=E_i+\hbar\Omega.\label{absorption}
\end{eqnarray}
من الواضح ان الجملة غير منحفظة. لكن اذا اخذنا الجملة الكلية  المشكلة من الجملة و الاضطراب الخارجي
$V(t)$
نجد انها جملة منحفظة كما يجب ان تكون.  الحالة الثانية
$(\ref{absorption})$
هي ممكنة فقط لما
$E_n>E_i$
اي لما تكون
$E_n$
عبارة عن حالة مثارة. هذا يوافق الامتصاص لان الجملة تتلقي طاقة
$\hbar\Omega$
من الاضطراب 
$V(t)$.
الحالة الاولي
$(\ref{stimulated})$
هي ممكنة فقط لما
$E_n<E_i$
اي لما تكون
$E_i$
هي الحالة المثارة. هذا يوافق الارسال المحفز لان الجملة تفقد طاقة
$\hbar\Omega$
للاضطراب. هذا الارسال يسمي محفز لان الاضطراب هو الذي تسبب فيه و لم يكن تلقائي. كمثال علي ذلك عندما نشع ضوء علي ذرة في الحالة المثارة
$E_i$ 
هذه الذرة يمكنها ان تقفز الي الحالة الادني
$E_n$.
بعبارة اخري الفوتون الوحيد الوارد علي الذرة يصبح فوتونين صادرين بنفس التواتر. هذا هو  بالضبط مبدأ التضخيم الذي يتحكم في اللايزر
\footnote{.${\rm LASER: light~amplification~by~stimulated~emission~of~radiation}$}.
نذكر هنا ان الارسال المحفز بالتفاعل الكهرومغناطيسي تنبأ به اولا اينشتاين. 

 قاعدة فيرمي الذهبية تكتب علي الشكل
 \begin{eqnarray}
w^{\rm stim-emis}_{i\longrightarrow [n]}=\frac{2\pi }{\hbar}\bigg[\overline{|V_{ni}|}^2\rho(E_n)\bigg]_{E_n= E_i-\hbar\Omega}.
\end{eqnarray} 
\begin{eqnarray}
w^{\rm abso}_{i\longrightarrow [n]}=\frac{2\pi }{\hbar}\bigg[\overline{|V^+_{ni}|}^2\rho(E_n)\bigg]_{E_n= E_i+\hbar\Omega}\label{abso-rate}.
\end{eqnarray} 
من النتيجة
$|V_{ni}|^2=|V^+_{in}|^2$
نحصل علي التوازن التفصيلي
\footnote{.${\rm detailed~balance}$}
الذي يعبر عن التناظر بين الامتصاص و الارسال المحفز. نكتب هذا التوازن التفصيلي علي الشكل
 \begin{eqnarray}
\frac{w^{\rm stim-emis}_{i\longrightarrow [n]}}{\rho(E_n)}=\frac{w^{\rm abso}_{n\longrightarrow [i]}}{\rho(E_i)}.
\end{eqnarray}
\subsection*{
الامتصاص و الارسال المحفز
}
\addcontentsline{toc}{subsection}{
الامتصاص و الارسال المحفز
} 
هاميلتونية شحنة
$q$
تتحرك تحت تأثير حقل كهربائي
$\vec{E}=-\vec{\nabla}\phi-\partial\vec{A}/\partial t$
و حقل مغناطيسي
 $\vec{B}=\vec{\nabla}x\vec{A}$ 
 تعطي ب
\begin{eqnarray}
H=\frac{1}{2m}(\vec{p}-q\vec{A})^2+q\phi.
\end{eqnarray}
ال
$\phi$ 
و
$\vec{A}$
هما الكمون السلمي و الكمون الشعاعي علي التووالي. نفرض الشرط المعياري لكولومب
\footnote{.${\rm Coulomb~gauge~condition}$}
الذي يعطي ب
\begin{eqnarray}
\vec{\nabla}\vec{A}=0.
\end{eqnarray}
يمكن كتابة الهاميلتونية علي الشكل
\begin{eqnarray}
H=\frac{\vec{p}^2}{2m}-\frac{q}{m}\vec{A}\vec{p}+\frac{q^2\vec{A}^2}{2m}+q\phi.
\end{eqnarray}
نعتبر حقل موجة مستوية وحيدة اللون
\footnote{.${\rm monochromatic}$}
الذي يعطي ب
\begin{eqnarray}
\phi=0~,~\vec{A}=2\hat{\epsilon}A_0\cos(k\hat{n}\vec{x}-\Omega t).
\end{eqnarray}
اتجاه الانتشار هو
$\hat{n}$
و اتجاه الاستقطاب هو
$\hat{\epsilon}$.
العدد الموجي هو
$k=\Omega/c$. 
الشرط المعياري
$\vec{\nabla}\vec{A}=0$
يعني ان الموجة هي عرضية اي ان
$\hat{\epsilon}\hat{n}=0$.
الهاميلتونية تصبح

\begin{eqnarray}
H=\frac{\vec{p}^2}{2m}+Ve^{i\Omega t}+V^+e^{-i\Omega t}.
\end{eqnarray}
\begin{eqnarray}
V=-\frac{qA_0}{m}\hat{\epsilon}\vec{p}e^{-ik\hat{n}\vec{x}}.
\end{eqnarray}
في المعادلة اعلاه اهملنا الحد الديامغناطيسي
\footnote{.${\rm diamagnetic}$}
 $q^2\vec{A}^2/2m$ 
 و استعملنا  النتيجة
 $\hat{\epsilon}\vec{p}\exp(ik\hat{n}\vec{x})=\exp(ik\hat{n}\vec{x})\hat{\epsilon}\vec{p}$. 

الحد
$e^{i\Omega t}V$
يوافق الارسال المحفز بينما يوافق الحد
$e^{-i\Omega t}V^+$ 
الامتصاص. فيما يلي سندرس الامتصاص بتفصيل اكبر. نحسب
(
مع
$q=e$)
\begin{eqnarray}
|V^+_{ni}|=\frac{e^2A_0^2}{m^2}|\hat{\epsilon}<n|\vec{p}e^{ik\hat{n}\vec{x}}|i>|^2.
\end{eqnarray}
نقوم بالتقريب التالي: طول موجة الاشعاع هو اكبر بكثير من بعد الذرات اي
$|k\hat{n}\vec{x}|=2\pi|\hat{n}\vec{x}/\lambda|<<1$.
 اذن يمكن ان نقرب الدالة الاسية ب
 $1$.
 هذا هو تقريب العزم الكهريبائي ثنائي القطبية
 \footnote{.${\rm elecrtic~dipole~approximation}$}
 .
 نحصل علي
 \begin{eqnarray}
|V^+_{ni}|=\frac{e^2A_0^2}{m^2}|\hat{\epsilon}<n|\vec{p}|i>|^2.
\end{eqnarray}
من العلاقة
$[x,H_0]=i\hbar p_x/m$ 
نحسب
 $<n|\vec{p}|i>=im\Omega_{ni}<n|\vec{x}|i>$.
 اذن نحصل علي
 \begin{eqnarray}
|V^+_{ni}|=A_0^2\Omega_{ni}^2|\hat{\epsilon}<n|\vec{P}|i>|^2.
\end{eqnarray}
الشعاع
$\vec{P}$ 
هو العزم الكهربائي ثنائي القطبية المعرف ب
\begin{eqnarray}
\vec{P}=e\vec{x}.
\end{eqnarray}
الحقل الكهربائي 
$\vec{E}$ 
خاصة الموجة المستوية وحيدة اللون اعلاه يعطي ب
$\vec{E}=-\hat{\epsilon}(2\Omega A_0)\sin(k\hat{n}\vec{x}-\Omega t)$. 
كثافة الطاقة (الطاقة في وحدة الحجم)
في موجة كهرومغناطيسية
هي
$u=(\epsilon_0E^2+B^2/{\mu}_0)/2=\epsilon_0E^2$. 
اذن المتوسط خلال دورة كاملة هو
$u=2\epsilon_0\Omega^2A_0^2$.
بالتالي نحصل باستعمال ايضا النتيجة
$\Omega_{ni}=\Omega$
علي
 \begin{eqnarray}
|V^+_{ni}|=\frac{u}{2\epsilon_0}|\hat{\epsilon}<n|\vec{P}|i>|^2.
\end{eqnarray}
نأخذ متوسط هذه العبارة علي جميع اتجاهات الورود
$\hat{n}$
و علي جميع اتجاهات الاستقطاب
$\hat{\epsilon}$. 
نعمل في الاحداثيات الكروية. الشعاعان
$\hat{n}$
و
$\hat{\epsilon}$
متعامدان. نختار المحور
$z$ 
علي طول محور الانتشار
 $\hat{n}$.
 نختار المحور
 $y$
 بحيث يكون الشعاع
 $<n|\vec{P}|i>$ 
 في المستوي
 $zy$.
 المحور
 $x$
 يصبح اذن مثبت. الزاوية بين
 $\hat{n}$
 و
 $<n|\vec{P}|i>$
 هي
 $\theta$
 و الزاوية بين المحور
 $x$ 
 و
 $\hat{\epsilon}$
 هي
 $\phi$.
 نحصل اذن علي
 \begin{eqnarray}
\hat{\epsilon}=\cos\phi\hat{i}+\sin\phi\hat{j}~,~<n|\vec{P}|i>=|<n|\vec{P}|i>|(\cos\theta \hat{k}+\sin\theta \hat{j}).
\end{eqnarray}
المتوسط يعطي اذن بالتكامل
\begin{eqnarray}
|V^+_{ni}|&=&\frac{u}{2\epsilon_0}|<n|\vec{P}|i>|^2\frac{1}{4\pi}\int \sin^2\phi\sin^2\theta \sin\theta d\theta d\phi\nonumber\\
&=&\frac{u}{6\epsilon_0}|<n|\vec{P}|i>|^2.
\end{eqnarray}
من الواضح ان
\begin{eqnarray}
<n|\vec{P}|i>|^2=<n|{P}_x|i>^2+<n|{P}_y|i>^2+<n|{P}_z|i>^2.
\end{eqnarray}
نكتب معدل الامتصاص
$(\ref{abso-rate})$
كالاتي
\begin{eqnarray}
w^{\rm abso}_{i\longrightarrow n}&=&\frac{2\pi }{\hbar} \overline{|V^+_{ni}|}^2\delta(E_n- E_i-\hbar\Omega)\rho(E_n)dE_n\nonumber\\
&=&\frac{\pi }{3\epsilon_0\hbar^2} |<n|\vec{P}|i>|^2\delta(\Omega_{ni}-\Omega)u \rho(E_n)dE_n.
\end{eqnarray} 
الموجة الكهرومغناطيسية ليست وحيدة اللون علي نحو كامل و بالتالي تأتي بعرض تواترات او ترددات محدود. نذكر ان
$\rho(E_n)dE_n$ 
هو عدد الحالات النهائية بطاقة بين
$E_n=\hbar(\Omega+E_i/\hbar)$
و
$E_n+dE_n=\hbar(\Omega+d\Omega+E_i/\hbar)$
حيث
$E_i$
نبقيه مثبت. اذن نري ان
$\rho(E_n)dE_n$ 
هو عدد
الانساق الكهرومغناطيسية
\footnote{.${\rm electromagnetic~modes}$}
التي لها تردد بين
$\Omega$
و
$\Omega+d\Omega$. 
كثافة الطاقة في النسق الكهرومغناطيسي ذو التردد
$\Omega$
هي
$u$. 
اذن
 $u \rho(E_n)dE_n$ 
 هي كثافة الطاقة في الانساق الكهرومغناطيسية التي لها تردد بين
 $\Omega$
 و
 $\Omega+d\Omega$. 
 نكتب هذا علي الشكل
\begin{eqnarray}
u \rho(E_n)dE_n=\rho_u(\Omega)d\Omega
\end{eqnarray} 
نحصل علي النتيجة النهائية
\begin{eqnarray}
w^{\rm abso}_{i\longrightarrow n}
&=&\frac{\pi }{3\epsilon_0\hbar^2} |<n|\vec{P}|i>|^2\delta(\Omega_{ni}-\Omega)\rho_u(\Omega)d\Omega.
\end{eqnarray} 
هذا يمكن كتابته علي الشكل
\begin{eqnarray}
w^{\rm abso}_{i\longrightarrow [n]}
&=&\frac{\pi }{3\epsilon_0\hbar^2} \bigg[|<n|\vec{P}|i>|^2\rho_u(\Omega)\bigg]_{\Omega=\Omega_{ni}}.
\end{eqnarray}

\newpage
\section*{
تمارين
}
\addcontentsline{toc}{section}{\bf
تمارين
} 


\paragraph{تمرين
$1$}
\begin{itemize}
\item[$(1$] 
نذكر بمعادلة تطور القيم المتوسطة التي تعطي ب
\begin{eqnarray}
i\hbar \frac{d}{d t}<\hat{Q}>=<[\hat{Q},\hat{H}]>.\nonumber
\end{eqnarray}
استعمل هذه المعادلة لتبرهن علي
\begin{eqnarray}
  \frac{d}{d t}<\hat{x}\hat{p}>=2<\hat{T}>-<\hat{x}\frac{\partial \hat{V}}{\partial {x}}(\hat{x})>.\nonumber
\end{eqnarray}
\item[$(2$] 
برهن علي النظرية الفيريالية
\footnote{.${\rm virial~theorem}$}
\begin{eqnarray}
 2<\hat{T}>=<\hat{x}\frac{\partial \hat{V}}{\partial \hat{x}}(\hat{x})>.\nonumber
\end{eqnarray}
\item[$(3$] 
من اجل ذرة الهيدروجين استعمل النظرية الفيريالية لتبين ان
\begin{eqnarray}
 <\hat{T}>=-\frac{<\hat{V}>}{2}=-E_n.\nonumber
\end{eqnarray}
\end{itemize}
\paragraph{تمرين
$2$}
\begin{itemize}
\item[$(1$] 
لتكن
$H$
هاميلتونية ماتتعلق بوسيط
$\lambda$ 
اي
$H=H(\lambda)$. 
القيم الذاتية و الاشعة الذاتية تتعلق بالتالي بالوسيط
$\lambda$
اي ان
$E_n=E_n(\lambda)$ و $|\psi_n>=|\psi_n(\lambda)>$.
برهن علي نظرية فايمان - هالمان

\begin{eqnarray}
\frac{\partial E_n}{\partial \lambda}=<\psi_n(\lambda)|\frac{\partial H}{\partial \lambda}|\psi_n(\lambda)>.\nonumber
\end{eqnarray}
\item[$(2$] 
الهاميلتونية الفعلية لدالة الموجة المدارية لذرة الهيدروجين تعطي ب
\begin{eqnarray}
H=-\frac{\hbar^2}{2m}\frac{d^2}{dr^2}+\frac{\hbar^2}{2m}\frac{l(l+1)}{r^2}-\frac{e^2}{4\pi\epsilon_0}\frac{1}{r}.\nonumber
\end{eqnarray}
طاقات بور تعطي ب
\begin{eqnarray}
E_n=-\frac{\alpha^2mc^2}{2(j_{\rm max}+l+1)^2}.\nonumber
\end{eqnarray}
استعمل نظرية فايمان - هالمان من اجل
 $\lambda=l$
لحساب
القيمة المتوسطة
$<1/r^2>$.
\end{itemize}
\paragraph{
تمرين
$3$}
بين ان المعادلة المدارية لذرة الهيدروجين يمكن ان تكتب علي الشكل
\begin{eqnarray}
\frac{d^2u}{dr^2}=\bigg[\frac{l(l+1)}{r^2}-\frac{2}{ar}+\frac{1}{n^2a^2}\bigg]u.\nonumber
\end{eqnarray}
نصف قطر بور معرف ب
 \begin{eqnarray}
a=\frac{4\pi\epsilon_0\hbar^2}{me^2}.\nonumber
\end{eqnarray}
استعمل المعادلة المدارية اعلاه من اجل ان تشتق علاقة كريمر
\begin{eqnarray}
\frac{s}{4}[(2l+1)^2-s^2]<r^{s-2}>-\frac{2s+1}{a}<r^{s-1}>+\frac{s+1}{n^2a^2}<r^s>=0.\nonumber
\end{eqnarray}
احسب القيمة المتوسطة
$<r^{-3}>$.
\paragraph{
تمرين
$4$}
جملة كمومية يمكن ان تتواجد في ثلاث حالات مستقلة خطيا. الهاميلتونية تعطي ب
\begin{eqnarray}
H^{\epsilon}=V_0\left(\begin{array}{ccc}
1-\epsilon &0 &0\\
0 &1 &\epsilon\\
0&\epsilon &2
\end{array}\right)~,~\epsilon<<1.\nonumber
\end{eqnarray}
\begin{itemize}
\item[$(1$]
حل مسالة القيم الذاتية للجملة غير المضطربة المعرفة ب
$\epsilon=0$.
\item[$(2$]
حل مسألة القيم الذاتية للجملة المضطربة من اجل اي قيمة ل
$\epsilon$.
\item[$(3$]
استعمل نظرية الاضطراب غير المنحلة من الرتبة الاولي و من الرتبة الثانية لايجاد التصحيح للقيمة الذاتية غير المنحلة ل
$H^0$.
قارن بالحل المضبوط.
\item[$(4$]
استعمل نظرية الاضطراب المنحلة من الرتبة الاولي لايجاد التصحيحات للقيمة الذاتية المضعفة الانحلال ل
$H^0$.
قارن بالنتيجة المضبوطة.
\end{itemize}
\paragraph{
تمرين
$5$
}

\begin{itemize}
\item[$(1$]
من اجل هزاز توافقي احادي البعد مؤثرات الموضع و كمية الحركة تعطي ب
\begin{eqnarray}
\hat{x}=\sqrt{\frac{\hbar}{2m\Omega}}(a^++a)~,~
\hat{p}=i\sqrt{\frac{\hbar m\Omega}{2}}(a^+-a).\nonumber
\end{eqnarray}
نعطي ايضا
\begin{eqnarray}
a|n>=\sqrt{n}|n-1>~,~a^+|n>=\sqrt{n+1}|n+1>.\nonumber
\end{eqnarray}
احسب
$<n^{'}|\hat{x}|n>$ و  $<n^{'}|\hat{x}^2|n>$.

\item[$(2$]
الهزاز التوافقي ثلاثي الابعاد معطي بالكمون
\begin{eqnarray}
V(r)=\frac{1}{2}m\Omega^2 ({x}^2+{y}^2+{z}^2).\nonumber
\end{eqnarray}
استعمل طريقة فصل المتغيرات من اجل حل معادلة شرودينغر المرافقة و عين الطاقات المسموح بها. عين انحلال كل مستوي طاقوي.
 
\item[$(3$]
ندخل الاضطراب
\begin{eqnarray}
\lambda H^1=\lambda x^2yz.\nonumber
\end{eqnarray}
استعمل نظرية الاضطراب غير المنحلة من الرتبة الاولي من اجل حساب تصحيح الحالة الاساسية.
\item[$(4$]
استعمل نظرية الاضطراب المنحلة من الرتبة الاولي من اجل ايجاد تصحيح الحالة المثارة الاولي.

\end{itemize}
\paragraph{
تمرين
$6$}
\begin{itemize}
\item[$(1$]
نعتبر جملة مشكلة من
ذرتين مستقطبتين تبعدان عن بعضهما البعض مسافة
$R$.
نأخذ كنموذج لهذه الجملة هزازان توافقيان مستقلان عن بعضهما البعض عبارة عن نوابض بثابت صلابة
 $k$.
نتصور الالكترونات ككتل نقطية
$m$
مرتبطة بهذه النوابض لكن الانوية ثقيلة الي الحد الذي يمكن ان نفترض معه انها ساكنة لا تتحرك في مراكز توازن النوابض. ازاحة الالكترونات تعطي ب
$x_1$ و $x_2$.
هاميلتونية الجملة تعطي ب 
\begin{eqnarray}
H^0=\frac{1}{2m}p_1^2+\frac{1}{2}k x_1^2+\frac{1}{2m}p_2^2+\frac{1}{2}k x_2^2.\nonumber
\end{eqnarray}
تفاعل كولومب بين الانوية يعطي بالكمون
$e^2/R$,
تفاعل كولومب بين النواة الاولي و الالكترون الثاني هو
$-e^2/(R+x_2)$,
تفاعل كولومب بين النواة الثانية و الالكترون الاول هو
$-e^2/(R-x_1)$ 
بينما تفاعل كولومب بين الالكترونات هو
$e^2/(R-x_1+x_2)$. 
التفاعل الكلي لكولومب
يعطي ب
 \begin{eqnarray}
H^1=\frac{1}{4\pi\epsilon_0}\bigg[\frac{e^2}{R}-\frac{e^2}{R-x_1}-\frac{e^2}{R+x_2}+\frac{e^2}{R-x_1+x_2}\bigg].\nonumber
\end{eqnarray}
بين انه من اجل
$|x_1|<<R$ و $|x_2|<<R$ 
فان
 \begin{eqnarray}
H^1=-\frac{e^2x_1x_2}{2\pi\epsilon_0R^3}.\nonumber
\end{eqnarray}
\item[$(2$] 
نقترح تغيير المتغيرات
\begin{eqnarray}
x_1=\frac{1}{\sqrt{2}}(x_++x_-)~,~x_2=\frac{1}{\sqrt{2}}(x_+-x_-).\nonumber
\end{eqnarray}
\begin{eqnarray}
p_1=\frac{1}{\sqrt{2}}(p_++p_-)~,~p_2=\frac{1}{\sqrt{2}}(p_+-p_-).\nonumber
\end{eqnarray}
احسب الطاقات المسموح بها و الحالات الذاتية المرافقة.
\item[$(3$] 
احسب الفرق
$\Delta V=E-E_0$ 
حيث
$E$ و $E_0$
هي طاقات الحالة الاساسية ب و بدون تفاعل كولومب.
\item[$(4$]
باعتبار
$H^0$ 
هي الهاميلتونية غير المضطربة و
$H^1$ 
هي الاضطراب احسب التصحيحات من الرتبة الاولي و الرتبة الثانية لطاقة الحالة الاساسية. ماذا تستنتج.
\end{itemize}
\paragraph{
تمرين
$7$}
ليكن
$\vec{J}$
العزم الحركي الكلي
$\vec{J}=\vec{L}+\vec{S}$.
الاشعة الذاتية
 $|jj_3>$ 
ل
$J^2$, $J_3$, $L^2$ و $S^2$ 
هي عبارة عن تركيب خطي للاشعة الذاتية
 $|lm>|s\sigma>$
ل
$L^2$, $L_3$, $S^2$ و $S_3$
بمعاملات
 $C_{jj_3}^{lms\sigma}$ 
تعرف باسم معاملات كلابش - جوردون.
احسب هذه المعاملات من اجل
$s=\frac{1}{2}$.
\paragraph{
تمرين
$8$}
\begin{itemize}
\item[$(1$]
اضطراب البنية الدقيقة لذرة الهيدروجين يعطي بالهاميلتونية
\begin{eqnarray}
H^1_{fs}&=&H^1_r+H^1_{so}\nonumber\\
&=&-\frac{p^4}{8m^3c^2}+(\frac{1}{2})(-\vec{\mu}_s.\vec{B}_{\rm int}).\nonumber
\end{eqnarray} 
المعامل 
 $1/2$
بين قوسين هو راجع لمداورة توماس.
العزم المغناطيسي المرفق بسبين الالكترون و الحقل المغناطيسي الداخلي الذي تولده الحركة المدارية يعطيان ب
\begin{eqnarray}
\vec{\mu}_s=-\frac{e}{m}\vec{S}~,~\vec{B}_{\rm int}=\frac{1}{4\pi\epsilon_0}\frac{e}{mc^2r^3}\vec{L}.\nonumber
\end{eqnarray}
احسب التصحيحات الناجمة عن هذا الاططراب لطاقات بور.
\item[$(2$]
نعتبر اضطراب اخر ناجم عن وجود حقل مغناطيسي خارجي غير منعدم. الهاميلتونية المرافقة
\begin{eqnarray}
H^1_Z=-(\vec{\mu}_s+\vec{\mu}_l).\vec{B}_{\rm ext}.\nonumber
\end{eqnarray} 
العزم المغناطيسي المرتبط بالعزم الحركي للالكترون هو
\begin{eqnarray}
\vec{\mu}_l=-\frac{e}{2m}\vec{L}.\nonumber
\end{eqnarray}
هذا الاضطراب يؤدي الي تأثير زيمان.

نعتبر اولا الجملة غير المضطربة المعرفة بهاميلتونية بور. حدد درجة انحلال المستوي الطاقوي
$E_2$
و اكتب الحالات الذاتية المقابلة. عبر عن الاشعة الذاتية
$|lsjj_3>$ 
بدلالة الاشعة الذاتية
$|lm>|s\sigma>$
باستعمال نتيجة المسألة السابقة.
\item[$(3$] 
احسب عناصر المصفوفة
$<\psi_{njj_3}|H^1_Z|\psi_{njj_3}>$. 
\item[$(4$] 
احسب عناصر المصفوفة
$<\psi_{njj_3}|H^1_{fs}|\psi_{njj_3}>$. 
\item[$(5$] 
عين مصفوفة الاضطراب الكلية
$W=H^1_{fs}+H^1_Z$ 
في الحالات التي لها
$n=2$. 
استعمل نظرية الاضطراب المنحلة لحساب التصحيحات من الرتبة الاولي للمستوي الطاقوي
$E_2$. 
\end{itemize}


\paragraph{
تمرين
$9$
}
قيم الطاقة و دوال الموجة لبئر كمون لا نهائي في بعد واحد تعطي ب
\begin{eqnarray}
E_n=\frac{n^2\pi^2\hbar^2}{2ma^2}~,~\psi_n(x)=\sqrt{\frac{2}{a}}\sin\frac{n\pi x}{a}~,~n=1,2,3,...\nonumber
\end{eqnarray}
 نؤثر باضطراب متعلق بالزمن خلال زمن
$T$
بحيث ان الكمون يصبح
\begin{eqnarray}
&&V(x)=V_0~,~0\leq x\leq \frac{a}{2}\nonumber\\
&&V(x)=0~,~\frac{a}{2}\leq x\leq 0\nonumber\\
&&V(x)=\infty~,~{\rm otherwise}.\nonumber
\end{eqnarray}
 توجد الجملة في اللحظة
$t=0$
في الحالة الاساسية
$n=1$.
ماهو الاحتمال ان تكون الجملة في اللحظة
$t=T$
قد قفزت الي الحالة المثارة الاولي
$n=2$.

\paragraph{
تمرين
$10$}
نضع هزاز توافقي احادي البعد تحت تأثير قوة منتظمة تتعلق بالزمن كالاتي
 \begin{eqnarray}
F(t)=\frac{F_0\tau}{2\pi\nu(\tau^2+t^2)}.\nonumber
\end{eqnarray}
يوجد الهزاز في الحالة الاساسية في اللحظة
$t=-\infty$.
احسب الاحتمال ان يكون الهزاز في اللحظة
$t=+\infty$ 
قد قفز الي الحالة المثارة الاولي.

\paragraph{
تمرين
$11$}
نعتبر جملة مشكلة من جسيمين سبينهما يساوي 
$1/2$.
هاميلتونية الجملة من اجل
$t<0$ 
تنعدم. من اجل
$t>0$
هاميلتونية الجملة تعطي ب

\begin{eqnarray}
H=\frac{4\Delta}{\hbar^2}\vec{S}_1\vec{S}_2.\nonumber
\end{eqnarray}
توجد الجملة في اللحظة 
$t<0$
في الحالة
$|+->$. 
\begin{itemize}
\item[$(1$]
احسب الاحتمال كدالة في الزمن ان نجد الجملة في الحالات
$|++>$, $|+->$, $|-+>$ و $|-->$ 
عن طريق حل المسألة بالضبط.
\item[$(2$]
احسب الاحتمال كدالة في الزمن ان نجد الجملة في الحالات
$|++>$, $|+->$, $|-+>$ و $|-->$ 
عن طريق حل المسألة باستعمال نظرية الاضطرابات المتعلقة بالزمن من الرتبة الاولي. قارن مع الحل المضبوط.
\end{itemize}
\paragraph{
تمرين
$12$}
نضع ذرة الهيدروجين في حقل كهربائي منتظم يتعلق بالزمن كالتالي

\[\vec{E}=0~,~t<0\]
\[\vec{E}=\vec{E}_0e^{-t/\tau}~,~\vec{E}_0=E_0\vec{k}~,~t\geq 0.\]
توجد الجملة من اجل
$t\leq 0$ 
في الحالة الاساسية 
$|\psi_{100}>$. 
استعمل نظرية الاضطرابات المتعلقة بالزمن من الرتبة الاولي لحساب الاحتمال كدالة في الزمن ان نجد ذرة الهيدروجين في الحالات
$|\psi_{200}>$ و $|\psi_{210}>$.
نعطي دوال الموجة 
\[\psi_{100}=\frac{1}{a^{1.5}}\frac{1}{\sqrt{\pi}}e^{-r/a}.\]
\[\psi_{200}=\frac{1}{a^{1.5}}\frac{1}{\sqrt{32}}\frac{1}{\sqrt{4\pi}}e^{-r/2a}(-{2r}/{a}+4).\]
\[\psi_{210}=\frac{1}{a^{1.5}}\frac{1}{\sqrt{24.12}}\frac{1}{\sqrt{4\pi}}\cos\theta e^{-r/2a}({6r}/{a}).\]

\paragraph{
تمرين
$13$}
\begin{itemize}
\item[$(1$]
نعتبر جملة غير مضطربة معطاة بذرة بور. اكتب القيم الذاتية للطاقة و الاشعة الذاتية المرافقة لها. ما هي درجة الانحلال.
\item[$(2$]نأخذ بعين الاعتبار سبينات الالكترون و البروتون. ماهي في هذه الحالة اشعة الحالات الذاتية للطاقة و درجة انحلالها.
\item[$(3$]يولد سبين الالكترون عزم مغناطيسي ثنائي. اكتب هاميلتونية الالكترون في حقل مغناطيسي 
$\vec{B}$.
\item[$(4$] 
سبين البروتون يوافق ايضا عزم مغناطيسي ثنائي معطي ب
\[\vec{\mu}_p=\frac{eg_p}{2m_p}\vec{S}_p~,~g_p=5.59.\]  
هذا العزم يولد في اي نقطة
$\vec{r}$
 حقل مغناطيسي
$\vec{B}_p$
معطي ب
\[\vec{B}_p=\frac{\mu_0}{4\pi r^3}\bigg[3(\vec{\mu}_p.\hat{r})\hat{r}-\vec{\mu}_p\bigg]+\frac{2\mu_0}{3}\vec{\mu}_p\delta^3(\vec{r}).\]  
اكتب هاميلتونية الالكترون في هذا الحقل. تسمي هذه الهاميلتونية هاميلتونية البنية فائقة الدقة لذرة بور.
\item[$(5$]احسب التصحيح الكمي من الرتبة الاولي للطاقة الناجم عن الحد الاول
في الحقل المغناطيسي
$\vec{B_p}$.
\item[$(6$]بين ان التصحيح الكمي من الرتبة الاولي للطاقة الناجم عن الحد الثاني في الحقل المغناطيسي
$\vec{B_p}$
 المتناسب مع دالة ديراك يأخذ الشكل
\[E^1_{hf}=\frac{\mu_0e^2g_p}{3m_em_p}<\sigma^{'}|<\sigma|\vec{S}_e\vec{S}_p|\sigma>|\sigma^{'}>|\psi_{nlm}(0)|^2.\]
\item[$(7$]ماهو التصحيح الكمي من الرتبة الاولي لطاقة المستوي الاساسي لذرة بور.
\item[$(8$]اكتب الحالات الذاتية الجيدة لذرة بور في هذه الحالة.

استعمل:
\[\int d\Omega (\vec{a}.\hat{r})(\vec{b}.\hat{r})=\frac{4\pi}{3}\vec{a}.\vec{b}.\]  
\[|\psi_{nlm}(0)|^2=\frac{1}{\pi a^3}.\]
\end{itemize}

\paragraph{
تمرين
$14$}
هاميلتونية ذرة الهيليوم تعطي ب
\begin{eqnarray}
H=\frac{\vec{p}_1^2}{2m}-\frac{1}{4\pi\epsilon_0}\frac{2e^2}{r_1}+\frac{\vec{p}_2^2}{2m}-\frac{1}{4\pi\epsilon_0}\frac{2e^2}{r_2}
+\frac{1}{4\pi\epsilon_0}\frac{e^2}{|\vec{r}_1-\vec{r}_2|}.
\end{eqnarray}
القيمة التجريبية لطاقة الحالة الاساسية لذرة الهيليوم تعطي ب
$E=-78.975 eV$.
استعمل نظرية الاضطرابات في الرتبة الاولي لحساب القيمة النظرية لطاقة الحالة الاساسية لذرة الهيليوم.

\paragraph{
تمرين
$15$}
يوجد هزاز توافقي احادي البعد في حالته الاساسية
$|0>$ 
 من اجل
 $t<0$.
نؤثر علي الهزاز بقوة منتظمة في الاتجاه
$x$
متعلقة بالزمن معطاة ب

\[F=F_0e^{-\frac{t}{\tau}}~,~t\geq 0.\]
احسب احتمال ايجاد الهزاز في اللحظة الزمنية
$t>0$
في الحالات المثارة 
$|n>$
باستعمال نظرية الاضطرابات
المتعلقة بالزمن من الرتبة الاولي.
احسب النهاية
$\tau\longrightarrow \infty$.
ماذا تلاحظ. استعمل 
\[<n^{'}|x|n>=\sqrt{\frac{\hbar}{4m\pi\nu}}(\sqrt{n}\delta_{n^{'},n-1}+\sqrt{n+1}\delta_{n^{'},n+1}).\]

\paragraph{
تمرين
$16$}
\begin{itemize}
\item[$(1$] 
نعتبر الكمون اللانهائي في ثلاث ابعاد المعطي ب
\begin{eqnarray}
&&V(x,y,z)=0~,~{\rm if}~0<x<a~,~0<y<a~,~0<z<a\nonumber\\
&&V(x,y,z)=\infty~,~{\rm otherwise}.\nonumber
\end{eqnarray}
اشتق قيم الطاقة المسموح بها و الدوال الموجية الذاتية المرافقة لها.

ملحوظة: قيم الطاقة المسموح بها و الدوال الموجية الذاتية المرافقة لها من اجل الكمون اللانهائي في بعد واحد تعطي ب
\[E_n=En^2~,~E=\frac{\pi^2\hbar^2}{2ma^2}.\]
\[\psi_n(x)=\sqrt{\frac{2}{a}}\sin\frac {n\pi}{a}x~,~\int_0^a dx~ \psi_n^*(x)\psi_m(x)=\delta_{nm}.\]
\item[$(2$] 
ندخل الاضطراب
\begin{eqnarray}
&&H^1=V_0~,~{\rm if}~0<x<\frac{a}{2}~,~0<y<\frac{a}{2}\nonumber\\
&&H^1=0~,~{\rm otherwise}.\nonumber
\end{eqnarray}
احسب التصحيح من الرتبة الاولي لطاقة الحالة الاساسية.
\item[$(3$]
احسب التصحيح من الرتبة الاولي للمستوي المثار الاول ثلاثي الانحلال.
\end{itemize}

\newpage
\section*{
حلول
}
\addcontentsline{toc}{section}{\bf
حلول
}

\paragraph{
تمرين
 $1$:}
\begin{itemize}
\item[$(1$] 
نحتار
$\hat{Q}=(\hat{x}\hat{p}+\hat{p}\hat{x})/2$
و نستعمل
$[\hat{x},\hat{p}]=i\hbar$
و
$[\hat{p},\hat{V}(\hat{x})]=-i\hbar\frac{\partial \hat{V}}{\partial \hat{x}}(\hat{x})$.
\item[$(2$]
القيم المنتظرة في الحالات المستقرة لا تتعلق بالزمن.
\item[$(3$]
اولا نعمم المبرهنة الفيريالية لثلاث ابعاد. ثم نستعمل
$\vec{r}.\vec{\nabla}V=r{\partial V}/{\partial r}$, $V=-{Ke^2}/r$ 
و
$<H>=E_n$
لنبين ان المبرهنة الفيريالية تصبح
$2<\hat{T}>=-<\hat{V}>$. 
القيم المنتظرة تحسب في الحالات
$|\psi_{nlm}>$.
\end{itemize}
\paragraph{
تمرين
 $2$:}
\begin{itemize}
\item[$(1$] 
نقوم بالنشر
\begin{eqnarray}
H(\lambda)=H(0)+\lambda\frac{\partial H}{\partial \lambda}|_{\lambda=0}+O(\lambda^2).
\end{eqnarray}
نعتبر الحد الاول الهاميلتونية غير المضطربة بينما نعتبر الحدود الاخري كاضطراب. مسألة القيم الذاتية غير المضطربة تعطي اذن ب
\begin{eqnarray}
H(0)|\psi_n(0)>=E_n(0)|\psi_n(0)>.
\end{eqnarray}
التصحيح من الرتبة الاولي يعطي اذن ب
\begin{eqnarray}
E_n^1=<\psi_n(0)|\bigg[\frac{\partial H}{\partial \lambda}|_{\lambda=0}+O(\lambda)\bigg]|\psi_n(0)>.
\end{eqnarray}
الطاقة
$E_n(\lambda)$
تعطي اذن ب
\begin{eqnarray}
E_n(\lambda)=E_n(0)+\lambda<\psi_n(0)|\bigg[\frac{\partial H}{\partial \lambda}|_{\lambda=0}+O(\lambda)\bigg]|\psi_n(0)>.
\end{eqnarray}
نستنتج ان
\begin{eqnarray}
\frac{\partial E_n(\lambda)}{\partial \lambda}|_{\lambda=0}=<\psi_n(0)|\frac{\partial H}{\partial \lambda}|_{\lambda=0}|\psi_n(0)>.
\end{eqnarray}
\item[$(2$]
نحسب 
$\partial H/\partial l$ 
و
$\partial E_n/\partial l$. 
نجد
\begin{eqnarray}
<\frac{1}{r^2}>=\frac{\alpha^2m^2c^2}{\hbar^2}\frac{1}{n^3(l+\frac{1}{2})}.
\end{eqnarray}
\end{itemize}
\paragraph{
تمرين
 $3$:}
 المعادلة المدارية لذرة الهيدروجين يمكن ان تكتب علي الشكل
\begin{eqnarray}
\frac{d^2u}{dr^2}=\bigg[\frac{l(l+1)}{r^2}-\frac{2}{ar}+\frac{1}{n^2a^2}\bigg]u.
\end{eqnarray}
باستعمال هذه المعادلة يمكن ان نحسب مباشرة
\begin{eqnarray}
\int u r^s \frac{d^2u}{dr^2}dr&=&\int u r^s\bigg[\frac{l(l+1)}{r^2}-\frac{2}{ar}+\frac{1}{n^2a^2}\bigg]u dr\nonumber\\
&=&l(l+1)<r^{s-2}>-\frac{2}{a}<r^{s-1}>+\frac{1}{n^2a^2}<r^s>.
\end{eqnarray}
بالتكامل بالتجزئة نحصل علي
\begin{eqnarray}
\int u r^s \frac{d^2u}{dr^2}dr&=&-\int r^s(\frac{du}{dr})^2 dr -s \int u r^{s-1}\frac{du}{dr} dr\nonumber\\
&=&\frac{2}{s+1}\int\frac{du}{dr}\frac{d^2u}{dr^2} r^{s+1}dr+ \frac{s(s-1)}{2}<r^{s-2}>.
\end{eqnarray}
نحسب ايضا
\begin{eqnarray}
\int\frac{du}{dr}\frac{d^2u}{dr^2} r^{s+1}dr&=&-\frac{l(l+1)(s-1)}{2}<r^{s-2}>+\frac{s}{a}<r^{s-1}>-\frac{s+1}{2n^2a^2}<r^s>.\nonumber\\
\end{eqnarray}
نضع كل شيئ معا نحصل
\begin{eqnarray}
\frac{s}{4}[(2l+1)^2-s^2]<r^{s-2}>-\frac{2s+1}{a}<r^{s-1}>+\frac{s+1}{n^2a^2}<r^s>=0.
\end{eqnarray}
من اجل
$s=-1$
نحصل علي
\begin{eqnarray}
-\frac{1}{4}[(2l+1)^2-1]<r^{-3}>+\frac{1}{a}<r^{-2}>=0.
\end{eqnarray}
بالتالي
\begin{eqnarray}
<r^{-3}>=\frac{1}{a l(l+1)}<r^{-2}>=\frac{\alpha^3m^3c^3}{\hbar^3 n^3l(l+1)(l+\frac{1}{2})}.
\end{eqnarray}
\paragraph{
تمرين
 $4$:}
 الحل مباشر.
\paragraph{
تمرين
$5$:}
\begin{itemize}
\item[$(1$]
نجد
\begin{eqnarray}
<n^{'}|\hat{x}|n>=\sqrt{\frac{\hbar}{2m\Omega}}\bigg(\sqrt{n+1}\delta_{n^{'},n+1}+\sqrt{n}\delta_{n^{'},n-1}\bigg).
\end{eqnarray}
\begin{eqnarray}
<n^{'}|\hat{x}^2|n>={\frac{\hbar}{2m\Omega}}\bigg(\sqrt{(n+1)(n+2)}\delta_{n^{'},n+2}+\sqrt{n(n-1)}\delta_{n^{'},n-2}+(2n+1)\delta_{n^{'},n}\bigg).
\end{eqnarray}
\item[$(2$]
في هذه الحالة تعطي معادلة شرودينغر ب
\begin{eqnarray}
\bigg(-\frac{\hbar^2}{2m}\frac{\partial^2}{\partial x^2}-\frac{\hbar^2}{2m}\frac{\partial^2}{\partial y^2}-
\frac{\hbar^2}{2m}\frac{\partial^2}{\partial z^2}+\frac{1}{2}m\Omega^2(x^2+y^2+z^2)\bigg)\Psi(x,y,z)=E\Psi(x,y,z).
\end{eqnarray}
فصل المتغيرات يعطي مباشرة الطاقات المسموح بها
\begin{eqnarray}
E_{n}=\hbar\Omega(n_x+n_y+n_z+\frac{3}{2})~,~n=n_x+n_y+n_z.
\end{eqnarray}
الحالات المقابلة تعطي ب
\begin{eqnarray}
\Psi(x,y,z)=<x|n_x><y|n_y><z|n_z>.
\end{eqnarray}
انحلال المستوي الطاقوي
$E_n$
حيث
$n=n_x+n_y+n_z$
يبقي مثبت يحسب كالتالي. اولا نثبت
 $n_x$ 
 اي
 $n_y+n_z=n-n_x$.
 من الواضح انه لدينا
 $n-n_x+1$
 امكانية من اجل الزوج
 $(n_y,n_z)$. 
 اذن انحلال
 $E_n$
 يعطي بالعلاقة
\begin{eqnarray}
d(n)=\sum_{n_x=0}^n(n-n_x+1)=\frac{n(n+1)}{2}.
\end{eqnarray}
\item[$(3$]
التصحيح من الرتبة الاولي لطاقة الحالة الاساسية
$E_{000}=(3\hbar\Omega)/2$ 
التي هي حالة غير منحلة يأخذ الشكل
\begin{eqnarray}
\lambda E^1&=&<0|<0|<0|\lambda x^2 y z|0>|0>|0>\nonumber\\
&=&\lambda <0|x^2|0><0|y|0><0|z|0>\nonumber\\
&=&0.
\end{eqnarray}
\item[$(4$]
المستوي الطاقوي المثار الاول
$E_1=(5\hbar\Omega)/2$ 
هو ثلاثي الانحلال. الحالات المقابلة هي
$|100>$, $|010>$ 
و
$|001>$.
حتي نحسب التصحيح من الرتبة الاولي نستعمل نظرية الاضطراب المنحلة من الرتبة الاولي. اذن يجب ان نجد القيم الذاتية لمصفوفة الاضطراب
$W_{i j}=<i|\lambda H^1|j>$
حيث
$i,j=100,010,001$.
نحسب
\begin{eqnarray}
W=\left(\begin{array}{ccc}
0 &0 &0\\
0 &0 &\epsilon\\
0&\epsilon &0
\end{array}\right)~,~\epsilon=\lambda\sqrt{2}(\frac{\hbar}{2m\Omega})^2.
\end{eqnarray}
القيم الذاتية هي
$0$, $+\epsilon$ 
و
$-\epsilon$
مع الاشعة الذاتية
$|100>$, $(|010>+|001>)/\sqrt{2}$
و
$(|010>-|001>)/\sqrt{2}$
علي التوالي.
\end{itemize}
\paragraph{
تمرين
 $6$:}
\begin{itemize}
\item[$(1$]
استعمل نشر تايلور.
\item[$(2$]
نجد
\begin{eqnarray}
E_{n_{+},n_{-}}=\hbar\Omega_{+}(n_{+}+\frac{1}{2})+\hbar\Omega_{-}(n_{-}+\frac{1}{2}).
\end{eqnarray}
\begin{eqnarray}
\Omega_{\pm}=\sqrt{\frac{k\mp\frac{e^2}{2\pi\epsilon_0R^3}}{m}}.
\end{eqnarray}
\item[$(3$]
نجد
\begin{eqnarray}
E=E_{0,0}=\hbar\frac{\Omega_{+}+\Omega_{+}}{2}.
\end{eqnarray}
\begin{eqnarray}
E_0=\hbar \Omega_{0}.
\end{eqnarray}
\begin{eqnarray}
\Omega_{0}=\sqrt{\frac{k}{m}}.
\end{eqnarray}
نحصل علي
\begin{eqnarray}
\Delta V=E-E_0=-\frac{\hbar}{8m^2\Omega_0^3}\bigg(\frac{e^2}{2\pi\epsilon_0R^3}\bigg)^2.
\end{eqnarray}
\item[$(4$]
التصحيح من الرتبة الاولي هو
\begin{eqnarray}
E^1=<0|<0|H^1|0>|0>=0.
\end{eqnarray}
التصحيح من الرتبة الثانية
\begin{eqnarray}
E^2=\sum_{m_1\neq 0}\sum_{m_2\neq 0}\frac{|<m_1|<m_2|H^1|0>|0>|^2}{E_0^0-E_m^0}.
\end{eqnarray}
نستعمل النتيجة
\begin{eqnarray}
H^1|0>|0>=-\frac{e^2}{2\pi\epsilon_0R^3}\hat{x}_1|0>\hat{x}_2|0>=-\frac{e^2}{2\pi\epsilon_0R^3}\frac{\hbar}{2m\Omega_0}|1>|1>.
\end{eqnarray}
اذن
(
مع
 $E_0^0=\hbar\Omega_0$
 و
 $E_1^0=3\hbar\Omega_0$
)
\begin{eqnarray}
E^2=\bigg(\frac{e^2}{2\pi\epsilon_0R^3}\bigg)^2\bigg(\frac{\hbar}{2m\Omega_0}\bigg)^2\frac{1}{E_0^0-E_1^0}=-\bigg(\frac{e^2}{2\pi
\epsilon_0R^3}\bigg)^2\frac{\hbar}{8m^2\Omega_0^3}.
\end{eqnarray}
\end{itemize}
\paragraph{
تمرين
 $7$:}
 لدينا
\begin{eqnarray}
|jj_3>&=&C_{jj_3}^{lj_3-\frac{1}{2}\frac{1}{2}\frac{1}{2}}|lj_3-\frac{1}{2}>|\frac{1}{2}\frac{1}{2}>+C_{jj_3}^{lj_3+\frac{1}{2}\frac{1}{2}-\frac{1}{2}}|lj_3+\frac{1}{2}>|\frac{1}{2}-\frac{1}{2}>\nonumber\\
&=&A|lj_3-\frac{1}{2}>|\frac{1}{2}\frac{1}{2}>+B|lj_3+\frac{1}{2}>|\frac{1}{2}-\frac{1}{2}>
\end{eqnarray}
يجب ان يكون لدينا
$|A|^2+|B|^2=1$. 
ايضا لدينا
\begin{eqnarray}
J^2=L^2+S^2+2L_3S_3+L_+S_-+L_-S_+.
\end{eqnarray}
نحسب
\begin{eqnarray}
J^2|lj_3-\frac{1}{2}>|\frac{1}{2}\frac{1}{2}>&=&\bigg[l(l+1)+j_3+\frac{1}{4}\bigg]|lj_3-\frac{1}{2}>|\frac{1}{2}\frac{1}{2}>\nonumber\\
&+&\sqrt{l(l+1)-j_3^2+\frac{1}{4}}|lj_3+\frac{1}{2}>|\frac{1}{2}-\frac{1}{2}>.
\end{eqnarray}
\begin{eqnarray}
J^2|lj_3+\frac{1}{2}>|\frac{1}{2}-\frac{1}{2}>&=&\bigg[l(l+1)-j_3+\frac{1}{4}\bigg]|lj_3+\frac{1}{2}>|\frac{1}{2}-\frac{1}{2}>\nonumber\\
&+&\sqrt{l(l+1)-j_3^2+\frac{1}{4}}|lj_3-\frac{1}{2}>|\frac{1}{2}\frac{1}{2}>.
\end{eqnarray}
الشرط
$J_3^2|jj_3>=j(j+1)|jj_3>$
يؤدي الي معادلتين متكافئتين في المجهولين
$A$
و
 $B$.
 المعادلة الاولي تأخذ الشكل
\begin{eqnarray}
A\bigg[l(l+1)+j_3+\frac{1}{4}\bigg]+B\sqrt{l(l+1)-j_3^2+\frac{1}{4}}=j(j+1)A.
\end{eqnarray}
من اجل
$j=l+\frac{1}{2}$ 
لدينا
\begin{eqnarray}
A=\sqrt{\frac{l+\frac{1}{2}+j_3}{2l+1}}~,~B=\sqrt{\frac{l+\frac{1}{2}-j_3}{2l+1}}.
\end{eqnarray}
من اجل
$j=l-\frac{1}{2}$
لدينا
\begin{eqnarray}
A=\sqrt{\frac{l+\frac{1}{2}-j_3}{2l+1}}~,~B=-\sqrt{\frac{l+\frac{1}{2}+j_3}{2l+1}}.
\end{eqnarray}

\paragraph{
تمرين
 $8$:}
\begin{itemize}
\item[$(1$]
انظر المحاضرة.
\item[$(2$]
من اجل
$n=2$ 
لدينا
$l=0$
و
$l=1$.
لما نجمع
$l=0$
و
$s=\frac{1}{2}$
نحصل علي
$j=\frac{1}{2}$
و لما نجمع
$l=1$
و
$s=\frac{1}{2}$ 
نحصل علي
$j=\frac{1}{2}$
و
$j=\frac{3}{2}$. 
مستوي بور الطاقوي 
$E_2$
هو ثماني الانحلال و هو يعطي ب
\begin{eqnarray}
E_2=-13.6 eV/4~,~-(\frac{\alpha}{4})^2E_2=(\frac{\alpha}{8})^213.6 eV=\gamma.
\end{eqnarray}
الحالات الذاتية الثمانية المقابلة هي
 $|\psi_{njj_3}>=|R_{nl}>|lsjj_3>$ 
 حيث
 $n=2$, $s=\frac{1}{2}$, $l=0,1$
 و
 $j=l+\frac{1}{2},l-\frac{1}{2}$.
 يجب ان نعبر عن الاشعة الذاتية
 $|lsjj_3>$
 بدلالة الاشعة الذاتية
 $|l m s\sigma>$. 
 نستعمل نتيجة التمرين السابق لايجاد
\begin{eqnarray}
&&|\psi_1>=|0\frac{1}{2}\frac{1}{2}\frac{1}{2}>=|00>|\frac{1}{2}\frac{1}{2}>\nonumber\\
&&|\psi_2>=|0\frac{1}{2}\frac{1}{2}-\frac{1}{2}>=|00>|\frac{1}{2}-\frac{1}{2}>.
\end{eqnarray}
\begin{eqnarray}
&&|\psi_6>=|1\frac{1}{2}\frac{1}{2}\frac{1}{2}>=\frac{1}{\sqrt{3}}|10>|\frac{1}{2}\frac{1}{2}>-\sqrt{\frac{2}{3}}|11>|\frac{1}{2}-\frac{1}{2}>\nonumber\\
&&|\psi_8>=|1\frac{1}{2}\frac{1}{2}-\frac{1}{2}>=\sqrt{\frac{2}{3}}|1-1>|\frac{1}{2}\frac{1}{2}>-\frac{1}{\sqrt{3}}|10>|\frac{1}{2}-\frac{1}{2}>.\nonumber\\
\end{eqnarray}
\begin{eqnarray}
&&|\psi_3>=|1\frac{1}{2}\frac{3}{2}\frac{3}{2}>=|11>|\frac{1}{2}\frac{1}{2}>\nonumber\\
&&|\psi_5>=|1\frac{1}{2}\frac{3}{2}\frac{1}{2}>=\sqrt{\frac{2}{3}}|10>|\frac{1}{2}\frac{1}{2}>+\frac{1}{\sqrt{3}}|11>|\frac{1}{2}-\frac{1}{2}>\nonumber\\
&&|\psi_7>=|1\frac{1}{2}\frac{3}{2}-\frac{1}{2}>=\frac{1}{\sqrt{3}}|1-1>|\frac{1}{2}\frac{1}{2}>+\sqrt{\frac{2}{3}}|10>|\frac{1}{2}-\frac{1}{2}>\nonumber\\
&&|\psi_4>=|1\frac{1}{2}\frac{3}{2}-\frac{3}{2}>=|1-1>|\frac{1}{2}-\frac{1}{2}>.
\end{eqnarray}
\item[$(3$]
اضطراب الحقل المغناطيسي الخارجي يعطي ب
\begin{eqnarray}
H^{1}_{Z}&=&-\vec{B}_{\rm ext}.(\vec{\mu}_l+\vec{\mu}_s)\nonumber\\
&=&\frac{e}{2m}B_{\rm ext}(L_3+2S_3)\nonumber\\
&=&\frac{\mu_BB_{\rm ext}}{\hbar}(L_3+2S_3)\nonumber\\
&=&\frac{\beta}{\hbar}(L_3+2S_3).
\end{eqnarray}
في المعادلة اعلاه افترضنا ان الحقل المعناطيسي هو في الاتجاه الثالث و ان
 $\mu_B$ 
 و
 $\beta$
 معرفان ب
\begin{eqnarray}
\mu_B=\frac{e\hbar}{2m}~,~\beta=\mu_BB_{\rm ext}.
\end{eqnarray}
نحسب
\begin{eqnarray}
&&(L_3+2S_3)|\psi_1>=\hbar|\psi_1>\nonumber\\
&&(L_3+2S_3)|\psi_2>=-\hbar|\psi_2>\nonumber\\
&&(L_3+2S_3)|\psi_3>=2\hbar|\psi_3>\nonumber\\
&&(L_3+2S_3)|\psi_4>=-2\hbar|\psi_4>\nonumber\\
&&(L_3+2S_3)|\psi_5>=\frac{2\hbar}{3}|\psi_5>+\frac{\sqrt{2}\hbar}{3}|\psi_6>\nonumber\\
&&(L_3+2S_3)|\psi_6>=\frac{\hbar}{3}|\psi_6>+\frac{\sqrt{2}\hbar}{3}|\psi_5>\nonumber\\
&&(L_3+2S_3)|\psi_7>=-\frac{2\hbar}{3}|\psi_7>+\frac{\sqrt{2}\hbar}{3}|\psi_8>\nonumber\\
&&(L_3+2S_3)|\psi_8>=-\frac{\hbar}{3}|\psi_8>+\frac{\sqrt{2}\hbar}{3}|\psi_7>.
\end{eqnarray}
اذن المركبات المصفوفية
 $<\psi_{njj_3}|H^1_{Z}|\psi_{nj^{'}j_3^{'}}>$ 
 يمكن تنظيمها في المصفوفة
\begin{eqnarray}
H^1_Z=\left(\begin{array}{cccccccc}
\beta &0 &0&0&0&0&0&0\\
0 &-\beta &0&0&0&0&0&0\\
0&0 &2\beta &0&0&0&0&0\\
0&0&0&-2\beta &0&0&0&0\\
0&0&0&0&\frac{2}{3}\beta &\frac{\sqrt{2}}{3}\beta &0&0\\
0&0&0&0&\frac{\sqrt{2}}{3}\beta &\frac{1}{3}\beta &0&0\\
0&0&0&0&0&0&-\frac{2}{3}\beta &\frac{\sqrt{2}}{3}\beta\\
0&0&0&0&0&0&\frac{\sqrt{2}}{3}\beta &-\frac{1}{3}\beta 
\end{array}\right).
\end{eqnarray}
\item[$(4$]
المركبات المصفوفية
$<\psi_{njj_3}|H^1_{fs}|\psi_{nj^{'}j_3^{'}}>$
تعطي ب
\begin{eqnarray}
<\psi_{njj_3}|H^1_{fs}|\psi_{njj_3}>&=&E^1_{fs}\delta_{j_3j_3^{'}}\delta_{jj^{'}}\nonumber\\
&=&\frac{E_n^2}{2mc^2}\bigg(3-\frac{4n}{j+\frac{1}{2}}\bigg)\delta_{j_3j_3^{'}}\delta_{jj^{'}}.
\end{eqnarray}
من اجل
$n=2$
لدينا صراحة
\begin{eqnarray}
H^1_{fs}=\left(\begin{array}{cccccccc}
-5\gamma &0 &0&0&0&0&0&0\\
0 &-5\gamma &0&0&0&0&0&0\\
0&0 &-\gamma &0&0&0&0&0\\
0&0&0&-\gamma &0&0&0&0\\
0&0&0&0&-\gamma &0 &0&0\\
0&0&0&0&0 &-5\gamma &0&0\\
0&0&0&0&0&0&-\gamma &0\\
0&0&0&0&0&0&0 &-5\gamma 
\end{array}\right).
\end{eqnarray}
\item[$(5$]
مصفوفة الاضطراب الكلية
$W=H^1_{fs}+H^1_Z$
تعطي ب
\begin{eqnarray}
W=\left(\begin{array}{cccccccc}
\beta-5\gamma &0 &0&0&0&0&0&0\\
0 &-\beta-5\gamma &0&0&0&0&0&0\\
0&0 &2\beta-\gamma &0&0&0&0&0\\
0&0&0&-2\beta-\gamma &0&0&0&0\\
0&0&0&0&\frac{2}{3}\beta-\gamma &\frac{\sqrt{2}}{3}\beta &0&0\\
0&0&0&0&\frac{\sqrt{2}}{3}\beta &\frac{1}{3}\beta-5\gamma &0&0\\
0&0&0&0&0&0&-\frac{2}{3}\beta-\gamma &\frac{\sqrt{2}}{3}\beta\\
0&0&0&0&0&0&\frac{\sqrt{2}}{3}\beta &-\frac{1}{3}\beta-5\gamma 
\end{array}\right).\nonumber\\
\end{eqnarray}
نلاحظ مباشرة القيم الذاتية
 $\pm \beta - 5\gamma$
 و
 $\pm 2\beta-\gamma$. 
 القيم الذاتية الاربعة الاخري تحل المعادلات المميزة
 \footnote{.${\rm characteristic~equations}$}
\begin{eqnarray}
x^2+(\beta-6\gamma)x+5\gamma^2-\frac{11}{3}\gamma\beta=0.
\end{eqnarray}
\begin{eqnarray}
y^2+(-\beta-6\gamma)y+5\gamma^2+\frac{11}{3}\gamma\beta=0.
\end{eqnarray}
نحصل علي الحلول
\begin{eqnarray}
x_{\pm}=3\gamma-\frac{\beta}{2}\pm\sqrt{4\gamma^2+\frac{\beta^2}{4}+\frac{2}{3}\beta\gamma}.
\end{eqnarray}
\begin{eqnarray}
y_{\pm}=3\gamma+\frac{\beta}{2}\pm\sqrt{4\gamma^2+\frac{\beta^2}{4}-\frac{2}{3}\beta\gamma}.
\end{eqnarray}
هذه القيم الذاتية هي التصحيحات من الرتبة الاولي للمستوي الطاقوي
$E_2$
الراجعة الي تأثيرات البنية الدقيقية و تأثير حقل مغناطيسي خارجي غير معدوم.
\end{itemize}


\paragraph{
تمرين
$9$:}
احتمال الانتقال يعطي ب

\begin{eqnarray}
P_{1\longrightarrow 2}(T)=|c_2^{(0)}(T)+c_2^{(1)}(T)+...|^2.
\end{eqnarray}
نحسب
\begin{eqnarray}
&&c_2^{(0)}(T)=\delta_{21}=0.
\end{eqnarray}
التصحيح من الرتبة الاولي يعطي ب
\begin{eqnarray}
c_2^{(1)}(T)
&=&\bigg(\frac{-i}{\hbar}\bigg)\int_{0}^Tdt e^{i\Omega_{21}t}V_{21}(t).
\end{eqnarray}

\begin{eqnarray}
\Omega_{12}=\frac{E_1-E_2}{\hbar}~,~V_{21}(t)=<2|V(t)|1>.
\end{eqnarray}
لان
$V$
ثابت خلال المجال الزمني
$T$
نحصل مباشرة علي

\begin{eqnarray}
c_2^{(1)}(T)
&=&V_{21}\frac{1-e^{i\Omega_{21}T}}{E_2-E_1}.
\end{eqnarray}
احتمال الانتقال من الرتبة الاولي يعطي اذن ب
\begin{eqnarray}
P_{1\longrightarrow 2}(T)&=&\frac{4|V_{21}|^2}{(E_2-E_1)^2}\sin^2\frac{(E_2-E_1)T}{2\hbar}\nonumber\\
&=&\bigg(\frac{4ma^2|V_{21}|}{3\pi^2\hbar^2}\sin\frac{3\pi^2\hbar T}{4ma^2}\bigg)^2.
\end{eqnarray}
يبقي ان نعين
$V_{21}$. 
لدينا
\begin{eqnarray}
V_{21}&=&<2|V|1>\nonumber\\
&=&\int_0^adx \psi_2^*(x)\psi_1(x)V(x)\nonumber\\
&=&V_0\int_0^{\frac{a}{2}}dx \psi_2^*(x)\psi_1(x)\nonumber\\
&=&\frac{V_0}{a}\int_0^{\frac{a}{2}}dx \bigg(\cos\frac{\pi x}{a}-\cos\frac{3\pi x}{a}\bigg)\nonumber\\
&=&\frac{4V_0}{3\pi}.
\end{eqnarray}
\paragraph{
تمرين
$10$:}
الحالة الاساسية و الحالة المثارة الاولي للهزاز التوافقي في بعد واحد تعطي بالدوال الموجية
\begin{eqnarray}
\psi_0(x)=\bigg(\frac{m\Omega}{\pi\hbar}\bigg)^{\frac{1}{4}}~e^{-\frac{m\Omega}{2\hbar}x^2}.
\end{eqnarray}
\begin{eqnarray}
\psi_1(x)=a^+\psi_0(x)=\bigg(\frac{m\Omega}{\pi\hbar}\bigg)^{\frac{1}{4}}\sqrt{\frac{2m\Omega}{\hbar}}~x~e^{-\frac{m\Omega}{2\hbar}x^2}.
\end{eqnarray}
الطاقات المقابلة تعطي ب
\begin{eqnarray}
E_0=\frac{\hbar\Omega}{2}~,~E_1=\frac{3\hbar\Omega}{2}.
\end{eqnarray}
 سعة احتمال الانتقال من الرتبة الصفر تعطي ب
\begin{eqnarray}
c_1^{(0)}(+\infty)=\delta_{10}=0.
\end{eqnarray}
سعة احتمال الانتقال من الرتبة واحد تعطي ب
\begin{eqnarray}
c_1^{(1)}(+\infty)&=&-\frac{i}{\hbar}\int_{-\infty}^{\infty}dt~e^{i\Omega_{10}t}V_{10}(t)\nonumber\\
&=&-\frac{i}{\hbar}\int_{-\infty}^{\infty}dt~e^{i\Omega t}<1|V(t)|0>.
\end{eqnarray}
القوة منتطمة في الفضاء و بالتالي فان الكمون يعطي ب
$V=-Fx$. 
نحسب
\begin{eqnarray}
<1|V(t)|0>=<1|(-Fx)|0>=-F\sqrt{\frac{\hbar}{2m\Omega}}<1|(a+a^+)|0>=-F\sqrt{\frac{\hbar}{2m\Omega}}.
\end{eqnarray}
اذن
\begin{eqnarray}
c_1^{(1)}(+\infty)
&=&\frac{i}{\sqrt{2m\Omega\hbar}}F_0\frac{\tau}{\Omega}\int_{-\infty}^{\infty}dt~\frac{e^{i\Omega t}}{\tau^2+t^2}\nonumber\\
&=&\frac{iF_0}{\Omega\sqrt{2m\Omega\hbar}}\int_{-\infty}^{\infty}dt~\frac{e^{i\Omega \tau t}}{1+t^2}.
\end{eqnarray}
ندخل تحويل لابلاس كالتالي
\begin{eqnarray}
\int_{-\infty}^{\infty} dt \frac{e^{i\Omega \tau t}}{1+t^2}&=&\int_{-\infty}^{\infty} dt \int_0^{\infty} d\alpha e^{-\alpha (1+t^2) +i\Omega\tau t}\nonumber\\
&=&\int d\alpha e^{-\alpha-\big(\frac{\Omega\tau}{2}\big)^2\frac{1}{\alpha}} \int_{-\infty}^{\infty}  dt e^{-\alpha \big(t-\frac{i\Omega\tau}{2\alpha}\big)^2}\nonumber\\
&=&\int d\alpha e^{-\alpha-\big(\frac{\Omega\tau}{2}\big)^2\frac{1}{\alpha}} \int_{-\infty}^{\infty}  dt e^{-\alpha t^2}\nonumber\\
&=&\int d\alpha e^{-\alpha-\big(\frac{\Omega\tau}{2}\big)^2\frac{1}{\alpha}} \sqrt{\frac{\pi}{\alpha}}\nonumber\\
&=&\sqrt{\frac{\pi\Omega\tau}{2}}\int \frac{d\alpha}{\alpha^{\frac{1}{2}}} e^{-\frac{\Omega\tau}{2}(\alpha+\frac{1}{\alpha})}\nonumber\\
&=&\sqrt{\frac{\pi\Omega\tau}{2}}2K_{-\frac{1}{2}}(\Omega\tau)\nonumber\\
&=&\sqrt{\frac{\pi\Omega\tau}{2}}2\sqrt{\frac{\pi}{2\Omega\tau}}e^{-\Omega\tau}\nonumber\\
&=&\pi e^{-\Omega\tau}.
\end{eqnarray}
اذن
\begin{eqnarray}
c_1^{(1)}(+\infty)
&=&\frac{iF_0}{\Omega\sqrt{2m\Omega\hbar}}\pi e^{-\Omega\tau}.
\end{eqnarray}
احتمال الانتقال يعطي اذن ب
\begin{eqnarray}
|c_1^{(1)}(+\infty)|^2
&=&\frac{F_0^2\pi^2}{2m\Omega^3\hbar} e^{-2\Omega\tau}.
\end{eqnarray}
 \paragraph{
تمرين
$11$:}
\begin{itemize}
 \item  $(1$
 علينا حل معادلة شرودينغر
 \begin{eqnarray}
H|\psi(t)>=i\hbar \frac{\partial}{\partial t}|\psi(t)>.
\end{eqnarray}
نكتب الحل علي الشكل
 \begin{eqnarray}
|\psi(t)>=\sum_n c_n\exp(-i E_n t/\hbar)|n>.
\end{eqnarray}
لدينا
\begin{eqnarray}
H=\frac{4\Delta}{\hbar^2}\vec{S}_1\vec{S_2}=2\Delta(s(s+1)-3/2),
\end{eqnarray}
حيث
$s$
هو السبين الكلي. اذن

\begin{eqnarray}
H|00>=E_0|00>~,~E_0=-3\Delta~,~|00>=\frac{1}{\sqrt{2}}(|+>|->-|->|+>).
\end{eqnarray}
\begin{eqnarray}
&&H|1m>=E_1|1m>~,~E_1=\Delta~,~\nonumber\\
&&|11>=|+>|+>~,~|10>=\frac{1}{\sqrt{2}}(|+>|->+|->|+>)~,~|1-1>=|->|->.\nonumber\\
\end{eqnarray}
دالة الموجة في اللحظة
$t$
تعطي اذن ب
\begin{eqnarray}
|\psi(t)>&=&c_{00}\exp(-i E_0 t/\hbar)|00>+c_{11}\exp(-i E_1 t/\hbar)|11>+c_{10}\exp(-i E_1 t/\hbar)|10>\nonumber\\
&+&c_{1-1}\exp(-i E_1 t/\hbar)|1-1>.
\end{eqnarray}
دالة الموجة الابتدائية هي
\begin{eqnarray}
|\psi(0)>=|+>|->=\frac{1}{\sqrt{2}}|00>+\frac{1}{\sqrt{2}}|10>.
\end{eqnarray}
اذن بالمقارنة

\begin{eqnarray}
c_{00}=c_{10}=\frac{1}{\sqrt{2}}~,~c_{11}=c_{1-1}=0.
\end{eqnarray}
دالة الموجة في اللحظة
$t$
تصبح اذن
\begin{eqnarray}
|\psi(t)>&=&\frac{1}{\sqrt{2}}\exp(-i E_0 t/\hbar)|00>+\frac{1}{\sqrt{2}}\exp(-i E_1 t/\hbar)|10>\nonumber\\
&=&\frac{1}{2}\big(\exp(-i E_0 t/\hbar)+\exp(-i E_1 t/\hbar)\big)|+>|->-\frac{1}{2}\big(\exp(-i E_0 t/\hbar)-\exp(-i E_1 t/\hbar)\big)|->|+>.\nonumber\\
\end{eqnarray}
نحصل علي الاحتمالات

\begin{eqnarray}
&&P(|+>|->\longrightarrow |+>|+>)=0\nonumber\\
&&P(|+>|->\longrightarrow |->|->)=0\nonumber\\
&&P(|+>|->\longrightarrow |+>|->)=\frac{1}{4}|\exp(-i E_0 t/\hbar)+\exp(-i E_1 t/\hbar)|^2\nonumber\\
&&P(|+>|->\longrightarrow |->|+>)=\frac{1}{4}|\exp(-i E_0 t/\hbar)-\exp(-i E_1 t/\hbar)|^2.\nonumber\\
\end{eqnarray}

 \item  $(2$
 الهاميلتونية غير المضطربة
 \begin{eqnarray}
H_0=0.
\end{eqnarray}
الاضطراب يعطي ب
\begin{eqnarray}
V=H=\frac{4\Delta}{\hbar^2}\vec{S}_1\vec{S_2}.
\end{eqnarray}
الحالة الابتدائية
\begin{eqnarray}
|i>=|+>|->.
\end{eqnarray}
نحسب عنصر المصفوفة
\begin{eqnarray}
V_{ni}=<n|V|i>&=&\frac{4\Delta}{\hbar^2}<n|\vec{S}_1\vec{S_2}|+>|->\nonumber\\
&=&-\frac{3}{\sqrt{2}}\Delta <n|00>+\frac{\Delta}{\sqrt{2}}<n|10>.
\end{eqnarray}
من اجل
\begin{eqnarray}
|n>=|->|+>=\frac{1}{\sqrt{2}}|10>-\frac{1}{\sqrt{2}}|00>,
\end{eqnarray}
نحصل علي
\begin{eqnarray}
V_{ni}=2\Delta.
\end{eqnarray}
نحسب الان الاحتمال
\begin{eqnarray}
P(|+>|->\longrightarrow |->|+>)&=&|c_n^{(0)}+c_n^{(1)}+...|^2\nonumber\\
&=&|0-\frac{i}{\hbar}\int_0^{t}dt_1\exp(i\Omega_{ni}t_1)V_{ni}(t_1)+...|^2\nonumber\\
&=&|-\frac{i}{\hbar}\int_0^{t}dt_1V_{ni}(t_1)+...|^2\nonumber\\
&=&\frac{4\Delta^2t^2}{\hbar^2}.
\end{eqnarray}
من اجل
\begin{eqnarray}
|n>=|+>|->=\frac{1}{\sqrt{2}}|10>+\frac{1}{\sqrt{2}}|00>,
\end{eqnarray}
نحصل علي
\begin{eqnarray}
V_{ni}=-\Delta.
\end{eqnarray}
نحسب الان الاحتمال
\begin{eqnarray}
P(|+>|->\longrightarrow |+>|->)&=&|c_n^{(0)}+c_n^{(1)}+...|^2\nonumber\\
&=&|1-\frac{i}{\hbar}\int_0^{t}dt_1\exp(i\Omega_{ni}t_1)V_{ni}(t_1)+...|^2\nonumber\\
&=&|1-\frac{i}{\hbar}\int_0^{t}dt_1V_{ni}(t_1)+...|^2\nonumber\\
&=&|1+\frac{i}{\hbar}\Delta t|^2\nonumber\\
&=&1+\frac{\Delta^2t^2}{\hbar^2}.
\end{eqnarray}
بنفس الطريقة نحسب
\begin{eqnarray}
P(|+>|->\longrightarrow |+>|+>)&=&0.
\end{eqnarray}
\begin{eqnarray}
P(|+>|->\longrightarrow |->|->)&=&0.
\end{eqnarray}
\end{itemize}
\paragraph{
تمرين
$12$:}
الحالة الابتدائية
\begin{eqnarray}
|i>=|\psi_{100}>.
\end{eqnarray}
الحالة النهائية
\begin{eqnarray}
|n>=|\psi_{2lm}>.
\end{eqnarray}
احتمال الانتقال او القفز يعطي ب
\begin{eqnarray}
P(|100>\longrightarrow|2lm>)=|c_2^{(0)}+c_2^{(1)}+...|^2.
\end{eqnarray}
لدينا
\begin{eqnarray}
c_2^{(0)}=0.
\end{eqnarray}
\begin{eqnarray}
c_2^{(1)}=-\frac{i}{\hbar}\int_0^t dt_1 \exp(i\Omega_{20}t_1)V_{20}(t_1).
\end{eqnarray}
نحسب
(
باستعمال
$E_n=E_1/n^2$
)
\begin{eqnarray}
\Omega_{20}=\frac{E_2-E_1}{\hbar}=-\frac{3E_1}{4\hbar}.
\end{eqnarray}
لدينا الكمون الكهربائي
(
$E=-\partial V/\partial z$
)
\begin{eqnarray}
V=-E z=-E r\cos\theta.
\end{eqnarray}
نحسب عنصر المصفوفة
\begin{eqnarray}
V_{20}&=&-E<2|r\cos\theta|0>\nonumber\\
&=&-E\int d^3x \psi_{2lm}^* r\cos\theta \psi_{100}
\end{eqnarray}
لدينا
\begin{eqnarray}
\psi_{100}=\frac{1}{\sqrt{\pi}}\frac{1}{a^{3/2}}\exp(-r/a).
\end{eqnarray}
لدينا حالتان. في الحالة الاولي نأخذ
\begin{eqnarray}
\psi_{200}=\frac{1}{\sqrt{128\pi}}\frac{1}{a^{3/2}}\exp(-r/2a)(-2r/a+4).
\end{eqnarray}
هذه الدالة لا تتعلق ب
$\theta$. 
اذن التكامل علي 
$\theta$
هو
صفر. اذن احتمال الانتقال ينعدم في هذه الحالة.

في الحالة الثانية نأخذ
\begin{eqnarray}
\psi_{210}=\frac{1}{\sqrt{8.36}\sqrt{4\pi}}\frac{1}{a^{3/2}}\cos\theta \exp(-r/2a)(6r/a).
\end{eqnarray}
التكامل علي
$\theta$
في هذه الحالة يعطي ب

\begin{eqnarray}
\int \sin\theta d\theta.\cos\theta.\cos\theta=\frac{2}{3}.
\end{eqnarray}
التكامل علي 
$\phi$
يعطي ب
\begin{eqnarray}
\int d\phi=2\pi.
\end{eqnarray}
التكامل علي
$r$
يعطي ب
\begin{eqnarray}
\int r^2dr.\exp(-r/2a)r.r.\exp(-r/a)=4! (2a/3)^5.
\end{eqnarray}
نحصل علي عنصر المصفوفة

\begin{eqnarray}
V_{20}=-\frac{E a}{\sqrt{8}}4!(\frac{2}{3})^6.
\end{eqnarray}
سعة الاحتمال تصبح
\begin{eqnarray}
c_2^{(1)}&=&-\frac{i}{\hbar}\int_0^tdt_1 \exp(i\Omega_{20}t_1)V_{20}(t_1)\nonumber\\
&=&\frac{i}{\hbar}E_0\frac{e^{(i\Omega_{20}-\frac{1}{\tau})t}-1}{i\Omega_{20}-\frac{1}{\tau}}\frac{a}{\sqrt{8}}4!(\frac{2}{3})^6.
\end{eqnarray}
احتمال الانتقال يعطي اذن ب

\begin{eqnarray}
P(|100>\longrightarrow |210>)&=&\frac{E_0^2}{\hbar^2}|\frac{e^{(i\Omega_{20}-\frac{1}{\tau})t}-1}{i\Omega_{20}-\frac{1}{\tau}}|^2\frac{a^2}{{8}}(4!)^2(\frac{2}{3})^{12}\nonumber\\
&=&\frac{E_0^2a^2}{\hbar^2}\frac{2^{15}}{3^{10}}\frac{1}{\Omega_{20}^2+\frac{1}{\tau^2}}\big[1+\exp(-2t/\tau)-2\cos\Omega_{20}t \exp(-t/\tau)\big].\nonumber\\
\end{eqnarray}
\paragraph{
تمرين
$13$}
\begin{itemize}
\item[$(1$]
واضح.
\item[$(2$]
الحالات تصبح
$|R_{nl}>|Y_{lm}>|s>|s^{'}>$
بدرجة انحلال
$4n^2$.
\item[$(3$]
واضح.
\item[$(4$] 
\begin{eqnarray}
 H_{\rm hf}&=&-\vec{mu}_p.\vec{B}_p\nonumber\\
 &=&\frac{\mu_0e^2g_p}{8\pi m_pm_e}\frac{1}{r^3}\big[3(\vec{S}_p\hat{r})(\vec{S}_e\hat{r})-\vec{S}_p\vec{S}_e\big]
 +\frac{\mu_0e^2g_p}{3 m_pm_e}\vec{S}_p\vec{S}_e\delta^3(\vec{r}).\nonumber\\
\end{eqnarray}
\item[$(5$]
التصحيح من الرتبة الاولي
\begin{eqnarray}
E_{\rm hf}^{(1)}&=&<R_{nl}|<Y_{lm}|<s|<s^{'}|H_{\rm hf}|R_{nl}>|Y_{lm}>|s>|s^{'}>\nonumber\\
&=&\int d^3\vec{r} R_{nl}^*(r)Y_{lm}^*(\theta,\phi)<s|<s^{'}|H_{\rm hf}|s>|s^{'}>Y_{lm}(\theta,\phi)R_{nl}(r).\nonumber\\
\end{eqnarray}
التكامل علي الزوايا من اجل
$l=m=0$
هو
\begin{eqnarray}
\int \sin\theta d\theta d\phi Y_{lm}^*(\theta,\phi)<s|<s^{'}|H_{\rm hf}|s>|s^{'}>Y_{lm}(\theta,\phi)&=&\frac{1}{4\pi}\int \sin\theta d\theta d\phi <s|<s^{'}|H_{\rm hf}|s>|s^{'}>.\nonumber\\
\end{eqnarray}
الحد الاول
\begin{eqnarray}
<s|<s^{'}|\frac{1}{4\pi}\int \sin\theta d\theta d\phi \big(3(\vec{S}_p\hat{r})(\vec{S}_e\hat{r})-\vec{S}_p\vec{S}_e\big)|s>|s^{'}>\frac{\mu_0e^2g_p}{8\pi m_pm_e}&=&\nonumber\\
<s|<s^{'}|\frac{1}{4\pi}\int \sin\theta d\theta d\phi \big(3\frac{4\pi}{3}\vec{S}_p\vec{S}_e-4\pi\vec{S}_p\vec{S}_e\big)|s>|s^{'}>\frac{\mu_0e^2g_p}{8\pi m_pm_e}&=&\nonumber\\
0.
\end{eqnarray}
\item[$(6$]
اذن التصحيح يصبح
\begin{eqnarray}
 E_{\rm hf}^{(1)}
&=&\int d^3\vec{r}\psi_{nl}^*(\vec{r})<s|<s^{'}|\frac{\mu_0e^2g_p}{3m_pm_e}\vec{S}_e\vec{S}_p\delta^3(\vec{r})|s>|s^{'}>\psi_{nlm}(\vec{r})\nonumber\\
&=&|\psi_{nl}(0)|^2\frac{\mu_0e^2g_p}{3m_pm_e}<s|<s^{'}|\vec{S}_e\vec{S}_p|s>|s^{'}>.
\end{eqnarray}
\item[$(7$]
لدينا
\begin{eqnarray}
\vec{S}_e\vec{S}_p=\frac{1}{2}((\vec{S}_e+\vec{S}_p)^2-\frac{3\hbar^2}{2}).
\end{eqnarray}
اذن
\begin{eqnarray}
\vec{S}_e\vec{S}_p=\frac{\hbar^2}{2}(-\frac{3}{2})~,~s=0.
\end{eqnarray}
\begin{eqnarray}
\vec{S}_e\vec{S}_p=\frac{\hbar^2}{2}(\frac{1}{2})~,~s=1.
\end{eqnarray}
اي
\begin{eqnarray}
 E_{\rm hf}^{(1)}
&=&\frac{1}{\pi a^3}\frac{\mu_0e^2g_p}{3m_pm_e}\hbar^2 (-\frac{3}{4})~,~s=0.
\end{eqnarray}
\begin{eqnarray}
 E_{\rm hf}^{(1)}
&=&\frac{1}{\pi a^3}\frac{\mu_0e^2g_p}{3m_pm_e}\hbar^2 (\frac{1}{4})~,~s=1.
\end{eqnarray}

\item[$(8$]
الحالات الذاتية تصبح
$|R_{nl}>|Y_{lm}>|SM>$
مع
$S=0,1$.

\end{itemize}

\paragraph{
تمرين
$14$}
الهاميلتونية 
غير المضطربة
تعطي ب
\begin{eqnarray}
 H_0=\frac{\vec{p}_1^2}{2m}-\frac{1}{4\pi\epsilon_0}\frac{2e^2}{r_1}+\frac{\vec{p}_2^2}{2m}-\frac{1}{4\pi\epsilon_0}\frac{2e^2}{r_2}.
\end{eqnarray}
الاضطراب يعطي ب
\begin{eqnarray}
V=\frac{1}{4\pi\epsilon_0}\frac{e^2}{|\vec{r}_1-\vec{r}_2|}.
\end{eqnarray}
الطاقة غير المضطربة للحالة الاساسية لذرة الهيليوم تاخذ الشكل
\begin{eqnarray}
E_1=E_1^{(1)}+E_1^{(2)}.
\end{eqnarray}
من اجل ذرة الهيدروجين
$E_1^{(1)}\propto (e^2)^2$.  
اذن من اجل ذرة الهيليوم غير المضطربة يجب ان يكون لدينا
$E_1^{(1)}\propto (ًًZe^2)^2$.
 بالتالي فان طاقات بور لذرة الهيليوم هي
 \begin{eqnarray}
{\cal E}_n=\frac{Z^2E_1}{n^2}~,~E_1=-\frac{m}{2\hbar^2}(\frac{e^2}{4\pi\epsilon_0})^2.
\end{eqnarray}
اي ان المستوي الاساسي لذرة الهيليوم له طاقة
\begin{eqnarray}
{\cal E}_1=Z^2E_1+Z^2E_1=8E_1.
\end{eqnarray}
الحالة الاساسية لذرة الهيدروجين
\begin{eqnarray}
\psi_{100}(r)=\frac{1}{a^{3/2}}\frac{1}{\sqrt{\pi}}\exp(-r/a)~,~a=\frac{\hbar}{\sqrt{-2mE_1}}.
\end{eqnarray}
اذن الحالة الاساسية لذرة الهيليوم نحصل عليها بالتعويض
$E_1\longrightarrow Z^2E_1$
او
$a\longrightarrow a/Z$
و بالتالي نحصل علي
\begin{eqnarray}
\psi_{100}(r)=\frac{Z^{3/2}}{a^{3/2}}\frac{1}{\sqrt{\pi}}\exp(-Zr/a).
\end{eqnarray}
الحالة الاساسية الكلية لذرة الهيليوم هي اذن
\begin{eqnarray}
\psi_{100}(r_1,r_2)=\frac{8}{\pi a^3}\exp(-2(r_1+r_2)/a).
\end{eqnarray}
التصحيح من الرتبة الاولي
\begin{eqnarray}
{\cal E}_1^{(1)}&=&<100|V|100>\nonumber\\
&=&\int d^3\vec{r}_1\int d^3\vec{r}_2\psi_{100}^*(r_1,r_2)\frac{1}{4\pi\epsilon_0}\frac{e^2}{|\vec{r}_1-\vec{r}_2|}\psi_{100}(r_1,r_2)\nonumber\\
&=&(\frac{8}{\pi a^3})^2\frac{e^2}{4\pi\epsilon_0}\int \exp(-4r_1/a)I_2(r_1)d^3\vec{r}_1.
\end{eqnarray}
التكامل علي
$\vec{r}_2$
يعطي ب
\begin{eqnarray}
I_2(r_1)=\int d^3\vec{r}_2\frac{\exp(-4r_2/a)}{|\vec{r}_1-\vec{r}_2|}.
\end{eqnarray}
نختار المحور
$z$
في اتجاه الشعاع
$\vec{r}_1$. 
نحصل علي
\begin{eqnarray}
I_2(r_1)&=&-\int r_2^2dr_2 d\cos\theta_2 d\phi_2 \frac{\exp(-4r_2/a)}{\sqrt{r_1^2+r_2^2-2r_1r_2\cos\theta_2}}\nonumber\\
&=&\frac{2\pi}{r_1}\int_0^{\infty} \exp(-4r_2/a) r_2dr_2\big(r_1+r_2-|r_1-r_2|\big)\nonumber\\
&=&\frac{\pi a^3}{8r_1}\big(1-(1+2r_1/a)\exp(-4r_1/a)\big).
\end{eqnarray}
اذن  التصحيح الطاقوي يصبح
\begin{eqnarray}
{\cal E}_1^{(1)}
&=&(\frac{8}{\pi a^3})\frac{e^2}{4\pi\epsilon_0}\int \exp(-4r_1/a)\frac{1}{r_1}\big(1-(1+2r_1/a)\exp(-4r_1/a)\big)d^3\vec{r}_1\nonumber\\
&=&\frac{5e^2}{16\pi\epsilon_0 a}\nonumber\\
&=&-\frac{5E_1}{2}.
\end{eqnarray}
طاقة الحالة الاساسية لذرة الهيليوم تصبح
\begin{eqnarray}
{\cal E}_1=8E_1-\frac{5E_1}{2}=\frac{11}{2}E_1=-74.8 eV.
\end{eqnarray}
\paragraph{
تمرين
$15$}
القوة منتظمة و بالتالي فان الكمون يعطي ب
\begin{eqnarray}
V=-F_0\exp(-t/\tau) x.
\end{eqnarray}
الحالة الابتدائية
\begin{eqnarray}
|i>=|0>.
\end{eqnarray}
احتمال الانتقال
\begin{eqnarray}
P_{0\longrightarrow n}=|c_n^{(0)}+c_n^{(1)}+...|^2.
\end{eqnarray}
\begin{eqnarray}
c_n^{(0)}=\delta_{n0}=0.
\end{eqnarray}
\begin{eqnarray}
c_n^{(1)}=-\frac{i}{\hbar}\int_0^t dt_1 \exp(i\Omega_{n0} t_1)V_{n0}(t_1).
\end{eqnarray}
نحسب
\begin{eqnarray}
V_{n0}=<n|V|0>=-F_0\exp(-t/\tau)\sqrt{\frac{\hbar}{2m\Omega}}\delta_{n1}.
\end{eqnarray}
اذن
\begin{eqnarray}
c_n^{(1)}=-\frac{i}{\hbar}(-F_0\sqrt{\frac{\hbar}{2m\Omega}}\delta_{n1})\frac{\exp(i\Omega_{n0}-1/\tau)t -1}{i\Omega_{n0}-1/\tau}.
\end{eqnarray}
من اجل
$n\neq 1$
الاحتمال ينعدم. من اجل
$n=1$
الاحتمال يعطي ب
(مع
$\Omega_{10}=\Omega$
)
\begin{eqnarray}
P_{0\longrightarrow n}=\frac{F_0^2}{2m\hbar \Omega}\frac{1}{\Omega^2+1/\tau^2}\big[1+\exp(-2t/\tau)-2\cos \Omega t \exp(-t/\tau)\big].
\end{eqnarray}
من اجل
$\tau\longrightarrow \infty$ 
لدينا
\begin{eqnarray}
P_{0\longrightarrow n}=\frac{F_0^2}{2m\hbar \Omega}\frac{1}{\Omega^2}.
\end{eqnarray}

\paragraph{
تمرين
$16$}
\begin{itemize}
 \item [$(1$]
 المستويات الطاقوية و دوالها الموجية
 \begin{eqnarray}
E_{n_xn_yn_z}=E(n_x^2+n_y^2+n_z^2)~,~\psi_{n_xn_yn_z}=(2/a)^{3/2}\sin n_x\pi x/a \sin n_y\pi y/a \sin n_z\pi z/a.
\end{eqnarray}
 
 \item [$(2$]
 الحالة الاساسية
 \begin{eqnarray}
E_{111}=3E~,~\psi_{111}=(2/a)^{3/2}\sin \pi x/a \sin \pi y/a \sin \pi z/a.
\end{eqnarray}
التصحيح من الرتبة الاولي للحالة الاساسية
  \begin{eqnarray}
E_{111}^{(1)}&=&<\psi_{111}|H^1|\psi_{111}>\nonumber\\
&=&V_0(2/a)^{3/2}\int_0^{a/2} dx (\sin \pi x/a)^2\int_0^{a/2} dy (\sin \pi y/a)^2\int_0^{a/2} dz (\sin \pi z/a)^2\nonumber\\
&=&\frac{V_0}{4}
\end{eqnarray}
 \item [$(3$]
 الحالات المثارة الاولي
 \begin{eqnarray}
E_{211}=E_{121}=E_{112}=6E~,~\psi_{211}=\psi_1~,~\psi_{121}=\psi_2~,~\psi_{112}=\psi_.
\end{eqnarray}
مصفوفة التصادم
\begin{eqnarray}
W_{ab}=<\psi_a|H^1|\psi_b>.
\end{eqnarray}
نحسب
\begin{eqnarray}
W_{11}=W_{22}=W_{33}=V_0/4~,~W_{13}=W_{23}=W_{31}=W_{32}=0.
\end{eqnarray}
\begin{eqnarray}
W_{12}=W_{21}=V_0K/4~,~K=(8/3\pi)^2.
\end{eqnarray}
مصفوفة التصادم تعطي صراحة ب
\begin{eqnarray}
W=\frac{V_0}{4}\left(\begin{array}{ccc}
1 &K &0\\
K &1 &0\\
0&0&1 
\end{array}\right).
\end{eqnarray}
القيم الذاتية نحصل عليها من المعادلة المميزة
\begin{eqnarray}
(1-\lambda)((1-\lambda)^2-K^2)=0.
\end{eqnarray}
نحصل علي
\begin{eqnarray}
&&\lambda_1=1~,~\lambda_2=1+K~,~\lambda_3=1-K.
\end{eqnarray}

\end{itemize}

\chapter*{
نظرية التصادم
}
\addcontentsline{toc}{chapter}{$3$
نظرية التصادم
}

\thispagestyle{headings}
\section*{
نظرية التصادم الكلاسيكية
}
\addcontentsline{toc}{section}{
{\bf
نظرية التصادم الكلاسيكية
}
} 
\subsection*{
المسائل المركزية
}
\addcontentsline{toc}{subsection}{
المسائل المركزية
} 

ليكن
$\vec{r}_1$ 
و
$\vec{r}_2$ 
شعاعي الموضع لكتلتين نقطتين
$m_1$ 
و
$m_2$ 
تتفاعلان فيما بينهما عبر قوة ناجمة عن الطاقة الكامنة
$U=U(\vec{r},\dot{\vec{r}})$
حيث
$\vec{r}$
هو شعاع الموضع النسبي المعرف ب
$\vec{r}=\vec{r}_2-\vec{r}_1$. 
يمكن ان نبين ان حركة هذين الجسمين حول مركز ثقلهما يمكن اختزالها الي حركة جسيم واحد كتلته
$\mu=m_1m_2/(m_1+m_2)$ 
حول مركز الثقل الذي يتواجد عند النقطة ذات شعاع الموضع
$\vec{R}=(m_1\vec{r}_1+m_2\vec{r}_2)/(m_1+m_2)$.
 الكتلة
$\mu=m_1m_2/(m_1+m_2)$ 
تسمي بالكتلة المختزلة. هذه النتائج تترتب من المعادلة
\begin{eqnarray}
\frac{1}{2}(m_1\dot{\vec{r}_1}^2+m_2\dot{\vec{r}_2}^2)=\frac{m_1+m_2}{2}\dot{\vec{R}}+\frac{1}{2}\mu\dot{\vec{r}}^2.
\end{eqnarray}
نعتبر اذن حركة جسيم واحد كتلته
$m$ 
تحت تأثير قوة منحفظة اي مشتقة من كمون 
$\vec{F}=-\vec{\nabla}V$ 
حيث ان الكمون
$V$
لا يتعلق الا بالمسافة القطرية
$r$.
هذه مسألة مركزية
\footnote{.${\rm central~problem}$}
لان القوة تقع بمحاذاة الشعاع
$\vec{r}$
و لا تتعلق الا بطويلة هذا الشعاع. من الواضح ان هذه المسألة متناظرة كروية وبالتالي فان العزم الحركي
$\vec{L}=\vec{r}x \vec{p}$
يجب ان يكون منحفظا و هذا الامر يمكن التحقق منه مباشرة عن طريق حساب المشتقة بالنسبة للزمن
$d\vec{L}/dt$. 
هذا يعني ايضا ان
$\vec{r}$
هو شعاع عمودي علي
اتجاه
$\vec{L}$
الذي هو اتجاه ثابت في الفضاء و بالتالي فان الحركة تقع في المستوي. الطاقة الحركية تعطي ب
\begin{eqnarray}
T=\frac{1}{2}m\dot{\vec{r}}^2=\frac{1}{2}m(\dot{r}^2+r^2\dot{\theta}^2).
\end{eqnarray}
اللاغرانجية تعطي اذن ب
\begin{eqnarray}
L=\frac{1}{2}m(\dot{r}^2+r^2\dot{\theta}^2)-V(r).
\end{eqnarray}
معادلة لاغرانج الاولي
\begin{eqnarray}
\frac{\partial L}{\partial \theta}-\frac{d}{dt}\frac{\partial L}{\partial\dot{\theta}}=0\Leftrightarrow \frac{d}{dt}(mr^2\dot{\theta})=0.
\end{eqnarray}
بعبارة اخري
\begin{eqnarray}
mr^2\dot{\theta}=l.\label{eq1}
\end{eqnarray}
العدد
$l$
هو طويلة العزم الحركي. ليس من الصعب ان نري ان
$r^2\dot{\theta}/2$
هي السرعة السطحية اي المساحة التي يمسحها الشعاع 
$\vec{r}$ 
في وحدة الزمن. اذن انحفاظ العزم الحركي هو مكافئ لقانون كيبلر 
\footnote{.${\rm Kepler}$}
الثاني الذي ينص علي ان الشعاع
$\vec{r}$ 
يمسح مساحات متساوية في ازمنة متساوية.

معادلة لاغرانج الثانية
\begin{eqnarray}
\frac{\partial L}{\partial r}-\frac{d}{dt}\frac{\partial L}{\partial\dot{r}}=0\Leftrightarrow m\ddot{r}-mr\dot{\theta}^2=-\frac{\partial V}{\partial r}=f(r).
\end{eqnarray}
لان القوة منحفظة فان الطاقة الكلية يجب ان تكون محفوظة. هذا يمكن رؤيته كالتالي. اولا نكتب معادلة الحركة اعلاه كالتالي
\begin{eqnarray}
m\ddot{r}&=&mr\dot{\theta}^2-\frac{\partial V}{\partial r}\nonumber\\
&=&\frac{l^2}{mr^3}-\frac{dV}{d r}\nonumber\\
&=&-\frac{d}{dr}(\frac{1}{2}\frac{l^2}{mr^2}+V).
\end{eqnarray}
بالضرب في
$\dot{r}$
نحصل علي
\begin{eqnarray}
m\dot{r}\ddot{r}
&=&-\frac{dr}{dt}\frac{d}{dr}(\frac{1}{2}\frac{l^2}{mr^2}+V).
\end{eqnarray}
مكافئ لهذه المعادلة المعادلة
\begin{eqnarray}
\frac{d}{dt}\bigg(\frac{1}{2}m\dot{r}^2+\frac{1}{2}\frac{l^2}{mr^2}+V\bigg)=0.
\end{eqnarray}
المقدار بين القوسين هو بالضبط الطاقة الكلية للجملة و واضح تماما انها محفوظة. نكتب
\begin{eqnarray}
E=\frac{1}{2}m\dot{r}^2+\frac{1}{2}\frac{l^2}{mr^2}+V.
\end{eqnarray}
الحل من اجل
$\dot{r}$
يعطي
\begin{eqnarray}
\dot{r}=\sqrt{\frac{2}{m}\bigg(E-V-\frac{l^2}{2mr^2})}.
\end{eqnarray}
اذن
\begin{eqnarray}
dt=\frac{dr}{\sqrt{\frac{2}{m}\bigg(E-V-\frac{l^2}{2mr^2})}}.\label{eq2}
\end{eqnarray}
بمكاملة كلا الطرفين من
 $t=0$ 
 حيث
 $r(0)=r_0$ 
 الي
 $t$
 حيث
 $r(t)=r$ 
 نحصل علي
\begin{eqnarray}
t=\int_{r_0}^r\frac{dr}{\sqrt{\frac{2}{m}\bigg(E-V-\frac{l^2}{2mr^2})}}.
\end{eqnarray}
هذا يعطي
$t=t(r)$.
عن طريق قلب هذه المعادلة نحصل علي
$r=r(t)$. 
الزاوية
$\theta$
يمكن اذن الحصول عليها من
\begin{eqnarray}
d\theta=\frac{ldt}{mr^2(t)}.
\end{eqnarray}
بالمكاملة من
$t=0$
حيث
$\theta(0)=\theta_0$
الي
$t$
حيث
$\theta(t)=\theta$
نحصل علي

 \begin{eqnarray}
\theta=\int_0^t\frac{ldt}{mr^2(t)}+\theta_0.
\end{eqnarray}
معادلة المدار
$r=r(\theta)$
يمكن الحصول عليها كالتالي. انطلاقا من معادلة الحركة
$(\ref{eq1})$
لدينا
 \begin{eqnarray}
m r^2d\theta=l dt.
\end{eqnarray}
بوضع هذه المعادلة في
$(\ref{eq2})$
نحصل علي
\begin{eqnarray}
d\theta=\frac{l dr}{m r^2\sqrt{\frac{2}{m}\bigg(E-V-\frac{l^2}{2mr^2})}}.
\end{eqnarray}
بالمكاملة من
$r_0$
حيث
$\theta(r_0)=\theta_0$ 
الي
$r$
حيث
$\theta(r)=\theta$
نحصل علي
\begin{eqnarray}
\theta&=&\int_{r_0}^r\frac{l dr}{m r^2\sqrt{\frac{2}{m}\bigg(E-V-\frac{l^2}{2mr^2})}}+\theta_0\nonumber\\
&=&\int_{r_0}^r\frac{dr}{r^2\sqrt{\frac{2m(E-V)}{l^2}-\frac{1}{r^2}}}+\theta_0.
\end{eqnarray}
من اجل قانون التربيع العكسي 
\begin{eqnarray}
V=-\frac{k}{r}~,~f=-\frac{k}{r^2}.
\end{eqnarray}
نعتبر ايضا التكامل غير المحدد
(مع
$u=1/r$
)
\begin{eqnarray}
\theta&=&\int \frac{dr}{r^2\sqrt{\frac{2m(E-V)}{l^2}-\frac{1}{r^2}}}\nonumber\\
&=&-\int \frac{du}{\sqrt{\frac{2mE}{l^2}+\frac{2mku}{l^2}-u^2}}.
\end{eqnarray}
نستعمل العلاقة
\begin{eqnarray}
\int \frac{dx}{\sqrt{\alpha+\beta x+\gamma x^2}}=\frac{1}{\sqrt{-\gamma}}\arccos -\frac{\beta +2\gamma x}{\sqrt{q}}~,~q=\beta^2-4\alpha\gamma.
\end{eqnarray}
اذن نحصل (
مع
ثابت تكامل
$\theta^{'}$
)
علي
\begin{eqnarray}
\theta&=&-\arccos\frac{\frac{l^2u}{mk}-1}{\sqrt{1+\frac{2l^2E}{mk^2}}}+\theta^{'}.
\end{eqnarray}
اي اننا نحصل علي
\begin{eqnarray}
\frac{1}{r}=C(1+e\cos(\theta-\theta^{'}))~,~C=\frac{mk}{l^2}~,~e=\sqrt{1+\frac{2El^2}{mk^2}}.\label{orbit}
\end{eqnarray}
المدار هو اذن مقطع مخروطي حيث ان احد المحرقين او البؤرتين تقع في نقطة المبدأ و 
$e$
هي الاختلاف المركزي
\footnote{.${\rm eccentricity}$}
. طبيعة المدار هي كالتالي
\begin{eqnarray}
&&e>1 \Leftrightarrow E>0~:~{\rm hyperbola}\nonumber\\
&&e=1\Leftrightarrow E=0~:~{\rm parabola}\nonumber\\
&&e<1\Leftrightarrow E<0~:~{\rm ellipse}\nonumber\\
&&e=0\Leftrightarrow E=-\frac{mk^2}{2l^2}~:~{\rm circle}.
\end{eqnarray}
\subsection*{
المقطع الفعال التفاضلي
}
\addcontentsline{toc}{subsection}{
المقطع الفعال التفاضلي
} 

نعتبر شعاع منتظم من الجسيمات التي لها نفس الكتلة و نفس الطاقة واردة علي مركز قوة. نفترض ان القوة تنعدم في اللانهاية و بالتالي فان مسار الجسيمات الواردة لما تكون بعيدة عن 
مركز القوة هو خط مستقيم. هذا المسار ينحرف عن مسار الورود المستقيم لما تقترب الجسيمات من مركز القوة و 
بعد المرور علي مركز القوة يستقيم المسار شيئا فشيئا الي ان يصبح خط مستقيم في اللانهاية معطي  بالضبط بمسار الجسيمات الواردة. زاوية التصادم
\footnote{.${\rm scattering~angle}$}
هي الزاوية بين محور الورود و محور الصدور.

نعرف محور الحضيض 
\footnote{.${\rm periapsis}$}
علي انه المحور الذي يمر عبر مركز القوة و عبر النقظة علي المسار الاقرب مسافة من مركز القوة. لتكن 
$\Psi$
الزاوية بين محور الورود و محور الحضيض. لان المسار متناظر حول محور الحضيض يجب ان يكون لدينا
\begin{eqnarray}
\Theta=\pi-2\Psi.
\end{eqnarray}
المسافة العمودية
$b$
بين محور الورود و مركز القوة تسمي وسيط الصدمة
\footnote{.${\rm impact~parameter}$}
. هذا الوسيط يعوض العزم الحركي 
 $l$
للجسيمات الواردة لان

\begin{eqnarray}
l=mv_0b=b\sqrt{2mE}.
\end{eqnarray} 
من الواضح ان الجسيمات الواردة ضمن مقطع مساحي متناه في الصغر
$d\sigma$ 
سوف تنتثر
\footnote{.${\rm scatter}$}
داخل زاوية صلبة
$d\Omega$. 

معامل التناسب يسمي المقطع الفعال التفاضلي
\footnote{.${\rm differential~cross~section}$}
و يعرف ب

\begin{eqnarray}
D(\Omega)=\frac{d\sigma}{d\Omega}.
\end{eqnarray} 
المعني الفيزيائي ل
$D(\Omega)$
هو كالتالي. ليكن
$I$
شدة او اضاءة الشعاع الوارد اي عدد الجسيمات الواردة في وحدة المساحة العمودية للشعاع في وحدة الزمن. عدد الجسيمات التي تعبر المساحة
$d\sigma$
في وحدة الزمن هي بالتالي
$dN=Id\sigma =ID(\Omega)d\Omega$. 
بعبارة اخري
\begin{eqnarray}
D(\Omega)=\frac{1}{I}\frac{dN}{d\Omega}.
\end{eqnarray}
هذا يعني ان
$ID(\Omega)d\Omega$
هو عدد الجسيمات المنتثرة داخل الزاوية الصلبة 
$d\Omega$ 
في وحدة الزمن. هذا يجب ان يكون مقدارا موجبا. من اجل الكمونات المركزية لدينا تناظر كامل حول محور الورود. اذن يمكن ان نكتب
$d\sigma=2\pi bdb$ 
و
$d\Omega=2\pi\sin\Theta d\Theta$. 
نحصل علي المقطع الفعال التفاضلي
\begin{eqnarray}
D(\Theta)=\frac{b}{\sin\Theta}|\frac{db}{d\Theta}|.
\end{eqnarray}
لان
$\Theta$
تتناقص لما يتزايد
$b$
فان المشتقة
$db/d\Theta$ 
هي سالبة و بالتالي اخذنا القيمة المطلقة للحفاظ علي ايجابية
$D(\Theta)$.
المقطع الفعال الكلي هو اذن

\begin{eqnarray}
\sigma=\int D(\Theta) d\Omega.
\end{eqnarray}
\subsection*{
تصادم رذرفورد
}
\addcontentsline{toc}{subsection}{
تصادم رذرفورد
} 
كمثال علي تصادم الجسيمات بواسطة مركز قوة نأخذ حالة قوة كولون المتنافرة بين شحنة مثبتة
$q_1=-Z e$
و شحنة
$q_2=-Z^{'}e$.
اذن
\begin{eqnarray}
k=-\frac{q_1q_2}{4\pi\epsilon_0}.
\end{eqnarray}
من اجل مسائل التصادم الطاقة
$E$ 
تكون موجبة و بالتالي
$e>1$.
باختيار
$\theta^{'}=\pi$
في
$(\ref{orbit})$
فان محور الحضيض يصبح موافق ل
$\theta=0$
لان ذلك يعطي اصغر قيمة ل
$r$.
 المدار يصبح
\begin{eqnarray}
\frac{1}{r}=\frac{mq_1q_2}{4\pi\epsilon_0l^2}(e\cos\theta -1).
\end{eqnarray}
معامل الاختلاف المركزي يصبح
\begin{eqnarray}
e=\sqrt{1+\bigg(\frac{8\pi\epsilon_0Eb}{q_1q_2}\bigg)^2}.
\end{eqnarray}
اتجاه الجسيمات الواردة يقابل
$\theta=\Psi$
لما
$r\longrightarrow\infty$.
بعبارة اخري
\begin{eqnarray}
\cos\Psi=\frac{1}{e}.
\end{eqnarray}
او بالمقابل
\begin{eqnarray}
\sin\frac{\Theta}{2}=\frac{1}{e}\Leftrightarrow \cot\frac{\Theta}{2}=\sqrt{e^2-1}=\frac{8\pi\epsilon_0Eb}{q_1q_2}.
\end{eqnarray}
بعبارة اخري
\begin{eqnarray}
b=\frac{q_1q_2}{8\pi\epsilon_0 E}\cot\frac{\Theta}{2}.
\end{eqnarray}
المقطع الفعال التفاضلي هو اذن
\begin{eqnarray}
D(\Omega)=\frac{1}{4}\bigg(\frac{q_1q_2}{8\pi\epsilon_0 E}\bigg)^2\frac{1}{\sin^4\frac{\Theta}{2}}.
\end{eqnarray}
المقطع الفعال الكلي
$\sigma=\int d\Omega D(\Omega)$
هو غير متناه. هذا راجع الي كون مدي التفاعلات الكهرومغناكيسية غير منته.
\thispagestyle{headings}
\section*{
نظرية التصادم الكمومية
}
\addcontentsline{toc}{section}{
{\bf
نظرية التصادم الكمومية
}
} 
\subsection*{
معادلة ليبمان - شوينغر
}
\addcontentsline{toc}{subsection}{
معادلة ليبمان - شوينغر
} 
نعتبر معادلة شرودينغر غير المتعلقة بالزمن المعطاة ب
\begin{eqnarray}
(H_0+V)|\psi>=E|\psi>.\label{eqsch}
\end{eqnarray}
الهاميلتونية
$H_0$
هي مؤثر الطاقة الحركية اي
\begin{eqnarray}
H_0=\frac{\vec{p}^2}{2m}.
\end{eqnarray}
 نفترض ان اطياف
$H_0$
و
$H_0+V$
 هي اطياف مستمرة. ليكن
 $|\phi>$
 الشعاع الذاتي ل
 $H_0$
 المرفق بالطاقة الذاتية 
 $E$ 
 اي
 \begin{eqnarray}
H_0|\phi>=E|\phi>.
\end{eqnarray}
الهدف هو ايجاد حل
$|\psi>$
ل
$(\ref{eqsch})$
مرفق بنفس قيمة الطاقة
$E$
بحيث
$|\psi>\longrightarrow |\phi>$
لما
$V\longrightarrow 0$.
هذا يوافق التصادم المرن لانه لا يقع اي تغيير لقيمة الطاقة. معادلة شرودينغر
$(\ref{eqsch})$
تأخذ الشكل
 \begin{eqnarray}
(E-H_0)|\psi>=(E-H_0)|\phi>+V|\psi>.
\end{eqnarray}
لدينا اذن الحل المنهجي
 \begin{eqnarray}
|\psi>=|\phi>+\frac{1}{E-H_0}V|\psi>.
\end{eqnarray}
حتي نحصل علي حل ذي معني يجب ان نتبني وصفة فايمان. نجعل
$E$
مركب قليلا. نحصل علي
\begin{eqnarray}
|\psi^{\pm}>=|\phi>+\frac{1}{E-H_0\pm i\epsilon}V|\psi^{\pm}>.
\end{eqnarray}
هذه هي معادلة ليبمان -  شوينغر
\footnote{.${\rm Lippmann-Schwinger~equation}$}. 
في اساس الموضع لدينا
\begin{eqnarray}
<\vec{x}|\psi^{\pm}>&=&<\vec{x}|\phi>+<\vec{x}|\frac{1}{E-H_0\pm i\epsilon}V|\psi^{\pm}>\nonumber\\
&=&<\vec{x}|\phi>+\int d^3x^{'}<\vec{x}|\frac{1}{E-H_0\pm i\epsilon}|\vec{x}^{'}><\vec{x}^{'}|V|\psi^{\pm}>.
\end{eqnarray}
اذن يجب حساب نواة هذا التكامل المعطاة ب
\begin{eqnarray}
G_{\pm}(\vec{x},\vec{x}^{'})&=&\frac{\hbar^2}{2m}<\vec{x}|\frac{1}{E-H_0\pm i\epsilon}|\vec{x}^{'}>\nonumber\\
&=&\frac{\hbar^2}{2m}\int d^3\vec{p}^{'}\int d^{3}\vec{p}^{''}<\vec{x}|\vec{p}^{'}><\vec{p}^{'}|\frac{1}{E-H_0\pm i\epsilon}|\vec{p}^{''}><\vec{p}^{''}|\vec{x}^{'}>.
\end{eqnarray}
نستعمل النتائج
\begin{eqnarray}
<\vec{x}|\vec{p}^{'}>=\frac{e^{\frac{i}{\hbar}\vec{p}^{'}\vec{x}}}{(2\pi\hbar)^{\frac{3}{2}}}~,~<\vec{p}^{''}|\vec{x}^{'}>=\frac{e^{-\frac{i}{\hbar}\vec{p}^{''}\vec{x}^{'}}}{(2\pi\hbar)^{\frac{3}{2}}}.
\end{eqnarray}
\begin{eqnarray}
<\vec{p}^{'}|\frac{1}{E-H_0\pm i\epsilon}|\vec{p}^{''}>=\frac{\delta^3(\vec{p}^{'}-\vec{p}^{''})}{E-\frac{\vec{p}^{'2}}{2m}\pm i\epsilon}.
\end{eqnarray}
اذن (
مع
$E=\hbar^2 k^2/(2m)$, $\vec{p}^{'}=\hbar \vec{q}$
و
$\vec{R}=\vec{x}-\vec{x}^{'}$
)
\begin{eqnarray}
G_{\pm}(\vec{x},\vec{x}^{'})&=&\frac{\hbar^2}{2m}\int \frac{d^3\vec{p}^{'}}{(2\pi\hbar)^3}\frac{e^{\frac{i}{\hbar}\vec{p}^{'}(\vec{x}-\vec{x}^{'})}}{E-\frac{\vec{p}^{'2}}{2m}\pm i\epsilon}\nonumber\\
&=&\int \frac{d^3\vec{q}}{(2\pi)^3}\frac{e^{i\vec{q}\vec{R}}}{k^2-q^2\pm i\epsilon}.
\end{eqnarray}
من هذه المعادلة من الواضح ان
$G_{\pm}(\vec{x},\vec{x}^{'})$
يحل معادلة هلمولتز
\footnote{.${\rm Helmoltz~equation}$}
بمنبع معطي بدالة دلتا
\footnote{.${\rm delta~function}$}
اي
\begin{eqnarray}
(\vec{\nabla}^2+k^2)G_{\pm}(\vec{x},\vec{x}^{'})&=&\delta^3(\vec{x}-\vec{x}^{'}).
\end{eqnarray}
الدالة
$G_{\pm}(\vec{x},\vec{x}^{'})$
هي دالة غرين 
\footnote{.${\rm Green's~function}$}
خاصة معادلة هلمولتز. هذه الدالة تقيس التجاوب مع المنبع المعطي بدالة دلتا.

الشعاع
$\vec{R}$
ثابت. يمكن ان نختار
المحور
$z$
بمحاذاة الشعاع 
$\vec{R}$.
اذن نحصل علي
\begin{eqnarray}
G_{\pm}(\vec{x},\vec{x}^{'})&=&\int \frac{q^2\sin\theta dqd\theta d\phi}{(2\pi)^3}\frac{e^{iqR\cos\theta}}{k^2-q^2\pm i\epsilon}\nonumber\\
&=&\frac{i}{8\pi^2 R}\int_{-\infty}^{\infty}\bigg[\frac{qe^{-iqR}dq}{k^2-q^2\pm i\epsilon}-\frac{qe^{iqR}dq}{k^2-q^2\pm i\epsilon}\bigg]\nonumber\\
&=&\frac{i}{8\pi^2 R}(-I_2+I_1).
\end{eqnarray}
\begin{eqnarray}
I_1&=&\int_{-\infty}^{\infty}\frac{qe^{iqR}dq}{-k^2+q^2\mp i\epsilon}.
\end{eqnarray}
\begin{eqnarray}
I_2&=&\int_{-\infty}^{\infty}\frac{qe^{-iqR}dq}{-k^2+q^2\mp i\epsilon}.
\end{eqnarray}
 يعطي القطبين بالشرط 
 $q_{\pm}^2=k^2\pm i\epsilon$.
 من اجل
 $G_+$ 
  يقع القطبين عند
  $q_{\pm}=\pm(k+i\epsilon^{'})$
  بينما من اجل
  $G_-$
  يقع القطبين عند
  $q_{\pm}=\pm(k-i\epsilon^{'})$.

سوف نحسب هذين التكاملين باستعمال صيغة كوشي التكاملية 
\footnote{.${\rm Cauchy's integral~formula}$}
المعطاة بالمعادلة
\begin{eqnarray}
\oint \frac{f(z)dz}{z-z_0}=2\pi i f(z_0).
\end{eqnarray}
اذن نحتاج ان نغلق محيط التكامل
\footnote{.${\rm contour~of~integration}$}
بدون تغيير قيمة التكامل. المعامل الاسي
$e^{iqR}$
الذي يظهر في
$I_1$ 
يقترب من الصفر اذا كانت
$q$
علي نصف دائرة ذات نصف قطر غير منته في النصف الاعلي من المستوي المركب. اذن نغلق محيط التكامل خاصة
$I_1$ 
باضافة نصف دائرة في اللانهاية في النصف الاعلي من المستوي المركب. من اجل
$G_+$
 فان القطب
$(k+i\epsilon^{'})$
هو الذي 
يقع داخل محيط التكامل بينما من اجل
$G_-$
فان القطب
$-(k-i\epsilon^{'})$ 
هو الذي يقع داخل محيط التكامل. اذن نحصل (مع
$f(z)=ze^{izR}/(z+z_0)$
و
$z_0=\pm (k\pm i\epsilon^{'})$)
علي

  \begin{eqnarray}
I_1&=&2\pi i f(z_0)=2\pi i(\frac{1}{2}e^{iz_0R})=\pi i e^{\pm i kR}.
\end{eqnarray}
بالمثل نغلق محيط التكامل خاصة
$I_2$
باضافة نصف دائرة في اللانهاية في النصف الاسفل من المستوي المركب. من اجل
$G_+$
فان القطب
$-(k+i\epsilon^{'})$
هو الذي يقع داخل محيط التكامل بينما من اجل
$G_-$
فان القطب
$(k-i\epsilon^{'})$ 
هو الذي يقع داخل محيط التكامل. اذن نحصل علي
(مع
$f(z)=ze^{-izR}/(z+z_0)$ 
و
$z_0=\mp (k\pm i\epsilon^{'})$
)
  \begin{eqnarray}
I_2&=&-2\pi i f(z_0)=-2\pi i(\frac{1}{2}e^{-iz_0R})=-\pi i e^{\pm i kR}.
\end{eqnarray}
اشارة الناقص تنجم عن اتجاهنا في اتجاه عقارب الساعة. نحصل اخيرا علي دالة غرين
\begin{eqnarray}
G_{\pm}(\vec{x},\vec{x}^{'})&=&-\frac{1}{4\pi}\frac{e^{\pm ikR}}{R}.
\end{eqnarray}
تصبح معادلة ليبمان - شوينغر في اساس الموضع معطاة ب
\begin{eqnarray}
<\vec{x}|\psi^{\pm}>
&=&<\vec{x}|\phi>-\frac{2m}{\hbar^2}\frac{1}{4\pi}\int d^3x^{'}\frac{e^{\pm ik|\vec{x}-\vec{x}^{'}|}}{|\vec{x}-\vec{x}^{'}|}<\vec{x}^{'}|V|\psi^{\pm}>.
\end{eqnarray}
من اجل الكمونات الموضعية
\footnote{.${\rm local~potentials}$}
لدينا
$<\vec{x}^{'}|V|\psi^{\pm}>=V(\vec{x}^{'})<\vec{x}^{'}|\psi^{\pm}>$ 
و بالتالي
\begin{eqnarray}
<\vec{x}|\psi^{\pm}>
&=&<\vec{x}|\phi>-\frac{m}{2\pi\hbar^2}\int d^3x^{'}\frac{e^{\pm ik|\vec{x}-\vec{x}^{'}|}}{|\vec{x}-\vec{x}^{'}|}V(\vec{x}^{'})<\vec{x}^{'}|\psi^{\pm}>.\label{lippmann}
\end{eqnarray}
الشعاع
$\vec{x}$
يعرف نقطة الملاحظة
$P$
اين نضع
الكاشف
\footnote{.${\rm detector}$}
. 
من اجل كمون ذو مدي منته فان جزء صغير فقط من الفضاء يعطي مشاركة غير منعدمة في التكامل علي
$\vec{x}^{'}$.
في مسائل التصادم نقطة الملاحظة
$P$
تكون  عموما بعيدة خارج مدي الكمون لانه لا يمكننا وضع الكاشف بالقرب من مركز التصادم. اذن يجب ان يكون لدينا 
$|\vec{x}|>>|\vec{x}^{'}|$. 
بالتالي
(مع
$|\vec{x}|=r$, $|\vec{x}^{'}|=r^{'}$, $\hat{r}=\vec{x}/|\vec{x}|$
و
$\alpha$ 
هي الزاوية بين
$\vec{x}$
و
$\vec{x}^{'}$)
 \begin{eqnarray}
|\vec{x}-\vec{x}^{'}|&=&\sqrt{r^2+r^{'2}-2rr^{'}\cos\alpha}\nonumber\\
&=&r\sqrt{1+\frac{r^{'2}}{r^2}-2\frac{\vec{x}^{'}\hat{r}}{r}}\nonumber\\
&=&r-\vec{x}^{'}\hat{r}+O(\frac{1}{r}).
\end{eqnarray}
اذن
\begin{eqnarray}
\frac{e^{\pm ik|\vec{x}-\vec{x}^{'}|}}{|\vec{x}-\vec{x}^{'}|}\simeq \frac{e^{\pm i kr} e^{\mp i\vec{k}^{'}\vec{x}^{'}}}{r}~,~\vec{k}^{'}=k\hat{r}.
\end{eqnarray}
الشعاع
$\hbar \vec{k}^{'}$
هو كمية حركة الجسيمات المنتثرة التي تصل الي نقطة الملاحظة
$P$.
كمية حركة الجسيمات الواردة هي
$\vec{p}_i=\hbar \vec{k}$ 
و شعاع الحالة المرافق هو
$|\phi>=|\vec{k}>$.
 يجب اذن ان يكون لدينا
\begin{eqnarray}
<\vec{x}|\phi>=<\vec{x}|\vec{k}>=\frac{e^{i\vec{k}\vec{x}}}{(2\pi)^{\frac{3}{2}}}.
\end{eqnarray}
اذن من اجل المسافات الكبيرة
$r$
معادلة ليبمان - شوينغر تصبح
\begin{eqnarray}
<\vec{x}|\psi_{\pm}>=\frac{1}{(2\pi)^{\frac{3}{2}}}\bigg[e^{i\vec{k}\vec{x}}+\frac{e^{\pm ikr}}{r}f(\vec{k}^{'},\vec{k})\bigg].
\end{eqnarray}
\begin{eqnarray}
f(\vec{k}^{'},\vec{k})=-\frac{m}{\hbar^2}\sqrt{2\pi}\int d^3x^{'} e^{\mp i\vec{k}^{'}\vec{x}^{'}}V(\vec{x}^{'})<\vec{x}^{'}|\psi^{\pm}>.
\end{eqnarray}
اذن التعلق الفضائي للحل الموجب (السالب) من اجل
$r$ 
كبير يعطي بمجموع الموجة المستوية الواردة و الموجة الكروية الصادرة (الواردة) التي يتسبب فيها مركز التصادم. في اغلب تجارب التصادم فان الحل الموجب هو المحقق في الواقع. في ما تبقي لنا سنعتبر فقط الحل الموجب.

لتكن
$v_0$
سرعة الشعاع الوارد. احتمال ان يعبر جسيم مقطع مساحي متناه في الصغر
$d\sigma$ 
عمودي لاتجاه الشعاع الوارد خلال زمن
$dt$
يعطي ب
\begin{eqnarray}
dP_{\rm incident}=|A|^2 (v_0dt d\sigma).
\end{eqnarray}
هذا الاحتمال يجب ان يكون مساو للاحتمال  $dP_{\rm scattered}$ في ان ينتثر الجسيم داخل الزاوية الصلبة
$d\Omega$
خلال الزمن
$dt$.
بعبارة اخري
$dP_{\rm scattered}$
هو احتمال ان يعبر الجسيم السطح
$r^2d\Omega$
خلال زمن
$dt$
اي
 
\begin{eqnarray}
dP_{\rm scattered}=\frac{|A|^2|f|^2}{r^2} (v_0dt ~r^2d\Omega).
\end{eqnarray}
اذن المقطع الفعال التفاضلي يعطي ب
\begin{eqnarray}
D(\Omega)=\frac{d\sigma}{d\Omega}=|f(\vec{k}^{'},\vec{k})|^2.
\end{eqnarray}
المعامل
$f(\vec{k}^{'},\vec{k})$
الذي يسمي سعة التصادم هو سعة احتمال التصادم في الاتجاه
$\Theta$
الذي هو الزاوية بين
$\vec{k}$
و
$\vec{k}^{'}$.

\subsection*{
تقريب بورن
}
\addcontentsline{toc}{subsection}{
تقريب بورن
} 

نعيد كتابة معادلة ليبمان - شوينغر و كذا سعة التصادم علي الشكل
\begin{eqnarray}
<\vec{x}|\psi_{+}>=\frac{1}{(2\pi)^{\frac{3}{2}}}\bigg[e^{i\vec{k}\vec{x}}+\frac{e^{+ ikr}}{r}f(k\hat{r},\vec{k})\bigg].
\end{eqnarray}
\begin{eqnarray}
f(k\hat{r},\vec{k})=-\frac{m}{\hbar^2}\sqrt{2\pi}\int d^3x^{'} e^{- ik\hat{r}\vec{x}^{'}}V(\vec{x}^{'})<\vec{x}^{'}|\psi^{+}>.
\end{eqnarray}
باستعمال معادلة ليبمان - شوينغر تأخذ سعة الاحتمال الشكل

\begin{eqnarray}
f(k\hat{r},\vec{k})=-\frac{m}{\hbar^2}\sqrt{2\pi}\int d^3x^{'} e^{- ik\hat{r}\vec{x}^{'}}V(\vec{x}^{'})\frac{1}{(2\pi)^{\frac{3}{2}}}
\bigg(e^{i\vec{k}\vec{x}^{'}}+\frac{e^{+ ikr^{'}}}{r^{'}}f(k\hat{r}^{'},\vec{k})\bigg).
\end{eqnarray}
يتلخص تقريب بورن الاول
\footnote{.${\rm first~Born~approximation}$}
في التخلص من الحد الاول بين القوسين اي
\begin{eqnarray}
f^{(1)}(k\hat{r},\vec{k})&=&-\frac{m}{\hbar^2}\sqrt{2\pi}\int d^3x^{'} e^{- ik\hat{r}\vec{x}^{'}}V(\vec{x}^{'})\frac{1}{(2\pi)^{\frac{3}{2}}}e^{i\vec{k}\vec{x}^{'}}\nonumber\\
&=&-\frac{m}{2\pi\hbar^2}\int d^3x^{'} e^{i(\vec{k}-k\hat{r})\vec{x}^{'}}V(\vec{x}^{'}).
\end{eqnarray}
هذا هو تحويل فورييه لكمون التفاعل. كمية الحركة المحولة خلال العملية هي
$\hbar \vec{q}$ 
حيث
$\vec{q}=\vec{k}-k\hat{r}$. 
بدلالة زاوية التصادم
$\Theta$
نحسب
\begin{eqnarray}
q=2k\sin\frac{\Theta}{2}.
\end{eqnarray}
نختار المحور
$z$
في اتجاه
$\vec{q}$.
اذن
\begin{eqnarray}
f^{(1)}(k\hat{r},\vec{k})
&=&-\frac{m}{2\pi\hbar^2}\int r^{'2}\sin\theta dr^{'}d\theta d\phi e^{iqr^{'}\cos\theta}V(\vec{x}^{'}).
\end{eqnarray}
من اجل كمون متناظر كرويا
$V(\vec{x}^{'})=V(r^{'})$ 
لدينا
\begin{eqnarray}
f^{(1)}(k\hat{r},\vec{k})
&=&\frac{m}{\hbar^2}\int r^{'2}dr^{'}d\cos\theta  e^{iqr^{'}\cos\theta}V(r^{'})\nonumber\\
&=&-\frac{2m}{q\hbar^2}\int r \sin qr V(r) dr.
\end{eqnarray}
حتي نفهم طبيعة تصحيحات بورن من الرتبة العليا نعود الي معادلة ليبمان - شوينغر في الشكل
$(\ref{lippmann})$.
ندخل الترميز
$\psi_+(\vec{x})=<\vec{x}|\psi_+>$, $\phi(\vec{x})=<\vec{x}|\phi>$
و
\begin{eqnarray}
G(\vec{x}-\vec{x}^{'})
&=&-\frac{m}{2\pi\hbar^2}\frac{e^{+ ik|\vec{x}-\vec{x}^{'}|}}{|\vec{x}-\vec{x}^{'}|}.
\end{eqnarray}
معادلة ليبمان - شوينغر تصبح
\begin{eqnarray}
\psi^{+}(\vec{x})
&=&\phi(\vec{x})+\int G(\vec{x}-\vec{x}_1)V(\vec{x}_1)\psi^{+}(\vec{x}_1)d^3x_1.
\end{eqnarray}
نحصل مباشرة علي السلسلة
\begin{eqnarray}
\psi^{+}
&=&\phi(\vec{x})+\int G(\vec{x}-\vec{x}_1)V(\vec{x}_1)\phi(\vec{x}_1)d^3x_1\nonumber\\
&+&\int G(\vec{x}-\vec{x}_1)V(\vec{x}_1)G(\vec{x}_1-\vec{x}_2)V(\vec{x}_2)\phi(\vec{x}_2)d^3x_1d^3x_2\nonumber\\
&+&\int G(\vec{x}-\vec{x}_1)V(\vec{x}_1)G(\vec{x}_1-\vec{x}_2)V(\vec{x}_2)G(\vec{x}_2-\vec{x}_3)V(\vec{x}_3)\phi(\vec{x}_3)d^3x_1d^3x_2d^3x_3+...\nonumber\\
\end{eqnarray}
هذا يعرف باسم سلسلة بورن
\footnote{.${\rm Born~series}$}
. في الحد رقم
$n$
دالة الموجة الواردة
$\phi$
تتفاعل
$n$
مرة
مع الكمون قبل ان تواصل انتشارها الي اللانهاية عبر دالة غرين
$G$.
 نعبر علي كل تفاعل بدلالة معامل عقدة
 \footnote{.${\rm vertex~factor}$}
  يساوي الي الكمون و بين كل تفاعلين متتاليين تنتشر دالة الموجة عن طريق دالة غرين
  $G$. 
  تعرف اذن دالة غرين بالمنتشر
  \footnote{.${\rm propagator}$}
  . 
  اذن يمكن فهم نشر بورن علي انه عبارة عن معاملات عقد
  $V$
  و منتشرات
  $G$
  موصولة فيما بينها لتشكل مخططات تعرف باسم مخططات فايمان
  \footnote{.${\rm Feynman~diagrams}$}
  .

\subsection*{
مؤثر الانتقال
}
\addcontentsline{toc}{subsection}{
مؤثر الانتقال
} 
سعة التصادم هي
\begin{eqnarray}
f(k\hat{r},\vec{k})&=&-\frac{m}{\hbar^2}\sqrt{2\pi}\int d^3x^{'} e^{- ik\hat{r}\vec{x}^{'}}V(\vec{x}^{'})<\vec{x}^{'}|\psi^{+}>\nonumber\\
&=&-\frac{m}{\hbar^2}(2\pi)^2<k\hat{r}|V|\psi^+>.
\end{eqnarray}
نعرف مؤثر الانتقال
$T$
ب
   \begin{eqnarray}
V|\psi^+>=T|\phi>~,~|\phi>=|\vec{k}>.
\end{eqnarray}   
اذن
  \begin{eqnarray}
f(k\hat{r},\vec{k})&=&-\frac{m}{\hbar^2}\sqrt{2\pi}\int d^3x^{'} e^{- ik\hat{r}\vec{x}^{'}}V(\vec{x}^{'})<\vec{x}^{'}|\psi^{+}>\nonumber\\
&=&-\frac{m}{\hbar^2}(2\pi)^2<k\hat{r}|T|\vec{k}>.
\end{eqnarray}
بضرب معادلة ليبمان - شوينغر ب
$V$ 
نحصل علي
\begin{eqnarray}
V|\psi^{+}>&=&V|\phi>+V\frac{1}{E-H_0+ i\epsilon}V|\psi^{+}>.
\end{eqnarray}  
بعبارة اخري
\begin{eqnarray}
T|\phi>&=&V|\phi>+V\frac{1}{E-H_0+ i\epsilon}T|\phi>.
\end{eqnarray}  
لان   الاشعة الذاتية لمؤثر كمية الحركة  
$|\phi>=|\vec{k}>$
هي مكتملة نحصل علي
\begin{eqnarray}
T&=&V+V\frac{1}{E-H_0+ i\epsilon}T\nonumber\\
&=&V+V\frac{1}{E-H_0+ i\epsilon}V+V\frac{1}{E-H_0+ i\epsilon}V\frac{1}{E-H_0+ i\epsilon}V+...
\end{eqnarray}

\section*{
طريقة الانسحابات الطورية
}
\addcontentsline{toc}{section}{\bf
طريقة الانسحابات الطورية
}

\subsection*{
معادلة شرودينغر في المنطقة
$V= 0$
}
\addcontentsline{toc}{subsection}{
معادلة شرودينغر في المنطقة
$V=0$
} 
نعتبر معادلة شرودينغر في كمون مركزي
$V(r)$
ذي مدي منته. الدوال الموجية تأخذ الشكل
\begin{eqnarray}
\psi(r,\theta,\phi)=R(r)Y_l^m(\theta,\phi).
\end{eqnarray}
المعادلة القطرية بالنسبة ل
 $u=rR$ 
 تعطي ب
\begin{eqnarray}
-\frac{\hbar^2}{2m}\frac{d^2u}{dr^2}+\bigg[V(r)+\frac{\hbar^2}{2m}\frac{l(l+1)}{r^2}\bigg]u=Eu.
\end{eqnarray}  
من اجل 
$r$
كبير, فيما يسمي بمنطقة الاشعاع, يمكن ان نهمل الكمون المركزي و كمون القوة الطاردة المركزية لنحصل علي
\begin{eqnarray}
\frac{d^2u}{dr^2}=-k^2u~,~k^2=\frac{2mE}{\hbar^2}.
\end{eqnarray}  
اذن
\begin{eqnarray}
u=Ae^{ikr}+Be^{-ikr}.
\end{eqnarray} 
من اجل موجة منتثرة يجب ان يكون لدينا
 $B=0$.
 اذن نحصل علي
\begin{eqnarray}
R=A\frac{e^{ikr}}{r}.
\end{eqnarray}  
في المنطقة الوسطي يمكن ان نهمل الكمون المركزي لنحصل علي 
\begin{eqnarray}
\frac{d^2u}{dr^2}-\frac{l(l+1)}{r^2}=-k^2u.
\end{eqnarray}  
الحل العام هو تركيب خطي لدوال بسال
\footnote{.${\rm spherical~Bessel~functions}$}
الكروية اي

\begin{eqnarray}
R=A j_l(kr)+B n_l(kr).
\end{eqnarray} 
دوال بسال الكروية من الرتبة
$l$
هي تعميم 
للدوال المثلثية
جب. هذه الدوال معرفة ب
\begin{eqnarray}
j_l(x)=(-x)^l(\frac{1}{x}\frac{d}{dx})^l\frac{\sin x}{x}~\longrightarrow \frac{2^l l!}{(2l+1)!}x^l~,~x\longrightarrow 0.
\end{eqnarray} 
دوال نيومن الكروية
\footnote{.${\rm spherical~Neumann~functions}$}
من الرتبة
$l$
هي تعميم للدوال المثلثية
تجب. هذه الدوال تعرف ايضا تحت مسمي دوال بسال من النوع الثاني و هي معرفة ب
\begin{eqnarray}
n_l(x)=-(-x)^l(\frac{1}{x}\frac{d}{dx})^l\frac{\cos x}{x}~\longrightarrow -\frac{(2l)!}{2^l l!}\frac{1}{x^{l+1}}~,~x\longrightarrow 0.
\end{eqnarray} 
تعميم الدوال الاسية
$e^{\pm ix}$
يعطي بما يسمي دوال هنكل
\footnote{.${\rm Hankel~functions}$}
.
دوال هنكل الكروية تعرف ب
\begin{eqnarray}
h_l^{(1)}(x)=j_l(x)+in_l(x)\longrightarrow \frac{(-i)^{l+1}}{x}e^{ix}~,~x\longrightarrow \infty.
\end{eqnarray} 
\begin{eqnarray}
h_l^{(2)}(x)=j_l(x)-in_l(x)\longrightarrow \frac{(i)^{l+1}}{x}e^{-ix}~,~x\longrightarrow \infty.
\end{eqnarray} 
الحل العام
$R$
يمكن اذن ان يكتب علي شكل تركيب خطي لدوال هنكل الكروية اي
\begin{eqnarray}
R=C h_l^{(1)}(kr)+D h_l^{(2)}(kr).
\end{eqnarray}
من التصرف الذي وجدناه من اجل
 $r$
 كبير نري ان
 $h_l^{(1)}$
 يوافق الموجة الكروية الخارجة بينما
 $h_l^{(2)}$ 
 يوافق الموجة الكروية الداخلة. بعبارة اخري يجب ان يكون لدينا
 $D=0$.
 نحصل علي
\begin{eqnarray}
R=C h_l^{(1)}(kr).
\end{eqnarray} 
دالة الموجة خارج منطقة التصادم اين
$V=0$
هي من الشكل
\begin{eqnarray}
\psi(r,\theta,\phi)=A\bigg[e^{ikz}+\sum_{l,m}C_{lm}h_l^{(1)}(kr)Y_l^m(\theta,\phi)\bigg].
\end{eqnarray} 
نختار المحور
$z$ 
في اتجاه الشعاع الوارد. الكمون المركزي هو كمون متناظر كرويا و بالتالي فان دالة الموجة لا يمكن ان تتعلق ب
$\phi$.
بعبارة اخري فقط الحدود
$m=0$
تشارك. نذكر

\begin{eqnarray}
Y_l^0(\theta,\phi)=\sqrt{\frac{2l+1}{4\pi}}P_l(\cos\theta).
\end{eqnarray} 
ندخل سعة الموجة الجزئية $a_l$ من الرتبة
$l$
بالمعادلة
$C_{l0}=i^{l+1}k\sqrt{4\pi(2l+1)} a_l$. 
نحصل علي
\begin{eqnarray}
\psi(r,\theta)&=&A\bigg[e^{ikz}+k\sum_{l}i^{l+1}(2l+1)a_lh_l^{(1)}(kr)P_l(\cos\theta)\bigg].\label{solutionoutside}
\end{eqnarray} 
من اجل
 $r$
 كبير نحصل علي
\begin{eqnarray}
\psi(r,\theta)
&\longrightarrow & A\bigg[e^{ikz}+f(\theta)\frac{e^{ikr}}{r}\bigg]~,~r\longrightarrow\infty.
\end{eqnarray} 
سعة التصادم تعطي بدلالة سعات الامواج الجزئية ب
\begin{eqnarray}
f(\theta)=\sum_{l}(2l+1)a_lP_l(\cos\theta).
\end{eqnarray} 
يبقي ان نعين سعات الامواج الجزئية عن طريق حل معادلة شرودينغر في منطقة التصادم او التناثر
اين
$V\neq 0$.

من النقاش اعلاه من الواضح انه لان الموجة الواردة
 $e^{ikz}$
 تحل معادلة شرودينغر من اجل
 $V=0$ 
 فانها يمكن كتابتها بدلالة دوال بسال الكروية و التوافقيات الكروية. النشر يأخذ الشكل
\begin{eqnarray}
e^{ikz}=\sum_{lm}\bigg(A_{lm} j_l(kr)+B_{lm} n_l(kr)\bigg)Y_l^m(\theta,\phi).
\end{eqnarray}
دوال نيومان تنفجر عند المبدأ و بالتالي يجب ان يكون لدينا
$B_{lm}=0$. 
بالاضافة الي هذا لان
$z=r\cos\theta$
فقط الحدود
$m=0$
 تشارك. اذن نحصل علي
\begin{eqnarray}
e^{ikz}=\sum_{l}i^lC_{l}(2l+1) j_l(kr)P_l(\cos\theta).
\end{eqnarray} 
التعبير التكاملي لدوال بسال يعطي ب
\begin{eqnarray}
j_l(q)=\frac{1}{2i^l}\int_{-1}^{+1}e^{iqx}P_l(x)dx.
\end{eqnarray} 
يمكن استعمال هذه العلاقة لنبين ان
$C_l=1$
و بالتالي
\begin{eqnarray}
e^{ikz}=\sum_{l}i^l(2l+1) j_l(kr)P_l(\cos\theta).\label{rayleigh}
\end{eqnarray} 
هذه العلاقة تعرف باسم علاقة رايلي
\footnote{.${\rm Rayleigh's~formula}$}
.
\subsection*{
الامواج المستوية و الكروية
}
\addcontentsline{toc}{subsection}{
الامواج المستوية و الكروية
} 
الاشعة الذاتية لهاميلتونية الجسيم الحر
$H_0$
هي اشعة حالة الامواج المستوية
$|\vec{k}>$ 
التي تحقق علاقة التعامد و التجانس
\begin{eqnarray}
<\vec{k}^{'}|\vec{k}>=\delta^3(\vec{k}-\vec{k}^{'}).
\end{eqnarray} 
الهاميلتونية
$H_0$
تتبادل مع مؤثرات العزم الحركي
 $L^2$
 و
 $L_3$.
 اذن الاشعة الذاتية ل
$H_0$ 
يمكن ايضا اخذها الاشعة الذاتية المشتركة ل
$H_0$ ,$L^2$
و
$L_3$
التي نرمز لها ب
$|Elm>$. 
هذه هي بالضبط اشعة حالة الامواج الكروية التي تحقق علاقة التعامد و التجانس
\begin{eqnarray}
<E^{'}l^{'}m^{'}|Elm>=\delta_{ll^{'}}\delta_{mm^{'}}\delta(E-E^{'}).
\end{eqnarray} 
من الواضح ان الدوال الموجية الكروية في فضاء كمية الحركة
$<\vec{k}|Elm>$
هي متناسبة مع 
$\delta(E-\frac{\hbar^2k^2}{2m})$. 
نكتب
\begin{eqnarray}
<\vec{k}|Elm>=\frac{\hbar}{\sqrt{mk}}\delta(E-\frac{\hbar^2 k^2}{2m})X_l^m(k,\hat{k}).
\end{eqnarray} 
نحسب
\begin{eqnarray}
\int d^3\vec{k} <E^{'}l^{'}m^{'}|\vec{k}><\vec{k}|Elm>=\delta(E-E^{'})\int d\Omega X_{l^{'}}^{m^{'}}(k,\hat{k})(X_{l}^{m}(k,\hat{k}))^{*}.
\end{eqnarray}
في المعادلة اعلاه
$d\Omega$
هي الزاوية الصلبة الموافقة لشعاع الوحدة
$\hat{k}$
و
$k$
هو بحيث 
$k=\sqrt{{2mE}/{\hbar^2}}$. 
يجب ان يكون لدينا
\begin{eqnarray}
\int d\Omega X_{l^{'}}^{m^{'}}(k,\hat{k})(X_{l}^{m}(k,\hat{k}))^{*}=\delta_{l^{'}l}\delta_{m^{'}m}.
\end{eqnarray} 
يعطي الحل مباشرة بالتوافقيات الكروية اي
\begin{eqnarray}
X_{l}^{m}(k,\hat{k})=X_{l}^{m}(\hat{k})=Y_l^m(\hat{k}).
\end{eqnarray} 
 يمكن نشراشعة حالة الامواج المستوية
$|\vec{k}>$ 
بدلالة اشعة حالة الامواج الكروية
$|Elm>$
كالاتي
 \begin{eqnarray}
|\vec{k}>&=&\sum_l\sum_m\int dE |Elm><Elm|\vec{k}>\nonumber\\
&=&\frac{\hbar}{\sqrt{mk}}\sum_l\sum_m |Elm>(Y_l^m(\hat{k}))^*~,~E=\frac{\hbar^2 k^2}{2m}.
\end{eqnarray} 
نحتاج الان ان نحسب
 \begin{eqnarray}
<\vec{x}|\vec{k}>
&=&\frac{\hbar}{\sqrt{mk}}\sum_l\sum_m <\vec{x}|Elm>(Y_l^m(\hat{k}))^*~,~E=\frac{\hbar^2 k^2}{2m}.
\end{eqnarray} 
الدوال الموجية الكروية في فضاء الموضع
هي
$<\vec{x}|Elm>$.
من الفقرة السابقة نعرف ان الامواج الكروية تعطي ب
 \begin{eqnarray}
<\vec{x}|Elm>=c_lj_l(kr)Y_l^m(\hat{r}).
\end{eqnarray} 
اذن
\begin{eqnarray}
\frac{e^{i\vec{k}\vec{x}}}{(2\pi)^{\frac{3}{2}}}
&=&\frac{\hbar}{\sqrt{mk}}\sum_l\sum_m c_l j_l(kr)Y_l^m(\hat{r})(Y_l^m(\hat{k}))^*\nonumber\\
&=&\frac{\hbar}{\sqrt{mk}}\frac{1}{4\pi}\sum_l(2l+1)c_l j_l(kr)P_l(\hat{k}\hat{r}).
\end{eqnarray} 
استعملنا اعلاه النتيجة
 \begin{eqnarray}
\sum_m Y_l^m(\hat{r})(Y_l^m(\hat{k}))^*=\frac{2l+1}{4\pi}P_l(\hat{k}\hat{r}).
\end{eqnarray}   
اذن
\begin{eqnarray}
e^{i\vec{k}\vec{x}}
&=&\frac{\hbar}{\sqrt{mk}}\sum_l\sum_m c_l j_l(kr)Y_l^m(\hat{r})(Y_l^m(\hat{k}))^*\nonumber\\
&=&{\hbar}{\sqrt{\frac{\pi}{2mk}}}\sum_l(2l+1)c_l j_l(kr)P_l(\hat{k}\hat{r}).
\end{eqnarray}
بمقارنة هذه العلاقة مع 
$(\ref{rayleigh})$
نحصل علي
  \begin{eqnarray}
c_l=\frac{i^l}{\hbar}\sqrt{\frac{2mk}{\pi}}.
\end{eqnarray}  
في الخلاصة الامواج الكروية في فضاء الموضع و فضاء كمية الحركة تعطي ب
\begin{eqnarray}
<\vec{k}|Elm>=\frac{\hbar}{\sqrt{mk}}\delta(E-\frac{\hbar^2 k^2}{2m})Y_l^m(\hat{k}).
\end{eqnarray} 
  \begin{eqnarray}
<\vec{x}|Elm>=\frac{i^l}{\hbar}\sqrt{\frac{2mk}{\pi}}j_l(kr)Y_l^m(\hat{r}).
\end{eqnarray}

\subsection*{
سعات الامواج الجزئية و الانسحابات الجزئية
}
\addcontentsline{toc}{subsection}{
سعات الامواج الجزئية و الانسحابات الجزئية
} 

ننطلق من سعة التصادم بدلالة مؤثر الانتقال التي تعطي ب
 \begin{eqnarray}
f(k\hat{r},\vec{k})
&=&-\frac{m}{\hbar^2}(2\pi)^2<k\hat{r}|T|\vec{k}>\nonumber\\
&=&-\frac{4\pi^2}{k}\sum_{ml}\sum_{m^{'}l^{'}}Y_{l^{'}}^{m^{'}}(\hat{r})(Y_l^m(\hat{k}))^*<El^{'}m^{'}|T|Elm>~,~E=\frac{\hbar^2k^2}{2m}.\nonumber\\
\end{eqnarray}
المؤثر
 $T$
 يتبادل مع
 $L^2$
 و 
 $L_3$ 
 بسبب التناظر تحت تأثير الدورانات. اذن
 $T$
 هو مؤثر سلمي و منه باستعمال مبرهنة
 فيجنر - ايكارت
 \footnote{.${\rm Wigner-Eckart~theorem}$}
 لدينا
\begin{eqnarray}
<El^{'}m^{'}|T|Elm>=T_l(E)\delta_{ll^{'}}\delta_{mm^{'}}.
\end{eqnarray}
بالتالي
 \begin{eqnarray}
f(k\hat{r},\vec{k})
&=&-\frac{4\pi^2}{k}\sum_{ml}Y_{l}^{m}(\hat{r})(Y_l^m(\hat{k}))^*T_l(E)\nonumber\\
&=&-\frac{\pi}{k}\sum_{l}(2l+1)T_l(E)P_l(\hat{k}\hat{r})\nonumber\\
&=&\sum_l(2l+1)a_lP_l(\hat{k}\hat{r}).
\end{eqnarray}
سعة الموجة الجزئية من الرتبة
$l$
تعطي ب
$a_l$
و هي معرفة ب
\begin{eqnarray}
a_l=-\frac{\pi}{k}T_l(E).
\end{eqnarray}
نكتب المعادلة اعلاه علي الشكل
\begin{eqnarray}
f(k\hat{r},\vec{k})
&=&\sum_l(2l+1)a_lP_l(\cos\theta).
\end{eqnarray}
نرجع الان الي معادلة ليبمان - شوينغر من اجل 
$r$ 
كبير التي تعطي ب
\begin{eqnarray}
<\vec{x}|\psi_{+}>=\frac{1}{(2\pi)^{\frac{3}{2}}}\bigg[e^{i\vec{k}\vec{x}}+\frac{e^{+ ikr}}{r}f(k\hat{r},\vec{k})\bigg].
\end{eqnarray}
لقد وجدنا ان
\begin{eqnarray}
e^{i\vec{k}\vec{x}}
&=&\sum_li^l(2l+1) j_l(kr)P_l(\cos\theta).\label{freewave}
\end{eqnarray}
ايضا لدينا التصرف
\begin{eqnarray}
j_l(kr)=\frac{1}{2kr}(-i)^{l+1}e^{ikr}+\frac{1}{2kr}i^{l+1}e^{-ikr}~,~r\longrightarrow \infty.
\end{eqnarray}
اذن
\begin{eqnarray}
<\vec{x}|\psi_{+}>=\frac{1}{(2\pi)^{\frac{3}{2}}}\sum_l(2l+1)\frac{P_l(\cos\theta)}{2ik}\bigg[\frac{e^{ikr}}{r}S_l-\frac{e^{-i(kr-l\pi)}}{r}\bigg].\label{larger}
\end{eqnarray}
\begin{eqnarray}
S_l=1+2ika_l.
\end{eqnarray}
عندما يكون المشتت اي مركز التصادم غائب اي لما
$V=0$
فان سعة الموجة الجزئية
$a_l$
تنعدم و نحصل علي جمع موجة كروية خارجة و موجة كروية داخلة من اجل كل
$l$. 
تأثير المشتت اذن هو ان يسحب معاملات الموجة الكروية الخارجة كالاتي
$1\longrightarrow 1+2ika_l$.


من قانون انحفاظ الاحتمال نعرف ان تدفق الجسيمات الواردة يجب ان يكون مساو لتدفق الجسيمات الخارجة. باستعمال قانون انحفاظ العزم
الحركي يجب ان يحدث هذا من اجل كل موجة جزئية علي حدة. هذا يعني ان سعات الامواج الواردة و الامواج الخارجة بنفس قيمة العزم الحركي
$l$ 
يجب ان تكون متساوية اي

\begin{eqnarray}
|S_l|=1.
\end{eqnarray}
هذه العلاقة تعرف باسم
علاقة الاحادية
\footnote{.${\rm unitarity~relation}$}
. 
بعبارة اخري فان الانسحاب في طور الموجة الخارجة ينجم بالضبط من عملية التصادم. الطور
$S_l$ 
هو العنصر القطري رقم
$l$
لمصفوفة التصادم
$S$
التي يجب ان تكون احادية بسبب قانون انحفاظ الاحتمال. نعرف الانسحاب الطوري
$\delta_l$
ب
\begin{eqnarray}
S_l=e^{2i\delta_l}.
\end{eqnarray}
اذن
\begin{eqnarray}
a_l=e^{i\delta_l}\frac{\sin\delta_l}{k}.
\end{eqnarray}
انطلاقا من هذه المعادلة نري ان
${\rm Re}(ka_l)=\sin 2\delta_l/2$ 
و
${\rm Im}(ka_l)-1/2=-\cos 2\delta_l/2$.
اذن
$({\rm Re}(ka_l))^2+({\rm Im}(ka_l)-1/2)^2=1/4$. 
بعبارة اخري 
$ka_l$
يقع علي دائرة نصف قطرها
$1/2$ 
و مركز
$(0,1/2)$
تعرف بالدائرة الاحادية.

اذن نحصل علي
\begin{eqnarray}
f(\theta)
&=&\frac{1}{k}\sum_l(2l+1)e^{i\delta_l} \sin\delta_l P_l(\cos\theta).
\end{eqnarray}
المقطع الفعال الكلي هو
\begin{eqnarray}
\sigma &=&\int d\Omega \frac{d\sigma}{d\Omega}\nonumber\\
&=&\int d\Omega |f(\theta)|^2\nonumber\\
&=&\frac{1}{k^2}\sum_{l,l^{'}}(2l+1)(2l^{'}+1)e^{i\delta_l-i\delta_{l^{'}}} \sin\delta_l \sin\delta_{l^{'}}\int d\Omega P_l(\cos\theta)P_{l^{'}}(\cos\theta).\nonumber\\
\end{eqnarray}
نستعمل العلاقة
\begin{eqnarray}
\int d\Omega P_l(\cos\theta)P_{l^{'}}(\cos\theta)=\frac{4\pi}{2l+1}\delta_{ll^{'}}.
\end{eqnarray}
اذن
\begin{eqnarray}
\sigma 
&=&\frac{4\pi}{k^2}\sum_{l}(2l+1) (\sin\delta_l)^2.
\end{eqnarray}
نلاحظ ان
\begin{eqnarray}
{\rm Im}f(0)&=&\frac{k}{4\pi}\sigma.
\end{eqnarray}
هذا يسمي بالمبرهنة الضوئية
\footnote{.${\rm optical~theorem}$}

يبقي ان نعين الانسحابات الطورية من اجل كمون معين
$V$.
مرة اخري نعتبر حالة كمون مركزي ذي مدي منته. هذه المرة نفترض ان الكمون ينعدم من اجل
$r>R$
حيث
$R$
هو مدي الكمون. من اجل
$r>R$ 
يجب ان يكون لدينا موجة كروية حرة. هذه يجب ان تكتب علي شكل تركيب خطي ل
$j_l(kr)P_l(\cos\theta)$
و
$n_l(kr)P_l(\cos\theta)$
حيث ان
$n_l$
هي  الان
مشمولة لان
المبدأ
$r=0$ 
مستبعد. بالمقابل
الموجة الكروية الحرة من اجل
$r>R$ 
يمكن كتابتها علي شكل تركيب خطي ل
 $h_l^{(1)}(kr)P_l(\cos\theta)$ 
 و
 $h_l^{(2)}(kr)P_l(\cos\theta)$.
 بعبارة اخري من اجل
 $r>R$
 يمكن ان نكتب
\begin{eqnarray}
<\vec{x}|\psi_+>
&=&\frac{1}{(2\pi)^{\frac{3}{2}}}\sum_li^l(2l+1) A_l(kr)P_l(\cos\theta).
\end{eqnarray}
\begin{eqnarray}
A_l(kr)=c_l^{(1)}h_l^{(1)}(kr)+c_l^{(2)}h_l^{(2)}(kr).
\end{eqnarray}
من اجل
$V=0$
هذا يختزل ل
$(\ref{freewave})$. 
هذه المعادلة مكافئة ل
$(\ref{solutionoutside})$.
هذا يمكن تبيانه كالاتي. من اجل
 $r$
 كبير نحصل علي

\begin{eqnarray}
<\vec{x}|\psi_{+}>=\frac{1}{(2\pi)^{\frac{3}{2}}}\sum_l(2l+1)\frac{P_l(\cos\theta)}{2ik}\bigg[\frac{e^{ikr}}{r}2c_l^{(1)}-2c_l^{(2)}\frac{e^{-i(kr-l\pi)}}{r}\bigg].
\end{eqnarray}
بالمقارنة مع
$(\ref{larger})$ 
نحصل علي
\begin{eqnarray}
c_l^{(1)}=\frac{1}{2}e^{2i\delta_l}~,~c_l^{(2)}=\frac{1}{2}.
\end{eqnarray}
اذن
\begin{eqnarray}
A_l(kr)=j_l(kr)+ika_lh_l^{(1)}(kr).
\end{eqnarray}
هذا يمكن ايضا كتابته علي الشكل
\begin{eqnarray}
A_l(kr)=e^{i\delta_l}\bigg[\cos\delta_l j_l(kr)-\sin\delta_l n_l(kr)\bigg].
\end{eqnarray}
نحسب
\begin{eqnarray}
\beta_l&=&\bigg(\frac{r}{A_l}\frac{d A_l}{dr}\bigg)_{r=R}\nonumber\\
&=&kR \frac{\cos\delta_l j_l^{'}(kR)-\sin\delta_l n_l^{'}(kR)}{\cos\delta_l j_l(kR)-\sin\delta_l n_l(kR)}.
\end{eqnarray}
من هذه المعادلة نستنتج
\begin{eqnarray}
\tan\delta_l=\frac{kRj_l^{'}(kR)-\beta_lj_l(kR)}{kRn_l^{'}(kR)-\beta_ln_l(kR)}.
\end{eqnarray}
اذن من اجل ايجاد
$\delta_l$ 
نحتاج ان نجد
$\beta_l$.
هذا الامر يمكن انجازه فقط عن طريق حل معادلة شرودينغر من اجل
$r<R$. 
بعبارة اخري نحتاج ان نعين
$A_l^{\rm inside}(r)={u_l(r)}/{r}$
حيث
$u$
تحقق المعادلة التفاضلية القطرية و الشرط الحدي المعطيان ب

\begin{eqnarray}
\frac{d^2u_l}{dr^2}+(k^2-\frac{2m}{\hbar^2}V-\frac{l(l+1)}{r^2})u_l=0.
\end{eqnarray}
\begin{eqnarray}
u_l(0)=0.
\end{eqnarray}
المشتقة الاولي لدالة الموجة يجب ان تكون مستمرة عند
$r=R$. 
اذن يجب ان يكون لدينا
\begin{eqnarray}
\beta_l^{\rm inside}\equiv \bigg(\frac{r}{A_l^{\rm inside}}\frac{d A_l^{\rm inside}}{dr}\bigg)_{r=R}=\beta_l^{}.
\end{eqnarray}
\newpage
\section*{
تمارين
}
\addcontentsline{toc}{section}{\bf 
تمارين
}

\paragraph{تمرين
$1$}
 تصادم جسيم مع كرة صلبة يعطي بالكمون
\begin{eqnarray}
V&=&0~,~r>R\nonumber\\
&=&\infty~,~r<R.\nonumber
\end{eqnarray}
\begin{itemize}
\item[$(1$]
عين زاوية التصادم, المقطع الفعال التفاضلي و المقطع الفعال الكلي من اجل تصادم كلاسيكي لجسيم عبر الكمون 
$V$.

\item[$(2$]
احسب التغير في الطور
$\delta_l$ 
من اجل تصادم كمي لجسيم عبر الكمون 
$V$.
تذكر ان دالة الموجة من اجل
$r>R$
تعطي ب
\begin{eqnarray}
<\vec{x}|\psi_+>
&=&\frac{1}{(2\pi)^{\frac{3}{2}}}\sum_li^l(2l+1) A_l(kr)P_l(\cos\theta).\nonumber
\end{eqnarray}

\begin{eqnarray}
A_l(kr)=e^{i\delta_l}\bigg[\cos\delta_l j_l(kr)-\sin\delta_l n_l(kr)\bigg].\nonumber
\end{eqnarray}

\item[$(3$]
عين التغير في الطور, المقطع الفعال التفاضلي و المقطع الفعال الكلي من اجل الطاقات الصغيرة
$kR<<1$. 
ماذا تستنتج.
\item[$(4$]
احسب المقطع الفعال الكلي من اجل الطاقات العليا
$kR>>1$.
ماذا تستنتج.
\end{itemize}
نعطي
\begin{eqnarray}
j_l(x)\longrightarrow \frac{2^l l!}{(2l+1)!}x^l~,~n_l(x)\longrightarrow -\frac{(2l)!}{2^l l!}\frac{1}{x^{l+1}}~,~x\longrightarrow 0.\nonumber
\end{eqnarray} 
\begin{eqnarray}
j_l(x)\longrightarrow \frac{1}{x}\sin(x-\frac{l\pi}{2})~,~
n_l(x)\longrightarrow -\frac{1}{x}\cos(x-\frac{l\pi}{2})~,~x\longrightarrow \infty.\nonumber
\end{eqnarray}
\paragraph{تمرين
$2$
}
نعطي كمون يوكاوا
 \begin{eqnarray}
V(r)=\beta\frac{e^{-r}}{r}.\nonumber
\end{eqnarray}
\begin{itemize}
\item[$(1$]
احسب سعة بورن من الرتبة الاولي.
\item[$(2$]
احسب المقطع الفعال التفاضلي.
\item[$(3$]
عين من النتيجة السابقة المقطع الفعال التفاضلي من اجل تصادمات كولومب. قارن مع نتيجة رذرفورد الكلاسيكية.  
\end{itemize}
\paragraph{تمرين
$3$
}
نعتبر تصادم الالكترونات مع ذرات الهيدروجين  ( الذي قد يكون مرن او غير مرن) كالاتي:
\begin{eqnarray}
e^{-}+H(~{\rm ground}~{\rm state})\longrightarrow e^{-}+H(~{\rm excited}~{\rm state}).\nonumber
\end{eqnarray}
\begin{itemize}
\item[$(1$]
اكتب الحالة الابتدائية
$|i>$
قبل التصادم, الحالة النهائية
$|f>$
 بعد التصادم, سعة بورن من الرتبة الاولي
$f^{(1)}$
 و كمون التفاعل
$V$.
 اذكر
الشروط التي يكون فيها التصادم مرنا و الشروط التي يكون فيها غير مرن.

 
\item[$(2$]
احسب تحويل فورييه لكمون كولومب.
\item[$(3$]
احسب عنصر المصفوفة
$<\vec{k}^{'}<n|V|\vec{k}>|1>$.
\item[$(4$]
احسب سعة بورن من الرتبة الاولي, المقطع الفعال التفاضلي و معامل الشكل من اجل تصادم مرن.
نذكر ان دالة موجة ذرة الهيدروجين في الحالة الاساسية هي
\begin{eqnarray}
\psi_{100}=\frac{1}{\sqrt{\pi a^3}}e^{-\frac{r}{a}}.\nonumber
\end{eqnarray}
\item[$(5$] 
احسب المقطع الفعال التفاضلي من اجل تصادم غير مرن انطلاقا من العلاقة
\begin{eqnarray}
\frac{d\sigma}{d\Omega}=\frac{k^{'}}{k}|f^{(1)}(\vec{k}^{'},n,\vec{k},1)|^2.\nonumber
\end{eqnarray}

\end{itemize}

\paragraph{تمرين
$4$}
نعتبر تصادم جسيمات عبر كمون دالة ديراك الذي يعطي ب
\[V(r)=\alpha\delta (r-a).\]
 سوف نعتبر هنا التصادمات ذات الطاقات المنخفضة اي 
$ka<<1$ 
و بالتالي فان الامواج 
$s$
هي التي تطغي علي التصادم. معادلة شرودينغر المدارية تعطي ب
\[-\frac{\hbar^2}{2m}\frac{d^2u}{dr^2}+\bigg[V(r)+\frac{\hbar^2}{2m}\frac{l(l+1)}{r^2}\bigg]u=Eu.\]
دالة الموجة تعطي ب
\[\psi(r,\theta,\phi)=\frac{u(r)}{r}Y_l^m(\theta,\phi).\]
نعتبر فقط التقريب الذي يمكننا فيه وضع
$l=0$
من البدأ.
\begin{itemize}
\item[$(1$]
أكتب حل معادلة شرودينغر المدارية من اجل
$r<a$.
\item[$(2$]
اكتب الحل العام من اجل
 $r>a$.
قارن مع تصرف الحل من اجل اي كمون كروي  لما
$r\longrightarrow\infty$
 الذي يعطي ب

\[\frac{1}{(2\pi)^{1.5}}\sum_l(2l+1)\frac{1}{2ik}\bigg(\frac{e^{ikr}}{r}S_l+e^{-ikr}{r}(-1)^{l+1}\bigg)P_l(\cos\theta).\]
\[S_l=e^{2i\delta_l}=1+2ika_l.\]
أستخرج عبارة للموجة الجزئية 
$a_0$
بدلالة
معاملات الامواج الواردة و المتصادمة. اكتب دالة الموجة في هذه المنطقة بدلالة التغير في الطور
$\delta_0$.
\item[$(3$]
دالة الموجة يجب ان تكون مستمرة
في 
$r=a$.
اكتب الشرط الحدي المقابل.

المشتقة الاولي لدالة الموجة يجب ان تكون مستمرة الا في النقاط التي يتباعد فيها الكمون. الشرط المقابل يكتب علي الشكل
\[\Delta(\frac{du}{dr})=\frac{2m\alpha}{\hbar^2}u(a).\]

استخدم هذين الشرطين الحديين من اجل تعيين الموجة الجزئية
$a_0$.
\item[$(4$]
احسب
$f(\theta)$, $d\sigma/d\Omega$
و
$\sigma$.
ماهو الجواب من اجل
$ka<<1$. 
عرف
\[\beta=\frac{2ma\alpha}{\hbar^2}.\]
\item[$(5$]
احسب التغير في الطور
$\delta_0$.
\end{itemize}

\paragraph{تمرين
$5$}
\begin{itemize}
\item[$(1$]
في الفعل الكهروضوئي الحالة الابتدائية للالكترون هي حالة ذرية مرتبطة ذات طاقة
$E_i<0$ 
في حين ان الحالة النهائية هي حالة حرة ذات طاقة مستمرة
حول
$E_n>0$.
اكتب قاعدة فيرمي الذهبية للفعل الكهروضوئي .
\item[$(2$]
احسب كثافة الحالات النهائية
$\rho(E_n)$.
استعمل تنظيم العلبة للامواج المستوية الذي يعطي 
ب

\begin{eqnarray}
<\vec{x}|\vec{k}>=\frac{1}{L^{\frac{3}{2}}}e^{i\vec{k}\vec{x}}.\nonumber
\end{eqnarray}
في هذه الحالة القيم المسموح بها ل
$k_x$, $k_y$ 
و
$k_z$ 
هي
\begin{eqnarray}
k_x=\frac{2\pi n_x}{L}~,~k_y=\frac{2\pi n_y}{L}~,~k_z=\frac{2\pi n_z}{L}.\nonumber
\end{eqnarray}
علاقة التعامد و التجانس  و علاقة الانغلاق تعطي ب
 \begin{eqnarray}
<\vec{k}^{'}|\vec{k}>=\delta_{\vec{k}^{'},\vec{k}}.\nonumber
\end{eqnarray}
 \begin{eqnarray}
\sum_{\vec{k}}|\vec{k}><\vec{k}|=1.\nonumber
\end{eqnarray}
\item[$(3$]
استخرج المقطع الفعال التفاضلي للتفاعل المعرف ب
\begin{eqnarray}
\frac{d\sigma}{d\Omega}=\frac{h\nu}{u}w^{\rm abso}_{i\longrightarrow n}.\nonumber
\end{eqnarray}
\end{itemize}

\paragraph{تمرين
$6$}
 يمكن تصور التصادم علي انه عملية انتقال من الحالة الحرة الابتدائية
$|\vec {k}>$ 
في اللحظة
 $t=-\infty$
(الماضي البعيد)
الي مجموعة من الحالات الحرة النهائية
 $|\vec{k}^{'}>$ 
في اللحظة
$t$
التي لها كمية حركة محتواة في الزاوية الصلبة
$d\Omega=d^3k^{'}/(k^{'2}dk^{'})$
.
سوف
نعتبر التصادمات المرنة فقط التي من اجلها
$k^{'}=k$. 
في عملية الانتقال هذه التفاعل منعدم في اللحظة
$t=-\infty$
 ثم يبدأ في التزايد ببطء (اي ادياباتيكلي) مع الزمن. الكمون المتعلق بالزمن يأخذ بالتالي  الشكل
\begin{eqnarray}
V(t)=Ve^{\eta t}.\nonumber
\end{eqnarray}
في هذه المعادلة 
$\eta$ 
هو
عدد حقيقي موجب يجب ارساله الي الصفر  عند نهاية الحساب.
\begin{itemize}
\item[$(1$]
برهن ان احتمال الانتقال  مازال يعطي بقاعدة فيرمي الذهبية.

\item[$(2$]
احسب معدل الانتقال الي مجموعة الحالات النهائية
$|\vec{k}^{'}>$ 
التي لها طاقة بين
$E_{{k}^{'}}$ و $E_{{k}^{'}}+dE_{{k}^{'}}$. 
استعمل كثافة الحالات النهائية التي تحصلنا عليها في المسألة الثانية.

\item[$(3$]
احسب التدفق الوارد الذي يعرف ب

\begin{eqnarray}
|\vec{j}|=\frac{\hbar}{m}|{\rm Im}(\psi^*\vec{\nabla}\psi)|~,~\psi=<\vec{x}|\vec{k}>.\nonumber
\end{eqnarray}
\item[$(4$]
استخرج المقطع الفعال التفاضلي. قارن النتيجة مع تقريب بورن من الرتبة الاولي.
 \end{itemize}

\newpage
\section*{
حلول
}
\addcontentsline{toc}{section}{\bf 
حلول
} 

\paragraph{
تمرين
$1$:}
\begin{itemize}
\item[$(1$]
جسيم كلاسيكي وارد بمعامل صدمة
$b$
اكبر من نصف قطر الكرة يخطئ الكمون بالكامل. في هذه الحالة اذن
$\theta=0$.
 في الحالة
 $b\leq R$
  ينعكس الجسيم علي سطح الكرة بزاوية
   $\alpha$
   حيث
   $2\alpha+\theta=\pi$.
   لدينا
\begin{eqnarray}
\frac{b}{R}=\sin\alpha=\cos\frac{\theta}{2}.
\end{eqnarray}
لدينا اذن
\begin{eqnarray}
\theta&=&0~,~b>R\nonumber\\
&=&2\cos^{-1}\frac{b}{R}~,~b<R.
\end{eqnarray}
المقطع الفعال التفاضلي هو اذن
\begin{eqnarray}
D(\theta)=\frac{b}{\sin\theta}|\frac{db}{d\theta}|=\frac{R^2}{4}.
\end{eqnarray}
المقطع الفعال الكلي هو
\begin{eqnarray}
\sigma=\int D(\theta)d\Omega=\pi R^2.
\end{eqnarray}
هذا هو المقطع الفعال الهندسي.


\item[$(2$]
يجب ان تنعدم دالة الموجة عند
$r=R$.
بالتالي يجب ان يكون لدينا
$A_l(kR)=0$
و بالتالي
\begin{eqnarray}
\tan\delta_l=\frac{j_l(kR)}{n_l(kR)}.
\end{eqnarray}
\item[$(3$]
من اجل الطاقات المنخفضة لدينا
$kR<<1$. 
اذن نحصل علي
\begin{eqnarray}
\tan\delta_l=-\bigg(\frac{2^ll!}{(2l)!}\bigg)^2\frac{(kR)^{2l+1}}{2l+1}.
\end{eqnarray}
لان
$kR<<1$ 
فان
$l=0$
هو فقط المهم. لدينا تصادم للامواج 
$s$. 
الانسحاب الطوري يعطي ب
\begin{eqnarray}
\delta_0=-kR.
\end{eqnarray}
المقطع الفعال التفاضلي يعطي ب
\begin{eqnarray}
\frac{d\sigma}{d\Omega}=|f(\theta)|^2&=&|\frac{1}{k}\sum_l(2l+1)e^{i\delta_l} \sin\delta_l P_l(\cos\theta)|^2\nonumber\\
&=&\frac{1}{k^2}\sin^2\delta_0\nonumber\\
&=&R^2.
\end{eqnarray}
المقطع الفعال الكلي يعطي ب
\begin{eqnarray}
\sigma &=&\int d\Omega \frac{d\sigma}{d\Omega}=4\pi R^2.
\end{eqnarray}
هذا اكبر باربع مرات من المقطع الفعال الهندسي و يساوي مساحة كرة نصف قطرها
$R$. 
اذن الجسيم الكمومي يحس بالكرة باكملها.

في هذه الحالة لدينا
\begin{eqnarray}
<\vec{x}|\psi_+>
&=&\frac{1}{(2\pi)^{\frac{3}{2}}}A_0(kr)=\frac{1}{(2\pi)^{\frac{3}{2}}}e^{-ikR}\frac{\sin k(r-R)}{kr}.
\end{eqnarray}
\item[$(4$]
من اجل التصادمات ذات الطاقات العليا لدينا
$kR>>1$.
 نحسب
\begin{eqnarray}
\sin^2\delta_l=\frac{\tan^2\delta_l}{1+\tan^2\delta_l}=\frac{j_l^2(kR)}{j_l^2(kR)+n_l^2(kR)}.
\end{eqnarray}
من الواضح ان
\begin{eqnarray}
\sin^2\delta_l=\sin^2(kR-\frac{l\pi}{2})
\end{eqnarray}
المقطع الفعال الكلي يعطي بالمعادلة
\begin{eqnarray}
\frac{k^2}{4\pi}\sigma &=&\sum_{l}(2l+1) (\sin\delta_l)^2.
\end{eqnarray}
نلاحظ العلاقة
$\sin^2\delta_{l+1}=1-\sin^2\delta_l$.
اذن من اجل القيم الزوجية ل
 $l$
 انطلاقا من
 $l=0$ 
 لدينا
 $\sin^2\delta_l=\sin^2kR$
 بينما من اجل القيم الفردية ل
 $l$
 انطلاقا من
 $l=1$
 لدينا
 $\sin^2\delta_l=\cos^2kR$.
 اذن نحصل علي

\begin{eqnarray}
\frac{k^2}{4\pi}\sigma &=&\frac{1}{2}\sum_{l}(2l+1) -\frac{1}{2}\cos 2kR\bigg(\sum_{l~{\rm even}}(2l+1)-\sum_{l~{\rm odd}}(2l+1)\bigg).\nonumber\\
\end{eqnarray}
نفترض الان ان العزوم الزاوية حتي
$l_{\rm max}=N-1$
حيث
$N=kR$ 
هي فقط التي تشارك. نحصل علي
\begin{eqnarray}
\frac{1}{4\pi R^2}\sigma &=&\frac{1}{2} -\frac{1}{2N}\cos 2kR.
\end{eqnarray}
الحد الثاني مهمل في نهاية الطاقات العليا اي لما
$N\longrightarrow\infty$.
 نحصل بالتالي علي
\begin{eqnarray}
\sigma &=&2\pi R^2.
\end{eqnarray}
هذا يساوي ضعف المقطع الفعال الهندسي.
\end{itemize}
\paragraph{
تمرين
$2$:}
\begin{itemize}
\item[$(1$]
من اجل كمون مركزي سعة بورن من الرتبة الولي تعطي ب
\begin{eqnarray}
f^{(1)}(k\hat{r},\vec{k})&=&-\frac{2m}{q\hbar^2}\int r\sin qr V(r)dr\nonumber\\
&=&\frac{im\beta}{q\hbar^2}\int_0^{\infty}\bigg[e^{(iq-\mu)r}-e^{-(iq+\mu)r}\bigg]dr\nonumber\\
&=&\frac{im\beta}{q\hbar^2}\frac{2iq}{\mu^2+q^2}\nonumber\\
&=&-\frac{2m\beta}{\hbar^2}\frac{1}{\mu^2+q^2}.
\end{eqnarray}
\item[$(2$]
المقطع الفعال التفاضلي يعطي ب
\begin{eqnarray}
\frac{d\sigma}{d\Omega}=|f^{(1)}(k\hat{r},\vec{k})|^2&=&\bigg(\frac{2m\beta}{\hbar^2}\bigg)^2\frac{1}{\bigg(\mu^2+4k^2\sin^2\frac{\theta}{2}\bigg)^2}.
\end{eqnarray}
استخدمنا النتيجة
\begin{eqnarray}
q=|\vec{k}-k\hat{r}|=2k\sin\frac{\theta}{2}.
\end{eqnarray}
\item[$(3$]
في النهاية
$\mu\longrightarrow 0$
نحصل علي كمون كولون مع المطابقة
\begin{eqnarray}
\beta=\frac{q_1q_2}{4\pi\epsilon_0}.
\end{eqnarray}
نحصل اذن علي
\begin{eqnarray}
\frac{d\sigma}{d\Omega}&=&\frac{1}{4}\bigg(\frac{q_1q_2}{8\pi\epsilon_0E}\bigg)^2\frac{1}{\sin^4\frac{\theta}{2}}.
\end{eqnarray}
هذه هي علاقة رذرفورد.
\end{itemize}
\paragraph{
تمرين
$3$:}
\begin{itemize}
\item[$(1$]
ليكن
$\vec{x}$
شعاع موضع الالكترون الوارد و
$\vec{x}_1$
شعاع موضع الالكترون المرتبط. الحالة الابتدائية هي
\begin{eqnarray}
|i>=|\vec{k}>|1>\Leftrightarrow <\vec{x}|<\vec{x}_1|\vec{k}>|1>=\frac{1}{(2\pi)^{\frac{3}{2}}}e^{i\vec{k}\vec{x}}\psi_1(\vec{x_1}).
\end{eqnarray}
الموجة المستوية تصف الالكترون الوارد بينما
$\psi_1(\vec{x}_1)$
هي دالة موجة الهيدروجين في الحالة الاساسية. الحالة النهائية هي
\begin{eqnarray}
|f>=|\vec{k}^{'}>|n>\Leftrightarrow <\vec{x}|<\vec{x}_1|\vec{k}^{'}>|n>=\frac{1}{(2\pi)^{\frac{3}{2}}}e^{i\vec{k}\vec{x}}\psi_n(\vec{x_1}).
\end{eqnarray}
ال
$\psi_n(\vec{x}_1)$
هي دالة موجة الهيدروجين في المستوي الطاقوي
$n$. 
من اجل التصادمات المرنة يجب ان يكون لدينا
$n=1$ 
و
$\vec{k}^{'}=k\hat{r}$.

سعة بورن من الرتبة الاولي في هذه الحالة هي
\begin{eqnarray}
f^{(1)}(\hat{k}^{'},n,\vec{k},1)=-\frac{m}{\hbar^2}(2\pi)^2<\vec{k}^{'}<n|V|\vec{k}>|1>.
\end{eqnarray}
الكمون هو
( مع
$r=|\vec{x}|$
)

\begin{eqnarray}
V=\frac{1}{4\pi\epsilon_0}\bigg(-\frac{e^2}{r}+\frac{e^2}{|\vec{x}-\vec{x}_1|}\bigg).
\end{eqnarray}
\item[$(2$]
تحويل فورييه لكمون كولون هو
\begin{eqnarray}
\int d^3x e^{i\vec{q}\vec{x}}\frac{1}{r}&=&-\frac{2\pi i}{q}{\rm Lim}_{\mu\longrightarrow 0}\int dr \bigg[e^{(iq-\mu)r}-e^{-(iq+\mu)r}\bigg]\nonumber\\
&=&-\frac{2\pi i}{q}.\frac{2iq}{q^2}\nonumber\\
&=&\frac{4\pi}{q^2}.
\end{eqnarray}
\item[$(3$]
نحسب 
(مع
$\vec{q}=\vec{k}-\vec{k}^{'}$
)
\begin{eqnarray}
<\vec{k}^{'}<n|V|\vec{k}>|1>=\frac{1}{(2\pi)^3}\int d^3x d^3x_1 e^{i\vec{q}\vec{x}}\frac{1}{4\pi\epsilon_0}\bigg(-\frac{e^2}{r}+\frac{e^2}{|\vec{x}-\vec{x}_1|}\bigg)\psi_n^*(\vec{x}_1)\psi_1(\vec{x}_1).
\end{eqnarray}
الحد الاول هو
\begin{eqnarray}
\frac{1}{(2\pi)^3}\int d^3x d^3x_1 e^{i\vec{q}\vec{x}}\frac{1}{4\pi\epsilon_0}\bigg(-\frac{e^2}{r}\bigg)\psi_n^*(\vec{x}_1)\psi_1(\vec{x}_1)&=&\frac{1}{(2\pi)^3}\delta_{n1}\frac{-e^2}{4\pi\epsilon_0}\int d^3x e^{i\vec{q}\vec{x}}\frac{1}{r}\nonumber\\
&=&\frac{1}{(2\pi)^3}\delta_{n1}\frac{-e^2}{4\pi\epsilon_0}\frac{4\pi}{q^2}.
\end{eqnarray}
الحد الثاني هو
\begin{eqnarray}
\frac{1}{(2\pi)^3}\int d^3x d^3x_1 e^{i\vec{q}\vec{x}}\frac{1}{4\pi\epsilon_0}\bigg(\frac{e^2}{|\vec{x}-\vec{x}_1|}\bigg)\psi_n^*(\vec{x}_1)\psi_1(\vec{x}_1)&=&\nonumber\\
\frac{1}{(2\pi)^3}\frac{e^2}{4\pi\epsilon_0}\int d^3x_1\psi_n^*(\vec{x}_1)\psi_1(\vec{x}_1)\int d^3x e^{i\vec{q}\vec{x}}\frac{1}{|\vec{x}-\vec{x}_1|}&=&\nonumber\\
\frac{1}{(2\pi)^3}\frac{e^2}{4\pi\epsilon_0}\int d^3x_1\psi_n^*(\vec{x}_1)\psi_1(\vec{x}_1)\int d^3x e^{i\vec{q}(\vec{x}+\vec{x}_1)}\frac{1}{r}&=&\nonumber\\
\frac{1}{(2\pi)^3}\frac{e^2}{4\pi\epsilon_0}F_n(\vec{q})\frac{4\pi}{q^2}.
\end{eqnarray}
الدالة
$F_n(\vec{q})$
تعرف بمعامل الشكل
من اجل الاثارة من
$|1>$
الي
$|n>$:
\begin{eqnarray}
F_n(\vec{q})&=&\int d^3x_1\psi_n^*(\vec{x}_1)\psi_1(\vec{x}_1)e^{i\vec{q}\vec{x}_1}\nonumber\\
&=&<n|e^{i\vec{q}\vec{x}_1}|1>.
\end{eqnarray}
نلاحظ ان
$e^{i\vec{q}\vec{x}_1}$ 
هو  تحويل فورييه لكثافة الالكترونات الموافقة للالكترون المرتبط الذي يوجد عند
$\vec{x}_1$
 والمعطاة ب
\begin{eqnarray}
\rho(\vec{x})=\delta^3(\vec{x}-\vec{x}_1).
\end{eqnarray}
نحصل اذن علي
\begin{eqnarray}
<\vec{k}^{'}<n|V|\vec{k}>|1>=\frac{1}{(2\pi)^3}\frac{e^2}{4\pi\epsilon_0}\frac{4\pi}{q^2}(-\delta_{n1}+F_n(\vec{q})).
\end{eqnarray}
\item[$(4$]
اذن من اجل التصادم المرن نحصل علي
\begin{eqnarray}
f^{(1)}(k\hat{r},n,\vec{k},1)=-\frac{2me^2}{\hbar^2q^2}\frac{1}{4\pi\epsilon_0}(-1+F_1(\vec{q})).
\end{eqnarray}
المقطع الفعال التفاضلي هو
\begin{eqnarray}
\frac{d\sigma}{d\Omega}=|f^{(1)}(k\hat{r},1,\vec{k},1)|^2=\bigg(\frac{2me^2}{\hbar^2q^2}\bigg)^2\bigg(\frac{1}{4\pi\epsilon_0}\bigg)^2(-1+F_1(\vec{q})).
\end{eqnarray}
معامل الشكل في هذه الحالة يعطي ب
\begin{eqnarray}
F_1(\vec{q})&=&\int d^3x_1\psi_1^*(\vec{x}_1)\psi_1(\vec{x}_1)e^{i\vec{q}\vec{x}_1}\nonumber\\
&=&\int r^2\sin\theta dr d\theta d\phi e^{-\frac{2r}{a}}e^{iqr\cos\theta}\nonumber\\
&=&-\frac{2i}{a^3q}\int rdr\bigg[e^{(iq-\frac{2}{a})r}-e^{-(iq+\frac{2}{a})r}\bigg]\nonumber\\
&=&-\frac{2i}{a^3q}\frac{\frac{8iq}{a}}{(q^2+\frac{4}{a^2})^2}\nonumber\\
&=&\frac{16}{(4+a^2q^2)^2}.
\end{eqnarray}
\item[$(5$]
من اجل التصادم غير المرن علاقة المقطع الفعال التفاضلي تتعدل الي
\begin{eqnarray}
\frac{d\sigma}{d\Omega}=\frac{k^{'}}{k}|f^{(1)}(\vec{k}^{'},n,\vec{k},1)|^2=\frac{k^{'}}{k}\bigg(\frac{2me^2}{\hbar^2q^2}\bigg)^2\bigg(\frac{1}{4\pi\epsilon_0}\bigg)^2(F_n(\vec{q})).
\end{eqnarray}
\end{itemize}
\paragraph{
تمرين
 $4$:}
\begin{itemize}
\item[$(1$]
من اجل
$r<a$
الكمون ينعدم. معادلة شرودينغر القطرية او المدارية تصبح
\begin{eqnarray}
 \frac{d^2u}{dr^2}=\big(\frac{l(l+1)}{r^2}-\frac{2mE}{\hbar^2}\big)u.
\end{eqnarray}
نضع
$l=0$. 
نحصل علي
\begin{eqnarray}
 \frac{d^2u}{dr^2}=-k^2u~,~k^2=\frac{2mE}{\hbar^2}.
\end{eqnarray}
الحل
\begin{eqnarray}
u=A\sin kr+B\cos kr.
\end{eqnarray}
تذكر ان دالة الموجة تعطي ب
\begin{eqnarray}
\psi=\frac{u}{r}Y_l^m.
\end{eqnarray}
اذن يجب ان نضع
$B=0$
لان
$\cos kr/r$
تنفجر عند
$r=0$
نحصل علي
\begin{eqnarray}
u=A\sin kr.
\end{eqnarray}

\item[$(2$]
من اجل منطقة الاشعاع
$r>a$
لدينا
$V(r)=0$
و بالتالي فان الحل هو ايضا من الشكل
\begin{eqnarray}
u=C\sin kr+D\cos kr.
\end{eqnarray}
نكتب هذا الحل ايضا علي الشكل
\begin{eqnarray}
u=C_1\exp(i kr)+D_1\exp(-ikr)~,~C_1=\frac{D-iC}{2}~,~D_1=\frac{D+iC}{2}.
\end{eqnarray}
الحل العام

\begin{eqnarray}
\sum_{lm}c_{lm}\bigg(\frac{C_1}{r}\exp(i kr)+\frac{D_1}{r}\exp(-ikr)\bigg)Y_{lm}(\theta,\phi).
\end{eqnarray}
التناظر الكروي يعني ان
$c_{lm}=c_{l0}\delta_{m0}$
نحصل اذن علي
\begin{eqnarray}
&&\sum_{l}c_{l0}\bigg(\frac{C_1}{r}\exp(i kr)+\frac{D_1}{r}\exp(-ikr)\bigg)Y_{l0}(\theta,\phi)=\nonumber\\
&&\sum_{l}\sqrt{\frac{2l+1}{4\pi}}c_{l0}\bigg(\frac{C_1}{r}\exp(i kr)+\frac{D_1}{r}\exp(-ikr)\bigg)P_l(\cos\theta).
\end{eqnarray}
بالمقارنة مع الحل العام من اجل اي كمون كروي لما
$r\longrightarrow\infty$
الذي يعطي ب
\begin{eqnarray}
\frac{1}{(2\pi)^{3/2}}\sum_{l}{\frac{2l+1}{2ik}}\bigg(\frac{S_l}{r}\exp(i kr)+\frac{(-1)^{l+1}}{r}\exp(-ikr)\bigg)P_l(\cos\theta).
\end{eqnarray}
نحصل علي
\begin{eqnarray}
\sqrt{\frac{2l+1}{4\pi}}c_{l0}C_1=\frac{1}{(2\pi)^{3/2}}{\frac{2l+1}{2ik}}S_l.
\end{eqnarray}
\begin{eqnarray}
\sqrt{\frac{2l+1}{4\pi}}c_{l0}D_1=\frac{1}{(2\pi)^{3/2}}{\frac{2l+1}{2ik}}(-1)^{l+1}.
\end{eqnarray}
اذن مباشرة
\begin{eqnarray}
\frac{C_1}{D_1}=\frac{S_l}{(-1)^{l+1}}.
\end{eqnarray}
من اجل
$l=0$
لدينا
$C_1/D_1=-S_0$
اي
\begin{eqnarray}
S_0=\frac{C+iD}{C-iD}.
\end{eqnarray}
لكن
$S_l=1+2ik a_l$
اذن
\begin{eqnarray}
ik a_0=\frac{-1}{1+i\frac{C}{D}}.
\end{eqnarray}
من جهة اخري الحل في المنطقة
$r>a$
يمكن ان يكتب علي الشكل
\begin{eqnarray}
u=C_1\exp(ikr)-\frac{C_1}{S_0}\exp(-ikr).
\end{eqnarray}
نستخدم
$S_0=\exp(2i\delta_0)$.
اذن
\begin{eqnarray}
u=2iC_1\exp(-i\delta_0)\sin(kr+\delta_0).
\end{eqnarray}
\item[$(3$]
دالة الموجة يجب ان تكون مستمرة عند
$r=a$:
\begin{eqnarray}
\frac{A}{D}\sin ka=\frac{C}{D}\sin ka+\cos ka
\end{eqnarray}
المشتقة الاولي لدالة الموجة يجب ان تكون مستمرة الا في النقاط التي يتباعد فيها الكمون. الشرط المقابل يكتب علي الشكل
\begin{eqnarray}
\Delta(\frac{du}{dr})=\frac{2m\alpha}{\hbar^2}u(a).
\end{eqnarray}

\begin{eqnarray}
k\frac{C}{D}\cos ka-k\sin ka-k\frac{A}{D}\cos ka=\frac{2m\alpha}{\hbar^2}\frac{A}{D}\sin ka.
\end{eqnarray}
من المعادلتين اعلاه نجد
\begin{eqnarray}
\frac{C}{D}\frac{2m\alpha}{k\hbar^2}\sin^2 ka=-1-\frac{2m\alpha}{k\hbar^2}\cos ka\sin ka.
\end{eqnarray}
لكن 
$ika_0=-1/(1+iC/D)$. 
اذن
\begin{eqnarray}
a_0k=\frac{\frac{2m\alpha}{k\hbar^2}i\sin^2ka}{\frac{2m\alpha}{k\hbar^2}\sin^2 ka -i-i\frac{m\alpha}{k\hbar^2}\sin 2ka}
\end{eqnarray}
\item[$(4$]
\begin{eqnarray}
f(\theta)=\sum_l (2l+1)a_lP_l(\cos\theta)\simeq a_0+...
\end{eqnarray}
المقطع الفعال التفاضلي
\begin{eqnarray}
\frac{d\sigma}{d\theta}=|f(\theta)|^2=|a_0|^2=\frac{1}{k^2}=\frac{\big(\frac{2m\alpha}{k\hbar^2}\sin^2ka\big)^2}{\big(\frac{2m\alpha}{k\hbar^2}\sin^2 ka\big)^2 +\big(1
+\frac{m\alpha}{k\hbar^2}\sin 2ka\big)^2}.
\end{eqnarray}
المقطع الفعال الكلي
\begin{eqnarray}
\sigma=4\pi |a_0|^2.
\end{eqnarray}
من اجل
$ka<<1$
نجد
\begin{eqnarray}
\frac{d\sigma}{d\theta}=\frac{\beta^2 a^2}{(1+\beta)^2}~,~\beta=\frac{2ma\alpha}{\hbar^2}.
\end{eqnarray}
\begin{eqnarray}
\sigma=4\pi \frac{\beta^2 a^2}{(1+\beta)^2}.
\end{eqnarray}
\item[$(5$]
لدينا
\begin{eqnarray}
a_l=\exp(i\delta_l)\frac{\sin \delta_l}{k}.
\end{eqnarray}
اذن
\begin{eqnarray}
ka_0=\cos\delta_0\sin\delta_0+i\sin^2\delta_0.
\end{eqnarray}
هذا من جهة. من جهة اخري
\begin{eqnarray}
ka_0=\frac{iv}{v-iw}=\frac{iv(v+iw)}{v^2+w^2}.
\end{eqnarray}
اي
\begin{eqnarray}
\frac{v^2}{v^2+w^2}=\sin^2\delta_0~,~-\frac{vw}{v^2+w^2}=\cos\delta_0\sin\delta_0\Rightarrow \cot\delta_0=-\frac{w}{v}.
\end{eqnarray}
اذن
\begin{eqnarray}
\delta_0=-\cot^{-1}\frac{w}{v}.
\end{eqnarray}
لكن نعرف ان
\begin{eqnarray}
v=\frac{2m\alpha}{k\hbar^2}\sin^2 ka\simeq ka\beta.
\end{eqnarray}
\begin{eqnarray}
w=1+\frac{m\alpha}{k\hbar^2}\sin 2 ka\simeq 1+\beta.
\end{eqnarray}
اذن
\begin{eqnarray}
\delta_0=-\cot^{-1}\big(\frac{ka}{\beta \sin^2 ka}+\cot ka\big).
\end{eqnarray}

\end{itemize}

\paragraph{
تمرين
 $5$:}
\begin{itemize}
\item[$(1$]
معدل الامتصاص بدون
$(1$
تقريب العزم الكهربائي
و بدون
$(2$ 
اخذ المتوسط علي جميع اتجاهات الورود
$\hat{n}$
و جميع اتجاهات الاستقطاب
$\hat{\epsilon}$
يعطي ب

\begin{eqnarray}
w^{\rm abso}_{i\longrightarrow n}
&=&\frac{\pi e^2 }{\hbar\epsilon_0 m^2(2\pi\nu)^2} |\hat{\epsilon}<n|\vec{p}e^{i\frac{2\pi\nu}{c}\hat{n}\vec{x}}|i>|^2\delta(E_n-E_i-
2\pi \hbar\nu)u \rho(E_n)dE_n.\nonumber\\
\end{eqnarray} 
هذا هو الاحتمال في وحدة الزمن لشحنة
$e$
تبدأ في الحالة الابتدائية
$|i>$
بطاقة
$E_i=2\pi\hbar\nu_i$
ان تقفز في اللحظة
$t$ 
نحو الحالة
$|n>$
بطاقة
$E_n=2\pi\hbar\nu_n$
تحت تأثير حقل كهرومغناطيسي (اضطراب متعلق بلزمن) ذي تواتر
$2\pi\nu$. 
اعلاه
$\vec{p}$
هو كمية حركة الشحنة,
 $u$ 
 هو كثافة الطاقة (الطاقة في وحدة الحجم) في الحقل و
 $\rho(E_n)dE_n$
 هو عدد الحالات النهائية بطاقة بين
 $E_n$ 
 و
 $E_n+dE_n$. 

\item[$(2$]
في الفعل الكهروضوئي الحالة الابتدائية للالكترون هي حالة مرتبطة ذرية بطاقة 
$E_i<0$ 
بينما الحالة النهائية هي حالة مستمرة حرة بطاقة
$E_n>0$. 
اذن
\begin{eqnarray}
E_n=\frac{\hbar^2\vec{k}^2}{2m}~,~|n>=|\vec{k}>. 
\end{eqnarray}
عدد الحالات بكمية حركة بين
$\hbar \vec{k}$
و
$\hbar(\vec{k}+d\vec{k})$ 
يساوي عدد الحالات بطاقة بين
$E_n$ 
و
$E_n+dE_n$
بكمية حركة في الاتجاه المعرف بالزاوية الصلبة
$d\Omega=d^3k/(k^2dk)$.
باستعمال تنظيم العلبة لحالات الامواج المستمرة لدينا
\begin{eqnarray}
<\vec{x}|\vec{k}>=\frac{1}{L^{\frac{3}{2}}}e^{i\vec{k}\vec{x}}.
\end{eqnarray}
في هذه الحالة القيم المسموح بها ل
 $k_x$, $k_y$
 و
 $k_z$
 هي
\begin{eqnarray}
k_x=\frac{2\pi n_x}{L}~,~k_y=\frac{2\pi n_y}{L}~,~k_z=\frac{2\pi n_z}{L}.
\end{eqnarray}
من الواضح انه لدينا حالة واحدة في مكعب حجمه
$(2\pi/L)^3$
في فضاء كمية الحركة. اذن عدد الحالات بطاقة بين
$E_n$
و
$E_n+dE_n$
حيث تعطي اتجاهات كمية الحركة بالزاوية الصلبة
$d\Omega=d^3k/(k^2dk)$ 
هو
\begin{eqnarray}
\rho(E_n)dE_n=\frac{k^2dkd\Omega}{(\frac{2\pi}{L})^3}=(\frac{L}{2\pi})^3\frac{km}{\hbar^2}dE_nd\Omega.
\end{eqnarray}
\item[$(3$]
نحصل اذن
\begin{eqnarray}
w^{\rm abso}_{i\longrightarrow n}
&=&\frac{\pi e^2 }{\hbar\epsilon_0 m^2(2\pi\nu)^2} |\hat{\epsilon}<\vec{k}|\vec{p}e^{i\frac{2\pi\nu}{c}\hat{n}\vec{x}}|i>|^2\delta(E_n-E_i-2\pi\hbar\nu)u 
(\frac{L}{2\pi})^3\frac{km}{\hbar^2}dE_nd\Omega.\nonumber\\
\end{eqnarray} 
بالمكاملة علي
 $E_n$
 نحصل علي
 (
 الان
 $k=\sqrt{2m(E_i+2\pi\hbar\nu)/\hbar^2}$
 )
 
\begin{eqnarray}
w^{\rm abso}_{i\longrightarrow n}
&=&\frac{\pi e^2 }{\hbar\epsilon_0 m^2(2\pi\nu)^2} |\hat{\epsilon}<\vec{k}|\vec{p}e^{i\frac{2\pi\nu}{c}\hat{n}\vec{x}}|i>|^2u (\frac{L}{2\pi})^3\frac{km}{\hbar^2}d\Omega.
\end{eqnarray}
نحسب
\begin{eqnarray}
\hat{\epsilon}<\vec{k}|\vec{p}e^{i\frac{2\pi\nu}{c}\hat{n}\vec{x}}|i>&=&\frac{1}{L^{\frac{3}{2}}}\hat{\epsilon}\int d^3x e^{-i\vec{k}\vec{x}+i\frac{2\pi\nu}{c}\hat{n}\vec{x}}\frac{\hbar}{i}\vec{\nabla}\psi_i(\vec{x})\nonumber\\
&=&\frac{\hbar}{L^{\frac{3}{2}}}\hat{\epsilon}\vec{k}\int d^3x e^{-i\vec{q}\vec{x}}\psi_i(\vec{x})~,~\vec{q}=\vec{k}-\frac{2\pi\nu}{c}\hat{n}.\nonumber\\
\end{eqnarray}
نذكر ان
$\psi_i(\vec{x})$
هي دالة موجة الالكترون المرتبط. اذن في حالة ذرة الهيدروجين
$\psi_i(\vec{x})$
هي دالة موجة الحالة الاساسية المعطاة ب

\begin{eqnarray}
\psi_i(\vec{x})=\frac{1}{\sqrt{\pi a^3}}e^{-\frac{r}{a}}.
\end{eqnarray}
نحصل علي
\begin{eqnarray}
\hat{\epsilon}<\vec{k}|\vec{p}e^{i\frac{2\pi\nu}{c}\hat{n}\vec{x}}|i>
&=&\frac{\hbar}{L^{\frac{3}{2}}}\hat{\epsilon}\vec{k}\frac{8\pi}{a\sqrt{\pi a^3}}\frac{1}{(q^2+\frac{1}{a^2})^2}.
\end{eqnarray}
نحصل في النهاية علي
\begin{eqnarray}
w^{\rm abso}_{i\longrightarrow n}
&=&\frac{u}{2\pi\hbar \nu}d\Omega\bigg(\frac{8 e^2 k}{\pi \epsilon_0 m (2\pi\nu)} (\hat{\epsilon}\vec{k})^2\frac{1}{a^5}\frac{1}{(q^2+\frac{1}{a^2})^4}\bigg).
\end{eqnarray}
نختار
$\hat{n}$
في الاتجاه
 $z$
 و
 $\hat{\epsilon}$ 
 في الاتجاه
 $x$. 
 الشعاع
 $\vec{k}$
 يعرف بالزاويتين
 $\theta$
 و
 $\phi$.
 اذن
\begin{eqnarray}
&&(\hat{\epsilon}\vec{k})^2=k_x^2=k^2\sin^2\theta\cos^2\phi,\nonumber\\
&&q^2=k^2+\frac{(2\pi\nu)^2}{c^2}-2\frac{2\pi\nu}{c}k_z=k^2+\frac{(2\pi\nu)^2}{c^2}-2\frac{2\pi\nu}{c}k\cos\theta.
\end{eqnarray}
\end{itemize}

\paragraph{
تمرين
$6$:}
\begin{itemize}

\item[$(1$]
احتمال الانتقال هو

\begin{eqnarray}
|<\vec{k}^{'}|U_I^{(1)}(t,-\infty)|\vec{k}>|^2&=&|\bigg(\frac{-i}{\hbar}\bigg)\int_{-\infty}^tdt_1<\vec{k}^{'}|V_I(t_1)|\vec{k}>|^2\nonumber\\
&=&|\bigg(\frac{-i}{\hbar}\bigg)\int_{-\infty}^tdt_1<\vec{k}^{'}|e^{\frac{i}{\hbar}H_0t_1}V(t_1)e^{-\frac{i}{\hbar}H_0t_1}|\vec{k}>|^2\nonumber\\
&=&|\bigg(\frac{-i}{\hbar}\bigg)\int_{-\infty}^tdt_1<\vec{k}^{'}|e^{\frac{i}{\hbar}H_0t_1}V e^{\eta t_1}e^{-\frac{i}{\hbar}H_0t_1}|\vec{k}>|^2\nonumber\\
&=&\frac{1}{\hbar^2}|<\vec{k}^{'}|V|\vec{k}>|^2\frac{e^{2\eta t}}{\frac{1}{\hbar^2}(E_{k^{'}}-E_k)^2\frac{1}{\hbar^2}+\eta^2}.
\end{eqnarray}
معدل الانتقال هو
\begin{eqnarray}
\frac{d}{dt}|<\vec{k}^{'}|U_I^{(1)}(t,-\infty)|\vec{k}>|^2
&=&\frac{1}{\hbar^2}|<\vec{k}^{'}|V|\vec{k}>|^2\frac{2\eta e^{2\eta t}}{\frac{1}{\hbar^2}(E_{k^{'}}-E_k)^2\frac{1}{\hbar^2}+\eta^2}.\nonumber\\
\end{eqnarray}
في النهاية
$\eta\longrightarrow 0$
يمكن ان نستعمل النتيجة
\begin{eqnarray}
{\rm Lim}_{\eta\longrightarrow 0}\frac{\eta}{\eta^2+x^2}=\pi\delta(x).
\end{eqnarray}
نحصل علي
\begin{eqnarray}
\frac{d}{dt}|<\vec{k}^{'}|U_I^{(1)}(t,-\infty)|\vec{k}>|^2
&=&\frac{2\pi}{\hbar}|<\vec{k}^{'}|V|\vec{k}>|^2\delta(E_{k^{'}}-E_k).
\end{eqnarray}
هذه هي  بالضبط قاعدة فيرمي الذهبية. 

\item[$(2$]
معدل الانتقال من مجموعة الحالات النهائية
$|\vec{k}^{'}>$
التي لها طاقة بين
$E_{{k}^{'}}$
و
$E_{{k}^{'}}+dE_{{k}^{'}}$
هو


\begin{eqnarray}
w&=&\frac{2\pi}{\hbar}|<\vec{k}^{'}|V|\vec{k}>|^2\delta(E_{k^{'}}-E_k)\rho(E_{k^{'}})dE_{k^{'}}.
\end{eqnarray}
لقد حسبنا سابقا ان
\begin{eqnarray}
\rho(E_{k^{'}})dE_{k^{'}}=(\frac{L}{2\pi})^3\frac{k^{'}m}{\hbar^2}dE_{k^{'}}d\Omega.
\end{eqnarray}
اذن
(
مع
$k^{'}=k$
)
\begin{eqnarray}
w&=&\frac{2\pi}{\hbar}|<\vec{k}^{'}|V|\vec{k}>|^2(\frac{L}{2\pi})^3\frac{km}{\hbar^2}d\Omega.
\end{eqnarray}
\item[$(3$]
التدفق الوارد
\begin{eqnarray}
|\vec{j}|=\frac{\hbar}{m}|{\rm Im}(\psi^*\vec{\nabla}\psi)|=\frac{\hbar}{m}|{\rm Im}\frac{e^{-i\vec{k}\vec{x}}}{L^{\frac{3}{2}}}\vec{\nabla}\frac{e^{i\vec{k}\vec{x}}}{L^{\frac{3}{2}}}|=\frac{\hbar k}{mL^3}.
\end{eqnarray}
\item[$(4$]
معدل الانتقال يجب ان يكون مساو للتدفق الوارد مضروب في المقطع الفعال المتناه في الصغر
$d\sigma$. 
اذن
\begin{eqnarray}
w&=&|\vec{j}|d\sigma.
\end{eqnarray}
بالتالي
\begin{eqnarray}
\frac{d\sigma}{d\Omega}&=&\frac{m^2L^6}{4\pi^2\hbar4}|<\vec{k}^{'}|V|\vec{k}>|^2\nonumber\\
&=&|\frac{1}{4\pi}\frac{2m}{\hbar^2}\int d^3x V(\vec{x})e^{i(\vec{k}-\vec{k}^{'})\vec{x}}|^2.
\end{eqnarray}
هذا هو تقريب بورن من الرتبة الاولي.
\end{itemize}


{\selectlanguage{english}


}

\end{document}